\documentclass[a4paper,12pt]{thesis}

\pdfoutput=1
 
\usepackage[dvips]{graphicx}
 
\usepackage{cmap}
\usepackage[english]{babel}
\usepackage{ucs}
\usepackage[utf8x]{inputenc}
\usepackage[T1]{fontenc}
 
\usepackage{microtype}
\usepackage{lmodern}

\usepackage{amsmath}
\usepackage{amstext}
\usepackage{amssymb}
\usepackage{mathbbol}
\usepackage{nicefrac}
\usepackage{slashed}
\usepackage{mathtools}
\usepackage{wasysym}
\usepackage{tensor}

\usepackage{enumerate}

\usepackage[dvipsnames]{xcolor}
 
\usepackage[breaklinks=true,colorlinks=true,linkcolor=blue,citecolor=magenta,urlcolor=Green]{hyperref}
 
\usepackage[hyperpageref]{backref}
\renewcommand*{\backref}[1]{}
\renewcommand*{\backrefalt}[4]{%
   \ifcase #1 %
      { {\scriptsize Not cited in the text.}}%
   \or
      { {\scriptsize Back to page} #2}%
   \else
      { {\scriptsize Back to page} #2}%
   \fi
}

\usepackage{cite}
\usepackage{comment}
\usepackage{rotating}
\usepackage{multirow}
\usepackage{hhline}

\let\oldsqrt\sqrt
\def\sqrt{\mathpalette\DHLhksqrt}
\def\DHLhksqrt#1#2{%
\setbox0=\hbox{$#1\oldsqrt{#2\,}$}\dimen0=\ht0
\advance\dimen0-0.2\ht0
\setbox2=\hbox{\vrule height\ht0 depth -\dimen0}%
{\box0\lower0.4pt\box2}}

\def\Tr#1{\mathrm{Tr}\left\{ #1 \right\}}

\def\up#1{^{\left( #1 \right) }}

\def\sc#1#2{\Phi_{#1}\up{#2}}
\def\fer#1#2{\Psi_{#1}\up{#2}}
\def\bos#1#2{B_{#1}\up{#2}}
\def\vecb#1#2{X_{#1}\up{#2}}

\def\aut{\,{\rm or}\,}

\newcommand{\WWee}{WWee}
\newcommand{\Wnue}{W \nu e}

\begin{document}

\frontmatter

\thispagestyle{empty}

\begin{center}


  \begin{figure}[h!]
  \center
  \includegraphics[width=12.21cm]{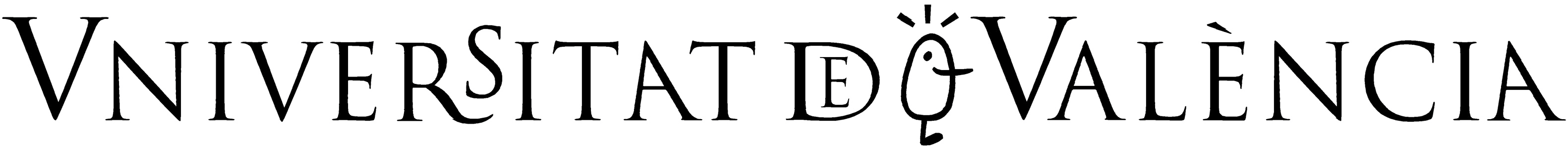}	
  \end{figure}
  
  \vspace{0.5 cm}
  
  {\bf \large Departament de Física Teòrica, IFIC -- CSIC}
  
  \vspace{0.6 cm}
  
  \begin{figure}[h!]
  \center
  \includegraphics[width=3.96cm]{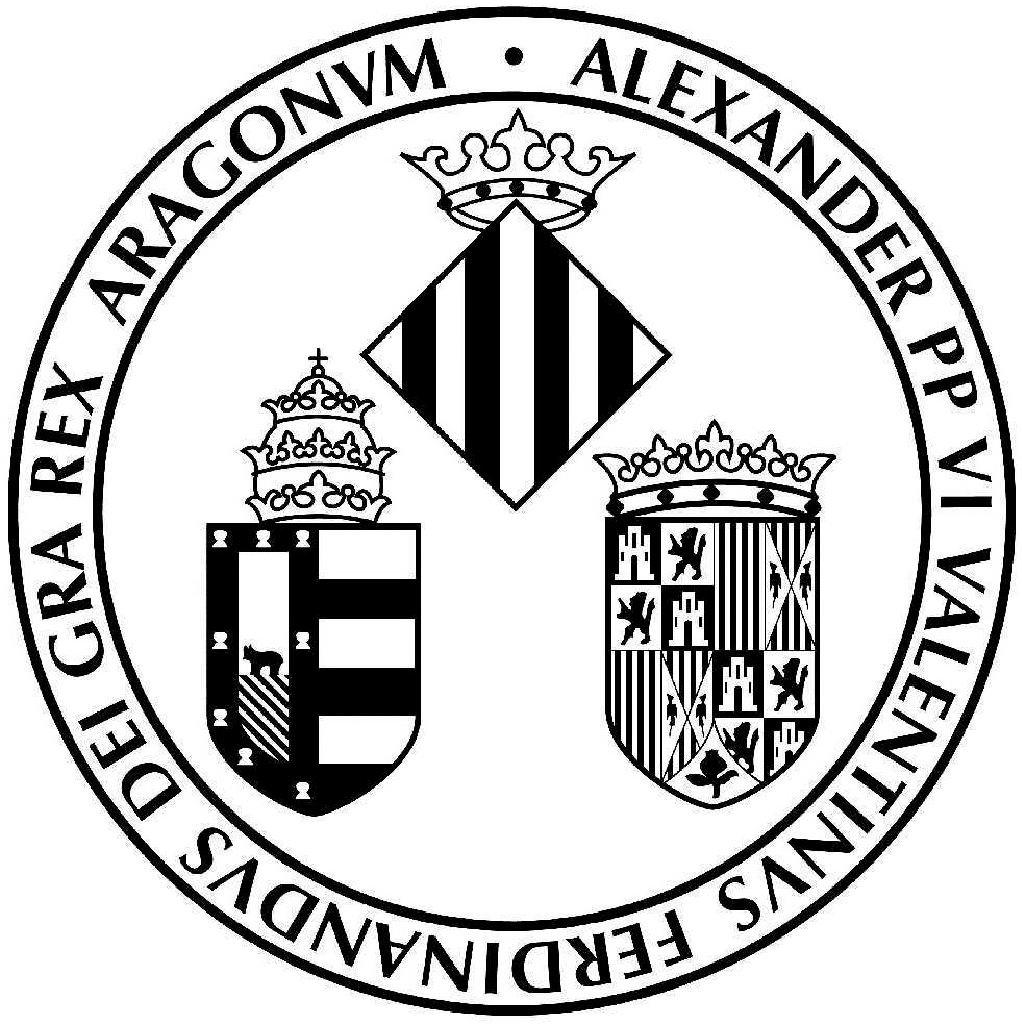}	
  \end{figure}

  \vspace{1. cm}

  {\bf \Large Exotic properties of neutrinos}\\
  \vspace{0.3 cm}
  {\bf \Large using effective Lagrangians}\\
  \vspace{0.3 cm}
  {\bf \Large and specific models}\\
  \vspace{1.3 cm}

\end{center}

\begin{flushleft}

  {\bf PhD dissertation by}\\
  \vspace{0.4 cm}

  {\bf \large Alberto Aparici Benages}\\
  \vspace{0.4 cm}

\end{flushleft}

\begin{flushright}

  {\bf Under the supervision of}\\
  \vspace{0.4 cm}
  {\bf \large Arcadi Santamaria i Luna}\\
  \vspace{1.1 cm}

\end{flushright}

\begin{center}

{\bf \large Valencia, December 16\textsuperscript{th} 2013}

\end{center}

\pagebreak

\thispagestyle{empty}

\begin{quote}
	\flushright
	\emph{There is no love of life without despair about life.}

	{\small \textsc{Albert Camus}, on PhD's.}
\end{quote}

\vspace{1cm}

\begin{quote}
	\flushright
	\emph{They wanted the end to be comforting, to say that it was not an end,
			but a beginning.
			But it's not a beginning, it is the real end; there will be 
			nothing afterwards, nothing.}

	{\small \textsc{Dmitri Shostakovich}, on postdoctoral positions.}
\end{quote}

\vspace{1cm}

\begin{quote}
	\flushright
	\emph{Of course it is happening inside your head, Harry, but why on earth
			should that mean that it is not} real\emph{?}

	{\small \textsc{Albus Dumbledore}, on renormalisation.}
\end{quote}

\vspace{1cm}

\begin{quote}
	\flushright
	\emph{I find your lack of faith disturbing.}

	{\small \textsc{Lord Darth Vader}, after presenting his latest 
			\mbox{neutrino mass model.}}
\end{quote}

\pagebreak

\thispagestyle{empty}

This doctoral dissertation is based on the research presented in the
following publications:

\vspace{1cm}

\begin{tabular}{ c p{0.8\textwidth} }
	\cite{Aparici:2009fh}  &
		\emph{Right-handed neutrino magnetic moments} \newline
		A. A., Kyungwook Kim, Arcadi Santamaria, José Wudka \newline
		\href{http://dx.doi.org/10.1103/PhysRevD.80.013010}
			{Phys.Rev.\textbf{D80} 013010 (2009)}
		\vspace{0.5cm}
	\\
	\cite{Aparici:2009mj}  &
		\emph{A model for right-handed neutrino magnetic moments} \newline
		A. A., Arcadi Santamaria, José Wudka \newline
		\href{http://dx.doi.org/10.1088/0954-3899/37/7/075012}
			{J.Phys.\textbf{G37} 075012 (2010)}
		\vspace{0.5cm}
	\\
	\cite{Aparici:2010zz}  &
		\emph{Right-handed neutrino magnetic moments} \newline
		A. A., Kyungwook Kim, Arcadi Santamaria, José Wudka \newline
		\href{http://dx.doi.org/10.1088/1742-6596/259/1/012089}
			{J.Phys.Conf.Ser.\textbf{259} 012089 (2010)}
		\vspace{0.5cm}
	\\
	\cite{delAguila:2011gr}  &
		\emph{A realistic model of neutrino masses with a large neutrinoless
			double beta decay rate} \newline
		Francisco del Águila, A. A., Subhaditya Bhattacharya, Arcadi Santamaria, 
			José Wudka	 \newline
		\href{http://dx.doi.org/10.1007/JHEP05(2012)133}
			{JHEP \textbf{1205} 133 (2012)}
		\vspace{0.5cm}
	\\
	\cite{delAguila:2012nu}  &
		\emph{Effective Lagrangian approach to neutrinoless double beta decay 
			and neutrino masses} \newline
		Francisco del Águila, A. A., Subhaditya Bhattacharya, Arcadi Santamaria, 
			José Wudka	 \newline
		\href{http://dx.doi.org/10.1007/JHEP06(2012)146}
		{JHEP \textbf{1206} 146 (2012)}
		\vspace{0.5cm}
	\\
	\cite{delAguila:2013zba}  &
		\emph{Neutrinoless double $\beta$ decay with small neutrino 
			masses} \newline
		Francisco del Águila, A. A., Subhaditya Bhattacharya, Arcadi Santamaria, 
			José Wudka	 \newline
		\href{http://arxiv.org/abs/arXiv:1305.4900}
			{PoS Corfu2012 028 (2013)}
\end{tabular}

\pagebreak

\addcontentsline{toc}{chapter}{Abbreviations}

\chapter*{Abbreviations}

This is a list of abbreviations used throughout the text:


\begin{itemize}
	\item[$\mathbf{0 \nu \beta \beta}$] neutrinoless double beta decay

	\item[\textbf{BAO}] baryon acoustic oscillations
	
	\item[\textbf{CHAMP}] charged massive particle

	\item[\textbf{CL}] confidence level

	\item[\textbf{CMB}] cosmic microwave background

	\item[\textbf{EFT}] effective field theory

	\item[\textbf{EWSB}] electroweak symmetry breaking
	
	\item[\textbf{LFV}] lepton flavour violation
	
	\item[\textbf{LFV-ing}] lepton-flavour-violating
	
	\item[\textbf{LN}] lepton number
	
	\item[\textbf{LNV}] lepton number violation
	
	\item[\textbf{LNV-ing}] lepton-number-violating
	
	\item[\textbf{NP}] new physics

	\item[\textbf{QCD}] quantum chromodynamics
	
	\item[\textbf{QED}] quantum electrodynamics
	
	\item[\textbf{QFT}] quantum field theory
	
	\item[\textbf{SM}] Standard Model

	\item[\textbf{SSB}] spontaneous symmetry breaking

	\item[\textbf{VEV}] vacuum expectation value

\end{itemize}


\tableofcontents

\mainmatter

\pagebreak

\chapter{Overview}

In the past fifteen years oscillation experiments have unveiled the massive 
character of neutrinos; their results, combined with the enduring efforts to 
directly measure neutrino \mbox{masses --not yet} \mbox{successful--, picture} 
an image of three extremely light neutrinos, lighter by several orders of
magnitude than the rest of the known particles. This mass gap calls for an
explanation, and suggests that neutrinos might hide other unexpected 
properties related to the physics that endows them with this 
very special personality.

In this work we study several nonstandard features of neutrinos, first 
focusing on their phenomenological consequences and then progressing
towards a more complete explanation. This scheme is realised in the
interplay between effective field theory and concrete models.
Effective interactions are powerful tools that allow to study the low-energy 
effects of heavy particles without having to specify the details of the
heavy physics; they provide, so, a natural way to carry out model-independent
analyses of exotic low-energy features. However, they are known to be limited:
whenever some knowledge of the high-energy physics is needed, effective theories
fail to yield predictions.
Models, on the other hand, 
provide a complete specification of the low- and high-energy physics,
thus allowing to calculate all and every feature of the theory; at the same time, 
and for this reason, they are limited: they must commit to one and
just one family of new particles and can only yield predictions about them.
Our aim is to combine the most powerful qualities of both approaches:
to use effective operators to spot and study new neutrino properties in a
model-independent way, and to complement that vision with concrete models
which provide sharp high-energy predictions and maybe additional low-energy 
features that the effective theory cannot predict.

The text is organised as follows: chapters \ref{chap:introduction}  and
\ref{chap:neutrinos} introduce the basic concepts on which our study lies:
the Standard Model, effective field theory and neutrino physics. Chapters
\ref{chap:nuR-magmo-eff} and \ref{chap:nuR-magmo-model} give a prominent role 
to right-handed neutrinos: we discuss in them effective operators that 
involve right-handed neutrinos as \emph{low}-energy fields. We find a 
dimension-five interaction that has remained essentially unnoticed to date
and which provides magnetic moments for the $\nu_{\mathrm{R}}$'s, opening
a plethora of scenarios where they might be relevant. In chapter 
\ref{chap:nuR-magmo-eff} we address the question from the effective theory
viewpoint, while in chapter \ref{chap:nuR-magmo-model} we discuss one
model that realises the magnetic moment interaction. Chapters 
\ref{chap:0nu2beta-eff}, \ref{chap:Wnue-model} and \ref{chap:WWee-model}
study a family of effective interactions that provide neutrinoless double 
beta decay and may overpower the usual contribution mediated by Majorana
neutrino masses. We find two alternative mechanisms: one involving 
one $W$ boson, one electron and one neutrino ($W \nu e$ mechanism) and
one involving two $W$'s and two electrons ($WWee$ mechanism), both with
explicit violation of lepton number. In chapter \ref{chap:0nu2beta-eff} we
study how to characterise these effective interactions, how they contribute
to $0 \nu \beta \beta$ and how they mediate the generation of neutrino masses.
We find a remarkable connection between the chirality of the leptons
involved in the effective interaction and the suppression of both
the interaction and the generated neutrino masses. In chapters 
\ref{chap:Wnue-model} and \ref{chap:WWee-model} we discuss models that
realise the two effective mechanisms: in chapter \ref{chap:Wnue-model} the
$W \nu e$ interaction is considered and chapter \ref{chap:WWee-model} is
devoted to the $WWee$ interaction. The latter model presents several
interesting phenomenological features and can provide signals in a
variety of different experiments.

\chapter{Introductory concepts}  \label{chap:introduction}

In this chapter we present the Standard Model and motivate some of its properties.
While we introduce the notation for the SM fields we will discuss the topic of
spontaneous symmetry breaking and the origin of mass in the Standard Model,
as it plays an important role in the following chapters.
We also introduce the concept of effective theory and explain how it applies 
to quantum field theory. The aim of these disquisitions is introductory, but the
interested reader will find throughout the text references to specialised
reviews where these topics are treated in more depth.

\section{Elements of the Standard Model: the weak interactions}

The Standard Model is the main framework of modern particle
physics. It is a quantum field theory that successfully describes the
electromagnetic, weak and strong interactions of the known particles by using
internal symmetries of the associated fields -- see in
\cite{Glashow:1961tr,Higgs:1964pj,Englert:1964et,Guralnik:1964eu,Higgs:1966ev,
Kibble:1967sv,Weinberg:1967tq,Salam:1968rm,
Han:1965pf,Fritzsch:1973pi,GellMann:1981ph,Fritzsch:1972jv}
the original work, and also for instance \cite{Pich:2012sx,Pich:1999yz} for modern
reviews. The fundamental fields include twelve fermions: six quarks, 
$u$, $d$, $s$, $c$, $b$ and $t$; three charged leptons, $e$, $\mu$ and 
$\tau$; and three neutrinos, $\nu_e$, $\nu_\mu$ and $\nu_\tau$. The number
of fundamental bosons may depend on how you count, but let's say we have
eleven gauge bosons: eight coloured gluons for QCD, a massless photon
for electromagnetism, and the $W$ and $Z$ for the weak interactions. And then
the Higgs boson. 

All these fields are organised in multiplets with 
well-defined internal symmetries; the symmetries are gauged, and this 
introduces naturally the gauge bosons into the theory and yields the right
interaction vertices. That is strictly true for the strong interactions,
and electromagnetism might in principle be accommodated.
The weak interactions, however, pose
a more difficult challenge because their associated bosons are known to be
massive and pure gauge theories yield strictly massless bosons; besides, 
the properties of the weak symmetries towards chirality
make it impossible to
construct a proper mass term for the fermions. These problems are solved at once
by introducing the Higgs field and making it break the weak symmetries
spontaneously. Let us explain how this happens, as we will deal mostly with the
weak interactions in the chapters that follow.

\subsection{The weak interactions: a broken $SU(2)$ symmetry}

The weak interactions are accurately described at low energies by the 
effective Fermi interaction, which couples two fermionic currents into
a dimension-six operator \cite{Fermi:1934sk,Fermi:1934hr,wilson:1150}.
The celebrated experimental work of Madame Wu, carried out in the fifties,
came to show that these interactions violate parity maximally \cite{Wu:1957my,
Lee:1956qn}, which has to be implemented in the effective interaction.
This can be done if the aforementioned fermionic currents only involve
one of the fermion chiralities; by convention, the left-handed chirality is
chosen to participate in the weak Fermi interactions whereas the right-handed
one is transparent to them.

The Fermi interaction, however good it is at low energies, cannot be the
final theory of the weak interactions: it is an effective interaction, and
as such it fails at high energies, where the underlying particles should
manifest. The sixties saw the development of several proposals for the 
high-energy content of the weak interactions, and finally it was the
theoretical framework of Weinberg, Glashow and Salam \cite{Glashow:1961tr,
Weinberg:1967tq,Salam:1968rm} which proved to be the most \mbox{successful -- as
of} summer 2013, some might find it even \emph{too} successful. This model 
gathers the
left-handed components of the fermions in doublets and invokes upon them
a $SU(2)$ gauge symmetry. Intuitively, we know from the Fermi interaction
that there's a high-energy interaction connecting electrons to electron
neutrinos and $u$ quarks to $d$ quarks, together with one or more heavy
particles; grouping the fields into multiplets and invoking a gauge symmetry
is a natural way to generate interactions among the fields
and, besides, provides new \mbox{particles --the gauge} \mbox{bosons-- that hold up}
the whole structure of new interactions. So, it seems a fairly good idea;
but why doublets and why precisely $SU(2)$? Indeed, originally it was not the
only option, and ultimately it was the experiment who established the better
proposal, but some valuable information can be extracted from an analysis of
the currents in the Fermi interaction, as the reader can find, for instance, 
in chapter 11 of \cite{Cheng:1985bj}. Let us here skip these details and focus
on the notation that we are to use along the text and on the process of
spontaneous symmetry breaking, which will recur over and over in the forthcoming
chapters; but as a necessary caveat, we propose to the reader the following
question: so, we are to group electrons and electron neutrinos into a doublet
and then we are to say the doublet is $SU(2)$-symmetric. But how? Would not that
imply that electrons and neutrinos should be indistinguishable? Indeed it would;
so we already know, from the start, that the story cannot end with $SU(2)$.
In due time this and other issues will be too strident to be ignored; but first 
things first: we promised some notation and here it comes.

\subsection{Doublets and singlets} \label{sec:intro-SM-doublets-singlets}

As we already argued, our theory of the weak interactions has the left-handed 
components of the fermions grouped into doublets which 
will be made $SU(2)$-invariant; we will denote the leptonic doublets by 
$\ell_{\mathrm{L}}$ and the quark doublets by $Q_{\mathrm{L}}$:
\begin{align}
	\ell_{\mathrm{L} e} &= \mathrm{P}_{\mathrm{L}} \begin{pmatrix}  \nu_e 
			\\ e \end{pmatrix}
		&
		\ell_{\mathrm{L} \mu} &= \mathrm{P}_{\mathrm{L}} \begin{pmatrix}  
			\nu_\mu \\ \mu \end{pmatrix}
		&
		\ell_{\mathrm{L} \tau} &= \mathrm{P}_{\mathrm{L}} \begin{pmatrix}  
			\nu_\tau \\ \tau \end{pmatrix}
		\nonumber
	\\ \label{intro-SM-doublets}
	Q_{\mathrm{L} 1} &= \mathrm{P}_{\mathrm{L}} \begin{pmatrix}  u \\ d \end{pmatrix}
			&
	Q_{\mathrm{L} 2} &= \mathrm{P}_{\mathrm{L}} \begin{pmatrix}  c \\ s \end{pmatrix}
			&
	Q_{\mathrm{L} 3} &= \mathrm{P}_{\mathrm{L}} \begin{pmatrix}  t \\ b \end{pmatrix}
			\, ,
\end{align}
where $\mathrm{P}_{\mathrm{L}}$ is the left-handed chiral projector, and the doublets
may bear a family index that we will denote in general by a greek letter, 
$\ell_{\mathrm{L} \alpha}$ or $Q_{\mathrm{L} \alpha}$; in most cases, though,
this index will be omitted in order to avoid overburdening the notation.
The charge carried by these doublets, and represented by their $SU(2)$ properties,
is called \emph{weak isospin}, and we will denote it by $T$. Hence, the doublets
carry $T^2 \, (\ell_{\mathrm{L}}) = T^2 \, ( Q_{\mathrm{L}}) = \nicefrac{1}{2}$, 
while the components possess a well-defined value for the isospin along the $z$-axis; 
for instance, we have $T_3 \, (e_{\mathrm{L}}) = T_3 \, (d_{\mathrm{L}} ) = - 
\nicefrac{1}{2}$ and $T_3 \, (\nu_{\mathrm{L} e}) = T_3 \, (u_{\mathrm{L}} ) = + 
\nicefrac{1}{2}$.

The right-handed components of the fermions are singlets, that is, chargeless 
under $SU(2)$; they will be denoted by $e_{\mathrm{R}}$, $\tau_{\mathrm{R}}$, 
$u_{\mathrm{R}}$ and so on, or sometimes as $e_{\mathrm{R} \alpha}$, 
$u_{\mathrm{R} \alpha}$, $d_{\mathrm{R} \alpha}$, $\nu_{\mathrm{R} \alpha}$, 
when we want to express them as components of a vector in flavour space. 
The existence of 
right-handed neutrinos is not proven yet, as we will discuss in more detail in 
section \ref{sec:neutrinos-dirac-or-majorana}, and usually they are not considered
part of the Standard Model. They can be added to the SM content in order to account
for the existence of neutrino masses, but that is not the only \mbox{option -- again,}
see section \ref{sec:neutrinos-dirac-or-majorana}.

The fermionic Lagrangian of the Standard Model, thus, considering only the 
weak $SU(2)$ symmetry, will read
\begin{multline}  \label{intro-SM-weakL}
	\mathcal{L}_{\mathrm{W}} = i \, \bar \ell_{\mathrm{L} \alpha} \, \gamma_\mu 
			D^\mu \,
			\ell_{\mathrm{L}}^\alpha \; + \; i \, \overline{Q_{\mathrm{L} \alpha}} \, 
			\gamma_\mu D^\mu \, Q_{\mathrm{L}}^\alpha \; +
			\\
			\vphantom{\raisebox{2ex}{a}} 
			+ \: i \, \bar e_{\mathrm{R} \alpha} \, \gamma_\mu \partial^\mu \,
			e_{\mathrm{R}}^\alpha \; + \; i \, \bar u_{\mathrm{R} \alpha} \, 
			\gamma_\mu \partial^\mu \, u_{\mathrm{R}}^\alpha \; + \; 
			i \, \bar d_{\mathrm{R} \alpha} \, \gamma_\mu \partial^\mu \,
			d_{\mathrm{R}}^\alpha \, , 
\end{multline}
where Einstein's convention is understood for family and Lorentz indices; 
from now on, in
simple expressions such as this one we will omit these indices. Note that 
there are no masses in Lagrangian \eqref{intro-SM-weakL}: since mass terms are of
the form $\bar e_{\mathrm{R}} e_{\mathrm{L}}$, they involve fields with 
different $SU(2)$ charges and therefore they're not gauge invariant. This 
is another issue that points to the necessity of breaking $SU(2)$.

But let us come back to this just in a moment; before we need to fix the 
notation for the gauge bosons: begin by noticing that, whereas the singlet 
right-handed fields have a
usual kinetic term with a spacetime derivative, the doublets, charged under
the gauge symmetry, need a covariant derivative that we denote by $D^\mu$.
This derivative contains a usual $\partial^\mu$ plus other terms involving
the gauge fields, that transform appropriately in order for $SU(2)$ to be
respected even locally. We write these terms as
\begin{displaymath}
	D_\mu = \partial_\mu \: - \: i g \, T_a W^a_\mu \, ,
\end{displaymath}
where $g$ is the coupling for the $SU(2)$ interactions, $T_a$ are the 
generators of $SU(2)$ (the Pauli matrices halved, $T_a = \frac{1}{2} 
\, \sigma_a$), and $W_\mu^a$ are three real vector fields, the gauge bosons
of $SU(2)$. The covariant kinetic terms for the doublets in \eqref{intro-SM-weakL} 
yield interactions among the fermions and the gauge bosons, of the form
\begin{align*}
	&g \, \bar \ell_{\mathrm{L}} \, \gamma^\mu T_a \, \ell_{\mathrm{L}} \, W_\mu^a
	& & &
	&g \, \overline{Q_{\mathrm{L}}} \, \gamma^\mu T_a \, Q_{\mathrm{L}} \, W_\mu^a 
		\, ,
\end{align*}
which are the embryos of the weak interaction vertices. The kinetic term for
the gauge bosons can be devised by looking at their transformation properties
under $SU(2)$ and writing the most general bilinear gauge-invariant term that
contains two derivatives, as it must be for bosonic fields. We will not enter
into the details, which can be consulted for instance in \cite{Pich:2012sx}, or
in chapter 8 of \cite{Cheng:1985bj}, but we provide the form of these kinetic 
terms:
\begin{align} \label{intro-SM-kingauge-su2}
	\mathcal{L}_{\mathrm{k} \, W} &= - \frac{1}{4} \, W^a_{\mu \nu} W^{\mu \nu}_a
		\, , &
	W^a_{\mu \nu} \equiv \partial_\mu W^a_\nu - \partial_\nu W^a_\mu + 
 		g \, \epsilon^a_{\phantom{a} b c} \, W^b_\mu W^c_\nu \, .
\end{align}

\subsection{The trouble with masses, and a solution} \label{sec:masses-SM}

It is important to note that the same gauge transformations that define
the gauge fields, as we see in \eqref{intro-SM-kingauge-su2}, explicitly forbid 
a mass term for them.
So, at this point of the story we are left with a symmetry that provides some
fields which quite resemble the particles we observe, but none of them
can have mass at all; besides, the symmetry renders the neutrino and the electron
undistinguishable, which certainly doesn't resemble at all the situation in the 
real world. One could conclude that a symmetry that yields the correct
elements for the model ends up restricting too much their properties.
The question is: is there a way to relax these
requirements so that we can reconcile the theory with reality?

The Higgs mechanism provides such a means; and, incidentally, one that  
remarkably matches the experimental data. The idea is to introduce a
scalar field, the Higgs field, which shifts the vacuum of the theory to a 
non-gauge-invariant
state; as a consequence, some fields with well-defined $SU(2)$ properties will
no longer represent physical excitations of the \mbox{theory -- for instance,} 
excitations
with definite mass. The fields that describe the actual particles of the 
theory will be combinations of the primordial fields, so we may say that the shift
in the vacuum induces a shift in the fields. The new fields will happen to
have mass terms, and will successfully describe the particles
we observe. At the end of the day it all happens \emph{as if} there was no 
$SU(2)$ symmetry at all, because fields that should be undistinguishable
end up being rather different, and terms which seemed forbidden appear
in a natural fashion. We say that the symmetry has been broken \emph{spontaneously},
for we have not introduced any term that breaks $SU(2)$; rather, it is 
the fact that we live in a noninvariant vacuum what makes us see the world
as if $SU(2)$ was not a symmetry. But the underlying laws of physics 
remain $SU(2)$-invariant, and the symmetry pervades most of the properties
of the particles and their interactions.

Let us tell this story in a somewhat more rigorous way. The Higgs field is 
introduced as a complex scalar field with $SU(2)$ charge; this last requirement
is crucial, for it must displace the vacuum to a state with definite $SU(2)$
properties. The simplest option is to choose the Higgs to be a doublet, like
the other $SU(2)$-charged fields in the theory. We write it as $\phi$, and
it contains four real-valued degrees of freedom; we can arrange them in several
ways,
\begin{equation} \label{intro-SM-def-phi}
	\phi (x) = \begin{pmatrix} \phi_1 (x) + i \, \phi_2 (x) \\ \phi_3 (x) + i \, 
			\phi_4 (x) \end{pmatrix} = e^{i \, \sqrt{2} \, T_a \, \theta^a (x) / f}
			\begin{pmatrix}	0 	\\ \rho (x) \end{pmatrix} \, .
\end{equation}
Maybe the right-hand side reordering proves to be the most useful%
\footnote{Of course there are infinitely many possible locations \mbox{for $\rho$ 
-- some sort} of ``modulus of $\phi$''. The convenience of this particular choice
will be seen in due course, when we discuss electric charge.
};
in it, $T_a$ are
the generators of $SU(2)$, $\rho$ and the $\theta_a$ are real scalar fields, and
$f$ is some mass scale, needed for dimensional consistency. 
We endow the Higgs field with a quartic self-coupling, 
$\lambda \, (\phi^\dagger \phi)^2$, and an `anomalous' mass term which apparently 
contains an imaginary mass, $- \mu^2 \, \phi^\dagger \phi$. Summarising, the
Lagrangian for the Higgs field is
\begin{equation} \label{intro-SM-HiggsL}
	\mathcal{L}_{\mathrm{Higgs}} = (D_\mu \phi)^\dagger D^\mu \phi +
			\mu^2 \, \phi^\dagger \phi - \lambda \, (\phi^\dagger \phi)^2 \, ,
\end{equation}
but of course $\phi$ cannot represent a real particle with such an unphysical 
mass. The introduction of this mass term forces us to look for the physical
fields somewhere else. Thinking a little bit semiclassically, one may notice that
$\phi = 0$ is not a minimum of the scalar potential of the theory, 
$V( \phi ) = \lambda \, (\phi^\dagger \phi)^2 - \mu^2 \phi^\dagger \phi$, and
this is due precisely to the anomalous sign of $\mu^2$. It is possible to prove
rigorously%
\footnote{Though we won't do so here; the interested reader
may look into \cite{Jackiw:1979wf,Jackiw:1974cv,Cornwall:1974vz}, chapter
2 of \cite{Georgi:weak}, and also chapter 5 of \cite{Coleman:aspects}.
}
that when the potential is minimised by nonzero values of some fields, then
these fields acquire finite vacuum expectation values; this, in turn, implies
that the vacuum of the theory corresponds to a state which is not invariant under some
of the symmetries felt by the relevant fields.
In other words: due to the sign of $\mu^2$, the symmetries under which $\phi$ is 
charged should
become spontaneously broken. 

Let's see how this affects the properties we're interested in: first, notice that
$V (\phi)$ depends solely on $\rho$, as defined in \eqref{intro-SM-def-phi},
and that $V (\rho)$ is minimised for $\rho = \sqrt{\nicefrac{\mu^2}{2 \lambda}}
\equiv v$;
this is the value of the VEV for $\rho$ and, hence, for $\phi$ we have
\begin{displaymath}
	\left< \phi \right> = \begin{pmatrix} 0 
			\\ v
			\end{pmatrix} \, .
\end{displaymath}
Then we may identify a field with a physically meaningful mass term by shifting
$\rho$ and defining a new field around the minimum:
\begin{align*}
	\rho (x) &= v + \frac{H (x)}{\sqrt{2}}
	&
	V (H) &= \mu^2 \, H^2 + \sqrt{2} \lambda v \, H^3 + \frac{\lambda}{4} \, H^4 \, ,
\end{align*}
where the $\sqrt{2}$ factor accounts for a proper normalisation of the kinetic term
of $H$, which is a real field. This way defined, the measured VEV of the 
Standard Model is $v \simeq 246 \; \mathrm{GeV}$ \cite{Beringer:1900zz}; 
some authors prefer to define the VEV with a $\sqrt{2}$ normalisation too, 
and then a value of 174 GeV is given.

The $H$ field, needless to say, is the now-superstar Higgs boson, whose mass
has been recently measured to $m_H = \mu \simeq 125 \; \mathrm{GeV}$ 
\cite{Aad:2012tfa,Chatrchyan:2012ufa}.
But the interesting story is what happens to the other
three degrees of freedom of the Higgs doublet: if we come back to equation
\eqref{intro-SM-def-phi}, what we have done is just
\begin{equation} \label{intro-SM-Higgs-with-VEV}
	\phi = e^{i \, \sqrt{2} \, T_a \, \theta^a / v}
		\begin{pmatrix}	0 	\\ v + \frac{H}{\sqrt{2}} \end{pmatrix} \, ,
\end{equation}
where we have fixed $f$ to the natural mass scale of the problem, $f = v$. 
If we substitute now this expression into the kinetic
term in \eqref{intro-SM-HiggsL}, a great deal of new pieces appear. We select
here some of them:
\begin{equation} \label{intro-SM-Goldstones-SSB}
	 (D_\mu \phi)^\dagger D^\mu \phi = \frac{1}{2} \, \partial_\mu \theta^a \, 
	 		\partial^\mu \theta_a 
			\: + \: \frac{g^2 v^2}{4} \, W_\mu^a \, W^\mu_a \: - \: 
			\frac{g v}{\sqrt{2}} \, W_\mu^a \, \partial^\mu \theta_a \: + 
			\: \ldots \, ,
\end{equation}
and from them we point out three important features:
first, the $\theta$'s are revealed as the Goldstone bosons associated to the
spontaneous breaking,
as they're massless, propagating scalar fields that transform in the adjoint 
representation of $SU(2)$. Second, the gauge bosons develop mass terms due to 
the shift of $\rho$ to $H$, or, equivalently, due to the shift of the vacuum 
to a non-$SU(2)$-invariant state. Third, the Goldstone bosons mix with the gauge 
bosons, thus providing the longitudinal degree of freedom needed to construct a 
consistent massive vectorial field%
\footnote{In fact, we can always, if we want, eliminate the Goldstone bosons from 
the description of the physics and consider just three vectorial bosons with 
definite nonzero mass. Formally, this is justified by the fact that the interactions
remain $SU(2)$-invariant and we can choose to express them in any $SU(2)$ gauge
we like; for instance, a gauge that cancels the $\theta$'s in every spacetime
point. Such gauge is called the \emph{unitary gauge}. For more information
on gauge fixing in spontaneously broken gauge theories the reader may refer
to \cite{Cheng:1985bj} or \cite{Abers:1973qs}.
}.
The terms in equation 
\eqref{intro-SM-Goldstones-SSB} prove how deep has been the introduction of that
anomalous sign in equation \eqref{intro-SM-HiggsL}.

Going a step further, the Higgs field can also help in giving masses to the
fermions. Notice that the $SU(2)$ charges allow to write Yukawa interactions
of the form $\bar \ell_{\mathrm{L}} Y_e e_{\mathrm{R}} \, \phi$, where $Y_e$
is a matrix in flavour space, and that after SSB these interactions provide,
among others, terms that exactly resemble a fermion mass term:
\begin{equation} \label{intro-SM-fermion-masses}
	\bar \ell_{\mathrm{L}} Y_e e_{\mathrm{R}} \, \phi = v \, \bar e_{\mathrm{L}}
			Y_e e_{\mathrm{R}} \: + \: \bar e_{\mathrm{L}} \frac{Y_e}{\sqrt{2}} 
			e_{\mathrm{R}} \, H \, ,
\end{equation}
where we have expressed the interactions after the breaking in the unitary gauge.
Through this mechanism, the Higgs field emerges as the necessary piece to understand
mass in the Standard Model: its VEV is the common scale for the
masses of the gauge bosons and the fermions; the complex spectrum we observe
is the consequence of the modulation of $v$ by the gauge and Yukawa couplings.
As a side effect,
the Higgs boson, the only excitation of the Higgs field that retains individuality,
inherits couplings to the fermions that are correlated to their mass spectrum:
the heavier the fermion mass, the larger the Yukawa coupling and the larger
the coupling to the Higgs boson, as we can see in \eqref{intro-SM-fermion-masses}.

The mechanism described above could be thought to provide masses only for the fields 
in the second component of the weak isodoublet, as the VEV is located in that 
position (the way we chose it, see equation \eqref{intro-SM-Higgs-with-VEV},
imposes a zero in the first component). This is not quite correct, for there
is one more way to couple the Higgs field to the fermion doublets. To see it,
remember that the
fundamental representation of $SU(2)$ is pseudoreal, that is, we can transform
isovectors in the fundamental representation to isovectors in the complex-conjugate
of the fundamental representation; to do so we only need an appropriate $SU(2)$
matrix, which happens to be $i \sigma_2$. We often call this matrix $\epsilon$,
as it is the totally antisymmetric $2 \times 2$ matrix. Then consider the Yukawa
couplings we can construct not with $\phi$, but with $\phi^*$: since we want
to couple it to fundamental-representation doublets, say $Q_{\mathrm{L}}$, 
we will need the mediation of $\epsilon$. As a result:
\begin{multline} \label{intro-SM-quark-mass-tilde}
	\overline{Q_{\mathrm{L}}} \, Y_u u_{\mathrm{R}} \, \epsilon \, \phi^* =
		\begin{pmatrix} \bar u_{\mathrm{L}} & \bar d_{\mathrm{L}} \end{pmatrix} \,
		Y_u \, u_{\mathrm{R}} \, \begin{pmatrix} 0 & 1 \\ -1 & 0 \end{pmatrix} \,
		\begin{pmatrix} 0 \\ v + \frac{H}{\sqrt{2}} \end{pmatrix} = 
		\\
		\vphantom{\raisebox{2.5ex}{a}} 
		= v \, \bar u_{\mathrm{L}} Y_u u_{\mathrm{R}} \: + \: \bar u_{\mathrm{L}} 
			\frac{Y_u}{\sqrt{2}} u_{\mathrm{R}} \, H \, ,
\end{multline}
again with the Goldstone bosons abstracted from the description of the system.
When writing down these terms it is common to use the more compact notation
$\tilde \phi \equiv \epsilon \, \phi^*$, and the same for the complex-conjugate
representation of the fermions, $\tilde \ell_{\mathrm{L}} \equiv \epsilon \, 
\ell_{\mathrm{L}}^{\, \mathrm{c}}$, with $\ell_{\mathrm{L}}^{\, \mathrm{c}}$ defined
as usual, $\ell_{\mathrm{L}}^{\, \mathrm{c}} \equiv C \, 
\bar \ell_{\mathrm{L}}^{\, \mathrm{T}}$.
In conclusion, we have a few more terms to add to our SM Lagrangian, which account
for the masses%
\footnote{As we said above, for the case of the neutrinos the use of right-handed
neutrinos and the Higgs mechanism is not the only option. In fact it is not the
most popular one. See section \ref{sec:neutrinos-dirac-or-majorana} for more
on this matter.
}
of the fermions and their interactions with the Higgs boson:
\begin{displaymath}
	\mathcal{L}_{\mathrm{Yukawa}} = \bar \ell_{\mathrm{L}} Y_e e_{\mathrm{R}} \, 
		\phi \: + \: \bar \ell_{\mathrm{L}} Y_\nu \nu_{\mathrm{R}} \, \tilde \phi
		\: + \: \overline{Q_{\mathrm{L}}} \: Y_u u_{\mathrm{R}} \, \tilde \phi
		\: + \: \overline{Q_{\mathrm{L}}} \: Y_d \, d_{\mathrm{R}} \, \phi \, .
\end{displaymath}

A final remark must be noted about these mass terms: both for leptons and 
quarks there are two mass matrices, one associated to $e_{\mathrm{R}}$ 
($d_{\mathrm{R}}$) and the other to $\nu_{\mathrm{R}}$ ($u_{\mathrm{R}}$).
All four matrices are independent, so in principle the $Y$'s are completely
general $3 \times 3$ complex matrices which we must diagonalise in order
to find the leptons (quarks) with well-defined masses. The diagonalisation
of a general complex matrix can be carried out through two unitary matrices,
one acting from the left and the other from the right; for instance, to
diagonalise $Y_e$ one only has to rotate adequately $e_{\mathrm{R}}$ and
$\ell_{\mathrm{L}}$ in flavour space, that is, one has to express
the charged leptons in the basis with well-defined mass. But then something
interesting happens when we try to do the same for the neutrinos: we can
rotate the $\nu_{\mathrm{R}}$'s, but we have already consumed the liberties
of $\ell_{\mathrm{L}}$ by finding the charged leptons with definite mass:
if we diagonalise $Y_\nu$ then we un-diagonalise $Y_e$ and vice-versa. This
issue implies that some of the non-diagonal terms in these matrices are
physical quantities that we cannot ignore: there is a mismatch between 
the charged lepton mass eigenstates and the neutrino mass eigenstates, or
as it is usually expressed, there is physical \emph{mixing} between the mass and
flavour states of the leptons. This is a direct consequence of $\ell_{\mathrm{L}}$
taking part in both $Y_e$ and $Y_\nu$, and the same happens for quarks.
Of course, there is no problem in working with quarks and leptons with
well-defined masses (though that doesn't have to be always convenient):
one can diagonalise at the same time $Y_e$ and $Y_\nu$ at the expense of
transporting the physical off-diagonal quantities to other interactions in 
the \mbox{Lagrangian -- for the SM,} to the weak charged currents. In that place
they will behave as mixing terms that may transform flavour-conserving
interactions into flavour-changing ones. More on the fundamental topic
of the dynamics of flavour can be found in \cite{Pich:2012sx} and in 
chapters 11 and 12 of \cite{Cheng:1985bj}; on the interesting matter
of identifying the physical flavour parameters we recommend 
\cite{Santamaria:1993ah}.

\subsection{Hypercharge}

At this point we might be tempted to think that the task is done: we have, indeed,
endowed our gauge theory with massive gauge bosons and we have found the way to
construct mass terms for the fermions, and all these things we've done without 
breaking any golden rule. Something, however, should make us suspect that the job 
is not quite finished yet: we know 
that the weak vector bosons are massive, but the $W$'s and the $Z$ have different 
masses, and equation \eqref{intro-SM-Goldstones-SSB} assigns the same mass to all of
them. Furthermore, some significant things happen related to the electric charge:
we know that electromagnetism is described by a $U(1)$ gauge theory; the photon
is, as far as we know, exactly massless, so the $U(1)$ symmetry is not broken.
Naively, we may think that this means that the symmetry-breaking Higgs field
must be electrically neutral. And that can be done; we can assign $Q (\phi) = 0$
and let all the electric charge rest on the fermionic fields, either singlets or
doublets. But this programme would encounter two problems: first, the doublets
must be assigned a charge as a whole, but we know that the neutrino is
neutral, while the electron is charged. Second, two of the Goldstone bosons
end up as the longitudinal components of the $W$'s, which are
electrically charged; these particular Goldstones, therefore, must be charged too, 
and as they
are degrees of freedom of the Higgs field, the picture of a neutral $\phi$ doesn't 
seem consistent. These two facts suggest some sort of `coupling' between $SU(2)$
and electric charge; this idea is reinforced if we note that, for all the
doublets in \eqref{intro-SM-doublets}, there's a difference of 1 between the charge 
of the up component and the down component.

To fix these issues \emph{hypercharge} is introduced; the idea is to put a new
gauge boson into the game in such a way that the photon emerges as a combination
of the $SU(2)$ bosons and this new one. If we get to do so, electric charge will
be a combination of the new charge and the $SU(2)$ charges, and we will have
$SU(2)$ multiplets with different electric charges for each component. As a side
effect we will also gain a new gauge coupling which will allow us to have
different masses for the $W$'s and the $Z$. Let us discuss how this happens: 
first we introduce a gauge phase symmetry ``of hypercharge'', that we denote as
$U(1)_Y$; this symmetry has just one generator and thus one associated gauge
boson, that we denote by $B_\mu$. The covariant derivative is then modified to
\begin{displaymath}
	D_\mu = \partial_\mu - i g \, T_a W^a_\mu - i g^{\, \prime} \, Y \, B_\mu
\end{displaymath}
for a general field with $SU(2)$ structure and hypercharge $Y$. Note that
$Y$ suffers from an ambiguity of definition: as the coupling, $g^{\, \prime}$,
and the charge itself always appear in the form of the product $g^{\, \prime} \, Y$,
the physics does not change if we rescale $Y$ by a factor $\kappa$ and 
$g^{\, \prime}$ by $\kappa^{-1}$. $g^{\, \prime}$ and $Y$, so, are ill-defined;
this is a common feature to all Abelian gauge 
symmetries. The degeneracy will be resolved when we choose a well-defined 
relation between hypercharge, $SU(2)$ charges and electric charge. To get to that
point let us first discuss the spontaneous breaking of this somewhat more
involved $SU(2) \otimes U(1)$ symmetry.

The scheme we follow is the same we have already practiced for plain $SU(2)$:
all the relevant features emerge from the Higgs field kinetic term, after we 
insert equation \eqref{intro-SM-Higgs-with-VEV}:
\begin{multline} \label{intro-SM-Higgs-kinetic-term-SU2-U1}
	(D_\mu \phi)^\dagger D^\mu \phi = \frac{v^2}{4} \left( g^2 \, W_\mu^a W_a^\mu + 
	 		4 {g^{\, \prime}}^2 \, Y_\phi^2  B_\mu B^\mu - 4 g g^{\, \prime} \, 
			Y_\phi B_\mu W^\mu_3 \right) \: + 
			\\
			+ \: v \sqrt{2} \left( g^{\, \prime} \, Y_\phi B_\mu \, \partial^\mu 
			\theta_3 - \frac{g}{2} \, W_\mu^a \, \partial^\mu \theta_a \right) 
			\: + \: \ldots 
\end{multline}
Here $Y_\phi$ represents the hypercharge assignment of the Higgs field.
We have selected two sets of pieces: 
the second one informs us that the hypercharge gauge boson couples only to the
third Goldstone, while the first yields the mass matrix for the vectorial bosons.
We note from this mass matrix that all the terms are diagonal save for one 
mixing between $W_3$ and $B$. $W_1$ and $W_2$ are degenerate, and their mass
is determined by the $SU(2)$ coupling constant; of course, any combination of
$W_1$ and $W_2$ will have the same mass as they have. Indeed, when we recover
electric charge we will see that a certain combination has definite electromagnetic
properties; we will identify such combination with the weak, charged $W$ boson.

For the remaining two gauge bosons, some discussion about their masses is in order.
It's easy to check that the mass matrix
\begin{equation} \label{intro-SM-neutral-gauge-boson-mass-matrix}
	\begin{pmatrix}  W^3_\mu  &  B_\mu  \end{pmatrix} \frac{v^2}{4}
	\begin{pmatrix}
		g^2   &   -2 g g^{\, \prime} \, Y_\phi
		\\
		\vphantom{\raisebox{2ex}{a}} 
		-2 g g^{\, \prime} \, Y_\phi   &   4 {g^{\, \prime}}^2 \, Y_\phi^2
	\end{pmatrix}
	\begin{pmatrix}   W_3^\mu  \\   \vphantom{\raisebox{2ex}{a}}
			B^\mu   \end{pmatrix}
\end{equation}
possesses a null eigenvalue, which has to correspond to the photon.
The rotation
\begin{displaymath}
	\begin{pmatrix}  W^3_\mu   \\   B_\mu \end{pmatrix}  =
	\begin{pmatrix}
		\cos \theta_W   &  \sin \theta_W  
		\\
		- \sin \theta_W   &   \cos \theta_W
	\end{pmatrix}
	\begin{pmatrix}   Z_\mu    \\   A_\mu   \end{pmatrix}
\end{displaymath}
with $\tan \theta_W \equiv  2 g^{\, \prime} Y_\phi / g$,
identifies the relevant combinations, and defines the important \emph{weak
angle}, $\theta_W$.
The photon is the only massless gauge boson that remains in the theory; it has
to be associated with an unbroken, one-generator subgroup of $SU(2) \otimes U(1)_Y$.
There is only one possibility if the subgroup is to have just one generator: a
$U(1)$ symmetry, which corresponds nicely with the known $U(1)_{\mathrm{e m}}$ of QED.
The conserved charge of $U(1)_{\mathrm{e m}}$ is, of course, electric charge;
we can identify it by isolating the photon in the  $SU(2) \otimes U(1)_Y$ covariant
derivative:
\begin{align*}
	D_\mu &= \partial_\mu - i g \, T_a W^a_\mu - i g^{\, \prime} \, Y \, B_\mu =
			\\
	&= \partial_\mu - i g \, \left( T_1 W^1_\mu + T_2 W^2_\mu \right) 
			- i \left( g \, \cos \theta_W \, T_3 - g^{\, \prime} Y \, \sin 
			\theta_W \right) Z_\mu -
			\\
	&\phantom{=} - i \, \frac{g g^{\, \prime}}{\sqrt{g^2 + 4 {g^{\, \prime}}^2 \, 
			Y_\phi^2}} \left( 2 Y_\phi \, T_3 + Y \right) A_\mu \, ,
\end{align*}
that allows us to define%
\footnote{Of course $U(1)_{\mathrm{em}}$ is Abelian too, and it is also affected
by a scaling ambiguity. It can be made apparent by substituting $Y \rightarrow 
\kappa \, Y$ and $g^{\, \prime} \rightarrow g^{\, \prime} / \kappa$, which also
relates it to the ambiguity in $U(1)_Y$. The definition for $Q$ in 
\eqref{intro-SM-def-e-and-Q-0}
is actually a choice, and fixes $\kappa = 1$. The reader can check that all the
discussion that follows can be carried out with $\kappa$ explicitly inserted in
the expressions and no argument can be found to constrain its value. Another
popular choice is $\kappa = 1 / 2$, in analogy to the Gell-Mann-Nishijima
formula that relates electric charge and strong isospin.
}:
\begin{align} \label{intro-SM-def-e-and-Q-0}
	e &\equiv \frac{g g^{\, \prime}}{\sqrt{g^2 + 4 {g^{\, \prime}}^2 Y_\phi^2}}
		&
	Q &\equiv 2 Y_\phi \, T_3 + Y \, .
\end{align}
These definitions are obscured by the presence of $Y_\phi$, which we should fix
to its numerical value. But before we must figure it out: $Y_\phi$, as we read
it in equation \eqref{intro-SM-def-e-and-Q-0}, is related to the charge
spacing between the up-components and the down-components of the doublets, which
we know to be of one unit. Indeed, if we apply the definition of electric charge
to the components of the Higgs field itself, should we write $\phi^\mathrm{T} = 
\begin{pmatrix} \phi_{\mathrm{up}} & \phi_{\mathrm{down}} \end{pmatrix}$, we obtain
\begin{align*}
	Q (\phi_{\mathrm{up}}) &= 2 Y_\phi
		\\ 
	Q (\phi_{\mathrm{down}}) &= 0 \, .
\end{align*}
There are two remarkable discoveries here: first, as we expected, the spacing
between the two components depends only on $Y_\phi$; in order to make it of $+ 1$
charge units, as we see in the lepton and quark doublets, we require $Y_\phi = + 
\nicefrac{1}{2}$. Second: the down-component has null
electric charge, irrespective of the value of $Y_\phi$; there is a good reason
for that: $\phi_{\mathrm{down}}$ is the component of the Higgs field that
gets a VEV, as we established in \eqref{intro-SM-Higgs-with-VEV}, and if the
photon exists and is massless it's because its associated symmetry is unbroken.
The theory ensures $U(1)_{\mathrm{em}}$ remains exact by assigning $Q = 0$ to the
relevant component of the symmetry-breaking field. 

Incidentally, let us digress for a moment to comment, as we promised, on the choice
we made in equation \eqref{intro-SM-Higgs-with-VEV} for the $SU(2)$ direction of 
the VEV. The reason is simple to state: electromagnetic interactions do not change 
flavour. Had the VEV been located in another direction, say 
$\left< \phi \right>^\mathrm{T} = \begin{pmatrix} \nicefrac{v}{\sqrt{2}} &
\nicefrac{v}{\sqrt{2}} \end{pmatrix}$, the $B$ boson would have mixed with some 
combination of the $W_a$, and the
photon would have been in part $W_1$ or $W_2$; consequently, $Q$ would have had
components along $T_1$ or $T_2$, which are nondiagonal and provide flavour-changing
interactions. This is not acceptable: a photon that matches phenomenology needs
$B$ to mix only with $W_3$. This is achieved by a handful of directions, of which
the one chosen in \eqref{intro-SM-Higgs-with-VEV} is the most commonly used.

Returning to our discussion of hypercharge, the result $Y_\phi = + 
\nicefrac{1}{2}$ greatly simplifies the expressions that describe the breakdown
of the $SU(2) \otimes U(1)_Y$ symmetry, starting with the coupling and the 
charge of the electromagnetic interaction, which become
\begin{align} \label{intro-SM-def-e-and-Q-1}
	e &= \frac{g g^{\, \prime}}{\sqrt{g^2 + {g^{\, \prime}}^2}}
		&
	Q &=  T_3 + Y \, .
\end{align}
This final expression for $Q$ allows us to find the hypercharges for all the
fields of the theory; we display them in table \ref{tab:intro-SM-hypercharges}
together with a reminder of their $SU(2)$ properties.
We can also identify the electrically charged $W$ boson: we know
it should appear out of a combination of $W_1$ and \mbox{$W_2$ --as it is} a complex
field it requires two real fields to be \mbox{assembled--, and we} know that any
field that creates a charge $q$ should verify $\left[ \, Q, \, \psi \, \right] 
= q \, \psi$;
as the $W$ bosons carry no hypercharge, it all reduces to finding a combination
of $W_1$ and $W_2$ such that $\left[ \, T_3, \, W^+ \, \right] = 
+ \, W^+$. The $SU(2)$ matrix which 
verifies this is $T_+ \equiv T_1 + i T_2$, 
and the associated combination of gauge bosons yields
\begin{displaymath}
	W^\pm \equiv \frac{1}{\sqrt{2}} \left( W_1 \mp i W_2 \right) \, .
\end{displaymath}
For practical purposes we will work, as usual, with a $W^\mu$ field and its
Hermitian conjugate. Let us define by convention $W \equiv W^-$ and 
$W^\dagger \equiv W^+$, meaning that the ``particle'' \mbox{states --those} 
created by the \mbox{operator $a^\dagger$-- will carry} charge $+1$.

\begin{table}[bt]
	\centering
	\[
	\begin{array}{| c | c | c | c | c | c | c | c |}
		\hline
		\vphantom{\raisebox{-1.5ex}{a}} \vphantom{\raisebox{1.5ex}{a}}
			 &  \ell_{\mathrm{L} \alpha} & e_{\mathrm{R} \alpha} & 
			 	\nu_{\mathrm{R} \alpha} &  Q_{\mathrm{L} \alpha}  &  
				u_{\mathrm{R} \alpha} &  d_{\mathrm{R} \alpha}  &  \phi
		\\
		\hline
		\vphantom{\raisebox{-1.5ex}{a}} \vphantom{\raisebox{1.5ex}{a}}
		SU(2)  &  \nicefrac{1}{2}  &  0  &  0  &  \nicefrac{1}{2}  &  0  &  0
			   &  \nicefrac{1}{2}
		\\
		\hline
		\vphantom{\raisebox{-1.5ex}{a}} \vphantom{\raisebox{1.5ex}{a}}
		Y  &  - \nicefrac{1}{2}  &  -1  &  0  & + \nicefrac{1}{6}  &  
			  + \nicefrac{2}{3}  &  - \nicefrac{1}{3}  &  + \nicefrac{1}{2}
		\\ 
		\hline
	\end{array}
	\]
	\caption{Electroweak charges of the non-gauge fields in the Standard Model.
			 For $SU(2)$ the representation is indicated; for hypercharge the
			 convention given in equation \eqref{intro-SM-def-e-and-Q-1} is
			 assumed. The doublets $\ell_{\mathrm{L}}$ and $Q_{\mathrm{L}}$ 
			 are detailed in equation \eqref{intro-SM-doublets}. The greek
			 indices represent components in a flavour vector, as the analogous
			 fields in different families have the same charge assignments.
			} \label{tab:intro-SM-hypercharges}
\end{table}

\subsection{Concluding remarks}

As for other properties of the gauge bosons, we can now write them down clearly.
For instance, we can read their masses from equation 
\eqref{intro-SM-Higgs-kinetic-term-SU2-U1}
and from the eigenvalues of \eqref{intro-SM-neutral-gauge-boson-mass-matrix}:
\begin{align*}
	m_W^2 &\equiv \frac{v^2}{4} \, g^2  = \frac{\pi \alpha}{\sin^2 \theta_W} \, v^2
	\\
	\vphantom{\raisebox{3ex}{a}} 
	m_Z^2 &\equiv \frac{v^2}{4} \left( g^2 + {g^{\, \prime}}^2 \right) =
			\frac{\pi \alpha}{\sin^2 \theta_W \, \cos^2 \theta_W} \, v^2 \, ,
\end{align*}
where we have expressed the masses first as a function of fundamental electroweak 
parameters and then in terms of the somewhat more practical $\alpha$ (the 
fine-structure constant) and $\theta_W$. We can also write the weak interaction
vertices in terms of the relevant fields after spontaneous symmetry breaking;
they arise from the fermion kinetic terms, which are of the form
\begin{displaymath}
	\mathcal{L}_{\mathrm{kin-ferm}} = i \, \overline{\Psi_{\mathrm{L}}} \gamma_\mu
		D^\mu \Psi_{\mathrm{L}} \: + \: i \, \overline{\psi_{\mathrm{R}}} \, 
		\gamma_\mu	D^\mu  \, \psi_{\mathrm{R}} \, ,
\end{displaymath}
where the covariant derivative acting on the left-handed fields contains $SU(2)$
structure, while the one acting on $\psi_{\mathrm{R}}$ carries, if anything, 
just the hypercharge part. Let us write explicitly the charged-current vertex
for a general doublet notated as $\Psi^\mathrm{T}_{\mathrm{L}} = \begin{pmatrix}
\Psi^\mathrm{up}_{\mathrm{L}}  &  \Psi^\mathrm{down}_{\mathrm{L}}  \end{pmatrix}$:
\begin{equation} \label{intro-SM-charged-currents}
	g \, \overline{\Psi_{\mathrm{L}}} \, \gamma_\mu \left( T^1 W_1^\mu + 
		T^2 W_2^\mu \right) \, \Psi_{\mathrm{L}} =
		\frac{e}{\sqrt{2} \, \sin \theta_W} \, 
		\overline{\Psi_{\mathrm{L}}^\mathrm{down}} \, \gamma_\mu
		\Psi_{\mathrm{L}}^\mathrm{up} \, W^\mu \: + \: \mathrm{H.c.}
\end{equation}
And then the somewhat more cumbersome neutral-current vertices, which get
stupendously simplified for the photon:
\begin{multline*}
	g \, T^3_\psi \,
	\overline{\psi_{\mathrm{L}}} \, \gamma_\mu \, \psi_{\mathrm{L}} 
	\, W_3^\mu \: + \: g^{\, \prime} (Q_\psi - T^3_\psi) \,
	\overline{\psi_{\mathrm{L}}} \, \gamma_\mu \, \psi_{\mathrm{L}} 
	\, B^\mu \: + \:
		g^{\, \prime} Q_\psi \, \overline{\psi_{\mathrm{R}}} \, \gamma_\mu \,
		\psi_{\mathrm{R}} \, B^\mu =
		\\
	\vphantom{\raisebox{4.5ex}{a}} 
	= e \, \bar \psi \, \gamma_\mu \left[ \frac{T^3_\psi}{\sin \theta_W
		\cos \theta_W} \, \mathrm{P}_{\mathrm{L}} - Q_\psi \tan \theta_W \right]
		\psi \, Z^\mu \: + \: e Q_\psi \, \bar \psi \gamma_\mu \psi \, A^\mu \, ,
\end{multline*}
and where $\psi_{\mathrm{L}}$ represents either $\Psi^\mathrm{up}_{\mathrm{L}}$ or
$\Psi^\mathrm{down}_{\mathrm{L}}$, with its corresponding isospin assignment being
$T^3_\psi$.

\section{Effective theories} \label{sec:intro-effective-theories}

A second element that will be present throughout the next chapters is the notion
of effective theory. Essentially%
\footnote{The reader interested in a more thorough discussion may look
into some of the many references available in the literature. Two classical
ones are \cite{Weinberg:1978kz,Georgi:1994qn}, but also interesting,
with different approaches, are \cite{Kaplan:1995uv,Burgess:1998nm,
Manohar:1996cq,Pich:1998xt,Wudka:1994ny}.}, 
the idea is that when a system has two
very separated scales the dynamics of each scale tends to proceed independently 
of that of the other. This principle is easy to understand in the specific case 
of particle physics, where the natural scale difference is \emph{energetic}:
there exist light particles and heavy particles; at low energies one
sees all the light particles, but none of the heavy ones; at high
energies one can produce the heavy states, but cannot discern the masses
of the light ones. One could think, thus, of devising specific theories
for working at low or high energies which would be simpler than the complete
model; such theories are called ``effective'' when some of the dynamical
degrees of \mbox{freedom --in the} case of particle physics, the fields
that describe the \mbox{particles-- are} completely erased in the process.
For instance one can think of simply eliminating the high-energy fields
from the Lagrangian in order to describe the dynamics at low energies;
of course, this would be just a first approximation, as the low-energy
behaviour is not completely independent of the heavy fields: they can mediate
process as virtual particles and generate effects that would not be 
observed if the light particles were the only ingredient. Therefore, a
second step is to incorporate all these effects \emph{without} reintroducing
the heavy fields. But one does not need to account for all the effects
of the heavy particles, as this would be equivalent to recovering the complete
model; part of the beauty of this game is to realise that some effects
are more important than others, and to identify the leading and the
subleading ones. In this way one can weigh how deep wants to dig into
the \mbox{theory --or,} analogously, how much precision one's calculation 
\mbox{requires-- and then} make an explicit cut, forget about the 
effects smaller than that. This sequence roughly describes the making
of an effective theory.

\begin{figure}[p]
	\centering
	\raisebox{4.5cm}{\emph{a)}}
	\includegraphics[width=0.65\textwidth]{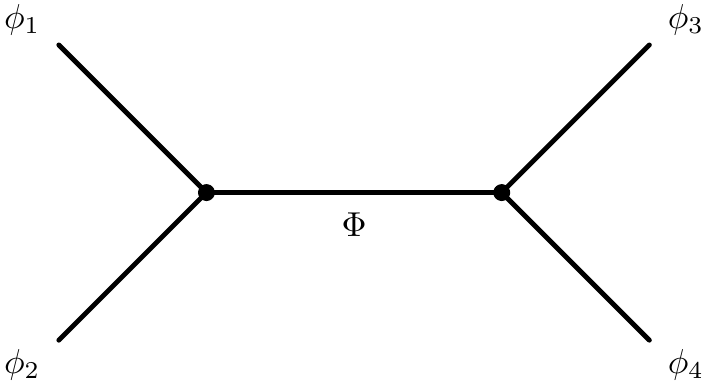}

	\vspace{0.7cm}
	\hspace{2cm}
	\includegraphics[width=0.05\textwidth]{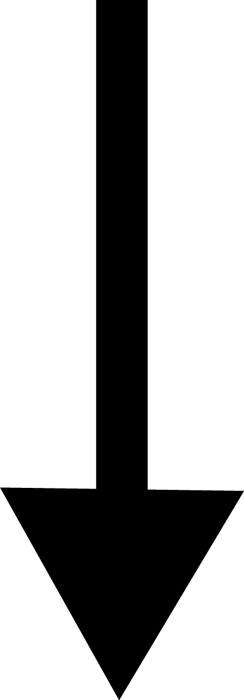}
	\raisebox{0.85cm}{$\quad m_\Phi \gg E$}
	\vspace{0.7cm}

	\raisebox{5.6cm}{\emph{b)}}
	\raisebox{3cm}{$\dfrac{- i}{m_\Phi^2} \; \times \quad$}
	\includegraphics[width=0.5\textwidth]{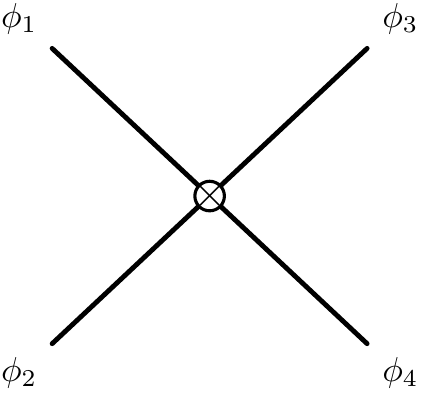}
	\caption{An example of how a heavy field is eliminated from the low-energy
			description of physics: the four light fields $\phi_1$, $\phi_2$, 
			$\phi_3$ and $\phi_4$ scatter with the mediation of a heavy field
			$\Phi$. If the typical energies of the $\phi_i$ system are much
			below the mass of $\Phi$, it can be reduced to a numerical factor
			that depends on its mass, and for the light fields there remains
			a $\phi_1 \phi_2 \phi_3 \phi_4$ effective vertex.
			} \label{fig:intro-EFT-diagrams}
\end{figure}

So, how do these ideas effectively crystallise in the everyday work 
of a particle physicist? Have a look at figure 
\ref{fig:intro-EFT-diagrams}\emph{a)}; it represents the transition
$\phi_1 \phi_2 \rightarrow \phi_3 \phi_4$ mediated by a $\Phi$ field
in the \emph{s} channel. Let us now assume that the fields $\phi_i$ are
light and $\Phi$ is heavy, and that the whole process proceeds with 
an energy sufficiently low so that a physical $\Phi$ cannot be produced by
any means; this transition, however, is allowed and will happen with 
some rate, as a $\Phi$ in an internal leg does not need to be a detectable
particle. We can calculate this rate: it only involves the internal propagator
of $\Phi$, which is just
\begin{equation} \label{intro-EFT-Phi-prop}
	\frac{i}{p^2 - m_\Phi^2} \, .
\end{equation}
Now, as we want to abstract $\Phi$ from our description of the dynamics of 
$\phi_1$,
$\phi_2$, $\phi_3$ and $\phi_4$, we could forget that \eqref{intro-EFT-Phi-prop}
comes at all from the contraction of a field; we could just consider it
a sort of `form factor', an addendum necessary for the calculation of 
$\phi_1 \phi_2 \rightarrow \phi_3 
\phi_4$. Moreover, if $m_\Phi$ is really much heavier than any energy scale
we want to consider, we can neglect the internal momentum. By doing this we are
left with $- i / m_\Phi^2$, a number; a dimensionful quantity too, but in any case
not a field nor any kind of operator. What remains then for the calculation
of the process? Just four external legs of light fields; we can represent the
situation after `forgetting' about $\Phi$ as the diagram in figure 
\ref{fig:intro-EFT-diagrams}\emph{b)}: the process has reduced to a four-field
vertex and a dimensionful factor.

This example is paradigmatic: heavy fields that only act as virtual particles
leave behind multi-field vertices that involve the light fields \emph{times}
dimensionful factors usually related to their masses. The next step is immediate:
if this four-field vertex has appeared as a byproduct of a four-field process,
other vertices, with arbitrary number of light fields, will emerge from
other, N-field processes. We can learn from here the way to construct our effective
theory, even if we know nothing about the underlying heavy physics: we will have
to consider every combination of light fields which can take part in a low-energy 
process; this essentially means any combination of fields that respects all the
conserved charges, as any physical process must do. The result will be
a list of operators with ever-growing number of fields, each of them
representing more and more complicated processes mediated by the heavy fields.
Of course, given a certain set of heavy fields not all the operators will be
equally important. It may even happen that some of them are not present at all,
if the symmetries of the heavy fields forbid the associated processes. But as
long as we are ignorant of the heavy physics we must consider all and
every one of them.

Let us write these ideas in a more systematic way. We have a quantum field
theory for which we know the relevant fields and symmetries, that is, the
Lagrangian, up to some energy. The theory shows no hint that further
physics is required: the Lagrangian is renormalisable, and so valid up to
arbitrarily high energies. At this stage we have simply stripped away the
heavy physics from the theory. Then we would like to introduce its virtual 
effects on the low-energy fields; for that aim, we proceed by
adding all the operators that respect Lorentz symmetry and any other symmetry
of the low-energy theory, irrespective of the number of fields. As the original
Lagrangian was renormalisable, it contained terms up to dimension four;
the effective terms, so, will begin at dimension five:
\begin{equation} \label{intro-EFT-Lagrangian}
	\mathcal{L}_{\mathrm{eff}} = \mathcal{L} + \sum_{n=5}^\infty \sum_i  
		\left( \frac{C^{(n)}_i}{\Lambda^{n-4}} \, \mathcal{O}\up n_i + 
		\mathrm{H.c.} \right)\, .
\end{equation}
In this expression $n$ indicates the dimensionality of the operator and
$i$ labels the different operators of the same dimension, whose number
will depend on the field content of $\mathcal{L}$. As a Lagrangian must
have dimensions of $E^4$, the effective operators must be accompanied
by dimensionful couplings with dimensions of inverse mass to an appropriate
power; this is a reflection of the fact that these operators are the result of
`freezing' one or more internal propagators from a physical process, much in
the way we display in figure \ref{fig:intro-EFT-diagrams}. The dimensionality
of the effective coupling is carried by the parameter $\Lambda$, which has 
dimensions of mass; the effective coefficients
$C \up n_i$ collect the dimensionless factors of the coupling.
The separation between $\Lambda$ and $C$ is somewhat arbitrary, and when one
knows nothing about the properties of the heavy particles it is customary
to take $C \up n_i = 1$.

The effective interactions are, by construction, nonrenormalisable. This
should not be a surprise, as we know that nonrenormalisable theories
are not valid to all energy scales: processes that involve a nonrenormalisable
vertex yield unacceptably large amplitudes at high energies, thus violating
the unitarity of the $S$ matrix. This violation occurs generically at 
energies of the order of $\Lambda$, and is considered an indication that
the scale of the new particles has been reached; one must then drop the effective 
theory and replace it by a new, renormalisable QFT that includes the 
heavy fields. For this reason, $\Lambda$ is usually termed the \emph{new physics
scale}, and for this reason $\Lambda$ is often considered the most important
parameter in the effective theory. When the underlying theory is simple it 
may coincide precisely with the mass of the new particles, but more commonly
it will be a combination, sometimes rather involved, of several dimensionful 
parameters; if these parameters are not too different, though, $\Lambda$ will
still point in the correct direction, indicating at least the order of magnitude
of the heavy masses. It may also happen, if the underlying theory presents
several mass scales, that different effective operators show different new physics
scales; again, if these are not too different the interpretation of $\Lambda$ is
safe. Moreover, equation \eqref{intro-EFT-Lagrangian} still holds, even in
this case, if we cleverly integrate inside the $C$'s the appropriate (dimensionless) 
combination of parameters so that $\Lambda$ is replaced, where needed,
by another $\Lambda^\prime$.

This far we have briefly explained how we construct an effective field theory. 
But once we have it, how exactly do we use it? 
Well, it depends on what we know about the underlying heavy physics.
If we know nothing about it we must consider the whole set of effective operators
and try to measure experimentally their couplings. That may be easier if we
select those operators that yield processes forbidden in the low-energy 
\mbox{theory -- for in}\-stance, some effective interactions may break a global 
symmetry that prevents certain process in the low-energy theory: that would be 
a good place to start. In the general case we would think that lower-dimensional
operators offer the best experimental shot, as they are suppressed by less 
powers of $\Lambda$, which is assumed to be large. As we dig deeper in the
details of the low-energy theory we gain access to higher-dimensional operators
and also to heavier masses in the lowest-dimensional ones. So, as we said 
above, details are relevant: the effective contribution may be hidden in the
fifth or sixth digit of a measurement, and this is also a way to investigate
the physics of heavy particles. When no positive experimental result is found,
at least we can place bounds on the new physics scale, which should roughly
correspond to the mass of the new particles.

If we know something about the underlying \mbox{theory --perhaps} that it breaks
certain global symmetry, or that it only involves scalar \mbox{particles--, we can}
probably restrict the spectrum of effective operators: we can know that some
of them don't exist at all, or that others are loop-generated and thus suppressed
respect to those generated at tree level.
These considerations help to focus on the correct operators
in order to look for the optimal experimental signatures.

Finally, we may be working in a theoretical framework; if we know absolutely the
physics of the heavy particles, our interest can be to deduce their low-energy
effects in order to simplify the calculations in that regime.
The procedure is to calculate in the underlying theory
the full process that generates the effective interaction, and then match this
calculation with the corresponding result in the effective theory.
We will do so at several points throughout this work, as it will allow us to
compare the sensitivity of direct searches for the new particles with 
that of precision low-energy measurements.

\chapter{Neutrinos}  \label{chap:neutrinos}

All the work presented here revolves around neutrinos and their properties. 
Due to their extremely
light masses neutrinos are ubiquitous in the universe, but their lack of electric
charge makes them very difficult to spot and study; as far as we know, they only
interact through purely weak interactions, that is, through processes mediated
by a $W$ or a $Z$, which are suppressed at low energies by $m_{W,Z}^{-2}$;
charged currents, for instance, which are responsible for processes such as
$\nu_e + X \rightarrow e + Y$, of uttermost experimental importance, are 
governed by the small Fermi effective coupling, $G_F \equiv \frac{\sqrt{2}}{8} \, 
\frac{g^2}{m_W^2} = (1.16637 \pm 0.00001) \times 10^{-5} \; \mathrm{GeV}^{-2}$.
It is, thus, interesting to investigate if neutrinos may feel additional 
interactions, which would provide new windows into their properties; this
investigation is especially relevant if the new interactions shed some
light on the neutrino properties that are still poorly known, such as 
their masses and their flavour structure. In this chapter we review some of these
features, focusing on those which will be of interest for the studies that
we are to present in the remaining of this work. The reader interested in a more
comprehensive review may consult any of the very good texts that
already exist in the literature, for instance \cite{Mohapatra:2005wg},
\cite{Mohapatra:1998rq,Giunti:2007ry,Xing:2011zza} or \cite{Raffelt:1996wa}.

\section{Neutrino masses}

The origin and nature of neutrino masses is one of the most enticing puzzles in 
modern particle physics. Their minute value, their flavour struc\-\mbox{ture --so 
dif}\-ferent from what we observe in the quark \mbox{sector--, or the} intriguing 
possibility that they open a gateway into the violation of lepton number are just some 
of the issues that attract our interest in this matter. In this section we review
several features of neutrino masses, focusing on those which will will be developed
in forthcoming chapters. For a more comprehensive review we refer the reader to
\cite{Mohapatra:1998rq,Mohapatra:2005wg,Strumia:2006db,GonzalezGarcia:2007ib}.

\subsection{Masses and the nature of fermionic particles} 
				  \label{sec:neutrinos-dirac-or-majorana}

One of the tough problems we face when studying neutrino masses is that 
we don't even know how to write them. We have learnt a lot about their structure,
and day after day we lay a better siege to their value, but the fact is that 
we still ignore their true character, their \emph{nature}. Fermions are known
to have two classes of mass terms: the first is derived from Paul Dirac's
relativistic quantum equation for the fermion \cite{Dirac:1928hu,Dirac:1928ej}; 
it describes the mixing 
of two fields, one of them left-handed under Lorentz transformations and the other
one right-handed:
\begin{displaymath}
	m \, \overline{\chi_{\mathrm{R}}} \, \psi_{\mathrm{L}} \, .
\end{displaymath}
This Dirac mass term yields fields with \emph{four} internal degrees of freedom,
and is suited for particles that carry internal charges such as electric charge,
for it can be naturally transmitted from $\psi_{\mathrm{L}}$ 
to $\chi_{\mathrm{R}}$.
The degrees of freedom distribute among two polarisations 
for an excitation which carries charge $+ Q$ and two for an excitation that carries
charge $- Q$: an \emph{antiparticle}.

The second option, proposed by Ettore Majorana \cite{Majorana:1937vz}, 
consists in changing one of the fields for the
\emph{charge-conjugate} of the other, for we know that the operation of 
charge conjugation changes the chirality of the affected field. The mass
term is written then as
\begin{displaymath}
	m \, \overline{\psi_{\mathrm{L}}^\mathrm{c}} \, \psi_{\mathrm{L}} \, ,
\end{displaymath}
and it describes a fermion field with just \emph{two} degrees of freedom;
besides, the field $\psi_{\mathrm{L}}$ can't be assigned an internal charge,
for it would be broken by the combination $\overline{\psi_{\mathrm{L}}^\mathrm{c}} 
\, \psi_{\mathrm{L}} \sim \psi_{\mathrm{L}} \, \psi_{\mathrm{L}}$. The result,
thus, resembles quite much a real field: having only the polarisations for one
fermionic excitation and being unable of carrying charges, it describes a 
fermion which is its own antiparticle. The Majorana field is a real fermionic 
field.

As we see, the matter between Dirac and Majorana fields is more about the available
degrees of freedom and the ability to carry charges than the mass term itself,
though these properties emerge of course from the symmetries of the mass term.
Most of the fundamental fermions in the Standard Model are electrically charged,
so they can only be Dirac fermions. And indeed, as we described in section
\ref{sec:masses-SM}, each left-handed field in the SM has a right-handed 
counterpart
which allows to construct Dirac mass terms. Neutrinos, however, are different:
they are neutral and they don't possess any other charge that we know of; they
might, therefore, be the only Majorana fermions on the landscape. Well, to
be absolutely truthful they \emph{could} actually carry one charge that we know
of: lepton number. Lepton number is a global symmetry felt only by the leptonic
fields of the theory that causes leptons to be created or annihilated
exclusively in pairs lepton-antilepton. It is defined by the transformations
\begin{align}
	\ell_{\mathrm{L} \alpha} &\rightarrow \: e^{i \, L \, \theta} \, 
		\ell_{\mathrm{L} \alpha}
	\nonumber
	\\ \label{intro-neutrinos-LN}
	e_{\mathrm{R} \alpha} &\rightarrow \: e^{i \, L \, \theta} \, e_{\mathrm{R} \alpha}
\end{align}
where the Lagrangian just requires that $L(\ell_{\mathrm{L} \alpha}) = 
L( e_{\mathrm{R} \alpha})$, but conventionally 
it is usually assumed that $L(\ell_{\mathrm{L} \alpha}) = L( e_{\mathrm{R} \alpha})
= + 1$. If we add right-handed neutrinos to our description of neutrino masses
they receive the same
lepton number assignment as $e_{\mathrm{R} \alpha}$. Lepton number would
be broken by the presence of a Majorana mass for the neutrinos;
its conservation, however, is experimentally established to an astonishing
level of precision \cite{Beringer:1900zz}. It may seem a bit
dull even to consider the possibility of neutrino masses being Majorana after
this statement, but something comes to the rescue: neutrino masses are
unbelievably small%
\footnote{See section \ref{sec:knowledge-mass-neutrino}.
}.
If they were the only source of lepton number violation then it would be only
very slightly violated; actually our experiments, being greatly precise,
would not have reached the precision required to record it. So, all in all,
Majorana masses are still an open possibility for neutrinos, and an exciting
one that points to a mass generation mechanism different from that of the
other fermions.

Let us follow this thread by reviewing some details about the generation of Dirac
or Majorana masses for the neutrinos. We begin with the known, more conventional,
Dirac masses. By an argument of economy it seems reasonable to think that they
are generated by the same Higgs mechanism that yields masses for the other
fermions. For this aim we need \mbox{three --at least-- right-}handed neutrinos
with charges as displayed in table \ref{tab:intro-SM-hypercharges}, 
and Yukawa couplings of the like
of \eqref{intro-SM-quark-mass-tilde}, which upon SSB provide
\begin{equation} \label{intro-numass-dirac-masses}
	\overline{\ell_{\mathrm{L}}} \, Y_\nu \, \nu_{\mathrm{R}} \: \tilde \phi 	
		\quad \longrightarrow \quad \overline{\nu_{\mathrm{L}}} \, m_\nu \, 
		\nu_{\mathrm{R}} + \ldots \, ,
\end{equation}
where $m_\nu = v \, Y_\nu$. From the theoretical point of view Dirac masses are
a simple and consistent choice: they require no further addition to a Standard
Model that we know to work fine; besides, they allow naturally for lepton
number conservation, as firmly suggested by experiment. However, the same
reasons that make Dirac masses acceptable also make them sort of disappointing:
in a pure-Dirac scenario Yukawa couplings are the only source for neutrino masses;
the same Yukawa couplings that yield a top quark of 173 GeV provide an electron
of 511 keV, and then three neutrinos with masses at a fraction of the eV. 
In the standard Higgs mechanism everything is encoded inside the Yukawa couplings,
of whose origin we know nothing. We obtain no explanation, no insight on 
the hierarchy of the fermion masses, and especially we learn nothing about the
enormous gap between the electron and the neutrinos. Is disappointment an argument
good enough to drop a physical theory? Probably not. But it is good enough to
entice imagination, and that's also part of what we do.

Can Majorana masses help in our aim to explain the smallness of neutrino masses?
Before considering this question we need to find a way to fit them within
the Standard Model. Constructing a Majorana mass for the neutrinos won't be
as simple as writing $m \: \overline{\nu_{\mathrm{L}}^\mathrm{c}} \, 
\nu_{\mathrm{L}}$: although neutrinos are electrically neutral they are not
chargeless. Neutrinos belong in the leptonic doublet $\ell_{\mathrm{L}}$, 
which carries $SU(2)$ charge as well as hypercharge; a combination like 
$\overline{\ell_{\mathrm{L}}} \, \tilde \ell_{\mathrm{L}}$ would break all
those charges, providing also a disastrous Majorana mass term for the charged
leptons. One can check that actually there's no renormalisable 
combination of the fields in the Standard Model that can yield a Majorana mass 
term for the neutrinos. It was Steven Weinberg in \cite{Weinberg:1979sa}%
\footnote{But see also the work by Weldon and Zee a year later, 
\cite{Weldon:1980gi}.}
who realised
that one can combine the operator $\tilde \phi^\dagger \, \ell_{\mathrm{L}}$,
which is hyperchargeless and $SU(2)$-singlet, with its complex-conjugate to
obtain a dimension-five interaction that upon SSB really provides
Majorana neutrino masses. This operator is usually known nowadays as the
\emph{Weinberg operator}:
\begin{equation} \label{intro-neutrinos-majorana-masses}
	\left( \overline{\tilde \ell_{\mathrm{L}} \vphantom{\raisebox{1.25ex}{a}}} \,
		\phi \right)  \xi  \left( \tilde \phi^\dagger \, \ell_{\mathrm{L}}
		\right)  \quad \longrightarrow \quad \overline{\nu_{\mathrm{L}}^\mathrm{c}}
		\, m_\nu \, \nu_{\mathrm{L}}  +  \ldots \, ,
\end{equation}
where we have $m_\nu = v^2 \, \xi$ and $\xi$ is an effective coupling with dimensions
of inverse mass, as it corresponds to a dimension-five operator; we can extract
the dimensionful part in the form of a new physics scale as we explained in
section \ref{sec:intro-effective-theories}, and write 
$\xi = C / \Lambda$, with $C$ some combination of dimensionless couplings 
which generate the operator in the underlying high-energy theory. 

Other operators of higher dimensionality also provide Majorana neutrino masses, 
but they are suppressed by higher powers of $\Lambda$ and their effect is subdominant;
the Weinberg operator is the primary source of Majorana masses for the neutrinos
in the Standard Model. This fact has two important consequences: first, if the
masses of the neutrinos are Majorana then there has to be new physics beyond
the Standard Model; this is straightforward, as they cannot be constructed 
with renormalisable, pure-SM interactions. Second, the resulting Majorana masses
have the form $m_\nu = C v^2 / \Lambda$, that is, if the new physics is much
heavier than the electroweak scale the masses can be strongly suppressed. In this
way, Majorana masses offer a twofold explanation: the masses of the neutrinos are
very small because they are related to very heavy physics, and we cannot explain
their smallness in the framework of the Standard Model because it is actually
related to something else. Phylosophically speaking this is much more rewarding
than ``they are very small because this is the way they are''.

Of course, from this starting point a wide world of possiblities has been explored
in the past thirty years: some Majorana mass models do not actually need very 
heavy new physics, some models mix Majorana and Dirac mass terms for the neutrinos,
and some others manage to construct naturally small Dirac \mbox{mass terms -- 
see,} just as an instance of the overwhelming variety, \cite{Mohapatra:1980yp,
Cheng:1980qt,Zee:1985id,SilvaMarcos:1999wy,Ma:2006km,Ibanez:2009du,
Aparici:2011nu,Peinado:2012tp} or the models described in part II of 
\cite{Mohapatra:1998rq}. We won't
review this vast and ever-changing world here, but in section \ref{sec:seesaw} 
we will briefly explain a framework for neutrino mass generation
that has become very popular and is taken by itself as a starting point for
proposing new ideas: the \emph{seesaw mechanism}. But let us first describe
what we know and what we don't about neutrino masses and establish some notation
for the forthcoming discussion.

\subsection{Neutrino oscillations} \label{sec:neutrino-oscillations}

Flavour mixing and oscillation phenomena constituted our initial gateway into
neutrino masses. Let us introduce in this section the intuitive notion of 
oscillation, which will be used in section \ref{sec:knowledge-mass-neutrino}
to describe our knowledge of neutrino masses.

As a starting point, note that neutrinos are always produced and 
detected in association with a charged \mbox{lepton -- actually,} 
what we see in many
experiments is indeed the charged lepton, and from it we infer the quality
of the unseen neutrino. For decades, the definitory fact for neutrino species
was flavour: we had one `electron', one `muon'
and one `tau' neutrino according to which lepton they were paired with in the
weak interactions. Equation \eqref{intro-SM-doublets} essentially follows 
this line 
of thought through the vertices in \eqref{intro-SM-charged-currents}.
However, as early as in the 1960's we learnt, by means of the Homestake experiment 
\cite{Davis:1964hf,Cleveland:1998nv}, that something odd was happening to 
these flavoured neutrinos: 
they seemed to get lost. The disappearance was first noted in solar neutrino 
experiments: these looked for electron neutrinos produced in the core of the Sun, 
and found only one third of the neutrino flux expected from the best solar models
available at the moment. Of course the first reactions were of astonishment: first
the theoretical solar calculations were blamed for the anomaly and then the 
experiment reliability was put under question; however, independent calculations
showed the theoretical expectations to be correct and subsequent solar 
neutrino searches \cite{Fukuda:1996sz,Hampel:1998xg,Fukuda:1998fd,Ahmad:2001an,
Abdurashitov:2002nt,Altmann:2005ix} confirmed the deficit. 
The advent of atmospheric neutrino experiments, though initially brought no
disagreement with the total expected flux \cite{Reines:1965qk,Achar1965196,
Aglietta:1988be,Daum:1994bf}, finally led to the identification of an 
anomaly \cite{BeckerSzendy:1992hq,Fukuda:1994mc}: 
these surveys looked for neutrinos produced in the upper atmosphere as a byproduct
of cosmic ray collisions; at low energies they expected to find twice as many 
muon neutrinos as
electron neutrinos, but the ratio showed to be just a little above unity. 
This result seemed to suggest again that muon neutrinos were getting lost
\mbox{somewhere -- or maybe} was it electron neutrinos who were appearing out of 
nowhere?

The situation was puzzling. Several mechanisms were proposed to explain the 
neutrino deficits, among them neutrino oscillations. The idea behind oscillations
was tricky, but it can be expressed in one sentence: if neutrinos do have masses
but the states with definite mass don't coincide with the states with definite
flavour, then flavour won't be conserved during propagation. In other words: there
are three neutrino flavours, and correspondingly one should have three neutrino
masses; however, there's no 
reason for these masses to be assigned to one flavour each. Maybe the 
quantum-mechanical
state with mass $m_1$ is neither $\nu_e$ nor $\nu_\mu$ nor $\nu_\tau$, but a 
combination of all three, and so on with the other two masses. Then the other
way around would also be true: a $\nu_e$ would be a combination of the states
with masses $m_1$, $m_2$ and $m_3$. If that were the case, then the components
in $\nu_e$ would propagate independently, for \mbox{propagation --time-}evolution, 
in \mbox{the end-- in quantum} mechanics deals with energies: it is the states
with definite energy which propagate unperturbed. $\nu_e$, which has no definite
\mbox{energy --no definite} \mbox{mass--, will be} perturbed during 
propagation: its three massive components will dephase and the mixture will
change, transforming the $\nu_e$ a little bit into a $\nu_\mu$ or a $\nu_\tau$.
When happening in a vacuum, this transformation will be cyclical: at some 
point during propagation the 
mixture will return to $\nu_e$ and the process will repeat over and
over again. That's why this bizarre behaviour got to be called \emph{oscillation}.
In the end, the effect is simple: neutrinos change flavour as they move. If
we didn't find all the solar neutrinos it was because we only looked for electron
neutrinos; the moment the SNO experiment also looked for the other two varieties 
\cite{Ahmad:2002jz} we found exactly what we expected. 

This, unfortunately, is no place to tell the full details about oscillations,
even though it is a beautiful and enlightening story. For more information we
refer the reader to specific texts on the matter, for instance 
\cite{Giunti:2000kw,Giunti:2008cf,Waltham:2003tf,GonzalezGarcia:2007ib}.
We will only go on to say that the fight to determine the mechanism behind the
several neutrino deficits was a tough one, and oscillations only emerged as a 
clear winner when the KamLAND experiment, which studied 
neutrinos produced in human-made nuclear reactors, allowed us to check 
that the disappearance probability of neutrinos is
energy- and distance-dependent in the way one should expect from 
\mbox{oscillations \cite{Araki:2004mb} -- see} the
references cited above, or simply check out equation 
\eqref{intro-oscillations-reactor}.

\subsection{Writing masses and mixings} 
			\label{sec:neutrinos-writing-masses-mixings}

Let us now take a small break in this section in order to establish the 
notation of neutrino masses and mixings as we will use throughout
the rest of the text. We start by writing
the fields that describe the flavour eigenstates; in a straightforward notation
we call them $\nu_e$, $\nu_\mu$, $\nu_\tau$ as in 
equation \eqref{intro-SM-doublets}.
Then we denote the fields with definite mass by $\nu_1$, $\nu_2$, $\nu_3$, and
their associated masses by $m_1$, $m_2$ and $m_3$, in good logic. 
It is important to note that ``1'', ``2'', ``3'' are just conventional labels;
they don't pretend to say anything about the values of the masses and, 
in particular, 
they \emph{don't}
imply $m_1 < m_2 < m_3$. In the following \mbox{section 
--\ref{sec:knowledge-mass-neutrino}-- we will} describe what we know
and what we don't about the parameters that we are about to define here.

Given the eigenfields with well-definite mass and flavour, let us now
consider the relation between them;
as usual in quantum mechanics, it can be described by a linear change of basis:
\begin{displaymath}
	\begin{pmatrix}  \nu_e  \\  \nu_\mu  \\  \nu_\tau  \end{pmatrix} =
			U_{\mathrm{PMNS}} \begin{pmatrix}  \nu_1  \\  \nu_2  \\  \nu_3  
			\end{pmatrix} \, ,
\end{displaymath}
where $U_{\mathrm{PMNS}}$ is usually called the Pontecorvo-Maki-Nakagawa-Sakata
matrix \cite{Pontecorvo:1957cp,Pontecorvo:1957qd,Maki:1962mu,Nakagawa:1963uw}; 
it contains the observable parameters that describe the mixing between
flavour and mass eigenstates \cite{Schechter:1980gr}. 
With three neutrino species three of these 
parameters are `mixing angles', that is, values which control the amount
of flavour violation in processes like oscillations or charged-current
interactions; one more parameter is a phase, a constant that triggers
CP-violating processes. If the neutrinos are Majorana particles, two more
phases will be present in the PMNS matrix. 
The reason for this is that the number and nature of the mixing parameters depends 
on the structure of the masses and the flavour symmetries available to 
the involved fields; initially, any parameter present in the mass matrices would 
be either a mass, a mixing or a phase, but some of them can be proven to be
physically irrelevant. The fields have flavour redundancies, they can be
rearranged in flavour space without changing the observable quantities of the
theory; if we can find one such rearrangement which eliminates from the 
Lagrangian a certain parameter, then that parameter was not physical, it could
not influence observable results. The process of identifying the physical
parameters is described in more detail in \cite{Santamaria:1993ah}, but
we sketch it here for the case of Dirac and Majorana masses in order to
clarify the differences in the PMNS matrix for these two scenarios.

If the masses of the neutrinos are of Dirac type, then the flavour matrix we're
interested in is $Y_\nu$, as written in equation \eqref{intro-numass-dirac-masses}.
This matrix is initially a completely general $3 \times 3$ complex matrix, and so it 
possesses 18 parameters, 9 of which are moduli and the remaining 9 are phases; 
out of the 9 moduli,
3 describe the masses themselves, and some of the remaining 6 can be physically
relevant mixings,
but we don't know how many; of the 9 phases we don't know either how many 
are physical. Then let us look at the relevant fields: the fields with
flavour structure in \eqref{intro-numass-dirac-masses} are $\ell_{\mathrm{L}}$ 
and $\nu_{\mathrm{R}}$. We should identify the flavour transformations that we
can apply upon them without changing the rest of the physics; that's easy to do
by looking at the kinetic terms in \eqref{intro-SM-weakL}: the weak interactions
remain invariant if we apply any unitary transformation in flavour space to 
$\ell_{\mathrm{L} \alpha}$ and $\nu_{\mathrm{R} \alpha}$. So, we could think
that we have two unitary $3 \times 3$ matrices to eliminate spurious parameters
from $Y_\nu$.

However, $\ell_{\mathrm{L}}$ not only contains neutrinos:
it also carries the left-handed components of the charged leptons, $e_{\mathrm{L}}$,
which in turn are involved in the charged lepton masses, whose
flavour structure is given by $Y_e$, as seen in equation 
\eqref{intro-SM-fermion-masses}. This means that any flavour transformation acting
upon $\ell_{\mathrm{L}}$ affects both $Y_\nu$ and $Y_e$: it might happen that
by a transformation on $\ell_{\mathrm{L}}$ we erase a parameter from $Y_\nu$
but it is transferred to $Y_e$ instead of being properly eliminated; such a 
parameter should count as physical. In consequence, it is not correct to look for
the physical parameters just in $Y_\nu$; the right approach is to consider the 
physical 
parameters in \emph{both} $Y_\nu$ and $Y_e$. And if $Y_e$ is involved then the
$e_{\mathrm{R} \alpha}$ are also involved. That makes \emph{two} completely
general $3 \times 3$ complex matrices, composed initially of 18 moduli and
18 phases, and to attack them we have \emph{three} unitary $3 \times 3$ matrices,
which provide 9 moduli and 18 phases.

These are our weapons. Now just one element remains to be unveiled: in the end, 
we aim to 
transform $Y_e$ and $Y_\nu$ in two diagonal matrices which describe Dirac masses
for the charged leptons and the neutrinos. These masses allow for one internal
symmetry, as we discussed earlier: lepton number; a global $U(1)$ symmetry 
acting simultaneoulsy on $\ell_{\mathrm{L}}$, $e_{\mathrm{R}}$ and $\nu_{\mathrm{R}}$.
For our purposes, a global phase in the unitary matrices that remains free, as we can 
see in equation \eqref{intro-neutrinos-LN}.
When we carry out our calculation we need to account for this liberty which in the
end has to remain.

The rest is arithmetics: we have 18 moduli and 18 phases; we can use the unitary
transformations to diagonalise the Yukawa couplings; that means setting 36 equations,
one for each degree of freedom in $Y_e$ and $Y_\nu$. All of them will be nonlinear,
and many can be difficult even to set, but that doesn't matter for our purposes: all
we want to know is how many parameters we can eliminate from the equations
by adjusting the
unitary matrices to the appropriate values. And that can be known by simple counting:
we have 9 free moduli from the unitary matrices, so we can neutralise 9 of the moduli;
we have 18 free phases, so we can eliminate 18 phases. But remember that one of the 
phases in the unitary matrices is to remain free, so we can only absorb 17 of the
phases in the Yukawas. Result: a Dirac mass pattern for the leptons yields
9 physical moduli and 1 physical phase. The final arrangement of these physical 
parameters is somewhat arbitrary, as it is also arbitrary which elements in the 
initial Yukawa matrices we choose to absorb. However, for a matter of physical
significance it is customary to take 6 of the physical parameters to be the masses
of the charged leptons and the neutrinos; the remaining 3 are the three mixing
angles in the PMNS matrix. The phase has become universally known as the 
\emph{Dirac phase} of the mixing matrix. The six masses remain of course in the
now-diagonal mass matrices, while the mixings and the phase are incorporated
into the PMNS matrix. Again there are many possible parametrisations, but it is
customary to use the following, in the form of three Euler angles with an added
phase:
\begin{multline} \label{intro-neutrinos-PMNS-Dirac}
	U_{\mathrm{PMNS}}^\mathrm{Dirac} = \begin{pmatrix}
				1  &  0  &  0  \\
				0  &  \cos \theta_{23}  &  \sin \theta_{23}  \\
				0  &  - \sin \theta_{23}  &  \cos \theta_{23} 
		\end{pmatrix} \times \vphantom{\raisebox{-5.5ex}{a}} 
		\\
		\times \begin{pmatrix}
				\cos \theta_{13}  &  0  &  \sin \theta_{13} \; e^{-i \, \delta}  \\
				0  &  1  &  0  \\
				- \sin \theta_{13} \; e^{i \, \delta}  &  0  &  \cos \theta_{13}
		\end{pmatrix}  \begin{pmatrix}
				\cos \theta_{12}  &  \sin \theta_{12}  &  0  \\
				- \sin \theta_{12}  &  \cos \theta_{12}  &  0  \\
				0  &  0  &  1
		\end{pmatrix}
\end{multline}

As for Majorana masses, all the matter is to repeat the same algorithm: now we
don't have $\nu_{\mathrm{R}}$'s, so we have one less unitary matrix to play
with; the neutrino mass matrix is determined by the $\xi$ couplings appearing
in \eqref{intro-neutrinos-majorana-masses}, which conform a complex \emph{symmetric}
$3 \times 3$ matrix; and the charged leptons' Yukawas are also involved in the
diagonalisation process, as they also involve the fields $\ell_{\mathrm{L}}$.
That makes 15 moduli and 15 phases in $\xi$ and $Y_e$, and 6 free moduli and 12 
free phases from the unitary transformations of $\ell_{\mathrm{L}}$ and 
$e_{\mathrm{R}}$. As lepton number is broken
due to the terms in \eqref{intro-neutrinos-majorana-masses}, none of the free
phases has to be preserved this time. The counting then yields 9 physical moduli 
(again) and 3 physical phases. The custom parametrisation is identical to that
for Dirac masses, but incorporating the two new phases as the phases of two of
the eigenvalues of the neutrino mass matrix. We call these two phases 
\emph{Majorana phases} and denote them by $\alpha_1$, $\alpha_2$; then we write
\begin{multline} \label{intro-neutrinos-PMNS-Majorana}
	U_{\mathrm{PMNS}}^\mathrm{Majorana} = \begin{pmatrix}
				1  &  0  &  0  \\
				0  &  \cos \theta_{23}  &  \sin \theta_{23}  \\
				0  &  - \sin \theta_{23}  &  \cos \theta_{23} 
		\end{pmatrix} \vphantom{\raisebox{-5.5ex}{a}} \begin{pmatrix}
				\cos \theta_{13}  &  0  &  \sin \theta_{13} \; e^{-i \, \delta}  \\
				0  &  1  &  0  \\
				- \sin \theta_{13} \; e^{i \, \delta}  &  0  &  \cos \theta_{13}
		\end{pmatrix} \times
		\\
		\times   \begin{pmatrix}
				\cos \theta_{12}  &  \sin \theta_{12}  &  0  \\
				- \sin \theta_{12}  &  \cos \theta_{12}  &  0  \\
				0  &  0  &  1
		\end{pmatrix}  \begin{pmatrix}
				e^{i \, \alpha_1 / 2}  &  0  &  0  \\
				0  &  e^{i \, \alpha_2 / 2}  &  0  \\
				0  &  0  &  1
		\end{pmatrix} \, .
\end{multline} 

We conclude this section by indicating an example of mixing pattern that
has gained some popularity in the recent years: the so-called tribimaximal mixing.
The great deal of data which was released in the late 90's and early
2000's hinted a mixing
structure with $\nu_3$ maximally mixed between $\nu_\mu$ and $\nu_\tau$ and
$\nu_2$ maximally mixed among all the flavours. The strict realisation of
this pattern was soon dubbed `tribimaximal mixing' \cite{Harrison:2002er}
due to the maximal `bi'-mixing of $\nu_3$ and the maximal `tri'-mixing of $\nu_2$; 
such realisation implied strong predictions for the angles:
the atmospheric angle was bound to be maximal, $\theta_{23} = 45^\circ$, 
the solar angle was to lead to the trimaximal mixing, $\theta_{12} \simeq 
35.3^\circ$, and the reactor angle was required to vanish. This latter condition
carried one further consequence: the mass-mixing system gained an additional
liberty and the Dirac phase ended up not being physical; it, this way, disappeared 
from the parametrisation. The final PMNS matrix of tribimaximal mixing looked like
\begin{displaymath}
	U_{\mathrm{PMNS}}^\mathrm{tbm} = \begin{pmatrix}
     		\sqrt{\nicefrac{2}{3}} & \vspace{0.2cm} \hspace{0.2cm} 
					\nicefrac{1}{\sqrt{3}} & 0 \hspace{0.2cm} 
			\\
			- \nicefrac{1}{\sqrt{6}} & \vspace{0.2cm} \hspace{0.2cm} 
					\nicefrac{1}{\sqrt{3}} & \nicefrac{1}{\sqrt{2}} \hspace{0.2cm} 
			\\
     		\hspace{0.1cm} \nicefrac{1}{\sqrt{6}} & \vspace{0.2cm} 
					- \nicefrac{1}{\sqrt{3}} & \nicefrac{1}{\sqrt{2}} \hspace{0.2cm}
	\end{pmatrix} \, .
\end{displaymath}
Nowadays tribimaximal mixing is ruled out as a realistic `final-word' description
of the leptonic mixing, as we can appreciate in table 
\ref{tab:intro-neutrino-bounds-and-measurements}, but
it can still be regarded as a fair approximation whose symmetries may hint 
the symmetries of the underlying flavour physics that generates neutrino masses.
As such, it's still worth to be borne in mind.

\section{Our knowledge of the mass sector} \label{sec:knowledge-mass-neutrino}

In this section we will review our knowledge about the neutrino mass and
mixing parameters. This means, as we discussed in the previous section, finding
out
the value of no less than 7 different parameters: three masses, three mixing
angles and one \mbox{phase -- plus two} more phases, were the neutrinos to be 
Majorana fermions. Needless to say, no single experiment can probe the 
whole set of parameters; rather, it's been the patient juxtaposition of
different experimental results what has provided our present knowledge 
about neutrino masses. Neutrino oscillation experiments, for instance,
are well-suited for measuring mixing angles and squared-mass differences between
the various massive neutrinos; they can also provide information about
the Dirac CP-violating phase, but they are not sensitive to the values of the 
masses themselves
or to the Majorana phases. Precise measurements of the $\beta$-decay 
spectrum of certain nuclides can yield information on the absolute scale
of neutrino masses, but they only probe a certain \emph{combination} of 
$m_1$, $m_2$ and $m_3$, not the three of them separately.
Cosmological observables are sensitive to a different combination of the \mbox{masses
--their} \mbox{sum-- and they} offer a fairly good constraint on it, but with the
drawback that it depends somewhat on our knowledge of the 
cosmological parameters.
Neutrinoless double beta decay, finally,
is an excellent probe of the Majorana/Dirac nature of the neutrinos,
but the interpretation of an eventual positive result must be taken with
care: while a positive $0 \nu \beta \beta$ signal immediately implies that
neutrinos are Majorana particles, the decay amplitude may or may not yield
numerical information on neutrino masses, depending on how the decay proceeds
\mbox{internally -- if} it is triggered by heavy, LNV-ing physics the amplitude
will inform about the couplings of such heavy physics, whereas if it's the active
neutrinos who provide the dominant contribution it will measure the $m_{ee}$
element of the mass matrix. Therefore, $0 \nu \beta \beta$ requires a synergy with 
other, possibly high-energy experiments in order to establish precisely what kind of
information we can extract from it.

As a general statement it can be said that the
struggle to measure the absolute masses of the neutrinos is a tough one;
when the day arrives it will require to combine the results of
several different experiments together with the best knowledge we can
gather about mixing angles and phases. Fortunately, the mixing sector
is reasonably well-known even today due to the fantastic precision 
achieved by the oscillation experiments; in fact, at the moment of writing 
only the Dirac
phase and the sign of one mass splitting remain to be measured. Let us 
begin our review without any further introduction by stating 
the results we can derive from neutrino oscillations.

\subsection{Oscillation experiments} \label{sec:oscillation-experiments}

Oscillation experiments probe, as we explained in section 
\ref{sec:neutrino-oscillations},
flavour nonconservation during neutrino free propagation. A review of the 
oscillation mechanism in full detail can be found for instance in 
\cite{Giunti:2000kw,Giunti:2008cf};
here, for the purposes we're interested in, let us consider the oscillation in 
vacuum of two neutrino species:
\begin{equation} \label{intro-oscillations-2species}
	P (\nu_\alpha \rightarrow \nu_\beta ) = \sin^2 (2 \theta_{ij}) \:
			\sin^2 \left( \frac{\Delta m_{ij}^2 \, L}{4 E} \right) \, .
\end{equation}
This expression describes the probability that a neutrino is detected with flavour
$\beta$ after a flight of length $L$ if it had flavour $\alpha$ at the moment
of production and it carries an energy $E$. We assume two mass eigenstates,
labeled as $i$ and $j$, to underlie the two flavour states 
$\alpha$ and $\beta$; 
the expression $\Delta m_{ij}^2$ refers to the difference of the square of
their masses, $\Delta m_{ij}^2 \equiv m_i^2 - m_j^2$; the angle $\theta_{ij}$
describes the mixing of $i$ and $j$ with $\alpha$ and $\beta$:
\begin{equation} \label{intro-oscillations-2species-mix}
	\begin{pmatrix}  \nu_\alpha  \\  \nu_\beta \end{pmatrix} =
			\begin{pmatrix}
				\cos \theta_{ij}  &  \sin \theta_{ij}
				\\
				- \sin \theta_{ij}  &  \cos \theta_{ij} 
			\end{pmatrix}  \begin{pmatrix}
				\nu_i  \\  \nu_j
			\end{pmatrix} \, .
\end{equation}
With these elements, we can read in \eqref{intro-oscillations-2species} two
important properties of oscillations: first, the amount of flavour violation
is directly dependent on the mixing angle, in such a way that if $\theta_{ij}=0$
a $\nu_\alpha$ remains $\nu_\alpha$ all the way \mbox{long -- that is,} 
looking at \eqref{intro-oscillations-2species-mix}: if the flavour and mass 
eigenstates coincide there are no oscillations. Second, the wavelength of
the oscillations depends exclusively on the difference of squared masses and
the energy, in such a way that if $\Delta m_{ij}^2 = 0$ no change in flavour
\mbox{occurs -- that is,} one can only observe oscillations if $m_i$ and $m_j$ are
\emph{different}.

Of course, expressions \eqref{intro-oscillations-2species} and
\eqref{intro-oscillations-2species-mix} do not represent the situation of 
real neutrinos, of which we have at least three species%
\footnote{For instance, two-species oscillations conserve CP, and indeed
there is no CP-violating phase in \eqref{intro-oscillations-2species}.
}, 
but they can help
in getting an idea of what is being observed in the experiments.
Essentially, oscillation experiments count the number of neutrinos of a certain
\mbox{species --sometimes} of several \mbox{species-- coming} from a known source, 
and look for discrepancies with the initial composition of the source. They also
measure the energy of the neutrinos, in order to look for an agreement
with spectra such as \eqref{intro-oscillations-2species}. There are
mainly four relevant sources of neutrinos: the Sun, the atmosphere, 
man-made nuclear reactors and \emph{ad-hoc} neutrino beams produced 
in particle accelerator facilities. Each source is sensitive in its own
particular way to a subset of the mixing parameters. Let us here sketch 
what information can be extracted from each family of experiments. A more
complete review can be found in \cite{GonzalezGarcia:2007ib}.

Solar neutrinos \cite{Antonelli:2012qu} originate in the inner core of the Sun 
as a byproduct of its fusion nuclear reactions \cite{Raffelt:1996wa}; 
significant processes 
are the proton-proton
reaction, $p + p \rightarrow d + e^+ + \nu_e$, which provides most of the solar 
neutrinos but only with low \mbox{energies --up to} \mbox{$400 \; \mathrm{keV}$--, 
and the} boron-8 process,
$\tensor[^8]{\mathrm{B}}{} \rightarrow {}^8\mathrm{Be} + e^+ + \nu_e$, 
which provides
only a small fraction of the total neutrino flux, but with energies 
above a few MeV's, that make them easier to detect. The neutrinos 
produced inside the
core travel through the Sun's plasma and oscillate in this environment. The
presence of matter can affect the oscillation process as described by the
Mikheyev–Smirnov–Wolfenstein \mbox{effect -- \cite{Wolfenstein:1977ue,
Wolfenstein:1979ni,Mikheev:1986wj,Mikheev:1986gs}, see} \cite{Smirnov:2004zv} 
for a review; in the case of the Sun, its density profile marks a change in the
quality of oscillations at $1-3 \; \mathrm{MeV}$: neutrinos
with energies below this \mbox{value --for instance} those coming from the 
$p$-$p$
\mbox{reactions-- oscillate} as they would do in vacuum, resulting in a
conversion of 30\% of the electron neutrinos into the other two species.
For energies above a few MeV, however, the conversion is enhanced by matter effects
and up to 70\% of the $\nu_e$'s are transformed. This satisfactorily accounts
for the discrepancies between theory and experiment that were recorded
during the $20^\mathrm{th}$ century \cite{Davis:1964hf,Fukuda:1996sz,
Hampel:1998xg,Cleveland:1998nv,Fukuda:1998fd,Ahmad:2001an,Abdurashitov:2002nt,
Altmann:2005ix}; the advent of the SNO experiment, which
was capable of measuring $\nu_e$'s as well as the other two flavours, confirmed
this interpretation \cite{Ahmad:2002jz}. Solar neutrino oscillations are 
mainly due to the mixing
between $\nu_1$ and $\nu_2$, and so these experiments allow to measure 
the angle $\theta_{12}$ and the squared-mass splitting $\Delta m_{21}^2$.
However, the uncertainties in our knowledge of the solar properties and
the eventful trip of the neutrinos from \mbox{the Sun --with} oscillations 
in matter along the solar interior, then oscillations in vacuum to reach Earth,
and then again oscillations in matter at least during the \mbox{nighttime--
casted} some degeneracy on the combination of oscillation parameters that 
could account for the observations. It was only by bringing together solar,
atmospheric and reactor results that the correct regime could be
identified (see section 3 of \cite{GonzalezGarcia:2007ib} for a more detailed 
discussion), yielding values of about $\theta_{12} \simeq 35^\circ$ and 
\mbox{$\Delta m_{21}^2 \simeq 10^{-4} \; \mathrm{eV}^2$ -- we present} the
current best values in table 
\ref{tab:intro-neutrino-bounds-and-measurements}.
Note that we have indicated here a full value for $\Delta m_{21}^2$, with a 
well-defined sign, even though the neutrino oscillation probabilities, like
\eqref{intro-oscillations-2species}, are insensitive to the signs of the
squared-mass splittings. The reason is that oscillations in matter, whose
equations we won't quote here, open up
the possibility of distinguishing the sign of the mass splitting; this is
an important feature of matter effects and it is expected to allow in the
near future to measure the sign of the remaining mass splitting.

Atmospheric neutrinos \cite{Kajita:2012vc} are produced several kilometers 
above the ground by 
the collision of incoming cosmic rays; the first cosmic-ray hit produces pions
and kaons,
which eventually decay to muons and, as lepton family number demands, muon 
(anti-)neutrinos; the muons then
subsequently decay and give one more muon antineutrino (neutrino) and one 
electron (anti-)neutrino.
We can summarise thus in a nutshell the most relevant processes for the 
production of 
atmospheric neutrinos; as a result of this scheme, if all the produced
muons get to decay before hitting the ground the flux of atmospheric
muon (anti-)neutrinos is expected to be roughly twice the flux of electron 
(anti-)neutrinos%
\footnote{This happens actually only for low-energy muons. Consequently, neutrinos
with $E_\nu \lesssim 1 \; \mathrm{GeV}$ show a ratio $\phi (\nu_\mu) / \phi(\nu_e)
\sim 2$, but as energy increases and less muons decay during their atmospheric
travel, less electron neutrinos are produced and this ratio also increases.
}.
Experiments to study this source of neutrinos have been 
carried out since the 1960's, but in these first stages no discrepancy was
found between the expected neutrino fluxes and the experimental data
\cite{Reines:1965qk,Achar1965196,Aglietta:1988be,Daum:1994bf}; later,
in the 80's, when the experiments became capable of discriminating between
electron-neutrino and muon-neutrino events, it was noticed that even if
the total flux was consistent with expectations, the \emph{relative} 
number of $\nu_\mu$'s respect to $\nu_e$'s was not in consonance
\cite{BeckerSzendy:1992hq,Fukuda:1994mc}. This was the first hint of
oscillations in atmospheric neutrinos. However, the big power of this
family of experiments is their capacity to observe neutrinos that
have travelled different distances, and so are in different points of
the oscillation curve \eqref{intro-oscillations-2species}. The trick is
that neutrinos can go through the whole planet without impediment, and so
one can select the flight length simply by recording the incidence
angle of the neutrino \mbox{event -- downward-}going neutrinos are to have been 
produced in the atmosphere directly above the experiment and they will have flown
about $15 \; \mathrm{km}$ until they reach the detector; upward-going neutrino
events, however, are to be associated with a neutrino produced at the other
side of the Earth, and they store information about a flight of approximately
$12800 \; \mathrm{km}$. With such a large range of available baselines
atmospheric neutrino experiments are very well suited for testing oscillation
curves, and indeed the first claim of observation of neutrino oscillations
came from Super-Kamiokande \cite{Fukuda:1998mi}; shortly after, other experiments 
confirmed the discovery \cite{Ambrosio:2001je,Sanchez:2003rb}.
Neutrino telescopes, like ANTARES \cite{AdrianMartinez:2012ph} and 
IceCube \cite{Aartsen:2013jza}, can also measure the parameters
of atmospheric oscillation.
Due to the hierarchies in
the mass splittings and the mixing angles, these experiments probe mainly 
the angle $\theta_{23}$ and its associated splitting, $\Delta m_{32}^2$;
for them, values around $\theta_{23} \simeq 45^\circ$ and $\left| \Delta 
m_{32}^2 \right| \simeq 10^{-3} \; \mathrm{eV}$ are found. 
Note that the sign of $\Delta m_{32}^2$ is not yet known. To determine it
we need experiments sensitive both to the atmospheric parameters and to
matter effects, which can discriminate the sign of the splitting. This can
be achieved with atmospheric neutrinos, especially if the experiments attain 
sensitivities in the energy range of 
$1-12 \; \mathrm{GeV}$, where oscillations in the mantle and the core of the 
Earth are enhanced by resonant matter effects \cite{Petcov:2005rv,Mena:2008rh}.
Such searches are already planned and will be carried out in the near future.

Reactor neutrinos \cite{Lasserre:2005qw} are neutrinos produced in nuclear 
reactors due to $\beta$ 
decay processes; they are, therefore, basically electron antineutrinos
with energies of the order of the MeV. With such a composition, reactor neutrino
fluxes resemble somewhat solar neutrinos, only with a different energy spectrum
as they are generated by different nuclear processes. And indeed reactor
neutrinos can yield information about the solar oscillation parameters: if we
write the 3-species electron antineutrino disappearance probability we can 
identify two competing terms,
\begin{multline} \label{intro-oscillations-reactor}
	P (\bar \nu_e \rightarrow \bar \nu_{\mu, \tau} ) = 4 \, \sin^2 \theta_{13}
			\, \cos^2 \theta_{13} \, \sin^2 \left( \frac{L}{4 E} \, \Delta m_{31}^2 
			\right) \: +  
			\\
			+ \: \cos^4 \theta_{13} \, \sin^2 ( 2 \theta_{12} ) \,
			\sin^2 \left( \frac{L}{4 E} \, \Delta m_{21}^2 \right) + \ldots \, ,
\end{multline}
plus other subleading contributions that include the interference of these
two terms and matter \mbox{effects -- one can} check that given our knowledge of 
the mass splittings and the possible baselines available on planet Earth those
contributions are indeed subdominant. Equation \eqref{intro-oscillations-reactor}
displays an interesting situation: the composition of two oscillations with
different charcteristic lengths; one of them is associated with solar oscillations,
and has $L_{\mathrm{char}} = \nicefrac{4 E}{\Delta m_{21}^2} \sim 200 
\; \mathrm{km}$
for the typical energies of reactor neutrinos. The second component is given
by the atmospheric scale, as $\Delta m_{32}^2 \gg \Delta m_{21}^2$ and so 
$\Delta m_{31}^2 \simeq \Delta m_{32}^2$; then, for the first term in 
\eqref{intro-oscillations-reactor} we have $L_{\mathrm{char}} \simeq \nicefrac{4 E}
{\Delta m_{32}^2} \sim 5 \; \mathrm{km}$. 
Therefore, the hierarchies of the neutrino mass splittings allow
to identify two regimes in electron antineutrino oscillations, and that in turn
opens up the possibility of designing two kinds of reactor neutrino experiments:
long-baseline ($L \sim 100 \; \mathrm{km}$) ones, which probe the solar 
parameters, $\theta_{12}$ and $\Delta m_{21}^2$, and short-baseline 
($L \sim 1 \; \mathrm{km}$) facilities, which should be sensitive to the third 
mixing angle, $\theta_{13}$, and the atmospheric scale, $\Delta m_{31}^2 \simeq
\Delta m_{32}^2$. The fact that $\theta_{13}$ is small compared to $\theta_{12}$
further helps to separate the two components in \eqref{intro-oscillations-reactor}.
The first reactor neutrino experiments were carried out with short or
intermediate baselines, and weren't sensitive enough to detect the
small antineutrino disappearance induced by $\theta_{13}$ 
\cite{Zacek:1986cu,Vidyakin:1994ut,Declais:1994su,Apollonio:1999ae,
Piepke:2002ju}. Then there came KamLAND, which operated with longer baselines
and was thus probing the solar neutrino parameters; its achievement was
a big one, because it firmly established the scenario of a large $\theta_{12}$
and $\Delta m_{12}^2$ above $10^{-6} \; \mathrm{eV}^2$ \cite{Eguchi:2002dm}.
Moreover, KamLAND provided definitive evidence that the mechanism underlying
neutrino flavour transformation was indeed oscillations by noticing that
the energy spectrum of the reactor neutrinos was distorted along 
\mbox{propagation
\cite{Araki:2004mb} -- as dif}\-ferent energies result in different survival
probabilities, see \eqref{intro-oscillations-reactor}. Finally, 
a new generation of short-baseline experiments has very recently provided
strong proof of a nonzero $\theta_{13}$ \cite{Abe:2011fz,An:2012eh,Ahn:2012nd}, 
opening thus the door to CP violation in the neutrino sector (see section 
\ref{sec:neutrinos-writing-masses-mixings}).
Due to the sensitivity of reactor experiments to $\theta_{13}$ it has grown
increasingly common to dub it the `reactor angle'; its value lies,
as of today, around $\theta_{13} \simeq 10^\circ$.

Finally, accelerator neutrinos are muon neutrinos produced in an accelerator
facility by colliding particles against a fixed target \cite{Feldman:2012qt}. 
The procedure has a 
similar effect to the cosmic ray collisions that generate atmospheric neutrinos:
pions are produced, and then these decay to muons and muon (anti-)neutrinos. The
difference is that for these experiments the muons are stopped as efficiently as 
possible, so that the electron neutrinos resulting from their decays are emitted in
random directions and the final neutrino beam contains mostly neutrinos of the
muon family. A detector is then located at some distance in the trajectory of the 
beam; often the experiments comprise two dectectors: one ``near'' detector 
close to the collision point whose function is to characterise the initial 
composition of the beam, and one ``far'' detector that allows to assess how
the beam has changed during its travel.
Such good knowledge of the beam properties allows to examine various channels 
of oscillation: the dominant $\nu_\mu 
\rightarrow \nu_\tau$ oscillations are usually searched for in terms of
``$\nu_\mu$ disappearance'', as most beams are not energetic enough to 
produce a $\tau$ (but see \cite{Agafonova:2012zz,DiMarco:2013qxa});
these searches, so, investigate the same physical process as atmospheric neutrino
experiments and provide independent measurements for the atmospheric oscillation
parameters \cite{Ahn:2006zza,deJong:2013/04/19pya}. Another interesting
process is the appearance of electron neutrinos out of the muon species, 
$\nu_\mu \rightarrow \nu_e$, a subdominant oscillation that is sensitive to 
the reactor angle and to the CP phase $\delta$; 
this channel has also been intensely investigated
\cite{Yamamoto:2006ty,Adamson:2011qu,Catanesi:2013fxa,Agafonova:2013xsk},
but with less sensitivity than the reactor experiments and with no luck yet
in what concerns CP violation. Future accelerator experiments will further
probe the $\nu_\mu \rightarrow \nu_e$ channel and plan to use matter effects
in the Earth's crust to elucidate the sign of the atmospheric mass splitting
\cite{Shanahan:2011zz}.

Accelerator neutrino beams have been also
used to design short-baseline experiments, located at the order of hundred
meters away from the collision point; these are sensitive to large 
mass-squared differences, and have been running to look for exotic oscillations
into more massive neutrino species, possibly sterile. The LSND experiment
reported a positive result in this direction \cite{Aguilar:2001ty}, but
subsequent independent searches seem to disfavour this claim 
\cite{Armbruster:2002mp,AguilarArevalo:2007it}.
The matter stays controversial as of autumn 2013 \cite{VandeWater:2012tsa}.

\begin{figure}[tb]
	\centering
	\includegraphics[width=0.7\textwidth]{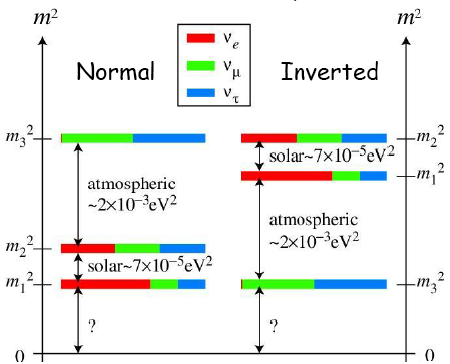}
	\caption{This picture displays the two possible mass orderings depending
			 on the sign of the atmospheric mass \mbox{splitting -- either} 
			 $\Delta m_{32}^2$ or $\Delta m_{31}^2$, for this matter.
			 In normal hierarchy the two mass eigenstates separated by the solar
			 mass splitting are lighter than the third one, whereas in inverted
			 hierarchy it is the other way around. Note that the sign of the
			 solar mass splitting is known, and consequently we know that $\nu_1$
			 is lighter that $\nu_2$. The colours represent the flavour content
			 of each mass eigenstate; the approximate tribimaximal mixing can
			 be appreciated in $\nu_2$ and $\nu_3$. The figure has been
			 obtained from \cite{King:2013eh}. 
			 } \label{fig:intro-neutrino-mass-hierarchy}
\end{figure}

Such a fantastic battery of experiments yields a healthy flux of independent
and cross-checkable results which can be combined in order to attain better
statistical significance. It is actually very common to do so, and the so-called
`global fits' have become the standard source of knowledge about neutrino 
parameters.
Table \ref{tab:intro-neutrino-bounds-and-measurements}, which summarises our 
latest information about neutrino 
masses and mixings, refers its oscillation data to the global 
fit in \cite{Tortola:2012te}. Some features can be noted from
these data: first, they are not anymore compatible with tribimaximal 
mixing; $\theta_{13}$ has been measured to be nonzero and $\theta_{23}$ shows
some preference for non-maximal values. The Dirac phase, thus, is physical, and
the experiments aiming to measure it now know that there's something to
measure indeed.

One more feature is noticeable from table 
\ref{tab:intro-neutrino-bounds-and-measurements}: as we already anticipated,
we have no clue on the sign of the atmospheric mass splitting, $\Delta m_{31}^2$
(or $\Delta m_{32}^2$, if we prefer). This means that we don't fully know the
ordering of the mass eigenstates. The reader may remember that we commented
in section \ref{sec:neutrinos-writing-masses-mixings} that the numbers in 
$m_1$, $m_2$, $m_3$ were mere
labels and they didn't imply an ordering; knowing the differences between them
we know the gaps that separate them, but not knowing the sign of those differences
we are left with a handful of possible ways to arrange the masses and the gaps.
Fortunately, we do know the sign of $\Delta m_{21}^2$, so only a twofold degeneracy
remains to be resolved. In figure \ref{fig:intro-neutrino-mass-hierarchy} we
depict the two possibilities; the configuration in which $m_1 < m_2 < m_3$ is 
conventionally called ``normal hierarchy'', whereas the second configuration,
that yields $m_3 < m_1 < m_2$ is dubbed ``inverted'' hierarchy. Note that
in table \ref{tab:intro-neutrino-bounds-and-measurements} some values of the 
best fit can be different if one or the other hierarchy is assumed.

\begin{table}[p]
	\centering
	\begin{tabular}{| c | c | c }
		\cline{1-2}
		\multicolumn{2}{| c |}{\textbf{Oscillation parameters} 
			\cite{Tortola:2012te}}  &  
			\vphantom{\raisebox{1.5ex}{a}} \vphantom{\raisebox{-1.5ex}{a}} 
		\\ \hhline{==~}
		$\Delta m_{21}^2$  &  $\left( 7.62 \pm 0.19 \right) \times 10^{-5} 
			\; \mathrm{eV}^2$  &  
			\vphantom{\raisebox{1.7ex}{a}} \vphantom{\raisebox{-1.3ex}{a}}
		\\ \cline{1-2} 
		\multirow{2}{*}{$\left| \Delta m_{31}^2 \right|$}  &  $\left( 2.55 \, 
			{+ 0.06 \atop
			- 0.09} \right) \times 10^{-3} \; \mathrm{eV}^2$  & {\small NH}
			\vphantom{\raisebox{1.8ex}{a}} \vphantom{\raisebox{-1.5ex}{a}}
		\\ \cline{2-2}
		  &  $\left( 2.43 \, {+ 0.07 \atop - 0.06} \right) \times 10^{-3} \; 
		  	\mathrm{eV}^2$  &  {\small IH}
			\vphantom{\raisebox{1.8ex}{a}} \vphantom{\raisebox{-1.5ex}{a}}
		\\ \cline{1-2}
		$\sin^2 \theta_{12}$  &  $0.320 \, {+ 0.016 \atop - 0.017}$  &  
			\vphantom{\raisebox{1.8ex}{a}} \vphantom{\raisebox{-1.5ex}{a}}
		\\ \cline{1-2}
		\multirow{3}{*}{\raisebox{-2ex}{$\sin^2 \theta_{23}$}}  &  $0.613 
			\, {+ 0.022 \atop 
			- 0.040}$  &  \multirow{2}{*}{\small \raisebox{-1ex}{NH}}
			\vphantom{\raisebox{1.8ex}{a}} \vphantom{\raisebox{-1.5ex}{a}}
	 	\\ \cline{2-2}
			&  $0.43 \pm 0.03$  &
			\vphantom{\raisebox{1.7ex}{a}} \vphantom{\raisebox{-1.3ex}{a}}
		\\ \cline{2-2}
			&  $0.600 \pm 0.03$  &  {\small IH}
			\vphantom{\raisebox{1.7ex}{a}} \vphantom{\raisebox{-1.3ex}{a}}
		\\ \cline{1-2}
		$\sin^2 \theta_{13}$  &  $0.025 \pm 0.003$  & 
			\vphantom{\raisebox{1.7ex}{a}} \vphantom{\raisebox{-1.3ex}{a}}
		\\ \cline{1-2}
		$\delta$  &  $0 - 2 \pi$  &
			\vphantom{\raisebox{1.7ex}{a}} \vphantom{\raisebox{-1.3ex}{a}}
		\\ \hhline{==~}
		\multicolumn{2}{| c |}{\textbf{Tritium $\beta$ decay}}  &  
			\vphantom{\raisebox{1.5ex}{a}} \vphantom{\raisebox{-1.5ex}{a}}
		\\ \cline{1-2}
		$m_\beta$  &  $< 2 \; \mathrm{eV}$ \quad 95\% CL &  \cite{Aseev:2011dq}
			\vphantom{\raisebox{1.7ex}{a}} \vphantom{\raisebox{-1.3ex}{a}}
		\\ \hhline{==~}
		\multicolumn{2}{| c |}{\textbf{Cosmological constraints}} & 
			\vphantom{\raisebox{1.5ex}{a}} \vphantom{\raisebox{-1.5ex}{a}}
		\\ \cline{1-2}
		$\sum_i m_i$  &  $< 0.23 \; \mathrm{eV}$ \quad 95\% CL & \cite{Ade:2013zuv}
			\vphantom{\raisebox{1.7ex}{a}} \vphantom{\raisebox{-1.3ex}{a}}
		\\ \hhline{==~}
		\multicolumn{2}{| c |}{\textbf{$0 \nu \beta \beta$ decay}}  &
			\vphantom{\raisebox{1.5ex}{a}} \vphantom{\raisebox{-1.5ex}{a}}
		\\ \cline{1-2}
		$m_{\beta \beta}$  &  $< 0.14 - 0.38 \; \mathrm{eV}$ \quad 90\% CL &  
			\cite{Auger:2012ar}
			\vphantom{\raisebox{1.7ex}{a}} \vphantom{\raisebox{-1.3ex}{a}}
		\\ \cline{1-2}
		$m_{\beta \beta}$  &  $< 0.2 - 0.4 \; \mathrm{eV}$ \quad 90\% CL & 
			\cite{Agostini:2013mzu}
			\vphantom{\raisebox{1.7ex}{a}} \vphantom{\raisebox{-1.3ex}{a}}
		\\ \cline{1-2}
	\end{tabular}
	\caption{A compilation of the present knowledge on the neutrino mass and mixing 
			parameters; each class of measurements is discussed in the
			corresponding section -- see \ref{sec:oscillation-experiments} for
			oscillations, \ref{sec:tritium-beta-decay} for $\tensor[^3]{\mathrm{H}}
			{}$ $\beta$ decay, \ref{sec:neutrinos-cosmology} for cosmology and
			\ref{sec:neutrinos-0nu2beta} for $0 \nu \beta \beta$.
			The displayed 
			uncertainties correspond to the $1 \sigma$ range.
			Note that the oscillation parameters are extracted from a global fit
			and in some cases the fit yields slightly different central 
			values for normal and inverted hierarchy. For the case of $\theta_{23}$
			in NH two different minima are obtained with similar statistical 
			significance.
			} \label{tab:intro-neutrino-bounds-and-measurements}
\end{table}

\subsection{Tritium beta decay and the magnitude of neutrino masses}
									  \label{sec:tritium-beta-decay}

Oscillation experiments allow to measure mixing angles and mass-squared 
differences, but offer no information on the absolute scale of the masses.
The lightest mass eigenstate could lie anywhere from zero to several
tenths of electronvolt%
\footnote{See section \ref{sec:neutrinos-cosmology} for the rationale of 
this upper bound.
}. 
A different experimental approach is needed in order to reveal the absolute
magnitude of the masses; unfortunately, we don't know yet of a procedure
to probe independently the values of $m_1$, $m_2$ and $m_3$
\cite{Drexlin:2013lha}. Among the
several possibilities, direct observation in $\beta$ decay spectra may be the 
closest to what we would naively
call ``measuring a mass''. Its main virtues are the fact that it involves very 
well-known physics, a reasonable independence from other measurements, and 
a low level of model-theoretical noise. The idea is simple: in a $\beta$ decay
process, a neutron is transformed into a proton by emission of an electron and
an electron antineutrino, $n \rightarrow p + e^- + \bar \nu_e$. The neutrino is
elusive, and only very rarely we would be able to detect it; the electron, however,
is easily traceable, and we can measure its energy and momentum. The available
energy for the proton-electron-antineutrino system is 
$Q = m_n - m_p - m_e - m_\nu$,
and it has to be shared among the three particles in a way that respects 
momentum conservation; as this share is not unique there are many energies 
available to the electron, and when we measure the energy of outgoing electrons 
from $\beta$ decay we observe a continuous spectrum. The spectrum has a lower cut
at \mbox{zero --for the} electrons emitted \mbox{at rest-- and an} upper cut 
\mbox{at $Q/2$ -- when} it is the proton or the neutrino who have been emitted at 
rest, and then the electron takes half the available energy and the remaining 
particle takes the other half. Thus, measuring the upper end point of the $\beta$
spectrum is as much as measuring $Q$, and if we know very accurately the masses
of the involved particles, we can measure the neutrino mass.

This far for the basics of the process. Of course a practical realisation of this
simple idea requires coping with many details and experimental challenges, as we
can read in \cite{Drexlin:2013lha}. 
We will be here interested in one of these details:
the fact that we aim to measure the mass of a $\bar \nu_e$. As we know, the 
flavour states don't have a definite mass; they are, rather, a superposition of
states with definite mass. For the electron (anti)neutrino, by looking at
\eqref{intro-neutrinos-PMNS-Dirac}, we see that
\begin{displaymath}
	\bar \nu_e = \cos \theta_{12} \cos \theta_{13} \, \bar \nu_1 + 
			\sin \theta_{12} \cos \theta_{13} \, \bar \nu_2 + \sin \theta_{13} \,
			e^{-i \delta} \, \bar \nu_3 \, .
\end{displaymath}
If we could just take a $\nu_e$ and measure its mass we would obtain as quantum
mechanics prescribes: a fraction $\cos^2 \theta_{12} \cos^2 \theta_{13}$ of the
measurements would yield $m_1$, a fraction $\sin^2 \theta_{12} \cos^2 \theta_{13}$
would yield $m_2$, and a fraction $\sin^2 \theta_{13}$, $m_3$. However, our
experimental setups are not yet sensitive enough to resolve $m_1$, $m_2$ and
$m_3$, so what we observe in the end is the \emph{average} of these three 
values, each weighed by its corresponding probability. We can regard this
observable as an effective mass; we denote it by $m_\beta$, and write
\begin{displaymath}
	m_\beta^2 = \cos^2 \theta_{12} \cos^2 \theta_{13} \, m_1^2 + 
			\sin^2 \theta_{12} \cos^2 \theta_{13} \, m_2^2 + \sin^2 \theta_{13} \,
			m_3^2 \, .
\end{displaymath}

The current best measurement of the effective electron neutrino mass was
carried out by the Troitsk experiment \cite{Aseev:2011dq}, not measuring neutron
\mbox{decay --this can be} cumbersome, as neutrons are difficult to store and their
half-life does not exceed \mbox{15 minutes--, but} $\beta$ decay of tritium 
($\tensor[^3]{\mathrm{H}}{}$). 
Their measurement could not distinguish the mass of the
neutrino but provided a bound, $m_\beta < 2 \; \mathrm{eV}$ at 95\% CL.

\subsection{Cosmological bounds} \label{sec:neutrinos-cosmology}

Neutrinos are abundant particles in the universe; the cosmic background of
neutrinos, though not yet directly observed, is expected to yield a number
density of 339 neutrinos/$\mathrm{cm}^3$, a figure comparable with that of
the photons of the cosmic microwave background. 
The cosmological observables, such as the
temperature anisotropies of the CMB or the matter power spectrum at large
scales, have been shaped by a long evolution of billions of years dominated
by gravitational interactions; the fact that neutrinos possess a mass is bound
to affect these observables, and they provide an independent way to 
probe their values. The analysis of cosmological observables usually implies
assuming a model for the evolution of the cosmos and comparing the observational
data with the results of \emph{in silico} simulations of the evolution of the
observables.
The details of these calculations, 
as well as the many subtleties that appear while examining each particular
observable,
fall out of the scope of this work; 
they can be consulted in specialised texts such as 
\cite{Lesgourgues:2012uu,Pastor:2013book}. 

Two features, however, of this family of constraints are worth being pointed out.
The first concerns the way in which cosmology is sensitive to neutrino masses: as
the cosmic neutrinos were produced during an epoch when the three leptonic families
were identical to a good \mbox{approximation --the energies} were so high that all 
lepton species could be regarded as \mbox{massless--, the back}\-ground neutrinos 
are evenly distributed among flavours; therefore, there must be also even proportions
of each mass species. As a consequence, the cosmological observables cannot 
distinguish
among the various neutrino masses: they are only sensitive to their total 
gravitational contribution, to the sum of the masses. It is a generic fact that
cosmological considerations only yield \mbox{bounds --or, eventual}\-ly, 
\mbox{a measurement-- for $\sum_i m_i$.}

The second aspect we would like to remark is the unavoidable parameter degeneracy
in cosmological arguments. 
The evolution of the cosmos is a complex phenomenon that 
involves various mechanisms and is driven by several sources. 
A certain observable effect of neutrino masses may in many cases
be mimicked by a change in other parameters of the cosmologicaol model. 
Some effects can be provided either
by neutrino masses or by some sort of nonstandard physics. The specific figures, 
finally, of a given bound may change 
depending on which parameters are allowed to vary and which are kept fixed 
when fitting the observed data.
These degeneracies are due in part to the complexity of cosmic dynamics, but
also
to the fact that we cannot \emph{experiment} with the cosmos, but only make 
observations and try to figure out how these observables came to be the way they
are. In the end, what we have is observable quantities and parameters, such
as neutrino masses, whose values we would like to extract from the observables;
some parameters may be derived with reasonable precision from a single
observable, but others can only be determined with precision when other parameters
have been fixed, so one needs a family of observables to do the job. The sum
of the masses of the neutrinos is a good example of this; it can be
probed quite satisfactorily just from the cosmic 
microwave background if it is above 1 eV, but the CMB
loses sensitivity below this level and one needs more observables to increase
the precision. We can see this effect in the
latest CMB data released by the Planck Collaboration 
\cite{Ade:2013zuv}: the reunion of a great deal of data about the CMB
by several \mbox{experiments --Planck} power spectrum plus polarisation 
information from WMAP plus information on high CMB multipoles
from the Atacama Cosmology Telescope and the South Pole \mbox{Telescope-- yields}
a bound on the sum of neutrino masses of
$\sum_i m_i < 0.66 \; \mathrm{eV}$ at 95\% CL; a good limit, which is tightened,
however, by a factor of three when we include in the analysis data on baryon
acoustic oscillations that allow to constrain the cosmological evolution
parameters.
The joint bound with CMB and BAO data reported in \cite{Ade:2013zuv} is
$\sum_i m_i < 0.23 \; \mathrm{eV}$, again at 95\% CL. 
This is considered the
state-of-the-art bound as of today, and as such we quote it in table
\ref{tab:intro-neutrino-bounds-and-measurements}. Future data releases by Planck,
such as information on the CMB polarisation, will probably improve this bound, but
it is remarkable that cosmological considerations begin to approach the inverted
hierarchy level, which lies around $\sum_i m_i \sim 0.1 \; \mathrm{eV}$.

\subsection{Neutrinoless double beta decay} \label{sec:neutrinos-0nu2beta}

Neutrinoless double beta decay is currently the most sensitive probe
to the possible Majorana nature of neutrinos. It is a rare nuclear decay process
that essentially consists of two simultaneous $\beta$ decays occurring to the
same nuclide with no outgoing neutrinos:
\begin{equation} \label{intro-0nu2beta-process}
	\tensor*[^A_Z]{\mathrm{X}}{} \longrightarrow  \tensor*[^A_{Z+2}]{\mathrm{Y}}{}
			+ 2 \, e^- \, ;
\end{equation}
as such, $0 \nu \beta \beta$ decay violates lepton number conservation by creating
two units of lepton number. 
A related but lepton-number-conserving process,
\begin{displaymath} 
	\tensor*[^A_Z]{\mathrm{X}}{} \longrightarrow  \tensor*[^A_{Z+2}]{\mathrm{Y}}{}
			+ 2 \, e^- + 2 \, \bar \nu \, ,
\end{displaymath}
dubbed simply `double beta decay', or in short $2 \nu \beta \beta$, occurs
with both Dirac and Majorana neutrinos, and it has already been observed
in about a dozen of nuclides \cite{Barabash:2005sh,Ackerman:2011gz}.
$2 \nu \beta \beta$ is a decay mode with very long 
half-life, about $10^{20}$ years, and so it is only relevant
for nuclides which have all other channels \mbox{closed -- the}
paradigmatic case is that of nuclides whose $\beta$-decay daughter has a 
higher energy than themselves but have an energetically available daughter at two 
$\beta$-decay steps.

$0 \nu \beta \beta$ was initially proposed as a means to enhance the lifetimes
of $2 \nu \beta \beta$ \cite{Furry:1939qr}: as the phase space of two particles
is less suppressed than that of four particles, $0 \nu \beta \beta$ was
expected to show higher rates. Nowadays we know that the possible 
sources of lepton number \mbox{violation --say,} \mbox{neutrino masses-- are also}
very suppressed, and $0 \nu \beta \beta$ is known to present even longer
half-lifetimes, were it to occur. But still, it is one of our best probes
of lepton number violation, and an independent window into the properties 
of the neutrino mass sector; as such, it retains a good deal of interest.
Let us briefly%
\footnote{For more extensive reviews see references \cite{Avignone:2007fu,
GomezCadenas:2011it,Bilenky:2012qi,Deppisch:2012nb}.
}
describe how $0 \nu \beta \beta$ can be used to investigate
Majorana neutrino masses.

\begin{figure}[tb]
	\centering
	\includegraphics[width=0.6\textwidth]{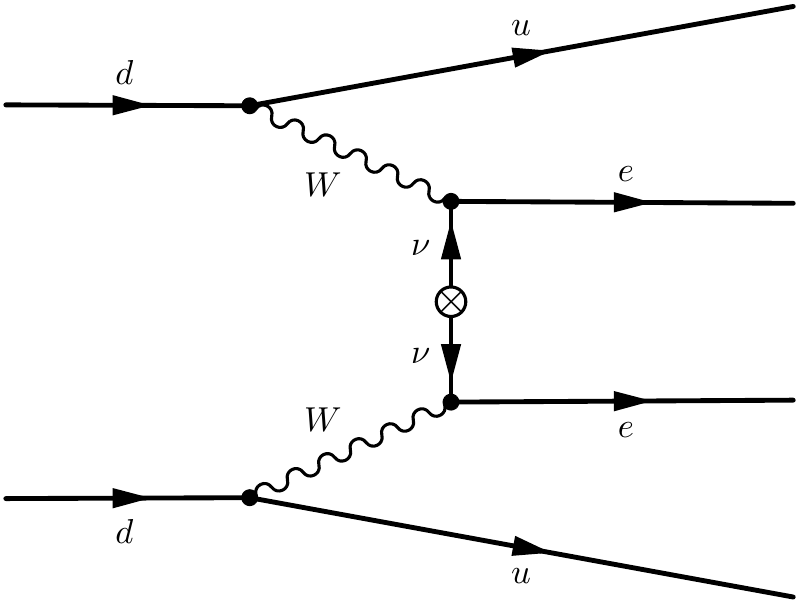}
	\caption{$0 \nu \beta \beta$ mediated by neutrino masses. The diagram displays
			 the process in terms of its fundamental \mbox{participants -- quarks,}
			 instead of nucleons or nuclei. The crossed circle represents a 
			 Majorana mass insertion necessary to change the helicity of the
			 neutrino and break lepton number.
			} \label{fig:intro-0nu2beta-mnu}
\end{figure}

\begin{figure}[p]
	\centering
	\raisebox{4.5cm}{\emph{a)}}
 	\includegraphics[width=0.5\textwidth]{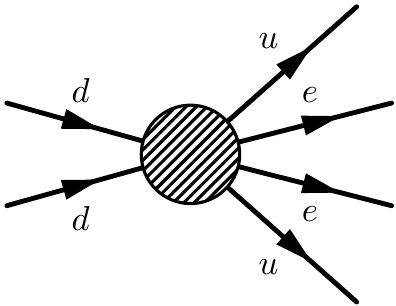}

	\vspace{0.5cm}
	\raisebox{4cm}{\emph{b)}}
 	\includegraphics[width=0.7\textwidth]{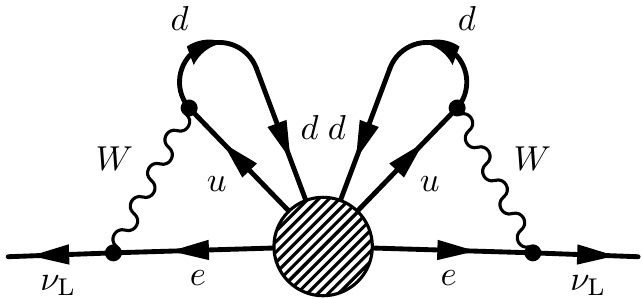}
	\caption{A diagrammatic representation of the Schechter-Valle theorem.
			 In diagram \emph{a)}, the most general form of generating $0 \nu 
			 \beta \beta$ is presented; the blob may be expanded in terms of any
			 lepton-number-violating physics that can accommodate two
			 ingoing $d$ quarks and two outgoing $u$ quarks and electrons.
			 In diagram \emph{b)} we show how the blob effective interaction
			 necessarily leads to the generation of a Majorana component for
			 the neutrino mass. This component can be small, as it's suppressed
			 by at least four loops, but the underlying theory \mbox{may --and in}
			 many cases \mbox{will-- produce} Majorana neutrino masses through 
			 other, less-suppressed processes.
			} \label{fig:intro-0nu2beta-Schechter-Valle}
\end{figure}

Neutrinoless double beta decay has been linked to neutrino masses right from
its first appearance in the scientific literature. It was initially conceived
to be mediated by lepton-number-violating neutrinos through the process depicted
in figure \ref{fig:intro-0nu2beta-mnu}, with two neutrons becoming two protons
by means of charged-current interactions. However, it is clear from 
\eqref{intro-0nu2beta-process} that it can be mediated by any interaction
that transforms two neutrons into protons and yields two electrons; the diagram
in figure \ref{fig:intro-0nu2beta-Schechter-Valle}\emph{a)} displays the most general
form of the interaction in terms of the fundamental fields of the Standard Model. 
One would say, then, that $0 \nu \beta \beta$ may or may not be related to
neutrino masses: inside the blob of figure 
\ref{fig:intro-0nu2beta-Schechter-Valle}\emph{a)}
there can be Majorana neutrinos and $W$'s, but also exotic new scalars, or 
grand-unifying fields. As far as figure 
\ref{fig:intro-0nu2beta-Schechter-Valle}\emph{a)}
is concerned, the neutrinos might well be Dirac particles and still we could see
$0 \nu \beta \beta$ if it's induced by a separate sector of lepton-number-violating
new physics. 

Wrong. A simple argument shows us so: if the interaction in 
figure \ref{fig:intro-0nu2beta-Schechter-Valle}\emph{a)} exists, we can always 
construct a Majorana mass term for the \mbox{neutrinos --at least} for the electron 
\mbox{neutrino-- by inserting} two $W$'s and closing the remaining legs in a
four-loop diagram; we can see it in figure 
\ref{fig:intro-0nu2beta-Schechter-Valle}\emph{b)}. This diagram will not be calculable 
in the effective theory, without knowledge of what's inside the blob, but once
we provide a renormalisable model that realises the blob it will become a 
perfectly regular diagram that provides a Majorana component for neutrino masses.
Indeed it won't be a large \mbox{Majorana mass -- after all,} it's at least four-loop
suppressed; but from that moment on, electron neutrinos will be Majorana particles
and will violate lepton number, if only by a small amount%
\footnote{Actually, in most models with lepton number violation Majorana neutrino
masses are generated much before the four-loop level. The diagram in figure
\ref{fig:intro-0nu2beta-Schechter-Valle}\emph{b)} will usually be just a 
renormalisation to the leading tree-level or one-loop contribution, and the 
masses of the neutrinos may be dominantly lepton-number-violating.
}. 
So, the correct
relationship between $0 \nu \beta \beta$ and neutrino masses is: if $0 \nu \beta 
\beta$ is observed, it may or may not be induced by Majorana neutrino masses, 
but neutrino masses \emph{will be} Majorana, though the lepton number violation
may be small. This result is known as the Schechter-Valle theorem, after the
authors who first figured it out \cite{Schechter:1981bd}.

Up to this point we have only discussed qualitative considerations about the
implications of $0 \nu \beta \beta$ regarding the nature of neutrinos. 
Let us now get more quantitative and see
how a measurement can inform us about the amount of lepton number violation
and about neutrino masses. The calculation of the width of a process like
\eqref{intro-0nu2beta-process} is in general complicated, as it involves 
nuclei, whose dynamics is not absolutely known. For this reason it's usual
to separate the width of a $0 \nu \beta \beta$ process in three parts:
\begin{equation} \label{intro-0nu2beta-halflifetime}
	\left[ T_{1/2}^{\, 0 \nu} \right]^{-1} = G_{0 \nu} \, \left| M_{0 \nu} \right|^2
			\, | \varepsilon |^2 \, ,
\end{equation}
where $G_{0 \nu}$ is a phase space factor that depends on the energy released
in the decay and the atomic number of the parent nucleus, $M_{0 \nu}$ is 
the nuclear matrix element between the initial and final nuclear states,
and $\varepsilon$ is the parameter or combination of parameters which describe
lepton number violation;
note, therefore, that ``$\varepsilon$'' must be understood as something of a
symbolic notation, and its concrete form can only be fixed when something about 
the underlying source of $0 \nu \beta \beta$ is specified.

Of these three elements, the phase space factors are the least problematic:
they don't depend on the new physics involved in the process, but just
on kinematic considerations and the electric charge distribution of the
nucleus; as such, they can be calculated with a reasonable precision and
don't introduce much uncertainty in $T_{1/2}^{\, 0 \nu}$. 
Unfortunately, we cannot say the same about the nuclear matrix elements:
their values can't be accessed through \mbox{experiment --that is,} through
other than a $0 \nu \beta \beta$ \mbox{experiment--, and they} must be evaluated
theoretically; but the calculation involves difficult many-body dynamics,
and it can only be attacked by means of models that attempt to capture the
nuclear structure and interaction. Popular models include the Interacting
Shell Model and the Quasiparticle Random Phase Approximation. The different
approaches essentially agree, but only within a range of a factor 2 or 3;
thus, the calculation of nuclear matrix elements becomes the main source
of uncertainty when interpreting the results of a $0 \nu \beta \beta$ experiment.
The interested reader may find more details on nuclear models and the calculation
of nuclear matrix elements for instance in \cite{Avignone:2007fu,
GomezCadenas:2011it} and references therein.

As for the third element, $\varepsilon$, it is the \emph{raison d'être} of
neutrinoless double beta decay experiments. If we see a positive $0 \nu \beta \beta$
signal it means that $\varepsilon$ is not zero, and thus we have proven that
lepton number is violated and, by the Schechter-Valle argument, that neutrinos
are Majorana particles. Now the thing is that the most general form of the
$0 \nu \beta \beta$ process is that of figure 
\ref{fig:intro-0nu2beta-Schechter-Valle}\emph{a)}, which \mbox{shows 
--neces}\-\mbox{sarily-- nothing} about the internal happenings of the process;
a means should be found to classify and parametrise the various possible
contributions. We find particularly useful the approach of references 
\cite{Pas:1999fc,Pas:2000vn}, where the new physics contributions are classified
according to the Lorentz structure of the resulting six-fermion effective vertex.
The parameter $\varepsilon$ is identified, for each different Lorentz arrangement
of the three fermionic bilinears, with the coupling of the corresponding 
effective operator, and bounds are derived for each coupling or for combinations
of them. This procedure allows to set bounds on large classes of models: for each
particular model one has just to identify the leading six-fermion operators
and express its couplings in terms of the model parameters, which allows to
translate the general bounds into limits specific for that model.

Let us now discuss in more detail the classic case of $0 \nu \beta \beta$ induced 
by Majorana neutrino masses. It is depicted in figure \ref{fig:intro-0nu2beta-mnu},
where we identify the lepton-number-violating element with a fermionic bilinear
that couples an electron neutrino with an electron antineutrino, that is to say,
a Majorana mass and in particular the element $m_{ee}$ of a possible neutrino
mass \mbox{matrix -- indeed,} observing $0 \nu \beta \beta$ only informs us about
violation of lepton number in the electron family, but we do know that $\nu_e$ is
not a mass eigenstate and its mixings are large, so LNV will be right away
transmitted to the other families.
Going back to equation \eqref{intro-0nu2beta-halflifetime}, we have just 
determined that for neutrino-mediated $0 \nu \beta \beta$ we have 
$\varepsilon = m_{ee}$. Let us then examine which knowledge $m_{ee}$ offers 
about the neutrino mass and mixing parameters; we have identified in equation 
\eqref{intro-neutrinos-PMNS-Majorana} the physical degrees of freedom of the
mixing sector for Majorana neutrinos; by properly convoluting $U_{\mathrm{PMNS}}$ 
with the diagonal neutrino mass matrix we can recover the flavour-basis mass 
matrix:
\begin{displaymath}
	(m_\nu)_{\mathrm{flavour}} = U_{\mathrm{PMNS}} \, \begin{pmatrix}
				m_1  &    &  \\
				  &  m_2  &  \\
				  &    &  m_3
			  \end{pmatrix}  \, U_{\mathrm{PMNS}}^\mathrm{T} \, ,
\end{displaymath}
where we find
\begin{multline*}
	m_{ee} = \cos^2 \theta_{12} \cos^2 \theta_{13} \: e^{i \, \alpha_1} \, m_1 \: +
			\\
			+ \: \sin^2 \theta_{12} \cos^2 \theta_{13} \: e^{i \, \alpha_2} \, m_2
				+ \sin^2 \theta_{13} \: e^{-i \, 2 \delta} \, m_3 \, .
\end{multline*}
Note, however, that for the purpose of neutrinoless double beta decay only the 
modulus of $\varepsilon$ matters, so the relevant $0 \nu \beta \beta$ observable
is finally defined as:
\begin{multline} \label{intro-0nu2beta-mbetabeta}
	m_{\beta \beta} = \left| \cos^2 \theta_{12} \cos^2 \theta_{13} \: e^{i \, 
			(\alpha_1 - 2 \delta)} \, m_1 \right. \: +
			\\
			+ \: \left. \sin^2 \theta_{12} \cos^2 \theta_{13} \: e^{i \, (\alpha_2 - 
				2 \delta)} \, m_2 + \sin^2 \theta_{13} \, m_3 \right| \, ,
\end{multline}
and sometimes dubbed the `effective Majorana mass' of neutrinoless double beta
decay.

Several features are worth being noted about this effective Majorana mass: 
first, it retains much
information about the phases; this far it is the only solid observable we have
that is sensitive to the Majorana phases, even if it is in the convoluted
combination of equation \eqref{intro-0nu2beta-mbetabeta}. Unfortunately, with 
three neutrino species there are two Majorana phases and just \emph{one} observable, 
so even in the case we attain a good knowledge of the masses, mixings, Dirac phase 
and nuclear matrix elements further experimental input would be needed to 
determine unambiguously $\alpha_1$ and $\alpha_2$; clearly, a synergy will be
required, possibly with CP-violating observables, if a positive $0 \nu \beta \beta$
signal is observed.

Second, the phase-dependence of $m_{\beta \beta}$ is good news, but it also opens
the door to undesired scenarios: an un-serendipitous combination of the parameters 
might yield $m_{\beta \beta} = 0$, leading to negative $0 \nu \beta \beta$ results 
even if the neutrinos are Majorana particles. This would require quite an amount
of fine tuning and may be regarded as unnatural, but it can be natural in the
framework of certain neutrino mass models: in chapter \ref{chap:WWee-model} 
we describe
a model which yields a naturally suppressed $m_{ee}$. However, even in this case
neutrinos are bound not to be the only source of lepton number violation: if
Majorana neutrino masses are generated by some high-energy physics these particles
are the ultimate source of LNV and may contribute by themselves to $0 \nu \beta
\beta$, as it is the case in the model of chapter \ref{chap:WWee-model}. All in all,
this reminds us that neutrinoless double beta decay results, either positive 
or null, cannot be taken as the last word about the nature of the neutrino: more
investigation is needed in order to determine the origin of neutrino masses and,
were the day to arrive, of a possible $0 \nu \beta \beta$ signal.

\begin{figure}[p]
	\centering
	\includegraphics[width=0.7\textwidth]{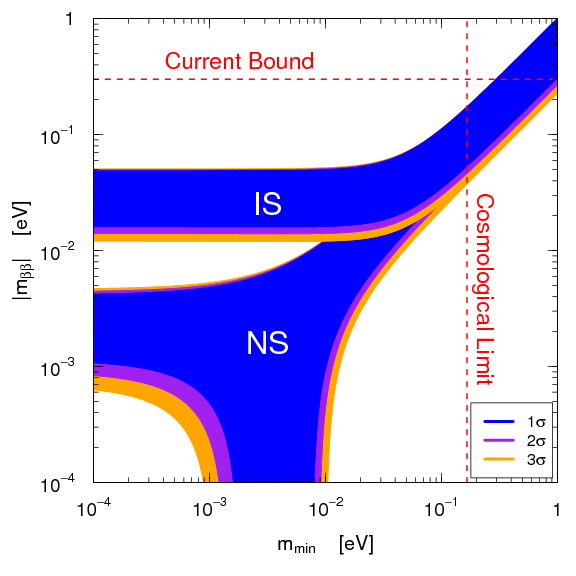}
	\caption{A graph that displays the range of allowed values for the effective
			 Majorana mass in terms of the mass of the lightest \mbox{neutrino -- 
			 either}
			 $m_1$ for normal hierarchy or $m_3$ for inverted hierarchy. The
			 variability along the $m_{\mathrm{min}}$ axis is the result of our
			 lack of knowledge about the absolute scale of neutrino masses,
			 whereas the width of the bands reflects the uncertainties in
			 phases, mixing angles and squared mass splittings. The allowed
			 range is very dependent on the mass hierarchy for small 
			 $m_{\mathrm{min}}$, yielding two bands, marked here ``NS'' for
			 ``normal spectrum'' and ``IS'' for ``inverted spectrum''; as
			 the spectrum grows heavier and the mass splittings become less
			 noticeable the two bands converge. The figure is taken from reference 
			 \cite{Bilenky:2012qi}.
			} \label{fig:intro-0nu2beta-elephant}
\end{figure}

Third, the combination of parameters in equation \eqref{intro-0nu2beta-mbetabeta}
is a rather complicated one, but we have measurements or strong constraints for many
of them and they can be used to evaluate the possible values of $m_{\beta \beta}$.
All mixings and mass splittings are known with a good level of precision, so
the main sources of uncertainty are the absolute scale of the masses, their
\mbox{ordering --either normal} or inverse hierarchy, see \mbox{section 
\ref{sec:oscillation-experiments}--, and} the phases, about which we know nothing.
This evaluation has been done and is well-known; we present it in figure 
\ref{fig:intro-0nu2beta-elephant}. Note the fact that cancellations among the
neutrino mass and mixing parameters can only
yield $m_{\beta \beta} = 0$ for normal hierarchy; note also, therefore, 
that attaining sensitivities
below $10 \; \mathrm{meV}$ without a positive result would mean almost
ruling out Majorana neutrinos with inverted hierarchy%
\footnote{The only way out in this scenario would be that some LNV-ing new
physics exists whose contribution to $0 \nu \beta \beta$ cancels exactly that
of light Majorana neutrinos, but this would require a certain amount of fine
tuning and could be regarded as unnatural.
}, 
but this achievement is
very unlikely, even for the next generation of experiments 
\cite{GomezCadenas:2011it}.

Let us, to conclude, present the most stringent bounds to date on the effective
Majorana mass. For more than a decade, the Heidelberg-Moscow experiment stayed
as the most sensitive experiment with an upper bound of $m_{\beta \beta} < 0.35 
\; \mathrm{eV}$ \cite{KlapdorKleingrothaus:2000sn} derived from the search
of $0 \nu \beta \beta$ in $\tensor[^{76}]{\mathrm{Ge}}{}$. A subset of the 
collaboration claimed a detection \cite{KlapdorKleingrothaus:2001ke} with 
$m_{\beta \beta} = 0.32 \pm 0.03 \; \mathrm{eV}$ \cite{KlapdorKleingrothaus:2006ff}, but this claim stays controversial and the
official position of the collaboration is a no-signal result with
$T_{1/2}^{\, 0 \nu} ( \tensor[^{76}]{\mathrm{Ge}}{} ) > 1.9 \times 10^{25} 
\; \mathrm{yr}$ at 90\% confidence level, from where the cited bound on
$m_{\beta \beta}$ is derived. Recently, the EXO collaboration has
released the first results of their searches in $\tensor[^{136}]{\mathrm{Xe}}{}$,
which offer no evidence of $0 \nu \beta \beta$ and allow to establish
$T_{1/2}^{\, 0 \nu} ( \tensor[^{136}]{\mathrm{Xe}}{} ) > 1.6 \times 10^{25} \; 
\mathrm{yr}$; the collaboration derives from there a bound on the effective neutrino 
Majorana mass of $m_{\beta \beta} < 0.14 - 0.38 \; \mathrm{eV}$
\cite{Auger:2012ar}, with the interval accounting for the uncertainties in
the nuclear matrix elements. Even more recently, the GERDA collaboration has
released their first searches of neutrinoless double beta decay 
\cite{Agostini:2013mzu}; they report a null result with a limit on the half life of 
$\tensor[^{76}]{\mathrm{Ge}}{}$ of $T_{1/2}^{\, 0 \nu} ( 
\tensor[^{76}]{\mathrm{Ge}}{} ) > 3.0 \times 10^{25} \; \mathrm{yr}$ at
90\% CL, which allows them to bound the effective mass at $m_{\beta \beta} <
0.2 - 0.4 \; \mathrm{eV}$. Both \cite{Auger:2012ar} and \cite{Agostini:2013mzu} 
clearly disfavour the claim in \cite{KlapdorKleingrothaus:2001ke}. 

As for the near future, $\tensor[^{136}]{\mathrm{Xe}}{}$ is
expected to draw some attention, with good prospects for EXO, NEXT and KamLAND-Zen;
the CUORE experiment, which investigates $\tensor[^{130}]{\mathrm{Te}}{}$, may
also attain good levels of sensitivity \cite{GomezCadenas:2011it}.

\section{Seesaw mechanism} \label{sec:seesaw}

Let us in this section review briefly the seesaw mechanism, one of the most
popular ways of explaining the smallness of neutrino masses. The basic idea is
simple to grasp: the masses of the neutrinos are very small because they are 
generated via interactions with very heavy particles; the mass-generating
process possesses an internal leg with one of the heavy particles, so the
heavier the new particles the lighter the generated masses.
The classic picture of the seesaw mechanism provides Majorana masses and generates
essentially the Weinberg operator, equation \eqref{intro-neutrinos-majorana-masses}; 
this means that these theories
are also models of violation of lepton number, with the violation originating in
the heavy sector and being transmitted to the neutrinos. Seesaw-like models
are usually classified into three types according to the identity of the
heavy particles: in type I seesaw, the neutrino masses are generated through
interactions with fermionic $SU(2)$ singlets, in type II the mediators are
scalar $SU(2)$ triplets, whereas in type III seesaw it is a heavy fermionic
triplet who aids to generate neutrino masses.

The mechanism was first proposed at the end of the 1970's by several authors;
the community assumes that the general knowledge of the time was ripe and ready 
to yield this idea and it blossomed in several frameworks: grand-unified theories
\cite{GellMann:1980vs,Ramond:1979py}, flavour symmetries 
\cite{Minkowski:1977sc,Yanagida:1979as} and left-right models 
\cite{Mohapatra:1979ia}. From these primeval 
forms the mechanism evolved and was developed into the three classes mentioned
above, inverse seesaw \cite{Wyler:1982dd,Mohapatra:1986aw,Mohapatra:1986bd}, 
Dirac-mass seesaw \cite{SilvaMarcos:1999wy} and many more. We will,
in the remaining of this section, discuss the three types of classic seesaw, which
are also the simplest realisations of the original idea.

\subsection{Type I Seesaw}

\begin{figure}[p]
	\centering
	\raisebox{3.75cm}{\emph{a)}}
	\includegraphics[width=0.5\textwidth]{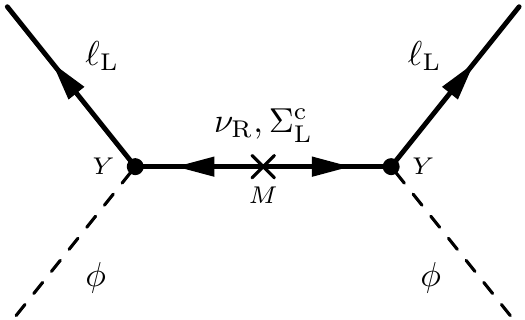}

	\vspace{0.7cm}
	\raisebox{5.3cm}{\emph{b)}} \phantom{a}
	\includegraphics[width=0.35\textwidth]{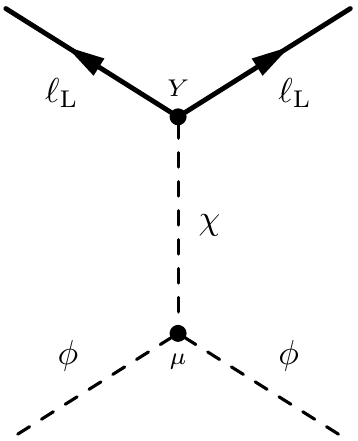}

	\caption{Feynman diagrams giving rise to the Weinberg operator 
			 \eqref{intro-neutrinos-majorana-masses} in the framework
			 of seesaws type I and III (diagram \emph{a)}) and seesaw type II
			 (diagram \emph{b)}). As emphasised in the text, the presence of
			 heavy particles in an internal propagator suppresses the coefficient
			 of the effective operator, and thus the magnitude of the resulting
			 neutrino masses.
			} \label{fig:intro-seesaw-diagrams}
\end{figure}

Type I seesaw was the first variety of seesaw being developed, back in the
end of the 70's \cite{GellMann:1980vs,Ramond:1979py,Minkowski:1977sc,
Yanagida:1979as,Mohapatra:1979ia}. In this class of seesaws
$n$ fermion fields, singlets under $SU(2)$ and with zero 
hypercharge, are added to the Standard Model; such singlets have the quantum 
numbers of right-handed neutrinos, so we denote them by $\nu_{\mathrm{R} \alpha}$. 
Note that to explain the neutrino data, which require at least two massive 
neutrinos, a minimum of two extra singlets are needed.
Having no charges under the SM, Majorana masses
for right-handed neutrinos are allowed by gauge invariance, so the 
new terms in the Lagrangian are:
\begin{equation} \label{intro-seesaw-lagrangian-type-I}
\mathcal{L}_{\nu_\mathrm{R}} = i \, \overline{\nu_\mathrm{R}} \, \gamma^\mu
	\partial_\mu \nu_\mathrm{R} - \left( \frac{1}{2} \:
	\overline{\nu_\mathrm{R}^\mathrm{c}} M \nu_\mathrm{R} +
	\overline{\ell_{\mathrm{L}}} \, Y \,\nu_\mathrm{R} \, \tilde \phi +
	\mathrm{H. c.} \right)  \, ,
\end{equation}
with $M$ a $n \times n$ symmetric matrix in flavour space, $Y$ a general 
$3 \times n$ flavour matrix, and with flavour indices omitted for simplicity.
We can recognise in equation \eqref{intro-seesaw-lagrangian-type-I} the
necessary interactions to generate the Weinberg operator through the diagram that
we display in figure \ref{fig:intro-seesaw-diagrams}\emph{a)}. 
As a $\nu_{\mathrm{R}}$ is propagated
in an internal leg, its mass will suppress the process, realising the seesaw
mechanism. It is also worth noting that the violation of lepton number 
is also originated in the $\nu_{\mathrm{R}}$ sector, in particular in their
Majorana masses, and is transmitted to the active neutrinos through the Weinberg 
operator.

The generation of neutrino masses can also be understood non-diagrammatically:
the interactions in \eqref{intro-seesaw-lagrangian-type-I} generate, upon
spontaneous symmetry \mbox{breaking --that is,} when the Higgs field develops a 
VEV, \mbox{$\langle \phi \rangle = \begin{pmatrix}  0  &  v_\phi  
\end{pmatrix}^\mathrm{T}$--, Dirac} mass terms for the neutrinos that can be 
arranged together with $M$ in a global `neutrino Majorana mass matrix' as follows:
\begin{equation} \label{intro-seesaw-type-I-mass-matrix}
\mathcal{L}_{\nu \: \mathrm{mass}} = - \frac{1}{2} \, 
	\begin{pmatrix} 
		\overline{\nu_\mathrm{L}} & \overline{\nu_\mathrm{R}^\mathrm{c}}
	\end{pmatrix} \, 
	\begin{pmatrix}
		0 & m_\mathrm{D} \\
		m_\mathrm{D}^\mathrm{T} & M
	\end{pmatrix} \,
	\begin{pmatrix}
		\nu_\mathrm{L}^\mathrm{c} \\
		\nu_\mathrm{R}
	\end{pmatrix} + \mathrm{H.c.} \, , 
\end{equation}
where $m_D = Y v_\phi$. The neutrino mass eigenstates will be the eigenvectors
of this matrix; unfortunately, if $m_{\mathrm{D}}$ and $M$ are arbitrary matrices
the diagonalisation is difficult to carry out analytically.
However, if one assumes $M \gg m_\mathrm{D}$, as the diagram in figure 
\ref{fig:intro-seesaw-diagrams}\emph{a)}
requires to obtain small masses for the $\nu_{\mathrm{L}}$'s, 
\eqref{intro-seesaw-type-I-mass-matrix} can be block-diagonalised perturbatively
and very approximate eigenvalues and eigenvectors can be obtained.
In particular, one finds $n$ heavy eigenvectors with \mbox{masses $\sim M$} which are
essentially combinations of the $\nu_{\mathrm{R}}$'s, and three light
eigenvectors which are mostly combinations of the $\nu_{\mathrm{L}}$'s with
a Majorana mass matrix
\begin{displaymath}
	m_\nu \simeq - m_\mathrm{D} \, M^{-1} \, m_\mathrm{D}^\mathrm{T} \, .
\end{displaymath}
This formula is now famous, and it 
naturally explains the smallness of light neutrino masses as a 
consequence of the presence of heavy SM singlet fermions.  

\subsection{Type II Seesaw}  \label{sec:seesaw-typeII}

The type II seesaw \cite{Konetschny:1977bn,Cheng:1980qt,Lazarides:1980nt,
Magg:1980ut,Schechter:1980gr} 
adds to the SM field content only one scalar $SU(2)$ triplet with hypercharge 1, 
that we will denote by $\chi$. It is useful, in order to deduce the interactions
of this triplet with the leptons, to find a notation that displays intuitively 
the composition of the doublet and triplet representations of $SU(2)$. A common
way of doing so is to write the triplets as $2 \times 2$ matrices that act on
vectors in the doublet representation. For the case of our hypercharge-1 $\chi$ 
field this matrix reads
\begin{displaymath}
	\chi =  \begin{pmatrix}
				\chi^+ / \sqrt{2}  &  \chi^{+ +} 
				\\
				\chi^0 & - \chi^+ / \sqrt{2} 
			\end{pmatrix} \, ,
\end{displaymath}
where $\chi^0$, $\chi^+$ and $\chi^{++}$ are the components of the triplet
with definite electric charge. This matrix form of $\chi$ transforms
under $SU(2)$ gauge transformations as $\chi \rightarrow U \chi 
U^\dagger$. It is easy then to see that gauge invariance allows for just one
coupling to the Standard Model fermions: a Yukawa 
interaction among $\chi$ and two lepton doublets,
\begin{equation} \label{intro-seesaw-type-II-yukawas}
	\mathcal{L}_{\chi \: \mathrm{Yukawas}} = - Y_{\alpha \beta} \: 
		\overline{\tilde \ell_\alpha} \chi \ell_\beta + \mathrm{H. c.} \, ,
\end{equation}
which allows to assign lepton number to the triplet, $L (\chi) = -2$. In 
\eqref{intro-seesaw-type-II-yukawas} the flavour indices are displayed
explicitly, and $Y$ is a $3 \times 3$ symmetric matrix.

The terms that need to be added to the SM Lagrangian include these Yukawa
vertices together with several pure-scalar interactions, both $\chi$ self-couplings
and interactions with the Higgs doublet. We display here two of them:
\begin{displaymath}
	\mathcal{L}_\chi = \mathcal{L}_{\chi \: \mathrm{Yukawas}} - m_\chi^2 \, 
			\mathrm{Tr}[ \chi \chi^\dagger] - \left( \mu \, \tilde \phi^\dagger
			\chi^\dagger \phi + \mathrm{H.c.} \right) + \ldots \, ;
\end{displaymath}
both are related to lepton number violation: $m_\chi^2$ would trigger a VEV
for $\chi$, were it to have the `wrong' sign, $m_\chi^2 < 0$. Lepton
number would then be spontaneously broken, as $\chi$ carries lepton number charge,
and Majorana neutrino masses should appear. Indeed,
\eqref{intro-seesaw-type-II-yukawas} would immediately induce Majorana masses
for the neutrinos, for the VEV of $\chi$ needs to have the form
\begin{displaymath}
	\langle \chi \rangle = \begin{pmatrix}
				0  &  0
				\\
				v_\chi  &  0
			\end{pmatrix}
\end{displaymath}
in order not to break electric charge. Unfortunately, this program is not 
phenomenologically viable, because a very small $| m_\chi^2 |$ would
be needed, and it would yield light charged scalars which have not been observed.
Besides, it would neither realise the seesaw mechanism nor explain the smallness 
of neutrino masses.

That's what the $\mu$ trilinear coupling is useful for. It violates explicitly
lepton number and allows to write the diagram in figure 
\ref{fig:intro-seesaw-diagrams}\emph{b)}, with
the scalar triplet running in an internal propagator. The $\mu$ interaction
induces a VEV for $\chi$ through the VEV of $\phi$,
irrespective of the sign of $m_\chi^2$; in particular,
when $m_\chi^2 > 0$ \mbox{and $m_\chi \gg v_\phi$ --as the} diagram in figure 
\ref{fig:intro-seesaw-diagrams}\emph{b)} seems \mbox{to hint-- we obtain}
\begin{displaymath}
	v_\chi \simeq \frac{\mu \, v_\phi^2}{m_\chi^2} \, .
\end{displaymath}
Then the Yukawa couplings in equation \eqref{intro-seesaw-type-II-yukawas} lead 
to a Majorana mass matrix for the left-handed neutrinos, with structure
\begin{displaymath}
	m_\nu = 2 \, Y v_\chi \simeq 2 \, \frac{\mu \, v_\phi^2}{m_\chi^2} \, 
			Y \, .
\label{eq:triplet-numass}
\end{displaymath}

Neutrino masses are thus proportional to both $Y$ and $\mu$, and
inversely proportional to $m_\chi^2$. The seesaw is realised, for the heavier
$\chi$ the lighter the $\nu_{\mathrm{L}}$'s, and lepton number is violated
by the simultaneous presence of $Y$, that assigns lepton number to
$\chi$, and $\mu$, that breaks it.

\subsection{Type III Seesaw}

In the seesaw mechanism of type III \cite{Foot:1988aq,Ma:2002pf}, 
the Standard Model 
is extended by fermion $SU(2)$ triplets, $\Sigma_{\mathrm{L} \alpha}$, with zero 
hypercharge. 
Like in type I, at least two fermion triplets are needed to have two nonvanishing 
light neutrino masses. As for the case of the type II seesaw (see 
section \ref{sec:seesaw-typeII}), the triplet $\Sigma_{\mathrm{L}}$ can be 
written as a $2 \times 2$ matrix that acts on doublets,
\begin{displaymath}
	\Sigma_{\mathrm{L}} = \begin{pmatrix}
				\Sigma_{\mathrm{L}}^0 / \sqrt{2}  &  \Sigma^+_{\mathrm{L}}
				\\
				\Sigma^-_{\mathrm{L}}  &  - \Sigma^0_{\mathrm{L}} / \sqrt{2}
		\end{pmatrix} \, , 
\end{displaymath}
with $\Sigma_{\mathrm{L}}^0$, $\Sigma_{\mathrm{L}}^+$ and $\Sigma_{\mathrm{L}}^{-}$
the components of $\Sigma_{\mathrm{L}}$ with definite electric charge.
The new terms in the Lagrangian are then given by
\begin{multline} \label{intro-seesaw-type-III-lagrangian}
	\mathcal{L}_{\Sigma_{\mathrm{L}}} = i \, \mathrm{Tr} \left[ 
		\overline{\Sigma_{\mathrm{L}}} \, \gamma^\mu D_\mu \Sigma_{\mathrm{L}} 
		\right] +
		\\ +
		\left( \frac{1}{2} \, M_{\alpha \beta} \, \mathrm{Tr} \left[ 
		\overline{\Sigma_{\mathrm{L} \alpha}} \, \widetilde 
		\Sigma_{\mathrm{L} \beta} \right] + \sqrt{2} \, 
		Y_{\alpha \beta} \, 
		\overline{\ell_{\mathrm{L} \alpha}} \, \widetilde \Sigma_{\mathrm{L} 
		\beta} \: \tilde \phi + \mathrm{H.c.} \right) \, ,
\end{multline}
where $\widetilde \Sigma_{\mathrm{L}}$ is defined analogously to $\tilde 
\ell_{\mathrm{L}}$, $\widetilde \Sigma_{\mathrm{L}} \equiv \epsilon \,
\Sigma_{\mathrm{L}}^\mathrm{c} \epsilon^\dagger$, in order to connect the
fundamental and complex-conjugate doublet representations of $SU(2)$.
The coupling content of the model is very similar to that of type I seesaw,
see equation \eqref{intro-seesaw-lagrangian-type-I}, and so is the diagram
that generates the Weinberg operator, 
figure \ref{fig:intro-seesaw-diagrams}\emph{a)}.
The difference is that the interactions are richer in content with a triplet:
the Yukawas provide mixing between the $\nu_{\mathrm{L}}$'s and 
$\Sigma_{\mathrm{L}}^0$, as expected, but also mix the charged leptons with the
charged components of $\Sigma_{\mathrm{L}}$; the `Majorana' mass term $M$ is
also richer than it may seem: it indeed yields a Majorana mass for 
$\Sigma_{\mathrm{L}}^0$, but also provides a Dirac mass for a Dirac fermion $E$
constructed with both $\Sigma_{\mathrm{L}}^+$ and $\Sigma_{\mathrm{L}}^-$, 
$E = \Sigma_{\mathrm{L}}^- + {\Sigma_{\mathrm{L}}^+}^\mathrm{c}$.
These features endow the model with a more varied phenomenology, though we will
not discuss this in detail here.

For what respects the generation of neutrino masses, everything proceeds in 
a similar fashion as for type I seesaw: the diagram in figure 
\ref{fig:intro-seesaw-diagrams}\emph{a)}
suggests that the heavier $M$, the lighter the active neutrino masses.
Indeed, if we select from \eqref{intro-seesaw-type-III-lagrangian} the bilinear
terms of the neutral leptons we can compose a Majorana mass matrix,
\begin{displaymath}
	\mathcal{L}_{\nu \: \mathrm{mass}} = - \frac{1}{2} \,
		\begin{pmatrix}
			\overline{\nu_\mathrm{L}}  &  \overline{\Sigma^0_{\mathrm{L}}}
		\end{pmatrix} \,
		\begin{pmatrix}
			0  &  m_{\mathrm{D}} 
			\\
			m_{\mathrm{D}}^{\mathrm{T}}  &  M
		\end{pmatrix} \,
		\begin{pmatrix}
			\nu_{\mathrm{L}}^{\mathrm{c}} \vphantom{\raisebox{-1.5ex}{a}}
			\\
			{\Sigma_{\mathrm{L}}^0}^\mathrm{c}
		\end{pmatrix} + \mathrm{H. c.}  
\end{displaymath}
with $m_{\mathrm{D}} = v Y$,
which is absolutely equivalent to that of type I seesaw. Therefore, once we
assume $M \gg m_{\mathrm{D}}$ it too
provides the classical seesaw formula for the light neutrino masses,
\begin{displaymath}
	m_\nu \simeq - m_{\mathrm{D}} \, M^{-1} \, m_{\mathrm{D}}^{\mathrm{T}} \, .
\end{displaymath}
As we already commented, the main difference between type I and type III seesaw
is the presence of charged components in $\Sigma_{\mathrm{L}}$; in what respects
neutrino masses, this leads to more stringent constraints upon $M$: as the new
charged leptons have not been observed one should require
$M_{\alpha \beta} \gtrsim 100 \; \mathrm{GeV}$.

\section{Neutrino magnetic moments}

The electromagnetic properties of neutrinos have been a matter of interest
for many years. From the experimental point of view neutrinos are known to have very
small couplings to photons \cite{Beringer:1900zz}, and indeed in the context
of the Standard Model their electric charge is usually assumed to be zero%
\footnote{Though that is not a fundamental requirement of the theory: they might
well have a tiny, but nonzero, electric charge, to the degree allowed by 
observations. 
The situation is different in grand unified theories: as they aim to explain
the relation between the charges of leptons and quarks they end up quantising
electric charge, and neutrinos are neutral by \mbox{construction -- see,}
for instance, \cite{Ross:1985ai} for more on this matter.
}. 
Other electromagnetic
couplings are even more free, and their magnitude depends
on the interactions of neutrinos with certain charged particles. In this section
we will discuss one such couplings, the magnetic dipole moment; we will motivate
its interest and will point out some features of phenomenological interest.

Magnetic and electric
dipole moments are dimension-five effective interactions that
couple a pair of fermions to a photon, with a characteristic Lorentz 
structure:
\begin{displaymath}
	\mathcal{L}_{\mathrm{dipole}} = - \frac{1}{2} \, \overline{\psi_{\mathrm{R}}} 
			\left( \mu + \epsilon \gamma_5 \right)  \sigma^{\alpha \beta} \,
			\chi_{\mathrm{L}} \, F_{\alpha \beta} + \mathrm{H. c.} 
\end{displaymath}
In this expression $\chi_{\mathrm{L}}$ and $\psi_{\mathrm{R}}$ are fermionic
fields with left and right chirality, and $F_{\alpha \beta}$ is the electromagnetic
field strength tensor, that involves derivatives of the photon field. $\mu$ is
then called a \emph{magnetic dipole moment} and $\epsilon$ an \emph{electric
dipole moment}. They can be induced by radiative processes which involve charged
particles inside a loop, such as those we show in figure \ref{fig:nuR-model-magmo}.
For the case of the neutrinos, being neutral particles, one can have two types of 
dipole moments: the fields $\chi_{\mathrm{L}}$ and $\psi_{\mathrm{R}}$ can be
the left- and right-handed neutrino fields, and then we have the 
lepton-number-conserving \emph{Dirac dipole moments},
\begin{displaymath}
	\mathcal{L}_{\textrm{D-moments}} = - \frac{1}{2} \, \overline{\nu_{\mathrm{R}}} 
			\left( \mu_D + \epsilon_D \gamma_5 \right)  \sigma^{\alpha \beta} \,
			\nu_{\mathrm{L}} \, F_{\alpha \beta} + \mathrm{H. c.} 
\end{displaymath}
But we can also construct this same structure with just one neutrino field, say,
a left-handed neutrino and its charge-conjugate, which is a right-handed field;
we have then \emph{Majorana dipole moments},
\begin{displaymath}
	\mathcal{L}_{\textrm{M-moments}} = - \frac{1}{4} \, 
			\overline{\nu_{\mathrm{L}}^\mathrm{c}} 
			\left( \mu_M + \epsilon_M \gamma_5 \right)  \sigma^{\alpha \beta} \,
			\nu_{\mathrm{L}} \, F_{\alpha \beta} + \mathrm{H. c.} \, ,
\end{displaymath}
which break every charge carried \mbox{by $\nu_{\mathrm{L}}$ -- in} particular, 
lepton number. 

The Dirac or Majorana nature of the dipole moments depends on the nature of the 
underlying interactions between the neutrinos and the charged particles, and very
importantly on the relation of such interactions with lepton number violation.
They also depend on whether right-handed neutrinos are allowed or not as standalone
fields in the picture of neutrino masses being considered. Actually, often there
is a very close relation between, specifically, neutrino magnetic moments and 
neutrino masses. For instance, in the Standard Model enlarged only with 
right-handed neutrinos Dirac magnetic moments are generated radiatively
if the $\nu_{\mathrm{R}}$'s participate in neutrino masses 
\cite{Fukugita:2003en,Kayser:1989iu};
the converse is also true: if the neutrinos are endowed with Dirac magnetic
moments, they immediately acquire a mass through radiative corrections 
\cite{Davidson:2005cs,Bell:2005kz}. This intercourse between neutrino
magnetic moments and masses makes them a very interesting
study case. For instance, the induced
magnetic moment in the Standard Model for Dirac neutrinos is tiny due 
to the smallness of neutrino masses; the exact value is
$\mu_\nu \simeq 3 \times 10^{-19} \, \left( \nicefrac{m_\nu}{1 \; \mathrm{eV}} 
\right)$, far below any foreseeable experimental sensitivity, but it is still
interesting to look for them in experiments, because were it to happen that
we observe a magnetic moment larger than that predicted by the SM,
we would possibly be probing the physics responsible for neutrino masses.

However, neutrino magnetic moments also have interest by themselves; 
in the past, neutrino 
magnetic moments were also a matter of interest because if they
allowed flavour transitions they might provide an explanation for the solar
neutrino puzzle (see section \ref{sec:neutrino-oscillations}). 
Today, after the great experiments of the 90's and the 2000's,
it is a well-established fact that the solar as well as atmospheric 
neutrino deficit
is due to oscillations induced by mass mixing between the different flavours 
(see section \ref{sec:oscillation-experiments} and references therein), 
but the existence of a large neutrino magnetic
moment still would be very relevant in a number of scenarios, of which
the most important are maybe astrophysical \cite{Raffelt:1999tx,Raffelt:1996wa}:
neutrinos are extremely weakly-interacting
particles; this means that they are difficult to trap inside any object, and so
they provide a major means for energy to escape from astrophysical bodies. Such
energy loss, usually referred to as the \emph{cooling} of the object, is in 
many cases
well-bounded, because a too large energy depletion might compromise the integrity
of the body, or drastically change its properties. 

A good example is the bound
extracted from the luminosity gap between helium-burning stars and 
hydrogen-burning
red giants; it is described in a beatifully simple paper written by 
Georg Raffelt in 1990 \cite{Raffelt:1990pj}. 
The idea is the following: the burning of helium
requires a temperature roughly 10 times higher than the burning of hydrogen; this 
temperature can only be achieved by gravitational pressure upon the helium
mass. While in its hydrogen-burning phase the star produces huge amounts of
helium that cannot be burnt yet, as the temperature inside the star is not
high enough; the helium falls toward the center and there it forms an inert core.
As the helium accumulates the star keeps burning hydrogen in a shell that
surrounds the helium core. The gravity of the core presses the shell, heating it
and enhancing the hydrogen burning rate. The star, so, shines brighter and 
brighter, and meanwhile it undergoes several structural transformations
which eventually lead it to the red giant phase. When finally the helium furnace
ignites the core expands, alleviating the pressure upon the hydrogen shell, which
reduces its burning rate. The helium burning is not energetic enough and cannot 
compensate for this decrease, so the overall energy emission of the star decreases:
a gap in luminosity appears between the stars that have just begun to burn helium
and those that are about to.

The interesting thing about this succession of facts is that energy losses in 
the helium core affect differently helium-burning and not-yet-helium-burning stars:
obviously, if the core is losing energy, say by emission of neutrinos, 
hydrogen-burning red giants will need more massive cores to attain the temperature
necessary to ignite helium; as this means more gravitational pressure on the
hydrogen shell, the stars will reach greater luminosities before the helium
flash; but at the same time, helium-burning stars will inherit more massive
cores, so they will also shine brighter. The trick is that our astrophysical models
inform us that an additional cooling mechanism enhances \emph{more} the luminosity of
hydrogen-burning red giants than that of their helium-burning relatives;
consequently, a measurement of the luminosity gap between these two populations
constitutes also a measurement of the anomalous cooling mechanisms acting on the 
stars. 

How can we relate this to neutrinos and their magnetic moments? It's simple:
neutrinos are accounted for as `standard' cooling sources when coming from the
nuclear reactions occurring inside the core; any additional source of neutrinos
is considered `anomalous'. And magnetic moments provide such an additional source,
because photons inside a plasma can decay directly to neutrinos via the 
process $\gamma \rightarrow \nu \bar \nu$ that would be forbidden in vacuum.
This is due to the fact that the electromagnetic interactions of the photons
with the charged particles in the plasma dress the photons and endow them with
an effective mass; such massive photons are called \emph{plasmons}, and we will
denote them by $\gamma_P$. The mass of plasmons depends on the temperature of the
plasma, and it can vary from roughly $10 \; \mathrm{keV}$ for red giant cores
to several MeV for neutron stars and tens of MeV for supernova cores. Using
experimental data on the luminosity gap before and after the helium flash, 
reference \cite{Raffelt:1990pj} provides a very stringent bound on magnetic 
moments for neutrino species with masses below $10 \; \mathrm{keV}$. 
We will use this bound in chapter \ref{chap:nuR-magmo-eff}.

\chapter{Right-handed neutrino magnetic moments}	\label{chap:nuR-magmo-eff}

In this chapter we will present and analyse a dimension-five effective operator 
that provides magnetic moments for the right-handed neutrino degrees of freedom.
Right-handed neutrinos are usually regarded as sterile particles, having no
gauge couplings at all as they do; an interaction like a magnetic
moment may change dramatically the physics of right-handed neutrinos, providing
production and decay mechanisms and opening the possibility that they
show up in collider experiments and astrophysical systems. We discuss
the major constraints that can be derived from the overall nonobservation
of right-handed neutrinos, and we also comment on the observation prospects
at the LHC and the possible consequences of these magnetic moments in the
early universe. Together with this discussion we also consider briefly
other dimension-five operators that arise if we consider as low-energy
fields the full SM plus right-handed neutrinos; at the dimension-five
effective level, the neutrino mass matrix receives additional contributions,
and one operator appears that couples the Higgs field to right-handed 
neutrinos, possibly providing new decay channels for the Higgs boson.
This work was carried out together with Arcadi Santamaria, José Wudka and
Kyungwook Kim.

\section{An effective theory capable of parametrising all neutrino mass scenarios}
											 \label{sec:nuR-eff-presentation}

When considering the low-energy effects of new, heavy particles
that are not directly probed, one can parametrise their low-energy effects
using a series of effective vertices that involve
only the light fields \cite{Weinberg:1980wa,Georgi:1994qn,Wudka:1994ny}.
These vertices are constrained only by the gauge invariance of the
light theory, irrespective of the new interactions that may appear in 
the complete model \cite{Veltman:1980mj}. Assuming that the heavy physics
is decoupling, their contributions will be suppressed by powers of the 
heavy scale $\Lambda$, as described in section \ref{sec:intro-effective-theories}.

In this work we aim to study the first-order effective corrections
to the Standard Model with a focus on neutrino physics. Some properties of the
neutrinos, however, are still poorly understood and this casts some
uncertainties on the definition
of a proper neutrino-oriented effective theory. Think, for instance,
on the identity of the low-energy degrees of freedom: neutrinos
have masses, but we don't know yet if they are Dirac-like or Majorana-like.
If neutrinos are Dirac particles then their right-handed counterparts
exist and must \mbox{be light -- at least} two of them, for one of the light
neutrinos is still allowed to be massless. If neutrinos are Majorana, then 
right-handed neutrinos are not necessary, but they are not barred; they might
even have masses below the electroweak scale and provide interesting
phenomenological features \cite{deGouvea:2006gz,Ibarra:2010xw}.
With all this considered, we think that it is prudent to include right-handed
neutrinos among the low-energy degrees of freedom: that allows the effective
theory to describe any \emph{a priori} possible pattern of neutrino masses,
and can be regarded as a conservative hypothesis. In this chapter we will
examine the consequences of retaining the $\nu_{\mathrm{R}}$'s among the 
low-energy fields.

The most general Lagrangian that includes the full Standard Model plus a number
of right-handed neutrinos will read
\begin{displaymath}
	\mathcal{L} = \mathcal{L}_{\mathrm{SM}} + \mathcal{L}_{\nu_{\mathrm{R}}} +
			\mathcal{L}_5 + \ldots \, ,
\end{displaymath}
where $\mathcal{L}_{\mathrm{SM}}$ represents the usual Standard Model with no
right-handed neutrinos,
\begin{equation} \label{nuR-eff-Lagrangian-SM}
	\mathcal{L}_{\mathrm{SM}}  =  i \, \overline{\ell_{\mathrm{L}}} \gamma_\mu 
		D^\mu \ell_{\mathrm{L}} + i \, \overline{e_{\mathrm{R}}} \gamma_\mu D^\mu 
		e_{\mathrm{R}} - (\overline{\ell_{\mathrm{L}}} \, Y_{e} \, e_{\mathrm{R}}
		\, \phi + \mathrm{H.c.}) + \ldots \, ,
\end{equation}
$\mathcal{L}_{\nu_{\mathrm{R}}}$ collects the renormalisable terms that appear
when $\nu_{\mathrm{R}}$'s are introduced,
\begin{equation} \label{nuR-eff-Lagrangian-nuR}
	\mathcal{L}_{\nu_{\mathrm{R}}} = i \, \overline{\nu_{\mathrm{R}}^{\, \prime}}
		\gamma_\mu \partial^\mu \nu_{\mathrm{R}}^{\, \prime} - \left( \frac{1}{2} \,
		\overline{{\nu_{\mathrm{R}}^{\, \prime}}^\mathrm{c}} M 
		\nu_{\mathrm{R}}^{\, \prime} + \overline{\ell_{\mathrm{L}}} \, Y_\nu \, 
		\nu_{\mathrm{R}}^{\, \prime} \, \tilde{\phi} + \mathrm{H.c.} \right) \, ,
\end{equation}
and $\mathcal{L}_5$ is the lowest-\mbox{order --i.e., dimension}-\mbox{five-- 
effective} operators that arise in such a theory,
\begin{equation} \label{nuR-eff-Lagrangian-5}
	\mathcal{L}_5 = \overline{{\nu_{\mathrm{R}}^{\, \prime}}^\mathrm{c}} \, \zeta \, 
		\sigma^{\mu \nu} \nu_{\mathrm{R}}^{\, \prime} \, B_{\mu \nu} + 
		\left(\overline{\tilde{\ell_{\mathrm{L}}}} \phi \right) \, \chi \, 
		\left(\tilde{\phi}^\dagger \ell_{\mathrm{L}} \right) - 
		\left(\phi^\dagger \phi \right) 
		\overline{{\nu_{\mathrm{R}}^{\, \prime}}^\mathrm{c}} \, \xi \, 
		\nu_{\mathrm{R}}^{\, \prime} + \mathrm{H.c.} \, .
\end{equation}
Notation in equations 
(\ref{nuR-eff-Lagrangian-SM}$\,$--$\,$\ref{nuR-eff-Lagrangian-5})
follows closely that described in sections \ref{sec:intro-SM-doublets-singlets}
and \ref{sec:masses-SM}, with the only difference of a prime ($\prime$) 
on some of the neutrino fields to denote that these are \emph{not} fields
with definite mass; we save the unprimed version for the fields 
representing light neutrinos with definite mass, equation 
\eqref{nuR-eff-mass-terms}. Consequently, the left-handed neutrinos in the 
flavour doublets also are understood to carry a prime: $\ell_{\mathrm{L}} 
\equiv {\nu_{\mathrm{L}}^{\, \prime} \choose e_{\mathrm{L}}}$. 
Apart from that, the $SU(2) \otimes U(1)$ charges of the SM fields
are the customary ones, see table \ref{tab:intro-SM-hypercharges};
the $B$ field, remember, represents the gauge boson associated
to $U(1)_Y$; and $\sigma^{\mu \nu}$ is a matrix in spinorial space
defined by $\sigma^{\mu \nu} \equiv \frac{i}{2} \, \left[ \gamma^\mu , \gamma^\nu
\right]$, with $\gamma^\alpha$ the Dirac matrices of the spinorial Clifford
algebra. The dots in equation \eqref{nuR-eff-Lagrangian-SM} indicate that
we have omitted the terms involving quarks, and also the kinetic and gauge-fixing
terms for the gauge fields, as they will not be used in our discussion.

As for the couplings with structure in generation space, their number and
complexity will depend on the number of families of $\nu_{\mathrm{R}}$'s
that we wish to consider. In the following we will assume that we have three
$\nu_{\mathrm{R}}^{\, \prime}$'s, as that is the minimum number needed to
give Dirac masses to the three light $\nu_{\mathrm{L}}^{\, \prime}$ families.
With this assumption, the Yukawa couplings $Y_e$ and $Y_\nu$ will be
completely general $3 \times 3$ matrices in flavour space; $M$, $\chi$,
and $\xi$ are complex symmetric $3 \times 3$ matrices, while $\zeta$
is a complex antisymmetric matrix that will ultimately provide magnetic
moments for the $\nu_{\mathrm{R}}$'s. Without loss of generality, $Y_e$
and $M$ can be taken diagonal with real, positive elements.

The term involving $M$ is the usual right-handed neutrino Majorana
mass. The term involving $\chi$ is known as the Weinberg operator and 
provides a Majorana mass for the left-handed neutrino fields plus
various lepton-number-violating neutrino-Higgs interactions, see
section \ref{sec:neutrinos-dirac-or-majorana}.
The term involving $\zeta$ has been mostly
ignored in the literature; it describes electroweak%
\footnote{Note that although we stress the generation of magnetic moments,
i.e., couplings to photons, the effective operator actually involves the
$B$ field, and consequently it will generate couplings to both $A$ and $Z$
after EWSB. The $Z$-moments are suppressed at low energies, but become 
important as one approaches energies of $\mathcal{O} (m_Z)$; they will thus
play a role in collider experiments and cosmological scenarios, as we discuss
in the corresponding sections.}
moment couplings for the right-handed neutrinos. We will dedicate a significant 
part of this chapter to the study of some of the consequences this operator
might have on various collider, astrophysical and cosmological observables.
Note that Dirac-type neutrino magnetic moments (involving $\ell_{\mathrm{L}}$
and $\nu_{\mathrm{R}}^{\, \prime}$) are generated by operators of dimension $\ge 6$,
while Majorana-type magnetic moments for left-handed neutrinos (involving
only $\ell_{\mathrm{L}}$) require operators of dimension $\ge 7$. One can
easily see that these effects are subdominant when compared to those
produced by the term containing $\zeta$ in $\mathcal{L}_5$.

The couplings $\chi$, $\xi$, $\zeta$ have dimension of inverse
mass, which is associated with the scale of the heavy physics responsible
for the corresponding operator. Though we will refer to this scale
generically as $\Lambda$ it must be kept in mind that different types
of new physics can be responsible for the various dimension-five operators,
and that the corresponding values of $\Lambda$ might be very different.
All these scales, of course, share that they must 
be much larger than the electroweak scale, $v \sim 0.25 \; \mathrm{TeV}$,
by consistency of the effective-theory approach. In the next section 
we discuss the possible
types of new physics that can generate these operators and the natural
size for the corresponding coefficients.

\section{Heavy physics content of the effective vertices} \label{sec:nuR-eff-NP}

In this section we will give a brief account of the combinations of heavy
particles that can provide the dimension-five interactions; for each operator
we will describe the fields that mediate its generation at the lowest possible order
in perturbation theory, and we estimate the size of the effective coupling.

\subsubsection{$\nu_{\mathrm{L}}^{\, \prime}$ Majorana mass term}

Using appropriate Fierz transformations we can rewrite the Weinberg operator
in \eqref{nuR-eff-Lagrangian-5}
as follows:
\begin{displaymath}
	\left( \overline{\tilde{\ell_{\mathrm{L} \alpha}}} \phi \right) 
		\left( \tilde{\phi}^\dagger \ell_{\mathrm{L} \beta} \right) = 
		\frac{1}{2} \left( \overline{\tilde{\ell_{\mathrm{L} \alpha}}} \, \vec 
		\sigma \, \ell_{\mathrm{L} \beta} \right) \cdot \left( \tilde{\phi}^\dagger
		\, \vec \sigma \, \phi \right) =
		- \left( \overline{\tilde{\ell_{\mathrm{L} \alpha}}} \, \vec \sigma \, 
		\phi \right) \cdot \left( \tilde{\phi}^\dagger \, \vec \sigma \,
		\ell_{\mathrm{L} \beta} \right) \, , 
\end{displaymath}
where $\alpha$ and $\beta$ are family indices and $\vec \sigma \equiv \left(
\sigma_1, \sigma_2, \sigma_3 \right)$ is a vector in the adjoint representation
of $SU(2)$ whose components are the $2 \times 2$ Pauli matrices. 
From these rearrangements we can
read the three different processes that generate the operator at tree-level:
in the first, a heavy fermion singlet with no hypercharge connects 
the two $SU(2)$-closed pairs; in the second, a hyperchargeless scalar triplet
is the mediator; in the third it is a fermion $SU(2)$ triplet, again with no
hypercharge. Of course these are the new physics additions that characterise
seesaws of type I, II and III, as we discussed in more detail in 
section \ref{sec:seesaw}. If perturbation
theory holds for the new physics the effective coupling should read,
roughly,
\begin{equation} \label{nuR-eff-chi-estimate}
	\chi \sim \frac{\lambda^2}{\Lambda} \, ,
\end{equation}
where $\Lambda$ in this simple tree-level case denotes the mass of the
heavy particle,
and $\lambda$ the coupling constants of the heavy fermions to $\phi \, \ell$,
or of the heavy scalar to $\phi \, \phi$ and $\ell \, \ell$.

\subsubsection{$\nu_{\mathrm{R}}^{\, \prime}$ Majorana mass term} 

The operator $\left( \phi^\dagger \phi \right) \, 
\overline{\nu_{\mathrm{R} \alpha}^{\, \prime \: \mathrm{c}}} \, 
\nu_{\mathrm{R} \beta}^{\, \prime}$ can also be generated at tree level.
By an argument similar to that in the previous section, the low-energy
fields can be grouped into $\phi \, \nu_{\mathrm{R}}$ pairs or in a 
$\phi^\dagger \, \phi$ 
and a $\nu_{\mathrm{R}} \, \nu_{\mathrm{R}}$ pair. The first groups can be connected
through a heavy fermion isodoublet of hypercharge $+ \, \nicefrac{1}{2}$; 
the second, via a hyperchargeless scalar singlet. And like in the previous section,
if the new physics is perturbative we expect
\begin{equation} \label{nuR-eff-xi-estimate}
	\xi \sim \frac{\lambda^2}{\Lambda} \, ,
\end{equation}
where $\Lambda$ again denotes the mass of the heavy particles, and
$\lambda$ the coupling of the heavy fermion to $\phi \, \nu_{\mathrm{R}}$ or the
heavy scalar to $\phi^\dagger \, \phi$ and $\nu_{\mathrm{R}} \, \nu_{\mathrm{R}}$.

\subsubsection{$\nu_{\mathrm{R}}^{\, \prime}$ electroweak coupling}

The electroweak moment operator, unlike the previous cases, cannot be generated
at tree level. The reason is not difficult to grasp; being its form
\begin{displaymath}
	\mathcal{L}_\zeta = \overline{\nu_{\mathrm{R} \alpha}^{\, \prime \: 
			\mathrm{c}}} \, \sigma^{\mu \nu} \, \nu_{\mathrm{R} \beta}^{\, \prime}
			\: B_{\mu \nu} \, ,
\end{displaymath}
it is clear that a charged internal leg is needed in order to insert the $B$
boson, as the $\nu_{\mathrm{R}}$'s are hyperchargeless. But if the 
$\nu_{\mathrm{R}}$'s are to be connected to charged particles 
there must be more than one of them so that their hypercharges cancel. In 
consequence, the topology of the generating diagram must include a loop 
of charged particles that closes to yield the external hyperchargeless legs.
At the lowest order, that is, at one loop, this can be realised by two different 
combinations of fields: first, via a 
scalar-fermion pair $\{ \omega, E \}$ which have opposite, 
nonzero hypercharges \emph{and}
present both interactions $\omega \, \overline{E} \, \nu_{\mathrm{R}}^{\, \prime}$ 
and $\omega \, \overline{E} \, {\nu_{\mathrm{R}}^{\, \prime}}^\mathrm{c}$.
Alternatively, the loop can also be provided by
a vector-fermion pair $\{ W_\mu^{\, \prime} , E \}$, again with opposite and 
nonzero hypercharges and again having couplings $W_\mu^{\, \prime} \, \overline{E} 
\gamma^\mu \nu_{\mathrm{R}}^{\, \prime}$ and $W_\mu^{\, \prime} \, \overline{E} 
\gamma^\mu {\nu_{\mathrm{R}}^{\, \prime}}^\mathrm{c}$.

In this class of models we will have, roughly, 
\begin{equation} \label{nuR-eff-zeta-estimate}
	\zeta \sim \frac{g^{\,\prime} y \lambda^2}{16 \pi^2} \, 
			\frac{m_{\mathrm{fermion}}}{\max (m_{\mathrm{fermion}}^2,
			m_{\mathrm{boson}}^2)} < \frac{g^{\, \prime} y \lambda^2}{16 \pi^2
			m_{\mathrm{fermion}}} \, ,
\end{equation}
where $\lambda$ denotes the coupling of the two heavy particles to
the $\nu_{\mathrm{R}}^{\, \prime}$, and $y$ the hypercharge of the heavy boson
or fermion. A specific example is provided in chapter \ref{chap:nuR-magmo-model}.

\subsubsection{Perturbativity of the heavy particle couplings}
						   \label{sec:nuR-eff-NP-strong-or-weak}

The reader should note that the estimates 
(\ref{nuR-eff-chi-estimate}$\,$--$\,$\ref{nuR-eff-zeta-estimate})
rely on the assumption that the heavy physics is connected to the low-energy
fields through weak, perturbative interactions. Were this not the case, we would
have no reliable way to guess the values of the effective couplings,
but we could still obtain estimates of their \emph{natural} values through
naive dimensional analysis (NDA) \cite{Manohar:1983md,Georgi:1992dw}; 
this would yield
\begin{align} \label{nuR-eff-NDA-estimates}
	\chi, \xi &\sim \frac{16\pi^2}{\Lambda}
	& & &
	\zeta &\sim \frac{1}{\Lambda} \, ,
\end{align}
where $\Lambda$ would be, in this case, the scale of the strong interactions.
However, a number of caveats apply even to these naturality estimates: 
\eqref{nuR-eff-NDA-estimates} assumes that \emph{all} the fields, including the
low-energy $\phi$, $\ell_{\mathrm{L}}$ and $\nu_{\mathrm{R}}$, participate 
in the strong interactions. Mixed scenarios, for instance with the new
particles having indeed nonperturbative couplings but being connected to 
the low-energy fields by weak interactions, could be more complicated. 
Anyway, it can be useful to have in mind estimates such as 
\eqref{nuR-eff-NDA-estimates} in order to weigh the influence of possible
nonperturbative scenarios. For practical purposes, in the remaining of this chapter
we will ignore these details and will label as ``$\Lambda$'' just the inverse
of the effective coupling; as we will discuss extensively the electroweak
moment interaction, most of the time it will be understood that
\begin{displaymath}
	\Lambda \equiv \frac{1}{\zeta} \, ,
\end{displaymath}
but when it has to be interpreted otherwise we will explicitly indicate so.
Note that this choice implies that
$\Lambda$ will have an immediate translation in terms of ``regime of strong
interactions of the heavy particles''; if the reader wants to recover the
values associated to a weakly-coupled theory it will be enough to substitute
$\Lambda \rightarrow 16 \pi^2 \, \Lambda$, as can be seen from equations
\eqref{nuR-eff-zeta-estimate} and \eqref{nuR-eff-NDA-estimates}.

\section{The Lagrangian in terms of mass eigenfields}
	  			 \label{sec:nuR-eff-mass-eigenfields}

In this section we will consider the possible mass structures of the neutrino sector,
and we will discuss the changes that allow to express the Lagrangian in terms of the 
fields with well-definite mass. For the representative case of three very light
neutrinos, mostly $\nu_{\mathrm{L}}^{\, \prime}$'s, and three relatively heavy 
ones, essentially $\nu_{\mathrm{R}}^{\, \prime}$'s, we present the effective
interactions of our interest in terms of the mass eigenfields.

The first step in this program is to construct the neutrino mass matrix and 
identify the degrees of freedom with definite mass.
From equations (\ref{nuR-eff-Lagrangian-SM}$\,$--$\,$\ref{nuR-eff-Lagrangian-5})
it is straightforward to obtain the neutrino and charged lepton
mass matrices after SSB. Replacing $\phi \to \langle \phi \rangle = \nicefrac{v}
{\sqrt{2}} \, (0,1)^\mathrm{T}$ yields the following mass terms for the leptons
\begin{equation} \label{nuR-eff-mass-Lagrangian}
	\mathcal{L}_{\mathrm{m}} = - \overline{e_{\mathrm{L}}} \, M_e \, e_{\mathrm{R}}
		- \overline{\nu_{\mathrm{L}}^{\, \prime}} \, M_D \, 
		\nu_{\mathrm{R}}^{\, \prime} - \frac{1}{2} \, 
		\overline{{\nu_{\mathrm{L}}^{\, \prime}}^\mathrm{c}} \, M_L \, 
		\nu_{\mathrm{L}}^{\, \prime} - \frac{1}{2} \, 
		\overline{{\nu_{\mathrm{R}}^{\, \prime}}^\mathrm{c}} \, M_R \, 
		\nu_{\mathrm{R}}^{\, \prime} + \mathrm{H.c.} \, ,
\end{equation}
where
\begin{align*}
	M_e &= \frac{v}{\sqrt{2}} \, Y_e  & M_D &= \frac{v}{\sqrt{2}} \, Y_\nu
	\\
	\vphantom{\raisebox{2ex}{a}}
	M_L &= \chi v^{2} & M_R &= M + \xi v^2  \, .
\end{align*}
It is worth noticing that, up to possible coupling-constant factors,
$M_D \sim v$ while $M_L \sim v^2 / \Lambda$. Various situations arise
depending on the hierarchy between $M_R$, $M_D$ and $M_L$:
if we have $M_R \gg M_D \gg M_L$ the theory describes a standard type I seesaw;
seesaws of types II and III would be indistinguishable in our dimension-five
Lagrangian, but would leave a footprint of the form $M_L \gg M_D^2 / M_R$
-- but beware, many other underlying models could leave the same mark.
In these seesaw-like cases there is no conserved or approximately conserved fermion
number and the mass eigenstates are Majorana fermions. In contrast,
when $M_D \gg M_R, M_L$ there is an approximately conserved fermion
number and the mass eigenstates will be Dirac fermions up to small
\mbox{admixtures -- the so-}called pseudo-Dirac fermions. Of course, the
details of the mass matrix diagonalisation vary tremendously from one scenario
to another. Throughout this chapter we will take as an example case the
seesaw scenario, $M_R \gg M_D \gg M_L$, which can accommodate rather easily 
our current phenomenological knowledge. In the following we proceed to describe the 
diagonalisation procedure and the mass eigenstates in this scenario. 

First of all, note that, as corresponds to a type I seesaw,
when $M_R \gg M_D \gg M_L$ the neutrino sector splits into two groups: the
light neutrinos, which are primarily $\nu_{\mathrm{L}}^{\, \prime}$'s, and the heavy
ones, which are composed mainly of $\nu_{\mathrm{R}}^{\, \prime}$'s. In such a case 
the mass matrices can be approximately diagonalised blockwise, leading to two 
$3 \times 3$ Majorana mass matrices,
\begin{align*}
	\mathrm{heavy}: \quad\! \mathcal{M}_N &\simeq M_R
	\\
	\mathrm{light}: \quad \mathcal{M}_\nu &\simeq M_L - M_D^* \, {M_R^\dagger}^{-1} 
			\, M_D^\dagger \, .
\end{align*}
These matrices can subsequently be diagonalised by using unitary
matrices $U_N$ and $U_\nu$, such that 
\begin{align*}
	M_N &\equiv U_N^\mathrm{T} \, \mathcal{M}_N \, U_N
	\\
	\vphantom{\raisebox{1.5ex}{a}} 
	M_\nu &\equiv U_\nu^\mathrm{T} \mathcal{M}_\nu \, U_{\nu} \, ,
\end{align*}
where $M_N$ and $M_\nu$ are diagonal matrices with positive elements (if one
chooses the $\nu_{\mathrm{R}}^{\, \prime}$ basis so that $M_R \simeq \mathcal{M}_N$
is diagonal then we have the further simplification that $U_N = \mathbb{1}$).
This way the mass Lagrangian \eqref{nuR-eff-mass-Lagrangian}
can be rewritten in terms of mass eigenfields as
\begin{equation} \label{nuR-eff-mass-terms}
	\mathcal{L}_{\mathrm{m}} = - \bar{e} \, M_e \, e - \frac{1}{2} \bar{\nu} \,
		M_\nu \, \nu - \frac{1}{2} \overline{N} \, M_N \, N \, ,
\end{equation}
where $\nu$ and $N$ are Majorana fields ($\nu^\mathrm{c} = \nu$ and $N^\mathrm{c} =
N$) that represent, respectively, the light and heavy mass eigenstates, 
and we assumed that the charged leptons are defined as the fields that
diagonalise $Y_e$.
This is enough to express the ``flavour'' spectrum of 
$\nu_{\mathrm{L}}^{\, \prime}$'s and $\nu_{\mathrm{R}}^{\, \prime}$'s 
in terms of the mass spectrum of $\nu$'s and $N$'s:
\begin{align}
	\nu_{\mathrm{L}}^{\, \prime} &= P_{\mathrm{L}} \, \left( U_\nu \, \nu + 
		\varepsilon \, U_N \, N + \ldots \right)
	\label{nuR-eff-nuLtonu}
	\\
	\nu_{\mathrm{R}}^{\, \prime} &= P_{\mathrm{R}} \, \left( U_N \, N - 
		\varepsilon^\mathrm{T} U_\nu \, \nu + \ldots \right) \, ,
	\label{nuR-eff-nuRtoN}
\end{align}
where $P_{\mathrm{L}, \mathrm{R}} = \frac{1}{2} (\mathbb{1} \mp \gamma_5)$ are the 
usual chiral projectors, and 
\begin{displaymath}
	\varepsilon \simeq M_D \, M_R^{-1} 
\end{displaymath}
is a $3 \times 3$ matrix that characterises the mixing between heavy and light
neutrinos. Note that barring cancellations in $\mathcal{M}_\nu$,
the elements of the mixing matrix $\varepsilon$ obey in most cases,
even when the light masses are not given by the seesaw mechanism, that
\begin{displaymath}
	| \varepsilon_{\alpha \beta} | \lesssim \sqrt{\frac{m_\nu}{m_N}},
\end{displaymath}
where $m_\nu$ represents a mass of the order of the light neutrino
masses and $m_N$ a mass of the order of the heavy neutrino masses. 
This is due to the fact that even when the seesaw is not the generating
mechanism for the light neutrino masses we need to ensure that seesaw-like
pieces don't spoil the mass \mbox{structure -- that is,} with too large
contributions. Therefore, the mixing is required to be at the level of
$\sqrt{\nicefrac{m_\nu}{m_N}}$ (the seesaw mixing) or below. The only
scenario that allows for large $\varepsilon$ is one where the right-handed
degrees of freedom have very light masses, of the order of the eV or
below; but this scenario is very constrained phenomenologically and
does not seem very \mbox{appealing -- see} the discussion in sections 
\ref{sec:nuR-eff-eebar-effects} and \ref{sec:nuR-eff-astro}.

Now we can use these expressions to write the effective operators in equation
\eqref{nuR-eff-Lagrangian-5} in terms of the mass eigenfields; this will
be of use when computing the amplitudes for processes in which the masses
are a relevant element. 
Substituting equation \eqref{nuR-eff-nuRtoN}
into the electroweak moment operator in \eqref{nuR-eff-Lagrangian-5} and using
the well-know expression for $B$ in terms of the photon and
the $Z$ fields, we obtain the magnetic and $Z$ moments in terms of the
mass eigenfields: 
\begin{multline} \label{nuR-eff-zeta-mass}
	\mathcal{L}_\zeta = \left( \overline{N} \, U_N^\dagger - \overline{\nu} \, 
		U_\nu^\dagger \, \varepsilon^* \right) \, \sigma^{\mu \nu} \, 
		\left( \zeta \, P_{\mathrm{R}} + \zeta^\dagger \, P_{\mathrm{L}} \right) 
		\times 
		\\
		\times \left( U_N \, N - \varepsilon^\mathrm{T} \, U_\nu \, \nu \right) \, 
		\left( c_W F_{\mu \nu} - s_W Z_{\mu \nu} \right) \, , 
\end{multline}
where $F_{\mu\nu}$ and $Z_{\mu\nu}$ are the Abelian field strengths
of the photon and the $Z$, and $c_W \equiv \cos \theta_W, s_W \equiv 
\sin \theta_W$, with $\theta_W$ the weak mixing angle. We see that the 
original $\nu_{\mathrm{R}}^{\, \prime}$ electroweak moment operator generates
a variety of couplings when expressed in terms of mass eigenstates:
both magnetic and $Z$-moments appear, and they connect either heavy-heavy,
light-heavy or light-light degrees of freedom with increasing suppression.
Note also that, as $\zeta$ is an antisymmetric matrix in flavour space,
\emph{all} the moments generated by $\mathcal{L}_\zeta$ are \emph{transition}
moments, even those linking heavy-heavy and light-light fields.

Similarly, if we use equation \eqref{nuR-eff-nuRtoN} to substitute in the last
term of \eqref{nuR-eff-Lagrangian-5}, 
we obtain, in addition to a contribution to the $N$ mass, a Higgs-heavy
neutrino interaction: 
\begin{equation} \label{nuR-eff-Higgs-NN-vertex}
	\mathcal{L}_\xi = -v \, H \, \overline{N} \left(\xi P_{\mathrm{R}} + 
		\xi^\dagger P_{\mathrm{L}} \right) \, N + \ldots \, ,
\end{equation}
where we took $U_N = \mathbb{1}$ and the dots represent other interactions
generated by this operator: $HHNN$ vertices, for instance, as well as $N-\nu$ and
$\nu-\nu$ interactions that are suppressed by the mixing $\varepsilon$;
these vertices are also generated by the neutrino Yukawa coupling
in $\mathcal{L}_{\nu_{\mathrm{R}}}$, where they're likewise suppressed.

Finally we should mention that when equation \eqref{nuR-eff-nuLtonu} is
introduced into the SM weak interactions $\overline{\nu_{\mathrm{L}}^{\, \prime}}
\gamma^\mu \nu_{\mathrm{L}}^{\, \prime} \, Z_\mu$ and $\overline{e_{\mathrm{L}}}
\gamma^\mu \nu_{\mathrm{L}}^{\, \prime} \, W_\mu$, one obtains $N - \nu -Z$ and 
$N - e - W$ couplings that, although suppressed by $\varepsilon$, are important 
for the decays of the lightest of the heavy neutrinos.

\section{Collider effects} \label{sec:nuR-eff-colliders}

\subsection{General considerations} \label{sec:nuR-eff-collider-prologue}

In this section we discuss the variety of effects that the dimension-five
operators can yield in a collider experiment. A first estimation of the importance
of such effects can be made as a result of the nonobservation of new particles
in LEP-II, Tevatron and the LHC \cite{Beringer:1900zz}: as the present bound is 
essentially $m_{\mathrm{NP}} >
100 \; \mathrm{GeV}$ or better for most classes of particles, we can infer that
the effective operators have at least $\Lambda > 100 \; \mathrm{GeV}$.
However, the bound affects differently the various effective interactions,
and the perturbativity or not of the heavy physics couplings can be an important 
factor. For instance, for the electroweak moments a weakly-coupled new physics
would imply $\Lambda = \nicefrac{1}{\zeta} \sim (4 \pi)^2 \, m_{\mathrm{NP}}$,
as seen in equation \eqref{nuR-eff-zeta-estimate},
and so $\Lambda > 15 \; \mathrm{TeV}$; this would most likely suppress
its effects beyond our present and near-future experimental sensitivities. 
A strongly-coupled scenario,
though more difficult to study theoretically, might significantly lower
the scales of the new interactions, putting them
at the reach of the LHC or the future linear collider;
in fact, if we have merely $\zeta = \nicefrac{1}{m_{\mathrm{NP}}}$ the effects
of this interaction would have been observable at LEP, as we consider later
in section \ref{sec:nuR-eff-eebar-effects}. 
In conclusion, a scenario with nonperturbative
new physics optimises the possibility of observing the effective interactions
in colliders; weakly-coupled new particles are not discarded, but in general 
direct searches seem a more efficient way to probe their existence.

Apart from the new physics responsible for the effective interaction, our
theory considers a number of new fermionic, neutral degrees of freedom associated 
to the $\nu_{\mathrm{R}}^{\, \prime}$. Our discussion below essentially focuses on
their production and observability in colliders, which can be achieved either
through mixing with the left-handed species or through the effective 
interactions. It is reasonable then to ask which is the relevant mechanism
for their collider phenomenology: would a positive experimental signal be
probing just mixing or the new effective interactions? We can easily argue
that in most cases it is the effective interactions who are in charge of
the collider effects: as we discussed in section 
\ref{sec:nuR-eff-mass-eigenfields}, 
if the new fermions are heavy the heavy-light mixing is 
$\varepsilon \lesssim \sqrt{\nicefrac{m_\nu}{m_N}}$,
a value too small to provide a significant production mechanism unless $m_N$ 
is rather small; but even in this case astrophysical considerations (see section
\ref{sec:nuR-eff-astro}) impose very stringent bounds on any production mechanism,
rendering any collider effect unobservable. Therefore, mixings can be safely
ignored for most purposes in collider phenomenology; a notable exception is
the decay of the neutral fermions, which is governed by the effective
magnetic moments except for the lightest of the $N$'s. 
In the following section we comment thoroughly on the phenomenology
of the neutral fermions' decays.

\subsection{Decay mechanisms for the new neutral fermions}
							   \label{sec:nuR-eff-N-decay}

Before discussing the impact of the new interactions in past and future
collider experiments, we would like to analyse the dominant decay modes
of the new neutral fermions associated to the $\nu_{\mathrm{R}}^{\, \prime}$'s.
These fermions can potentially be produced in colliders through the electroweak
moments or Higgs decays, as described in the remaining passages of this section;
once produced they will decay, and the decays will provide a means to identify
the relevant interactions. Let us here, so, consider such decay mechanisms in
some detail. Though in principle we could have three or more right-handed 
neutrinos, for this discussion it will be enough to consider two of them:
the lightest heavy state, that we will label $N_1$, and the next-to-lightest,
identified \mbox{as $N_2$ -- of course} with $m_1 < m_2$. The decays of all
the heavier particles will proceed as those of $N_2$ to $N_1$.

The heavy neutrinos have essentially two mechanisms to decay: through mixing,
involving the SM weak vertices, and through the electroweak moment interactions%
\footnote{A third option, namely $N_2 \rightarrow N_1 H$ through the $\xi$ operator,
will not be considered in this analysis.}.
The latter allows decays from one of the $N$'s to another with emission of a 
photon or $Z$, and are suppressed by the unknown new physics scale $\Lambda$;
the first link the heavy neutrinos to a gauge boson and a light lepton, and are
in turn suppressed by the mixing $\varepsilon$. As these mixings are expected to 
be small, the electroweak moments are expected to dominate the decays of the
heavier states, if they are observable at all. The $N_1$, however, has no
heavy state to decay to, and so its decays probe the heavy-light mixing.

We begin by describing the decay modes mediated by the electroweak moments.
If they are strong enough to produce the new neutral fermions%
\footnote{See sections \ref{sec:nuR-eff-eebar-effects} and 
\ref{sec:nuR-eff-production-LHC} for a discussion of the production of heavy
neutrinos in colliders.
}, 
the dominant 
decay channel for $N_2$ will be $N_2 \rightarrow N_1 \gamma$, and also 
$N_2 \rightarrow N_1 Z$ if $N_2$ is heavy enough. The most prominent experimental 
signature of these decays should be the presence of a hard photon, which will 
appear if $N_2$ is heavy \mbox{enough --say,} \mbox{$m_2 > 10 \; \mathrm{GeV}$-- 
and the} mass splitting with $N_1$ is large enough%
\footnote{It could happen that $N_2$ and $N_1$ are almost perfectly degenerate; 
then this sort of decays will be suppressed by kinematics, and other channels to 
SM particles, like $N_2 \rightarrow \nu \gamma$, or $N_2 \rightarrow \ell W, 
\, \nu Z, \, \nu H$, although suppressed by $\varepsilon$, could be relevant.}.
The lifetime of the heavy neutrinos will be very short if these modes are available;
for instance, we find that for a $N_2$ produced at center of mass energies ranging 
from $100-1000 \; \mathrm{GeV}$ the associated decay lengths are well below 
$10 \; \mathrm{nm}$ unless $m_2 \simeq m_1$. The reader can find the expressions
for the decay widths collected in section \ref{sec:nuR-eff-appendix} 
at the end of the chapter.

\begin{figure}[tb]
	\centering
	\includegraphics[width=0.8\textwidth]{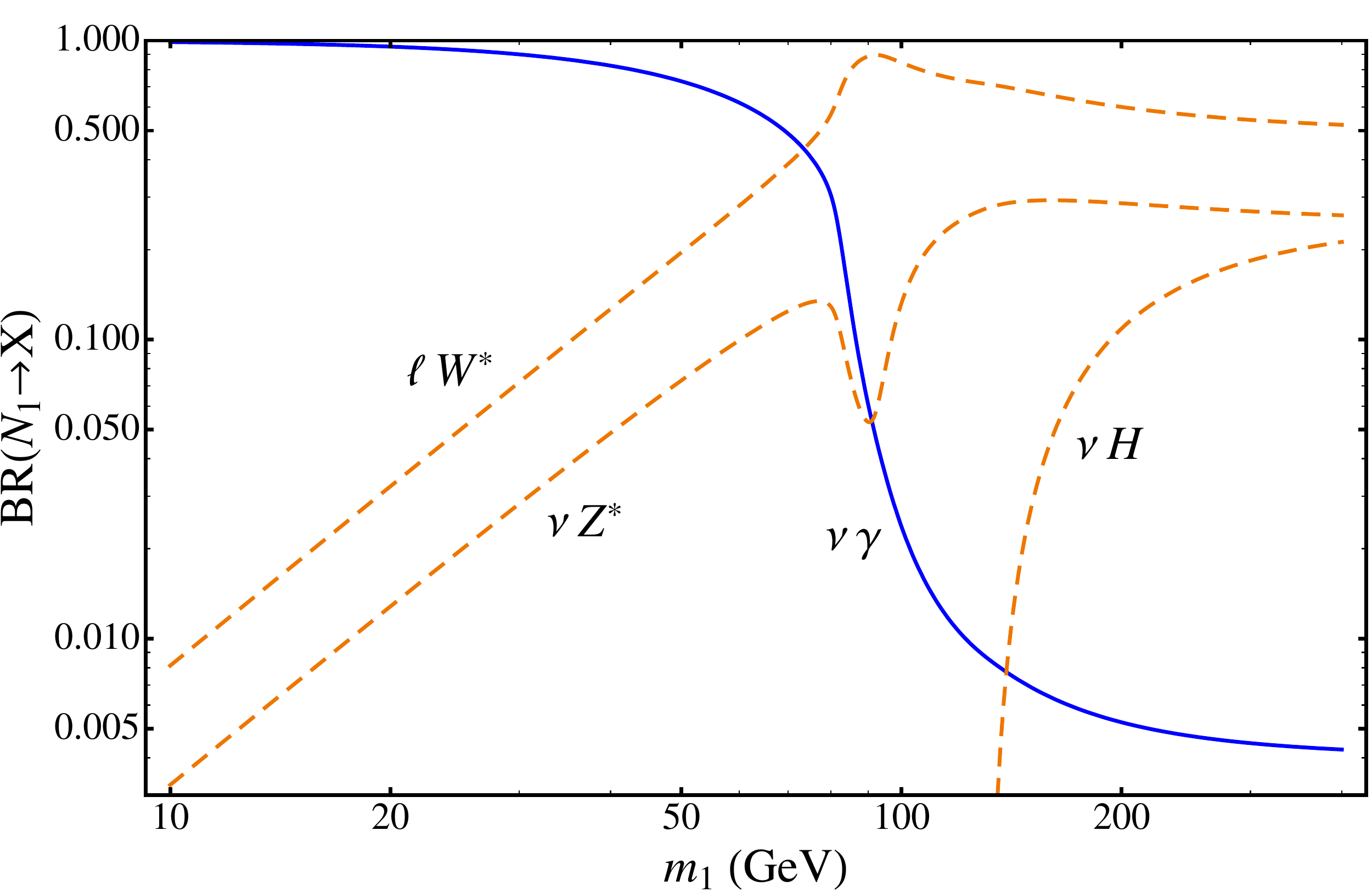}
	\caption{Decay branching ratios for $N_1$. The solid line corresponds to
			the channel mediated by the magnetic moments \emph{and} mixing,
			$N_1 \rightarrow \nu \gamma$, while the dashed lines represent
			the various decays to SM particles through SM vertices and
			mixing. For this figure we took $\varepsilon = 10^{-6}$,
			$\Lambda = 10 \; \mathrm{TeV}$, $m_H = 130 \; \mathrm{GeV}$.
			} \label{fig:nuR-eff-N1-BR}
\end{figure}

The decays of the lightest heavy neutrino are richer, as it decays to a variety of
SM particles. As discussed above, this means that the $N_1$ decays
will always be suppressed by the mixing parameter $\varepsilon$, and the 
corresponding decay lengths will be much longer. In figure \ref{fig:nuR-eff-N1-BR}
we present the dominant decay modes of $N_1$ for a set of typical values of
the parameters. As we can see, the electroweak moments provide a channel 
suppressed by mixing, whose products are a light neutrino and a photon; the
rest of the modes are mediated by SM vertices plus mixing. 
Since all the decay widths are proportional to $\varepsilon$,
the branching ratios will depend weakly on the heavy-light mixing
parameters; they will, however, be sensitive to the strength of the
new magnetic moment interaction. For light $N_1$ masses, under the threshold
of $W$'s and $Z$'s, the electroweak moments prevail over the SM modes;
the magnitude of this prevailance depends mainly on the value of $\Lambda$,
and could serve to measure it. Note, in fact, that the decay width 
$\Gamma (N_1 \rightarrow \nu \gamma)$ is suppressed by $\Lambda^{-2}$, while the 
rates to gauge bosons are not; thus, for relatively small $\Lambda$, say 
$\Lambda \sim 1 \; \mathrm{TeV}$, the mode $N_{1}\rightarrow\nu\gamma$ could 
also be relevant even above the threshold of production of the weak gauge bosons.

For $m_1$ above $m_W$ the decays are dominated by the two-body
mode $N_1 \rightarrow \ell W$ and for masses above $m_Z$ the
decay $N_1 \rightarrow \nu Z$ is also important.
If $m_1 > m_H$, the $N_1$ can also decay into a physical Higgs
boson; the main difference with the gauge boson channels is that the
Higgs width varies tremendously along its allowed mass range; for
light Higgs masses, like the one considered in figure \ref{fig:nuR-eff-N1-BR},
the width is very small and therefore virtual production is suppressed
and the branching ratio drops rapidly for $m_1 \lesssim m_H$. 
Notice that for $m_1 \gg m_H$ the decay widths $\Gamma (N_1 \rightarrow \nu Z)$ 
and $\Gamma (N_1 \rightarrow \nu H)$ converge, and equal one half of
$\Gamma (N_1 \rightarrow \ell W)$, as required by the equivalence theorem
\cite{Lee:1977eg,Cornwall:1974km}; see also, for more on this, the discussion in 
section \ref{sec:nuR-eff-appendix}.

\begin{figure}[tb]
	\centering
	\includegraphics[width=0.8\textwidth]{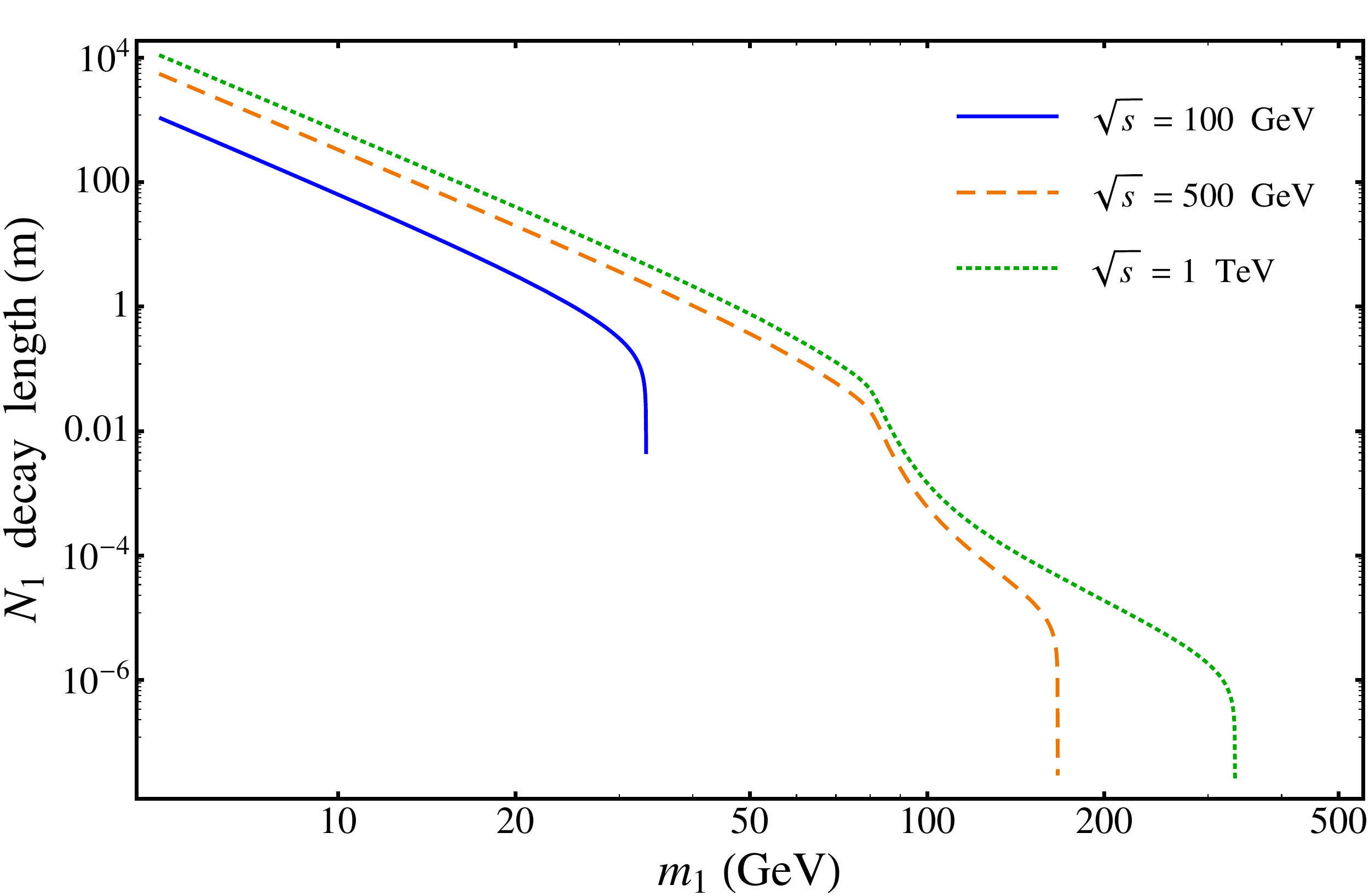}
	\caption{$N_1$ decay lengths for the case of a pair-production of $N_1$ and
			$N_2$ with several center-of-mass energies. The solid line corresponds 
			to
			$100 \; \mathrm{GeV}$, the dashed line to $500 \; \mathrm{GeV}$,
			and the dotted one to $1 \; \mathrm{TeV}$. For this calculation
			we chose $m_2 = 2 m_1$, $\Lambda = 10 \; \mathrm{TeV}$ and
			$\varepsilon = 10^{-6}$.
			} \label{fig:nuR-eff-N1-decaylength}
\end{figure}

The experimental signatures of the $N$ decays are very well-defined,
but they involve photons that can be differently boosted, or single gauge bosons
plus unobservable \mbox{neutrinos -- that is,} missing energy. This means that
the $N$ detection will be affected by a significant background of SM processes.
It would be nice if we had some other feature that can be easily separated 
from the standard background. The $N_1$'s may provide such a feature, for their
decays are suppressed by the mixing factor $\varepsilon$ which can be very small,
and so they might decay away from the collision point and provide a displaced
vertex. Such vertices have been considered as the experimental signatures of
theories with long-lived neutral particles
\cite{Terwort:2008ii,Zalewski:2007up,Prieur:2005xv}, and the same studies
might be applicable to right-handed neutrinos endowed with magnetic 
moments. In order to know if our theory yields such a feature, we have
estimated the decay length for an $N_1$ produced in an accelerator 
with present or near-future energies. The results are not conclusive, because
the decay lenght depends on the mixing factor $\varepsilon$, which can vary
along several orders of magnitude, and on the boost of the $N_1$, which in
turn depends on the production mechanism and the properties of the collider
in which it is produced. Besides, in order to observe a displaced vertex we
need the decay length not to be very short, but we need also that the $N_1$
decays inside the detector; this leaves an experimentally observable range
between some tenths of a millimeter and several meters.
In figure \ref{fig:nuR-eff-N1-decaylength} we present some estimates of the
$N_1$ decay lengths as a function of its mass. Several important assumptions
have been made: the $N_1$ is produced through the electroweak moment interaction
together with an $N_2$ (remember that the electroweak moments are all of them
transition moments, so they cannot produce a $N_1$ pair); the $N_1$ later
decays into the allowed decay channels, all suppressed by $\varepsilon$.
Decay lengths are presented as a function of the
$N_1$ mass for different values of the energy in the center-of-mass frame, and 
assuming $m_2 = 2 m_1$,
$\Lambda = 10 \; \mathrm{TeV}$ and $\varepsilon=10^{-6}$. We observe
that the decay lengths of the $N_1$ will be very small for masses
above $100 \; \mathrm{GeV}$. However, for masses below this value
the decay lengths could range from a few millimeters to a few kilometers,
depending on the $N_1$ and the $N_2$ masses, the heavy-light
mixing, the electroweak coupling and the kinematical configuration
of the experiment. In conclusion, it is possible that the $N_1$'s provide
displaced vertices in collider experiments, but a scenario with too short or
too long decay lengths is also perfectly plausible.

\subsection{Heavy neutrinos in $e^+ e^-$ colliders}
			  	  \label{sec:nuR-eff-eebar-effects}

The electroweak moments provide an optimal mechanism for the production of the
heavy neutrinos, as they couple to photons and $Z$'s; pair-production through
a photon or a $Z$ in the \mbox{$s$ channel --the so-}called Drell-Yan 
\mbox{process--
has become} paradigmatic for charged particles in all sorts of colliders, and
thanks to the electroweak moments it would be also applicable to the chargeless
heavy neutrinos. But precisely because of this, the nonobservation of heavy
neutral leptons in the previous generation of accelerators must place bounds
on the strength of the electroweak interactions and the masses of the heavy
neutrinos. In particular they were not observed at a $Z$ factory such as 
LEP1 \cite{Abreu:1996pa,Adriani:1992pq,Akrawy:1990zq,Dittmar:1989yg}
and its successor LEP2 \cite{Abdallah:2003np,Achard:2003tx,Abbiendi:1998yu}.
The most conservative bounds are obtained
by assuming that all the produced neutrinos escape undetected; but remember
that the electroweak moments are antisymmetric, and thus the heavy neutrinos
are produced in dissimilar pairs. If an $N_1$ is produced, it is relatively
likely that it escapes, because it can only decay
through heavy-light mixing and, as discussed above, the corresponding
decay length could be very large. The heavier states, however, will decay
rapidly into $N_1$ and $\gamma$, with the energetic photon providing a
potentially clear signature. For that reason, it should be possible to
set bounds stronger than those that merely assume that all neutrinos escape,
but these bounds would show some dependence on the neutrino \mbox{spectrum -- for 
instance,} if the $N_1$ and $N_2$ are almost degenerate the
photon could be too soft to provide a viable signal. 
Instead of providing an exhaustive description of all possible scenarios
we will focus on the more diaphanous bound on the invisible decays
of the $Z$ boson, leaving for the end of the section a brief comment on the
bounds derived from visible heavy neutrino decays.

Data from LEP has established with a remarkable precision
that $Z$'s decay essentially to three species of invisible particles, that we 
identify with the three families of active neutrinos. The invisible width of
the $Z$ reads \cite{Beringer:1900zz}
\begin{displaymath}
	\Gamma_{\mathrm{inv}} = 499.0 \pm 1.4 \; \mathrm{MeV} \, ,
\end{displaymath}
whereas we can construct the standard contribution of the three active
neutrinos by using the experimental $Z$ width to charged leptons, 
$\Gamma_{\ell^+ \ell^-} = 83.984 \pm 0.086 \; \mathrm{MeV}$, and the theoretical 
SM calculation for the corresponding widths, $\Gamma_{\bar{\nu} \nu}^{\, \mathrm{th}}
\, / \, \Gamma_{\ell^+ \ell^-}^{\, \mathrm{th}} = 1.991 \pm 0.001$. Then
\begin{displaymath}
	\Gamma_{\nu \bar \nu}^\mathrm{SM} = 3 \, \left( \frac{\Gamma_{\nu \bar 
		\nu}}{\Gamma_{\ell^+ \ell^-}} \right)^\mathrm{th} \, 
			\Gamma_{\ell^+ \ell^-} = 501.6 \pm 0.6 \; \mathrm{MeV} \, ,
\end{displaymath}
which is fairly compatible with the experimental result. A small room, though,
remains for other contributions to the invisible width, provided they are 
rare enough. If we assume that these nonstandard contributions are entirely
given by decays to heavy neutrinos, $\Gamma_{\mathrm{inv}} = \Gamma_{\nu \bar 
\nu}^\mathrm{SM} + \Gamma (Z \rightarrow N_1 N_2)$, we have
\begin{equation} \label{nuR-eff-GammaZNN-rough}
	\Gamma (Z \rightarrow N_1 N_2) = -2.6 \pm 1.5 \; \mathrm{MeV} \, .
\end{equation}
But of course the width is a positive-definite quantity, and this result has to
be interpreted as ``$\Gamma (Z \rightarrow N_1 N_2)$ is compatible with zero''.
The question then is how large can $\Gamma (Z \rightarrow N_1 N_2)$ be, given
the statistical significance of \eqref{nuR-eff-GammaZNN-rough}. We calculate
this value by using the Feldman \& Cousins prescription \cite{Feldman:1997qc}
and obtain
\begin{displaymath}
	\Gamma(Z \rightarrow N_1 N_2) < 0.72 \; \mathrm{MeV} \qquad 95\% \: 
			\mathrm{CL} 
\end{displaymath}
Now, if we look at the theoretical expression for $\Gamma(Z \rightarrow N_1 N_2)$,
given in section \ref{sec:nuR-eff-appendix}, we deduce
\begin{displaymath}
	\Lambda = \frac{1}{|\zeta_{12}|} > 7 \, \sqrt{f_Z (m_Z, m_1, m_2)} \; 
			\mathrm{TeV} \, , 
\end{displaymath}
where $f_Z (m_Z, m_1, m_2)$ is a phase space factor defined in equation
\eqref{nuR-eff-fZ} and normalised in such a way that $f_Z (m_Z, 0, 0) = 1$. 
For example, $\Lambda > 1.9 \; \mathrm{TeV}$ if $m_1 = m_2 = 35 \; \mathrm{GeV}$.

If the right-handed neutrino electroweak moment is large enough to
allow significant production of $N_1, N_2$ pairs at LEP energies,
the dominant decay of $N_2$ will be also mediated by the electroweak moments,
$N_2 \rightarrow N_1 \gamma$, unless the mass of $N_1$ is very close to that of
$N_{2}$. Then, the resulting photons could be detected and separated from the
background if $E_\gamma > 10 \; \mathrm{GeV}$. In fact, searches for
this type of proc\-\mbox{esses --note} that some searches for excited neutrinos 
probe exactly this sort \mbox{of signal-- have} been
conducted at LEP1 \cite{Abreu:1996pa,Abreu:1996vd,Adriani:1992pq,Akrawy:1990zq}
and at LEP2 \cite{Abdallah:2003np,Achard:2003tx,Abbiendi:1998yu}.
If the mass of the heavy neutrino is $\lesssim 90 \; \mathrm{GeV}$,
one typically obtains upper bounds on the production branching ratio,
$BR(Z \rightarrow N_1 N_2)$, of the order of $2-8 \times 10^{-6}$
\cite{Abreu:1996vd,Abreu:1996pa} depending on the masses of $N_1$ and $N_2$.
These bounds usually rely on several assumptions on the properties and spectrum
of the heavy neutrinos; for instance, the limit we just quoted for the branching
ratio assumes that $BR (N_2 \rightarrow N_1 \gamma) = 1$ and that 
$m_2 > 5 \; \mathrm{GeV}$. But this procedure allows to 
set much stronger bounds. For instance, assuming that
$m_1 = 0$ and $m_2$ is relatively light, $10 \; \mathrm{GeV} < m_2 < m_Z$,
we can use the conservative limit $BR (Z \rightarrow N_1 N_2) < 8 \times 10^{-6}$
to obtain $\Lambda > 40 \; \mathrm{TeV}$. Data from LEP2 can also be used to 
place limits on the couplings for masses up to $200 \; \mathrm{GeV}$
\cite{Abdallah:2003np,Achard:2003tx,Abbiendi:1998yu}. For typical
values of $m_1, m_2$ one can set upper bounds on the production cross
section of the order of $0.1 \; \mathrm{pb}$ (this number calculated for 
$\sqrt{s}= 207 \; \mathrm{GeV}$), which translate into bounds on $\Lambda$ of 
the order of a few TeV. In general these bounds depend 
strongly on the spectrum of $N_1$ \mbox{and $N_2$ --for in}\-stance, they are 
completely lost if \mbox{$m_2 - m_1 \lesssim 10 \; \mathrm{GeV}$--, but they} could
be important if some signal of this type is seen at the LHC.

\begin{figure}[tb]
	\centering
	\includegraphics[width=0.8\textwidth]{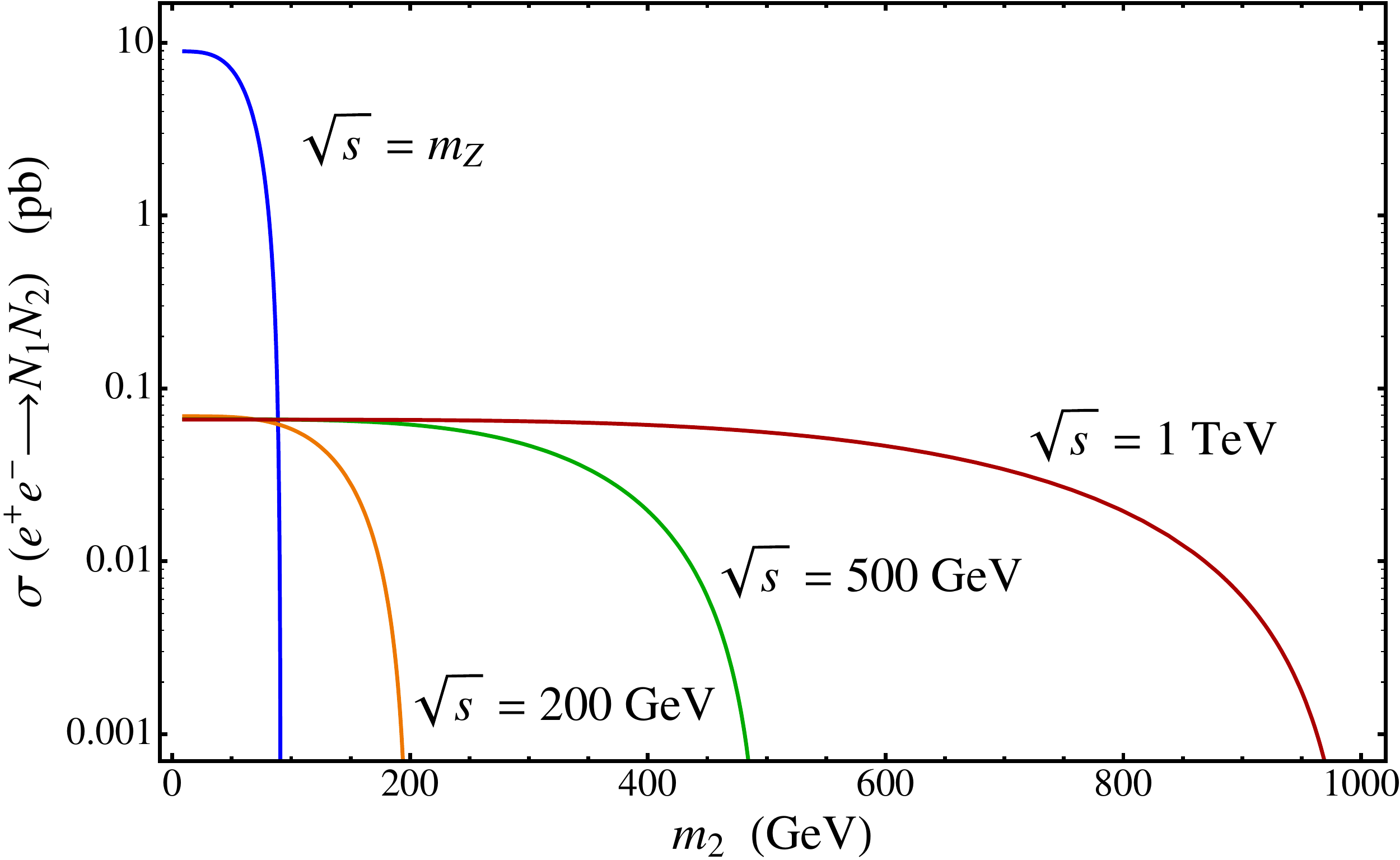}
	\caption{$e^+ e^- \rightarrow N_1 N_2$ as a function of the heavy neutrino
			mass, $m_2$, for different center-of-mass energies. For producing
			this plot we took $m_1 = 0$ and $\Lambda = 10 \; \mathrm{TeV}$.
			} \label{fig:nuR-eff-eebar-production}
\end{figure}

Finally, it is a straightforward calculation to obtain the production
cross section of the heavy neutrinos in a $e^+ e^-$ collider.
We show it in section \ref{sec:nuR-eff-appendix}, and in figure 
\ref{fig:nuR-eff-eebar-production} we plot it as a function of $m_2$ for 
the center-of-mass 
energies of LEP1 and LEP2. We also included results for $\sqrt{s} = 500 \; 
\mathrm{GeV}$ and $\sqrt{s} = 1000 \; \mathrm{TeV}$, in view of the proposals 
for future $e^+ e^-$ colliders like the International Linear Collider (ILC). 
We see that, except for collisions at the $Z$ peak, which are enhanced by about 
two orders of magnitude, or close to the threshold of production, which are 
suppressed by phase space, the cross sections are quite independent of the 
center-of-mass energy
and are determined essentially by the magnitude of the electroweak moments;
for the example value of $\Lambda = 10 \; \mathrm{TeV}$ the cross sections
are of the order of $0.1 \; \mathrm{pb}$.

\subsection{Neutral heavy lepton production at the LHC}
					 \label{sec:nuR-eff-production-LHC}

\begin{figure}[tb]
	\centering
	\includegraphics[width=0.8\columnwidth]{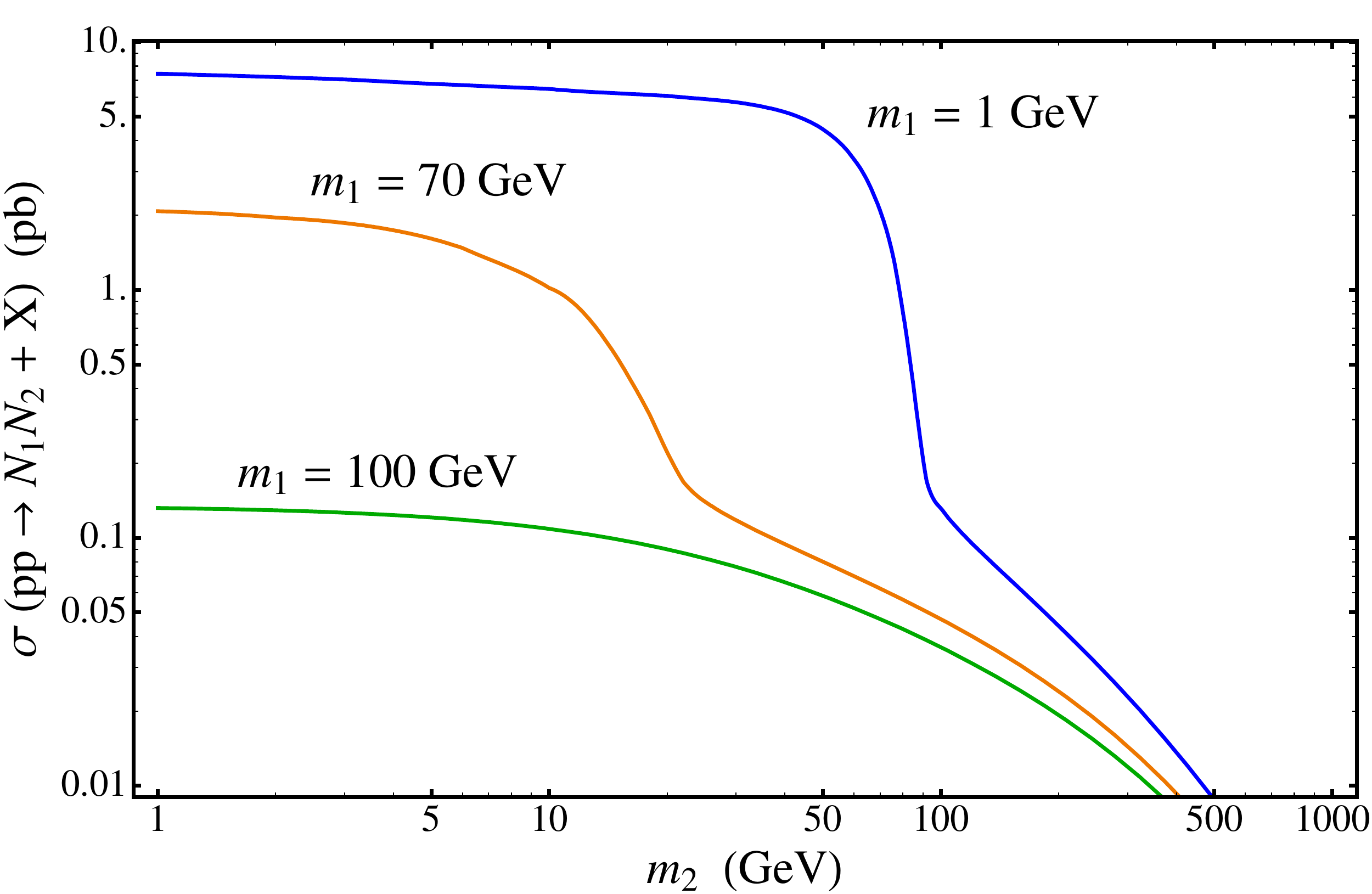}
	\caption{$pp \rightarrow N_1 N_2 + X$ cross section at the LHC 
			as a function of the mass of $N_2$ assuming the
			nominal center-of-mass energy of $\sqrt{s} = 14 \; \mathrm{TeV}$.
			For this plot we chose $\Lambda = 10 \; \mathrm{TeV}$
			and drew three curves for a few representative values of the mass of 
			$N_1$.
			} \label{nuR-eff-LHC-production}
\end{figure}

The right-handed electroweak moment can mediate the production of the heavy
neutrinos at hadron colliders. In particular, at the LHC the main production
mechanism will be the Drell-Yan mechanism, that is, through an $s$-channel 
$Z$ or photon. The differential cross section for proton-proton collisions can 
be computed in terms of the the partonic cross sections as follows%
\footnote{See \cite{Campbell:2006wx} for a very clear review on the matter.
}:
\begin{multline*}
	\mathrm{d} \sigma (pp \rightarrow N_1 N_2 + X) = \sum_{q} \int_{0}^{1}
			\mathrm{d}x_1 \int_{0}^{1} \mathrm{d}x_2 \, \bigl( f_q (x_1, \hat s) 
			f_{\bar q} (x_2, \hat s) + 
			\\
			+ (q \leftrightarrow \bar q) \bigr) \:
			\mathrm{d} \hat \sigma (q \bar q \rightarrow N_1 N_2, \hat s) \, ,
\end{multline*}
where $\hat s = x_1 x_2 s$ is the invariant mass in the partonic center-of-mass
reference frame, $\hat \sigma$ is the partonic cross section, and 
$f_q (x_1, \hat s)$, $f_{\bar q} (x_2, \hat s)$ are the parton distribution 
functions for the proton. Using the partonic cross sections given in section
\ref{sec:nuR-eff-appendix} and performing the convolution over the parton 
distribution functions%
\footnote{We used, in particular, the CTEQ6M parton distribution 
sets \cite{Pumplin:2002vw}, and checked our results against the CompHEP 
software \cite{Pukhov:1999gg,Boos:2004kh}.
}
we find the total cross section as a function of the heavy neutrino
masses, as displayed in figure \ref{nuR-eff-LHC-production}.
The cross section depends on the masses and the coupling $\zeta_{12} = 
1 / \Lambda$. In figure \ref{nuR-eff-LHC-production} we represent the total 
cross section in terms of $m_2$ and offer several curves for representative
values of $m_1$. The main conclusion is that large cross sections can be obtained, 
but mainly for $m_1 + m_2 \lesssim m_Z$, where the LEP bound applies.
The reason for this enhancement is precisely that for light masses
the heavy neutrino pairs can be produced in the decay of a real $Z$, whose
resonant production dominates the Drell-Yan process. But since the decay
$Z \rightarrow N_1 N_2$ has not been observed at LEP, it seems unlikely
that we can observe such enhanced production of heavy neutrinos at the LHC.
For larger masses the cross section decreases rapidly, down to still observable
but less favourable levels.

\begin{figure}[tb]
	\centering
	\includegraphics[width=0.8\columnwidth]{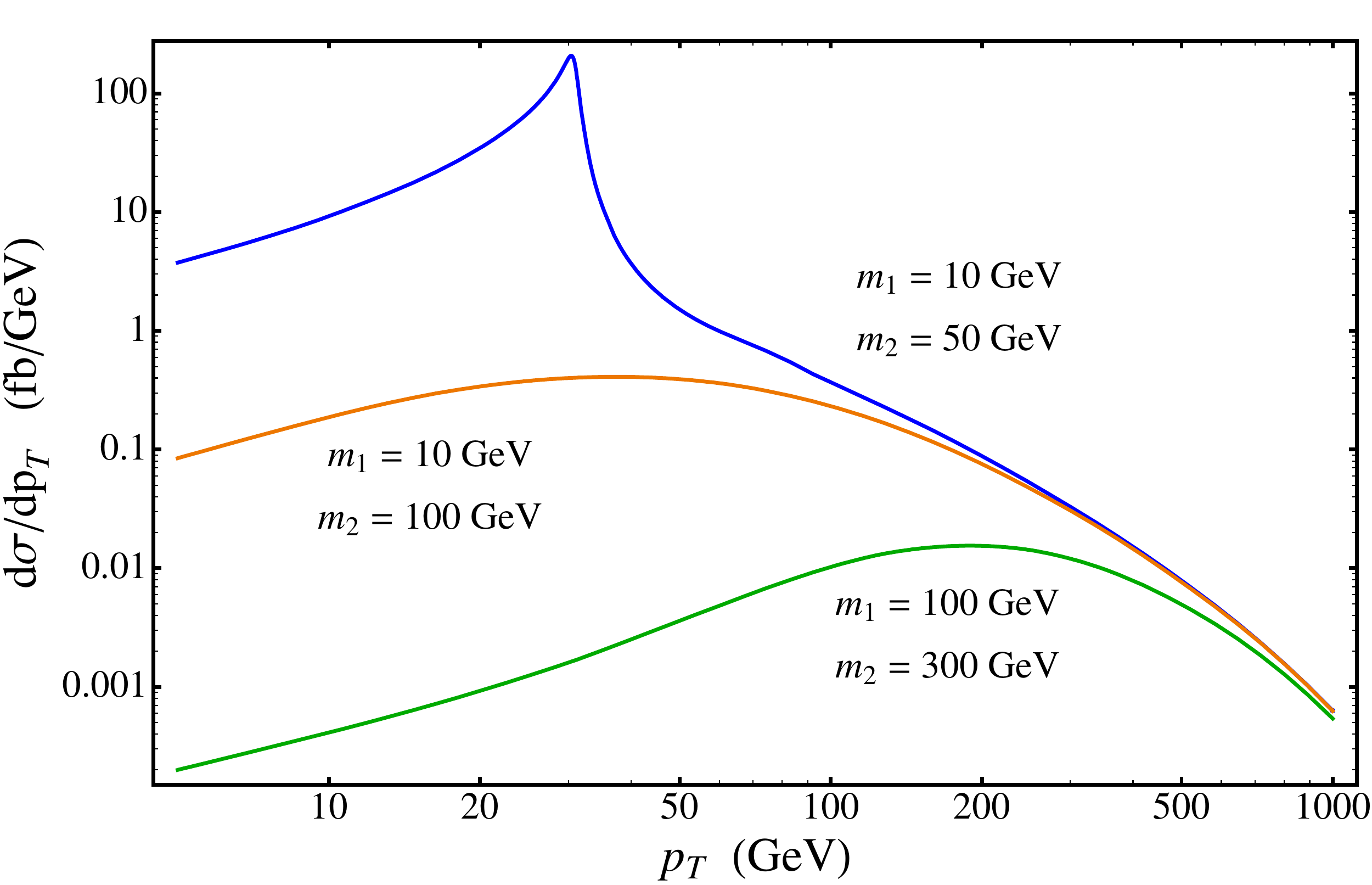}
	\caption{Transverse momentum distribution of the process $pp \rightarrow 
			N_1 N_2 + X$ for different sets of heavy neutrino masses. When the
			sum of the masses allows for it, the production process is dominated
			by a resonant intermediate state with a real $Z$ which shows
			up as a peak roughly at $p_T^\mathrm{res} = m_Z - m_1 - m_2$.			
			} \label{fig:nuR-eff-LHC-differential-sigma}
\end{figure}

Finally, in figure \ref{fig:nuR-eff-LHC-differential-sigma} we present the
differential cross section for the process $pp \rightarrow N_1 N_2 + X$
for different sets of heavy neutrino masses. For $m_1 + m_2 \lesssim m_Z$ we observe
a peak in transverse momentum for the pairs produced through an on-shell $Z$.

\subsection{Higgs decays into heavy neutrinos} \label{sec:nuR-eff-Higgs-to-NN}

In our discussion of the dimension-five Lagrangian we are focusing mainly on
the right-handed neutrino electroweak moments, which are phenomenologically
richer than the other two operators. However, there is one scenario in which
the effective Majorana mass for the $\nu_{\mathrm{R}}^{\, \prime}$ becomes
relevant: as commented in section \ref{sec:nuR-eff-mass-eigenfields}, the $\xi$
operator yields interactions between the physical Higgs boson and the
heavy neutrinos which could be relevant for Higgs searches at the LHC/ILC. 
In particular, it yields a $H NN$ vertex that induces 
new additional decays of the Higgs into heavy neutrinos, if allowed by
kinematics. Let us briefly discuss the possible effects of this operator.

\begin{figure}[tb]
	\centering
	\includegraphics[width=0.8\columnwidth]{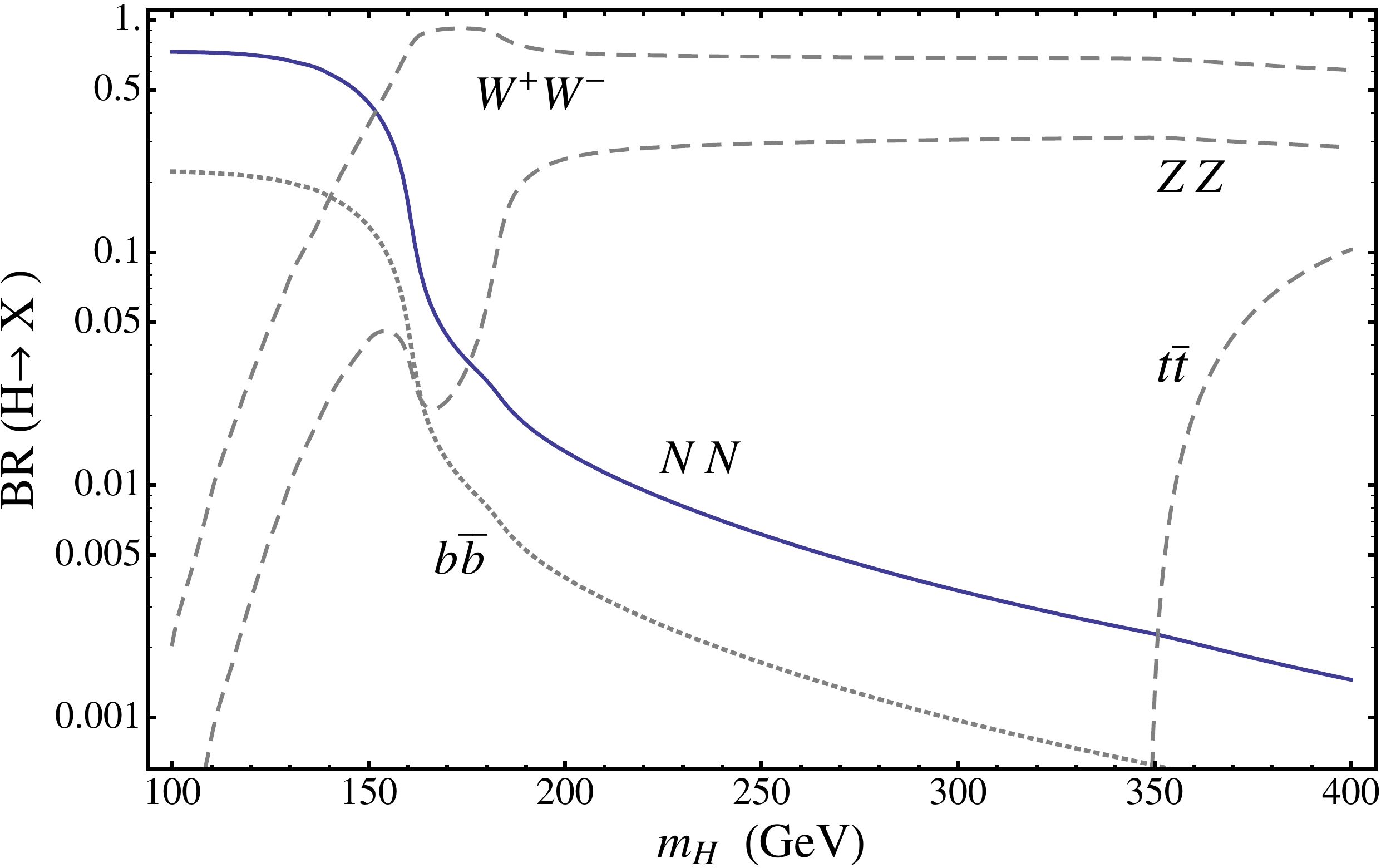} 
	\caption{Estimated branching ratios for Higgs decays with a new physics scale
			of $1 / \xi = 10 \; \mathrm{TeV}$. The solid line corresponds
			to the new channel to heavy neutrinos, which dominates below the
			$WW$ threshold; the usual SM channels are indicated as dashed lines,
			except for the $b \bar b$ channel, which is displayed as a dotted
			line. For this calculation we have neglected the masses of the 
			heavy neutrinos with respect to the Higgs boson mass.
			} \label{fig:nuR-eff-Higgs-BR}
\end{figure}

Equation \eqref{nuR-eff-Higgs-NN-vertex} describes the vertex that is relevant 
for our discussion; several other
interactions, involving heavy-light or light-light pairs of neutrinos have
been omitted, as they are suppressed by the heavy-light mixing, which is 
expected to be small. As a first approach we will ignore also here those
interactions and will focus on the heavy-heavy vertex, parametrised just 
by the effective coupling $\xi$. From \eqref{nuR-eff-Higgs-NN-vertex}
we compute the decay width of the Higgs boson into two heavy neutrinos,
see equation \eqref{nuR-eff-Higgs-to-NN-width}.
Then we can compare with the usual SM decay rates of the Higgs boson; in
figure \ref{fig:nuR-eff-Higgs-BR} we represent the decay branching ratios
into the different channels for a new physics scale of $\Lambda \equiv 1 / \xi = 
10 \; \mathrm{TeV}$, and neglecting the masses of the heavy neutrinos respect to
the mass of the Higgs boson. Notice that, as this interaction can be 
generated at tree 
level, $\Lambda$ can be identified roughly with the masses of the new particles that
generate the \mbox{operator -- see} equation \eqref{nuR-eff-xi-estimate},
but remember that it relies on the new physics being perturbative, and the
couplings $\lambda$ might be small.

Looking at figure \ref{fig:nuR-eff-Higgs-BR} we see that if $m_H$ lies below 
the $WW$ threshold the heavy neutrinos can dominate the decays of the Higgs, 
In fact, for low enough $\Lambda$ these decays could be significant even
when the $WW$ and $ZZ$ channels are open. This has an immediate consequence:
if the heavy neutrinos escape undetected, the detection of the Higgs boson
becomes a challenge; other important modes, such as $H \rightarrow \gamma \gamma$,
become additionally suppressed, and this might even lead to an ``invisible Higgs''
scenario \cite{Ghosh:2012ep,Belanger:2013kya}. 
However, the effect of this new interaction
needs not be troublesome; once produced, the $N$'s have to decay.
If the magnetic moment interaction of right-handed neutrinos is also
present the heaviest neutrinos can decay into lighter ones and photons,
and those photons could be detected. Moreover, the lightest of the
heavy neutrinos will decay into light neutrinos and photons. As discussed
in section \ref{sec:nuR-eff-N-decay}, this is suppressed by the heavy-light mixing,
therefore the $N_1$ could be rather long-lived and produce non-pointing
photons which could be detected. If the magnetic moment interaction
is not present, the heavy neutrinos will have three-body decays of the type
$N_1 \rightarrow W^* \nu$ or $N_1 \rightarrow Z^* \nu$, suppressed by the 
heavy-light mixing, as well as a one-loop-suppressed $N_1 \rightarrow \nu \gamma$
decay. 

In conclusion, the introduction of the $\xi$ operator opens new possibilities for 
the phenomenology of the Higgs boson that can be explored in the near future at 
the LHC, though the recent discovery of the Higgs boson \cite{Aad:2012tfa,
Chatrchyan:2012ufa} and the first measurements of its decay branching ratios
\cite{Chatrchyan:2013zna,Aad:2013wqa} seem to disfavour that the exotic
couplings play a prominent role.

\section{Astrophysical and cosmological considerations} \label{sec:nuR-eff-astro-cosmo}

In this section we consider several astrophysical and cosmological
systems and processes that may be affected by the presence of an extra magnetic
coupling of the neutrinos. The potential number is of course vast; we don't
pretend here to exhaust the possibilities, but rather we present some of the
most interesting scenarios in which the magnetic moment can play a role.

\subsection{Astrophysical effects} \label{sec:nuR-eff-astro}

Among the various astrophysical processes that are affected by neutrino
magnetic couplings, the cooling of red giant stars plays a prominent
role because it provides a very tight bound on the magnitude of the
magnetic \mbox{moments -- provided} the masses of the neutrinos involved
are sufficiently small. This limit is based on the observation that photons
in a plasma acquire a temperature-dependent mass, being
then referred to as \emph{plasmons} ($\gamma_P$); any electromagnetic neutrino 
coupling will then open a decay channel for the plasmon into a neutrino pair,
unless kinematically forbidden. If produced, the neutrinos leave fast the
star, resulting in an additional cooling mechanism that is very sensitive
to the size of the magnetic moment 
\cite{Castellani:1993hs,Catelan:1996-461,Haft:1993jt,Raffelt:1989xu,Raffelt:1990pj,Raffelt:1992pi,Heger:2008er};
this can be used to impose a stringent upper limit on this moment.

The electroweak moment affects in principle only the right-handed degrees of
freedom; this translates into a preferential coupling to the heavy neutrinos,
as we described in section \ref{sec:nuR-eff-mass-eigenfields}. From equation
\eqref{nuR-eff-zeta-mass}, and taking already $U_N = \mathbb{1}$, we write
the magnetic coupling to the heavy species as
\begin{displaymath}
	\mathcal{L}_{\zeta \mathrm{-heavy}} = c_W \, \overline{N} \sigma^{\mu \nu} 
			\left(\zeta P_{\mathrm{R}} + \zeta^{\dagger} P_{\mathrm{L}} \right) 
			N \, F_{\mu \nu} \, .
\end{displaymath}

In a nonrelativistic nondegenerate plasma the emissivity of neutrinos
is dominated by transverse plasmons \cite{Raffelt:1996wa}, which
have an effective mass equal to the plasma frequency, $m_P = \omega_P$.
A calculation \cite{Raffelt:1996wa} shows that the decay width of these 
plasmons into two
neutrino species with definite mass, labeled by $i$ and $j$ and satisfying 
$m_i + m_j < m_P$, is
\begin{displaymath}
	\Gamma( \gamma_P \rightarrow N_i N_j) = \frac{2 c_W^2}{3 \pi} \, 
			\frac{m_P^4}{\omega} \, \left| \zeta_{ij} \right|^2 \,
			f_Z (m_P, m_i, m_j) \, ,
\end{displaymath}
where $\omega$ is the plasmon energy in the plasma rest frame, and $f_Z$
has been defined in equation \eqref{nuR-eff-fZ}. The total decay rate is then
\begin{gather*}
	\Gamma (\gamma_P \to NN) = \frac{\mu_{\mathrm{eff}}^2}{24 \pi} \, 
			\frac{m_P^4}{\omega} \, ,
	\\
	\vphantom{\raisebox{2.5ex}{a}} 
	\mu_{\mathrm{eff}}^2 = 16 c_W^2 \, \sum_{\mathrm{all}} \left| \zeta_{ij} 
			\right|^2 \, f_Z (m_P, m_i, m_j) \, ,
\end{gather*}
where the sum runs over all allowed channels, with $i>j$ and such that 
$m_i + m_j < m_P$. The observational limits from red giant cooling then 
imply \cite{Raffelt:1996wa}
\begin{displaymath}
	\mu_{\mathrm{eff}} < 3 \times 10^{-12} \, \mu_B \, ,
\end{displaymath}
where $\mu_B$ is the Bohr magneton. This translates into a bound
on the couplings $\zeta_{ij}$ \emph{provided} the sum of the associated
neutrino masses lies below $m_P$; for instance, for $\zeta_{ij}$ real
and assuming $m_i, m_j \ll m_P \simeq 8.6 \; \mathrm{keV}$ we have
\begin{displaymath}
	\left| \zeta_{ij} \right| < 8.5 \times 10^{-13} \, \mu_{B} 
		\qquad \Longleftrightarrow \qquad
	\Lambda > 4 \times 10^6 \; \mathrm{TeV} \; .
\end{displaymath}
This bound is somewhat degraded when the neutrino masses are comparable to $m_P$.

This far we have just considered the plasmon decay into the heavy species of
neutrino (those which are primarily $\nu_{\mathrm{R}}^{\, \prime}$); 
it is clear from equation \eqref{nuR-eff-zeta-mass} that the plasmon
can also decay into $N$-$\nu$ and $\nu$-$\nu$ pairs, but with the amplitude 
suppressed by $\varepsilon$ and $\varepsilon^2$, respectively, which are small 
numbers unless the $N$'s are very light (see the discussion in section 
\ref{sec:nuR-eff-mass-eigenfields}). But this case, $m_N \sim m_\nu$, is 
already covered by our discussion above and the `mixed' vertices offer no new
insight. Another option is to consider the case 
in which the plasmon decay
to $N$'s is kinematically forbidden, $m_N > m_P \sim 10 \; \mathrm{keV}$,
and then set bounds on the process $\gamma_P \rightarrow \nu \nu$ 
that proceeds through mixing;
the bound depends in that case on $m_N$, and is generally weak. For instance,
taking $\varepsilon$ at its maximum, $\varepsilon^2 \sim m_\nu / m_N$, and
$m_N$ at its minimum, $m_N = 10 \; \mathrm{keV}$, the bound reduces to just
$\Lambda > 40 \; \mathrm{TeV}$ for a large value of the light mass, $m_\nu = 0.1 \; 
\mathrm{eV}$. With $m_N$ around several hundreds of keV's the bound drops 
already below the TeV.

The same line of reasoning can be applied to other astrophysical objects.
This might be of interest because the corresponding plasma frequency, and
thus $m_P$, will be larger in denser objects, and the corresponding
limits will apply to heavier neutrino states. Unfortunately, the limits
themselves are much poorer. As an example, we consider the case of
a neutron star, whose plasma frequency in the crust is $\omega_P \sim 1 \; 
\mathrm{MeV}$. This could allow us to extend the magnetic moment bounds to 
higher neutrino masses; however, the much weaker limit, $\mu_{\mathrm{eff}} < 
5 \times 10^{-7} \, \mu_B$ \cite{Iwamoto:1994zd} implies $\Lambda \gtrsim 23 
\; \mathrm{TeV}$ when $m_i, m_j \lesssim 1 \; \mathrm{MeV}$,
which is not competitive with bounds derived below from $\gamma + \nu \to N$
in supernovae, which also apply in this range of masses. Limits derived
from plasmon decays in the sun and supernovas are also not competitive
\cite{Raffelt:1996wa,Raffelt:1999gv}.

The neutrino electromagnetic coupling would also affect other interesting
processes. For example, it generates a new supernova cooling mechanism
through $\gamma + \nu \to N$, when kinematically allowed, with the
$N$ escaping. Limits on this anomalous cooling \cite{Raffelt:1996wa}
imply that the effective magnetic moment must lie below $3 \times 10^{-12} \mu_B$
provided the heavy neutrino mass lies \mbox{below $\sim 30 \; \mathrm{MeV}$ --
the or}\-der of the maximum energy of the neutrinos in the supernova core. 
As the process involves a $\nu - N$ transition, the coupling is suppressed
by an $\varepsilon$ factor; this leads to a bound of the form
\begin{displaymath}
	\Lambda \gtrsim 4 \times 10^6 \times \sqrt{m_\nu / m_N} \; \mathrm{TeV}
\end{displaymath}
at its best, i.e., using the most constraining value for the mixing.
Taking, for example, $m_\nu \sim 0.1 \; \mathrm{eV}$ we obtain $ \Lambda
> 1.5 \times 10^4 \; \mathrm{TeV}$ for $m_N = 10 \; \mathrm{keV}$ and 
$\Lambda > 390 \; \mathrm{TeV}$ for $m_N = 10 \; \mathrm{MeV}$. 
These limits are interesting in the region $10 \; \mathrm{keV} < m_N < 30 \;
\mathrm{MeV}$, where red giant bounds do not apply.

\subsection{The electroweak moments as a source of CP asymmetry in the 
			early universe}	  \label{sec:nuR-eff-lepto}

The electroweak moments are also of interest because they violate lepton number
and may contribute to the baryon asymmetry of the universe. As 
is well known, the baryonic content of the observable universe is 
greatly dominated by matter, with only trace amounts of antimatter
\cite{Steigman:1976ev,Cohen:1997ac}; Sakharov proposed in the sixties
that this matter domination needs not to be an initial condition of the
universe, but can be a result of the dynamics of the particle species
if certain conditions are satisfied \cite{Sakharov:1967dj}. 
For this aim, the Standard Model seems to lack
some of the necessary ingredients; in particular it cannot provide 
a strong enough violation of CP so to produce the observed baryon
asymmetry \cite{Gavela:1993ts,Gavela:1994dt,Huet:1994jb}.
The electroweak moments that we are discussing provide several new
phases that are potentially CP-violating, therefore it is worth to check if
they can help in this matter. As the theory presents
a number of heavy neutrinos it is natural to consider a scheme of 
baryogenesis through leptogenesis \cite{Fukugita:1986hr,Flanz:1996fb}. 

In the leptogenesis scenario%
\footnote{It would go beyond the scope of this work to heartily review the
intricate details of baryogenesis and leptogenesis. To the interested
reader we recommend the good reviews \cite{Dolgov:1991fr,Davidson:2008bu}.
} 
the baryon asymmetry is not generated in the
quark sector, but rather it originates in the dynamics of leptons and
then it is transferred to the baryons through certain nonperturbative
electroweak processes \cite{Rubakov:1996vz}. The way in which the lepton
asymmetry is first produced depends on the particular realisation of
leptogenesis, 
but a standard mechanism is the CP-violating decays of heavy netrinos. 
These processes are affected by \mbox{CP-odd phases --in 
our case,} the physical phases that appear in the electroweak \mbox{moments-- 
and by} phases that do not change sign under CP transformations, these ones 
appearing in the one-loop diagrams of the $N$ decay as imaginary parts of
the loop integral. The interference between the two classes of phases
generates in the end different rates for the two CP-conjugated channels,
$N \rightarrow \ell_{\mathrm{L}} \phi$ and $N \rightarrow \bar \ell_{\mathrm{L}} 
\bar \phi$, and thus a different number of leptons and antileptons.
The electroweak moments do not change substantially this panorama:
they just provide a new contribution to the one-loop correction, with a $B$
gauge boson running inside the loop and $\zeta$ delivering some of the CP-odd
phases.

\begin{figure}[p]
	\centering
	\includegraphics[width=0.4\textwidth]{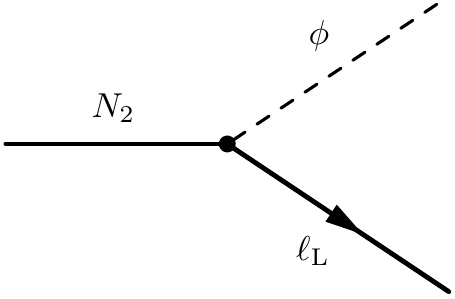}

	\vspace{1cm}
	\includegraphics[width=0.55\textwidth]{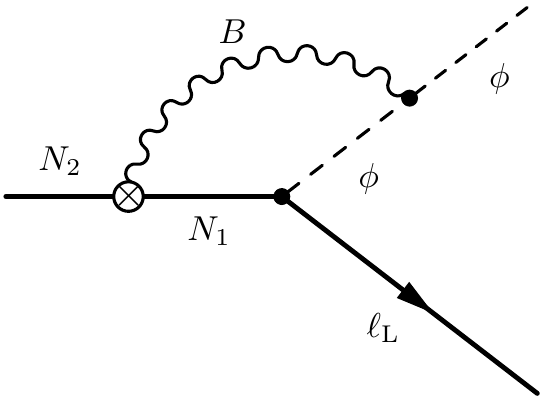}

	\vspace{0.4cm}
	\includegraphics[width=0.55\textwidth]{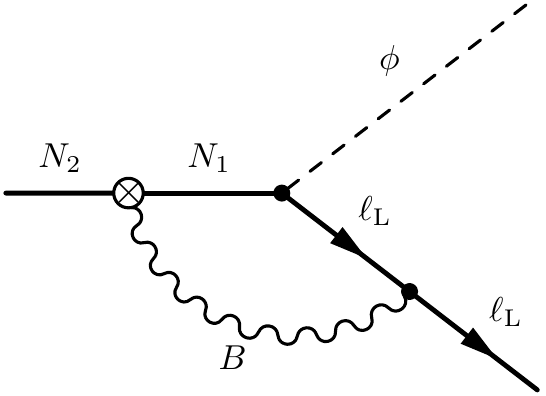}
	\caption{The tree-level and one-loop diagrams that allow to generate a lepton
			asymmetry through the electroweak moments. The effective vertices
			are marked as crossed circles.
			} \label{fig:nuR-eff-lepto}
\end{figure}

The relevant diagrams are presented in figure \ref{fig:nuR-eff-lepto}. Note that
the one-loop corrections require a second heavy neutrino due to the antisymmetry
of the electroweak moments; and as we need these graphs to yield an imaginary part,
by the optical theorem the `inner' heavy neutrino must be lighter than the outer 
one. This scenario is not free of problems: if $m_1 < m_2$ then the channel 
$N_2 \rightarrow N_1 \gamma$ is open, and it will in most cases dominate the
decays of $N_2$, leaving a negligible number of CP-violating leptonic decays.
The only situation in which the $\zeta$-mediated mechanism can be relevant
is when $N_2$ and $N_1$ are nearly degenerate; this scenario is usually 
referred to as \emph{resonant leptogenesis}
\cite{Pilaftsis:1997jf,Pilaftsis:2003gt}.

We now go on to calculate the lepton asymmetry emerging from the processes in
figure \ref{fig:nuR-eff-lepto}. We will assume that $m_N \gg v$ so that we can 
neglect electroweak symmetry breaking and assume that all gauge bosons, leptons 
and scalars are massless except for the heavy neutrinos, which have a 
Majorana mass 
term. Also, for simplicity, we neglect the Yukawa couplings of the charged leptons; 
as the electroweak moments provide new sources of CP violation, the Yukawas
need not be an essential piece in producing a lepton asymmetry. The relevant
parts of the Lagrangian are discussed in sections \ref{sec:nuR-eff-presentation}
and \ref{sec:nuR-eff-mass-eigenfields}, but we recapitulate them here:
\begin{align*}
	\mathcal{L}_N  = \; & \frac{i}{2} \: \overline{N} \gamma_\mu \partial^\mu N 
			\: - \: \frac{1}{2} \, \overline{N} M_N N \: - \: 
			\left( \overline{\ell_{\mathrm{L}}}	\, Y_\nu \, P_{\mathrm{R}} N \, 
			\tilde{\phi} \: + \mathrm{H.c.} \right) \: + 
			\\
		\vphantom{\raisebox{2.5ex}{a}} 
		&+ \: \overline{N} \sigma^{\mu \nu} (\zeta P_{\mathrm{R}} + \zeta^\dagger
			P_{\mathrm{L}} ) \, N \, B_{\mu \nu} \, ,
\end{align*}
where $N$ are Majorana fields and $M_N$ is their mass matrix which,
without loss of generality, can be taken diagonal. Since we ignore
the charged lepton Yukawa couplings we can rotate the doublet fields
$\ell$ so that $Y_\nu$ is Hermitian; there are no other possible
field redefinitions, so $\zeta$ is, in general, antisymmetric and
complex. For $n$ generations both $Y_\nu$ and $\zeta$ contain
$n (n-1) / 2$ phases; in particular, for $n=3$ we will have a total
of $6$ phases. But even for $n = 2$ we have two phases since both
$Y_{12}$ and $\zeta_{12}$ can be complex. This is important because
$CP$-violating observables should depend on those couplings; it
also means that we can make our estimates in a model with just two
generations, as we will do for simplicity.

Assuming two generations with $N_2$ the heavier of the heavy
neutrinos, we consider the lepton-number-violating decays $N_2 \to \ell_e \phi$
and $N_2 \to \bar \ell_e \bar \phi$. At tree level the amplitudes are simply
\begin{align*}
	\mathcal{A}_0 (N_2 \to \ell_e \phi) &= Y_{e2} \: \bar{u}(p_e) P_R u(p_2) 
	\\
	\vphantom{\raisebox{2.5ex}{a}} 
	\mathcal{A}_0 (N_2 \to \bar \ell_e \bar \phi) &= Y_{e2}^* \:
			\bar{v}(p_2) P_L v(p_e) = - Y_{e2}^* \: \bar{u}(p_e) P_L u(p_2) \, ,
\end{align*}
where we used $v(p)=u^{c}(p)$. The one-loop corrections that we show in
figure \ref{fig:nuR-eff-lepto} require a tedious but otherwise straightforward
calculation. Naive application of the optical theorem indicates that the loop
should develop an imaginary part if $m_1 < m_2$; this expectation is confirmed,
and such CP-even phase allows the process to yield CP-dissimilar amplitudes.
We compute such asymmetry using the standard $\epsilon$ parameter,
\begin{displaymath}
	\epsilon \equiv \frac{\Gamma (N_2 \rightarrow \ell_e \phi) - 
		\Gamma (N_2 \rightarrow \bar \ell_e \bar \phi)}{\Gamma (N_2 \rightarrow 
		\ell_e \phi) + \Gamma (N_2 \rightarrow \bar \ell_e \bar \phi)} \, ,
\end{displaymath}
which in our case results
\begin{displaymath}
	\epsilon = - \frac{g^\prime}{2 \pi} \, (m_2^2 - m_1^2) \, \frac{m_1}{m_2^3} \,
		\mathrm{Im}\left[ \frac{Y_{e2} Y_{e1}^*}{|Y_{e2}|^2} \, 
		\left( \zeta_{12}^* m_2 + \zeta_{12} m_1 \right) \right] \, .
\end{displaymath}
Now, particularising to $m_1 \ll m_2$ we obtain
\begin{displaymath}
	\epsilon = - \frac{g^\prime}{2 \pi} \, \frac{m_1}{\Lambda} \, 
		\mathrm{Im}\left[ \frac{Y_{e2} Y_{e1}^*}{|Y_{e2}|^2} \, e^{-i \delta_{12}}
		\right] \, ,
\end{displaymath}
where we defined the complex $\zeta_{12}$ as $\zeta_{12} \equiv e^{i \delta_{12}} /
\Lambda$, with $\Lambda$ real.

In conclusion, we find that the Majorana electroweak moments do generate 
additional
contributions to CP-violating asymmetries in heavy neutrino decays.
These, however, are relevant only for the decays of the heavier neutrinos
and could be relevant for leptogenesis only when $m_1$ and $m_2$
are relatively close; in this limit the amplitude is suppressed by a factor 
$m_2^2 - m_1^2$. In the following section we estimate the region in 
parameter space where this source of CP violation can be relevant for
leptogenesis.

\section{Summary of bounds and conclusions}
						\label{sec:nuR-eff-conclusions}

As can be seen from the previous sections, the dimension-five operators
involving right-handed neutrinos open up observable effects in several
scenarios of interest. The electroweak moment operator provides the richest
phenomenology, but there are other interesting possibilities. We review in this
section the phenomenology of these operators. 

The $\xi$ operator in equation \eqref{nuR-eff-Lagrangian-5} is most of all 
prominent for its ability to affect Higgs boson decays.
After spontaneous symmetry breaking, this operator gives rise
to several interaction vertices involving heavy neutrinos and
the physical Higgs boson, the strongest being a simple $H \: N_i N_j$ trilinear,
which provides new decay channels of the Higgs to $N$'s, if such
a process is kinematically allowed. These decays could dramatically
change the branching ratios for the Higgs, see figure \ref{fig:nuR-eff-Higgs-BR},
especially in the region $100 \; \mathrm{GeV} < m_H < 160 \; \mathrm{GeV}$,
where the gauge boson channels are still closed. The new decays could
result in an invisible Higgs, if the heavy neutrinos cannot be detected,
or in new, enhanced detection channels if they can be seen through their own 
decay channels, for instance $N_2 \rightarrow N_1 \gamma$,
or $N_1 \rightarrow \nu \gamma$ and $N_1 \rightarrow e W$ with a
displaced vertex. The first measurements of the Higgs boson branching
ratios at the LHC \cite{Chatrchyan:2013zna,Aad:2013wqa} seem to indicate
that the decays to heavy neutrinos are not dominant; a precise determination
of the $b \bar b$ branching ratio, which is the main SM decay channel for a Higgs
mass of $125 \; \mathrm{GeV}$, will reveal if there is room for these
exotic decays. Subsequent measurements of the whole set of SM branching
ratios will allow to strongly constrain this effective interaction.

\begin{figure}[tb]
	\centering
	\includegraphics[width=0.8\columnwidth]{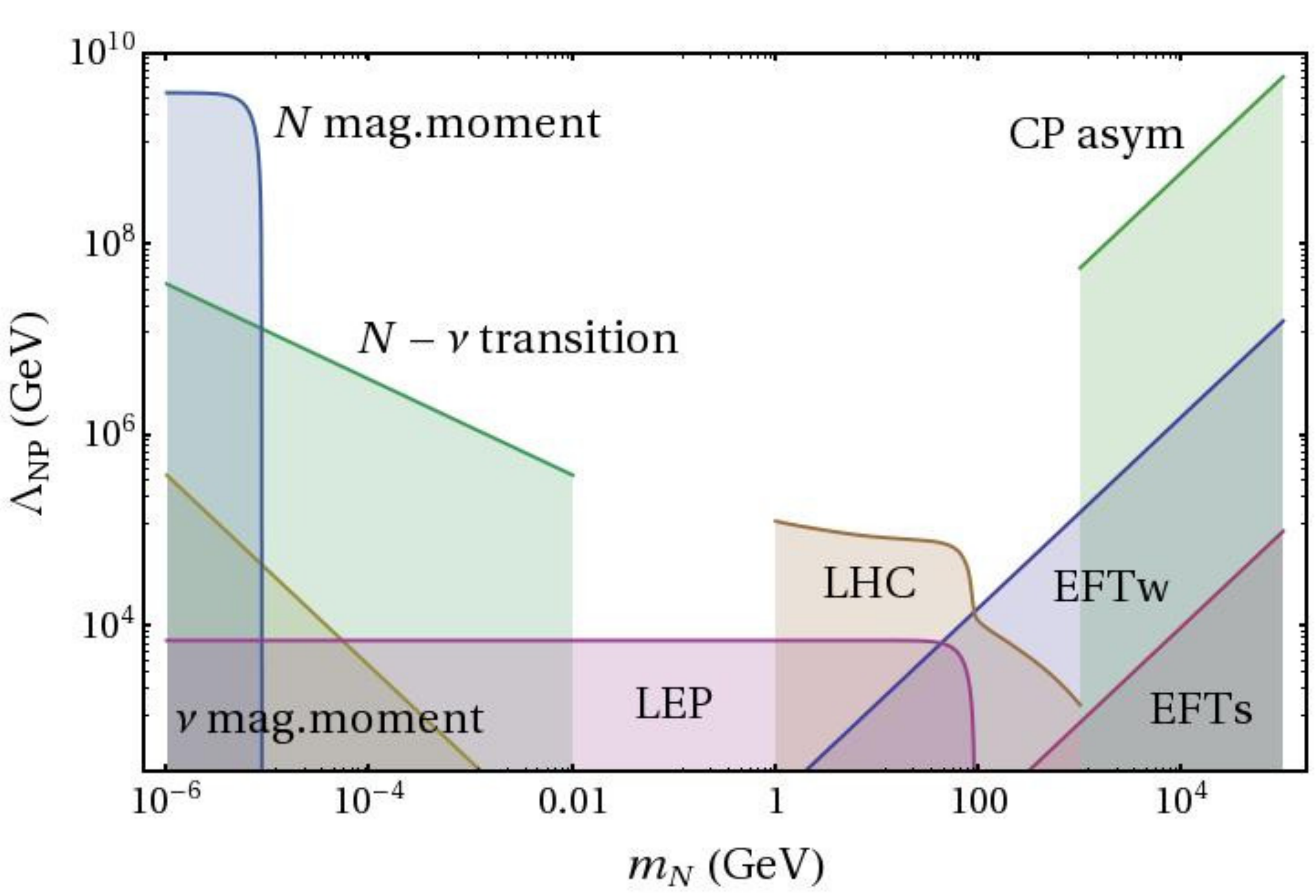}
	\caption{Summary of bounds and prospects for the electroweak moment operator.
			All the shaded areas represent exclusion regions, except for
			the ones marked as ``LHC'' and ``CP asym'' which are regions of
			interest. For the construction of this plot we assumed a maximal,
			seesaw-like mixing, $\varepsilon \sim \sqrt{m_\nu / m_N}$ with
			$m_\nu = 0.1 \; \mathrm{eV}$; for the processes where two heavy 
			neutrinos are in play we assumed $m_N \equiv m_2$ and neglected $m_1$.
			More details on the interpretation of the several regions can be
			found in the main text.
			} \label{fig:nuR-eff-summary-bounds}
\end{figure}

Most of this chapter has been devoted to the electroweak moment operator;
and indeed it presents a variety of observable effects that make it very 
interesting from the phenomenological point of view.
When expanded in terms of mass eigenstates the unique electroweak
moment operator generates $N-N$, $N-\nu$ and $\nu-\nu$ magnetic
moments, and $N-N$, $N-\nu$ and $\nu-\nu$ tensor couplings to the
$Z$-bosons, as we show in equation \eqref{nuR-eff-zeta-mass}. All the
phenomenology of these interactions is controlled essentially by three
parameters: the effective \mbox{coupling --or, equiv}\-alently, the new physics
\mbox{scale--, $\zeta \equiv 1 / \Lambda$,} the heavy-light mixing, $\varepsilon$, 
and the mass of the heavy neutrino, $m_N$. In figure 
\ref{fig:nuR-eff-summary-bounds} we summarise, in terms of $\Lambda$ and
$m_N$, the constraints discussed in the various sections of this chapter.
For that aim, several assumptions are made: first, the mixing is supposed
to have a seesaw-like structure, $\varepsilon \sim \sqrt{m_\nu / m_N}$;
then the light neutrino masses are assumed to lie at $m_\nu = 0.1 \; \mathrm{eV}$;
as for the heavy neutrino masses, they are identified with $m_N$ except in the
cases where two heavy neutrinos are involved, in which we take $m_N \equiv m_2$
and we neglect $m_1$.

With these assumptions, figure \ref{fig:nuR-eff-summary-bounds} shows 
the regions in the $\Lambda - m_N$ plane forbidden by the red
giant bound on the $N$ and $\nu$ magnetic moments, by the supernova
bound on the transition magnetic moment $N - \nu$ and by the LEP bound
from the invisible $Z$ boson decay width. Two more ``exclusion'' regions
are displayed: the areas labeled as ``EFTw'' and ``EFTs'' represent the
regions where the effective theory would be no longer valid in the
weak and strong coupling regime, respectively; inside these regions our
calculations make no sense by construction, and particular models should be 
used to examine the heavy neutrino interactions. We assume that the failure
of the effective approach occurs when the low-energy degrees of \mbox{freedom
--the heavy} neutrinos, in this \mbox{case--
become} heavier than the new physics excitations; this happens in the
strong regime for $m_N > \Lambda$ and in the weak regime for $m_N > \Lambda / 
(4 \pi)^2$, see section \ref{sec:nuR-eff-NP-strong-or-weak}. As these
are only order-of-magnitude estimations, the EFTw and EFTs regions should
be taken with a grain of salt.

Apart from these excluded regions, figure \ref{fig:nuR-eff-summary-bounds}
presents two more which could be described as ``regions of interest'': one in
which the electroweak moments can produce observable signals at the LHC, 
and other where a sizable lepton asymmetry can be generated in the decays
of the heavy neutrinos. There is some uncertainty as to how to define the
LHC region of interest: to test the new interactions at the LHC one should first 
produce the heavy neutrinos and after that one should be able to detect them. 
The analysis of the detection is complicated and depends on the details of the 
spectrum and the capabilities of the detectors, so we choose to make a 
conservative requirement: we demand that the heavy neutrinos are produced
with rates large enough. In particular, we plot the region where the cross 
section of $pp \rightarrow N_1 N_2 + X$ is at least $100 \; \mathrm{fb}$. 
The area is interrupted below $1 \; \mathrm{GeV}$ because our calculation,
which uses partons and the proton parton distribution functions, becomes
unreliable; for low $N$ masses another approach, maybe involving chiral
perturbation theory, should be used.

As for the second region of interest, we have discussed in section 
\ref{sec:nuR-eff-lepto} that the electroweak moments contain new sources of CP
nonconservation which can modify the standard leptogenesis scenarios.
In particular we found that the electroweak moment operator gives
additional contributions to the CP asymmetry in heavy neutrino decays.
These could be relevant in leptogenesis if $\epsilon \sim (g^\prime / 2\pi) \, 
m_N / \Lambda > 10^{-6}$ and $m_N > 1 \; \mathrm{TeV}$. We represent this
region in figure \ref{fig:nuR-eff-summary-bounds} and label it as ``CP asym''.

Figure \ref{fig:nuR-eff-summary-bounds} draws the following conclusions about
the electroweak moments:
\begin{itemize}
\item[\emph{i)}] Red giants cooling imposes very tight bounds for $m_N \lesssim 
		10 \; \mathrm{keV}$, strong enough as to require $\Lambda > 4 \times 10^9 
		\; \mathrm{GeV}$; in this region of the parameter space, obviously, any 
		effect of the electroweak moment coupling would be totally negligible for
		any present or planned collider experiment.

\item [\emph{ii)}] For $10 \; \mathrm{keV} \lesssim m_N \lesssim 10 \; \mathrm{MeV}$
		supernova cooling produced by the magnetic moment transitions $\gamma + 
		\nu \rightarrow N$ provides strong bounds. These bounds, however, depend on 
		the assumptions made on the heavy-light mixing parameter, $\varepsilon$.
		For this mass range there are also limits from red giant cooling, but they
		are derived from plasmon decay into a $\nu$ pair, which is proportional
		to $\varepsilon^2$ and yields less restrictive constraints.

\item [\emph{iii)}] For $m_N \lesssim m_Z$, the invisible $Z$ decays require
		$\Lambda \gtrsim 7 \times 10^3 \; \mathrm{GeV}$, depending on the
		details of the heavy neutrino spectrum.

\item [\emph{iv)}] For $m_N \sim 1-200 \; \mathrm{GeV}$, and roughly
		$7 \; \mathrm{TeV} < \Lambda < 100 \; \mathrm{TeV}$, heavy neutrinos
		could be produced at the LHC with large cross sections. The production
		mechanism necessarily yields two different mass eigenstates due to
		the antisymmetry of the Majorana electroweak moments. 
		The heaviest neutrino in the pair would decay rapidly to hard photons which
		could be detected. The lightest one is quite long-lived and, in a fraction
		of the parameter space, would produce non-pointing photons which could
		be detected.

\item [\emph{v)}] For masses above the TeV the electroweak moments can potentially
		interfere with the standard leptogenesis processes, even providing
		by themselves a significant lepton asymmetry.
\end{itemize}

A caveat is in order about figure \ref{fig:nuR-eff-summary-bounds} and the 
conclusions drawn above. As commented in section 
\ref{sec:nuR-eff-NP-strong-or-weak}, throughout this chapter we have defined
the new physics scale as $\Lambda \equiv 1 / \zeta$.
Since our operator is a magnetic-moment-like operator, this scale
corresponds directly to the mass of new particles only in a nonperturbative
context. If the interaction is generated by perturbative physics it arises at one
loop and one expects $\zeta \sim 1 / ((4\pi)^2 \, m_{\mathrm{NP}})$, where 
$m_{\mathrm{NP}}$ are the masses of the particles running inside the loop.
Thus, for the perturbative case all the constraints discussed
above apply, but can be reexpressed in terms of the masses of the new
particles using $m_{\mathrm{NP}} = \Lambda / (4\pi)^2$. A particularly interesting
instance for this change is the region of interest for the LHC; in that case,
the relevant range $\Lambda \sim 10 - 100 \; \mathrm{TeV}$ translates
into $m_{\mathrm{NP}} \sim 100 - 1000 \; \mathrm{GeV}$. For such low masses the
effective theory cannot be applied at LHC energies and one should
use the complete theory that gives rise to the right-handed neutrino electroweak
moments. Those models should contain new particles with nonzero hypercharges
and masses around $100 - 1000 \; \mathrm{GeV}$ which should be
produced in the LHC via, for instance, Drell-Yan processes. Most likely, 
weakly-coupled theories providing these electroweak moments will be probed 
more efficiently by looking for the new particles than they would be by observing
the effective interactions.

\section{Appendix: Decay rates and cross sections} 
					  \label{sec:nuR-eff-appendix}

In this section we collect the formulas that describe the decay rates and the
cross sections referred to in the text. Let us first introduce some notation
that will be useful to simplify the presentation of the expressions. As usual,
$s_W \equiv \sin \theta_W$ and $c_W \equiv \cos \theta_W$ will represent
the sine and cosine of the weak mixing angle. We will denote by $q_f$ the 
electric charge of a fermion $f$, and its vector and axial couplings will
be denoted by $v_f = t_3(f) \, \left( 1 - 4 |q_f| s_W^2 \right)$ and 
$a_f = t_3(f)$, respectively, with $t_3(f) = + 1/2 \, (-1/2)$
for up-type (down-type) fermions. When required, we will separate modulus
and phase of the effective electroweak coupling as $\zeta_{ij} \equiv 
|\zeta_{ij}| \, e^{i \delta_{ij}}$. At several moments we will use Källen's
lambda function, which reads
\begin{displaymath}
	\lambda(a,b,c) = a^2 + b^2 + c^2 - 2ab - 2ac - 2bc \, . 
\end{displaymath}

\subsubsection{$Z \rightarrow N_i N_j$}

The decay width of the $Z$ boson into heavy neutrinos is 
\begin{displaymath}
	\Gamma (Z \rightarrow N_i N_j) = \frac{2}{3 \pi} \, 
			s_W^2 |\zeta_{ij}|^2 m_Z^3 f_Z (m_Z, m_i, m_j) \, ,
\end{displaymath}
where $f_Z(m_Z, m_i, m_j)$ is a kinematic factor that reads
\begin{multline} \label{nuR-eff-fZ}
	f_Z (m_Z, m_i, m_j) = \frac{\sqrt{\lambda (m_Z^2, m_i^2, m_j^2)}}{m_Z^6} \, 
			\Bigl[ m_Z^2 \, \left(m_Z^2 + m_i^2 + m_j^2 - \right.
			\\
			\left. - 6 m_i m_j \cos 2 \delta_{ij} \bigr) - 2 \, 
			\left( m_i^2 - m_j^2 \right)^2 \right] \, .
\end{multline}
Note that $f_Z$ is defined in such a way that $f_Z (m_Z,0,0) = 1$.

\subsubsection{$N_2$ decay rates}

If the electroweak interaction is strong enough, the dominant decays of the
heavier neutral leptons will proceed to photons or $Z$'s plus one of the 
lighter heavy neutrinos. In our simplified spectrum with just two heavy 
neutrinos we would have
\begin{align*}
	\Gamma( N_2 \rightarrow N_1 \gamma) &= \frac{2}{\pi} \, c_W^2 |\zeta_{12}|^2
			m_2^3 \, \left( 1 - \nicefrac{m_1^2}{m_2^2} \right)^3
	\\
	\vphantom{\raisebox{3.5ex}{a}} 
	\Gamma( N_2 \rightarrow N_1 Z) &= \frac{2}{\pi} \, s_W^2 |\zeta_{12}|^2
			m_2^3 \, f_2 (m_Z, m_1, m_2) \, ,
\end{align*}
where 
\begin{displaymath}
	f_2 (m_Z, m_1, m_2) = - \frac{m_Z^6}{2 m_2^6} \, f_Z (m_Z, m_1, m_2) \, , 
\end{displaymath}
again normalised in such a way that $f_2 (0, 0, m_2) = 1$.

\subsubsection{$N_1$ decay rates} 

The lightest of the heavy neutrinos, $N_1$, can decay only due
to mixing with the SM sector. The mixing between left- and right-handed
neutrinos induces heavy-light transition interactions in the SM weak vertices,
as described at the end of section \ref{sec:nuR-eff-mass-eigenfields}. This 
opens up three SM-like decay channels, whenever kinematically allowed: two of them
are neutral, with emission of light neutrino plus a $Z$ or a Higgs boson, and 
one charged, involving a charged lepton and a $W$ in the final state. Actually, 
these decays are a consequence of the $\nu_{\mathrm{R}}^{\, \prime}$ couplings
to the four degrees of freedom of the Higgs doublet; as three of those degrees
of freedom end up inside the weak gauge bosons, most of them proceed with a 
gauge boson in the final state. We will later come back to this point.

Provided $m_1 > m_W$, the charged channel is to occur with the following rate:
\begin{equation} \label{nuR-eff-N1-to-W}
	\Gamma (N_1 \rightarrow \ell_\beta W) = \frac{1}{16} \, 
			\left| \varepsilon_W^{\beta 1} \right|^2 \alpha \, \frac{m_1^3}
			{s_W^2 m_W^2} \, \left( 1 - \frac{m_W^2}{m_1^2} \right)^2 
			\left(1 + 2 \, \frac{m_W^2}{m_1^2} \right) \, ,
\end{equation}
where $\alpha$ is the fine-structure constant and $\varepsilon_W$ characterizes
the mixing of heavy-light neutrinos in $W$ boson couplings, which
is order $\sqrt{m_\nu / m_N}$ or smaller.

For the first neutral channel, with a $Z$ in the final state, we obtain
\begin{equation} \label{nuR-eff-N1-to-Z}
	\Gamma ( N_1 \rightarrow \nu_\beta Z ) = \frac{1}{16} \, 
			\left| \varepsilon_Z^{\beta 1} \right|^2 \alpha \, \frac{m_1^3}
			{s_W^2 \, c_W^2 m_Z^2} \, \left( 1 - \frac{m_Z^2}{m_1^2} \right)^2
			\left( 1 + 2 \, \frac{m_Z^2}{m_1^2} \right)
\end{equation}
whenever $m_1 > m_Z$, with $\varepsilon_Z$ the mixing factor that 
corresponds to the $Z$ couplings.
Notice that since $m_W = c_W m_Z$ the two decay widths are equal up to phase 
space factors and differences in the mixing factors $\varepsilon_Z$
and $\varepsilon_W$. However, we have two decay channels into $W$'s,
$N_1 \rightarrow e^- W^+$ and $N_1 \rightarrow e^+ W^-$,
and only one into $Z$'s. This is an indication that there are two charged 
degrees of freedom in $\phi$, but only one of the neutral ones goes to the $Z$.

If $m_1 > m_H$ the second neutral channel is also open, and its decay rate is
\begin{equation} \label{nuR-eff-N1-to-Higgs}
	\Gamma (N_1 \rightarrow \nu_\beta H ) = \frac{1}{32 \pi} \, 
			|Y_\nu^{\beta 1}|^2 \: m_1 \, \left( 1 - \frac{m_H^2}{m_1^2} \right)^2 \, .
\end{equation}
This looks at first sight quite different from the other two SM-like channels; 
however, if we
use that $\varepsilon \simeq M_D M_N^{-1}$, $M_D = Y_\nu \, v / \sqrt{2}$
and $\alpha / (s_W^2 m_W^2) = 1 / (\pi v^2)$ we can rewrite 
\begin{displaymath}
	|\varepsilon|^2 \alpha \, \frac{m_1^3}{s_W^2 m_W^2} \sim |Y_\nu|^2 
			\frac{m_1}{2\pi} \, ,
\end{displaymath}
and so, in the limit $m_1 \gg m_H, m_W, m_Z$ the three decay widths are 
identical. This is required by the \emph{equivalence theorem} 
\cite{Lee:1977eg,Cornwall:1974km}, which states that at energies much above
the symmetry breaking scale any calculation can be performed using the unbroken
theory; in that theory all the fields except for the $N$'s are massless, there is 
no heavy-light mixing and the $N$'s decay into the doublet of leptons and the 
Higgs scalar doublet through the SM Yukawa couplings. Therefore, the decay rates
to all four degrees of freedom in the Higgs doublet must be equal. However,
for moderate $m_1$ the theory begins to `perceive' the broken symmetry, and
since $m_H > m_Z > m_W$ the phase space factors become important.
The equivalence of equations (\ref{nuR-eff-N1-to-W} -- \ref{nuR-eff-N1-to-Higgs})
in the limit $m_1 \gg m_H, m_W, m_Z$, which is apparent in figure
\ref{fig:nuR-eff-N1-BR}, shows that it is the scalar degrees of freedom which are 
in play in these decays.

Up to this point we have discussed the SM-like decay channels above
the kinematical threshold for producing the electroweak bosons. The three
of them can also proceed below their threshold through a virtual heavy boson,
yielding suppressed three-body decays. These virtual modes, however, are 
immediately in competition with any other weaker interaction at hand. 
And indeed, it so happens that the same electroweak moments
provide such an interaction: as we see in equation \eqref{nuR-eff-zeta-mass}, 
with one mixing insertion a channel to photon plus light neutrino is open. 
This channel is not competitive when the SM-like modes are fully open, but it 
can be important if $m_1 < m_W$. Its decay rate reads
\begin{displaymath}
	\Gamma (N_1 \rightarrow \nu_\beta \gamma ) = \frac{2}{\pi} \, 
			\left| \varepsilon_\gamma^{\beta 1} \right|^2 \, c_W^2 \, m_1^3 \, ,
\end{displaymath}
where $\varepsilon_{\gamma}$ is a parameter that comprises the effective
coupling together with the mixing factors as seen in equation 
\eqref{nuR-eff-zeta-mass}; it is of order $(1 / \Lambda) \,
\sqrt{\nicefrac{m_\nu}{m_N}}$. The relevance of this channel for $N_1$ decays 
is further discussed in section \ref{sec:nuR-eff-N-decay}.

\subsubsection{$e^+ e^- \rightarrow N_1 N_2$ cross section}

By neglecting the heavy-light mixing, the LEP and ILC heavy neutrino production
is given entirely by the electroweak couplings. The cross section then reads
\begin{displaymath}
	\sigma (e^+ e^- \rightarrow N_1 N_2) = \frac{2}{3} \, \alpha 
			\left| \zeta_{12} \right|^2 \, f_Z (\sqrt{s}, m_1, m_2) \, 
			\eta_{e}(s) \, ,
\end{displaymath}
with $f_Z$ defined in equation \eqref{nuR-eff-fZ} and $\eta_e$ carrying
information about the involved diagrams. For a general fermion we have
\begin{equation} \label{nuR-eff-etaf}
	\eta_f (s) = 4 c_W^2 \, q_f^2 - 4 \, q_f v_f \, \mathrm{Re} \left[ \chi (s)
			\right] + \frac{v_f^2 + a_f^2}{c_W^2} \, | \chi(s) |^2 \, ,
\end{equation}
where $\chi (s)$ represents the Lorentzian-like structure generated by the 
one-loop-corrected $Z$ propagator,
\begin{displaymath}
	\chi(s) = \frac{s}{s - m_Z^2 + i m_Z \, \Gamma_Z} \, .
\end{displaymath}
Of course, for the case of an $e^+ e^-$ collider it is enough to take $f = e$,
but these general expressions will be of use shortly.

\subsubsection{Partonic cross sections for $pp \rightarrow N_1 N_2 + X$}

To compute the $pp \rightarrow N_1 N_2 + X$ cross section we proceed by computing
the collisions among the partons inside the proton and then convoluting over the 
parton distribution functions of the proton. So, diagrammatically we just compute
the cross section for $q \bar{q} \rightarrow N_1 N_2$ proceeding through a photon 
or a $Z$, which is the main production process that the partons can provide.
The cross section reads
\begin{multline*}
	\frac{\mathrm{d} \hat{\sigma}}{\mathrm{d} \Omega} (q \bar{q} \rightarrow N_1
			N_2) = \frac{\alpha}{6 \pi} \, \left| \zeta_{12} \right|^2 \eta_{q}
			(\hat{s}) \, \frac{\sqrt{\lambda \left(\hat{s}, m_1^2, m_2^2 \right)}}
			{\hat{s}^3} \times
	\\
	\times \Bigl[ (m_1^2 + m_2^2) \, \left( \hat{s} + 2 \hat{t} \right) - 
			2 \hat{t} \, \left(\hat{s} + \hat{t} \right) - m_1^4 - m_2^4 - 
	\\		
	- 2 \, \hat{s} \, m_1 m_2 \, \cos 2 \delta_{12} \Bigr] \, ,
\end{multline*}
with $\hat{s}$ and $\hat{t}$ the Mandelstam variables for the partonic
collision in the center-of-mass frame of the quarks, and $\eta_q$
the function defined in equation \eqref{nuR-eff-etaf} but with the appropriate
quantum numbers for the relevant quarks.

\subsubsection{Higgs boson decays into heavy neutrinos, $H \rightarrow N_i N_j$}

Up to now we have only discussed cross sections and decays induced by
the electroweak moment interaction or by SM interactions plus heavy-light mixing. 
The $\xi$ term in equation \eqref{nuR-eff-Lagrangian-5} also has interesting
consequences, in particular if the $N$'s are light enough it can induce new 
decay modes for the Higgs boson. We found
\begin{multline} \label{nuR-eff-Higgs-to-NN-width}
	\Gamma (H \rightarrow N_i N_j) = \frac{v^2}{2 \pi \, m_H^3} \,
			|\xi_{ij}|^2 \, \sqrt{\lambda(m_H^2, m_i^2, m_j^2)} \times
	\\		
	\times \left[ m_H^2 - m_i^2 - m_j^2 - 2 \, m_i m_j \, 
			\cos 2 \delta_{ij}^{\, \prime} \right] \, ,
\end{multline}
where $\xi_{ij} = |\xi_{ij}| \, e^{i \delta_{ij}^{\, \prime}}$. Note that, as
$\xi$ is a symmetric matrix, there's no need that $i \neq j$, as it happened
with the electroweak moments. A full 
discussion of these decays can be found in section \ref{sec:nuR-eff-Higgs-to-NN}.

\chapter{A model for right-handed neutrino magnetic moments} 
								\label{chap:nuR-magmo-model}

In this chapter we present a very simple model that gives rise to right-handed
neutrino electroweak moments. In the previous chapter  we 
discussed the phenomenological consequences of such an interaction; maybe the
most prominent would be the fact that right-handed neutrinos cease to be
`sterile' if the coupling is large enough. We also concluded that if the
underlying physics is weakly coupled the observation of the electroweak
moments would be more difficult than the observation of the new particles
themselves. Motivated by this, we present here an extension of the Standard
Model in which the right-handed neutrino electroweak moments can arise, and
we analyse part of its phenomenology. The model includes, in addition to the 
SM fields
and the right-handed neutrinos, a charged scalar singlet and a charged,
$SU(2)$-singlet, vector-like fermion. The electroweak moments are generated at one
loop, and their couplings are calculable and well-defined. In the simplest
version of the model the new charged particles are stable due to a global 
symmetry and they can constitute charged dark matter, which is strongly
disfavoured by cosmological and astrophysical considerations.
To avoid such problems we extend minimally the model by allowing a soft 
breaking of the symmetries, which is enough to induce CHAMP decays.
The resulting decays proceed mainly to leptons of the third family.
We conclude that in collider experiments the model should be searched
for in a similar fashion as one does with heavy charged leptons.
This work was carried out together with Arcadi Santamaria and José Wudka.

\section{The model} \label{sec:nuR-model-presentation}

As discussed extensively in the previous chapter, the dimension-five interactions
in a low-energy scenario consisting of the Standard Model and a number of 
right-handed neutrinos fall into three classes:
\begin{equation} \label{nuR-model-dim-5-lagrangian}
	\mathcal{L}_5 = \overline{\nu_{\mathrm{R}}^\mathrm{c}} \, \zeta \,
			\sigma^{\mu \nu} \nu_{\mathrm{R}} \: B_{\mu \nu} + 
			\left( \overline{\tilde \ell_{\mathrm{L}}} \phi \right) \, \chi \,
			\left( \tilde \phi^\dagger \ell_{\mathrm{L}} \right) - 
			\left(\phi^\dagger \phi \right) \: 
			\overline{\nu_{\mathrm{R}}^\mathrm{c}} \, \xi \, \nu_{\mathrm{R}} 
			+ \mathrm{H.c.} \, ,
\end{equation}
with the notation described in detail in section \ref{sec:nuR-eff-presentation}. 
The reader will notice
a difference between equation \eqref{nuR-model-dim-5-lagrangian} and the 
corresponding expressions in chapter \ref{chap:nuR-magmo-eff}; there we used 
$\nu_{\mathrm{R}}^{\, \prime}$ to
denote the flavour-basis right-handed neutrino fields, in order to distinguish
them from the mass eigenfields. In this chapter we will omit this notation, as we 
will not be interested in the details of neutrino mass diagonalisation.
The couplings $\chi$, $\xi$, $\zeta$ in equation \eqref{nuR-model-dim-5-lagrangian}
have dimension of inverse mass, which is
associated with the scale of heavy physics responsible for the corresponding
operator. The $\chi$ term is the well-known Weinberg operator
\cite{Weinberg:1979sa} that yields Majorana masses for the left-handed neutrinos;
the term involving $\xi$ gives a contribution to the right-handed neutrino
Majorana masses, but also provides interactions that could induce
invisible Higgs decays, as commented in section \ref{sec:nuR-eff-Higgs-to-NN}; 
the $\zeta$ operator was the main character of chapter \ref{chap:nuR-magmo-eff}, 
as it induces magnetic-moment-like
interactions for the $\nu_{\mathrm{R}}$'s which have many phenomenological
conse\-\mbox{quences -- see} sections \ref{sec:nuR-eff-colliders} and 
\ref{sec:nuR-eff-astro-cosmo}.
Regarding their flavour structure, $\chi$ and $\xi$ are in general complex symmetric 
matrices in flavour space, while $\zeta$ is complex and antisymmetric.
The number of $\nu_{\mathrm{R}}$'s is not fixed from first principles; let us
consider, throughout this chapter, three right-handed neutrino families,
in analogy to the three known lepton families.

In this chapter we will describe a model that gives rise to the electroweak
moments $\zeta$.
As we argued in section \ref{sec:nuR-eff-NP} of the previous chapter, this has 
to occur at least at the one-loop level, and the models necessarily
involve either a scalar-fermion pair with opposite, nonzero hypercharges
and having Yukawa couplings to both $\nu_{\mathrm{R}}$ and 
$\nu_{\mathrm{R}}^\mathrm{c}$, or a vector-fermion pair with the same properties. 
Here we will focus on the first possibility, since a massive vector field
suffers from well-known renormalisation problems that are not related to
the generation of electroweak moments. Therefore, we enlarge the SM by adding
a negatively charged scalar singlet, $\omega$, and one negatively charged 
vector-like%
\footnote{This fermion should contain the two chiralities if it is to couple
to $\nu_{\mathrm{R}}$ and $\nu_{\mathrm{R}}^\mathrm{c}$. Of course one can
imagine more complicated scenarios with two or more fermionic additions,
but we will stick to this minimal version of the mechanism.
}
fermion, $E$; formally, we write their charges under $SU(3)_{\scriptscriptstyle C}
\otimes SU(2)_L \otimes U(1)_Y$ as
\begin{displaymath}
	\omega \sim (0, 0, -1), 
	\qquad \qquad 
	E_{\mathrm{L}} \, , \, E_{\mathrm{R}} \sim (0, 0, -1) \, ,
\end{displaymath}
where it is understood that $Q = Y + T_3$.

The Lagrangian of the model contains all the renormalisable terms allowed
by gauge invariance. To describe it, let us first separate the standard pieces
from the new physics additions:
\begin{displaymath}
	\mathcal{L} = \mathcal{L}_{\mathrm{SM}} + \mathcal{L}_{\mathrm{NP}} \, .
\end{displaymath}
We include in $\mathcal{L}_{\mathrm{NP}}$ all the terms that involve the
new particles, including the right-handed neutrinos. $\mathcal{L}_{\mathrm{SM}}$,
thus, comprises just the `pure' Standard Model terms; among all of them, we
will use throughout our discussions the following:
\begin{displaymath}
	\mathcal{L}_{\mathrm{SM}} = i \, \overline{\ell_{\mathrm{L}}} \gamma_\mu D^\mu
		\ell_{\mathrm{L}} + i \, \overline{e_{\mathrm{R}}} \gamma_\mu D^\mu 
		e_{\mathrm{R}} + \left( \overline{\ell_{\mathrm{L}}} \, Y_e \, 
		e_{\mathrm{R}} \, \phi + \mathrm{H.c.} \right) + \ldots \, ,
\end{displaymath}
with $Y_e$ the Yukawa couplings of the charged leptons, which are completely
general $3 \times 3$ matrices in flavour space. The omitted terms include
the SM gauge boson, Higgs boson and quark kinetic terms, quark Yukawa
interactions and the SM Higgs potential. 

We now go on to describe the new physics contribution;
in order to present the different terms, we split $\mathcal{L}_{\mathrm{NP}}$
into four parts:
\begin{displaymath}
	\mathcal{L}_{\mathrm{NP}} = \mathcal{L}_{\mathrm{K}} + \mathcal{L}_Y - 
			V_{\mathrm{NP}} + \mathcal{L}_{\mathrm{extra}} \, .
\end{displaymath}
$\mathcal{L}_{\mathrm{K}}$ describes the kinetic and mass terms of the  
new particles,
\begin{multline}
	\mathcal{L}_{\mathrm{K}} = D_\mu \omega^\dagger \, D^\mu \omega + 
		i \, \bar E \, \gamma_\mu D^\mu E - m_E \, \bar E E \: + 
	\\
	+ i \, \bar \nu_{\mathrm{R}}
		\gamma_\mu \partial^\mu \nu_{\mathrm{R}} - \left( \frac{1}{2} \: 
		\overline{\nu_{\mathrm{R}}^\mathrm{c}} \, M_R \nu_{\mathrm{R}} + 
		\mathrm{H.c.} \right) \, ,
\end{multline}
with $M_R$ the Majorana mass term of right-handed neutrinos, which
is a complex symmetric matrix in flavour space. 

$\mathcal{L}_Y$ contains the standard Yukawa interactions of right-handed 
neutrinos and the new Yukawa couplings needed to generate the electroweak moments:
\begin{equation}
	\mathcal{L}_Y = \overline{\ell_{\mathrm{L}}} \, Y_\nu \nu_{\mathrm{R}} \,
		\tilde \phi + \overline{\nu_{\mathrm{R}}^\mathrm{c}} \: h^\prime E \,
		\omega^\dagger + \overline{\nu_{\mathrm{R}}} \, h \, E \,\omega^\dagger
		+ \mathrm{H.c.} 
\end{equation}
$Y_\nu$ is a general $3 \times 3$ complex matrix and, if there is
just one generation of $E$'s, $h$ and $h^\prime$ are three-component
vectors in flavour space. Note that lepton number is explicitly broken in this 
model even in the absence of a Majorana mass for the $\nu_{\mathrm{R}}$'s, due
to the joint action of the $h$ and $h^{\, \prime}$ couplings, that assign
opposite lepton number charges to $E$ or $\omega$. Indeed, the lepton-number-violating
$\zeta$ and $\xi$ interactions will proceed through one $h$ and one 
$h^{\, \prime}$ insertion, as seen in figures \ref{fig:nuR-model-magmo} and
\ref{fig:nuR-model-2Higgs}.

The SM scalar potential is enlarged with several terms involving $\omega$; we
collect them inside $V_{\mathrm{NP}}$:
\begin{displaymath}
	V_{\mathrm{NP}} = {m_\omega^\prime}^2 \, \omega \omega^\dagger + 
		\lambda_\omega \, \left( \omega \omega^\dagger \right)^2 + 
		2 \lambda_{\omega \phi} \: \omega \omega^\dagger \, \phi^\dagger \phi \, . 
\end{displaymath}
Note that $m_\omega^\prime$ is \emph{not} the physical mass of the charged scalar;
after the breaking of electroweak symmetry, an additional $\omega \omega^\dagger$
term will emerge from the quartic coupling $\lambda_{\omega \phi}$. If we have
$\left< \phi^0 \right> = v / \sqrt{2}$, the final physical mass will be
$m_\omega^2 = {m_\omega^\prime}^2 + \lambda_{\omega \phi} \, v^2$.
To $V_{\mathrm{NP}}$ we will just demand that it yields a meaningful theory:
we need $\lambda, \lambda_\omega > 0$ and $\lambda \lambda_\omega > 
\lambda_{\omega \phi}^2$ in order for the potential to be bounded from below,
and $m_\omega^2 > 0$ so that $U(1)_{\mathrm{em}}$ remains unbroken.
It is important to remark that
with only one Higgs doublet there cannot be trilinear couplings between
the doublet and the singlet. Therefore, the potential has two
independent $U(1)$ global symmetries, one for the singlet and one for the
doublet.

Apart from those already commented, the SM symmetries allow for the following 
Yukawa couplings and mass terms that involve the new particles:
\begin{equation} \label{nuR-model-lagrangian-extra}
	\mathcal{L}_{\mathrm{extra}} = \overline{E_{\mathrm{L}}} \, \kappa e_{\mathrm{R}}
		+ \overline{\ell_{\mathrm{L}}} \, Y_E E_{\mathrm{R}} \, \phi + 
		\overline{\tilde{\ell_{\mathrm{L}}}} \, f \, \ell_{\mathrm{L}} 
		\omega^\dagger + 
		\overline{e_{\mathrm{R}}} \, f^{\, \prime} \, \nu_{\mathrm{R}}^\mathrm{c} 
		\omega + \mathrm{H.c.} \, ,
\end{equation}
which can be set to zero by imposing a discrete symmetry that affects
only the new  particles,
\begin{align*}
	E &\rightarrow -E  
	& 
	\omega &\rightarrow -\omega \, ,
\end{align*}
in which case \emph{all} low-energy effects of the new particles will be 
loop-generated. In fact, the resulting Lagrangian will have
a larger continuous symmetry,
\begin{align}
	E &\rightarrow e^{i \alpha} \, E 
	& 
	\omega &\rightarrow e^{i \alpha} \, \omega \, , \label{nuR-model-global-symmetry}
\end{align}
which is not anomalous, and therefore there will be a charge, carried only
by $E$ and $\omega$, which will be exactly conserved. In such case, the
lightest of the $E$ or $\omega$ will be completely stable becoming
a CHAMP, which could create serious problems in the 
standard cosmology scenarios. We will discuss this issue in section 
\ref{sec:nuR-model-CHAMPs}.

\section{Generation of the dimension-five operators} 

\subsection{The right-handed neutrino electroweak moments}

\begin{figure}[tb]
	\centering
	\includegraphics[width=0.6\textwidth]{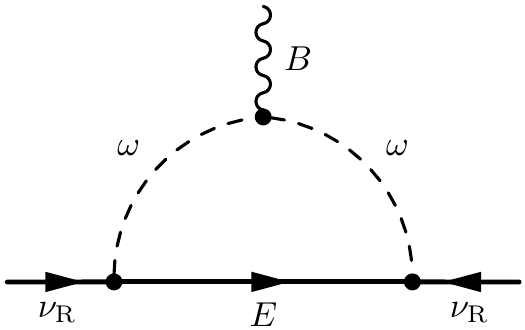}

	\vspace{0.5cm}
	\includegraphics[width=0.6\textwidth]{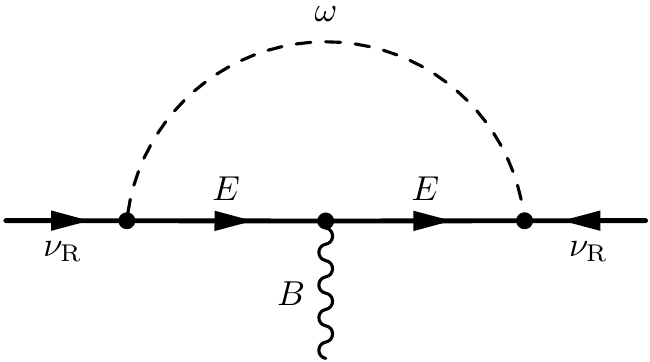}
	\caption{The diagrams that contribute to the right-handed neutrino Majorana 
			electroweak moments at leading order.
			} \label{fig:nuR-model-magmo}
\end{figure}

The model is constructed in such a way that the generation of electroweak moments
is straightforward: the $h$ Yukawas allow to open up the $\nu_{\mathrm{R}}$'s into a 
loop of $E$ and $\omega$ which violates lepton number if closed through 
a $h^{\, \prime}$
coupling; then, to obtain electroweak moments one has only to attach a $B$ 
to the $E$ or $\omega$ legs. The relevant diagrams are presented in figure
\ref{fig:nuR-model-magmo}. The calculation is somewhat cumbersome, but it can
be simplified by assuming $M_R \ll m_E, m_\omega$ and then neglecting all
external momenta and masses. Then we obtain
\begin{displaymath}
	\zeta_{ij} = \frac{g^{\, \prime}}{(4\pi)^{2}} \, \left( h_i^{\, \prime} h_j^* - 
			h_j^{\, \prime} h_i^* \right) \, \frac{f(r)}{4 m_E} \, ,
\end{displaymath}
where $r \equiv (m_\omega / m_E)^2$, and $g^{\, \prime}$ is, as usual, the 
$U(1)_Y$ gauge coupling. $f(r)$ is a loop function which reads
\begin{equation} \label{nuR-model-fr}
	f(r) = \frac{1}{1-r} + \frac{r}{(1-r)^2} \, \log r \quad \rightarrow \quad
		\left\{ \begin{array}{ll}
						1 \, ,   			  & r \ll 1 
						\\
						1/2 \, , 			  & r = 1 
						\\
						(\log r - 1) / r \, , & r \gg 1 
				\end{array} \right. \, .
\end{equation}
Note that only a small subset of all the interactions provided by the model
are in play in the generation of electroweak moments. In particular, the 
moments violate lepton number, but the Majorana mass of the $\nu_{\mathrm{R}}$'s
is not needed at all; in fact, these sort of one-loop diagrams can give us clues
about the natural value for $M_R$ in this model, see section 
\ref{sec:nuR-model-nuR-mass}. Notice, too, that none of the `rejected' interactions 
in equation \eqref{nuR-model-lagrangian-extra} participate in the electroweak 
moments; therefore, the moments offer no clue, at least at leading order, about 
the stability of the new particles.

We can use these expressions to estimate the magnitude of the generated 
electroweak moments; take, for instance, $m_\omega = m_E$, and
$\left(h_i^{\, \prime} h_j^* - h_j^{\, \prime} h_i^* \right) = 0.5$,
while $g^{\, \prime} = \sqrt{4\pi \, \alpha} / c_W \simeq 0.35$; then 
$\zeta \simeq 10^{-4} / m_E$. In terms of $\Lambda \equiv 1 / \zeta$ we would have 
$\Lambda = 10^4 \, m_E$, meaning that the bounds derived for the electroweak
moment will be roughly 10,000 times weaker than those derived from direct
search of the new charged particles. This is in agreement with our conclusions
from chapter \ref{chap:nuR-magmo-eff} for weakly-coupled theories, see for instance 
section \ref{sec:nuR-eff-collider-prologue}. For this particular example
the LEP and Tevatron bound of $m_E > 100 \; \mathrm{GeV}$ translates into 
$\Lambda > 10^6 \; \mathrm{GeV}$; as we can see in figure 
\ref{fig:nuR-eff-summary-bounds}, this bound is way better than those derived
on the electroweak moments from collider experiments. Only in the regime
of light $\nu_{\mathrm{R}}$'s, where astrophysical bounds apply, we find
such stringent limits on the electroweak moments that we can extract stronger
bounds on the masses of the new particles. For example, for very light right-handed
neutrinos, where the red giant cooling applies, we have $\Lambda \gtrsim 4 \times
10^9 \; \mathrm{GeV}$ and so $m_E \gtrsim 400 \; \mathrm{TeV}$, a stupendously
demanding bound.

\subsection{The effective Higgs boson interaction}
					  \label{sec:nuR-model-2Higgs}

\begin{figure}[tb]
	\centering
	\includegraphics[width=0.6\textwidth]{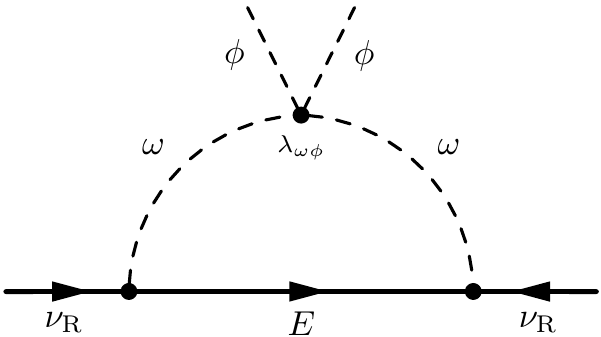}
	\caption{The diagram that yields the $\left(\phi^\dagger \phi \right) \: 
			\overline{\nu_{\mathrm{R}}^\mathrm{c}} \, \xi \, \nu_{\mathrm{R}}$
			interaction at one loop.} \label{fig:nuR-model-2Higgs}
\end{figure}

In addition to the electroweak moments, this model also provides at one loop
the dimension-five operator $\left(\phi^\dagger \phi \right) \: 
\overline{\nu_{\mathrm{R}}^\mathrm{c}} \, \xi \, \nu_{\mathrm{R}}$,
which yields several lepton-number-violating interactions among right-handed
neutrinos and the Higgs boson; in section \ref{sec:nuR-eff-Higgs-to-NN}
of the previous chapter we pointed out that some of these interactions could
enhance the decays of the Higgs boson to $\nu_{\mathrm{R}}$'s, maybe providing
a mechanism for invisible Higgs decays. The process that generates the $\xi$
interaction is very similar to those for the electroweak moment, but with
one insertion of the quartic scalar coupling $\lambda_{\omega \phi}$; we
present the relevant diagram in figure \ref{fig:nuR-model-2Higgs}.
A simple calculation yields
\begin{equation} \label{nuR-model-xi}
	\xi_{ij} = \frac{\lambda_{\omega\phi}}{(4\pi)^2} \, 
			\left(h_i^{\, \prime} h_j^* + h_j^{\, \prime} h_i^* \right) \, 
			\frac{f_\phi (r)}{4 m_E} \, ,
\end{equation}
where we have again $r \equiv (m_\omega / m_E)^2$, and $f_\phi (r)$ can be written 
in terms of $f(r)$, defined in equation \eqref{nuR-model-fr}, as follows: 
$f_\phi (r) \equiv 4 \, f (1/r) / r$.

As we did with the electroweak moments, we can use now equation 
\eqref{nuR-model-xi} to estimate the magnitude of the effective interaction.
Take, say, $\lambda_{\omega\phi} = \left(h_i^{\, \prime} h_j^* + h_j^{\, \prime} 
h_i^* \right) = 0.5$ and $m_E = m_\omega$; then we will have $\xi \simeq 10^{-3} \,
m_E$ or, expressed in terms of the new physics scale, $\Lambda \equiv 1 / \xi = 
10^3 \, m_E$. In section \ref{sec:nuR-eff-Higgs-to-NN} we showed that a new physics 
scale of a few tens of TeV might still interfere with the Higgs boson decays. 
The interaction generated by the model seems, thus, too weak; this is not a surprise, 
as this is a weakly-coupled model, and the effective interaction has to compete
with the nonsuppressed renormalisable Higgs couplings.

\section{Phenomenological analysis} 

\subsection{Considerations on the mass of the right-handed neutrinos}
						 			   \label{sec:nuR-model-nuR-mass}

In this model, the tree-level Majorana mass for the right-handed neutrinos, $M_R$,
is in principle arbitrary. However, there are several one-loop processes that
interfere with the tree-level Majorana masses; in this short
section we describe these mechanisms and discuss the information they provide
about the scale of Majorana masses within this model.

The first contribution emerges from the $\xi$ interaction discussed in section
\ref{sec:nuR-model-2Higgs}; after the spontaneous breaking of electroweak
symmetry this operator generates several terms, and among them there is
a Majorana mass for the $\nu_{\mathrm{R}}$'s. This mass  will be present
even if $M_R$ is set to zero in the Lagrangian, and so it can serve as a 
sort of `basal value' for the final Majorana mass in the model; indeed, the 
only way in which the theory could yield smaller Majorana masses in the end
is by finely cancelling this contribution, a possibility that we regard as
scarcely natural and we prefer to dismiss. It is important to note that
even though this mass is generated at one \mbox{loop --see fig}\-ure 
\ref{fig:nuR-model-2Higgs}, and replace the Higgs doublets \mbox{by VEV's--
it does \emph{not}} renormalise the tree-level Majorana mass $M_R$,
as it results from the $\xi$ operator, which is different in essence
from a bare Majorana mass. The effect of renormalisation is different and
subtler, and is to be discussed below.

We see, by looking at equations \eqref{nuR-model-dim-5-lagrangian} and
\eqref{nuR-model-xi}, that the Majorana mass provided by the $\xi$ operator
reads roughly
\begin{displaymath}
	M_R^{\mathrm{(\xi)}} \sim \frac{\lambda_{\omega\phi}}{(4\pi)^2} \, h^{\, \prime} 
			h \, \frac{v^2}{4 m_E} \, .
\end{displaymath}
Therefore we can estimate a lower bound on the final Majorana mass of the 
$\nu_{\mathrm{R}}$'s as
\begin{displaymath}
	M_R^\mathrm{(f)} \gtrsim \frac{0.1 \; \mathrm{GeV}^2}{m_E} \sim 1 \; 
			\mathrm{MeV} \, ,
\end{displaymath}
where the superindex ``(f)'' indicates that this is the ``final'' Majorana mass,
resulting from the sum of all the effective contributions.
For this estimate we took first $h^{\, \prime} = h = \lambda_{\omega \phi} = 0.1$
and then $m_E \sim 100 \; \mathrm{GeV}$. Smaller values for the coupling constants,
or heavier $E$ masses would yield milder bounds.

The second consideration about the right-handed neutrino Majorana masses regards
the naturality of the renormalisation procedure. The same one-loop diagrams
that generate the electroweak moments, see figure \ref{fig:nuR-model-magmo},
yield a renormalisation of the $\nu_{\mathrm{R}}$ Majorana mass if we remove
the $B$ leg. The diagrams, of course, are divergent and their contribution
is only meaningful after we regularise the theory and remove the infinities;
moreover, their contribution is added to the tree-level Majorana mass and it
is only this combination which is physically observable. Therefore, rigorously
speaking, we should have nothing to discuss about these one-loop corrections, as
they're not physical; however, if we trust that perturbation theory and
quantum field theory are capturing some features of reality, we would be inclined 
to think that the one-loop corrections should be \emph{corrections} indeed, and so
smaller than the tree-level mass. This argument appeals to the naturality and
consistency of the theory, and though it is not strictly phenomenological it can
provide an intuition of a `reasonable' scale for the parameters. Applied to the 
Majorana masses, it yields the following result: the one-loop diagrams are
logarithmically divergent, and their corrections will be of order
\begin{displaymath}
	\delta M_R \sim \frac{h^{\, \prime} h}{(4\pi)^2} \, m_E \, ,
\end{displaymath}
suppressed by an additional factor $(m_E / m_\omega)^2$ if the scalar is
much heavier than the vector-like fermion. 
It is then natural to require $M_R \gtrsim \delta M_R$, which is estimated to
be $M_R \gtrsim 5 \; \mathrm{MeV}$ for $h = h^{\, \prime} = 0.1$ and 
$m_E = 100 \; \mathrm{GeV}$.

\subsection{E or $\omega$ as CHAMP's} \label{sec:nuR-model-CHAMPs}

The model as described so far contains only the couplings necessary
to generate the right-handed neutrino Majorana electroweak moments.
But it is clear that the trilinear vertices $\bar \nu_{\mathrm{R}} E 
\omega^\dagger$ and $\bar \nu_{\mathrm{R}}^\mathrm{c} E \omega^\dagger$ 
alone cannot induce decays
for both the $E$ and the $\omega$. The lightest of the two will
remain stable and could then accumulate in the galaxy clusters, appearing
as electrically charged dark matter. The idea that dark matter could
be composed mostly of charged massive particles was proposed in
\cite{DeRujula:1989fe,Dimopoulos:1989hk}
and it is strongly constrained from very different arguments
\cite{Basdevant:1989fh,Hemmick:1989ns,Yamagata:1993jq,Gould:1989gw,
Chivukula:1989cc}.
One might still consider the possibility of having massive stable
$E$ or $\omega$ particles within the reach of the LHC, but with
a cosmic abundance lower than the one required for dark matter. Unfortunately,
such scenario seems also to be excluded, at least under the most usual 
assumptions: if one considers, as in 
\cite{DeRujula:1989fe},
that the $E$'s and $\omega$'s were produced in the early universe
through the standard freeze-out mechanism \cite{Wolfram:1978gp},
the bounds from interstellar calorimetry \cite{Chivukula:1989cc}
and terrestrial searches for super-heavy nuclei 
\cite{Hemmick:1989ns,Yamagata:1993jq}
completely close the window of under-TeV CHAMP abundances.

There is, however, a way to escape all these bounds. A recent paper
\cite{Chuzhoy:2008zy} notes that CHAMP's, if very massive or carrying
very small charges, are expelled from the galactic disk by the magnetic
fields. That situation prevents any terrestrial or galactic detection
and leaves room for CHAMP's to exist. The bound specifically states
that particles with $100 \, (Q/e)^2 \; \mathrm{TeV} \lesssim m \lesssim 10^8 \, 
(Q/e) \; \mathrm{TeV}$
are depleted from the disk, and in fact our model, if we forbid the
terms in equation \eqref{nuR-model-lagrangian-extra}, does not fix 
the hypercharge of
$E$ and $\omega$, so they can be millicharged. Unfortunately, this
situation is not interesting for our purposes, for this kind of CHAMP's
would give rise to very small neutrino magnetic moments and wouldn't
show up in the future accelerators, either due to their heavy masses
or to their small couplings.

In conclusion, we need an additional mechanism for $E$ or $\omega$
decays. The easiest way to accomplish this is by allowing one or more
of the couplings in equation \eqref{nuR-model-lagrangian-extra}, which 
can be taken small, if needed, by arguing that 
\eqref{nuR-model-global-symmetry} is an almost
exact symmetry. We discuss one of the possibilities in section
\ref{sec:nuR-model-CHAMP-decay}.
The scenario of decaying CHAMP's has, on its own, a number of advantages
and drawbacks. Some recent papers \cite{Jedamzik:2007qk,Jedamzik:2007cp}
have pointed out that the presence of a massive, charged and colourless
particle during the process of primordial nucleosynthesis might lead
to an explanation for the cosmic lithium problem. Also, the decay
of massive particles during nucleosynthesis could have a dramatic
influence in the final abundances of primordial elements, which provides
us with  bounds on the lifetime and abundance of CHAMP's that could
be useful.

\subsection{Allowing for CHAMP decays} \label{sec:nuR-model-CHAMP-decay}

If the new particles have to decay the global symmetry 
\eqref{nuR-model-global-symmetry}
needs to be broken, and for that aim it is enough to allow some of the terms
in equation \eqref{nuR-model-lagrangian-extra}. 
For the sake of simplicity, we will consider
only the case in which the symmetry is softly broken by mixings between $E$
and the charged leptons; that is, we will add to our Lagrangian the term
\begin{displaymath}
	\mathcal{L}_\kappa = \overline{E_{\mathrm{L}}} \, \kappa e_{\mathrm{R}} + 
			\mathrm{H.c.} \, ,
\end{displaymath}
where $\kappa$ is, if there is just one generation of $E$'s, a three-component
vector in flavour space with dimensions of mass.
The choice of a soft breaking is motivated, as usual, by the fact that the
model remains closed under the renormalisation group flow; with only 
$\mathcal{L}_\kappa$
added, none of the other terms in \eqref{nuR-model-lagrangian-extra} needs to
be introduced for the model to remain renormalisable.

The $\kappa$ term will induce decays of $E$ into SM particles much like the
heavy neutrino decays in seesaw models, since only this mixing links
$E$ to the SM degrees of freedom. After diagonalisation of the
charged lepton mass matrix one obtains interactions that connect 
$E$ to $W \nu$, $Z \ell$ and $H \ell$%
\footnote{The $\gamma \ell$ channels are subdominant: since $U(1)_{\mathrm{em}}$ 
remains unbroken, flavour-changing vertices involving photons are forbidden at 
tree level; the corresponding processes occur, but proceed necessarily
through loops and are thereby suppressed.
}.
As the current
bound on heavy charged leptons requires that $m_E > 100 \; \mathrm{GeV}$,
the $W$ and $Z$ will be produced on-shell; the Higgs channel may
or may not be open depending on the value of $m_E$. 
$\omega$, on the other hand, can only decay through the Yukawa
$\bar E \, \nu_{\mathrm{R}} \omega$ vertices, and kinematics can complicate
the scenario if the masses of $\omega$, $E$ and the $\nu_{\mathrm{R}}$'s 
are comparable. For the sake of simplicity, in the following we will
consider $m_\omega > m_E \gg M_R$, so that the $\omega$'s decay rapidly to
on-shell $E$ and $\nu_{\mathrm{R}}$ and then the $E$ decays through the
channels described above.

\begin{figure}[p]
	\centering
	\raisebox{6.3cm}{\emph{a)}}
	\includegraphics[width=0.8\textwidth]{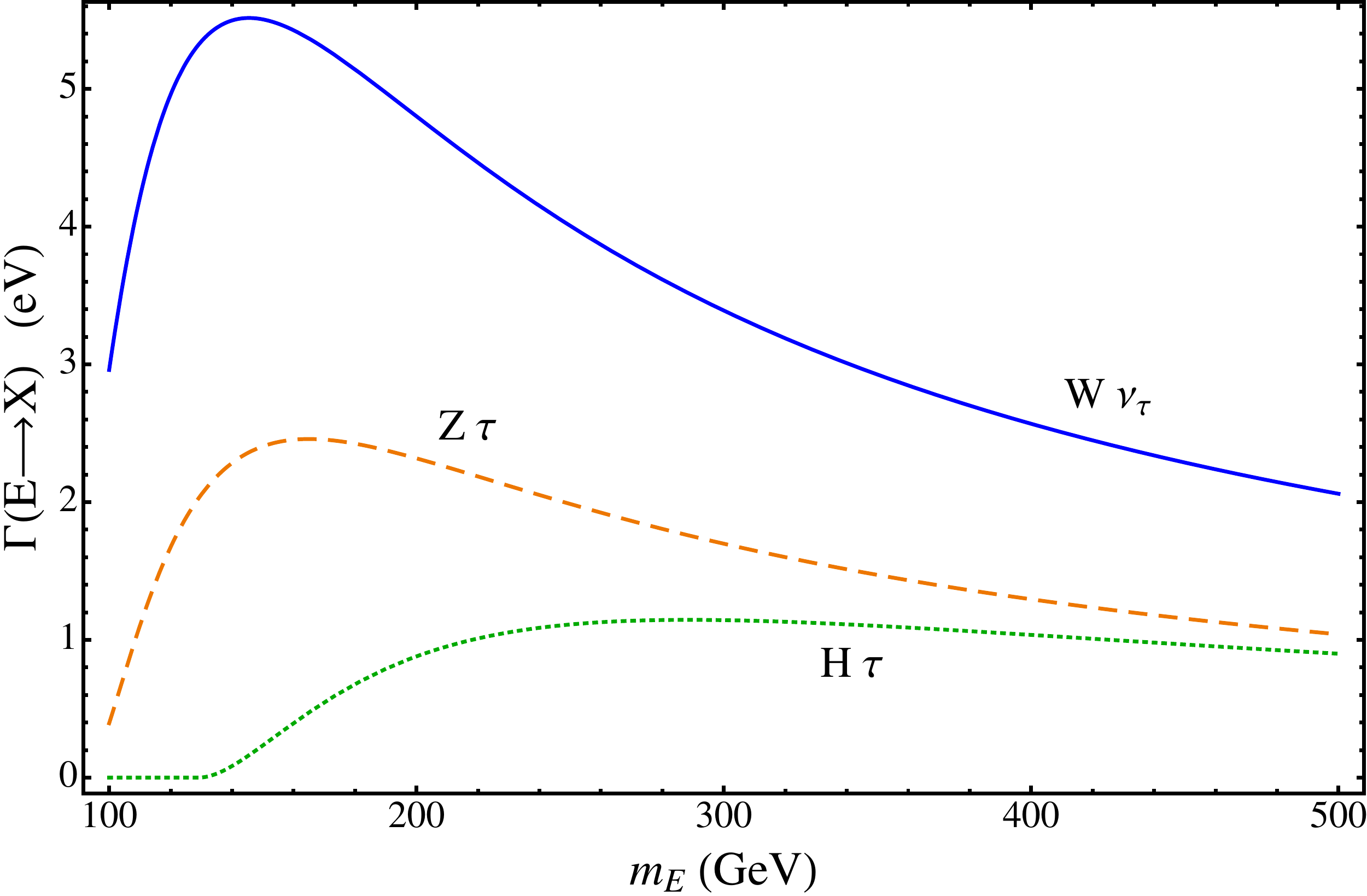}

	\vspace{0.75cm}
	\raisebox{6.3cm}{\emph{b)}}
	\includegraphics[width=0.81\columnwidth]{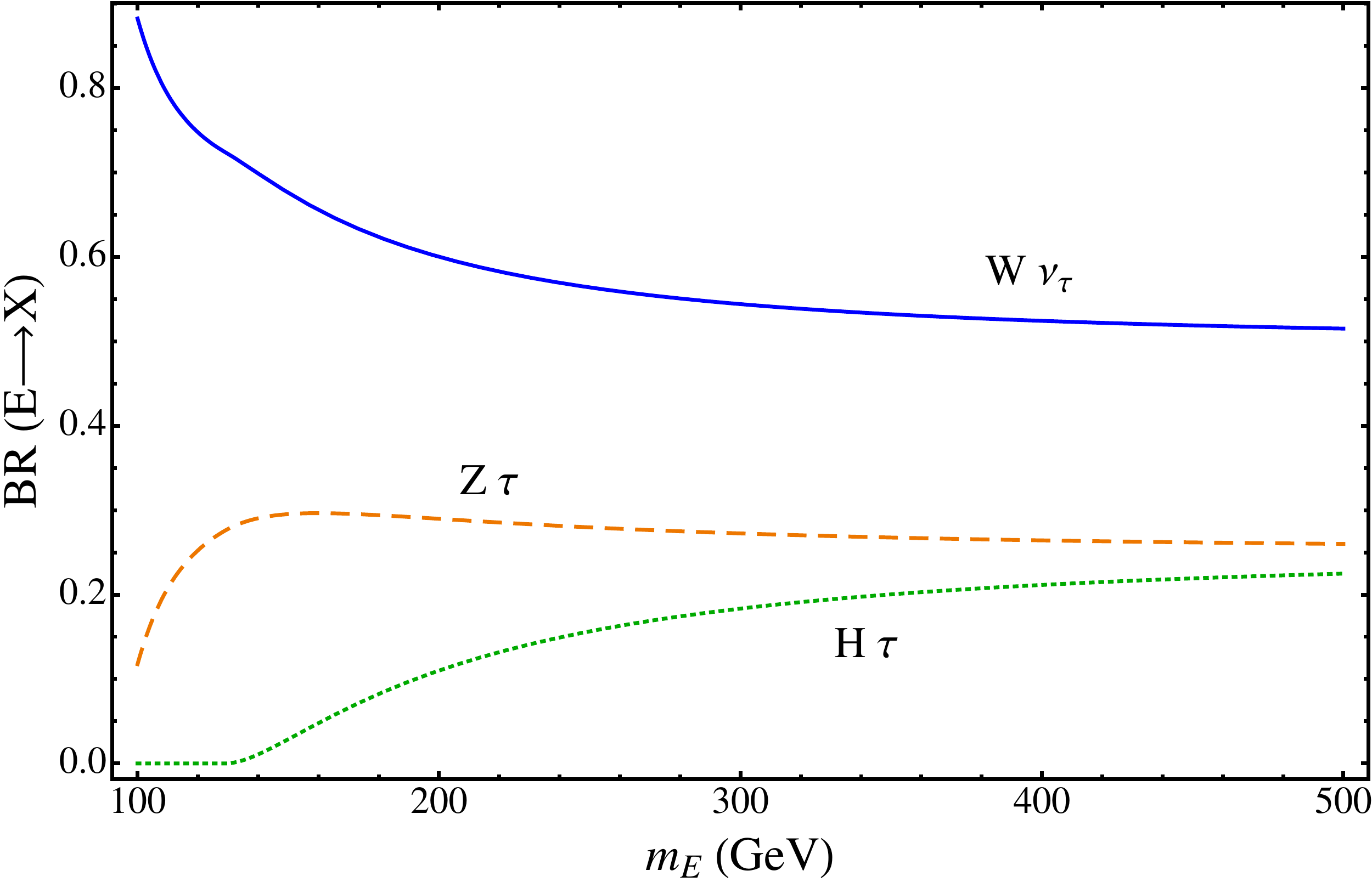}
	\caption{Decay widths and branching ratios for the dominant decay channels
			of the vector-like fermion $E$. The Higgs boson mass has been
			fixed to $m_H = 129 \; \mathrm{GeV}$;
			in \emph{a)} we have taken for the mixing $\kappa_\tau = 1 \;
			\mathrm{GeV}$.
			} \label{fig:nuR-model-E-decays}
\end{figure}

In figure \ref{fig:nuR-model-E-decays} we offer the widths and branching
ratios for the leading decay channels of $E$. Several features can be noted, 
all related to the fact that these are mixing-mediated decays. First, the
decay widths are suppressed by powers of $m_\ell / m_E$, where $m_\ell$
is the mass of the charged lepton in the final state; therefore, if the
$\kappa$'s are all of the same order $E$ will decay mainly to leptons of the 
third family. Figure \ref{fig:nuR-model-E-decays} reflects this fact and
only considers channels with $\tau$ or $\nu_\tau$ in the final state.
Second, in plot \emph{a)} of figure \ref{fig:nuR-model-E-decays} we
notice that the rates decrease when $m_E$ increases, even though the phase
space increases with $m_E$; again, this is due to the $m_\tau / m_E$ factor,
which compensates for the increase in phase space. Thirdly, we note in figure
\ref{fig:nuR-model-E-decays}\emph{b)} that the branching ratios of the
$Z \tau$ and $H \tau$ channels converge as $m_E$ increases; interestingly,
the branching ratio of the $W \nu_\tau$ channel converges to twice the value
of the other two. This is again a mixing effect: the decays are controlled
by mass couplings, and the bosons involved are essentially the components
of the scalar Higgs doublet; even when the decay proceeds to $W$ or $Z$, it is
the longitudinal component which dominates the process. Therefore,
and accordingly with the equivalence \mbox{theorem --see} section 
\ref{sec:nuR-eff-appendix} for more on this \mbox{topic--, as we} progress to high
energies the channels reproduce the statistics of the decays to components of
$\phi$, that is: half the decays to $\phi^\pm \nu_\tau$ ($W^\pm \nu_\tau$), 
one quarter to $\phi^0_p \, \tau$ ($Z \tau$), and one quarter to $\phi^0_s 
\, \tau$ ($H \tau$).

As for the possible cosmological effects of the new particles and their decays,
for the example value of $\kappa_\tau = 1 \; \mathrm{GeV}$ the decay
widths of $E$ are of the order of the eV. With such lifetimes there will be
no $E$'s at the epoch of primordial nucleosynthesis and it won't be affected.
Note, however, that the decay rates depend upon $\kappa_\tau^2$, and 
$\kappa_\tau$ is relatively free, so the decay rates can vary in several orders 
of magnitude depending on the value of $\kappa_\tau$. For $\kappa_\tau < 10^{-7}
\; \mathrm{GeV}$ the CHAMP's will affect nucleosynthesis and might
help to solve the cosmic lithium problem \cite{Jedamzik:2007qk,Jedamzik:2007cp}.
They might also affect the CMB by distorting the blackbody spectrum and
by modifying the optical depth; these two observables are very sensitive and
could impose tight limits on the value of $\kappa$.
For this discussion we will take a conservative approach and we will just 
require that the relic CHAMP's have decayed by the present time;
this means considering $\kappa_\tau > 10^{-16} \; \mathrm{GeV}$.

\subsection{Lepton Flavour Violating processes}

For general $\kappa$'s and $Y_e$'s family lepton
flavour is not conserved. It is then mandatory to determine if processes
such as $\mu \rightarrow 3e$, $\mu \rightarrow e\gamma$,
or $\tau \rightarrow 3\mu$ introduce any restrictions on the parameters of the 
model. In this section we discuss this possibility.

In order to calculate the amplitudes for these low-energy processes we choose to 
integrate out the heavy particles that mediate them and use an effective
Lagrangian which describes the low-energy interactions of the light leptons.
We carry out the integration%
\footnote{For a detailed example of the integration procedure in the case of a 
singly-charged scalar see \cite{Bilenky:1993bt}.
} 
by using the equations of motion for $E$ and expanding in powers of $1 / m_E$. 
One then obtains 
\begin{displaymath}
	\mathcal{L}_{\mathrm{LFV}} = - \frac{i}{m_E^4} \, \overline{e_{\mathrm{R}}} \,
			\kappa \kappa^\dagger ( \gamma_\mu D^\mu)^3 e_{\mathrm{R}} + 
			\ldots \, ,
\end{displaymath}
which, after spontaneous symmetry breaking and using the equations of motion,
leads to a lepton-flavour-violating interaction of the $Z$ with the
left-handed component of the charged leptons,
\begin{align}
	\mathcal{L}_{\mathrm{LFV}} &= \frac{e}{2 \, s_W c_W} \, Z_\mu \, 
		\overline{e_{\mathrm{L}}} \, C_{\mathrm{LFV}} \gamma^\mu \, e_{\mathrm{L}} 
		\, ,
		& 
		C_{\mathrm{LFV}} &\simeq \frac{v^2}{2 m_E^4} \, Y_e \kappa 
			\kappa^\dagger Y_e^\dagger \, .
		\label{nuR-model-LFV-Lagrangian}
\end{align}
$C_{\mathrm{LFV}}$ is a matrix in flavour space which is not, in general,
diagonal; therefore, the interaction \eqref{nuR-model-LFV-Lagrangian} will induce 
processes such as $\mu \rightarrow 3e$ and $\tau \rightarrow 3 \mu$. Without loss
of generality we can take $Y_e$ diagonal with elements proportional
to the charged lepton masses; then we can estimate the branching ratio
for $\mu \rightarrow 3e$ as
\begin{displaymath}
	\mathrm{BR} (\mu \rightarrow 3e) = \frac{\Gamma (\mu \rightarrow 3e)}
		{\Gamma (\mu \rightarrow e \nu \bar \nu)} \simeq 
		\frac{\left| m_e ( \kappa \kappa^\dagger )_{e \mu} m_\mu
		\right|^2}{m_E^8} \, .
\end{displaymath}
Our effective Lagrangian is an expansion in powers of $1 / m_E$ which
could be compensated, in part, by $\kappa \kappa^\dagger$ factors
in the numerator; thus, for consistency, we should require $\kappa < m_E$,
which allows us to establish an upper bound on the branching ratio.
Recalling also that the present limit on the mass of charged heavy
leptons is $m_E > 100 \; \mathrm{GeV}$ \cite{Beringer:1900zz}, we obtain
\begin{displaymath}
	\mathrm{BR} (\mu \rightarrow 3e) < \left( \frac{m_\mu m_e}{(100 \; 
			\mathrm{GeV)^2}} \right)^2 < 10^{-16} \, ,
\end{displaymath}
to be compared with present bounds, which are of the order of $10^{-12}$
\cite{Beringer:1900zz}.
As we see, the model contribution to $\mu \rightarrow 3 e$ is still far from
being observable, even taking large values of $\kappa$.

We can use the same reasoning to calculate the model contribution to
$\tau \rightarrow 3 \mu$; it is enough to notice that once $\kappa$ is fixed
$\mu \rightarrow 3 e$ and $\tau \rightarrow 3 \mu$ only differ in a kinematical
factor of $(m_\tau / m_\mu)^2$:
\begin{displaymath}
	\mathrm{R} (\tau \rightarrow 3 \mu) \equiv \frac{\Gamma (\tau \rightarrow 
			3 \mu)}{\Gamma (\tau \rightarrow \mu \nu \bar \nu)} < 
			\left( \frac{m_\tau m_\mu}{(100 \; \mathrm{GeV)^2}} \right)^2
			< 10^{-10} \, ,
\end{displaymath}
which is, too, under the present sensitivity for this process, fixed at
about $10^{-7}$ \cite{Beringer:1900zz}.

Another very restrictive process is $\mu\rightarrow e\gamma$, whose
branching ratio is bounded at $5.7 \times 10^{-13}$ by the 
MEG experiment \cite{Adam:2013mnn}.
The collaboration expects to improve this limit to roughly $10^{-13}$
\cite{Nishiguchi:2011zz}. However, this process
only arises in our model at one loop, and therefore
we do not expect stringent bounds from it. A further possibility would 
be the electroweak oblique parameters, but as the new particles are singlets
under $SU(2)$ they only enter in the $W$ self-energies through mixing, 
suppressed by powers of the masses of the light leptons; therefore, their 
contributions are
too small to be observed at the currently available precision.

Finally there is $\mu - e$ conversion in nuclei, which also provides strong limits;
for instance, $\mu - e$ conversion in titanium yields $\sigma (\mu^-
\mathrm{Ti} \rightarrow e^- \mathrm{Ti}) / \sigma (\mu^- \mathrm{Ti} \rightarrow 
\mathrm{capture}) < 4.3 \times 10^{-12}$ \cite{Dohmen:1993mp}. 
In our model, the process is induced by 
exactly the same interaction \eqref{nuR-model-LFV-Lagrangian} that gives 
$\mu \rightarrow 3e$, and we again do not expect, at the present, a strong bound.
However, given the future plans to improve the limits in several orders
of magnitude \cite{Kuno:2013mha,Pasternak:2013ksa}, then perhaps 
$\mu - e$ conversion will provide the
best bound for LFV-ing processes in this model. In any case, current data
on LFV cannot constrain this mechanism for $E$ decays.

\subsection{The model at colliders}

\begin{figure}[tb]
	\centering
	\includegraphics[width=0.8\textwidth]{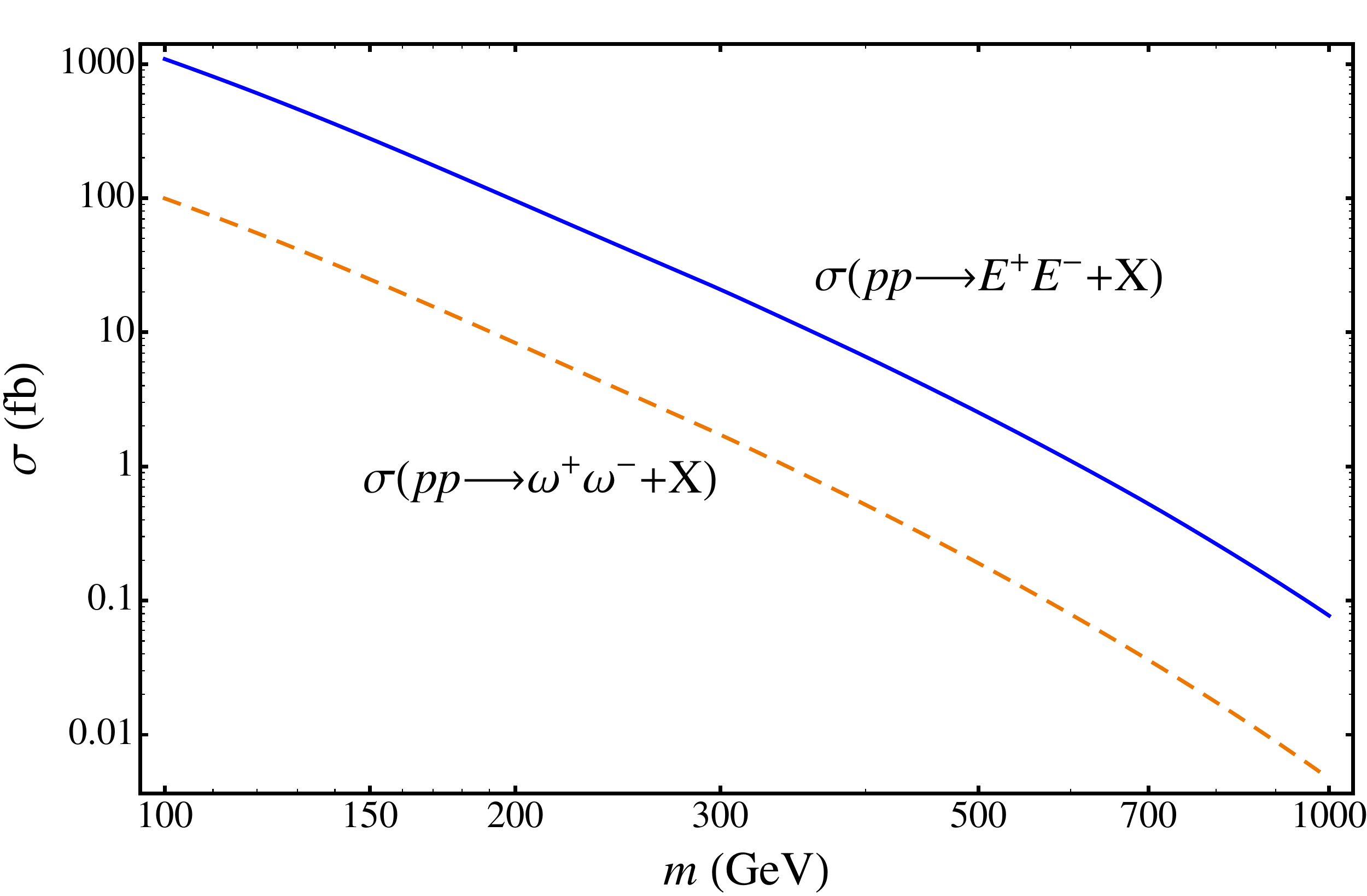}
	\caption{Production cross sections of the charged particles at the LHC, 
			calculated at its nominal center-of-mass energy, $\sqrt{s} = 14 \; 
			\mathrm{TeV}$. $m$ represents either $m_E$ or $m_\omega$, as it
			corresponds for each curve. $X$ represents that in a proton-proton
			collider we expect to generate together with the heavy particle
			pair other hadronic of leptonic products.
			} \label{fig:nuR-model-sigma-collider}
\end{figure}

Despite the fact that the new particles are $SU(2)$ singlets
and only have Yukawa couplings to right-handed neutrinos, they have
electric charge and can be copiously produced at the LHC, if light enough, 
through the Drell-Yan process. In figure \ref{fig:nuR-model-sigma-collider}
we show the production cross sections for a proton-proton collision at the 
LHC. To compute these cross sections we calculated the corresponding
quark-antiquark collision and convoluted its cross section with the
parton distribution functions of the proton%
\footnote{See, for a very clear review on the matter, \cite{Campbell:2006wx}.
In our calculation we used the CTEQ6M parton distribution sets 
\cite{Pumplin:2002vw} and we checked our results against the CompHEP 
software \cite{Boos:2004kh,Pukhov:1999gg}.
}.
Since the  particles are produced by $\gamma$ and $Z$ exchange, there are no 
unknown free
parameters except for the masses of the particles. We see that cross sections
from $1 \; \mathrm{fb}$ to $1 \; \mathrm{pb}$ are easily obtained for
the production of $E^+ E^-$ pairs with $100 \; \mathrm{GeV} < m_E < 700 \; 
\mathrm{GeV}$. For equal masses the production cross section
for $\omega$ is roughly one order of magnitude smaller. 

Once pair-produced, the  particles have to be detected and identified.
The characteristic signatures for this identification are very different
depending on the lifetimes of the  particles, mostly because if the
$E$ and $\omega$ are long-lived they can be tracked directly in
the detectors or, at least, be identified through a displaced decay
vertex. The relevant parameter for this behavior is $\kappa$, the
$E-e$ mixing. For $\kappa \lesssim 1 \; \mathrm{MeV}$, the $E$'s will have 
decay lengths roughly over $1$~centimeter%
\footnote{Note that there's room in the parameter space for this kind of effects
even if one requires that CHAMP's do not affect the primordial nucleosynthesis,
for if $\kappa > 100 \; \mathrm{eV}$ all the $E$'s will have decayed
before nucleosynthesis.
}; 
additionally, for $\kappa < 0.2 \; \mathrm{MeV}$, they will go through the
detector and behave as a heavy ionizing particle. A lot of work has
been carried out to analyse the signatures of CHAMP's inside the \mbox{detectors
--see,} for instance, \cite{Fairbairn:2006gg}, and \cite{Chen:2009gu}
for a recent \mbox{improvement--, and also} displaced vertices have been 
\mbox{discussed -- see,} for example, \cite{Franceschini:2008pz,deCampos:2008re}.
If $\kappa > 1 \; \mathrm{MeV}$ the $E$'s will decay near the collision
point and behave roughly as a fourth-generation charged lepton. 

Discovering the $\omega$'s can be much harder, because they will
be produced at a significantly lower rate and the signatures of their
decays depend strongly on the details of the model. In the $m_\omega > m_E$
scenario, they will decay quickly into an $E$ and a heavy \mbox{neutrino
--at least} if we want $h$ and $h^{\, \prime}$ large enough to have significant
electroweak \mbox{moments-- and then} one has to rely again on the detection
of $E$'s unless the heavy neutrino provides a cleaner signal, which
is unlikely. In any case, we think that the $E$'s, produced in a
much greater number, should be considered the signature of this model,
and perhaps the doorway to understand the $\omega$ and heavy neutrino
decays.

\chapter{Neutrinoless double beta decay and neutrino masses} \label{chap:0nu2beta-eff}

Neutrino oscillation and tritium $\beta$-decay experiments offer us a picture
of stupendously small neutrino masses that claim for an explanation. At the
same time, the search for neutrinoless double beta decay 
is our best shot to elucidate the Dirac or Majorana character of neutrino masses.
Indeed, in the classic picture of the process $0 \nu \beta \beta$ is induced
by neutrino \mbox{masses -- see,} for example, figure \ref{eff-0nu2beta-5}.
If we interpret the present and near future $0 \nu \beta \beta$ experiments 
in this terms, then only degenerate mass regime will be \mbox{explored --
it is} possible that we enter 
the inverse hierarchy regime,
but probably we won't be able to completely explore it (see section
6.11 in \cite{GomezCadenas:2011it}).
However, one can imagine other scenarios: maybe $\nu$ masses are indeed Majorana
but they don't constitute the dominant contribution to $0 \nu \beta \beta$;
were this the case, a signal in $0 \nu \beta \beta$ would not
necessarily pin neutrino masses to the degenerate or inverse hierarchies.
In this chapter we examine a class of effective operators
that provide such sort of a situation; these operators
involve at low energies only the leptons and bosons of the Standard Model.
Here we will analyse their relevance in a model-independent way by using
effective field theory, whereas in chapters \ref{chap:Wnue-model} and
\ref{chap:WWee-model} we will discuss particular models that realise 
the scenarios described here.
This work was carried out together 
with Arcadi Santamaria, José Wudka, Francisco del Águila and Subhaditya
Bhattacharya.

\section{Violating lepton number through leptons and bosons} 
		\label{sec:0nu2beta-eff-Violating-LN}

The relationship between neutrino masses and neutrinoless double beta decay 
is a deep and well-known one; we can state for sure that if $0 \nu \beta \beta$
is observed then neutrino masses are of the Majorana kind, as it was uncovered
in 1982 in a celebrated paper by Schechter and Valle \cite{Schechter:1981bd}. 
Beyond
this, however, we can say nothing, for the mechanisms generating neutrino masses
and, were it the case, $0 \nu \beta \beta$, are unknown. 
We want to use effective field theory to investigate this relationship in 
a model-independent way and, in particular, to examine scenarios where
$0 \nu \beta \beta$ is enhanced with respect to neutrino masses; EFT is a good
choice for this aim, as both $0 \nu \beta \beta$
and $\nu$ masses occur at low energies and in most scenarios the physics 
responsible for
LNV lies above the TeV scale. 

The exploration of $0 \nu \beta \beta$ and neutrino masses through EFT
has been carried out in the past \cite{Babu:2001ex,Choi:2002bb,Engel:2003yr,
deGouvea:2007xp}, usually including operators with quark fields
but excluding those which involve gauge bosons; the survey by Babu and 
Leung \cite{Babu:2001ex} specifically states ``It may be more difficult 
to generate such operators at tree level from an underlying renormalisable
theory, making them perhaps less interesting for the generation of 
neutrino masses''. We will focus on the converse situation: operators
without quarks but explicitly including the SM gauge bosons; we will 
explore this scenario and will prove that large families of models
can generate such operators.

\begin{figure}[tb]
	\centering
	\includegraphics[width=0.7\textwidth]{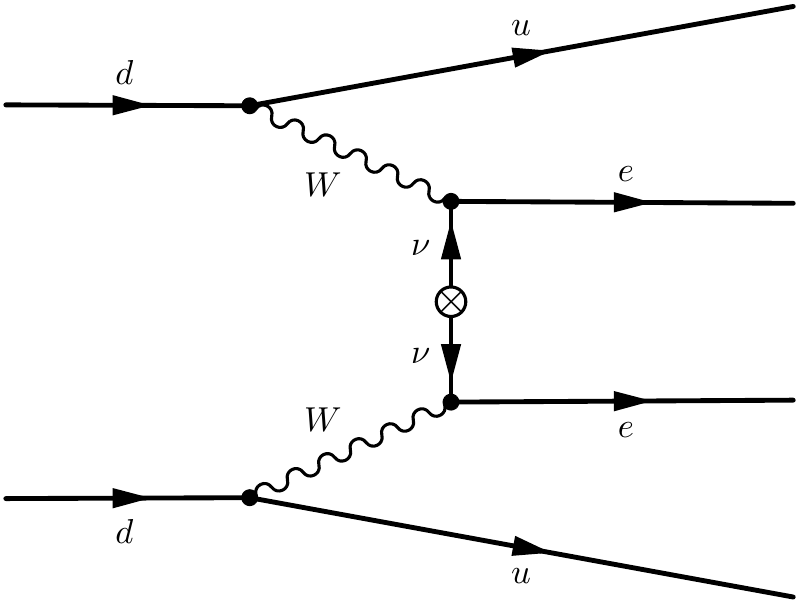}

	\caption{The `classical' mechanism for $0 \nu \beta \beta$, with
				a Majorana mass insertion in the internal neutrino
				propagator; the effective interaction is realised
				through the well-known Weinberg operator, depicted
				here as a crossed circle. This can be considered
				the first and most widely known of three ways of
				generating $0 \nu \beta \beta$ with effective operators
				that don't involve quarks.
				} \label{eff-0nu2beta-5}
\end{figure}

\begin{figure}[p]
	\centering
	\raisebox{6.2cm}{\emph{a)}}
	\includegraphics[width=0.7\textwidth]{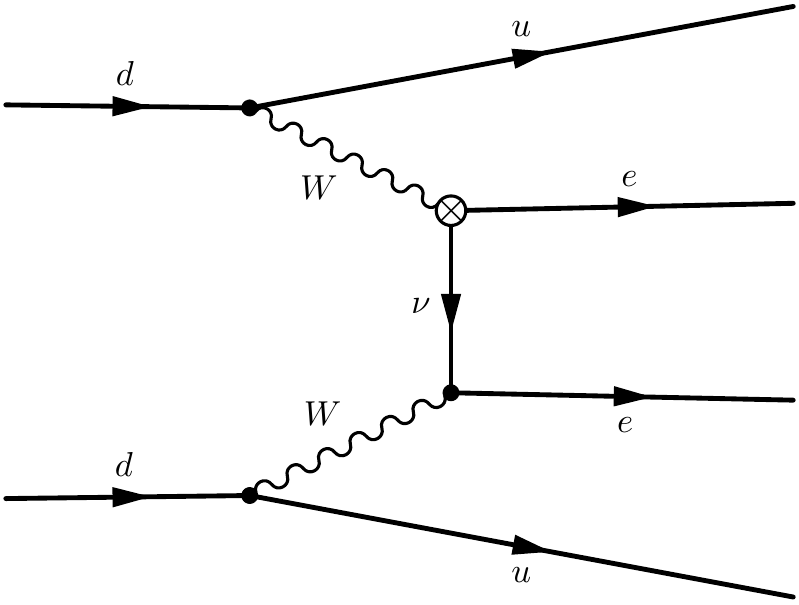}

	\vphantom{\raisebox{6ex}{a}} 
	\raisebox{6cm}{\emph{b)}}
	\includegraphics[width=0.7\textwidth]{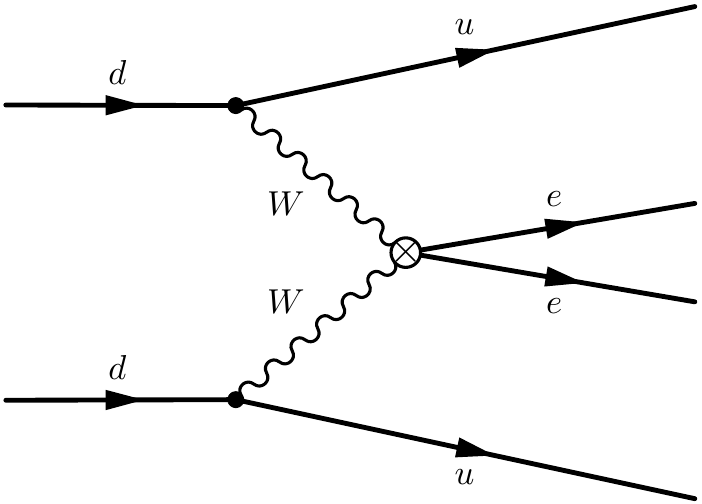}
	\caption{The two ways of generating $0 \nu \beta \beta$
			with effective vertices that involve $W$'s but not quarks:
			in \emph{a)}, through a $W \nu e$ vertex; in \emph{b)},
			through a $WWee$ interaction.
			} \label{fig:0nu2beta-eff-7-and-9}
\end{figure}

As a first step, let us briefly (and rather naively) consider how 
this class of operators might provide $0 \nu \beta \beta$ and 
neutrino masses. Consider first $0 \nu \beta \beta$; the `traditional'
mechanism, with all LNV accounted for by neutrino Majorana masses,
is displayed in figure \ref{eff-0nu2beta-5}. This scenario would fit
among the ones we wish to explore, as no quarks are
involved in the effective interaction, but just \mbox{leptons -- and}
scalar Higgs bosons, which get a VEV, see section 
\ref{sec:neutrinos-dirac-or-majorana}. However, this mechanism
has been extensively discussed and investigated and we won't stay on it
for long; we will rather take it as a sure reference point and maybe a
scenario we'd like to avoid, as we will argue later on.
Anyway, how else can $0 \nu \beta \beta$ be generated through effective
interactions that only involve leptons and gauge bosons? If we want
it to be generated at tree level in the effective \mbox{theory --and that's}
desirable, as we would like large rates \mbox{of $0 \nu \beta \beta$-- we
are} left with just two more scenarios: in the first one, depicted in 
figure \ref{fig:0nu2beta-eff-7-and-9}\emph{a)}, the effective interaction 
involves one $W$ and one electron, and leaves one neutrino which acts as a 
$t$-channel mediator and is absorbed in a common weak interaction vertex. Note
that all the LNV occurs in the effective vertex, and the neutrino
does not need to be Majorana to provide $0 \nu \beta \beta$.
This interaction, that we present here in a post-EWSB fashion, can
arise from several gauge-invariant nonrenormalisable operators; let us 
refer from now on to any such kind of vertex as a ``$W \nu e$ interaction'';
let us also call ``$W \nu e$ mechanism'' to the several processes that 
yield $0 \nu \beta \beta$ and neutrino masses through the mediation of a
$W \nu e$ vertex.

The second option is displayed in figure 
\ref{fig:0nu2beta-eff-7-and-9}\emph{b)}, and
involves a single LNV-ing vertex with two $W$'s and two electrons.
There are no more possibilities, as two $W$'s and two electrons
are necessarily involved in the \mbox{process --the first} due to the fact
that the quarks must transform through the SM weak interaction, and
the latter because they're the observable products of the \mbox{reaction--
and the} emission of any other particle in the effective interaction
would imply the closing of a loop, which we wish to avoid.
We will label from now on this interaction as a ``$W W e e$ vertex'', and 
the associated processes, the ``$WWee$ mechanism''. Correspondingly, to
the Weinberg operator that appears in figure \ref{eff-0nu2beta-5} we will
refer sometimes as a ``$\nu \nu$ interaction'', and to the generation
of neutrino masses at tree level and figure \ref{eff-0nu2beta-5} itself,
the ``$\nu \nu$ mechanism''.

\begin{figure}[t]
	\centering
	\raisebox{3.5cm}{\emph{a)}}
	\includegraphics[width=0.7\textwidth]{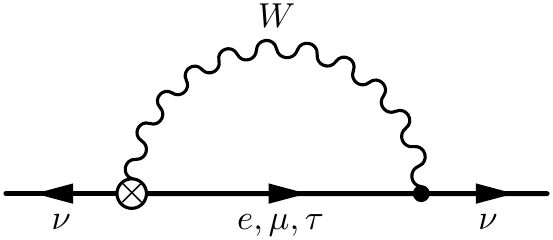}

	\vphantom{\raisebox{3ex}{a}} 
	\raisebox{4cm}{\emph{b)}}
	\includegraphics[width=0.7\textwidth]{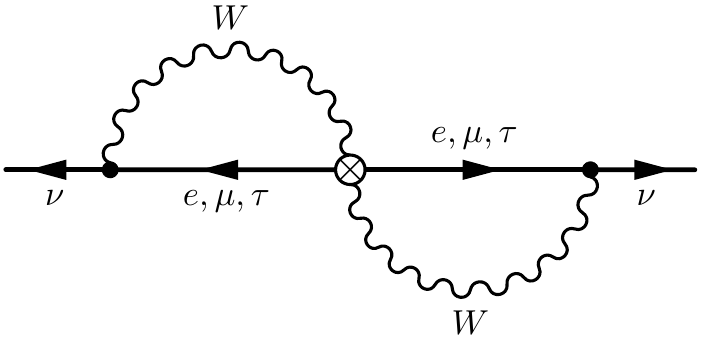}
	\caption{The diagrams that illustrate the generation of neutrino 
			mas\-\mbox{ses \emph{a)} at one loop,} 
			in effective theories with a $W \nu e$ vertex, 
			\mbox{and \emph{b)} at two} loops, in effective theories that 
			include a $WWee$ interaction.
			} \label{fig:0nu2beta-eff-numass-7-and-9}
\end{figure}

On the other hand we have neutrino masses; we want to know what sort of
Majorana masses can be associated with the three effective operators we have
just described.
The first one is trivial: it is already describing a Majorana
neutrino mass, which is generated at tree level and that's all.
More interesting are the other two: it is apparent at first sight that every 
$W-e$ pair can produce a neutrino by closing a loop. Thus, we would
expect the $W \nu e$ interaction to generate neutrino masses at one loop
and the $W W e e$ interaction to generate them at two loops, as seen in 
figure \ref{fig:0nu2beta-eff-numass-7-and-9}. Note, however, that in the 
effective theory these radiative neutrino masses are not truly calculable:
as the theory is nonrenormalisable, of the loops depicted in figure 
\ref{fig:0nu2beta-eff-numass-7-and-9} we can only compute the
running part; the matching, as is only logical, is lost when we abstract
from the underlying, high-energy theory, and cannot be recovered from
the low-energy operators. In other words: a complete calculation of the 
loop-generated neutrino masses requires that we stick with a particular
model. In chapters \ref{chap:Wnue-model} and \ref{chap:WWee-model} we offer 
two such models that allow to specify the loops in figure 
\ref{fig:0nu2beta-eff-numass-7-and-9} with full detail. In section
\ref{sec:0nu2beta-eff-neutrinomasses} we will comment further on
what the effective theory can tell us about neutrino masses.

It's time now, after this somewhat qualitative introduction, to proceed
to a formal deduction of these effective operators in an explicitly
gauge-invariant form that will allow for practical calculations.
In the next section we will work out such a deduction, and in the
meantime we will discover a remarkable connection between the chirality
of the leptons involved in the effective interaction and the leading
mechanism for $0 \nu \beta \beta$; this will have consequences for
both model-building and radiative generation of neutrino masses.

\section{Identification of the relevant operators. The role of chirality} 

\subsection{Preliminary remarks} \label{sec:0nu2beta-eff-preliminary-remarks}

The first step for the design of an effective theory is the choice of
the low-energy fields; for this role we will take the whole Standard
Model with the usual doublet Higgs field and no further addition, neither
new light fermions (in particular we won't consider light right-handed
neutrinos) nor any extension of the scalar sector\footnote{But see
chapter \ref{chap:Wnue-model}, where we will present a model with a second
light Higgs doublet which will simplify our discussion.}.
This election seems reasonable, as the physics responsible for LNV is
deemed to lie above the TeV scale. 
The effective theory will then be constructed by listing all the operators
comprised of the SM fields and compatible with the SM local symmetries,
including those of dimension greater than four. The effects of new physics,
which are necessarily virtual at low energies, 
are described, or rather parametrised,
by the coefficients $C\up{n}_i$ which accompany the tower
of nonrenormalisable operators. Explicitly, we write
\begin{displaymath} 
	\mathcal{L} = \mathcal{L}_{\mathrm{SM}} + \sum_{n=5}^\infty \sum_i  
		\left( \frac{C^{(n)}_i}{\Lambda^{n-4}} \, \mathcal{O}\up n_i + 
		\mathrm{H.c.} \right)\, ,
\end{displaymath}
where $n$ denotes the canonical dimension of the operator, 
$i$ labels the several independent operators of the same dimension,
and $\Lambda$ represents the NP scale. When the NP contains several scales
$\Lambda > \Lambda^\prime > \ldots$ the coefficients $C$ may contain
powers of $\Lambda / \Lambda^\prime$. For this analysis we'll be
considering all along the new physics to be weakly coupled and decoupling.

Operators contributing to $0 \nu \beta \beta$ decay 
must involve two leptons of either chirality, $\ell_{\mathrm{L}}$ 
or $e_{\mathrm{R}}$, 
and a number of scalar doublets, $\phi$, to make the product invariant under 
$SU(2)_L \times U(1)_Y$ transformations. Besides, they can have covariant 
derivative ($D_\mu$) insertions, which do not change the field quantum numbers. 
It is useful to note that operators differing in 
an even number of covariant derivatives,
say $\mathcal{O}^{\, \prime} \sim D^2 \mathcal{O}$, are generated 
simultaneous\-\mbox{ly --i.e.,} at the same loop \mbox{level-- in a} 
given model. This can be understood if 
we think of the two covariant derivatives as two extra gauge bosons in
the effective vertex; if a model can generate $\mathcal{O}$ at some loop
level, then to generate $\mathcal{O}^{\, \prime}$ one only needs to emit
a gauge boson at some point in the Feynman diagram and immediately afterwards
revert the change with the emission of a second gauge boson. There's a 
particularly interesting application of this principle for the case we are
considering: the no-$W$ operator depicted in figure \ref{eff-0nu2beta-5} is
nothing but a Majorana mass for the neutrinos, and so the Weinberg
operator, $(\overline{\tilde{\ell}_{\mathrm{L}}} \phi) \, ( \tilde \phi^\dagger
\ell_{\mathrm{L}})$. This operator involves two lepton doublets, and so 
two left-handed leptons, specifically $\nu_{\mathrm{L}} \nu_{\mathrm{L}}$; 
by this argument, there will be one operator which will be generated along
with the Weinberg operator and which only differs from it in two
covariant derivatives. This operator will yield an interaction of the
form $W W e_{\mathrm{L}} e_{\mathrm{L}}$, exactly of the $W W e e$ kind
depicted in figure \ref{fig:0nu2beta-eff-7-and-9}\emph{b)}. Of course, the 
$W W e_{\mathrm{L}} e_{\mathrm{L}}$ interaction will be suppressed with
two additional powers of $\Lambda$ respect to the Weinberg operator and
will be subdominant for what $0 \nu \beta \beta$ concerns. But the
lesson we must extract from this is that if we want to model $0 \nu
\beta \beta$ according to the $W W e e$ mechanism \emph{we cannot rely
on an operator with two left-handed electrons} for it will be generated
along with the Weinberg operator, which will take control of the double
beta decay. However, if we look for a model generating an operator
of the type $W W e_{\mathrm{R}} e_{\mathrm{R}}$, which is not directly
related to the Weinberg operator, maybe we'll have the option
of relegating it to a higher loop level and obtain a $WW ee$-dominated 
$0 \nu \beta \beta$.

We take this idea as an inspirational starting point: in what remains
of this section we will prove that chirality can be used to isolate
qualitatively different operators that correspond to the three mechanisms
described in the previous section. As they \emph{are} qualitatively 
different, they can be used to devise models in which each of the
three mechanisms dominates $0 \nu \beta \beta$ and neutrino masses.

\subsection{Notation}

The deduction of effective operators for the SM, and especially this class
that includes covariant derivatives, can be a little bit cumbersome due
to the need to correctly close Lorentz and $SU(2)$ indices and to the several
Fierz reorderings that can be carried out. In order to simplify the process
let us introduce some composite fields with definite lepton number
and hypercharge and no $SU(2)$ charge. As we want to construct operators
with just two leptons we will only need a handful of them. First let us present
the composite fields with one lepton:
\begin{align}
	N_{\alpha} &= \phi^\dagger \tilde \ell_{\mathrm{L} \alpha} &
		\Psi^\mu_{\alpha} &= \phi^\dagger D^\mu \tilde \ell_{\mathrm{L} \alpha} 
		\nonumber \\
	Y(N_\alpha) &= 0   &		Y(\Psi^\mu_\alpha) &= 0  \label{composite-1f} \\
	L(N_\alpha) &= -1  &		L(\Psi^\mu_\alpha) &= -1 \nonumber \\
	\mathrm{dim} (N_\alpha) &= \nicefrac{5}{2}	&
		\mathrm{dim}(\Psi^\mu_\alpha) &= \nicefrac{7}{2} \nonumber
\end{align} 
where the usual notation has been used for the \mbox{SM fields -- see} sections
\ref{sec:intro-SM-doublets-singlets} and \ref{sec:masses-SM}.
Note that we have selected hyperchargeless combinations of fields in order 
to make even easier the construction of the effective operators. These fields
need not to have a definite physical signification, as they're just artifacts
that we are going to use according to their charges and canonical dimension.
But anyway, someone might insist; in such a case
$N$ could be regarded as a ``left-handed, $SU(2)$-singlet lepton'', 
and $\Psi$ maybe
as a ``left-handed, singlet lepton plus a gauge boson''. Indeed, if we let the
electroweak symmetry to break spontaneously we obtain
\begin{align} \label{N-Psi-after-SSB}
	N_\alpha &= - v \, \nu_{\mathrm{L} \alpha}^{\mathrm{c}}  + \ldots \, , 
		\\
	\Psi^\mu_\alpha &= - v \left[ \left( \partial^\mu + \frac{i}{2} \, 
		\frac{g}{\cos \theta_W} Z^\mu \right) \nu_{\mathrm{L} \alpha}^{\mathrm{c}} +
		\frac {i g}{\sqrt{2}} \, W^\mu \, e_{\mathrm{L} \alpha}^{\mathrm{c}} \right] 
		+ \ldots \, , \nonumber
\end{align}
where the ellipsis denote the terms involving the physical Higgs boson, which
are not of interest for our study. Note that $N$ represents in the end just
a left-handed neutrino, whereas $\Psi$ is a more complicated combination of
fields. Note also that we have not defined a composite
field involving the right-handed $SU(2)$ singlets; as they only carry hypercharge
and lepton number it's not difficult to add them directly to the effective
operators when needed.

Let us now go ahead to composite fields with two leptons; as we are interested
in two-lepton operators, such composite fields must comprise all the leptonic
content in the final operator; and as we want the operators to violate lepton
number in two units, so must do the composite field. It so happens that there
is only one field combination that fullfils all these conditions and cannot
be built from $N$ or $\Psi$ through Fierz rearrangements. It is this one
here:
\begin{align}
	J^\mu_{\alpha \beta} &= \bar \ell_{\mathrm{L} \alpha} D^\mu 
			\tilde \ell_{\mathrm{L} \beta} \nonumber \\
	Y(J^\mu_{\alpha \beta}) &= 1   \label{composite-J} \\
	L(J^\mu_{\alpha \beta}) &= -2   \nonumber \\
	\mathrm{dim}(J^\mu_{\alpha \beta}) &= 4 \, ,  \nonumber 
\end{align}
which might be described as a ``LNV-ing leptonic current plus a gauge boson'';
indeed after SSB it takes the somewhat convoluted form
\begin{multline} \label{J-after-SSB}
	J^\mu_{\alpha \beta} = \overline{\nu_{\mathrm{L} \alpha}} 
			\left[ {\stackrel \leftrightarrow{\partial^\mu}} -
			i g \left(\frac{\cos 2 \theta_W}{2 \cos \theta_W} \, Z^\mu + 
			\sin \theta_W A^\mu \right) \right] 
			e_{\mathrm{L} \beta}^{\mathrm{c}} + \\
		+ \frac{ig}{\sqrt{2}} \, \left( {W^\mu}^\dagger \,
			\overline{\nu_{\mathrm{L} \alpha}} \nu_{\mathrm{L} \beta}^{\mathrm{c}} -
			W^\mu \overline{e_{\mathrm{L} \alpha}} e_{\mathrm{L} \beta}^{\mathrm{c}}
			\right) - 
			\frac{i g}{2} \, \frac{1}{\cos \theta_W} \, Z^\mu \, 
			\overline{e_{\mathrm{L} \alpha}} \nu_{\mathrm{L} \beta}^\mathrm{c} \, .
\end{multline}

This exhausts the list of leptonic pieces that we are to use. Now only one more
element is needed: if we look at \eqref{composite-J} we see that $J$ has
hypercharge 1; we don't have, however, any other piece that can cancel
such hypercharge, were we to use $J$ at some point: $N$ and $\Psi$ are
hyperchargeless, the leptonic right-handed singlets would yield too many
leptons in the operator, and the combination $\phi^\dagger \tilde \phi$
vanishes with just one scalar doublet. We need, so, one more bosonic
composite field; it must be analogous to $\phi^\dagger \tilde \phi$, 
but nonzero. The only option is to add a vector boson:
\begin{align}
	\mathcal{W}^\mu &= \phi^\dagger D^\mu \tilde \phi \nonumber \\
	Y(\mathcal{W}^\mu) &= -1   \label{composite-W} \\
	L(\mathcal{W}^\mu) &= 0   \nonumber \\
	\mathrm{dim}(\mathcal{W}^\mu) &= 3 \, .  \nonumber 
\end{align}
This $\mathcal{W}$ operator is a sort of ``hidden gauge boson'' plus other 
terms with scalar fields, as we can see by letting 
$\phi$ take a VEV:
\begin{equation} \label{W-after-SSB}
	\mathcal{W}^\mu = -i \, \frac{g}{\sqrt{2}} \, v^2 \, W^\mu + \ldots \, .
\end{equation}

With all these elements we are ready to begin our task. Just remember a few
rules that must be observed while constructing the operators:
\begin{enumerate}[\it i)]
	\item \textbf{Only two leptons:} We only want two leptons in our effective
			operator, in order to generate $0 \nu \beta \beta$ with no loops. 
			This rapidly exhausts the possible combinations of fields; 
			for example, if $J$ is used there's no room for more leptons.

	\item \textbf{Right-handed singlets:} The composite fields
			$N$, $\Psi$, $J$ and $\mathcal{W}$ contain only left-handed
			leptons; the right-handed singlets $e_{\mathrm{R} \alpha}$ 
			must be added by hand whenever convenient. Recall their charge 
			assignments and dimensionality:
			\begin{displaymath}
				Y(e_{\mathrm{R} \alpha}) = -1 \qquad \qquad
				L(e_{\mathrm{R} \alpha}) = 1  \qquad \qquad
				\mathrm{dim} (e_{\mathrm{R} \alpha}) =  \nicefrac{3}{2}
			\end{displaymath}

	\item \textbf{Watch out for hypercharge:} Hypercharge constrains considerably
			the possible combinations: $N$ and $\Psi$ are hyperchargeless,
			and $J$, $\mathcal{W}$ and $e_{\mathrm{R}}$ have unit hypercharge
			(with various signs), but $J$ and $e_{\mathrm{R}}$ cannot be
			combined.

	\item \textbf{Close Lorentz indices:} $\Psi_\mu$, $J_\mu$ and 
			$\mathcal{W}_\mu$ have a Lorentz four-vector index; 
			they can be closed with one another, but also by using
			any Dirac matrix, like $\gamma_\mu$ or $\sigma_{\mu \nu}$,
			or additional derivatives. Derivatives acting upon the composite
			fields or lepton singlets must be adapted to their charge content;
			for example, $N$ and $\Psi$ are hyperchargeless singlets, so
			they require just common partial derivatives, $\partial_\mu$.
			All the pieces we are to use are $SU(2)$ singlets, so these
			`external' derivatives will never include the $SU(2)$ 
			generators or gauge bosons.

			Finally, remember that $N$, $\Psi$ and $e_{\mathrm{R}}$ have
			spinorial indices that must also be closed in a consistent way.

\end{enumerate}

\subsection{Operators with two left-handed leptons}

This class of operators comprises, as we already anticipated, the Weinberg
operator and several other higher-order operators. They must be constructed
without using the right-handed singlets $e_{\mathrm{R} \alpha}$, and remember
that they must violate lepton number in two units, $\Delta L = \pm 2$.
The lowest-order operator can be built using only the $N$ fields;
it is of dimension five, and reads
\begin{equation} \label{Weinberg-(5)}
	\mathcal{O}_{\alpha \beta}\up 5 = 
		\overline{N_\alpha} N_\beta^\mathrm{c} = 
		- \left( \overline{\tilde \ell_{\mathrm{L} \alpha}} \phi \right) \,
		\left( \tilde \phi^\dagger \ell_{\mathrm{L} \beta} \right) =
		v^2 \, \overline{\nu_{\mathrm{L} \alpha}^\mathrm{c}} 
		\nu_{\mathrm{L} \beta} \, .
\end{equation}
Of course, as expected this operator is nothing but the Weinberg operator,
and generates tree-level neutrino masses suppressed by just one power of
the new physics scale. 

At dimension 6 we don't find any operator of our interest:
all LNV-ing operators at this level involve quarks and also violate baryon number
in one unit; they are, thus, not useful for providing $0 \nu \beta \beta$
and neutrino masses without additional new physics. At dimension 7, however,
several possibilities appear: first we find an operator which is made
of $\mathcal{O} \up 5$ plus a SM-singlet pair of scalar doublets,
$\left( \phi^\dagger \phi \right)$; but this operator does not provide
a new mechanism for $0 \nu \beta \beta$ or neutrino masses, and
we will ignore it. The interesting dimension-7 operators are those 
which are structurally different from $\mathcal{O} \up 5$, and it's
easy to check that there're only three of them:
\begin{align*}
	\mathcal{O}_{\alpha \beta} \up{7 \, \textsc{i}} &= 
			\overline{N_\alpha^\mathrm{c}} \, \partial_\mu \Psi_\beta^\mu =
			- \left( \overline{\ell_{\mathrm{L} \alpha}} \, \tilde \phi \right) 
			\partial_\mu \left( \phi^\dagger D^\mu \tilde
			\ell_{\mathrm{L} \beta} \right) 
			\\
	\mathcal{O}_{\alpha \beta} \up {7 \, \textsc{ii}} &= 
			\overline{\Psi_\alpha^\mathrm{c}}_\mu \Psi^\mu_\beta =
			- \left( \overline{ D_\mu \ell_{\mathrm{L} \alpha}} \, \tilde \phi
			\right) \left( \phi^\dagger D^\mu \tilde \ell_{\mathrm{L} \beta}
			\right)  
			\\
	\mathcal{O}_{\alpha \beta} \up {7 \, \textsc{iii}} &=
			J^\mu_{\alpha \beta} \, \mathcal{W}_\mu =
			\left( \overline{\ell_{\mathrm{L} \alpha}} \, D^\mu
			\tilde \ell_{\mathrm{L} \beta} \right)
			\left( \phi^\dagger D_\mu \tilde \phi \right) 
\end{align*}
By looking at \eqref{N-Psi-after-SSB}, \eqref{J-after-SSB} and \eqref{W-after-SSB}
we see that all three operators yield vertices of the type 
$W \nu_{\mathrm{L}} e_{\mathrm{L}}$;
additionally, $\mathcal{O} \up {7 \, \textsc{iii}}$ also provides a 
$W W e_{\mathrm{L}} e_{\mathrm{L}}$ interaction. 
We could think that these operators can be a source
for the mechanisms described in section \ref{sec:0nu2beta-eff-Violating-LN},
but they are not: they differ from $\mathcal{O} \up 5$ in two covariant
derivatives, and so they will be produced always together with it.
$\mathcal{O} \up 5$ will yield neutrino masses at tree level and
the $\mathcal{O} \up 7$'s will provide them through loops; 
$\mathcal{O} \up 5$ will give $0 \nu \beta \beta$ suppressed by
$\Lambda^{-1}$, whereas the $\mathcal{O} \up 7$'s will be suppressed
\mbox{by $\Lambda^{-3}$ -- with the} $\Lambda$ scale identical for
one and the others, as all of them are generated by the same physics.
All effect of these left-left $\mathcal{O} \up 7$'s, therefore, will
be subdominant; they cannot provide leading and observable
$W \nu e$- and $W W e e$-like mechanisms. To produce such mechanisms
we need to resort to operators with one or more right-handed lepton.
Let us in the next two sections derive such operators.

\subsection{Operators with one left- and one right-handed lepton}

For this class of operators we need the usual two-unit violation of lepton number
and one leptonic singlet, $e_{\mathrm{R}}$; the field $J$ will be therefore
forbidden, as its use would yield one lepton too many. If we study the
possible combinations of the remaining fields we easily see that there's only one 
lowest-order operator, and it is of dimension 7:
\begin{equation} \label{0nu2beta-eff-O7}
	\mathcal{O}_{\alpha \beta} \up 7 = \overline{e_{\mathrm{R} \alpha}}
			\gamma^\mu N_\beta \, \mathcal{W}_\mu =
			\overline{e_{\mathrm{R} \alpha}} \gamma^\mu \left( \phi^\dagger
			\tilde \ell_{\mathrm{L} \beta} \right) \left( \phi^\dagger
			D_\mu \tilde \phi \right) \, .
\end{equation}
We label it just ``$\mathcal{O} \up 7$'', as it is the only dimension 7
operator that we will be considering from now on. By looking at 
\eqref{N-Psi-after-SSB} and \eqref{W-after-SSB} we identify that this 
$\mathcal{O} \up 7$ provides a vertex
\begin{equation} \label{O7-after-SSB}
	\mathcal{O}_{\alpha \beta} \up 7 = i \, \frac{g}{\sqrt{2}} \, v^3 \,
			W_\mu \, \overline{e_{\mathrm{R} \alpha}} \gamma^\mu 
			\nu_{\mathrm{L} \beta}^\mathrm{c} + \ldots \, ,
\end{equation}
plus many others that involve the physical Higgs boson. This vertex will be
the starting point for the $W \nu e$ mechanism. We will not discuss here any 
other higher-order operator of this class, even if they could eventually
prove independent from $\mathcal{O} \up 7$ and thus provide their own sources
for the $W \nu e$ interaction; as $\mathcal{O} \up 7$ is the least suppressed
of this family of operators we will concentrate our analysis on it.

\subsection{Operators with two right-handed leptons}

For this class of operators all leptons are provided by the $SU(2)$ singlets
$e_{\mathrm{R}}$; hence, of the composite fields only $\mathcal{W}$ will be
available. As happened with the left-right class, we soon realise that 
with these elements there is only one lowest-order operator violating LN
in two units; it appears at dimension 9, and reads
\begin{equation} \label{0nu2beta-eff-O9}
	\mathcal{O}_{\alpha \beta} \up 9 = \overline{e_{\mathrm{R} \alpha}}
			e_{\mathrm{R} \beta}^\mathrm{c} \, \mathcal{W}^\mu
			\mathcal{W}_\mu = \overline{e_{\mathrm{R} \alpha}}
			e_{\mathrm{R} \beta}^\mathrm{c} \left( \phi^\dagger
			D^\mu \tilde \phi \right) \left( \phi^\dagger
			D_\mu \tilde \phi \right) \, .
\end{equation}
As in the previous case, this one will be the only dimension 9 operator we
will be considering, so we label it just ``$\mathcal{O} \up 9$''. By letting
electroweak symmetry to break we find this operator to yield an interaction
\begin{displaymath}
	\mathcal{O}_{\alpha \beta} \up 9 = - \frac{g^2}{2} \, v^4 \, W^\mu W_\mu \,
			\overline{e_{\mathrm{R} \alpha}} e_{\mathrm{R} \beta}^\mathrm{c} +
			\ldots
\end{displaymath}
plus others involving the physical Higgs boson; this interaction will be the 
basis for the $W W e e$ mechanism. As with the left-right operators we won't
consider higher-order interactions of this class, as $\mathcal{O} \up 9$
is the least suppressed of all; such investigation might be interesting
if other, higher-order operators can be generated idenpendently of 
$\mathcal{O} \up 9$.

\subsection{A note to the model builder} \label{sec:0nu2beta-eff-note-model-builder}

In this section we have derived the gauge-invariant form of the lowest-order
effective vertices that generate the $\nu \nu$, $W \nu e$ and $WW ee$ interactions.
We have observed that the chirality of the involved leptons allows to 
select a set of operators which dominantly contribute to just one of the mechanisms;
in other words, chirality helps not to mix the three mechanisms at the effective 
theory level. But at the level of the underlying model the situation can be 
different: if lepton number is broken, all three $\mathcal{O} \up 5$,
$\mathcal{O} \up 7$ and $\mathcal{O} \up 9$ are to be generated; this is inevitable,
and in general the three mechanisms will compete and mix. As model builders,
in many cases we will want this not to happen. This issue can be addressed by
imposing restrictions on the model, usually in the form of additional 
symmetries; the point will be to relegate the generation of the unwanted 
operators to a higher loop level and, equally important, to make sure 
that they are generated through the same physics as the one we want to 
focus on. 

For instance, consider this rather paradigmatic case: we want to
produce a model where $0 \nu \beta \beta$ and the neutrino masses are
dominated by the $WWee$ mechanism. We have a model that generates 
$\mathcal{O} \up 9$ at tree level, that's all right, but it also generates
$\mathcal{O} \up 5$ at tree level. This poses a double trouble: first,
$\mathcal{O} \up 5$ is going to compete with $\mathcal{O} \up 9$ for 
the preeminence in $0 \nu \beta \beta$, but as $\mathcal{O} \up 5$ is 
suppressed by $\Lambda^{-1}$ and $\mathcal{O} \up 9$ by $\Lambda^{-5}$,
the $\nu \nu$ mechanism is bound to win. Consequently, we devise a
suitable symmetry which forbids some of the couplings that yield
$\mathcal{O} \up 5$; it's gone, and problem solved. But lepton number
is broken, and it reappears at one loop through the mediation of other
couplings; it is not anymore a threat to $0 \nu \beta \beta$, because
it's loop-suppressed, but it yields one-loop neutrino masses which
compete with the two-loop masses generated by $\mathcal{O} \up 9$.
We apply the same principle again: look for the way to forbid any of
the relevant couplings for the one-loop generation of $\mathcal{O} \up 5$.
And again it goes away. But at two loops\dots $\;$here it is once more.
But of course it is: $\mathcal{O} \up 9$ generates neutrino masses at two
loops; we realise that the diagrams generating $\mathcal{O} \up 5$ are 
nothing but the same that yield neutrino masses through $\mathcal{O} \up 9$.
So we can just forget about $\mathcal{O} \up 5$: the $WW ee$ mechanism
explains it all, and if we control $\mathcal{O} \up 9$ we control the
$WW ee$ mechanism. Moral: the problem is \emph{not} the presence of 
$\mathcal{O} \up 5$; you cannot, in fact, avoid its presence. 
The problem is, rather, if it arises
from some physics independent of the operator you're interested in; then it competes,
and it probably spoils some of your plans. If you can control the generation
of the unwanted operators to the point that they arise from the operator
you want to dominate, then everything will probably be all right. 
We will find, in chapters \ref{chap:Wnue-model} and \ref{chap:WWee-model},
particular examples of how to deal with these issues.

\section{Implications for neutrinoless double beta decay}

The fundamental parameter of an effective theory is the new physics scale, 
$\Lambda$. It is easy just by intuition to understand that $0 \nu \beta \beta$
can provide a \emph{lower} bound for it: the higher $\Lambda$,
the lower the $0 \nu \beta \beta$ rate; we have upper bounds on this
rate, so they must translate into lower bounds on $\Lambda$. This is the main
purpose of this section. For that aim we will use the limits on
$0 \nu \beta \beta$ in $\tensor[^{76}]{\mathrm{Ge}}{}$ 
\cite{Barabash:2011fg}; as the traditional
picture for $0 \nu \beta \beta$ is that mediated by neutrino mas\-\mbox{ses
--what we} have called the $\nu \nu$ \mbox{mechanism--, the limit} is usually 
expressed in such terms, and written
\begin{equation} \label{76Ge-bound}
	m_{\beta \beta} \equiv \left| m_{ee} \right| < 0.24  - 0.5\ \; 
				\mathrm{eV} \, .
\end{equation}
But we would prefer a more versatile expression, one that can be easily applied
to the $W \nu e$ and $W W e e$ mechanisms, as they do not directly involve neutrino
masses. To derive such an expression we
first estimate the amplitude of the $\nu \nu$ process by looking at figure
\ref{eff-0nu2beta-5},
\begin{equation} \label{A-nunu}
	\left| \mathcal{A}_{0\nu\beta\beta} \up {\nu \nu} \right| \sim
			\frac{G_F^2}{p_{\mathrm{eff}}^2} \, |m_{ee}| \, .
\end{equation}
Here $p_{\mathrm{eff}}$ represents the energy scale of $0 \nu \beta \beta$,
$p_{\mathrm{eff}} \sim m_\pi$, which is the dominant scale in the neutrino
propagators. From \eqref{A-nunu} we construct a dimensionless quantity
that we can bound by using \eqref{76Ge-bound}:
\begin{equation} \label{A-0nu2beta-bound}
	\frac{p_{\rm eff}}{G^2_F} \, \left| \mathcal{A}_{0 \nu \beta \beta} 
			\up {\nu \nu} \right| \sim
			\frac{|m_{ee}|}{p_{\rm eff}} < 5 \times 10^{-9} \, ,
\end{equation}
and then we note that the amplitudes for the $W \nu e$ and $W W e e$ mechanisms
have the same dimensionality as $\mathcal{A}_{0 \nu \beta \beta} \up {\nu \nu}$
(look for example at figures \ref{fig:0nu2beta-O7} and \ref{fig:0nu2beta-O9}, 
and notice that each mechanism removes from
the diagram one neutrino propagator but adds a $v/\Lambda^2$ factor); so,
the quantity displayed in \eqref{A-0nu2beta-bound} is also dimensionless for 
the other mechanisms, and can be used directly to derive bounds for them.

\subsubsection{Bounds on $\mathcal{O} \up 5$}

\begin{figure}[tb]
	\centering
	\includegraphics[width=0.7\textwidth]{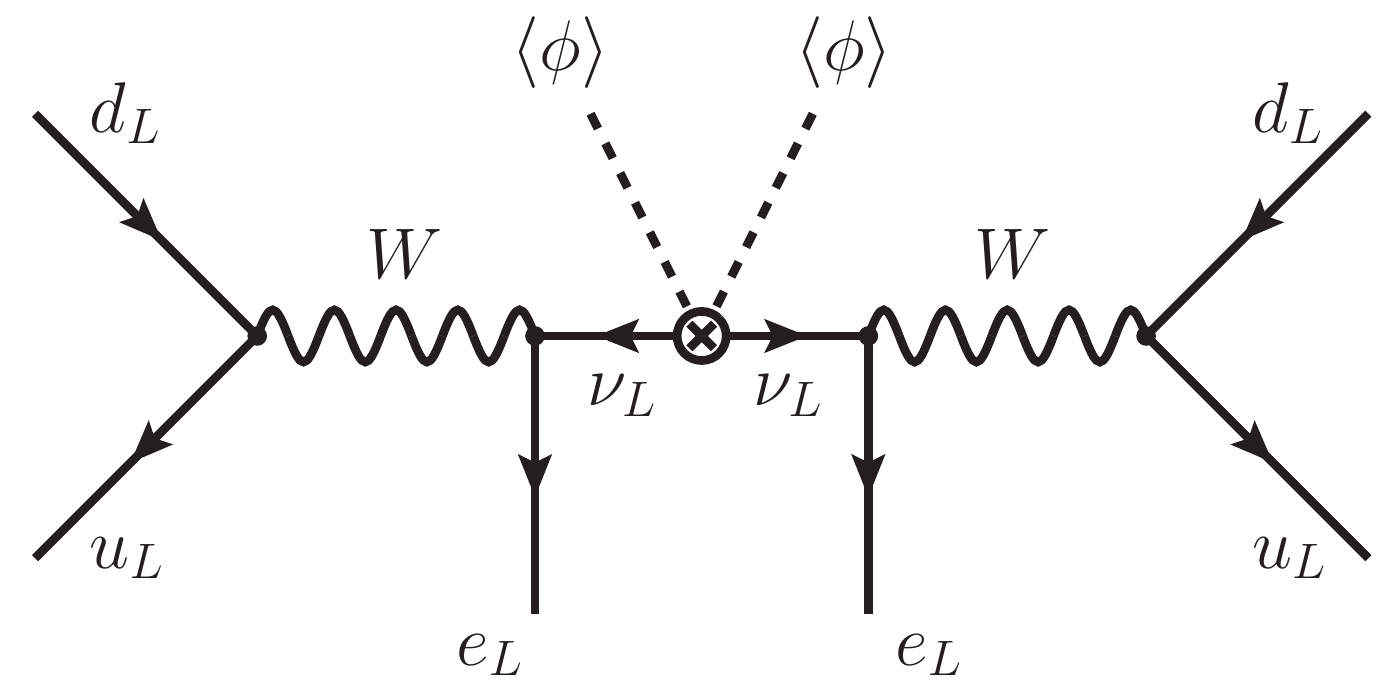}
	\caption{The process of $0 \nu \beta \beta$ through the $\nu \nu$
			mechanism, particularised here to the $\mathcal{O} \up 5$ operator.
			} \label{fig:0nu2beta-O5}
\end{figure}

The previous discussion essentially gives us all the elements: when the 
$\nu \nu$ mechanism is realised by $\mathcal{O} \up 5$, neutrinoless double
beta decay looks as in figure \ref{fig:0nu2beta-O5}; then we have
\begin{equation} \label{0nu2beta-eff-amplitude-5}
	\left| \mathcal{A}_{0 \nu \beta \beta} \up 5 \right| \sim
			\frac{G_F^2}{p_{\mathrm{eff}}^2} \, \frac{v^2}{\Lambda} \,
			\left| C_{ee} \up 5 \right| \, ,
\end{equation}
and so, by applying \eqref{A-0nu2beta-bound} we obtain a terrific bound,
\begin{equation} \label{bound-0nu2beta-O5}
	\frac{\Lambda}{| C_{ee} \up 5 |} > 10^{11} \; \mathrm{TeV} \, .
\end{equation}
This result would put the particles responsible for the $\nu \nu$ mechanism 
way beyond the reach of any present or planned experiment, unless $C_{e e} \up 5$
proves to be extremely small. This is the main reason why we turn to the 
other two mechanisms, that yield less pessimistic prospects from the experimental
point of view.

\subsubsection{Bounds on $\mathcal{O} \up 7$}

\begin{figure}[tb]
	\centering
	\includegraphics[width=0.7\textwidth]{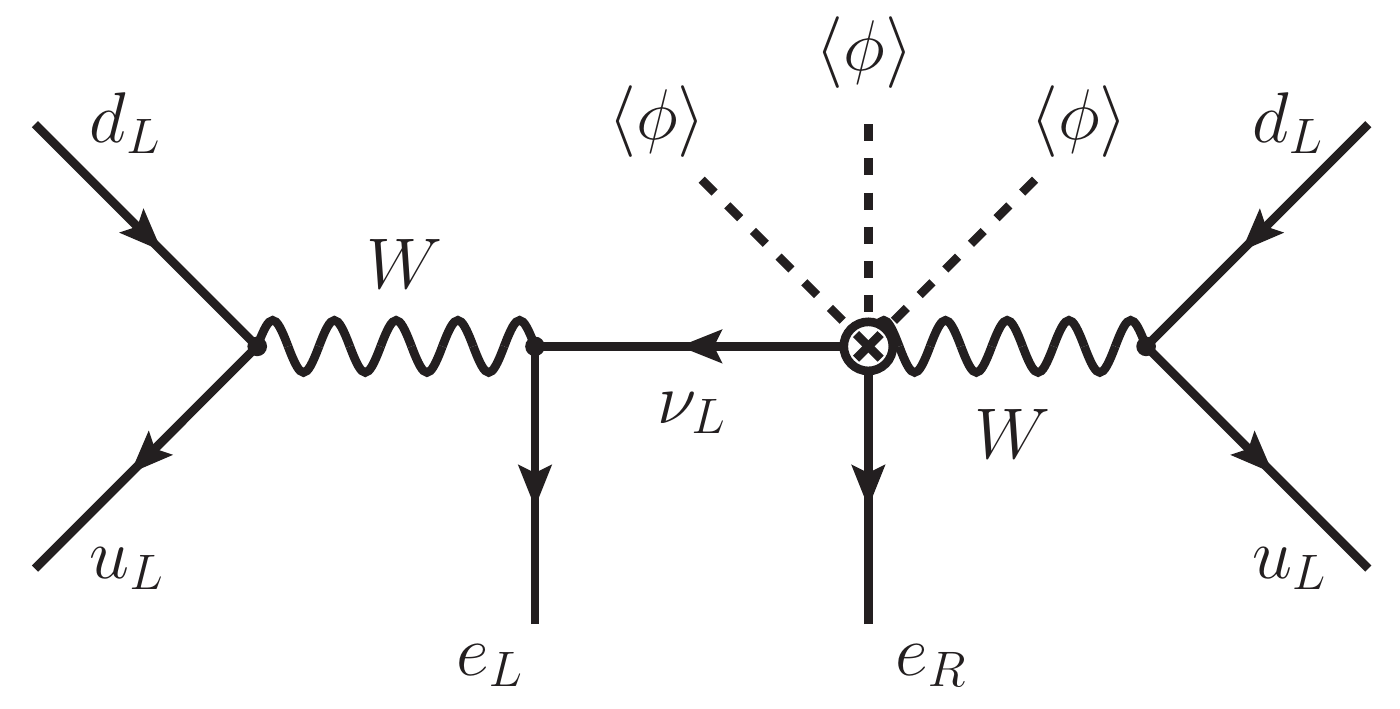}
	\caption{The process of $0 \nu \beta \beta$ as generated by the
			$W \nu e$ mechanism, particularised here to the $\mathcal{O} \up 7$
			operator.} \label{fig:0nu2beta-O7}
\end{figure}

Figure \ref{fig:0nu2beta-O7} depicts $0 \nu \beta \beta$ mediated by 
$\mathcal{O} \up 7$; from there we can estimate
\begin{equation} \label{0nu2beta-eff-amplitude-7}
	\left| \mathcal{A}_{0 \nu \beta \beta} \up 7 \right| \sim
		\frac{G_F^2}{p_{\mathrm{eff}}} \, \frac{v^3}{\Lambda^3} \,
			\left| C_{ee} \up 7 \right| \, ,
\end{equation}
and then from \eqref{A-0nu2beta-bound}
\begin{displaymath}
	\frac{\Lambda}{| C_{ee} \up 7 |^{\nicefrac{1}{3}}} > 
			100 \; \mathrm{TeV} \, .
\end{displaymath}

These are order-of-magnitude estimates, but one can also use detailed nuclear 
matrix elements available in the literature 
to derive bounds on new physics providing $0 \nu \beta \beta$
\cite{Muto:1989cd,Pas:1999fc}. The $W \nu e$ interaction 
induced by the operator $\mathcal{O} \up 7$
can be expressed as a modification of the standard weak interaction,
$W_\mu \overline{e} \gamma^\mu \left((1-\gamma_5) + \eta (1+\gamma_5) \right) \nu$, 
where $\nu=\nu_L+\nu_L^c$ is a Majorana field. Then, the strong limit on $\eta$ 
derived using
detailed nuclear matrix elements calculations, $|\eta| < 4.4\times 10^{-9}$ 
(note that $\eta$, in \cite{Pas:1999fc}'s notation, reads $\epsilon^{V+A}_{V-A}$),
implies in our case
\begin{displaymath}
	|\eta| = \frac{v^3}{\Lambda^3} \left| C_{ee} \up 7 \right| < 
			4.4 \times 10^{-9} \, ,
\end{displaymath}
as can be seen by comparing the definition of $\eta$ with \eqref{O7-after-SSB}.
This now translates into a bound which is very close to our estimate:
\begin{equation} \label{bound-0nu2beta-O7}
	\frac{\Lambda}{| C_{ee} \up 7 |^{\nicefrac{1}{3}}} > 
			106 \; \mathrm{TeV}\, .
\end{equation}

\subsubsection{Bounds on $\mathcal{O} \up 9$}

\begin{figure}[tb]
	\centering
	\includegraphics[width=0.6\textwidth]{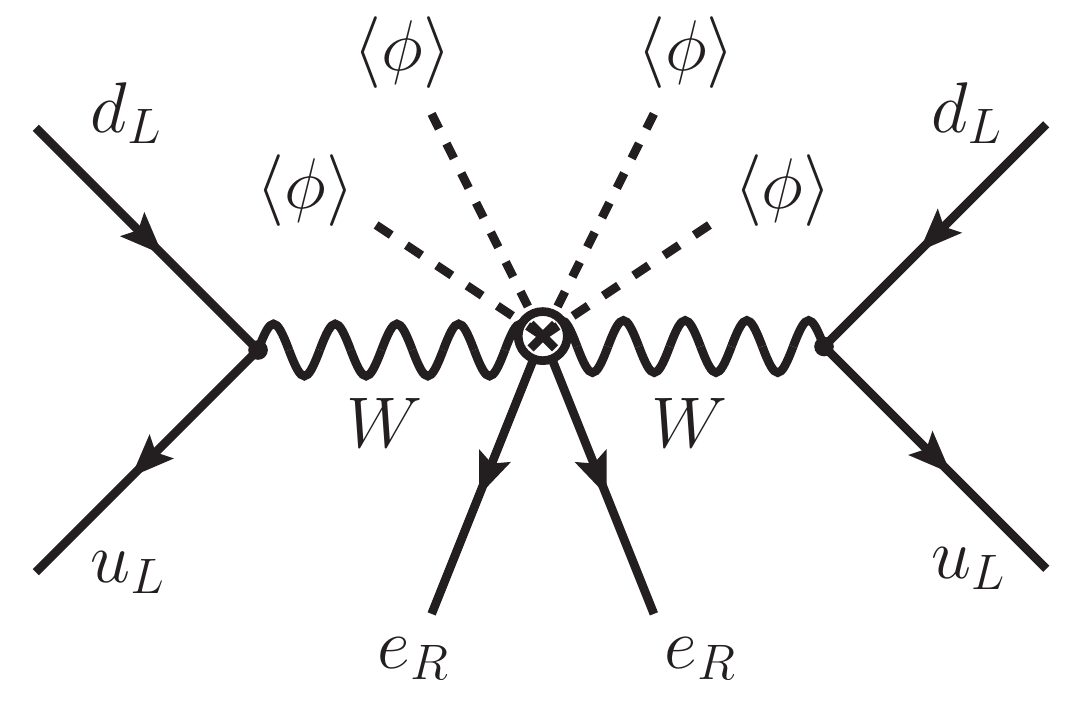}
	\caption{The process of $0 \nu \beta \beta$ as generated by the
			$W W e e$ mechanism, particularised here to the $\mathcal{O} \up 9$
			operator.} \label{fig:0nu2beta-O9}
\end{figure}

A first estimate of the bound can be obtained by looking at figure 
\ref{fig:0nu2beta-O9}; we can write then
\begin{equation} \label{0nu2beta-eff-amplitude-9}
	\left| \mathcal{A}_{0\nu\beta\beta} \up 9 \right| \sim
			G_F^2 \, \frac{v^4}{\Lambda^5} \, \left| C_{ee} \up 9 \right| \, ,
\end{equation}
and again by comparing with \eqref{A-0nu2beta-bound} we obtain
\begin{displaymath}
	\frac{\Lambda}{| C_{ee} \up 9 |^{\nicefrac{1}{5}}} > 
			2 \; \mathrm{TeV} \, ,
\end{displaymath}
which is a rather hopeful bound, suggesting that the physics responsible
for the $W W e e$ mechanism might be probed at the LHC. It is thus
even more interesting to check the more solid bounds that can be
obtained by using detailed nuclear matrix element calculations
\cite{Pas:2000vn}. In this work the authors derive bounds for the
coefficients of effective six-fermion interactions that provide $0 \nu
\beta \beta$; the six-fermion interaction
induced by $\mathcal{O} \up 9$, with \cite{Pas:2000vn}'s notation 
and normalisation, reads
\begin{displaymath}
	\mathcal{L}_{0\nu\beta\beta} = \frac{G_{F}^{2}}{2m_{p}} \, \epsilon_3 \,
			\left( \bar{u} \gamma^{\mu} (1-\gamma_{5}) d \right) \,
			\left( \bar{u} \gamma_{\mu} (1-\gamma_{5}) d \right) \,
			\bar{e} (1-\gamma_{5}) e^{\mathrm{c}} \, ,
\end{displaymath}
where $m_p$ denotes the proton mass, and for our case
\begin{displaymath}
	\epsilon_3 = -2 m_p \, \frac{v^4}{\Lambda^5} \, C_{ee} \up 9 \, .
\end{displaymath}
Now we read from \cite{Pas:2000vn} the bound for this particular type of 
new physics, $|\epsilon_3| < 1.4 \times 10^{-8}$ at 90\% CL%
\footnote{Actually there is a misprint in reference \cite{Pas:2000vn}. We 
are very grateful to the authors for providing us with the correct limit 
on $\epsilon_3$.
}, 
and from there we derive
\begin{equation} \label{bound-0nu2beta-O9}
	\frac{\Lambda}{| C_{ee} \up 9 |^{\nicefrac{1}{5}}} > 
			2.7 \; \mathrm{TeV} \, ,
\end{equation}
which is very close to our estimate, and emphasises the interest of
$\mathcal{O} \up 9$ from the experimental point of view, as the new particles
that generate it could be at the reach of the LHC; see, for more on this,
the model we discuss in chapter \ref{chap:WWee-model}.

\section{Implications for neutrino masses} \label{sec:0nu2beta-eff-neutrinomasses}

Once the effective theory generates one of the LNV-ing operators that 
produce $0 \nu \beta \beta$, neutrinos will get a mass at some loop order, 
even if there is no other independent source of neutrino \mbox{masses -- see}
figure \ref{fig:0nu2beta-eff-numass-7-and-9};
the three operators $\mathcal{O} \up 5$, $\mathcal{O} \up 7$ and 
$\mathcal{O} \up 9$ not only stand for new physics at quite 
different mass scales, but also result in different neutrino mass structures. 
In this section we want to discuss the details and restrictions that arise 
for neutrino mass generation in each mechanism.

First note that at the effective theory level we can obtain nothing but
estimates. The diagrams in figure \ref{fig:0nu2beta-eff-numass-7-and-9}
involve one or two loops, and nonrenormalisable theories do not provide
unambiguous calculations of divergent diagrams without resorting to matching, 
that is, without specifying a concrete high-energy model. As our aim in
this chapter is to draw general conclusions, we will stick to the estimates;
later, in chapters \ref{chap:Wnue-model} and
\ref{chap:WWee-model} we will discuss particular realisations of the
effective mechanisms and we will check that these estimates indeed yield
reasonable results.

Second, we need a quantity that we can impose to the generated neutrino mass
matrices, something that more or less captures `the correct neutrino mass
scale'. Unfortunately, our lack of knowledge about the neutrino mass
\mbox{eigenvalues --see} section \ref{sec:tritium-beta-decay} for a brief 
\mbox{discussion-- makes it} difficult to state even the order of magnitude
of particular elements of the mass matrix. However, we can devise an
argument that is reasonable at least for the families of models that we
are considering now: first note that even if we don't know the absolute
scale of neutrino masses, the atmospheric mass \mbox{splitting --see} 
\mbox{table \ref{tab:intro-neutrino-bounds-and-measurements}-- points to} 
a scale of at least 0.05 eV. Second, note from the diagrams in figure
\ref{fig:0nu2beta-eff-numass-7-and-9} that we expect the elements
of the generated mass matrices to be suppressed by powers of the 
corresponding charged lepton \mbox{mass -- as a} chirality flip is needed to 
close the loop, see figures \ref{fig:numass-eff-O7} and \ref{fig:numass-eff-O9}.
Thus, we can expect generically that the electron and muon elements of the
neutrino mass matrix are rather small, but still some pieces of the mass
matrix must account for a $\sim 0.05 \; \mathrm{eV}$ scale; the natural
candidates are the tau elements. In conclusion, we will take as a reasonable
assumption for the neutrino mass matrix, sensible at least for these
classes of models, that $| m_{\tau \tau} | \sim 0.1 \; \mathrm{eV}$.
From this assumption we will derive order-of-magnitude estimates for the
scales of new physics of the effective interactions.

\subsubsection{Neutrino masses from $\mathcal{O} \up 5$}

In the $\nu \nu$ mechanism neutrino masses are generated trivially at tree
level once the electroweak symmetry breaks spontaneously. To estimate
the scale of new physics for this mechanism we just observe from
\eqref{Weinberg-(5)} that
\begin{displaymath}
	m_{\tau \tau} = - 2 {C_{\tau \tau} \up 5}^* \, \frac{v^2}{\Lambda} \, .
\end{displaymath}
Then, imposing $| m_{\tau \tau} |  \sim 0.1 \; \mathrm{eV}$ we conclude
\begin{displaymath}
	\frac{\Lambda}{| C_{\tau \tau} \up 5 |} \sim 6 \times 10^{11} \; \mathrm{TeV} \, ,
\end{displaymath}
which is of the same order as the bound we obtained from $0 \nu \beta \beta$ 
for this mechanism, equation \eqref{bound-0nu2beta-O5}. Beware, this is by no
means an indication that the $\nu \nu$ mechanism is about to yield a signal
in $0 \nu \beta \beta$. It is just a reflection that both neutrino masses
and $0 \nu \beta \beta$ are tree-level processes for this mechanism; thus, 
since the bound on $m_{e e}$ from $0 \nu \beta \beta$ and the bound on $m_{\nu_e}$
from direct searches are of the same order, the two estimates yield similar
results.

\subsubsection{Neutrino masses from $\mathcal{O} \up 7$}

\begin{figure}[tb]
	\centering
	\includegraphics[width=0.6\textwidth]{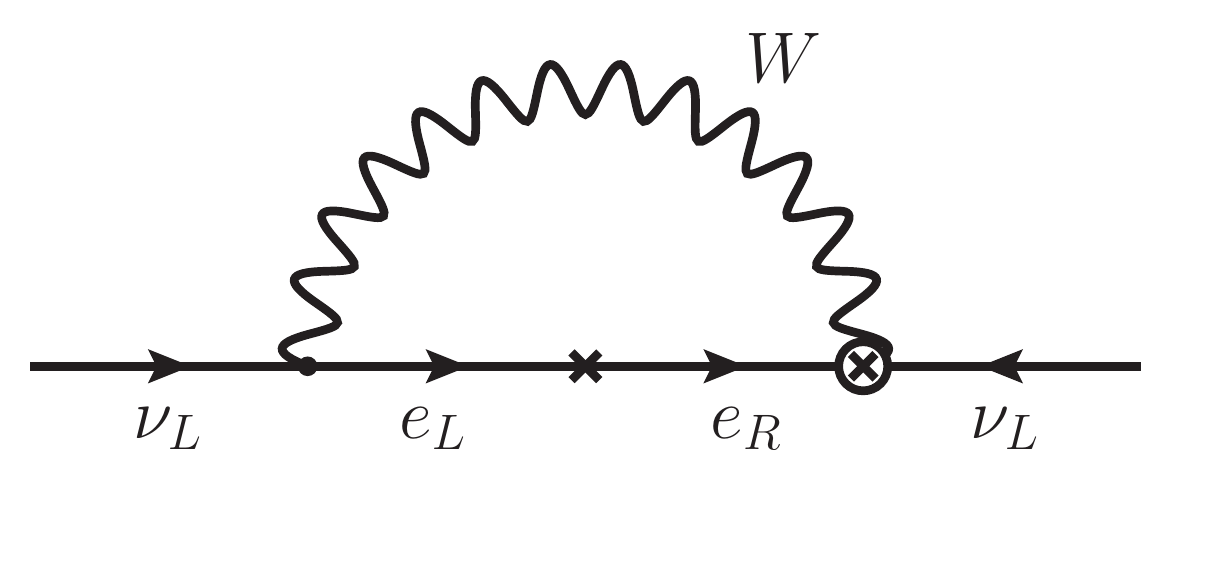}
	\caption{The diagram that generates neutrino masses at one loop with
			one $\mathcal{O} \up 7$ insertion.} 
			\label{fig:numass-eff-O7}
\end{figure}

In this case the neutrino masses are generated at one loop,
as can only be expected given that the effective interaction yields one neutrino
and one charged lepton; we need thus to close a loop with the $W$ and the
$e_{\mathrm{R}}$ to produce a second left-handed neutrino, as we see in figure
\ref{fig:numass-eff-O7}. Notice also that as the charged lepton is right-handed
a chirality flip, i.e., a mass insertion, is needed; the generated neutrino
masses, thus, will be additionally suppressed by one power of the charged 
lepton masses. When considering particular models this mechanism for 
providing neutrino masses will happen one loop above the generation of
$\mathcal{O} \up 7$; 
in chapter \ref{chap:Wnue-model} we present a model that realises the 
least-suppressed scenario: $\mathcal{O} \up 7$ is generated
at tree level, and so neutrino masses at one loop.

What can we say about these masses from the EFT point of
view? The effective theory can provide logarithmic contributions to
the loop depicted in figure \ref{fig:numass-eff-O7}; these contributions
are calculable and are obtained in 
standard effective-field-theoretical fashion using dimensional regularisation
and a renormalisable gauge. We obtain for them
\begin{displaymath}
	\delta m_{\alpha \beta} \sim \frac{v^3}{16 \pi^2 \Lambda^3} \, 
			\left(m_\alpha C_{\alpha \beta} \up 7 + 
			m_\beta C_{\beta \alpha} \up 7 \right) \, 
			\ln \frac{\Lambda}{v} \, ,
\end{displaymath}
where the $m_\alpha$ are understood to be charged lepton masses, whereas
$m_{\alpha \beta}$ represents an element of the neutrino mass matrix.
These logarithmic contributions, however, are subleading; 
the dominant contributions
to the loop come from the matching of the effective theory with the underlying
model, which cannot of course be calculated in general. We can nonetheless
estimate them using dimensional analysis, and we obtain
\begin{equation} \label{numass-matching-O7}
	m_{\alpha \beta} \sim \frac{v}{16 \pi^2 \Lambda} \, 
			\left(m_\alpha C_{\alpha \beta} \up 7 + m_\beta 
			C_{\beta \alpha} \up 7 \right) \, .
\end{equation}
Notice, however, that even for this estimation it is important
to use a renormalisable gauge, for in the unitary gauge spurious
positive powers of $\Lambda$ may appear. Note too that the matching estimate
is only suppressed by one power of $\Lambda$; this is logical, as the Weinberg
operator, of dimension five, describes the leading contribution to neutrino 
masses in the absence of $\nu_{\mathrm{R}}$. That is to say: the matching
estimate is describing a contribution to the Weinberg operator that is
generated one loop obove $\mathcal{O} \up 7$; even if a dimension seven operator
is involved, some powers of the heavy masses will be \mbox{compensated 
--dimen}\-sionally compensated, \mbox{we mean-- by dimen}\-sionful couplings
or by the loop integrals in such a way that the final result will be
effectively suppressed by just one power of heavy scales. Of course,
as always, this $\Lambda$ needs not to be the mass of an actual particle
of the theory; if the theory presents several mass scales it can be 
a combination of them with dimensions of mass. In the model presented
in chapter \ref{chap:WWee-model} this is in fact the case; look at equation
\eqref{WWee-model-generated-numass}, where all $m_\kappa$, $m_\chi$ and
$\mu_\kappa$ are heavy scales of the model.

Now we can use equation \eqref{numass-matching-O7} to estimate the new 
physics scale for the $W \nu e$ models. Requiring once more that
$| m_{\tau \tau} |  \sim 0.1 \; \mathrm{eV}$ we get
\begin{equation} \label{0nu2beta-eff-numass-bound-O7}
	\frac{\Lambda}{| C_{\tau \tau} \up 7 |} \sim 4 \times 10^7 \; \mathrm{TeV} \, ,
\end{equation}
which is consistent with the bound from $0 \nu \beta \beta$, equation 
\eqref{bound-0nu2beta-O7}, but points to a much heavier scale. Note that equation
\eqref{bound-0nu2beta-O7} marks the scales that are about to be explored in 
the forthcoming $0 \nu \beta \beta$ experiments, and it would be nice to obtain
a signal there. Equation \eqref{0nu2beta-eff-numass-bound-O7} informs us
that in the framework of the $W \nu e$ mechanism only models that provide 
very suppressed $C \up 7$ coefficients, with $| C \up 7 | \sim 10^{-9}$,
can provide a signal in the next round of $0 \nu \beta \beta$ experiments.
This is the case of the model we discuss in chapter \ref{chap:Wnue-model}.

\subsubsection{Neutrino masses from $\mathcal{O} \up 9$}

\begin{figure}[tb]
	\centering
	\includegraphics[width=0.6\textwidth]{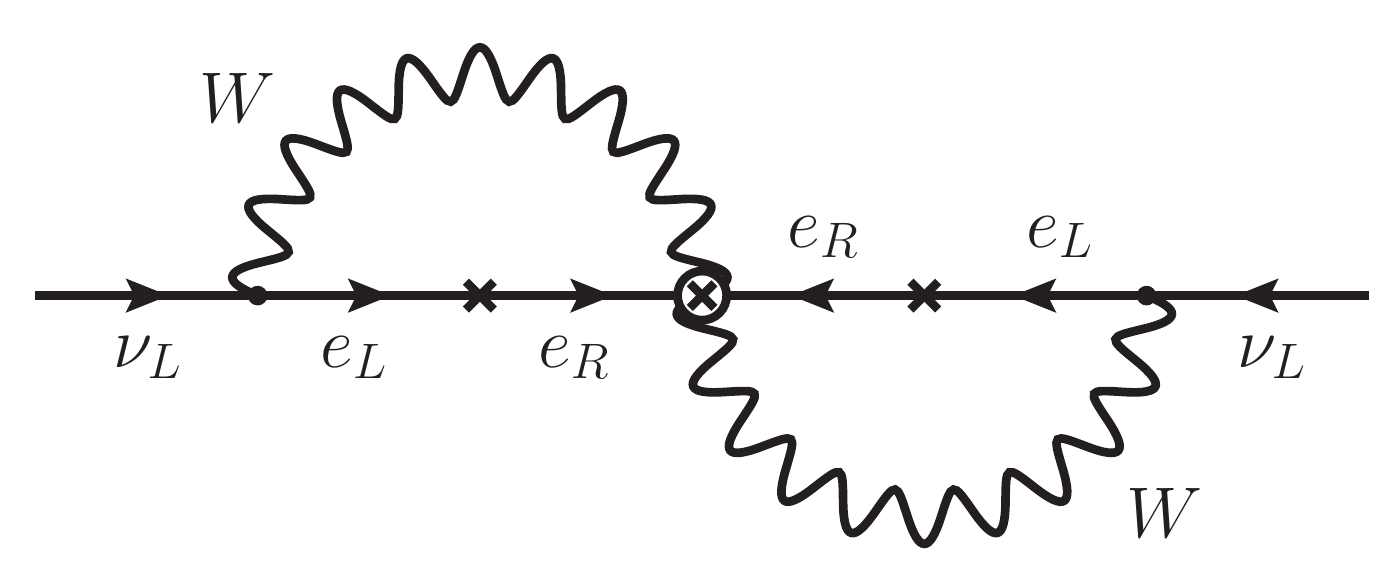}
	\caption{The diagram that generates neutrino masses at one loop with
			one $\mathcal{O} \up 9$ insertion.} 
			\label{fig:numass-eff-O9}
\end{figure}

The $\mathcal{O} \up 9$ operator provides neutrino masses at two loops, as
two charged leptons must be turned into neutrinos; since both leptons are 
right-handed, two powers of the charged lepton masses will additionally suppress 
the neutrino masses. The process is shown in figure \ref{fig:numass-eff-O9}. 
As in the previous section we distinguish between the calculable (logarithmic)
contributions,
\begin{displaymath}
	\delta m_{\alpha \beta} \sim \frac{v^4}{(16 \pi^2)^2 \, \Lambda^5} \,
			m_\alpha C_{\alpha \beta} \up 9 m_\beta \, 
			\ln \frac{\Lambda}{v} \, , 
\end{displaymath}
which are however subleading, and the dominant contributions from matching,
which we cannot but just estimate:
\begin{equation} \label{numass-matching-O9}
	m_{\alpha \beta} \sim \frac{1}{(16 \pi^2)^2 \, \Lambda} \,
			m_\alpha C_{\alpha \beta} \up 9 m_\beta \, . 
\end{equation} 

Again, by using \eqref{numass-matching-O9} we can estimate 
the new physics scale for the models that provide the $WW ee$ mechanism.
Recalling our reasonable requirement $| m_{\tau \tau} |  \sim 0.1 \; \mathrm{eV}$
we obtain
\begin{equation} \label{bound-numass-O9}
		\frac{\Lambda}{| C_{\tau \tau} \up 9 |} \sim 10^3 \; \mathrm{TeV} \, ,
\end{equation}
pointing once more to a heavier scale than \eqref{bound-0nu2beta-O9}, 
which marks the
current limit of observability for the $0 \nu \beta \beta$ effects of
$\mathcal{O} \up 9$. This implies that only models with $| C \up 9 | \sim 10^{-4}$
can yield a signal in near-future $0 \nu \beta \beta$ experiments; 
for models with $| C_{\tau \tau} \up 9 | \sim 1$, the
$WW ee$ mechanism will 
leave no trace of $0 \nu \beta \beta$, nor will the new particles show up at 
the LHC or other
planned accelerator experiments.
Fortunately, many underlying models can 
provide small $C \up 9$ coefficients, which would lower the scale in
\eqref{bound-numass-O9}. In chapter \ref{chap:WWee-model} we discuss a model
that, although tightly constrained, proves to be able to yield 
signals in a variety of different experiments while generating acceptable
neutrino masses.

\section{Preeminence of the $W \nu e$ and $WWee$ mechanisms in neutrinoless
			double beta decay}
		\label{sec:0nu2beta-eff-preeminence}

In the previous two sections we described how the $W \nu e$ and $WWee$
mechanisms can yield both $0 \nu \beta \beta$ and neutrino masses. Our final
aim will be to devise models where these mechanisms lead the generation of
these two processes. Neutrino masses, as commented in section 
\ref{sec:0nu2beta-eff-note-model-builder}, can generally be dealt with
by imposing suitable symmetries that forbid the interfering interactions,
but $0 \nu \beta \beta$ presents an additional 
complication: neutrino masses are themselves a source of $0 \nu \beta \beta$,
through the $\nu \nu$ mechanism. Consequently, it might happen that the
new mechanisms generate perfectly acceptable neutrino masses and then these
take control of $0 \nu \beta \beta$. It would be useful to have an indication
of under which circumstances our model will be affected by this inconvenience.
In this section we will expose a general argument
that allows to identify the regime where the new mechanisms overpower
their own generated masses in the competition for $0 \nu \beta \beta$.

First let us consider a theory that provides $0 \nu \beta \beta$ and neutrino
masses only through the $W \nu e$ mechanism. At the model level, $0 \nu \beta \beta$
will be produced by some process and neutrino masses will be generated by other,
related process with one additional loop. From the effective theory point of
view what we will have is one $\mathcal{O} \up 7$ operator that will yield
$0 \nu \beta \beta$ at tree level and one $\mathcal{O} \up 5$ operator that 
will yield neutrino masses at \mbox{tree level -- remember,} the loop diagrams
in figure \ref{fig:0nu2beta-eff-numass-7-and-9} are not calculable in the 
effective theory; the $\mathcal{O} \up 5$ operator accounts at any rate for the 
loop-generated neutrino masses, with the loop suppression encoded in the 
coefficient $C \up 5$. Both $\mathcal{O} \up 5$ and $\mathcal{O} \up 7$ 
contribute to $0 \nu \beta \beta$, and we can read their contributions
from equations \eqref{0nu2beta-eff-amplitude-5} and \eqref{0nu2beta-eff-amplitude-7}:
\begin{align} 
	\mathcal{A}_{0 \nu \beta \beta} \up 5 &\sim
			\frac{G_F^2}{p_{\mathrm{eff}}^2} \, \frac{v^2}{\Lambda} \,
			C_{ee} \up 5 
	\nonumber
	\\ \vphantom{\raisebox{3.5ex}{a}} \label{0nu2beta-eff-0nu2beta-5-and-7}
	\mathcal{A}_{0 \nu \beta \beta} \up 7 &\sim
		\frac{G_F^2}{p_{\mathrm{eff}}} \, \frac{v^3}{\Lambda^3} \,
			C_{ee} \up 7 \, ,
\end{align}
where the new physics scale is the same for both contributions, as they all
come from the $W \nu e$ mechanism. The total contribution to $0 \nu \beta \beta$, 
then, will be simply
$\mathcal{A}_{0 \nu \beta \beta} = \mathcal{A}_{0 \nu \beta \beta} \up 5 + 
\mathcal{A}_{0 \nu \beta \beta} \up 7$, and our question is under which
circumstances $\mathcal{A}_{0 \nu \beta \beta} \up 7$ is the leading piece.
However, the presence of the effective coefficients $C \up 5$ and $C \up 7$
makes it difficult to compare the two terms. We can eliminate them by
recalling that neutrino masses connect $\mathcal{O} \up 7$ and $\mathcal{O} \up 5$;
the neutrino masses in the effective theory emerge from $\mathcal{O} \up 5$,
and they are, by equation \eqref{Weinberg-(5)},
\begin{equation} \label{0nu2beta-eff-numass-5-preem}
	m_{\alpha \beta} = -2 \, \frac{v^2}{\Lambda} \, {C_{\alpha \beta} \up 5}^* \, ;
\end{equation}
but at the same time by hypothesis neutrino masses are provided by the
$\mathcal{O} \up 7$ interaction, and though we can't calculate them we 
can estimate them. By equation \eqref{numass-matching-O7}:
\begin{equation} \label{0nu2beta-eff-numass-7-preem}
	m_{\alpha \beta} \sim \frac{v}{16 \pi^2 \Lambda} \, 
			\left(m_\alpha C_{\alpha \beta} \up 7 + m_\beta 
			C_{\beta \alpha} \up 7 \right) \, .
\end{equation}
Equations \eqref{0nu2beta-eff-numass-5-preem} and \eqref{0nu2beta-eff-numass-7-preem}
can now be used to eliminate $C_{ee} \up 5$ and $C_{ee} \up 7$ in favour of 
$m_{ee}$. Neglecting phases, we have
\begin{equation} \label{0nu2beta-eff-preem-A-total}
	\mathcal{A}_{0 \nu \beta \beta} \sim \left( \frac{G_F}{p_{\mathrm{eff}}}
		\right)^2 \frac{m_{e e}}{2} \, \left[ 1 + \frac{p_{\mathrm{eff}}}{m_e} 
		\, (4 \pi)^2 \, \frac{v^2}{\Lambda^2} \right] \, ,
\end{equation}
and so the condition for the preeminence of $\mathcal{O} \up 7$ is just
\begin{equation} \label{0nu2beta-eff-preem-scale}
	\Lambda \lesssim 4 \pi \, v \, \sqrt{\frac{p_{\mathrm{eff}}}{m_e}} 
			\simeq 35 \; \mathrm{TeV} \, .
\end{equation}
In conclusion: if we want the $W \nu e$ mechanism to dominate the generation
of both neutrino masses and $0 \nu \beta \beta$ we need the new physics scale
to be relatively light. This is good news, for it means that the mechanism
can be probed, and eventually ruled out, by direct search of the new particles.
Remember, however, that $\Lambda$ needs not to be the mass of a particle, and
so equation \eqref{0nu2beta-eff-preem-scale} is only giving us a rough idea 
of the scale where the generation of $0 \nu \beta \beta$ reverts from one 
mechanism to the other.

As for the case of the $WW ee$ mechanism, the result is the same as that of the
$W \nu e$ mechanism, equation \eqref{0nu2beta-eff-preem-scale}. The reason is
that by working out equations \eqref{0nu2beta-eff-amplitude-9} and
\eqref{numass-matching-O9} we see that the contribution of $\mathcal{O} \up 9$ 
to equation \eqref{0nu2beta-eff-preem-A-total} is the same as for 
$\mathcal{O} \up 7$ but squared. As we compare with 1, the result is identical. 
We conclude, so, that we can have preeminence of the $W \nu e$ and
$WW ee$ mechanisms for both neutrino masses and $0 \nu \beta \beta$, and to 
achieve that we don't need to renounce the observability of the new particles.

\section{Modeling the lepton-boson mechanisms} \label{sec:0nu2beta-eff-newphysics}

In this section we will survey the possible combinations of new particles
that can generate the operators $\mathcal{O} \up 5$, $\mathcal{O} \up 7$
and $\mathcal{O} \up 9$. We will do so just by deducing their spin and
gauge charges, with no reference to possible new symmetries or interactions,
which may be required if one wants to enhance a particular mechanism.
So, the listings offered here can be regarded as a first step for the construction
of a model. Bear in mind, however, that we will restrict this discussion to 
sets of particles that generate the relevant operators \emph{at tree level},
in order to obtain large effects in neutrinoless double beta decay.
We will assume that the underlying theory is weakly coupled
and contains only renormalisable vertices;
we also assume the NP respects all the gauge symmetries of the SM. 
In listing the heavy particles we denote by
$\vecb TY$, $\fer TY$ and $\sc TY$
a heavy vector, fermion or scalar with isospin
$T$ and hypercharge $Y$, respectively. 
When the heavy particles can be either a heavy
vector or heavy scalar with the same
isospin and hypercharge, we use $\bos TY$
to denote both possibilities.

\subsubsection{New physics to generate $\mathcal{O} \up 5$}

\begin{figure}[tb]
	\centering
	\includegraphics[width=0.48\textwidth]{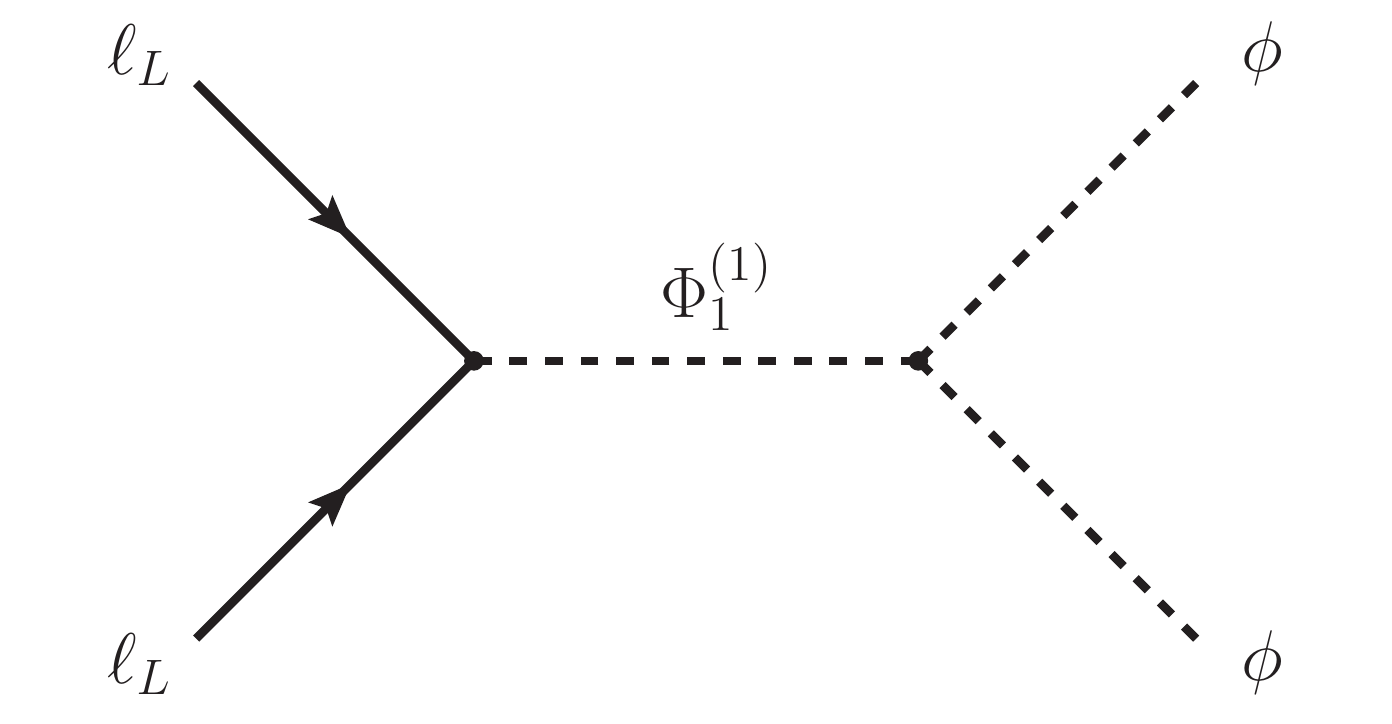}
	\includegraphics[width=0.48\textwidth]{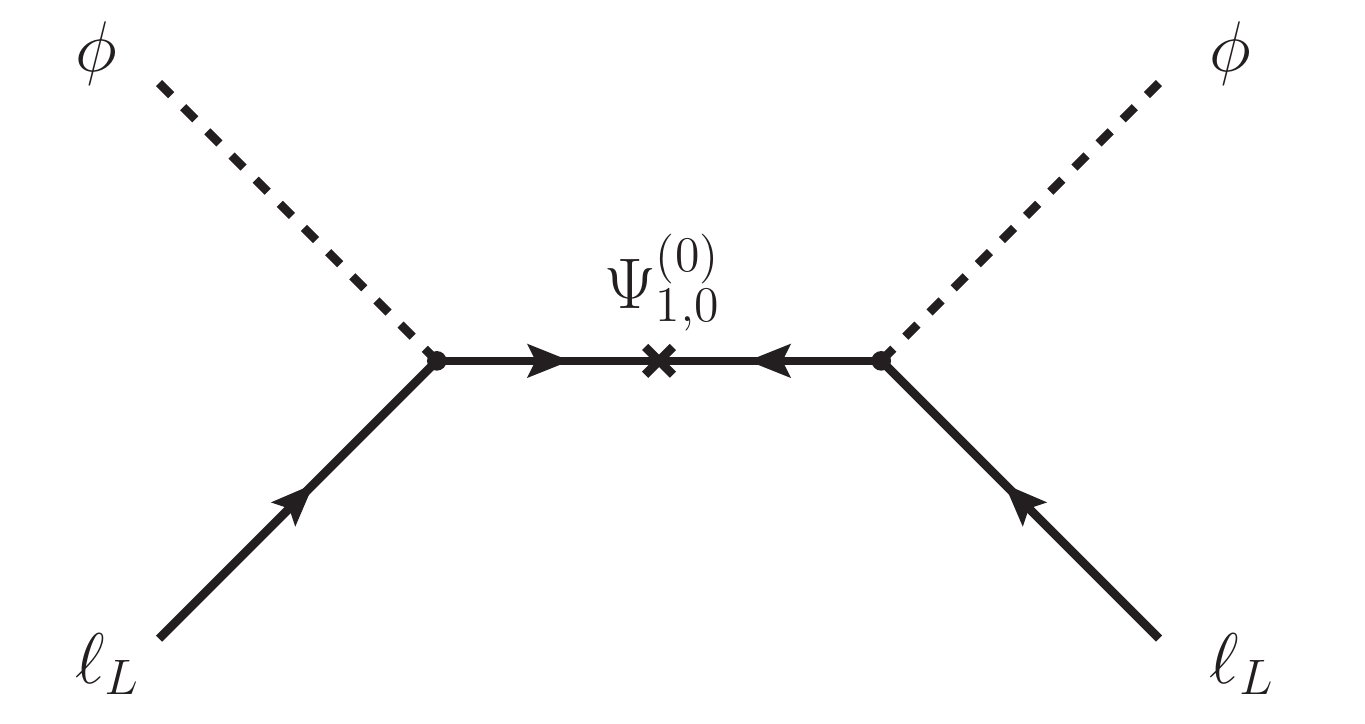}
	\caption{Topologies relevant for the generation of $\mathcal{O} \up 5$.
			} \label{fig:topologies-O5}
\end{figure}

The Weinberg operator, $\mathcal{O} \up 5$, comprises two 
leptonic doublets and two scalar doublets; we could represent it by
$\ell_{\mathrm{L}}^2 \phi^2$, omitting any reference to the closing of spinorial 
and $SU(2)$ indices. That essentially means that the Feynman diagram generating
$\mathcal{O} \up 5$ must have four external legs, two with $\ell_{\mathrm{L}}$'s 
and two with $\phi$'s; there're not so many ways to achieve this. At tree
level, and with no nonrenormalisable interactions, there are only two:
with two leptons at one side and two scalars at the other, united by a heavy 
mediator, or with a pair lepton-scalar at each side, united by a heavy 
mediator. We depict these two possible topologies in figure 
\ref{fig:topologies-O5}; with this in hand, the rest is computing spin
and charge sums: in the first case the mediator must be a hypercharge-1,
$SU(2)$-triplet scalar, and in the second a fermion, either $SU(2)$ singlet or
triplet, with hypercharge 0. We collect these results in table 
\ref{tab:O5-topologies} in a compact and systematic fashion; this will
be useful when considering higher-order operators with more complicated 
topologies. Note also that the diagrams in figure \ref{fig:topologies-O5} also
generate the subleading $\mathcal{O} \up {7 \, \textsc{i} - \textsc{iii}}$
just by inserting the appropriate number of $W$'s into the legs that are not
$SU(2)$ singlets.

\begin{table}[tb]
$$
\begin{array}{|c|c|c|c|c|}
\hline
\multicolumn{5}{|c|}{\includegraphics[width=1.5in]{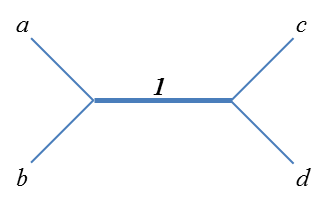}} \cr\hline
1                  & a    & b    & c    & d    \cr \hline\hline
\sc11 \vphantom{\raisebox{-1ex}{a}} \vphantom{\raisebox{1.7ex}{a}} 
				   & \ell_{\mathrm{L}} & \ell_{\mathrm{L}} & \phi & \phi \cr\hline
\fer00 \aut \fer10 \vphantom{\raisebox{-1ex}{a}} \vphantom{\raisebox{1.7ex}{a}}
				   & \ell_{\mathrm{L}} & \phi & \ell_{\mathrm{L}} & \phi \cr\hline
\end{array}
$$
\caption{Symbolic representation of the diagrams that can generate
		$\mathcal{O} \up 5$ at tree level. As we see, roman letters are used
		to label external (i.e., SM) legs and numbers are used to label
		internal (i.e., new physics) legs.
		} \label{tab:O5-topologies}
\end{table}

To summarise, the new physics additions that can produce $\mathcal{O} \up 5$
are
\begin{equation} \label{NP-for-O5}
 \fer{0}0 \qquad \qquad \sc{1}1 \qquad \qquad \fer10 \; ,
\end{equation}
that is to say, exactly the new particle content of so-called seesaws type I,
II and III, extensively studied in neutrino mass (i.e. Weinberg operator)
generation.

\subsubsection{New physics to generate $\mathcal{O} \up 7$}

\begin{table}[tb]
$$
\begin{array}{|c|c|c|c|c|c|c|}
\hline
\multicolumn{7}{|c|}{\includegraphics[width=1.5in]{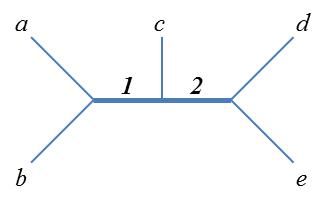}} 
\cr\hline
1              & 2     & a & b & c    & d    & e  \cr\hline\hline
\vecb{1/2}{3/2} \vphantom{\raisebox{-1ex}{a}} \vphantom{\raisebox{1.7ex}{a}}
		& \vecb01 \aut \vecb11 & \ell_{\mathrm{L}} & e_{\mathrm{R}}    & \phi 
		& \phi & \phi \cr\hline
\vecb{1/2}{3/2} \vphantom{\raisebox{-1ex}{a}} \vphantom{\raisebox{1.7ex}{a}}
		& \sc11   & \ell_{\mathrm{L}} & e_{\mathrm{R}}    & \phi & \phi 
		& \phi \cr\hline
\fer10 \vphantom{\raisebox{-1ex}{a}} \vphantom{\raisebox{1.7ex}{a}}        
		& \vecb11  & \ell_{\mathrm{L}} & \phi & e_{\mathrm{R}}    & \phi 
		& \phi \cr\hline
\fer10 \vphantom{\raisebox{-1ex}{a}} \vphantom{\raisebox{1.7ex}{a}}    
		& \sc11   & \ell_{\mathrm{L}} & \phi & e_{\mathrm{R}}    & \phi 
		& \phi \cr\hline
\fer00 \vphantom{\raisebox{-1ex}{a}} \vphantom{\raisebox{1.7ex}{a}}        
		& \vecb01   & \ell_{\mathrm{L}} & \phi & e_{\mathrm{R}}    & \phi 
		& \phi \cr\hline
\fer{1/2}{1/2} \vphantom{\raisebox{-1ex}{a}} \vphantom{\raisebox{1.7ex}{a}}
		& \vecb01 \aut \vecb11 & e_{\mathrm{R}}    & \phi & \ell_{\mathrm{L}} 
		& \phi & \phi \cr\hline 
\fer{1/2}{1/2} \vphantom{\raisebox{-1ex}{a}} \vphantom{\raisebox{1.7ex}{a}}
		& \sc11   & e_{\mathrm{R}}    & \phi & \ell_{\mathrm{L}} & \phi 
		& \phi \cr\hline 
\fer{1/2}{1/2} \vphantom{\raisebox{-1ex}{a}} \vphantom{\raisebox{1.7ex}{a}}
		& \fer00 \aut \fer10 & e_{\mathrm{R}}    & \phi & \phi & \phi 
		& \ell_{\mathrm{L}} \cr\hline
\end{array}
$$
\caption{Representantion of the diagrams generating $\mathcal{O} \up 7$
		at tree level, together with the new physics additions that can 
		realise them.
		} \label{tab:O7-topologies}
\end{table}

The presence of a covariant derivative makes the generation of
$\mathcal{O} \up 7$ somewhat more involved, as it consists of several pieces
with different gauge boson species. In order to simplify the
process it is enough to consider the term of $\mathcal{O} \up 7$ with no
gauge bosons:
\begin{equation} \label{eq:O7-no-gauge-bosons}
	\mathcal{O}\up{7} = \bar e_{\mathrm{R}}  \gamma^\mu \left( \phi^\dagger
				\tilde \ell_{\mathrm{L}} \right) \left(\phi^\dagger 
				\partial_\mu \tilde \phi \right) + \ldots \, ,
\end{equation}
as any model which generates this interaction will necessarily, by gauge
invariance, generate also the complete operator. This can be diagrammatically
understood by taking a graph that generates \eqref{eq:O7-no-gauge-bosons}
and attaching the appropriate number of light gauge bosons
to the internal heavy propagators, wherever allowed by the quantum numbers.
The relevant interaction, so, consists of one
$\ell_{\mathrm{L}}$, one $e_{\mathrm{R}}$ and three $\phi$'s\footnote{Actually,
as we see in \eqref{eq:O7-no-gauge-bosons}, $\mathcal{O} \up 7$ comprises the 
\emph{charge conjugates} of all these fields, but we prefer not to increase
the cumbersomeness of our notation. After all, the physics that generates 
$\mathcal{O} \up 7$ is the same that generates 
${\mathcal{O} \up 7}^\dagger$ but charge-conjugated, so our conclusions remain
intact.}; this means five external legs in our generating diagram, which
can be realised only with one topology: that depicted in table 
\ref{tab:O7-topologies}.
A second topology with all three $\phi$'s in one vertex and the two leptons
at the other side is forbidden at tree level, for a scalar alone cannot couple 
to the
combination $\overline{\tilde \ell_{\mathrm{L}}} \gamma^\mu e_{\mathrm{R}}$
and a vector cannot couple to $\phi^3$ if the vertex is to be of dimension
4 or lower.

In summary, the new physics we may be interested in is necessarily grouped
in pairs, and the possible couples are:
\begin{align}
	\bigl\{ \vecb{1/2}{3/2} &, \vecb{0,1}1 \bigr\}  &
			&  &
			\bigl\{ \vecb{1/2}{3/2} &, \sc{1}1 \bigr\} 
	\nonumber
	\\ \label{0nu2beta-eff-models-O7-content}
	\bigl\{ \fer{1/2}{1/2} &, \vecb{0,1}1 \bigr\} &
			\bigl\{ \fer{1/2}{1/2} &, \sc{1}1 \bigr\} &
		 	\bigl\{ \fer{1/2}{1/2} &, \fer{0,1}0 \bigr\}
	\\
	\bigl\{ \fer00 &, \vecb01 \bigr\}    &  
			&  &		
			\bigl\{ \fer10 &, \bos11 \bigr\} \,.
	\nonumber
\end{align}
It is worth noting that in the last two possibilities the
heavy fermions $\fer{0,1}0$ necessarily have couplings
that would also generate $\mathcal{O} \up 5$ at tree level.
In contrast, models containing $\sc{0,1}1$ may or may not generate them, 
depending on whether the heavy scalars couple to 
$\overline{\ell_{\mathrm{L}}} \tilde \ell_{\mathrm{L}}$,
which may be forbidden by the symmetries of the underlying theory.
From this brute-force list, so, one can select classes of models
more or less adequate to enhance the $W \nu e$ mechanism.
In chapter \ref{chap:Wnue-model} we consider a model that realises one
of these possibilities; as we discuss there, for the model to be
realistic several other additions are needed, including a second
light scalar doublet.

\subsubsection{New physics to generate $\mathcal{O} \up 9$}

\begin{table}[tb]
$$
\begin{array}{|c|c|c||c|c|c|c|c|c|}
\hline
\multicolumn{9}{|c|}{\includegraphics[width=1.7in]{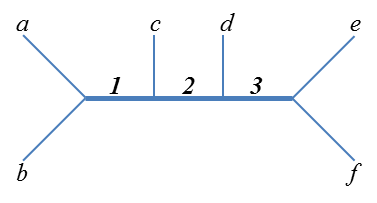}} \cr\hline
1  & 2  & 3 & a & b & c & d & e  & f    \cr\hline\hline
\sc02   \vphantom{\raisebox{-1ex}{a}} \vphantom{\raisebox{1.7ex}{a}}   
		& \bos{1/2}{3/2}   & \sc11   & e_{\mathrm{R}} & e_{\mathrm{R}} & \phi 
		& \phi & \phi & \phi \cr\hline
\sc02  \vphantom{\raisebox{-1ex}{a}} \vphantom{\raisebox{1.7ex}{a}}        
		& \bos{1/2}{3/2}  & \vecb11 \aut \vecb01 & e_{\mathrm{R}} & e_{\mathrm{R}} 
		& \phi & \phi & \phi & \phi \cr\hline
\fer{1/2}{1/2} \vphantom{\raisebox{-1ex}{a}} \vphantom{\raisebox{1.7ex}{a}}
		& \bos{1/2}{3/2}   & \sc11  & e_{\mathrm{R}} & \phi & e_{\mathrm{R}} 
		& \phi & \phi & \phi \cr\hline
\fer{1/2}{1/2} \vphantom{\raisebox{-1ex}{a}} \vphantom{\raisebox{1.7ex}{a}}
		& \bos{1/2}{3/2}  & \vecb11 \aut \vecb01 & e_{\mathrm{R}} & \phi 
		& e_{\mathrm{R}} & \phi & \phi & \phi \cr\hline
\fer{1/2}{1/2} \vphantom{\raisebox{-1ex}{a}} \vphantom{\raisebox{1.7ex}{a}}
		& \fer10  & \sc11  & e_{\mathrm{R}} & \phi & \phi & e_{\mathrm{R}} 
		& \phi & \phi \cr\hline
\fer{1/2}{1/2} \vphantom{\raisebox{-1ex}{a}} \vphantom{\raisebox{1.7ex}{a}}
		& \fer10   & \vecb11   & e_{\mathrm{R}} & \phi & \phi & e_{\mathrm{R}} 
		& \phi & \phi \cr\hline
\fer{1/2}{1/2} \vphantom{\raisebox{-1ex}{a}} \vphantom{\raisebox{1.7ex}{a}}
		& \fer00  & \vecb01  & e_{\mathrm{R}} & \phi & \phi & e_{\mathrm{R}} 
		& \phi & \phi \cr\hline
\fer{1/2}{1/2} \vphantom{\raisebox{-1ex}{a}} \vphantom{\raisebox{1.7ex}{a}}
		& \fer00 \aut \fer10 & \fer{1/2}{1/2}  & e_{\mathrm{R}} & \phi & \phi 
		& e_{\mathrm{R}} & \phi & \phi \cr\hline
\bos11  \vphantom{\raisebox{-1ex}{a}} \vphantom{\raisebox{1.7ex}{a}}       
		& \fer10  & \sc11  & \phi & \phi & e_{\mathrm{R}} & e_{\mathrm{R}} 
		& \phi & \phi \cr\hline
\vecb11  \vphantom{\raisebox{-1ex}{a}} \vphantom{\raisebox{1.7ex}{a}}       
		& \fer10  & \vecb11  & \phi & \phi & e_{\mathrm{R}} & e_{\mathrm{R}} 
		& \phi & \phi \cr\hline
\vecb01  \vphantom{\raisebox{-1ex}{a}} \vphantom{\raisebox{1.7ex}{a}}       
		& \fer00  & \vecb01  & \phi & \phi & e_{\mathrm{R}} & e_{\mathrm{R}} 
		& \phi & \phi \cr\hline
\end{array}
$$
\caption{The first topology that can generate $\mathcal{O} \up 9$ at tree 
		level, with a detailed list of the new heavy particles that must
		run in the internal propagators.
		} \label{tab:O9-topologies-1}
\end{table}
\begin{table}[tb]
$$
\begin{array}{|c|c|c||c|c|c|c|c|c|}
\hline
\multicolumn{9}{|c|}{\includegraphics[width=1.5in]{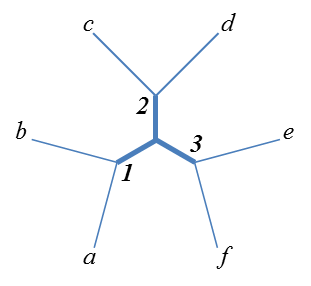}} \cr\hline
1 & 2  & 3  & a & b & c    & d    & e   & f    \cr\hline\hline
\sc02  \vphantom{\raisebox{-1ex}{a}} \vphantom{\raisebox{1.7ex}{a}}        
		& \sc11    & \sc11   & e_{\mathrm{R}} & e_{\mathrm{R}} & \phi & \phi 
		& \phi & \phi \cr\hline
\sc02  \vphantom{\raisebox{-1ex}{a}} \vphantom{\raisebox{1.7ex}{a}}        
		& \sc11   & \vecb11  & e_{\mathrm{R}} & e_{\mathrm{R}} & \phi & \phi 
		& \phi & \phi \cr\hline
\sc02  \vphantom{\raisebox{-1ex}{a}} \vphantom{\raisebox{1.7ex}{a}}        
		& \vecb11   & \vecb11 & e_{\mathrm{R}} & e_{\mathrm{R}} & \phi & \phi 
		& \phi & \phi \cr\hline
\sc02  \vphantom{\raisebox{-1ex}{a}} \vphantom{\raisebox{1.7ex}{a}}        
		& \vecb01   & \vecb01  & e_{\mathrm{R}} & e_{\mathrm{R}} & \phi & \phi 
		& \phi & \phi \cr\hline
\fer{1/2}{1/2} \vphantom{\raisebox{-1ex}{a}} \vphantom{\raisebox{1.7ex}{a}}
		& \sc11  & \fer{1/2}{1/2} & e_{\mathrm{R}} & \phi & \phi & \phi
		& \phi & e_{\mathrm{R}} \cr\hline
\fer{1/2}{1/2} \vphantom{\raisebox{-1ex}{a}} \vphantom{\raisebox{1.7ex}{a}}
		& \vecb11 \aut \vecb01  & \fer{1/2}{1/2}  & e_{\mathrm{R}} & \phi 
		& \phi & \phi & \phi & e_{\mathrm{R}} \cr\hline
\end{array}
$$
\caption{The second diagram topology to generate $\mathcal{O} \up 9$ at
		tree level.
		} \label{tab:O9-topologies-2}
\end{table}
\begin{table}[tb]
$$
\begin{array}{|c|c||c|c|c|c|c|c|}
\hline
\multicolumn{8}{|c|}{\includegraphics[width=1.5in]{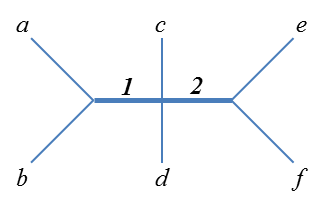}} \cr\hline
1   & 2    & a & b & c    & d    & e    & f    \cr\hline\hline
\sc02  \vphantom{\raisebox{-1ex}{a}} \vphantom{\raisebox{1.7ex}{a}}        
		& \sc11 & e_{\mathrm{R}} & e_{\mathrm{R}} & \phi & \phi & \phi 
		& \phi \cr\hline
\end{array}
$$
\caption{An analysis of the third topology that provides $\mathcal{O} \up 9$
		at tree level.
		} \label{tab:O9-topologies-3}
\end{table}

As in the previous case, we will discuss here the new physics additions
that can generate the piece of $\mathcal{O} \up 9$ with no gauge bosons;
by gauge invariance, those same models will also provide the complete
operator. In this case the relevant piece is
\begin{displaymath}
	\mathcal{O} \up 9 = \overline{e_{\mathrm{R}}} e_{\mathrm{R}}^\mathrm{c} \,
			\left( \phi^\dagger \partial_\mu \tilde \phi \right) 
			\left( \phi^\dagger \partial^\mu \tilde \phi \right) + 
			\ldots \, ,
\end{displaymath}
which comprises two $e_{\mathrm{R}}$'s and four $\phi$'s\footnote{As in 
the previous discussion, for simplicity we omit here the
complex-conjugates that should be in order for the component fields.}.
There are three different topologies that can provide
the corresponding six-legged diagram; we depict them in tables 
\ref{tab:O9-topologies-1}, \ref{tab:O9-topologies-2} and \ref{tab:O9-topologies-3},
together with the various combinations of new particles required for
their consistent generation.

Summarising all these possible additions we conclude we can generate
$\mathcal{O} \up 9$ whenever our model provides at least one of the
following combinations:
\begin{align}
	\bigl\{ \bos11 &, \sc02 \bigr\}  & \bigl\{ \bos11 &, \fer10 \bigr\}
			& \bigl\{ \bos11 &, \fer{1/2}{1/2} \bigr\}
			\nonumber
	\\
	\bigl\{ \vecb01 &, \sc02 \bigr\}  & \bigl\{ \vecb01 &, \fer00 \bigr\}
			& \bigl\{ \vecb01 &, \fer{1/2}{1/2} \bigr\}
			\nonumber
	\\
	& & \bigl\{ \fer{1/2}{1/2} &, \fer{0,1}0 \bigr\} & &  \label{NP-for-O9}
	\\
	\bigl\{ \bos{1/2}{3/2} , & \sc02 , \bos11 \bigr\}
			&  &
			& \bigl\{ \bos{1/2}{3/2} , & \sc02 , \vecb01 \bigr\}
			\nonumber
	\\
	\bigl\{ \bos{1/2}{3/2} , & \fer{1/2}{1/2} , \bos11 \bigr\}
			&  &
			& \bigl\{ \bos{1/2}{3/2} , & \fer{1/2}{1/2} , \vecb01 \bigr\}
			\nonumber
\end{align}
Despite the presence of $\sc11$ and $\fer{0,1}0$ in some of these
options, these heavy particles need not have the same vertices as the ones
leading to $\mathcal{O} \up 5$. If they do, $\mathcal{O} \up 9$ would
have only subdominant effects; but this is in general not the case.
In chapter \ref{chap:WWee-model} we discuss a model that includes essentially
a doubly-charged isosinglet $\sc02$ and a scalar triplet of type $\sc11$;
imposing additional symmetries the couplings of this model can be selected
in a way that suppresses $\mathcal{O} \up 5$.

\chapter{A model realising the $\protect \Wnue$ mechanism} \label{chap:Wnue-model}

In this chapter we will present a short account of a model that realises the
$W \nu e$ mechanism. The model will be presented and its main 
features will be discussed only briefly; we have not yet developed a full
analysis of this model, but it should come about in the near future.
As we will see, the model presents a deviation from the plan described in
section \ref{sec:0nu2beta-eff-preliminary-remarks}: it contains a second
light scalar doublet with the same quantum numbers as the Higgs doublet.
This variation is convenient in order to have neutrino masses dominantly 
produced by the $W \nu e$ mechanism while retaining tree-level $0 \nu \beta \beta$.
Thus, we could regard this model as an extension
of the popular Two Higgs Doublet Model \cite{Lee:1973iz,Branco:2011iw}. 
As we will see, LNV in this model can be originated either in spontaneous
breaking of lepton number or in explicit breaking in the scalar potential;
these two `paths to LNV' lead in fact to different effective operators,
either involving the usual Higgs doublet or the second scalar doublet.
In our discussion we try to determine which is the dominant mechanism
and how this dominance depends on the parameters of the model.
From our analysis we deduce bounds on the VEV of the second doublet,
which lies around the electroweak scale, and on the VEV of the triplet,
which is forced to be very small unless compensated by very heavy
particles or very small Yukawas.
Lepton flavour violation is not directly addressed in our discussion,
but the model can accommodate flavour symmetries that effectively
suppress all LFV-ing effects.
All in all, the model succeeds in realising the $W \nu e$ mechanism,
and the parameters can be arranged in such a way that both 
$0 \nu \beta \beta$ and neutrino masses are dominated by this mechanism,
with some of the new particles at the reach of the LHC.
More investigation is still needed, but the model seems promising and
its most interesting version will be probed in the next round of
high- and low-energy experiments. The work presented in this chapter
was carried out together 
with Arcadi Santamaria, José Wudka, Francisco del Águila and Subhaditya
Bhattacharya.

\section{A model with spontaneous and explicit breaking of lepton number} 

We begin the construction of our model by considering the couple 
\mbox{$\bigl\{ \fer{1/2}{1/2}, \sc11 \bigr\}$\vphantom{\raisebox{1.5ex}{a}}\vphantom{\raisebox{-1.5ex}{a}}} 
as read in equation \eqref{0nu2beta-eff-models-O7-content}. Table 
\ref{tab:O7-topologies}
informs us that a Yukawa interaction connecting the heavy and light fermion
doublets will be needed, 
\mbox{$\overline{\fer{1/2}{1/2}} \, \sc11 \, \ell_{\mathrm{L}}$\vphantom{\raisebox{1.5ex}{a}}},
as well as a cubic coupling between the scalars, 
$\phi^\dagger \, \sc11 \, \tilde \phi$, that will induce a VEV for $\sc11$;
the hypercharge assignments ensure that such terms are present. Unfortunately,
plain hypercharge enforcement also provides a type-II-seesaw-like coupling,
\mbox{$\overline{\tilde \ell_{\mathrm{L}}} \, \sc11 \, \ell_{\mathrm{L}}$\vphantom{\raisebox{1.7ex}{a}}}, 
which yields
masses for the neutrinos at tree level independently of the $W \nu e$ mechanism.
That is inconvenient. To forbid this coupling we call for a $Z_2$ discrete symmetry 
under which the heavy fields are odd and the usual SM fields are even. This
of course eliminates the unwanted Yukawas, but also forbids the necessary 
cubic scalar coupling. In order to overcome this difficulty we introduce a 
new scalar doublet, $\phi^{\, \prime}$, with the same $SU(2) \otimes U(1)$ 
assignments as the Higgs doublet but odd under $Z_2$; then the vertex 
${\phi^\dagger \, \sc11 \, \tilde \phi^{\, \prime}}$ is allowed. But by looking
at the diagram in table \ref{tab:O7-topologies} we see that the $\phi^{\, \prime}$
occupies an external leg, therefore it cannot be a heavy field. Our setup,
thus, requires a second \emph{light} scalar doublet that might show up
at the LHC in the near future.

\begin{table}[bt]
	\centering
	\[
	\begin{array}{| c | c | c | c | c |}
		\hline
		\multicolumn{1}{|l}{ } &  & T & Y & Z_2 \\ \hline
		\sc11 \vphantom{\raisebox{-1ex}{a}} \vphantom{\raisebox{1.7ex}{a}}
				& \chi & 1 & 1 & - 
		\\ \hline
		{\fer{1/2}{1/2}}^\mathrm{c} \vphantom{\raisebox{-1ex}{a}} 
				\vphantom{\raisebox{1.7ex}{a}}
				& L_{\mathrm{L}} &  \nicefrac{1}{2} & - \nicefrac{1}{2} & - 
		\\ \hline
			&  L_{\mathrm{R}} & \nicefrac{1}{2} & - \nicefrac{1}{2} & - 
			\vphantom{\raisebox{-1.2ex}{a}} \vphantom{\raisebox{1.5ex}{a}}
		\\ \hline
			&  \phi^{\, \prime} & \nicefrac{1}{2} & \nicefrac{1}{2} & - 
			\vphantom{\raisebox{-1.2ex}{a}} \vphantom{\raisebox{1.5ex}{a}}
		\\	\hline
	\end{array}
	\]
	\caption{The new fields of the model and their charge assignments. The
			leftmost column presents the fields in the notation of section
			\ref{sec:0nu2beta-eff-newphysics}; a new, more wieldy naming
			is introduced in the next column. The $\phi^{\, \prime}$ doublet is 
			not a heavy field, but a second doublet with a mass of the
			order of the electroweak scale.
			} \label{tab:Wnue-model-particles}
\end{table}

We summarise the new physics additions of our model in table 
\ref{tab:Wnue-model-particles},
where we also introduce a simpler notation to ease the discussion;
essentially we will write $\sc11 \equiv \chi$ and $\fer{1/2}{1/2} \equiv
L_{\mathrm{L}}^\mathrm{c}$, so that $L_{\mathrm{L}}$ has the same hypercharge
as the SM doublets $\ell_{\mathrm{L}}$. In order to guarantee the decoupling of the
heavy fields we take the $L$ fermions to be vector-like, and therefore
we introduce a right-handed counterpart with the same quantum numbers.
We will also allow for several families of the heavy fermions; this is not 
strictly necessary \emph{a priori}, but can be greatly helpful when dealing with 
LFV constraints. With all these considerations accounted for, we are ready to 
write the Lagrangian of the model; the pieces involving the heavy fermions
read
\begin{equation} \label{Wnue-model-yukawas}
	\mathcal{L}_L = \overline{L_\alpha} (i \, \slashed{D} - m_{L_\alpha}) L_\alpha 
		+ \left[ y^e_{\alpha \beta} \, \overline{L_{\mathrm{L} \alpha}} \, 
		e_{\mathrm{R} \beta} \, \phi^{\, \prime} + y^\nu_{\alpha \beta} \, 
		\overline{\tilde{L}_{\mathrm{L} \alpha}} \, \chi \, \ell_{\mathrm{L} \beta} 
		+ \mathrm{H.c.} \right] \, ,
\end{equation}
where it is understood that 
$L_\alpha = L_{\mathrm{L} \alpha} + L_{\mathrm{R} \alpha}$ and the usual 
\mbox{notation --see} sections \ref{sec:intro-SM-doublets-singlets} 
\mbox{and \ref{sec:masses-SM}-- is used} for the SM
fields. The mass matrix $m_L$ can be assumed diagonal without loss of generality;
the Yukawas $y^e$ and $y^\nu$, however, are general matrices whose number 
of physical parameters depends on the number of heavy families. Let's say 
there are three of them, $\alpha = 1,2,3$; then $y^e$ and $y^\nu$ are 
$3 \times 3$ general matrices from which six phases can be \mbox{removed -- overall} 
from the two of them; if it's convenient we can choose to take six phases
from one Yukawa matrix and none from the other.

As for the pure-scalar part of the Lagrangian, we can write it as a scalar 
potential:
\begin{align} 
	V  =  &- m_{\phi}^{2} \, \phi^\dagger \phi - {m_{\phi}^\prime}^{2} \, 
		{\phi^{\, \prime}}^\dagger \phi^{\, \prime}
		+ m_{\chi}^{2} \, \Tr{\chi^{\dagger} \chi}  + \lambda_{\phi} \, 
		\left( \phi^\dagger \phi \right)^2 \: +
		\nonumber 
		\\
	&+ \lambda_{\phi}^\prime \, \left( {\phi^{\, \prime}}^\dagger 
		\phi^{\, \prime} \right)^2 + 
		\lambda_{\chi} \, \left( \Tr{\chi^{\dagger} \chi} \right)^{2} + 
		\bar \lambda_{\chi} \, \Tr{\left(\chi^{\dagger}\chi\right)^{2}} \: +
	 	\vphantom{\raisebox{2.5ex}{a}} 
		\nonumber 
		\\
  &+ \lambda_{\phi \phi} \, \left( \phi^\dagger \phi^{\, \prime} \right)^2 +
		\lambda_{\phi\chi} \, \phi^\dagger \phi \, \Tr{\chi^{\dagger} \chi} +
 		\bar \lambda_{\phi\chi} \, \phi^{\dagger} \chi^{\dagger} \chi \, \phi \: +
		\vphantom{\raisebox{2.5ex}{a}} 
		\label{Wnue-model-scalar-potential}
		\\
  &+ \lambda_{\phi \chi}^\prime \, {\phi^{\, \prime}}^\dagger \phi^{\, \prime} \, 
  		\Tr{\chi^{\dagger} \chi} + \bar \lambda_{\phi\chi}^\prime \, 
		{\phi^{\, \prime}}^{\dagger} \chi^{\dagger} \chi \, \phi^{\, \prime} \: +
		\vphantom{\raisebox{2.5ex}{a}} 
		\nonumber 
		\\
  &+ \left[ \mu \, \phi^\dagger \chi \, \tilde \phi^{\, \prime} + \mathrm{H.c.} 
  		\right] \, .
		\vphantom{\raisebox{2.5ex}{a}} 
		\nonumber
\end{align}
Note that in this expression we already disposed the signs so that $m^2_\phi,
{m_\phi^\prime}^2 < 0$ and $m_\chi^2 > 0$. The reason for this election is
straightforward: the triplet VEV is phenomenologically forced to be small 
(see section \ref{sec:WWee-model-vevs} in the next chapter), 
but the physical scalars associated
to the triplet must be heavier than 100 GeV; if the triplet VEV is to be induced 
by its mass term, then it should be small too. A simple way to reconcile these
two conditions is to allow for a large, positive $m^2_\chi$ and let the triplet
VEV to be induced by the VEV's of the doublets plus the trilinear interaction
$\phi^\dagger \chi \, \tilde \phi^{\, \prime}$. Then, if we write $\left< \phi 
\right> \equiv v_\phi$ and $\left< \phi^{\, \prime} \right> \equiv v_\phi^{\, 
\prime}$ we will have
\begin{displaymath}
	\left< \chi \right> \equiv v_\chi \simeq - \mu^* \, \frac{v_\phi \,
		v_\phi^{\, \prime}}{ m_\chi^2 } \, ,
\end{displaymath}
which can be small either if $\mu$ is small or
if $m_\chi$ is large. 
As for other features of the scalar potential, we note that
a sufficient condition for it to be bounded from below, and so allowing for
a meaningful vacuum, is that all the quartic couplings are real and positive;
no more elaboration on this matter will be needed for our purposes.

Consider now the situation of lepton number in this model; from equations
\eqref{Wnue-model-yukawas} and \eqref{Wnue-model-scalar-potential} we see
that it's explicitly broken. LN is transmitted by the Yukawas to the 
$\phi^{\, \prime}$ and $\chi$ scalars, and this assignment is honored by the
trilinear term $\mu$; but then we have the quartic term $\lambda_{\phi \phi}$,
and the usual Higgs doublet, $\phi$, carries no lepton number because it couples
to quarks. Consequently, lepton number is not a good symmetry for this model,
due to the joint action of the $y^\nu$, $y^e$, $\mu$ and $\lambda_{\phi \phi}$ 
interactions. One might therefore think that any LNV-ing amplitude should contain all
these couplings, but in the model there is an additional source of lepton
number violation: the vacuum; 
even if some of the couplings needed to violate lepton number were set to zero,
lepton number would be broken spontaneously 
when $\chi$ and $\phi^{\, \prime}$ take a VEV.
And indeed, the interactions of the model are arranged
in such a way that a curious competition between explicit and spontaneous
breaking arises: when we examine a LNV-ing process we find contributions which 
require spontaneous violation through the VEV of $\phi^{\, \prime}$, and
others that involve the VEV of $\phi$ and explicit violation. It so happens
that in this model the latter contributions appear suppressed by one loop
with respect to the former ones; one might think that they are subdominant, but the 
first involve $\left< \phi^{\, \prime} \right>$, which may be small.
An interesting contest, thus, is taking place between the two sources of LNV.
In sections \ref{sec:Wnue-model-O7} and \ref{sec:Wnue-model-numass-0nu2beta}  
we present examples of this feature for the processes we are interested in.

%

\section{Generation of $\mathcal{O} \up 7$} \label{sec:Wnue-model-O7}

The model that we are discussing has two light scalar doublets, and this fact 
affects the analysis carried out in section \ref{sec:0nu2beta-eff-newphysics}:
new effective operators have to be considered that involve the second doublet.
As $\phi$ and $\phi^{\, \prime}$ have the same $SU(2) \otimes U(1)$ quantum 
numbers, these operators are mainly versions of those already discussed
but with some $\phi$'s replaced by $\phi^{\, \prime}$'s, as long as $Z_2$
is respected; one important addition is introduced: that the $Y = -1$ combination
$\phi^\dagger \tilde \phi^{\, \prime}$ is nonzero with more than one doublet, 
but this won't be very relevant for the processes we want to analyse. Following
our aim of examining the $W \nu e$ mechanism, we will focus in this section 
on the generation of the $\mathcal{O} \up 7$ operator, as seen in equation  
\eqref{0nu2beta-eff-O7}. But as we just commented, the presence of 
$\phi^{\, \prime}$ increases the number of $\mathcal{O} \up 7$-like operators;
in fact, the $Z_2$ symmetry reduces the possibilities to two: one with three
$\phi$ doublets and another with two $\phi^{\, \prime}$'s and one $\phi$.
The gauge-invariant form of these operators reads
\begin{align} 
	\mathcal{O}_{\alpha \beta} \up 7 &= \overline{e_{\mathrm{R} \alpha}} 
			\gamma^\mu \left( \phi^\dagger \tilde \ell_{\mathrm{L} \beta} \right) 
			\left( \phi^\dagger D_\mu \tilde \phi \right)
	\nonumber \\  \label{Wnue-model-O7s}
	{\mathcal{O}_{\alpha \beta} \up 7}^\prime &= \overline{e_{\mathrm{R} \alpha}} 
			\gamma^\mu \left( {\phi^{\, \prime}}^\dagger 
			\tilde \ell_{\mathrm{L} \beta} \right) \left( \phi^\dagger D_\mu 
			\tilde \phi^{\, \prime} \right) \, .
\end{align}

\begin{figure}[tb]
	\centering
	\raisebox{5.5cm}{\emph{a)}}
	\includegraphics[width=0.65\textwidth]{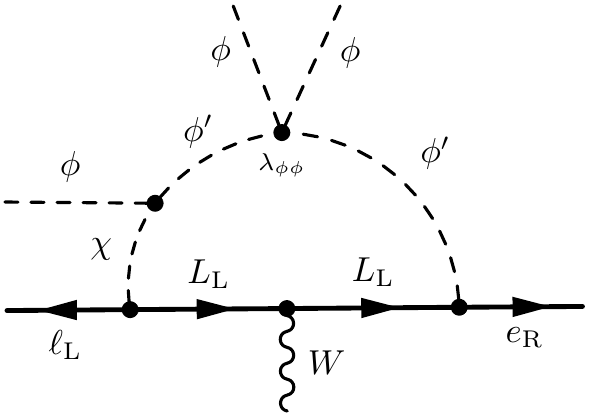}

	\vspace{0.4cm}
	\raisebox{4.5cm}{\emph{b)}}
	\includegraphics[width=0.65\textwidth]{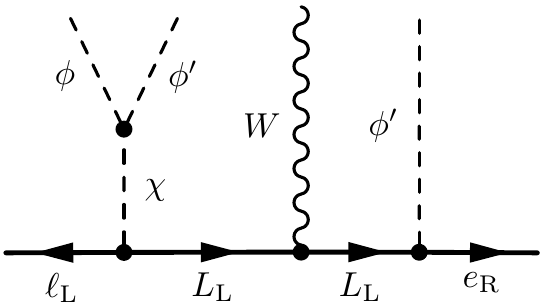}

	\caption{The two main diagrams generating the $\mathcal{O} \up 7$ and
			${\mathcal{O} \up 7}^\prime$ operators seen in equation
			\eqref{Wnue-model-O7s}.
			} \label{fig:Wnue-model-O7s}
\end{figure}

We present the diagrams that yield the main contributions to the operators in 
figure \ref{fig:Wnue-model-O7s}.
As we commented in the previous section, the operator with two $\phi^{\, \prime}$
legs is generated at tree level, and becomes LNV-ing once these $\phi^{\, \prime}$
acquire a VEV. The pure-$\phi$ operator is generated at one loop and requires
the LNV-ing interaction $\lambda_{\phi \phi}$, as $\phi$ carries no lepton number.
In principle it seems reasonable to think that ${\mathcal{O} \up 7}^\prime$
is responsible for the leading contribution to the $W \nu e$ mechanism, as 
the $W \nu e$ vertex generated by
$\mathcal{O} \up 7$ is suppressed by a factor that is roughly 
$\frac{\lambda_{\phi \phi}}{(4 \pi)^2} \, \frac{v_\phi^2}{{v_\phi^{\, \prime}}^2}$.
However, it's necessary to point out that the value of $v_\phi^{\, \prime}$ is not 
known, and it could
be forced to be small by phenomenological constraints; the preeminence of
either $\mathcal{O} \up 7$ or ${\mathcal{O} \up 7}^\prime$ is therefore not 
firmly decided. In the following we will focus our discussion on the effects of 
${\mathcal{O} \up 7}^\prime$ and related processes,
and we will check that a first phenomenological inspection allows for a 
reasonably large value of $v_\phi^{\, \prime}$ that ensures that it dominates
the $W \nu e$ mechanism.

The calculation of the effective coupling can be carried out just by inspection
of the diagram in figure \ref{fig:Wnue-model-O7s}\emph{b)}. The further assumption
that all the masses in the $\chi$ multiplet are equal allows to write the
result in terms of the triplet VEV:
\begin{equation} \label{Wnue-model-C7-prime}
	\frac{{C_{\alpha \beta} \up 7}^\prime}{\Lambda^3} \simeq -i  \mu \, \sum_\eta
		\frac{(y^e_{\eta \alpha})^* \, (y^\nu_{\eta \beta})^*}
		{m^2_\chi m^2_{L_\eta}} \simeq 
		-i \, \frac{v_\chi}{v_\phi v_\phi^{\, \prime}} \, 
		\sum_\eta \frac{(y^e_{\eta \alpha})^* \, (y^\nu_{\eta \beta})^*}
		{m^2_{L_\eta}}	\, .
\end{equation}
The presence \mbox{of $v_\chi$ --or,} equivalently, the trilinear 
\mbox{coupling $\mu$--, together} with the Yukawas $y^e$ and $y^\nu$, pinpoints
the violation of lepton number in the model. We will find this same combination
in the expressions for the neutrino masses that we derive in the next section,
and we can find it too in the processes that involve explicit violation of
LN through the $\lambda_{\phi \phi}$ interaction. In fact, the relation between
$v_\chi$ and LNV can be used to derive bounds on $v_\chi$ from the limits
on $0 \nu \beta \beta$ and neutrino masses; these bounds, that we discuss in 
section \ref{sec:Wnue-model-pheno}, can be very stringent, especially if we 
don't want the new particles to be very heavy. And there are good reasons not 
to want them to, as we discuss in the next section.

%
%

\section{Neutrino masses and neutrinoless double beta decay} 
		\label{sec:Wnue-model-numass-0nu2beta}

At this point, the generation of $0 \nu \beta \beta$ is straightforward: 
it suffices to insert either $\mathcal{O} \up 7$ or ${\mathcal{O} \up 7}^\prime$ 
into the appropriate combination of
tree-level SM processes as seen in figure \ref{fig:0nu2beta-eff-7-and-9}\emph{a)}.
Consequently, the corresponding bounds on the parameters of the model can be 
simply adapted from \eqref{bound-0nu2beta-O7} by using equation
\eqref{Wnue-model-C7-prime}; we leave this discussion to the next section,
where we will gather all phenomenological constraints together. In this section
our concern will be, rather, which is the mechanism that actually mediates
$0 \nu \beta \beta$; as we discussed in section \ref{sec:0nu2beta-eff-preeminence}
of the previous chapter, the $W \nu e$ mechanism contributes to $0 \nu \beta \beta$,
but it also yields masses for the neutrinos which can take over the generation
of $0 \nu \beta \beta$. In that section we obtained a generic bound on 
the new physics scale of $\Lambda \lesssim 35 \; \mathrm{TeV}$; in order to see
how that bound applies to this particular model, let us examine the generation
of neutrino masses.

Analogously to the case of the $\mathcal{O} \up 7$ operator discussed in the
previous section,
the neutrino masses in this model can arise either through spontaneous or
explicit violation of lepton number. These two possibilities correspond to
the two dimension-five operators that we find in the effective theory 
with a light $\phi^{\, \prime}$:
\begin{align*} 
	\mathcal{O}_{\alpha \beta} \up 5 &= - \left( \overline{\tilde 
		\ell_{\mathrm{L} \alpha}} \phi \right) \, \left( \tilde \phi^\dagger 
		\, \ell_{\mathrm{L} \beta} \right) 
	\\
	{\mathcal{O}_{\alpha \beta} \up 5}^\prime &= - \left( \overline{\tilde 
		\ell_{\mathrm{L} \alpha}} \phi^{\, \prime} \right) \,
		\left( {\tilde \phi^{\, \prime} \vphantom{b}}^\dagger \,
		\ell_{\mathrm{L} \beta} \right) 
		\, .
\end{align*}

\begin{figure}[tb]
	\centering
	\raisebox{4cm}{\emph{a)}}
	\includegraphics[width=0.65\textwidth]{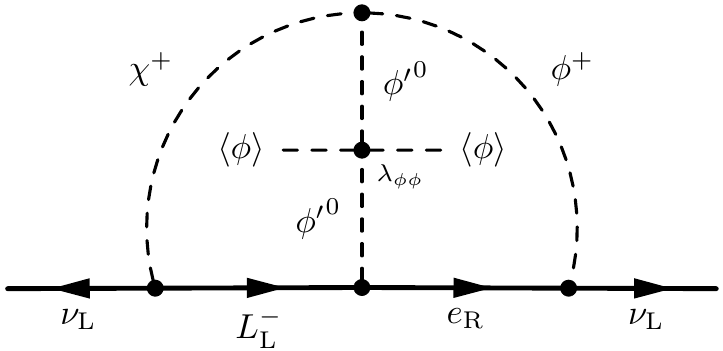}

	\vspace{0.4cm}
	\raisebox{6cm}{\emph{b)}}
	\includegraphics[width=0.65\textwidth]{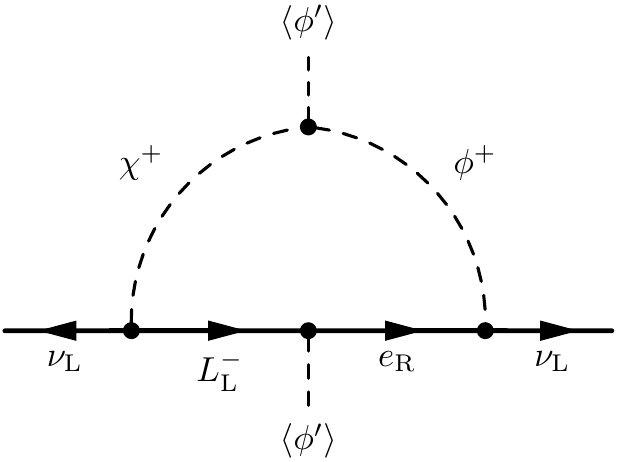}

	\caption{The two leading diagrams that yield neutrino masses, \emph{a)}
			through the $\mathcal{O} \up 5$ operator, and \emph{b)} through
			the ${\mathcal{O} \up 5}^\prime$ operator.
			} \label{fig:Wnue-model-O5s}
\end{figure}

The leading diagrams contributing to these operators are depicted in figure
\ref{fig:Wnue-model-O5s}. In the same way as in the previous section, the 
masses generated by $\mathcal{O} \up 5$ are suppressed by a factor of roughly 
$\frac{\lambda_{\phi \phi}}{(4 \pi)^2} \, \frac{v_\phi^2}{{v_\phi^{\, \prime}}^2}$
with respect to those generated by ${\mathcal{O} \up 5}^\prime$. This means that, 
barring other cancellations from the loop functions or the Yukawas,
 the spontaneous breaking of lepton number 
is the main source for neutrino masses unless $v_\phi^{\, \prime}$ is small.

Proceeding with our idea of considering spontaneous breaking as the main
source of LNV, the neutrino masses are generated by one-loop processes among which
the one presented in figure \ref{fig:Wnue-model-O5s}\emph{b)} is the leading
contribution. From this diagram we estimate the neutrino masses to be
\begin{equation} \label{Wnue-model-numass-0}
	(m_\nu)_{\alpha \beta} \simeq \frac{{v_\phi^{\, \prime}}^2 \mu}
			{2 (4 \pi)^2 \, v} \sum_\eta \frac{m_\alpha (y^e_{\eta \alpha})^* 
			(y^\nu_{\eta \beta})^* + m_\beta (y^e_{\eta \beta})^* 
			(y^\nu_{\eta \alpha})^*}{m_{L_\eta}^2 - m_\chi^2} \,
			\log \frac{m_{L_\eta}^2}{m^2_\chi} \, ,
\end{equation}
where we have assumed that $m_{L_\eta}$ and $m_\chi$ are much greater than all 
other masses. This expression is very detailed, even a bit too much for our 
purposes; a more handy form can be obtained if we assume the masses in the heavy 
multiplets to blend into one common heavy scale, $m_{L_\eta} \sim m_\chi \sim M$, 
and thus $\log \sim 1$:
\begin{equation} \label{Wnue-model-numass}
	(m_\nu)_{\alpha \beta} \sim \frac{v_\phi^{\, \prime} \, v_\chi}{(4 \pi)^2 \, 
			v_\phi^2} \sum_\eta \left[ m_\alpha (y^e_{\eta \alpha})^* 
			(y^\nu_{\eta \beta})^* + m_\beta (y^e_{\eta \beta})^* 
			(y^\nu_{\eta \alpha})^* \right]
\end{equation}

We are now prepared to estimate the contributions of neutrino masses and the
$W \nu e$ vertex to $0 \nu \beta \beta$. Proceeding as in equation
\eqref{0nu2beta-eff-0nu2beta-5-and-7}, and using \eqref{Wnue-model-numass}
and \eqref{Wnue-model-C7-prime}:
\begin{align*} 
	\mathcal{A}_{0 \nu \beta \beta} \up 5 &\sim
		\frac{G_F^2}{p_{\mathrm{eff}}^2} \, \frac{(m_\nu)_{e e}}{2} \sim
		\frac{G_F^2}{p_{\mathrm{eff}}^2} \, \frac{v_\phi^{\, \prime} \, v_\chi}
		{(4 \pi)^2 \, v_\phi^2} \, m_e \sum_\eta (y^e_{\eta e})^* (y^\nu_{\eta e})^*
	\\ \vphantom{\raisebox{3.5ex}{a}} 
	\mathcal{A}_{0 \nu \beta \beta} \up 7 &\sim
		\frac{G_F^2}{p_{\mathrm{eff}}} \, \frac{v_\phi^{\, \prime} \, v_\chi}
			{M^2} \sum_\eta (y^e_{\eta e})^* (y^\nu_{\eta e})^* \, ,
\end{align*}
where we dropped all phases and assumed again that $m_{L_\eta} \sim m_\chi 
\sim M$. The full $0 \nu \beta \beta$ amplitude is the coherent sum of these
two contributions,
\begin{displaymath}
	\mathcal{A}_{0 \nu \beta \beta} \sim \frac{G_F^2}{p_{\mathrm{eff}}^2} \,
		v_\phi^{\, \prime} \, v_\chi \sum_\eta (y^e_{\eta e})^* (y^\nu_{\eta e})^*
		\, \left[ \frac{m_e}{(4 \pi)^2 \, v_\phi^2} + \frac{p_{\mathrm{eff}}}{M^2}
		\right] \, ,
\end{displaymath}
and so the $W \nu e$ vertex dominates if
\begin{align*}
	\frac{p_{\mathrm{eff}}}{M^2} &> \frac{m_e}{(4 \pi)^2 \, v_\phi^2}
		& \Longrightarrow& & M &< 35 \; \mathrm{TeV} \, .
\end{align*}

So, for the case of this model, the requirement that $0 \nu \beta \beta$ is
mediated by the heavy particles has a simple meaning: it is not a bound on a 
complicated combination of scales, but merely on the masses of the $L$ fermions 
and the $\chi$ scalars, which should be at the level of the TeV or at most a few 
tens of TeV. This will be a useful constraint for the next section, where we
analyse the allowed values for the parameters of the model.

\section{Phenomenological constraints on the model parameters} 
		\label{sec:Wnue-model-pheno}
		
For the processes we are interested in, the model presents four groups of
relevant parameters: the masses of the heavy particles, the Yukawa couplings
$y^e$ and $y^\nu$, the VEV of the second doublet, $v_\phi^{\, \prime}$, and the 
trilinear coupling $\mu$. The aim of this section is to characterise a set
of viable values for these parameters from a first inspection focused mainly 
on LNV-ing observables. For this purpose we will prefer to write the $\mu$ coupling
in terms of the somewhat more physical triplet VEV, $v_\chi$; throughout this 
section, too, we will omit the details of the heavy particles' spectrum, and
we will assimilate all heavy masses to a common heavy scale, $m_{L_\eta} \sim 
m_\chi \sim M$.

The Yukawas are a source for family-number-violating processes:
once $\phi^{\, \prime}$ and $\chi$ acquire VEV's, the light $e_{\mathrm{R}}$ and
$\ell_{\mathrm{L}}$ leptons mix with the $L_\eta$; 
the low-energy effects of such mixings will be proportional to 
$y^e_{\alpha \beta} v_\phi^{\, \prime} / m_{L_\alpha}$
or $y^\nu_{\alpha \beta} v_\chi / m_{L_\alpha}$, and can be made as small 
as experimentally required either by increasing the heavy masses 
$m_{L_\alpha}$ or by reducing the Yukawas $y^{e, \nu}_{\alpha \beta}$ or
the VEV's. The first option doesn't seem convenient, since by the argument
exposed in section \ref{sec:Wnue-model-numass-0nu2beta} increasing $m_{L_\alpha}$
would lead to $0 \nu \beta \beta$ dominated by the light neutrino masses;
lowering the VEV's is also troublesome, for it would significantly suppress
the neutrino masses and possibly the $0 \nu \beta \beta$ rate.
As it seems, the only option appears to be to choose small Yukawas.
However, we may note that the number of heavy fermion families is not fixed in 
this model, and if there happened to be three or more we could choose 
the Yukawas to be aligned with those of the light leptons. And in fact we have
and argument in favour of more than one $L$ family: the light neutrino masses;
by looking at \eqref{Wnue-model-numass-0} we see that with only one generation
of $L$'s $m_\nu$ will have at most rank two, that is to say, the lightest 
neutrino will be forced to be massless. This would lead to a tight
neutrino mass spectrum that would further constrain the model parameters.
Two $L$ generations allow for three
massive light neutrinos, but, for a slightly higher prize, three $L$ families,
mimicking those of the chiral light leptons, open the possibility of imposing
a flavour symmetry that would efficiently suppress all LFV-ing effects. Of course
this could be sort of arbitrary in a model like this one, not concerned
with lepton family hierarchies, but it may be natural in a larger model.
In conclusion, increasing the number of heavy lepton families allows us
not to worry with the LFV constraints, with the $y^e$ and $y^\nu$ possibly
being of order 1 as long as they are almost-diagonal in the same basis in which 
$Y_e$ is diagonal%
\footnote{For
an estimation of how much alignment we would need to fulfill the LFV constraints
we can use the results obtained for other models with a similar lepton mixing
structure. For instance, several analyses focused on Littlest Higgs models
with T-parity \cite{Blanke:2007db,delAguila:2008zu,Goto:2010sn,delAguila:2010nv}
yield that the heavy flavours must be aligned with the light charged leptons 
with a precision better than $1-10$ \% for heavy masses of $\mathcal{O} 
(\mathrm{TeV})$.
}.

With this idea in mind, we move to LNV-ing observables: from equation 
\eqref{bound-0nu2beta-O7} we learn what $0 \nu \beta \beta$ imposes, and
in section \ref{sec:0nu2beta-eff-neutrinomasses} we establish 
$(m_\nu)_{\tau \tau} \sim 0.1 \; \mathrm{eV}$ as a reasonable requirement
for neutrino masses. Looking at equation \eqref{Wnue-model-C7-prime} we
see that we can identify ${C_{\alpha \beta} \up 7}^\prime \equiv -i 
\sum_\eta (y^e_{\eta \alpha})^* \, (y^\nu_{\eta \beta})^*$ and 
$\Lambda^3 \equiv v_\phi v_\phi^{\, \prime} M^2 / v_\chi$; if we assume
$y^e, y^\nu \sim 1$, \eqref{bound-0nu2beta-O7} translates into
\begin{equation} \label{Wnue-model-0nu2beta-bound-0}
	\frac{v_\phi v_\phi^{\, \prime} \, M^2}{v_\chi} > 10^6 \; \mathrm{TeV}^3 \, .
\end{equation}
On the other hand, if we demand $(m_\nu)_{\tau \tau} \sim 0.1 \; \mathrm{eV}$
to \eqref{Wnue-model-numass} under the same assumptions we obtain
\begin{equation} \label{Wnue-model-numass-bound-0}
	\frac{v_\phi^{\, \prime}}{v_\phi} \, \frac{v_\chi}{v_\phi} \sim 10^{-8} \, ,
\end{equation}
which we can use to eliminate $v_\phi^{\, \prime} / v_\phi$ from 
\eqref{Wnue-model-0nu2beta-bound-0} and then we get
\begin{equation} \label{Wnue-model-vchi-bound}
	\frac{v_\chi}{v_\phi} \lesssim 3 \times 10^{-8} \; \frac{M}{\mathrm{TeV}} \, .
\end{equation}
Finally, using \eqref{Wnue-model-vchi-bound} in return on 
\eqref{Wnue-model-numass-bound-0} we obtain for $v_\phi^{\, \prime}$:
\begin{equation} \label{Wnue-model-vphip-bound}
	\frac{v_\phi^{\, \prime}}{v_\phi} \gtrsim 0.3 \; \frac{\mathrm{TeV}}{M} \, .
\end{equation}

From these constraints we learn that $0 \nu \beta \beta$ demands a very small
VEV for the triplet, unless we consent that the new particles are very heavy,
thus losing the $W \nu e$ contribution to $0 \nu \beta \beta$. With the new
particles around the TeV scale, $v_\chi$ shouldn't be much above the keV scale;
this is quite a significant suppression, that we may think calls for 
further justification. We can, for example, invoke a global symmetry in this
fashion:
\begin{align*}
	L_{\mathrm{L}} &\rightarrow e^{i \alpha} \, L_{\mathrm{L}}
		& \phi^{\, \prime} &\rightarrow e^{i \alpha} \, \phi^{\, \prime}
	\\
	L_{\mathrm{R}} &\rightarrow e^{i \alpha} \, L_{\mathrm{R}}
		& \chi &\rightarrow e^{- i \alpha} \, \chi
\end{align*}
which is broken by the trilinear coupling $\mu$; if this symmetry is just 
approximate then we can understand that $\mu$ is small, and in turn \mbox{so 
$v_\chi$ is.} We note, additionally, that this symmetry would also suppress the 
$\lambda_{\phi \phi}$ coupling that violates LN explicitly; noticing that both
$\mu$ and $\lambda_{\phi \phi}$ are involved in the processes with explicit
LNV, we may conclude that spontaneous breaking dominates more easily in this
scenario.

As for $\phi^{\, \prime}$, equation \eqref{Wnue-model-vphip-bound} yields
an interesting conclusion: the second doublet is light, but it cannot be
\emph{too} light -- at least it cannot if its masses and VEV are to be of the
same order, as naturality would require. If the new particles have masses 
around the TeV, then the two doublets lie roughly at the electroweak scale;
this is consistent with the fact that no fundamental charged scalars have
been observed yet. Of course we could make the new particles heavier and
we would have room for lighter $v_\phi^{\, \prime}$, but at the expense of
losing some of the features of the model, and possibly creating a conflict
between naturality and direct searches. This situation should be studied with
care, considering all contributions to the masses of the charged scalars.
Whichever the case, $\phi^{\, \prime}$ appears to live between two 
somewhat close boundaries:
not too light as required by $0 \nu \beta \beta$ and neutrino masses, and
not too heavy if it is to remain as a low-energy excitation of the theory.

In a different line of thought, \eqref{Wnue-model-vphip-bound} also informs
us about the preeminent mechanism of LNV: with $\nicefrac{v_\phi^{\, \prime}}
{v_\phi} \sim 0.3$, lepton number would be mainly violated by spontaneous
breaking, but taking a slightly heavier $M$ would set a draw which could be
decided by
the value of the coupling $\lambda_{\phi \phi}$. Explicit breaking may 
even dominate if the heavy mass scale goes farther towards higher tens of
TeV or maybe $\mathcal{O} (100 \; \mathrm{TeV} )$, with the relevant $W \nu e$ 
diagrams gradually losing importance for the generation of $0 \nu \beta \beta$.
All in all, dominance of spontaneous breaking seems to be favoured, but 
a mixed scenario with a significant contribution of explicit breaking 
cannot be excluded, at least from our present analysis. 

Remember, to conclude, that the present discussion relies on several 
simplifications and order-of-magnitude assumptions; we have not considered,
for instance, the role of the Yukawas, which can be large if family
symmetries are invoked, but they can also be suppressed for different reasons.
Additional suppression from $y^e$ or $y^\nu$ would generally result in 
looser bounds for $v_\chi$ and $v_\phi^{\, \prime}$. We think that a good
conclusion to draw from this analysis is that the model is consistent with
\mbox{\emph{(i)} $0 \nu \beta \beta$ gen}\-erated through the $W \nu e$ vertex, 
possibly with large rates observable in the next round of experiments;
\mbox{\emph{(ii)} new par}\-ticles at the reach of the LHC, especially charged
scalars from the $\phi^{\, \prime}$ doublet; \mbox{\emph{(iii)} a variety} of
new LFV-ing signals mediated by the heavy leptons and possibly observable
in near future experiments.

\section{Collider effects} 

This model contains, besides the scalar triplet of unit 
hypercharge $\chi$, a second scalar isodoublet and several vector-like lepton
doublets. Direct evidence for the $W \nu e$ mechanism in an accelerator 
experiment would require the
discovery of the new particles together with a demonstration
of the presence of LNV-ing interactions. This latter goal seems difficult;
as we have seen throughout the discussion, violation of LN in this model involves 
several \mbox{couplings --typically} expected to be \mbox{small--, which yields}
small LNV-ing production and decay rates. In general, the dominant production 
mechanisms are standard and LN-conserving, or otherwise the production
itself is suppressed.
As for decays, not all the decay channels of the new particles produce 
LNV-ing signals, 
though in some cases they might be dominant. Generally, the idea is  
to look first for the new resonances in the most sensitive channels and 
only afterwards to address the observation of LNV-ing events. 

The scalars of the theory offer maybe the most promising detection prospects.
The triplet, for instance, includes a doubly-charged scalar that has fixed 
couplings to photons and is produced at colliders with known cross sections.
As for the decays, the triplet couples to a light and a heavy lepton, and the 
latter decays into a light lepton and a $W$, a $Z$, or a Higgs boson; this 
may result in a four-lepton decay. Similarly, the extra scalar isodoublet 
can also undergo decays with four fermions in the final state. Instances
of these processes can be
\begin{align*}
	pp \rightarrow \chi^{+ +} + X \rightarrow \ell_\alpha^+ L_\beta^+
		&\rightarrow \ell_\alpha^+ \ell_\eta^+ Z \rightarrow \ell_\alpha^+ 
		\ell_\eta^+ \ell_\lambda^+ \ell_\lambda^-
	\\
	\vphantom{\raisebox{2.5ex}{a}}
	pp \rightarrow {\phi^{\, \prime}}^{\, 0} + X \rightarrow \ell_\alpha^+ 
		{L_\beta^-}^* &\rightarrow \ell_\alpha^+ \ell_\eta^- Z \rightarrow
		\ell_\alpha^+ \ell_\eta^- \ell_\lambda^+ \ell_\lambda^-
\end{align*}
Moreover, the new particles can also be pair-produced via the Drell-Yan 
mechanism, resulting in final states with four charged leptons + jets, or even
eight leptons or more. The signal can be 
striking due to the large number of charged leptons, but there are many open 
channels and it may be difficult to resolve the different samples. 

The heavy vector-like lepton doublets are also mainly produced in pairs. They
violate the Glashow-Iliopoulos-Maiani mechanism \cite{Glashow:1970gm}, and can 
decay through a flavour-changing neutral current into a light lepton and a $Z$ 
or Higgs boson \cite{delAguila:1989rq}; as for the case of sequential fermions, 
they can 
also decay into a lepton and a $W$ boson through the usual charged current 
interaction. In these cases the final states have at least six fermions, so that 
the heavy leptons will be relatively easy to find if light enough, 
$m_{L_\eta} \lesssim \mathrm{TeV}$ for a center-of-mass energy of 
$14 \; \mathrm{TeV}$ and an integrated luminosity of $100 \; \mathrm{fb}^{-1}$
\cite{delAguila:1989rq,AguilarSaavedra:2009ik,delAguila:2010es}.

\chapter{A model realising the $\protect \WWee$ mechanism} \label{chap:WWee-model}

This chapter is dedicated to the study of a model that realises the $W W e e$
mechanism for $0 \nu \beta \beta$ and neutrino masses. 
It is but one of a family of models that give rise to the operator
$\mathcal{O} \up 9$ while suppressing the generation of $\mathcal{O} \up 5$
\mbox{and $\mathcal{O} \up 7$ -- for more} information on these operators and the
mechanisms they induce see chapter \ref{chap:0nu2beta-eff}, or, more plainly,
equations \eqref{Weinberg-(5)}, \eqref{0nu2beta-eff-O7} and 
\eqref{0nu2beta-eff-O9}. The $W W e e$ mechanism
is particularly interesting, as it can provide observable signals in a variety
of experiments, ranging from $0 \nu \beta \beta$ to high-energy
accelerators; we will give here an account of the several points of 
phenomenological interest and evaluate their importance for the viability
of the model. As we will see, the model is tightly constrained from various
directions: not only phenomenology, but also our desire to produce a signal
in the next round of $0 \nu \beta \beta$ experiments, and even the perturbative
consistence of the theory, tend to push the new particles to a rather narrow
region in the mass plane; all in all, though, there still seems to be room 
for them. Another interesting aspect of this model is the structure of
the generated neutrino mass matrix, which shows a remarkable suppression in
several
elements; as a consequence, the model predicts a nonzero value for the
reactor mixing angle, $\theta_{13}$, and a definite correlation between this
value and that of the CP-violating Dirac phase, $\delta$. The allowed range
for $\theta_{13}$ is consistent with the recent Daya Bay, RENO and Double
Chooz measurements.
This work was carried out in collaboration 
with Arcadi Santamaria, José Wudka, Francisco del Águila and Subhaditya
Bhattacharya.

\section{A model with lepton number softly broken} 
					\label{sec:WWee-model-presentation}

\begin{table}[bt]
	\centering
	\[
	\begin{array}{| c | c | c | c | c |}
		\hline
		\multicolumn{1}{|l}{ } &  & T & Y & Z_2 \\ \hline
		\sc02 \vphantom{\raisebox{-1ex}{a}} \vphantom{\raisebox{1.7ex}{a}}
				& \kappa &  0 & 2 & + \\ \hline
		\sc11 \vphantom{\raisebox{-1ex}{a}} \vphantom{\raisebox{1.7ex}{a}}
				& \chi & 1 & 1 & - \\ \hline
		  &  \sigma & 0 & 0 & - \\ \hline
	\end{array}
	\]
	\caption{The new scalars of the model and their charge assignments. The
			leftmost column presents the fields in the notation of section
			\ref{sec:0nu2beta-eff-newphysics}; a new, more wieldy naming
			is introduced in the next column. The $\sigma$ field is taken
			as real to avoid the creation of a Majoron when it develops
			a VEV.
			} \label{tab:WWee-model-particles}
\end{table}

For the construction of this model we will elaborate upon the 
$\bigl\{ \sc02 , \sc11 \bigr\}$ combination found in \eqref{NP-for-O9}. 
This combination includes two scalars: a doubly-charged singlet and a 
unit-hypercharge triplet, and is, by itself, sufficient to generate the 
$\mathcal{O} \up 9$ operator. However, in order to have a viable
and interesting model we need some extra work: to begin with,
if we want the $WWee$ mechanism to be dominant 
we need to suppress the competing operators $\mathcal{O} \up 5$
and $\mathcal{O} \up 7$, both of lower dimensionality and generating
neutrino masses at lower loop order. Note, for in\-\mbox{stance --see} 
\mbox{\eqref{NP-for-O5}--, that} the presence of $\sc11$ alone is enough to 
provide $\mathcal{O} \up 5$ at tree level; but by looking at table
\ref{tab:O5-topologies} we also see that the coupling 
$\overline{\tilde \ell_{\mathrm{L}}} \, \sc11 \ell_{\mathrm{L}}$
is fundamental to realise the generation of $\mathcal{O} \up 5$; if we
are clever enough to suppress this coupling we could relegate $\mathcal{O} 
\up 5$ to a subdominant role. This can be simply achieved by imposing
a discrete symmetry, for example a $Z_2$ under which all SM
fields are even and $\sc11$ is odd. We could then ask which can be the 
$Z_2$ assignment for $\sc02$; by looking at tables 
\ref{tab:O9-topologies-2} and \ref{tab:O9-topologies-3}, which describe
the topologies that generate $\mathcal{O} \up 9$, we see that a
$\overline{e_{\mathrm{R}}^\mathrm{c}} e_{\mathrm{R}} \, \sc02$ coupling
is needed. $\sc02$, thus, must be $Z_2$-even. But then, again
looking at tables \ref{tab:O9-topologies-2} and \ref{tab:O9-topologies-3},
we see that a $\phi^\dagger \sc11 \tilde \phi$ coupling is needed, and
it is forbidden by our $Z_2$ assignments! There are two options at this 
point: either we introduce this interaction by hand, explicitly breaking
the $Z_2$ symmetry, or we introduce a new scalar field that breaks 
spontaneously $Z_2$ and allows for this vertex to exist. We choose to
follow the second path, but look at section \ref{sec:WWee-model-drawbacks},
for the first one can be convenient on its own merits. The new scalar 
field must be a hyperchargeless singlet, odd under $Z_2$; we will take
it real, for reasons that will become clear in a moment.

The properties of the new scalars are summarised in table 
\ref{tab:WWee-model-particles}, where a new, simpler notation is defined for 
the fields. The most general scalar potential compatible with these assignments is
\begin{align} \label{WWee-model-scalar-potential}
V  =  &- m_{\phi}^{2} \, \phi^\dagger \phi - m_{\sigma}^{2} \, \sigma^{2} 
		+ m_{\chi}^{2} \, \Tr{\chi^{\dagger} \chi} + m_{\kappa}^{2} \,
		\kappa^\dagger \kappa \: + 
		\nonumber \\
  &+ \lambda_{\phi} \, \left( \phi^\dagger \phi \right)^2 + \lambda_{\sigma} \, 
  		\sigma^{4} + \lambda_{\kappa} \, \left( \kappa^\dagger \kappa \right)^2 +
		\lambda_{\chi} \, \left( \Tr{\chi^{\dagger} \chi} \right)^{2} + 
	 	\vphantom{\raisebox{2.5ex}{a}} \nonumber \\
  &+ \lambda_{\chi}^{\prime} \, \Tr{\left(\chi^{\dagger}\chi\right)^{2}} +
 		\lambda_{\phi \sigma} \, \phi^\dagger \phi \: \sigma^{2} + 
		\lambda_{\phi \kappa} \, \phi^\dagger \phi \: \kappa^\dagger \kappa \: + 
		\vphantom{\raisebox{2.5ex}{a}} \\
  &+ \lambda_{\phi\chi} \, \phi^\dagger \phi \, \Tr{\chi^{\dagger} \chi} +
 		\lambda_{\phi\chi}^{\prime} \, \phi^{\dagger} \chi^{\dagger} \chi \, \phi +
		\lambda_{\sigma\kappa} \, \sigma^{2} \: \kappa^\dagger \kappa \: + 
		\vphantom{\raisebox{2.5ex}{a}} \nonumber \\
  &+ \lambda_{\kappa\chi} \, \kappa^\dagger \kappa \, \Tr{\chi^{\dagger}\chi} +
		\lambda_{\sigma\chi} \, \sigma^{2} \, \Tr{\chi^{\dagger}\chi} + 
		\vphantom{\raisebox{2.5ex}{a}} \nonumber \\
  &+ \left[ \mu_{\kappa} \, \kappa \: \Tr{\chi^{\dagger} \chi^{\dagger}} -
		\lambda_{6} \, \sigma \: \phi^{\dagger} \chi \, \tilde{\phi} + 
		\mathrm{H.c.} \right] \, ; 
		\vphantom{\raisebox{2.5ex}{a}} \nonumber
\end{align}
among all these interactions%
\footnote{The reader may notice that 
\eqref{WWee-model-scalar-potential} lacks terms such as $\Tr{\chi^2} \, 
\Tr{{\chi^\dagger}^2}$, $\Tr{\chi^2 \, {\chi^\dagger}^2}$ and 
$\phi^\dagger \chi \chi^\dagger \phi$. These terms are not independent from
the ones presented and can be related to them by using the fact that any two
traceless $2 \times 2$ matrices $A$ and $B$ verify $\{ A, B \} = \Tr{A B} \,
\mathbb{1}$.
}, 
the most important ones for our purposes are the latter
two. Note that by an appropriate rephasing of the fields
we can always choose $\mu_\kappa$ and $\lambda_6$ to be real of either sign; 
for simplicity, we will take in the following $\mu_\kappa > 0$ and
$\lambda_6 > 0$. Notice, however, the minus sign assigned to the $\lambda_6$
term, whose purpose is to ensure that the VEV of $\chi$ is positive;
indeed, precision electroweak measurements
such as the $\rho$ parameter constrain the VEV of any triplet scalar to be
small. This could be accomplished by taking a (negative) small $m_\chi^2$, but
that would result in very light charged scalars that have not been observed.
Therefore, our strategy will be to let $\phi$ and $\sigma$ to take a VEV
on their own and transmit it to $\chi$ via the $\lambda_6$ interaction,
thus resulting in a naturally small $v_\chi$. Note that the sign assignments
for the mass terms in \eqref{WWee-model-scalar-potential} already prefigure
this program, allowing us to consider all $m^2 > 0$. 
A complete discussion on the breakdown of electroweak symmetry in this model
can be found in section \ref{sec:WWee-model-vevs}.

Also of interest are the Yukawa couplings allowed in the Lagrangian,
\begin{equation} \label{WWee-model-yukawas}
	\mathcal{L}_{\mathrm{Y}} = \overline{\ell_{\mathrm{L}}} \, Y_{e} 
			e_{\mathrm{R}} \,\phi + \overline{e^{\mathrm{c}}_{\mathrm{R}}} \,
			g \, e_{\mathrm{R}} \, \kappa + \mathrm{H.c.} \, ,
\end{equation}
where $Y_e$ is a $3 \times 3$ matrix that can be taken diagonal with positive
eigenvalues without loss of generality; $g$ is a complex symmetric $3 \times 3$
matrix with only three physical phases.

Note from \eqref{WWee-model-scalar-potential} and \eqref{WWee-model-yukawas}
the intrincate situation of lepton number in this model: considering that
$L (\ell_{\mathrm{L}}) = L (e_{\mathrm{R}}) = -1$, $\kappa$ is assigned
$L (\kappa) = +2$ by the $g$ Yukawas. Were $\sigma$ to be a complex field,
$\chi$ would be assigned $L (\chi) = +1$ by the $\mu_\kappa$ coupling in 
\eqref{WWee-model-scalar-potential} and transmit this charge to $\sigma$
via the $\lambda_6$ coupling, with $L (\sigma) = -1$. But then, as $\sigma$
takes a VEV a Majoron would be created that would additionally constrain the
model; even though these constraints wouldn't possibly be very tight, as the
Majoron would be mainly a singlet with small couplings to matter, we prefer
to choose $\sigma$ to be real, and therefore unable to carry
\mbox{charges -- but} see section \ref{sec:WWee-model-drawbacks} for more 
comments on the possibility of 
a complex $\sigma$. With $\sigma$ real, $\chi$ suffers from opposed LN assignments:
the $\mu_\kappa$ interaction would allot $L (\chi) = +1$, but the $\lambda_6$
vertex requires $L (\chi) = 0$; lepton number is, so, explicitly broken
if $\sigma$ is real. Moreover, as $\mu_\kappa$ is dimensionful it is 
\emph{softly} broken and the model will not suffer from infinity sickness
in radiative corrections. Notice also that none of the LN assignments for
$\chi$ is $+2$, which is the one associated to the 
$\overline{\tilde \ell_{\mathrm{L}}} \, \chi \ell_{\mathrm{L}}$ coupling
that we wish to suppress: that interaction is correctly forbidden by
the $Z_2$ symmetry and when it is loop-generated, after the $\sigma$ VEV breaks
$Z_2$, the resulting neutrino masses will be finite and calculable.

To finish this section, a further remark on lepton number and its role on 
the processes we are
interested in: note that in the limit of vanishing $Y_e$ one can define two
different ``lepton numbers'', one associated to the doublets $\ell_{\mathrm{L}}$
and the other to the singlets $e_{\mathrm{R}}$. The first, for 
$Y_e \rightarrow 0$, is uncommunicated to the scalar sector, and thus
conserved; as a consequence, any sort of $0 \nu \beta \beta$ to left-handed leptons
is forbidden, and Majorana neutrino masses are barred too. The second is
transmitted to the scalars via the $g$ Yukawas, and then broken in the scalar
potential by the $\mu_\kappa$ and $\lambda_6$ couplings. Therefore, 
$0 \nu \beta \beta$ to right-handed leptons is allowed in any case, and will
be proportional to $g_{\alpha \beta}$, $\mu_\kappa$ and $\lambda_6$, whereas
neutrino masses require these three elements together with $Y_e$. We will find
both results in the corresponding sections, \ref{sec:WWee-model-0nu2beta}
and \ref{sec:WWee-model-neutrino-masses}.

\section{The scalars of the theory} \label{sec:WWee-model-scalars}

The model includes three new scalar fields that yield nine degrees of freedom,
to sum to the four provided by the usual SM doublet. Their
properties are profoundly intertwined as a consequence of the interactions in
the scalar potential and the nontrivial breakdown of electroweak symmetry,
which involves three vacuum expectation values. The final physical scalars,
with definite masses and electric charges, are combinations of these 
pre-EWSB building blocks. In this section we present a possible pattern
of symmetry breaking that simplifies this potentially intrincate panorama
and fulfills all phenomenological constraints; we also describe the final
spectrum of physical scalars and remark on some couplings that will be
of use in the following sections.

\subsection{Boundedness of the potential} \label{sec:WWee-model-boundedness}

A necessary condition for any QFT to make sense is that the scalar potential
is bounded from below; this is what allows masses to be defined properly
and in turn leads to the notion of vacuum expectation value. 
The scalar potential presented in equation 
\eqref{WWee-model-scalar-potential} is a complicated one and involves many
different interactions. Even though most of them will not be of interest for
our forthcoming discussion let us briefly analyse the conditions under which
the model will present a potential bounded from below, if only to satisfy a drive
of professional zeal.

The condition of below-boundedness can be translated essential\-\mbox{ly --for a} 
regular potential, with no \mbox{poles-- to the} requirement that the potential
tends to $+ \infty$ in all directions when the fields, taken as C-numbers, diverge.
As it is a condition at infinity, it suffices to consider the highest-degree
part of the potential, that will be dominant in that re\-\mbox{gime -- in our} 
case, we just need 
the quartic terms. Specifically, let us call
\begin{align} \label{WWee-model-quartic-pieces}
	V \up 4 \equiv 
  & \phantom{+} \lambda_{\phi} \, \left( \phi^\dagger \phi \right)^2 + 
  		\lambda_{\sigma} \, 
  		\sigma^{4} + \lambda_{\kappa} \, \left( \kappa^\dagger \kappa \right)^2 +
		\lambda_{\chi} \, \left( \Tr{\chi^{\dagger} \chi} \right)^{2} + 
		\vphantom{\raisebox{2.5ex}{a}} \nonumber \\
  &+ \lambda_{\chi}^{\prime} \, \Tr{\left(\chi^{\dagger}\chi\right)^{2}} +
 		\lambda_{\phi \sigma} \, \phi^\dagger \phi \: \sigma^{2} + 
		\lambda_{\phi \kappa} \, \phi^\dagger \phi \: \kappa^\dagger \kappa \: + 
		\vphantom{\raisebox{2.5ex}{a}} \nonumber  \\
  &+ \lambda_{\phi\chi} \, \phi^\dagger \phi \, \Tr{\chi^{\dagger} \chi} +
 		\lambda_{\phi\chi}^{\prime} \, \phi^{\dagger} \chi^{\dagger} \chi \, \phi +
		\lambda_{\sigma\kappa} \, \sigma^{2} \: \kappa^\dagger \kappa \: + 
		\vphantom{\raisebox{2.5ex}{a}} \\
  &+ \lambda_{\kappa\chi} \, \kappa^\dagger \kappa \, \Tr{\chi^{\dagger}\chi} +
		\lambda_{\sigma\chi} \, \sigma^{2} \, \Tr{\chi^{\dagger}\chi} - 
		\vphantom{\raisebox{2.5ex}{a}} \nonumber \\
  &- \lambda_{6} \left( \sigma \: \phi^{\dagger} \chi \, \tilde{\phi} + 
		\sigma \: \tilde \phi^{\dagger} \chi^\dagger \, \phi \right) \, ; 
		\vphantom{\raisebox{2.5ex}{a}} \nonumber
\end{align}
note that most terms are biquadratic, meaning that the
the field combinations always yield a positive C-number, and the only negative
contributions may come from a possibly negative coefficient. Were $V \up 4$
strictly biquadratic, a sufficient condition for below-boundedness would be
that all the coefficients are positive. Unfortunately, we have also the $\lambda_6$
term, which we want to be negative and is linear in $\chi$ and $\sigma$. This
term forces us into a more convoluted discussion.

To our knowledge, there is no general method for finding the conditions of 
below-boundedness in a quartic form. 
We do know, however, of a very straightforward procedure to do so in a 
\emph{quadratic} form, and we can adapt it to our case here; 
it is a well-known method, consisting in diagonalising the quadratic polynomial and
requiring that the coefficients of the diagonalised \mbox{form --which} only accompany
perfect-squared variables, an so positive quanti\-\mbox{ties-- are} positive. 
Those coefficients are the eigenvalues of the matrix associated to the quadratic
form. Therefore, we will proceed to express our quartic form as a quadratic 
polynomial, making use of its almost-biquadratic form and implementing
variable changes of the form $\phi^2 \rightarrow x$. The rest will be to 
properly collect the matrix of the `new' quadratic polynomial and imposing 
positive eigenvalues for it.

The components of the quartic form we're interested in are the scalar fields
of our model or, more specifically, the degrees of freedom contained inside them.
This adds up to thirteen elements, but not all of them are relevant for
the present discussion. In order to identify the relevant degrees of freedom
let us express the involved fields in a new notation. Begin first with the 
triplet, which as we know encloses three complex fields with definite 
electric \mbox{charge -- see} for instance section \ref{sec:seesaw-typeII} and
references therein.
That makes six degrees of freedom, but as any piece of the potential
is $SU(2) \otimes U(1)$ gauge-invariant we can work in an $SU(2)$ gauge which
eliminates three of \mbox{them -- remem\-}ber, $SU(2)$ is a three-parameter 
Lie group. Let us then
choose the gauge that shapes the triplet, in the doublet representation, as
\begin{displaymath}
	\chi = \begin{pmatrix}
			0 & \sqrt{b} \, e^{-i \mu} \\
			\sqrt{a} \, e^{i \mu} & 0
	\end{pmatrix} ,
\end{displaymath}
with $a$ and $b$ positive real C-numbers and $\mu$ an ordinary phase that, we
will check, can be abstracted from this discussion. The doublet, in that same 
gauge, will present its most general form,
\begin{displaymath}
	\phi = \begin{pmatrix}
			\sqrt{p} \, e^{i \alpha}  \\
			\sqrt{q} \, e^{i \beta}
	\end{pmatrix} ,
\end{displaymath}
with again $p$ and $q$ real positive C-numbers and $\alpha$, $\beta$ irrelevant
phases. As for the singlets, notice that the pieces in 
\eqref{WWee-model-quartic-pieces} only involve the hyperchargeless combination
$\kappa^\dagger \kappa$, and so a `phase' of $\kappa$ can also be ignored 
here; let us define
\begin{displaymath}
	\kappa^\dagger \kappa \equiv c \, ,
\end{displaymath}
and $\sigma$, for `dimensional consistency',
\begin{displaymath}
	\sigma^2 \equiv d \, ,
\end{displaymath}
with both $c$ and $d$ real positive C-numbers.

Expressing \eqref{WWee-model-quartic-pieces} in this new notation yields
\begin{displaymath}
	V \up 4 = X^\mathrm{T} \mathcal{V} \, X - 
			2 \lambda_6 \left[ q \sqrt{a d} \, \cos (\mu - 2 \beta ) -
			p \sqrt{b d} \, \cos (\mu - 2 \alpha ) \right] \, ,
\end{displaymath}
where $X^\mathrm{T} \mathcal{V} \, X$ is already a quadratic form with
{\scriptsize
\begin{displaymath} 
	\mathcal{V} = \begin{pmatrix}
		\lambda_\chi + \lambda_\chi^\prime & \lambda_\chi 
			& \lambda_{\kappa \chi} / 2  & \lambda_{\sigma \chi} / 2 
			& (\lambda_{\phi \chi} + \lambda_{\phi \chi}^\prime ) / 2
			& \lambda_{\phi \chi} / 2 
			\\
		\lambda_\chi & \lambda_\chi + \lambda_\chi^\prime 
			& \lambda_{\kappa \chi} / 2 & \lambda_{\sigma \chi} / 2
			& \lambda_{\phi \chi} / 2 
			& (\lambda_{\phi \chi} + \lambda_{\phi \chi}^\prime )/2
			\\
		\lambda_{\kappa \chi} / 2 & \lambda_{\kappa \chi} / 2 
			& \lambda_{\kappa} & \lambda_{\sigma \kappa} / 2 
			& \lambda_{\phi \kappa} / 2 & \lambda_{\phi \kappa} / 2
			\\
		\lambda_{\sigma \chi} / 2 & \lambda_{\sigma \chi} / 2 
			& \lambda_{\sigma \kappa} / 2 & \lambda_{\sigma}
			& \lambda_{\phi \sigma} / 2 & \lambda_{\phi \sigma} / 2
			\\
		( \lambda_{\phi \chi} + \lambda_{\phi \chi}^\prime ) / 2
			& \lambda_{\phi \chi} / 2 & \lambda_{\phi \kappa} / 2 
			& \lambda_{\phi \sigma} / 2 & \lambda_\phi & \lambda_\phi
			\\
		\lambda_{\phi \chi} / 2 
			& ( \lambda_{\phi \chi} + \lambda_{\phi \chi}^\prime ) / 2
			& \lambda_{\phi \kappa} / 2 & \lambda_{\phi \sigma} / 2
			& \lambda_\phi & \lambda_\phi
	\end{pmatrix}
\end{displaymath}
}
$\!\!$and
\begin{displaymath}
	X^\mathrm{T} = \begin{pmatrix}
		a & b & c & d & p & q
	\end{pmatrix} \, ,
\end{displaymath}
and the $\lambda_6$ terms break this structure, as we already knew. The
goal is now to integrate these terms into the quadratic form in a way that
preserves the significant information about below-boundedness. That can happen;
observe that the sign of these terms is not fixed, as they contain cosines
of in principle arbitrary phases. Note too that these terms are a problem
for below-boundedness as far as they \emph{subtract} from the quadratic
polynomial, and so pull to $- \infty$ two of the directions. Let us now
place ourselves in the \emph{worst} of such situations: that where the phases 
conspire to yield
\begin{multline*}
	- 2 \lambda_6 \left[ q \sqrt{a d} \, \cos (\mu - 2 \beta ) -
		 p \sqrt{b d} \, \cos (\mu - 2 \alpha ) \right] =
	\\
	\vphantom{\raisebox{2ex}{a}}
	= - 2 \lambda_6 \left[ q \sqrt{a d} + p \sqrt{b d} \right] \, .
\end{multline*}
Then consider the general inequality $\sqrt{x y} \leq (x + y) / 2$; by using
it we can obtain quadratic-like terms that are always \emph{worse} than the
original ones,
\begin{displaymath}
	- 2 \lambda_6 \left[ q \sqrt{a d} + p \sqrt{b d} \right] \geq
		- \lambda_6 \left[ q (a + d) + p (b + d) \right] \, .
\end{displaymath}
Then we can construct a new quadratic form
\begin{displaymath}
	\bar V \up 4 \equiv X^\mathrm{T} \mathcal{V} \, X -
		 \lambda_6 \left[ q (a + d) + p (b + d) \right] \equiv
		 X^\mathrm{T}  \bar{\mathcal{V}} \, X
\end{displaymath}
with a slightly modified matrix
{\scriptsize
\begin{displaymath}
	\bar{\mathcal{V}} = \begin{pmatrix}
		\lambda_\chi + \lambda_\chi^\prime & \lambda_\chi 
			& \lambda_{\kappa \chi} / 2  & \lambda_{\sigma \chi} / 2 
			& (\lambda_{\phi \chi} + \lambda_{\phi \chi}^\prime ) / 2
			& \bar \lambda_{\phi \chi} / 2 
			\\
		\lambda_\chi & \lambda_\chi + \lambda_\chi^\prime 
			& \lambda_{\kappa \chi} / 2 & \lambda_{\sigma \chi} / 2
			& \bar \lambda_{\phi \chi} / 2 
			& (\lambda_{\phi \chi} + \lambda_{\phi \chi}^\prime )/2
			\\
		\lambda_{\kappa \chi} / 2 & \lambda_{\kappa \chi} / 2 
			& \lambda_{\kappa} & \lambda_{\sigma \kappa} / 2 
			& \lambda_{\phi \kappa} / 2 & \lambda_{\phi \kappa} / 2
			\\
		\lambda_{\sigma \chi} / 2 & \lambda_{\sigma \chi} / 2 
			& \lambda_{\sigma \kappa} / 2 & \lambda_{\sigma}
			& \bar \lambda_{\phi \sigma} / 2 & \bar \lambda_{\phi \sigma} / 2
			\\
		( \lambda_{\phi \chi} + \lambda_{\phi \chi}^\prime ) / 2
			& \bar \lambda_{\phi \chi} / 2 & \lambda_{\phi \kappa} / 2 
			& \bar \lambda_{\phi \sigma} / 2 & \lambda_\phi & \lambda_\phi
			\\
		\bar \lambda_{\phi \chi} / 2 
			& ( \lambda_{\phi \chi} + \lambda_{\phi \chi}^\prime ) / 2
			& \lambda_{\phi \kappa} / 2 & \bar \lambda_{\phi \sigma} / 2
			& \lambda_\phi & \lambda_\phi
	\end{pmatrix} \, ,
\end{displaymath}
}
$\!\!$where $\bar \lambda_{\phi \chi, \phi \sigma} \equiv \lambda_{\phi \chi,
\phi \sigma} - \lambda_6$.
It is sufficient condition for $V \up 4$ to be bounded from below that
$\bar V \up 4$ is; so, below-boundedness holds if the eigenvalues of
$\bar{\mathcal{V}}$ are all positive.

As we anticipated, this long discussion ends with a not-so-spectacular conclusion:
essentially, we learn that we can proceed to the analysis of the
model under assumptions usually considered `reasonable'; 
we don't really need to calculate the
eigenvalues of $\bar{\mathcal{V}}$, involving as they do many parameters that
are not of interest to our discussion. It is enough to take 
all the $\lambda$'s positive, including \mbox{$\lambda_6$ --that we need--,}
as far as we keep posi\-\mbox{tive --or at} least \mbox{balanced-- the values} of
$\lambda_{\phi \chi, \phi \sigma} - \lambda_6$.

\subsection{Vacuum expectation values} \label{sec:WWee-model-vevs}

The model possesses three neutral scalars susceptible of developing a VEV:
one comes from the doublet, one from the triplet and the third from the 
neutral singlet $\sigma$. We need at least the doublet to acquire a VEV, in order
to break electroweak symmetry. $0 \nu \beta \beta$ and neutrino masses need
lepton number to be broken, but as it is explicitly violated in the potential
(see section \ref{sec:WWee-model-presentation} for more details) the triplet
and the singlet can remain VEV-less at the price of having $0 \nu \beta \beta$
at one \mbox{loop --instead} of tree-\mbox{level-- and} neutrino masses at 
three \mbox{loops -- instead} of two. As we want to enhance
the rate of $0 \nu \beta \beta$ we prefer the scenario with three VEV's, which 
presents additional difficulties.

The VEV of the triplet is problematic because we have very strong bounds
from electroweak observables. In particular, the $\rho$ parameter is highly
sensitive to the $SU(2)$ charge of the fields that break the electroweak symmetry,
and it is consistent with ``breaking by a doublet'' to a good degree of
precision; specifically, a global fit to electroweak precision data 
\cite{Beringer:1900zz} yields
\begin{equation} \label{WWee-model-rho-exp}
	\rho = 1.0004 {\,}^{+0.0023}_{-0.0011}
\end{equation}
at $2 \sigma$, very close to $\rho = 1$, which would indicate pure doublet-mediated
electroweak symmetry breaking. This means that any triplet VEV must be small
enough not to compete with the standard Higgs doublet for the dominant role 
in EWSB. We can calculate to which extent \eqref{WWee-model-rho-exp} constrains 
the triplet VEV. The theoretical expression for $\rho$ at tree level is
\begin{displaymath}
	\rho = \frac{\Sigma_j \, v_j^2 \, \left[ T_j (T_j + 1) - Y_j^2 \right]}
			{\Sigma_j \, 2 v_j^2 \, Y_j^2} \, , 
\end{displaymath}
where the sums run over all the scalars of the theory, $T_j$ represents their
isospin, $Y_j$ their hypercharge and $v_j$ their VEV\footnote{Throughout 
this section we are already implementing the notation presented in equation
\eqref{WWee-model-neutral-and-vevs} in the next section.}. 
For the case under 
consideration we obtain a simple expression that can be simplified even
further if we assume that $v_\chi \ll v_\phi$:
\begin{equation} \label{WWee-model-rho-theo}
	\rho = \frac{v_\phi^2 + 2 v_\chi^2}{v_\phi^2 + 4 v_\chi^2}
		\simeq 1- \frac{2 v_\chi^2}{v_\phi^2} \, .
\end{equation}
As we see, the VEV of the triplet contributes negatively to the $\rho$ parameter,
while \eqref{WWee-model-rho-exp} presents a value slightly greater than 1. 
We can thus derive an upper bound on the value of $v_\chi$, that reads
\begin{displaymath}
	v_\chi < 3 \; \mathrm{GeV} \qquad \! \! 95\% \; \mathrm{CL}
\end{displaymath}
for $v_\phi = 174 \; \mathrm{GeV}$. This bound is similar to those obtained
from more comprehensive surveys, for example a global fit including
explicitly the scalar triplet effects \cite{delAguila:2008ks} which
yields $v_\chi < 2 \; \mathrm{GeV}$ at 90\% CL.

However, a more complete analysis should also include the radiative corrections 
to the $\rho$ parameter induced by the scalar triplet itself, which can be
positive \cite{Einhorn:1981cy}. For instance, in the triplet Majoron model
loop corrections provide a positive contribution that cancels partially the 
tree-level one, $\Delta \rho = \frac{1 - \ln 2}{2 \pi^2 \sqrt{2}} \, G_F \, 
m_{\chi^{++}}^2$, with $m_{\chi^{++}}$ the mass of the doubly-charged 
scalar \cite{Golden:1986jn}.
Since these contributions depend on the mass splitting of the triplet components, 
which in our model is not fixed, when we want to consider a conservative
bound on $v_\chi$ we will take
\begin{equation} \label{WWee-model-vchi-bound}
	v_\chi < 5 \, \mathrm{GeV} \, ,
\end{equation}
which is in the line of another recent work \cite{Kanemura:2012rs} that, including
one-loop corrections, points to a relaxation of the bound of up to 
$v_\chi < 6 \; \mathrm{GeV}$ at 95\% CL.

As for the VEV of the singlet, it is essentially unconstrained, as it does
not contribute to EWSB. The issues with this VEV are more related to its 
qualitative consequeces, such as the creation of domain walls due to the
breaking of $Z_2$, which we address in section \ref{sec:WWee-model-drawbacks},
or the necessary induction of a VEV for the triplet, irrespective of the
sign of $m_\chi^2$. Indeed, once $\phi$ and $\sigma$ take a VEV the $\lambda_6$
interaction drives $\chi$ into a nontrivial minimum, bringing to the focus
the restrictions on $v_\chi$ that we just described. We could think of
dropping the singlet VEV, but as commented before that would mean relegating
$0 \nu \beta \beta$ to the one-loop level, which would spoil part of the
interest of the model. Let us, so, examine this nontrivial minimum that
involves the three neutral fields and see if it can accommodate the
phenomenological restrictions on $\chi$.

The minimisation of $V$ can be carried out in a simple way: just substituting
in \eqref{WWee-model-scalar-potential} the neutral components $\phi^0$,
$\chi^0$ and $\sigma$ by $v_\phi$, $v_\chi$ and $v_\sigma$ while putting zeros
in place of the charged components, and then looking for a minimum
of this three-variable function. To simplify the procedure, we will just look
for an extremum (the point where the three partial derivatives vanish) and
later we will check that it is indeed a minimum; for that aim it is enough
to require that the resulting scalar masses are physical, i.e., $m^2 \ge 0$.
See section \ref{sec:WWee-model-scalar-spectrum} for a discussion on this
matter.

Several solutions are obtained by following this program; most of them
 present vanishing VEV's for either the doublet or the singlet.
We select the solution that yields nonzero values for all three VEV's,
and further simplify it by assuming $v_\chi \ll v_\phi$;
this is not a very grave loss of generality: given the bound 
\eqref{WWee-model-vchi-bound} this condition is merely factual and must
be imposed at some point. Already in this approximation, the VEV's
at the minimum read
\begin{gather} 	
	v_\phi^2 \simeq \frac{2 \lambda_\sigma \, m_\phi^2 - 
		\lambda_{\phi\sigma} \, m_\sigma^2}
		{4 \lambda_\sigma \lambda_\phi - \lambda_{\phi\sigma}^2}  
	\qquad
	v_\sigma^2 \simeq \frac{2 \lambda_\phi \, m_\sigma^2 -
		\lambda_{\phi\sigma} \, m_\phi^2}
		{4 \lambda_\sigma \lambda_\phi - \lambda_{\phi\sigma}^2}
	\nonumber \\ 
	\vphantom{\raisebox{4ex}{a}}
	v_\chi \simeq \frac{\lambda_6 \, v_\sigma v_\phi^2}
		{m_\chi^{2} + \lambda_{\phi\chi} \,v_\phi^2 + 
		\lambda_{\sigma \chi} \, v_\sigma^2} \; .
	\label{WWee-model-vevs-minimum}
\end{gather}
Notice that $v_\phi$ and $v_\sigma$ are real and positive for all $\lambda_\phi,
\lambda_\sigma, \lambda_{\phi \sigma} > 0$, in consistency with the conditions
derived in section \ref{sec:WWee-model-boundedness} for the potential to be 
bounded from below. As for $v_\chi$, note that with the present sign assignment
$\lambda_6 > 0$ is the right phase choice to make $v_\chi$ real and positive;
note also that the condition $v_\chi \ll v_\phi$ somehow suggests that
$m_\chi \gg v_\phi, v_\sigma$, which looks convenient, as the new charged
scalars provided by the triplet have not yet been observed. Anyway, we will
make no assumptions about $m_\chi$; the phenomenological constrains themselves
will yield the appropriate restrictions \mbox{for it -- see} 
section \ref{sec:WWee-model-constraints-parameters}.

\subsection{Physical scalar spectrum} \label{sec:WWee-model-scalar-spectrum}

In this section we describe the final spectrum of physical scalars of the model.
As we have already mentioned, the theory possesses thirteen scalar degrees
of freedom that are arranged into two complex fields with electric charge $Q=2$,
two more complex fields but with unit electric charge, and five neutral
real fields, of which three are proper scalars and the remaining two are 
pseudoscalars. Let us establish some notation: we will label the fields provided 
by the triplet as
\begin{equation} \label{WWee-model-chi-Q-explicit}
	\chi = \begin{pmatrix}
				\chi^+ / \sqrt{2} & \chi^{+ +}  \\
				\chi^0 & - \chi^+ / \sqrt{2}
		   \end{pmatrix} \, ,
\end{equation}
whereas those present in the doublet we will call
\begin{equation} \label{WWee-model-phi-Q-explicit}
	\phi = \begin{pmatrix}
				\phi^+ \\
				\phi^0
	       \end{pmatrix} \, .
\end{equation}
In addition, for the purposes of this section we will display explicitly the 
electric charge of the singlet $\kappa$, namely by writing $\kappa^{+ +}$. The 
neutral components of the various fields, moreover, are separated into
their scalar and pseudoscalar components; besides, they 
are allowed to take VEV's. We define all these elements as follows:
\begin{align} \label{WWee-model-neutral-and-vevs}
	\phi^0 &= v_\phi + \frac{1}{\sqrt{2}} \, \left( \phi_s + i \, \phi_p \right) ,
	&
	\chi^0 &= v_\chi + \frac{1}{\sqrt{2}} \, \left( \chi_s + i \, \chi_p \right) ,
	&
	\sigma &= v_\sigma + \sigma_s,
\end{align}
with an ``$s$'' subscript representing ``scalar'' and a ``$p$'' subscript 
representing ``pseudoscalar''.

After electroweak symmetry breaking fields with equal electric charge
and parity mix among themselves; this happens irrespective of
their $SU(2)$ lineage and so the scalars are sorted into four groups, each
of them endowed with its own mass and mixing matrix. These matrices can
easily be obtained by substituting equations \eqref{WWee-model-chi-Q-explicit},
\eqref{WWee-model-phi-Q-explicit} and \eqref{WWee-model-neutral-and-vevs}
into \eqref{WWee-model-scalar-potential}. If we arrange the fields as
\begin{multline*}
	\mathcal{L}_{\mathrm{mass}}  =  - \begin{pmatrix}
					\kappa^{- -} & \chi^{- -} 
			\end{pmatrix} M_{+ +}^2 \begin{pmatrix}
					\kappa^{++} \\
					\chi^{++}
			\end{pmatrix} - \begin{pmatrix}
				\phi^- & \chi^-
			\end{pmatrix} M_{+}^2 \begin{pmatrix}
				\phi^{+} \\
				\chi^{+}
			\end{pmatrix} -
	\\
	- \frac{1}{2} \begin{pmatrix}
					\phi_s & \chi_s & \sigma_s 
			\end{pmatrix} M_s^2 \begin{pmatrix}
					\phi_s \\
					\chi_s \\
					\sigma_s
			\end{pmatrix} - \frac{1}{2} \begin{pmatrix}
				\phi_p & \chi_p
			\end{pmatrix} M_p^2 \begin{pmatrix}
				\phi_p \\
				\chi_p
			\end{pmatrix} \, ,
\end{multline*}
then the mass and mixing matrices\footnote{Beware
that the ``2'' superindex is part of the `proper name'
of the matrix, and does not at all represent a matricial product. The $M^2$ 
matrices are not ``mass matrices'', but ``squared-mass matrices''.}
take the following form:
\begin{equation} \label{WWee-model-doubly-charged-mass-matrix}
	M_{+ +}^2 = \begin{pmatrix}
			\bar m_\kappa^2  &  0   
			\\
			0   & \bar m_\chi^2 + \lambda_{\phi \chi}^\prime \, v_\phi^2 
		 \end{pmatrix} 
		 + \; \frac{v_\chi}{v_\phi} \; \begin{pmatrix}
		 	0  & 2 \mu_\kappa \, v_\phi
			\\
			2 \mu_\kappa \, v_\phi & 0
		 \end{pmatrix}
		 + \mathcal{O} \left( \nicefrac{v_\chi^2}{v_\phi^2} \right) \, ,
		 \vphantom{\raisebox{-3ex}{a}} 
\end{equation}
\begin{equation} \label{WWee-model-singly-charged-mass-matrix}
	M_{+}^2 \simeq \begin{pmatrix}
					0   &   0
					\\
					0   &   \bar m_\chi^2
				\end{pmatrix}  
				+ \; \frac{v_\chi}{v_\phi} \; \begin{pmatrix}
					0  &  - \sqrt{2} \, \bar m_\chi^2 
					\\
					- \sqrt{2} \, \bar m_\chi^2   &  0
				\end{pmatrix}
			+ \mathcal{O} \left( \nicefrac{v_\chi^2}{v_\phi^2} \right) \, ,
		 		\vphantom{\raisebox{-3.5ex}{a}} 
\end{equation}
\begin{equation} \label{WWee-model-scalars-mass-matrix}
	M_s^2 = \begin{pmatrix}
			4 \lambda_\phi \, v_\phi^2  &  0 
			&  2 \sqrt{2} \, \lambda_{\phi \sigma} \, v_\phi v_\sigma
			\\
			0  &  \bar m_\chi^2
			&  - \sqrt{2} \, \lambda_6 \, v_\phi^2
			\vphantom{\raisebox{2ex}{a}} \vphantom{\raisebox{-2ex}{a}} 
			\\
			2 \sqrt{2} \, \lambda_{\phi \sigma} \, v_\phi v_\sigma
			&  - \sqrt{2} \, \lambda_6 \, v_\phi^2
			&  8 \lambda_\sigma \, v_\sigma^2
		 \end{pmatrix} 
		 + \mathcal{O} \left( \nicefrac{v_\chi}{v_\phi} \right) \, ,
\end{equation}
\begin{equation} \label{WWee-model-pseudoscalars-mass-matrix}
	M_p^2 = \begin{pmatrix}
		 	0   &   0
			\\
			0  &  \bar m_\chi^2
		 \end{pmatrix} 
		 + \; \frac{v_\chi}{v_\phi} \; \begin{pmatrix}
		 	0  &  -2 \, \bar m_\chi^2
			\\
			-2 \, \bar m_\chi^2  &  0
		 \end{pmatrix}
		 + \mathcal{O} \left( \nicefrac{v_\chi^2}{v_\phi^2} \right) \, ,
		   \vphantom{\raisebox{3.5ex}{a}} 
\end{equation}
where we used the notation $\bar m_j^2 \equiv m_j^2 + \lambda_{\phi j} \, 
v_\phi^2 + \lambda_{\sigma j} \, v_\sigma^2$ bearing in mind that in the most 
interesting regions of the
parameter space $m_j^2$ will be the dominant term in such expressions. We
have simplified the output by using \eqref{WWee-model-vevs-minimum}
to eliminate the doublet and singlet squared masses in favour of the somewhat
more physical $v_\phi$ and $v_\sigma$.
We have also separated explicitly the order-0 and order-1 terms in powers of
$\nicefrac{v_\chi}{v_\phi}$, the small parameter of the model; this separation
is significant as far as the various $\lambda$'s are of the same order and
don't introduce a new low scale in the model. Finally, we have 
neglected in the expression of $M_+^2$ several terms of order 
$\nicefrac{v_\phi^2}{\bar m_\chi^2}$.

By looking at the mass matrices (\ref{WWee-model-doubly-charged-mass-matrix}
-- \ref{WWee-model-pseudoscalars-mass-matrix}) we note several important
features: first, the mixing of the two doubly-charged scalars, dominated 
by the cubic dimensionful coupling $\mu_\kappa$: unless $\mu_\kappa$ is
much above $m_\kappa$ and \mbox{$m_\chi$ --and 
this} may not be natural, see sec\-\mbox{tion 
\ref{sec:WWee-model-unitarity-naturality}--, the} 
mixing is bound to be small, and so one of the physical doubly-charged
states will be mainly $\kappa$, and the other mainly $\chi^{+ +}$. Secondly,
the somewhat seesaw-like structure of $M_{+}^2$ and $M_p^2$, particularly realised 
if $m_\chi \gg v_\phi, v_\sigma$; this is consistent with the fact that 
these two matrices must yield one zero eigenvalue each, for to those
eigenvalues will correspond the two would-be Goldstone states 
that have to provide the longitudinal degree of freedom for the $W$ and $Z$
bosons. Actually, as we anticipated in section \ref{sec:WWee-model-vevs}, to 
(\ref{WWee-model-doubly-charged-mass-matrix}
-- \ref{WWee-model-pseudoscalars-mass-matrix}) we still have to require that they
only yield nonnegative eigenvalues, in order for \eqref{WWee-model-vevs-minimum}
to represent a true minimum of the potential. These conditions can be
easily implemented in the mass matrices, and we note that they are particularly
easy to implement in the limit of large $m_\chi$.
Indeed, this limit seems especially convenient: it both ensures a 
phenomenologically viable triplet VEV (see the discussion at the end of 
section \ref{sec:WWee-model-vevs})
and greatly simplifies the scalar spectrum, by
fixing one definite-mass state in each group to an almost-triplet condition.
As we have already stated, we will not restrict our discussion to this 
sce\-\mbox{nario --for it} would be appealing to have the triplet scalars 
at the reach of \mbox{the LHC--, but} we will keep it in mind for the sake of 
having at hand a simple picture of the scalar panorama.

Moving on to a more detailed discussion, let us fix some notation for the
definite-mass scalar states. We will call $\kappa_1$ and $\kappa_2$
the two massive doubly-charged states; they are the result of a certain
mixture of $\kappa$ and $\chi^{+ +}$,
\begin{align}
	\kappa_1 &= \cos \theta_{+ +} \: \kappa^{+ +} + \sin \theta_{+ +} \: \chi^{++} 
	\nonumber \\
	\kappa_2 &= -\sin \theta_{+ +} \: \kappa^{+ +} + \cos \theta_{+ +} \: \chi^{++} 
	\, , 
	\label{WWee-model-doubly-charged-mass-eigenfields}
\end{align}
with $\theta_{+ +}$ given by
\begin{displaymath}
	\tan 2\theta_{+ +} \simeq \frac{4 \, \mu_\kappa v_\chi}{m_\chi^2 - 
			m_\kappa^2 + \lambda_{\phi \chi}^\prime \, v_\phi^2} \, .
\end{displaymath}
Note that this mixing is small for almost the entire parameter space: the 
denominator is at least of the order of the electroweak scale squared, while the
numerator contains $v_\chi$, which is smaller by a factor 100, and $\mu_\kappa$,
that can be large, but if it's much heavier than $m_\chi$ or $m_\kappa$ causes
the theory to lose naturality in loop corrections (see section 
\ref{sec:WWee-model-unitarity-naturality} for more on this matter). 
In conclusion, $\kappa_1$ is mainly composed of $\kappa$ for most of the
interesting scenarios, whereas $\kappa_2$ is essentially $\chi^{+ +}$; 
consequently, $\kappa_1$ will prefer singlet-like couplings and will have
small triplet-like couplings, while the opposite will be true for $\kappa_2$.
See section \ref{sec:WWee-model-couplings-pheno} for specific examples of 
this feature.

We turn now to the pseudoscalar sector. The matrix 
\eqref{WWee-model-pseudoscalars-mass-matrix} describes the masses and
mixings for two real pseudoscalar fields; one of them must be massless,
as the would-be Goldstone boson that provides the 
longitudinal part of the $Z$ is a pseudoscalar. 
One can check%
\footnote{This check requires the pieces of $\mathcal{O} \left( 
\nicefrac{v_\chi^2}{v_\phi^2} \right)$ in 
\eqref{WWee-model-pseudoscalars-mass-matrix}, as that is the order of
the independent term in the characteristic equation of $M_p^2$.
}
that this immediately happens if the fields are shifted to
the minimum of the potential, given by \eqref{WWee-model-vevs-minimum}.
Moreover, the would-be 
Goldstone must be mainly a doublet, as argued in section 
\ref{sec:WWee-model-vevs}, and indeed this is the case: let us define
the definite-mass pseudoscalar fields as
\begin{align*}
	G^0 &= \cos \theta_p \: \phi_p + \sin \theta_p \: \chi_p
	\\
	A &= - \sin \theta_p \: \phi_p + \cos \theta_p \: \chi_p \, ,
\end{align*}
where $G^0$ is the would-be Goldstone, $A$ the remaining physical 
pseudoscalar, and $\theta_p$ the mixing angle between the doublet contribution,
$\phi_p$, and the triplet one, $\chi_p$. This angle has to be small, as the
electroweak symmetry is broken mainly by the doublet and $G^0$ must be
mainly $\phi_p$. And that's what happens, as can be seen rather easily
from \eqref{WWee-model-pseudoscalars-mass-matrix}:
\begin{displaymath}
	\tan 2 \theta_p \simeq - 4 \, \frac{v_\chi}{v_\phi} \, .
\end{displaymath}
It is sort of elegant that everything can be 
rewritten in a form that involves only symmetry-breaking-related parameters:
as soon as $v_\chi$ is small compared to $v_\phi$, $G^0$ will be indeed 
mainly doublet.

If we look at the singly-charged scalars we find a similar situation to that
of the pseudoscalars: the matrix \eqref{WWee-model-singly-charged-mass-matrix}
must have one zero eigenvalue corresponding to the would-be Goldstone that 
provides the longitudinal part \mbox{of $W$ -- note} that this time the would-be
Goldstone is a complex field, as $W$ is. 
The would-be Goldstone, again, has to be essentially a doublet, and we can
check that this is precisely the case. Expressing the definite-mass fields as
\begin{align}
	G^+ &= \cos \theta_{+} \: \phi^{+} + \sin \theta_{+} \: \chi^{+}
	\nonumber \\	
	\omega^+ &= - \sin \theta_{+} \: \phi^{+} + \cos \theta_{+} \: \chi^{+} \, ,
	\label{WWee-model-singly-charged-mass-eigenfields}
\end{align}
where $G^+$ is the would-be Goldstone and $\omega^{+}$ is the remaining 
physical singly-charged scalar, the mixing angle can be calculated as
\begin{displaymath}
	\tan 2 \theta_{+} \simeq -2 \sqrt{2} \: \frac{v_\chi}{v_\phi}  \, ,
\end{displaymath}
and we observe again that in the limit of $v_\chi \ll v_\phi$ the Goldstone
will be mainly a doublet, as expected.

As for the neutral scalar sector of the model, their mass matrix is not so 
constrained. One could obtain a few more constraints from requiring that all
three eigenvalues are positive, but as our main goal is not to thoroughly
characterise the scalar parameters we will avoid that discussion. We will
just say that by looking at \eqref{WWee-model-scalars-mass-matrix} we see
that in the limit of large $m_\chi$ one of the physical states is mainly a
triplet, as expected, with mass approximately $m_\chi$. The other two
neutral scalars depend greatly on the balance of the relevant parameters:
$v_\phi$, $v_\sigma$ and the quartic couplings involved. If $v_\phi$ and
$v_\sigma$ are of the same order and $\lambda_{\phi \sigma}$ is not very
small, we can expect $\phi_s$ and $\sigma_s$ to be pretty much mixed
as part of at least two of the physical neutral scalars of the model.

\subsection{Some couplings of phenomenological interest}
							\label{sec:WWee-model-couplings-pheno}

In this section we list some couplings that will be of interest during
the remainder of our discussion. To begin with, it can be instructive
to express the $\kappa$ Yukawas to right-handed electrons in terms of
the mass eigenfields:
\begin{equation} \label{WWee-model-yukawas-kappa1-kappa2}
	\overline{e_{\mathrm{R}}^\mathrm{c}} \, g \, e_{\mathrm{R}} \, \kappa = 
	\overline{e_{\mathrm{R}}^\mathrm{c}} \, g \, e_{\mathrm{R}} \,
	\left(\cos \theta_{+ +} \: \kappa_1 - \sin \theta_{+ +} \: \kappa_2 \right) \, ,
\end{equation}
checking indeed that \mbox{$\kappa_1$ --mainly} \mbox{singlet-- retains} most of 
the coupling, while $\kappa_2$, mainly triplet, has this interaction suppressed
by the sine of the small angle $\theta_{+ +}$.

Also important for $0 \nu \beta \beta$ and neutrino masses will be the gauge
interaction of the doubly-charged component of the triplet with two $W$'s,
namely
\begin{equation} \label{WWee-model-gauge-couplings-kappa1-kappa2}
	g^2 \, W_\mu W^\mu \, {\chi^0}^\dagger \chi^{++} = 
		g^2 v_\chi \, W_\mu W^\mu \left( \sin \theta_{+ +} \: \kappa_1 +
		\cos \theta_{+ +} \: \kappa_2 \right) + \ldots \, ,
\end{equation}
which presents the opposite feature: $\kappa_1$, mainy singlet, has this 
triplet-like interaction suppressed by $\sin \theta_{+ +}$. The dots above
represent the terms that involve $\chi_s$ and $\chi_p$, which we will not use.

We present also explicitly here some trilinear couplings that we will use at
some point:
\begin{displaymath}
	- \mu_\kappa \, \kappa^{++} \chi^- \chi^- - 2 \mu_\kappa \, 
			\kappa^{++} \chi^{--} {\chi^0}^\dagger + \lambda_6 \, \sigma \, 
			\chi^{++} \phi^- \phi^- + \mathrm{H.c.} \, ;
\end{displaymath}
of course, these couplings can also be expressed in terms of VEV's and mass 
eigenfields by using equations \eqref{WWee-model-neutral-and-vevs}, 
\eqref{WWee-model-doubly-charged-mass-eigenfields} and
\eqref{WWee-model-singly-charged-mass-eigenfields}.

Finally, let us write explicitly the Yukawas that change charge and chirality:
\begin{displaymath}
	\overline{\nu_{\mathrm{L}}} \, Y_e \, e_{\mathrm{R}} \, \phi^{+} = 
			\overline{\nu_{\mathrm{L}}} \, Y_e \, e_{\mathrm{R}} \,
			\left( \cos \theta_{+} \: G^{+} - \sin \theta_{+} \: \omega^{+} \right)
			\, .
\end{displaymath}

\section{Neutrinoless double beta decay} \label{sec:WWee-model-0nu2beta}

$0 \nu \beta \beta$ through the $WWee$ mechanism requires couplings to 
right-handed electrons and to $W$ bosons; in this model, the first are provided
by the singlet $\kappa$ and the second by the doubly-charged component of the
triplet. As the two of them mix new massive states arise, $\kappa_1$ and 
$\kappa_2$, which participate of both kinds of interactions, as described in sections
\ref{sec:WWee-model-scalar-spectrum} and \ref{sec:WWee-model-couplings-pheno}. 
These particles are the necessary mediators for $0 \nu \beta \beta$.
In this section we calculate their contribution to the process of double beta 
decay and set the appropriate constraints upon the involved parameters to fit 
the current experimental bounds; we also deduce the range in which the relevant 
parameters should lie if $0 \nu \beta \beta$ is to be observed in the next round
of experimental searches.

We can deduce the model contribution to $0 \nu \beta \beta$ even before
thinking about the particular physical process that generates it. Assuming that 
$m_\kappa, m_\chi \gg v_\phi, v_\sigma$ and
integrating out the heavy $\kappa$ and $\chi$ modes we find, after 
a straightforward calculation, that the effective Lagrangian contains the term
\begin{equation} \label{eq:derived.O9}
\mathcal{L}_9 = 4 \: \frac{(\lambda_6 v_\sigma)^2 \, \mu_\kappa}
				{m_\kappa^2 m_\chi^6} \, \left(\overline{e_{\mathrm{R} \alpha}}
				\, g_{\alpha \beta}^* \, e_{\mathrm{R} \beta}^\mathrm{c} \right)
				\left(\phi^\dagger D^\mu \tilde{\phi} \right)
				\left(\phi^\dagger D_\mu \tilde{\phi} \right)
				+ \mathrm{H.c.} \, ,
\end{equation}
as expected, because the model contains the necessary ingredients to generate
the $\mathcal{O} \up 9$ operator (see section \ref{sec:WWee-model-presentation}
for more details). 
The explicit $WWee$ vertex is revealed once the electroweak symmetry breaks
spontaneously and looks
\begin{equation} \label{eq:Lagrangian-WWee}
\mathcal{L}_9 = - 2 g^2 \: \frac{\mu_\kappa v_\chi^2}
				{m_\kappa^2 m_\chi^2} \, 
				\overline{e_{\mathrm{R} \alpha}} \, g_{\alpha \beta}^* \, 
				e_{\mathrm{R} \beta}^\mathrm{c} \: W^\mu W_\mu + 
				\mathrm{H.c.} + \ldots \, ,
\end{equation}
where we have used the expression for $v_\chi$ in the limit of large $m_\chi$,
$v_\chi \simeq \lambda_6 v_\sigma v_\phi^2 / m_\chi^2$, derived from
\eqref{WWee-model-vevs-minimum}. It must be emphasised that in this model 
the interactions violating LN in two units, equation 
\eqref{eq:derived.O9}, happen to be proportional to $v_\chi^2$;
this is qualitatively different from other \mbox{models --for} example, seesaw 
\mbox{type II-- where} the LNV-ing interactions are linear in $v_\chi$, indicating
that a different mechanism is in action here. This is a consequence of having
chosen carefully our fundamental vertices so that $\chi$ cannot be assigned lepton
number 2, thus qualitatively separating our model from those that generate
$\mathcal{O} \up 5$ at tree level.

\begin{figure}[tb]
	\centering
	\includegraphics[width=0.75\columnwidth]{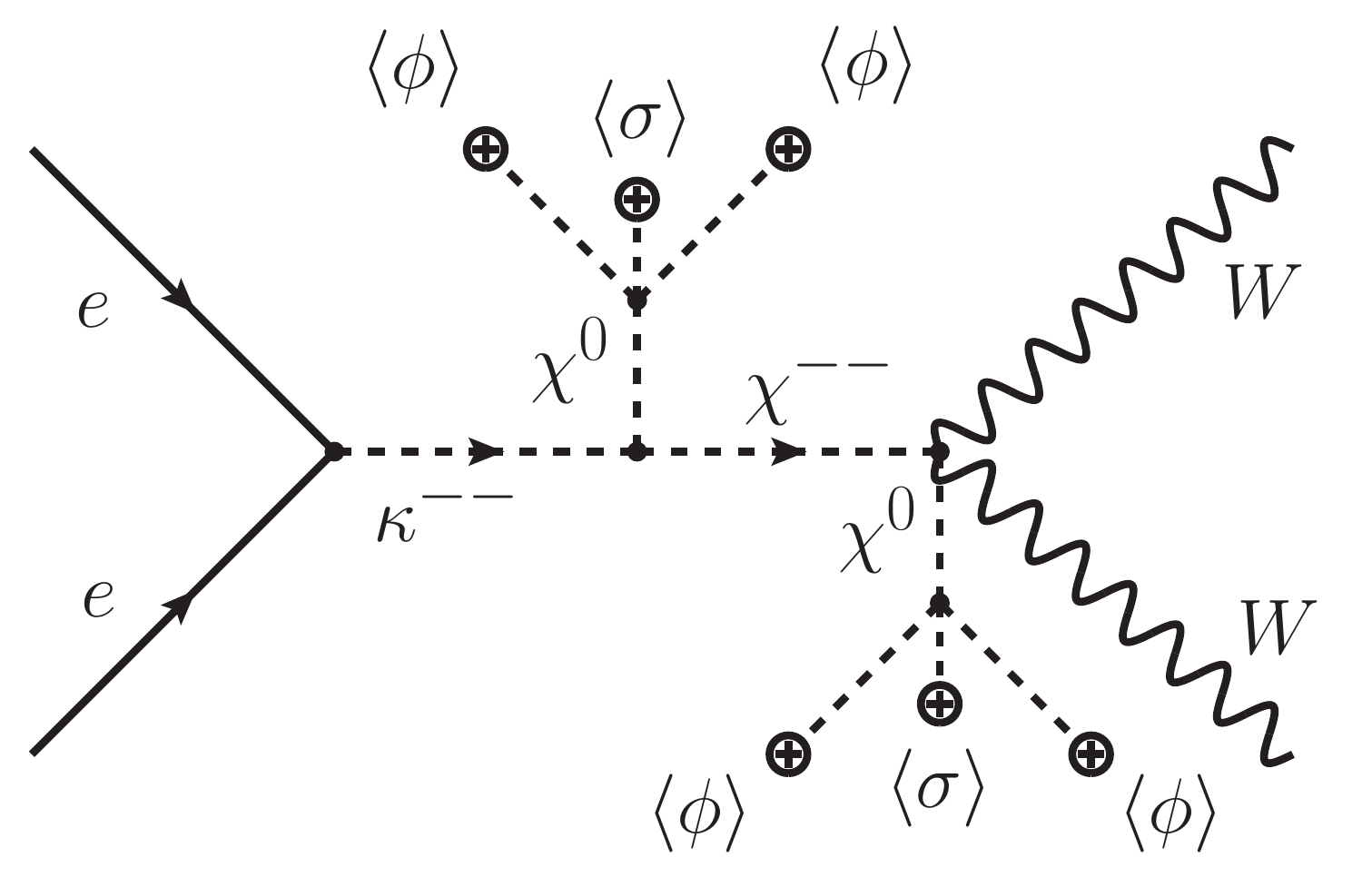}
	\caption{Dominant tree-level diagram contributing to the effective 
			neutrinoless double beta decay operator.
			} \label{fig:WWee-model-0nu2beta}
\end{figure}

One can better understand  the form of \eqref{eq:derived.O9} by
considering the physical process that gives rise to it. The dominant diagram is
shown in figure \ref{fig:WWee-model-0nu2beta}, where the different couplings 
and VEV's involved are displayed explicitly. We note the importance of the mixing 
between $\kappa$ and $\chi^{+ +}$, represented by the presence of $\mu_\kappa$; 
we also note that two VEV's of $\chi$ are present, but they can be expressed either 
as such, like in \eqref{eq:Lagrangian-WWee}, or as combinations of $v_\phi$ and 
$v_\sigma$ with $\lambda_6$ insertions, as in \eqref{eq:derived.O9}.
This same process could be displayed in terms of the physical massive states 
$\kappa_1$ and $\kappa_2$ by using their couplings 
\eqref{WWee-model-yukawas-kappa1-kappa2} and 
\eqref{WWee-model-gauge-couplings-kappa1-kappa2}; we would see then an apparently 
much cleaner diagram, with just $\kappa_1$ or $\kappa_2$ mediating between the 
pair of electrons and the pair of $W$'s. In this view many of the couplings and 
VEV's that we see in \eqref{eq:derived.O9} would be codified inside the 
$\theta_{+ +}$ mixing; to avoid this `hiding' we prefer the pre-EWSB vision
of figure \ref{fig:WWee-model-0nu2beta}, to which we will refer when necessary.

Let us now derive bounds on the model parameters from the experimental limits
on $0 \nu \beta \beta$. For that aim we will use the very general analysis 
carried out in \cite{Pas:2000vn}, where the authors used detailed nuclear 
matrix element calculations to set bounds on all the six-fermion interactions
that can provide $0 \nu \beta \beta$. In our model,
the interaction \eqref{eq:Lagrangian-WWee} leads to a six-fermion vertex
of the form
\begin{equation} \label{WWee-model-0nu2beta-six-fermion}
\mathcal{L}_{0 \nu \beta \beta} = \frac{G_F^2}{2m_p} \: \epsilon_3 \: 
		\left[ \bar{u} \gamma^\mu (1 - \gamma_5) d \right] 
		\left[ \bar{u} \gamma_\mu (1 - \gamma_5) d \right] \:
		\bar{e} (1 - \gamma_5) e^\mathrm{c} \, ,
\end{equation}
where $m_p$ denotes the proton mass and $\epsilon_3$, which corresponds to
\cite{Pas:2000vn}'s notation, reads
\begin{displaymath} 
	\epsilon_3 = - 8 \, \frac{m_p \mu_\kappa v_\chi^2}{m_\kappa^2 m_\chi^2} 
				\: g^*_{ee} \, .
\end{displaymath}
All that is left, so, is to read from \cite{Pas:2000vn} the bound that applies to
operators like \eqref{WWee-model-0nu2beta-six-fermion}; it is just
$\epsilon_3 < 1.4 \times 10^{-8}$ at 90\% CL%
\footnote{Actually
there is a misprint in \cite{Pas:2000vn}. We are very grateful to the
authors for providing us with the correct result.
}.
But we will want something else: we would like the model to yield a signal in
the next round of $0 \nu \beta \beta$ experiments. In order to implement this
requirement we turn again to \cite{Pas:2000vn} and see that their bounds are
derived from the Heidelberg-Moscow limit%
\footnote{This experiment remained for more than a decade the most sensitive one
to the effective neutrino Majorana mass, $m_{\beta \beta}$, but has been
recently superseded by the EXO bound for $\tensor[^{136}]{\mathrm{Xe}}{}$ 
\cite{Auger:2012ar} and the GERDA bound for $\tensor[^{76}]{\mathrm{Ge}}{}$
\cite{Agostini:2013mzu}. We, nevertheless, stick to the numbers in \cite{Pas:2000vn}.
},
$T_{1/2} > 1.9 \times 10^{25} \, \mathrm{years}$ for ${}^{76}\mathrm{Ge}$ 
\cite{KlapdorKleingrothaus:2000sn}. These numbers are reported to be improved 
in the next years up to
lifetimes of $6 \times 10^{27}$ years \cite{Barabash:2011fg}, that is to say, an 
increase in a factor of 20 in sensitivity. Gathering together all these ideas 
we obtain the following requirement for the parameters of our model:
\begin{equation} \label{WWee-model-0nu2beta-bounds}
	8.75 \times 10^{-11} \, \stackrel{\mathrm{{\scriptstyle Next}}}{<} \,
			\frac{m_p \mu_\kappa v_\chi^2}{m_\kappa^2 m_\chi^2} \; 
			|g_{ee}| \, < \, 1.75 \times 10^{-9} \quad (90\% \,\mathrm{CL}) \,,
\end{equation}
where the upper bound corresponds to the Heidelberg-Moscow experimental limit,
whereas the ``Next'' superindex in the lower bound indicates that it's not
an experimental requirement, but only the limit in sensitivity to be attained
by the next round of experiments. These conditions will prove to be
rather restrictive because their range of variation is relatively narrow;
see section \ref{sec:WWee-model-constraints-parameters} for a discussion of
the constraints that this and other phenomenological requirements set on
the new particles of the model.

\section{Lepton flavour violation constraints}
						\label{sec:WWee-model-lfv}

Lepton flavour violation constitutes a very relevant probe to test the 
phenomenological viability of the model. The $g$ Yukawas introduce a new
flavour hierarchy, independent of that of the Higgs Yukawas $Y_e$, and
so they can induce all sorts of LFV-ing processes. Here we will look into the 
most restrictive of them: the tree-level-mediated three-body decays of the 
form $\ell_\alpha^- \rightarrow \, \ell_\beta^+ \ell_\gamma^- \ell_\eta^-$.
Other LFV-ing processes yield milder constraints: $\ell_\alpha^- \rightarrow
\ell_\beta^- \gamma$ proceeds at one loop and the bound on the model 
parameters is weaker due to the loop factor; similarly happens for
$\mu - e$ conversion in nuclei; the bounds for $\mu^+ e^- \leftrightarrow
\mu^- e^+$ (muonium-antimuonium conversion), though tree-level, are
also less restrictive%
\footnote{For 
a more comprehensive discussion of LFV mediated by doubly-charged scalar
singlets the reader might find interesting \cite{Nebot:2007bc,Raidal:1997hq}, 
where models with similar features are considered.
}.

\begin{figure}[tb]
	\centering
	\includegraphics[width=0.5\textwidth]{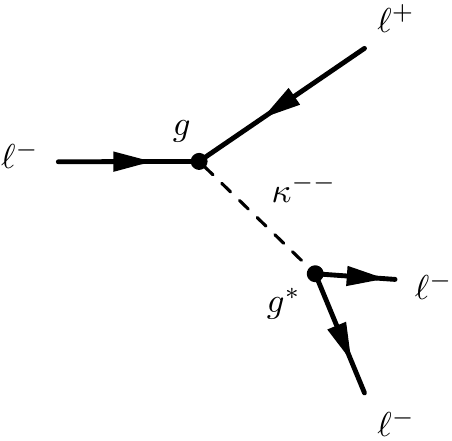}
	\caption{The tree-level diagram that yields LFV-ing three-body decays through
			the $g$ Yukawas.} \label{fig:diagram-l-3l}
\end{figure}

So, let us consider the three-body leptonic decays $\ell_\alpha^- \rightarrow 
\, \ell_\beta^+ \ell_\gamma^- \ell_\eta^-$; in our model these are mediated
by the doubly-charged singlet $\kappa$, as depicted in figure 
\ref{fig:diagram-l-3l}.
Of course, in the mass-eigenstate basis the $\kappa$ couplings get split 
between the two massive eigenfields $\kappa_1$ and $\kappa_2$, as seen in 
\eqref{WWee-model-yukawas-kappa1-kappa2}, and this is reflected in the
expression for the branching ratio of the process,
\begin{equation} \label{WWee-model-BR-l-3l}
	\mathrm{BR} (\ell_\alpha^- \rightarrow \ell_\beta^+ \ell_\gamma^- 
			\ell_\eta^-) = \frac{1}{2 (1 + \delta_{\gamma \eta})} \, 
			\left| \frac{g_{\alpha \beta} g_{\gamma \eta}^*}{G_F \, 
			\tilde{m}_\kappa^2} \right|^2 \: \mathrm{BR} (\ell_\alpha^- 
			\rightarrow \ell_\beta^- \nu \bar{\nu}) \, , 
\end{equation}
where $\tilde m_\kappa$ is a compact way of representing the contribution
of the two mass eigenstates:
\begin{displaymath}
	\frac{1}{\tilde{m}_\kappa^2} \equiv \frac{\cos^2 \theta_{+ +}}
			{m_{\kappa_1}^2} + \frac{\sin^2 \theta_{+ +}}{m_{\kappa_2}^2} \, ,
\end{displaymath}
and $\delta_{\gamma \eta}$ accounts for the statistics of identical particles,
were it the case that $\gamma = \eta$. \mbox{Note --see the} details in 
sec\-\mbox{tions
\ref{sec:WWee-model-scalar-spectrum} and \ref{sec:WWee-model-unitarity-naturality}--
that} if naturality is required then $\theta_{+ +}$ is rather small,
and we will have $\tilde m_\kappa \simeq m_{\kappa_1}$,
recovering the case in which only the \mbox{singlet --or its} direct 
off\-\mbox{spring, $\kappa_1$-- mediates} the process.

Let us now adapt the expression \eqref{WWee-model-BR-l-3l} to the cases that 
will be more
important for our description of the model. The most constrained decay of the
type $\ell \rightarrow 3 \ell$ is $\mu \rightarrow 3 e$; for it we have the
upper bound $\mathrm{BR} (\mu^- \rightarrow e^+ e^- e^-) < 1.0 \times 10^{-12}$ 
\cite{Beringer:1900zz}, which translates into
\begin{equation} \label{WWee-model-muto3e-bound}
	|g_{\mu e} g_{ee}^*| < 2.3 \times 10^{-5} \, (\tilde{m}_{\kappa} / 
			\mathrm{TeV})^2 \, .
\end{equation}
As we will see, for our purposes this constitutes mainly a constraint on 
$g_{\mu e}$ because we want $g_{e e}$ as large as possible in order to 
enhance $0 \nu \beta \beta$.

As it couldn't be but expected, the LFV bounds essentially favour small 
$g$ couplings and large scalar masses. This sets a conflict with other 
phenomenological features of the model, and especially with neutrino masses,
which are rather suppressed and would prefer 
\emph{large} values of the $g$ Yukawas.
Amidst this discus\-\mbox{sion --see} sections 
\ref{sec:WWee-model-neutrino-masses} and 
\ref{sec:WWee-model-constraints-parameters} for more \mbox{details--, 
one par}\-ticular LFV-ing process happens to cast a special tension:
it's $\tau^- \rightarrow e^+ \mu^- \mu^-$, whose experimental branching ratio
is bounded to be $\mathrm{BR} (\tau^- \rightarrow e^+ \mu^- \mu^-) < 
1.7 \times 10^{-8}$ \cite{Beringer:1900zz}. 
This means that for our model we must impose
\begin{equation} \label{WWee-model-tautoe2mu-bound}
	|g_{e \tau} g_{\mu \mu}^*| < 0.007 \, (\tilde m_\kappa / \mathrm{TeV})^2 \, .
\end{equation}

As for other three-body decays, most add interesting constraints or suggest that
the model could be just one step beyond the current experimental sensitivity,
at least if it is to fulfill requirements such as ``yield a signal in the next
round of $0 \nu \beta \beta$ experiments'' or ``leave the new particles at the
reach of the LHC''. As a general statement we can say that the $g$ couplings 
involving only heavy leptons, such as $g_{\mu \mu}$ or $g_{\mu \tau}$, are
required not to be very small due to constraints on the neutrino mass matrix;
were the new particles not very heavy, these processes could be a good probe
of the interactions of the model.

\section{Constraints from naturality and perturbative unitarity}
							\label{sec:WWee-model-unitarity-naturality}

In the discussion of $0 \nu \beta \beta$ carried out in section 
\ref{sec:WWee-model-0nu2beta} we found that its rate in this model
is proportional to the combination of parameters $\mu_\kappa v_\chi^2 g_{e e}$; 
that happened somehow as expected, for all these couplings are needed in order 
to violate lepton number, as argued at the end of section 
\ref{sec:WWee-model-presentation}. We will find equal\-\mbox{ly --see 
sec}\-\mbox{tion \ref{sec:WWee-model-neutrino-masses}-- that neu}\-trino masses
depend on the same combination of parameters. And in both cases we would like
it to be large: for $0 \nu \beta \beta$ we want to know if the rate can be
large enough to yield a signal in near-future experiments; 
for neutrino masses, we want to overcome
the suppression introduced by the charged-lepton Yukawas. Therefore, at this
point the question of how large some couplings can be becomes relevant. It is
well known that very large couplings make a quantum field theory unaccessible
to perturbative ex\-\mbox{ploration -- and so} most of our calculations meaningless.
It is also generally acknowledged that a theory requiring fine tuning in the
process of renormalisation cannot be considered natural; this happens, for
example, if two mass scales related by loop corrections are largely dissimilar.
In this section we will use these considerations to set upper bounds on the
values of the couplings of the model.

Let us first discuss perturbative unitarity. As we know, the scattering matrix
that we commonly use to evaluate physical processes is nothing but a reexpression
of the time-evolution operator, and the unitary character of this operator can be
used to set bounds on the elements of the \mbox{S-matrix -- see, for} example
\cite{Weinberg:1995mt}. But to calculate such elements we
use perturbation theory, so these bounds can be better expressed as ``unitarity
bounds as long as perturbation theory is valid''. We can use those bounds
for our case. Consider the electron-electron scattering mediated by the 
doubly-charged sin\-\mbox{glet $\kappa$ -- or more} precisely by the massive
states that participate from it, $\kappa_1$ and $\kappa_2$. At high-energy,
tree-level unitarity requires
$| g_{e e} | < \sqrt{4 \pi}$ \cite{Lee:1977yc,Lee:1977eg,Chanowitz:1978mv}. 
Exactly the same is obtained 
for the analogous processes involving other leptons. In conclusion, we won't 
be considering values of the $g$ Yukawas above $\sqrt{4 \pi}$,
\begin{equation} \label{WWee-model-g-bound-unitarity}
	|g_{\alpha \beta}| \stackrel{\mathrm{{\scriptstyle Unit}}}{<} \sqrt{4 \pi} \, .
\end{equation}

Tree-level unitarity at high energy does not give useful information on dimensionful
parameters like $\mu_\kappa$ because amplitudes involving these couplings 
decrease with energy. To bound this scale we turn to naturality arguments,
noting that the one-loop self-energies of $\kappa$ and $\chi$ involve loops
with two powers of $\mu_\kappa$ which correct $m_\kappa^2$ and $m_\chi^2$.
These diagrams are of course divergent, but after renormalisation yield
corrections that roughly read $\delta m_{\kappa, \chi}^2 \sim \mu_\kappa^2 /
(4 \pi)^2$. For the theory not to be fine-tuned it is enough to require that
the one-loop corrections are below the tree-level contribution,
that is,
\begin{equation} \label{WWee-model-muk-bound-naturality}
	\mu_\kappa \stackrel{\mathrm{{\scriptstyle Nat}}}{<} 4 \pi \, 
			\min (m_{\kappa_1}, m_{\kappa_2}) \, ,
\end{equation}
where we bound with respect to $m_{\kappa_1}, m_{\kappa_2}$ rather than 
$m_\kappa, m_\chi$, as those are the physical masses of the theory. Note that
other alternative requirements for naturality, such as the width of the 
new particles $\kappa_1$ and $\kappa_2$ not being larger than their masses, 
are also met if 
\eqref{WWee-model-g-bound-unitarity} and \eqref{WWee-model-muk-bound-naturality} 
hold.

One must be aware, however, that all these limits are estimates 
which depend on what is our idea of a `natural theory'. 
Thus, although at the price of fine tuning, one might
decide to fix the model parameters outside the range defined by these limits; 
moreover, there can be model extensions where those values are natural.  At any rate, 
we use equations \eqref{WWee-model-g-bound-unitarity} and 
\eqref{WWee-model-muk-bound-naturality} in the text 
to illustrate that the allowed regions in parameter space 
are at a large extent bounded if the perturbative theory must stay 
natural.

\section{A very constrained neutrino spectrum}
						\label{sec:WWee-model-neutrino-masses}

\begin{figure}[p]
	\centering
	\raisebox{3.5cm}{\emph{a)}} 
	\includegraphics[width=0.6\textwidth]{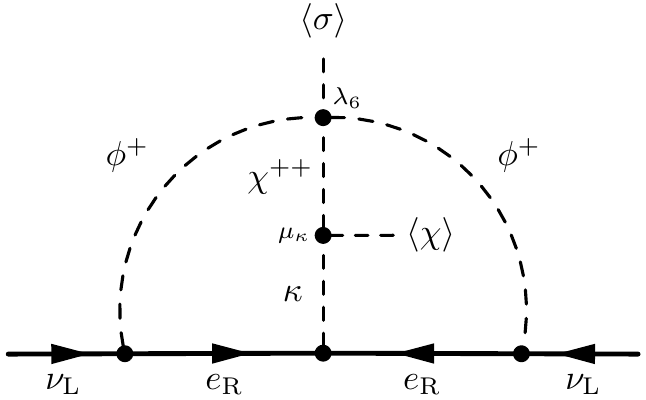}

	\vspace{0.5cm}
	\raisebox{3.5cm}{\emph{b)}} 
	\includegraphics[width=0.6\textwidth]{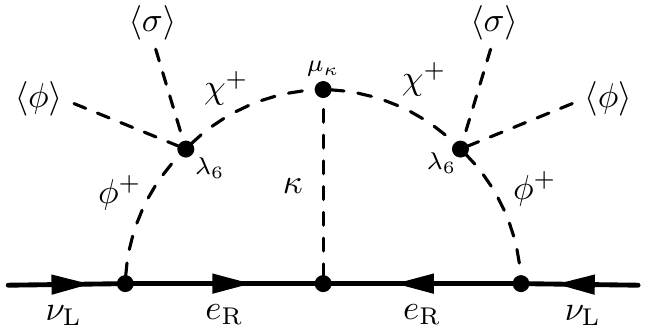}

	\raisebox{5cm}{\emph{c)}} 
	\includegraphics[width=0.6\textwidth]{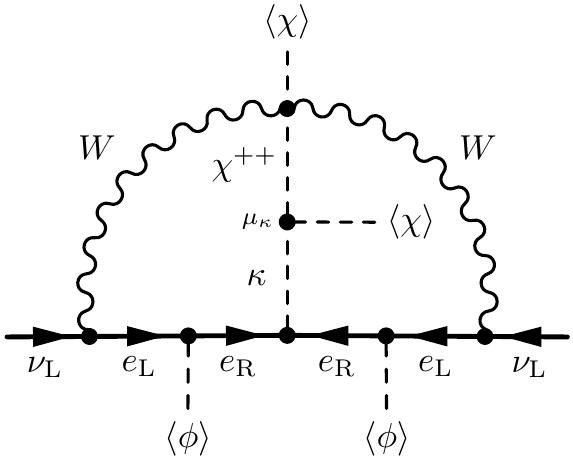}

	\caption{Three of the two-loop processes that provide neutrino masses in
			this model, displayed in the mass insertion approach. The reader
			can check that the complete list includes many more, mainly
			combining several sorts of $W - \phi^+$, $W - \chi^+$ or 
			$\phi^+ - \chi^+$ mixing in the scalar legs. However, among all
			of them, \emph{a)} and \emph{b)} represent the dominant
			contributions.
			} \label{fig:WWee-model-numass-3diag}
\end{figure}

Lepton number is violated in this model through the conjoint action of the
$g$ Yukawas and the $\mu_\kappa$ and $\lambda_6$ scalar interactions; 
the violation affects at this level the leptonic right-handed singlets,
and that's why $0 \nu \beta \beta$ proceeds mainly to two $e_{\mathrm{R}}$'s.
But the breaking is right away transmitted to the left-handed doublets 
through the charged-lepton Yukawas $Y_e$, and once this happens nothing
can help the left-handed neutrinos from developing Majorana masses.
This occurs at the two-loop level, as was expected for the
$WWee$ mechanism (see section \ref{sec:0nu2beta-eff-Violating-LN} for
more on the rationale of the mechanism), and the generated masses are finite
and calculable, as none of the mass-generating interactions appear 
at lower orders. The complete picture of neutrino mass generation,
however, is complicated: many interactions are involved, and the nontrivial
vacuum structure of the model yields a variety of vertices that connect
the several degrees of freedom of the theory. In order to illustrate
this situation we present in figure \ref{fig:WWee-model-numass-3diag} three of
the diagrams that provide neutrino masses; they are presented in the mass
insertion picture so that the role played by the vacuum expectation values
can be explicitly identified. As we see, the VEV's enter the diagrams in a 
variety of ways, and the $SU(2)$ charges of the involved degrees of freedom
are widely dissimilar. It would appear, at first glance, that these processes
represent very different contributions to neutrino masses. To make the 
situation even more convoluted, figure \ref{fig:WWee-model-numass-3diag}
hardly represents a small fraction of the total amount of contributing
diagrams: the several mixing terms between $W$, $\phi^+$ and $\chi^+$ 
yield a fauna of mixed graphs with varying coupling content and suppression
factors. All in all, diagrams \emph{a)} and \emph{b)} represent the
leading contributions in this approach, but it's hard to estimate the
importance of the remaining graphs, surely subdominant but also
much more numerous.

Fortunately, there's more than differences between the graphs presented in
figure \ref{fig:WWee-model-numass-3diag}; a number of common features
can be observed that will prove to be revealing:
the loop topology, for instance, is the same for all of them; the electric charge of
the particles in each line remains the same; and most important: after
maybe some reworking by using equations \eqref{WWee-model-vevs-minimum},
\emph{all} the diagrams are shown to be proportional 
\mbox{to $\mu_\kappa v_\chi^2 Y_e^2
g$ -- where} $g$ represents not the weak gauge coupling but one of
the $\kappa$ Yukawas. 
All these are features which suggest that a simpler picture must exist for
the generation of neutrino masses. Most likely one can find a basis for the 
bosons of the theory where this complicated set of interactions and mixings is
encoded inside the couplings of just a few relevant degrees of freedom. 
A way to explore the possible rearrangements of the bosonic sector is
by choosing a particular gauge. One could think initially that the unitary
gauge is a good candidate: after all, all the fields there represent 
physical excitations with definite mass, and only two singly-charged bosons
are present: the massive $W$ and the scalar $\omega$. However, in this gauge
other type of difficulties arise: the $W$ propagators happen to render some 
quadratically-divergent pieces in the loops, and we know they must cancel
because the neutrino masses are finite, but this makes the actual calculation 
sort of tricky and cumbersome. This is why we looked for a different gauge which
suceeds in simplifying the mass generation picture without yielding
further complications. Fortunately, such a gauge does indeed exist.

\subsection{Calculation of the two-loop neutrino masses}

\begin{figure}[tb]
	\centering

	\includegraphics[width=0.6\textwidth]{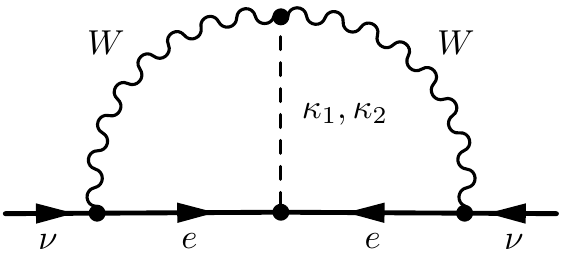}
	
	\vspace{0.5cm}
	\includegraphics[width=0.6\textwidth]{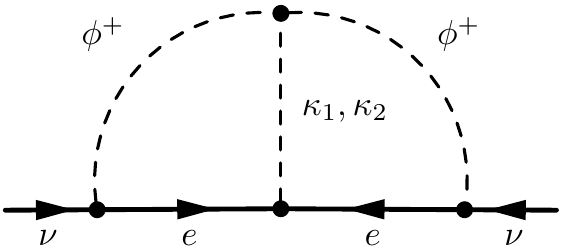}

	\caption{The two diagrams that yield neutrino masses at two loops in the
			gauge with no $W - \phi$, $W - \chi$ and $\chi - \phi$ mixing.
			} \label{fig:WWee-model-numass-WudkaGauge}
\end{figure}

The difficulty that we want to address with the gauge choice is the large number
of mass-generating diagrams, which is mainly due to the extensive mixing between
the bosonic fields, both the scalars and the gauge bosons. Therefore, we
seek a gauge which cancels most of the mixings between the various
singly-charged degrees of
freedom; that can be achieved by a gauge fixing of the form
\begin{displaymath}
	\mathcal{L}_{\mathrm{g.f.}} = a \, \left( \partial^{\, \mu} \, W_\mu^\dagger + 
			b \, \chi^+ + c \, \phi^+ \right) \left( \partial^{\, \mu} \, 
			W_\mu^\dagger + b \, \chi^+ + c \, \phi^+ \right)^\dagger \, ,
\end{displaymath}
where the constants $a, b, c$ represent
\begin{align*}
	a &= - \frac{m_W^2}{m_\omega^2} 
	&  
	b &= - i \, \frac{m_\omega^2}{m_W^2} \, \sqrt{1 - \rho}
	& 
	c &= - i \, \frac{m_\omega^2}{m_W^2} \, ,
\end{align*}
with $m_\omega$ the combination of parameters
\begin{equation} \label{WWee-model-momega-definition}
	m_\omega^2 \equiv m_\chi^2 +  \lambda_{\sigma \chi} \, v_\sigma^2 + 
			\left( \lambda_{\phi \chi} + \frac{1}{2} \, \lambda_{\phi \chi}^\prime
			\right) \, v_\phi^2 \, ,
\end{equation}
which roughly corresponds to the mass of the physical singly-charged scalar
of the theory, and $\rho$ the 
electroweak $\rho$ parameter yielded by the model, equation 
\eqref{WWee-model-rho-theo}. The reader can check that in this gauge there are
no $W \chi^+$, $W \phi^+$ or $\phi^+ \chi^-$ mixing terms; this means that
there are just two mass-generating diagrams: one with $W$'s and one with $\phi^+$'s,
which we display in figure \ref{fig:WWee-model-numass-WudkaGauge}. Without
mixing no other two-loop diagram can be closed, because
$\chi^+$ doesn't couple to fermions and there is no mixed 
$W \phi^+ \kappa_{1,2}$ \mbox{vertex -- not just} in this gauge, there is no such 
an interaction in this model. Other relevant features of this gauge include the 
fact that $\phi^+$ is massive with mass $m_\omega$, and a modified propagator
for the $W$ bosons,
\begin{displaymath}
	i D_{\mu \nu} (k) = \frac{- i}{k^2 - m_W^2} \, \left[ \eta_{\mu \nu} -
			\left( 1 - \frac{m_\omega^2}{m_W^2} \right) \, \frac{k_\mu k_\nu}
			{k^2 - m_\omega^2} \right] \, ,
\end{displaymath}
which has no quadratic terms, and so does not yield divergent pieces
in the loop integrals.

With all these ingredients at hand, we are now prepared to calculate the neutrino
masses. Let us define the neutrino mass matrix as
\begin{displaymath}
	\mathcal{L}_{\mathrm{mass}} = -\frac{1}{2} \, \overline{\nu_{\mathrm{L}}} \,
			m \, \nu^\mathrm{c}_{\mathrm{L}} + \mathrm{H.c.} \, ,
\end{displaymath}
and note that in the following we will use $m_{\alpha \beta}$ to denote the
elements of the neutrino mass matrix and $m_\alpha$ to represent the mass
of the charged lepton of the family $\alpha$.
By working out the diagrams in figure \ref{fig:WWee-model-numass-WudkaGauge}
we obtain
\begin{equation} \label{WWee-model-generated-numass}
	m_{\alpha \beta} = \frac{1}{2 \, (2 \pi)^4} 
			\frac{\mu_\kappa \, v_\chi^2}{v_\phi^4} \, 
			m_\alpha g^*_{\alpha \beta} m_\beta \, I_\nu \, ,
\end{equation}
where it is worth noting the dependence on the combination of 
parameters $\mu_\kappa \, 
v_\chi^2$, which we already found while calculating the post-SSB form of the 
$WWee$ interaction, equation \eqref{eq:Lagrangian-WWee};
its presence here is just an indication that this is
again the relevant physics in play for neutrino mass generation.
In contrast, models which rely on the triplet Yukawas 
$\overline{\tilde{\ell}_{\mathrm{L}}} \chi \ell_{\mathrm{L}}$ to yield
neutrino masses, like for instance type-II seesaw, show a linear dependence 
in $v_\chi$. In our model the $Z_2$ symmetry forbids the triplet Yukawas and 
neutrino masses arise from the $WWee$ mechanism.

Some discussion is also in order concerning the function $I_\nu$ which 
appears in \eqref{WWee-model-generated-numass}. $I_\nu$ is a factor provided
by the loop integrals, and as such it is a function of the parameters of
the model, especially of the masses running inside the loops. For the case
of the gauge we are considering, $I_\nu$ comprises two contributions, one
from each of the diagrams in figure \ref{fig:WWee-model-numass-WudkaGauge}.
We will write $I_\nu = I_W + I_\phi$, and we will have, after appropriate 
rescaling to fit the factors in equation \eqref{WWee-model-generated-numass},
\begin{multline} \label{WWee-model-IW}
	I_W = -2 (4\pi)^4 m_W^4 \, \cos^4 \theta_{+} \: \times
		\\
		\times \int \frac{\mathrm{d}^4 k \, \mathrm{d}^4 q}{k^2 \, (k^2 - m_W^2) \, 
		q^2 \, (q^2 - m_W^2) \, \left[ (k - q)^2 - m_{\kappa_1}^2 \right] 
		\left[ (k - q)^2 - m_{\kappa_2}^2 \right]} \times
		\\
		\times \Bigg[ 4 - \left(1 - \frac{m_\omega^2}{m_W^2} \right) 
		\left( \frac{k^2}{k^2 - m_\omega^2} + \frac{q^2}{q^2 - m_\omega^2} \right) +
		\\
		+ \left(1 - \frac{m_\omega^2}{m_W^2} \right)^2 \, 
		\frac{\left( k \cdot q \right)^2}{(k^2 - m_\omega^2) \, 
		(q^2 - m_\omega^2)} \Bigg] \, , 
\end{multline}
\begin{multline} \label{WWee-model-Iphi}
	I_\phi = (4\pi)^4 \, \left( m_\omega^2 - \frac{1}{2} \, 
		\lambda_{\phi \chi}^\prime \, v_\phi^2 \right) \times 
	\\
	\times \int \frac{\mathrm{d}^4 k \, \mathrm{d}^4 q \; \: k \cdot q}{k^2 \, 
		(k^2 - m_\omega^2) \, q^2 \, (q^2 - m_\omega^2) \, 
		\left[ (k-q)^2 - m_{\kappa_1}^2 \right] \, \left[ (k-q)^2 - 
		m_{\kappa_2}^2 \right]} \, , 
\end{multline}
where we have used $m_\omega$ as defined in equation 
\eqref{WWee-model-momega-definition}. Note that each of the two diagrams in figure
\ref{fig:WWee-model-numass-WudkaGauge} represents in turn two processes: one 
with $\kappa_1$ and other with $\kappa_2$ running inside the loop. Some reworking,
which involves the doubly-charged mixing $\theta_{+ +}$, is needed in order to
express the total result as seen in equations (\ref{WWee-model-generated-numass}
-- \ref{WWee-model-Iphi}); essentially, the unitarity of the doubly-charged
mixing matrix is used to join the $\kappa_1$ and $\kappa_2$ contributions into 
one single integral that contains both $m_{\kappa_1}$ and $m_{\kappa_2}$ and
which is manifestly finite.

\begin{figure}[tb]
	\centering
	\includegraphics[width=0.7\textwidth]{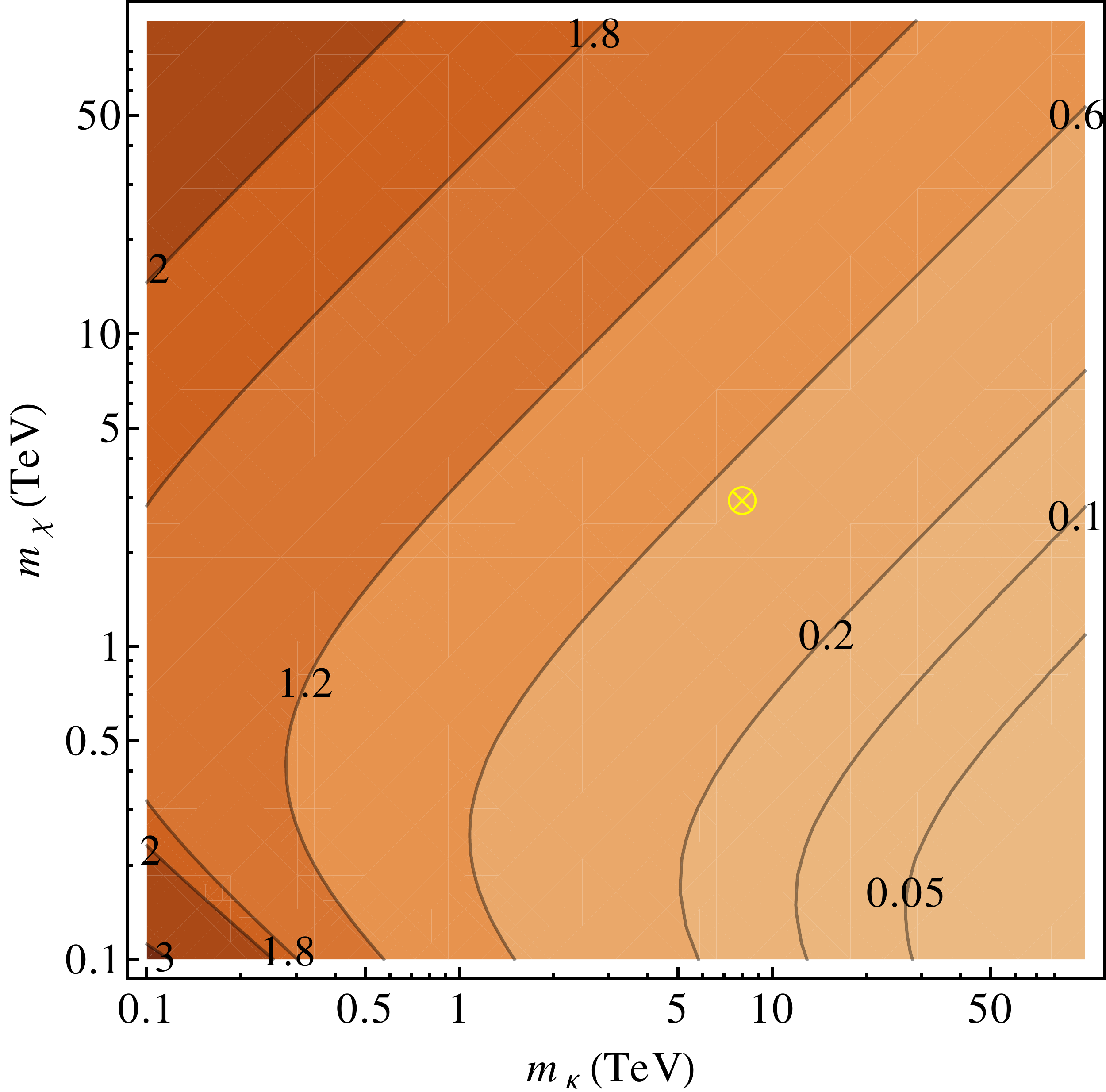}
	\caption{Contour plot for the loop integral $I_\nu$ in the limit $v_\chi \ll
			v_\phi$, as a function of the masses $m_\kappa \simeq m_{\kappa_1}$ 
			and $m_\chi \simeq m_{\kappa_2}$, and assuming $m_\omega = m_\chi$
			and $m_W = 80 \; \mathrm{GeV}$. The cross marks the reference
			point presented in equation \eqref{WWee-model-reference-point}.
			} \label{fig:WWee-model-Inu-contourplot}
\end{figure}

In general, $I_\nu$ is a complicated function of $m_W, m_{\kappa_1}, m_{\kappa_2}, 
m_\omega$ and the ratio $v_\chi / v_\phi$, which enters into the mixing angles.
However, a few reasonable assumptions can be made that very much simplify
the scenario: first, it is a phenomenological requirement that $v_\chi \ll v_\phi$,
which ensures that $\kappa_1$ and $\kappa_2$ are not very much mixed; this means
that $m_{\kappa_1} \simeq \bar m_\kappa$ and $m_{\kappa_2} \simeq \bar m_\chi$, 
in the notation used for equation \eqref{WWee-model-doubly-charged-mass-matrix}.
Second, we assume that $\bar m_\kappa \simeq m_\kappa$ and $\bar m_\chi, m_\omega
\simeq m_\chi$, which is true either if $m_\kappa, m_\chi \gg v_\phi, v_\sigma$ 
or if the $\lambda$ couplings involved are small. With these assumptions,
$I_\nu$ becomes a function only of $m_\kappa^2 / m_W^2$ and $m_\chi^2 / m_W^2$,
which we compute numerically. In figure \ref{fig:WWee-model-Inu-contourplot}
we plot the contours of constant $I_\nu$ as a function of the masses 
$m_{\kappa_1} \simeq m_\kappa$ and $m_{\kappa_2} \simeq m_\chi$, 
with the $W$ mass fixed to $m_W = 80 \; \mathrm{GeV}$. From the figure we appreciate
that $I_\nu$ is $\mathcal{O} (1)$ for a large region of the parameter space,
only producing small values when $m_\kappa \gg m_\chi$. We conclude, so,
that for most cases
$I_\nu$ can be ignored in equation \eqref{WWee-model-generated-numass}
for order-of-magnitude estimates, with the remaining involved parameters
adequately capturing the physics of neutrino mass generation.

\subsection{The structure of the neutrino mass matrix}

Equation \eqref{WWee-model-generated-numass} presents a neutrino mass matrix
highly influenced by the charged leptons' masses. The factor $m_\alpha m_\beta /
v_\phi^2$ poses an extra suppression which is
element-dependent, and so induces a well-definite hierarchy in the mass matrix
that is only modulated by the Yukawas $g_{\alpha \beta}$ and by global factors. 
From our present knowledge of neutrino parameters the mass matrix is by no means
arbitrary, so it is natural to wonder whether this induced hierarchy is admissible or
if it dooms the model to exclusion. In this section we discuss the range
of variation of the most tightly pressed elements, while in the next one we examine
if this hierarchy can accommodate a phenomenologically viable pattern of neutrino 
masses. 

Let us begin with the matrix element $m_{e e}$; being proportional to $m_e^2$, 
it is most surely forced to be small. How quantitative can we make this assertion? 
If we take $\mu_\kappa \sim 10 \; \mathrm{TeV}$, $v_\chi \sim 2 \; \mathrm{GeV}$, and 
$I_{\nu}, |g_{ee}| \sim 1$, equation \eqref{WWee-model-generated-numass} yields
$| m_{ee} | \sim 3.7 \times 10^{-6} \; \mathrm{eV}$, a certainly small
\mbox{value -- compare} with the other known neutrino mass scales: the atmospheric 
scale,
$\sqrt{\Delta m_{\mathrm{atm}}^2} = 0.05 \; \mathrm{eV}$, and the solar scale,
$\sqrt{\Delta m_{\astrosun}^2} = 0.009 \; \mathrm{eV}$. How much can $m_{e e}$
vary? Making it smaller would only require to take smaller $| g_{e e} |$,
$\mu_\kappa$ \mbox{or $v_\chi$ -- of course} at the prize of reducing the 
$0 \nu \beta \beta$ rate, which we do not desire, but in this way we could easily
have $| m_{ee} | \rightarrow 0$. However, how \emph{large} can it be?
How much can we push the limit to the more interesting `heavy' scenario? 
In cooking a large $| m_{ee} |$ we will end up facing the upper limit
on $v_\chi$, the unitarity limit on $| g_{e e} |$ and the naturality limit
for $\mu_\kappa$ (see equations \eqref{WWee-model-vchi-bound}, 
\eqref{WWee-model-g-bound-unitarity} and \eqref{WWee-model-muk-bound-naturality},
respectively, and the discussion upon them). None of these is a strict and
inescapable bound, but rather they're order-of-magnitude estimates that
could be stretched, or might change when detailed radiative corrections are 
considered. We implement them in \eqref{WWee-model-generated-numass} in order
to get an estimate of the upper bound on $| m_{ee} |$; for $v_\chi$
we prefer the 2 GeV bound that does not require loop corrections. With all this
considered, we obtain
\begin{displaymath}
	| m_{ee} | < 1.6 \times 10^{-5} \, \left( \frac{\min (m_{\kappa_1}, 
			m_{\kappa_2})}{\mathrm{TeV}} \right) \; \mathrm{eV} \, .
\end{displaymath}
Now we would like an upper bound on the lightest of $m_{\kappa_1}$ and 
$m_{\kappa_2}$; as larger masses imply less observable effects, this cannot
be obtained from an experimental limit. Rather, we turn to the requirement 
that a signal is observed in the next round of $0 \nu \beta \beta$ experiments,
equation \eqref{WWee-model-0nu2beta-bounds}, which naturally disfavours very 
heavy masses; from there we find that $\min ( m_{\kappa_1}, m_{\kappa_2} ) \sim
10 \; \mathrm{TeV}$, and this translates into
\begin{equation} \label{WWee-model-mee-upperbound}
	| m_{ee} | < 1.6 \times 10^{-4} \; \mathrm{eV} \, .
\end{equation}
Alternatively, we could also translate the experimental limits on 
$0 \nu \beta \beta$ into bounds on $| m_{ee} |$, but for large scalar masses 
this limit is less stringent than \eqref{WWee-model-mee-upperbound}. 
In either case, one can say that $| m_{ee} |$ is typically less than 
$10^{-4}$.

Another very constrained matrix element is $m_{e \mu}$; this one is 
proportional to $g_{e \mu}$, which is tightly bounded by \mbox{$\mu \rightarrow 3 e$
--see} equation \mbox{\eqref{WWee-model-muto3e-bound}--, especial}\-ly if
one desires to have large $0 \nu \beta \beta$ rates, and so large values of
$| g_{e e} |$. By substituting \eqref{WWee-model-muto3e-bound} into 
\eqref{WWee-model-generated-numass}, one gets
\begin{displaymath}
	| m_{e \mu} | < 2.3 \times 10^{-5} \, \left( \frac{m_\kappa}{\mathrm{TeV}}
			\right)^{2} \, \frac{\mu_\kappa \, v_\chi^2}{2 (2\pi)^4 \, v_\phi^4} \,
			\frac{m_e m_\mu}{| g_{ee} |} \, I_{\nu} \, ,
\end{displaymath}
from where we can eliminate $| g_{e e} |$ by imposing the observation of 
$0 \nu \beta \beta$ in the next generation of experiments, and then again
pushing the parameters to their maximum values:
\begin{displaymath}
	| m_{e \mu} | < 4.3 \, \frac{\mu_\kappa^2}{m_\chi^2} \,
			\frac{v_\chi^4}{v_\phi^4} \, I_\nu \; \mathrm{eV} <
			1.2 \times 10^{-5} \; \mathrm{eV} \, ,
\end{displaymath}
where we took again $I_\nu \sim 1$ and $v_\chi \sim 2 \; \mathrm{GeV}$,
and used the naturality bound on $\mu_\kappa$ together with the observability
of $0 \nu \beta \beta$.

In conclusion, the neutrino mass matrix generated by this model must have 
tiny $ee$ and $e \mu$ elements; more precisely, $| m_{e e} |, 
| m_{e \mu} | \lesssim 10^{-4} \; \mathrm{eV}$. This follows 
from two assumptions:
\emph{(i)} that $0 \nu \beta \beta$ is at the reach of the next round of experiments, 
and \emph{(ii)} that the theory is perturbative and free of unnatural fine tuning up 
to several tens of TeV. These limits could be somewhat relaxed: in the 
$| m_{ee} |$ case by making doubly-charged scalar masses larger, and for
$| m_{e \mu} |$ by allowing a smaller $|g_{ee}|$. In both cases the model
would lose some interesting feature, and we prefer to stick with these requirements
and see if the general picture can make sense anyway. The next section 
addresses this question.

\subsection{Consequences for the neutrino mass spectrum} 
					\label{sec:WWee-model-neutrino-spectrum}

In the previous section we discussed that the neutrino spectrum generated by
this model must present very small $m_{e e}$ and $m_{e \mu}$.
The question now becomes whether it is possible to accommodate the
observed pattern of neutrino masses and mixings into this particular
structure. To answer this question we need to construct the neutrino
mass matrix, because experiments don't probe directly mass matrix elements
but rather magnitudes such as mixing angles, phases and squared-mass splittings.
Let us use the parametrisation of neutrino masses described in section
\ref{sec:neutrinos-writing-masses-mixings},
\begin{equation} \label{WWee-model-neutrino-mass-construction}
	m_\nu = U D_\nu \, U^\mathrm{T} \; , \quad \mathrm{with} \; \; 
			D_\nu = \mathrm{diag} (m_1, m_2, m_3) 
\end{equation}
and $U$ the PMNS matrix as seen in equation \eqref{intro-neutrinos-PMNS-Majorana}.
The current best values for the neutrino mass and mixing parameters
can be consulted in table \ref{tab:intro-neutrino-bounds-and-measurements};
remember that the phases, the sign of $\Delta m_{31}^2$ and the 
absolute scale of neutrino masses are not yet known (see section 
\ref{sec:knowledge-mass-neutrino} for more details). 
%

The values in table \ref{tab:intro-neutrino-bounds-and-measurements}
can be used, together with equations 
\eqref{WWee-model-neutrino-mass-construction} and 
\eqref{intro-neutrinos-PMNS-Majorana},
to obtain the elements of the neutrino mass matrix in the flavour basis.
First thing to note is the importance of the hierarchy choice; indeed, it is 
easy to
check that for inverted hierarchy the experimental
values of the parameters require $| m_{e e} | > 10^{-2} \; \mathrm{eV}$,
which is incompatible with the texture generated by our model. Normal hierarchy,
however, allows for small $m_{e e}$ and $m_{e \mu}$. Therefore,
a first conclusion is that this class of models, if phenomenologically
viable at all, yields neutrino masses in normal hierarchy, that is,
with $\Delta m_{31}^2 > 0$.

The next step should be to explore the neutrino parameter space and see if
one can find phenomenologically allowed matrices that agree with the structure
provided by \eqref{WWee-model-generated-numass}. A possible course of action
would be to generate \emph{in silico} sets of random values for the 
neutrino parameters within their $1\sigma$ allowed ranges, as read in 
table \ref{tab:intro-neutrino-bounds-and-measurements},
and obtain the corresponding neutrino mass matrix elements 
$m_{\alpha \beta}$.
Then one can use equation \eqref{WWee-model-generated-numass} to solve for
$g_{\alpha \beta}$ in terms of $\mu_\kappa v_\chi^2  I_\nu$
and check if the phenomenologically allowed values for \mbox{$\mu_\kappa, v_\chi,
m_\kappa, m_\chi$ --as de}\-scribed in sections \ref{sec:WWee-model-0nu2beta},
\ref{sec:WWee-model-lfv} \mbox{and \ref{sec:WWee-model-unitarity-naturality}--
yield} phenomenologically allowed values for $g_{\alpha \beta}$. One would find
tensions between the various phenomenological requirements: while neutrino masses
could favour large values for $g_{\alpha \beta}$, lepton flavour violation
constraints would prefer small values; $v_\chi$ is bounded to be small, but
the largest possible values would favour the observability of $0 \nu \beta \beta$;
and so on. But let us as a first approach
to face a simplified version of the problem: roughly, what we are wondering
is whether $|m_{e e}|, |m_{e \mu}| \sim 0$ is consistent with 
neutrino oscillation 
data. We can try and explore this question independently of
our model: the task will be just to look if there is
room in the \emph{neutrino} parameter space for a mass matrix with exactly
vanishing $m_{e e}$ and $m_{e \mu}$. If the answer is yes, then
we will know that the model is viable, and we
can turn to the other interesting question of in which way small 
$m_{e e}$'s and $m_{e \mu}$'s constrain the parameters of the model.

This new, simpler task can be carried out analytically. 
In order to see how this comes about it is useful to go through
a straightforward parameter-counting exercise:
$m$ is a $3 \times 3$  complex and symmetric matrix specified
by 12 real numbers: 3 of these are unphysical and can be absorbed 
by rephasing the neutrino fields, and 5 of the remaining 9 are \mbox{measured --
2 mass} differences and 3 mixing angles, including $\theta_{13}$.
If we now impose $m_{ee} = m_{e \mu} = 0$, meaning
4 additional (real) constraints, only a discrete set of points 
will be consistent. In fact, there might be no allowed values at all! 
Fortunately, we checked that this is not the case: for most of the neutrino
parameter space at least one solution exists and thus the model is 
viable, although possibly confined to a narrow region of neutrino parameters.
Which is not bad news; that, rather, means that the $WWee$ mechanism yields
very sharp predictions for neutrino masses that can be tested experimentally.

Let us now give an example of neutrino parameters that realise
the simplified conditions: 
using the central values of the global fit in \cite{Schwetz:2011zk}
we have obtained the following for the unmeasured parameters:
\begin{align}
	\alpha_1 &= 0.65 \; \mathrm{rad} & \alpha_2 &= -2.32 \; \mathrm{rad}
	\nonumber
	\\ \label{WWee-model-example-point}
	\delta &= -0.78 \; \mathrm{rad} & m_1 &= 0.0036 \; \mathrm{eV} \, .
\end{align}
These values, when substituted in \eqref{WWee-model-neutrino-mass-construction}
and \eqref{intro-neutrinos-PMNS-Majorana} produce the following mass matrix:
\begin{equation} \label{WWee-model-mass-matrix-example}
	m \simeq \begin{pmatrix}
			0  &  0  &  0.59 + 0.58 i \\
 			0  &  2.47 - 0.2 i  &  2.64 + 0.2 i \\
 			0.59 + 0.58 i  &  2.64 + 0.2 i  &  2.12 - 0.21 i
		\end{pmatrix} \times 10^{-2} \; \mathrm{eV} \, ,
\end{equation}
which presents a recognisable texture. In our model we would interpret
this hierarchy as the effect of the charged lepton masses, but note that 
\eqref{WWee-model-mass-matrix-example} has had no interaction with 
the model; the texture appears as a consequence of the requirement 
that $m_{ee} = m_{e \mu} = 0$, but the result is consistent 
with the structure suggested by equation \eqref{WWee-model-generated-numass}.

\begin{figure}[p]
	\centering
	\includegraphics[width=0.7\textwidth]{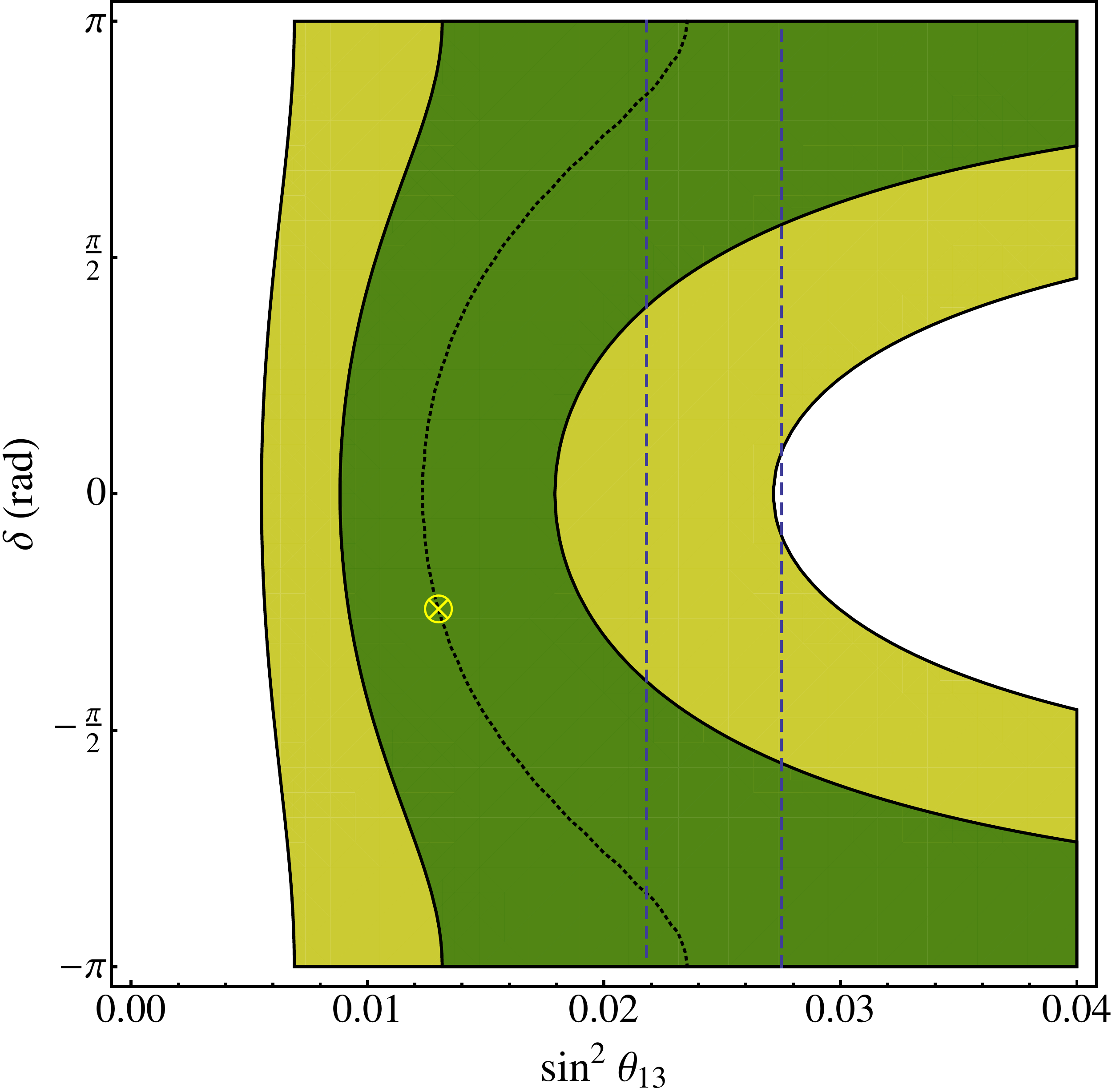}
	\caption{Allowed $\sin^2 \theta_{13} - \delta$ region if
			$m_{ee} = m_{e \mu} = 0$ is to have solution.
			The green, darker region represents the positive $\theta_{13} - \delta$
			values when $s_{12}, s_{23}, \Delta m_{21}^2, \Delta m_{31}^2$ are 
			allowed to vary within their $1 \sigma$ range; the yellow, lighter
			zone is obtained when the neutrino parameters vary throughout their
			$3 \sigma$ band. The middle dotted curve corresponds 
			to the central values of the neutrino masses and mixings in the global 
			fit performed in \cite{Schwetz:2011zk}. For comparison, we draw 
			(vertical dashed lines) the $1\sigma$ band for $\sin^2 \theta_{13}$
			from the global fit in \cite{Tortola:2012te}, which incorporates
			the recent measurements by Daya Bay, RENO and Double Chooz.
			The cross stands for the reference point presented
			in equation \eqref{WWee-model-reference-point}.
			} \label{fig:WWee-model-graf-delta-s13}
\end{figure}

The main source of uncertainty in calculations such as 
\eqref{WWee-model-example-point} is the value of $\theta_{13}$,
which was only hinted at the time of this analysis%
\footnote{Barely a few months before the first Daya Bay data were 
released\dots $\;$We consider, however, that this discussion is still useful and
it can be interpreted in the light of the new measurements, as we do below.
}.
If we repeat the exercise 
for different values \mbox{of $\sin^2 \theta_{13}$ --away from} the central 
\mbox{point in \cite{Schwetz:2011zk}--, we find} a surprising result:
$m_{ee} = m_{e \mu} = 0$ only has solution for 
\mbox{$0.012 < \sin^2 \theta_{13} < 0.024$,} at least for 
$s_{12}, s_{23}, \Delta m_{21}^2,
\Delta m_{31}^2$ fixed to their central values. It would seem, so, that the
texture induced by the charged lepton masses forces $\theta_{13}$ to be nonzero
in this class of models. How much does this result depend on the values of
the measured neutrino parameters? Is it robust if we vary $s_{12}, s_{23}, 
\Delta m_{21}^2, \Delta m_{31}^2$ within their allowed ranges? We calculated 
the values of $\theta_{13}$ and $\delta$ that allow for a solution to
$m_{ee} = m_{e \mu} = 0$ letting the rest of neutrino parameters 
take values in their $1 \sigma$ and $3 \sigma$ bands, and we display
the results in figure \ref{fig:WWee-model-graf-delta-s13}. As we can see,
the lower bound on $\theta_{13}$ is robust, and interesting correlations 
are observed with the allowed values of the phase $\delta$: large
values of $\theta_{13}$ tend to prefer $\delta$'s around $\pi$, while
$\delta$'s around 0 roughly imply $\theta_{13}$ near the minimum of its 
allowed range. Actually, now that we have strong evidence for a nonzero 
$\theta_{13}$ \cite{Abe:2011fz,An:2012eh,Ahn:2012nd} we can interpret
this evidence in the framework of our model: in figure 
\ref{fig:WWee-model-graf-delta-s13} we draw two vertical lines representing
the $1\sigma$ band for $\sin^2 \theta_{13}$ from the global fit in
\cite{Tortola:2012te}, which includes the new measurements. From the 
correlations between $\theta_{13}$ and $\delta$ we would deduce that our
model prefers a Dirac phase around $\pi$, rather than around 0.

\begin{figure}[p]
	\centering

	\raisebox{5.75cm}{\emph{a)}}
	\includegraphics[width=0.75\textwidth]{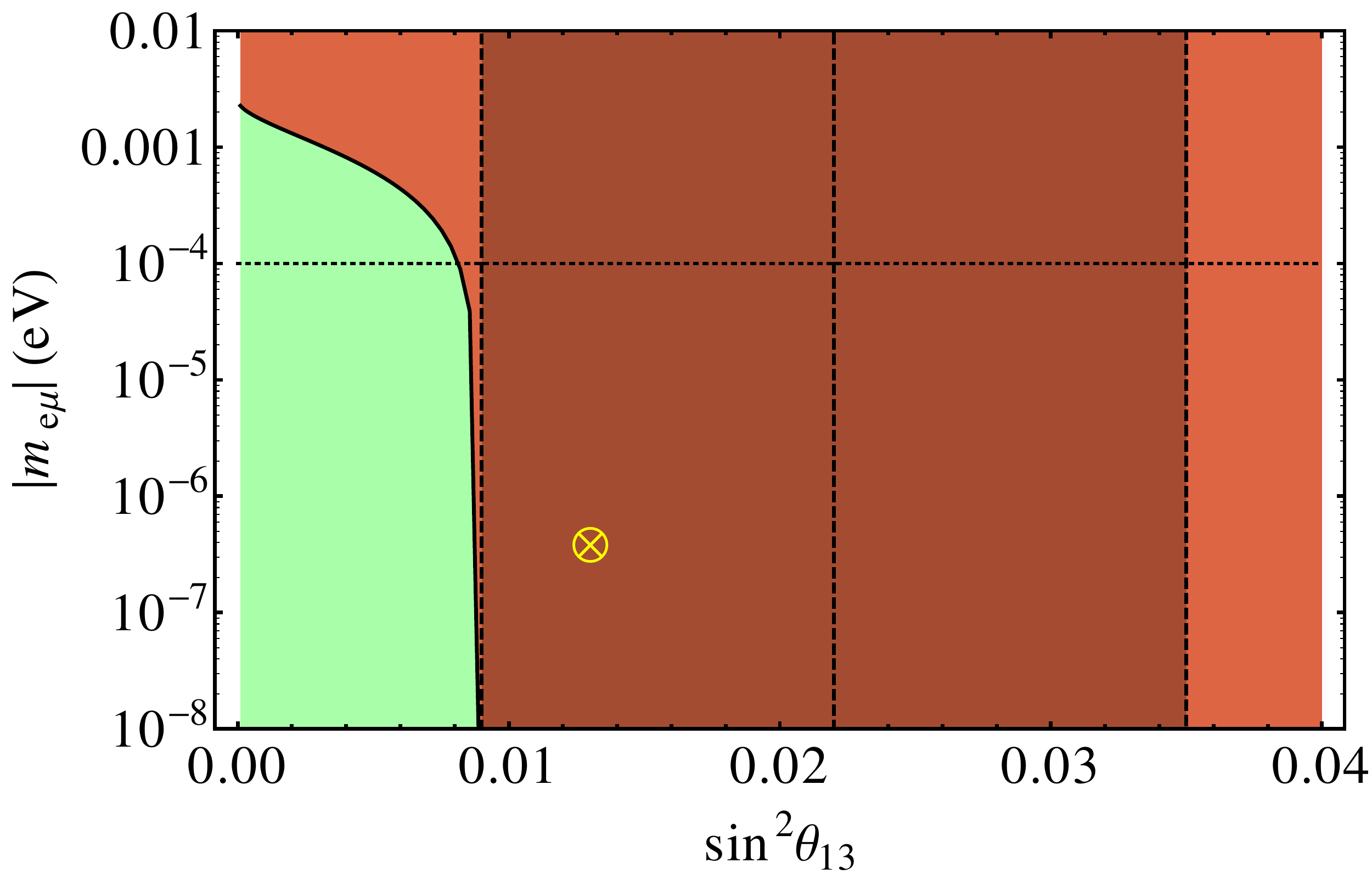}

	\vphantom{\raisebox{0.5cm}{a}} 
	\raisebox{5.5cm}{\emph{b)}}
	\includegraphics[width=0.75\textwidth]{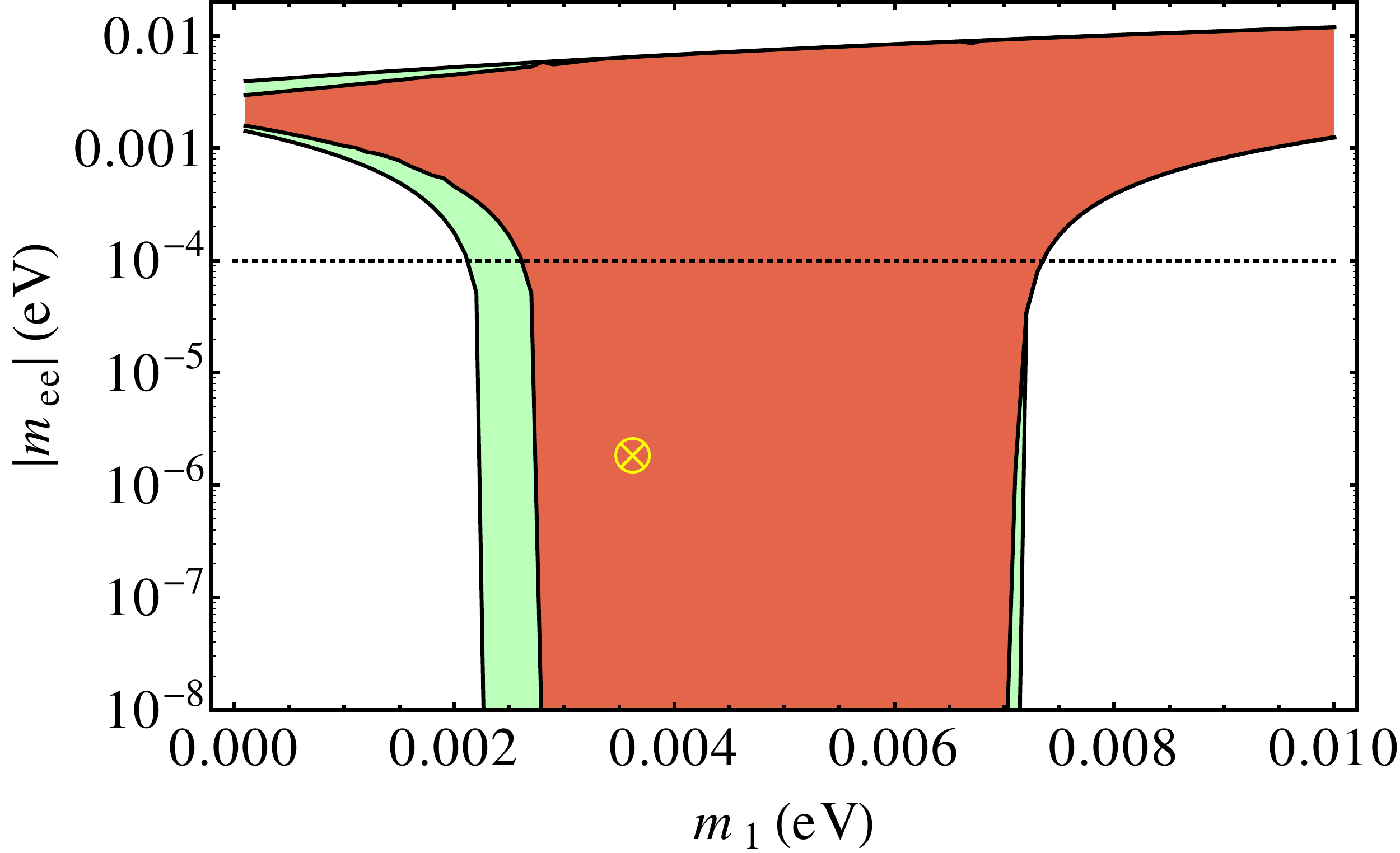}
	\caption{These plots show in green the allowed values for $| m_{e \mu} |$ (up)
			and	$| m_{e e} |$ (down) if we let the neutrino parameters vary within
			their $1\sigma$ experimental band. $| m_{e \mu} |$ is 
			shown in terms of $\sin^2 \theta_{13}$, with the red 
			region displaying the additional constraint that $| m_{e \mu} | 
			> | m_{e e} |$ (see the main text for an explanation).
			$| m_{ee} |$ is shown in terms of $m_1$, and this time the
			red area displays the opposite condition, $| m_{e e} | 
			> | m_{e \mu} |$. In both graphs we have drawn the estimate
			upper bound on these two matrix elements, $10^{-4} \; \mathrm{eV}$. 
			The yellow crosses represent
			the reference point presented in equation 
			\eqref{WWee-model-reference-point}.
			} \label{fig:WWee-model-full-neutrino-masses}
\end{figure}

Once checked that the model is viable and indeed can yield interesting predictions,
it's compulsory to examine the full case with $| m_{ee} |, | m_{e \mu} | 
\lesssim 10^{-4} \; \mathrm{eV}$. In figure 
\ref{fig:WWee-model-full-neutrino-masses}
we present two plots that illustrate the situation: in \emph{a)} we display the
values of $| m_{e \mu} |$ in terms of $\sin^2 \theta_{13}$ if we let the
remaining neutrino parameters vary within their $1 \sigma$ band. The green, light
area which extends to the whole panel informs us that for a given value
of $\theta_{13}$ one can obtain \emph{any} value for $| m_{e \mu} |$ by
correctly selecting the rest of the parameters. But in plotting this we're not
constraining $| m_{e e} |$, and it could be large for some of such 
combinations of parameters. The red, darker region attempts to constrain 
$| m_{e \mu} |$ and $| m_{e e} |$ simultaneously: it represents
the values of $| m_{e \mu} |$ in terms of $\sin^2 \theta_{13}$ wherever
it is fulfilled that $| m_{e \mu} | > | m_{e e} |$. Therefore,
in the red area we can be sure that we are keeping both elements under control;
and we see that, as we obtained for the $| m_{e e} | = | m_{e \mu} | = 0$
case, small values of $\theta_{13}$ are forbidden, meaning that we cannot
produce both $| m_{e e} |$ and $| m_{e \mu} |$ small with small
$\theta_{13}$\footnote{One
could worry about the possibility that $| m_{e e} |$ is greater than 
$| m_{e \mu} |$ but still under the $10^{-4} \; \mathrm{eV}$ 
\mbox{bound -- say,} $| m_{e \mu} | \sim 10^{-6} \; \mathrm{eV}$ and 
$| m_{e e} | \sim 10^{-5} \; \mathrm{eV}$. But it's possible to check, 
for instance with a plot of
$| m_{e e} |$ against \mbox{$\sin^2 \theta_{13}$ --not shown} \mbox{here--, 
that} such configurations cannot be achieved with small $\theta_{13}$ within 
$1 \sigma$ variation of the neutrino parameters.}. 
Quantitatively, we see that in this realistic case the upper bound for $\theta_{13}$
disappears, whereas the lower bound is somewhat relaxed, but still clear; we 
could write that
\begin{equation} \label{WWee-model-theta13-bound}
	\sin^2 \theta_{13} \gtrsim 0.008
\end{equation}
if the upper bound on $| m_{e e, e \mu} |$ is fixed in $10^{-4} \; 
\mathrm{eV}$. We observe that if we relax this upper bound to $10^{-3}$ the
lower bound upon $\theta_{13}$ also decreases, until it completely disappears
a little bit above $10^{-3} \; \mathrm{eV}$. 

Figure \ref{fig:WWee-model-full-neutrino-masses}\emph{b)} is similar to \emph{a)},
but illustrates the bounds that the neutrino mass texture renders upon $m_1$,
the lightest neutrino mass. Here we represent the element $| m_{e e} |$
in terms of $m_1$. We could expect, from the well-known plots of $0 \nu \beta 
\beta$ sensitivity (see figure \ref{fig:intro-0nu2beta-elephant})
that if we are to have very small 
$m_{e e}$ we should be lying in the `elephant leg' of $10^{-3} \; 
\mathrm{eV} \lesssim m_1 \lesssim 10^{-2} \; \mathrm{eV}$. And indeed that is
what we find in figure \ref{fig:WWee-model-full-neutrino-masses}\emph{b)}: the
green, lighter region shows just $| m_{e e} |$ in terms of $m_1$ varying
the rest of the neutrino parameters within their $1 \sigma$ range, and the
red, darker area verifies the additional condition that $| m_{e e} | > 
| m_{e \mu} |$. Both regions show a clear upper and lower bound for $m_1$,
which is constrained to be $\mathcal{O} (10^{-3} \; \mathrm{eV})$, or, more
precisely,
\begin{equation} \label{WWee-model-m1-bound}
	0.002 \; \mathrm{eV} \lesssim m_1 \lesssim 0.007 \; \mathrm{eV} \, .
\end{equation}

The model, so, by virtue of the texture induced in the neutrino mass matrix
by the charged lepton masses, yields very definite predictions about some 
still unmeasured neutrino parameters. These predictions, summarised in equations
\eqref{WWee-model-theta13-bound} and \eqref{WWee-model-m1-bound}, can be used to 
probe the model and eventually falsify it,
particularly in the light of the recent measurements of $\theta_{13}$.

\section{Constraints on the parameters of the model}  
				\label{sec:WWee-model-constraints-parameters}

In the previous sections we have listed a number of theoretical and 
phenomenological features that constrain in several ways the parameters 
of the model. Not all of them are equally critical: the experimental bounds
on $0 \nu \beta \beta$ and LFV-ing processes, sections \ref{sec:WWee-model-0nu2beta}
and \ref{sec:WWee-model-lfv}, and also the requirement that
the neutrino mass matrix fits with the known neutrino parameters, section
\ref{sec:WWee-model-neutrino-spectrum}, are inescapable demands; the
unitarity and naturality bounds, section \ref{sec:WWee-model-unitarity-naturality},
and the bound on the triplet VEV, equation \eqref{WWee-model-vchi-bound},
are to be fulfilled, but might be stretched a bit if the appropriate 
conditions are met; finally, the requirement that $0 \nu \beta \beta$ is
observed in the next round of experiments is something that we would like to
happen, and we impose it in order to find out if that's possible. In this section
we will consider all these constraints together and see how much room they
leave for the model; as we will see, it is possible to fulfill all of them, 
but not without tensions and a little bit of high-wire walking.

\begin{figure}[p]
	\centering
	\raisebox{6cm}{\emph{a)}}
	\includegraphics[width=0.5\textwidth]{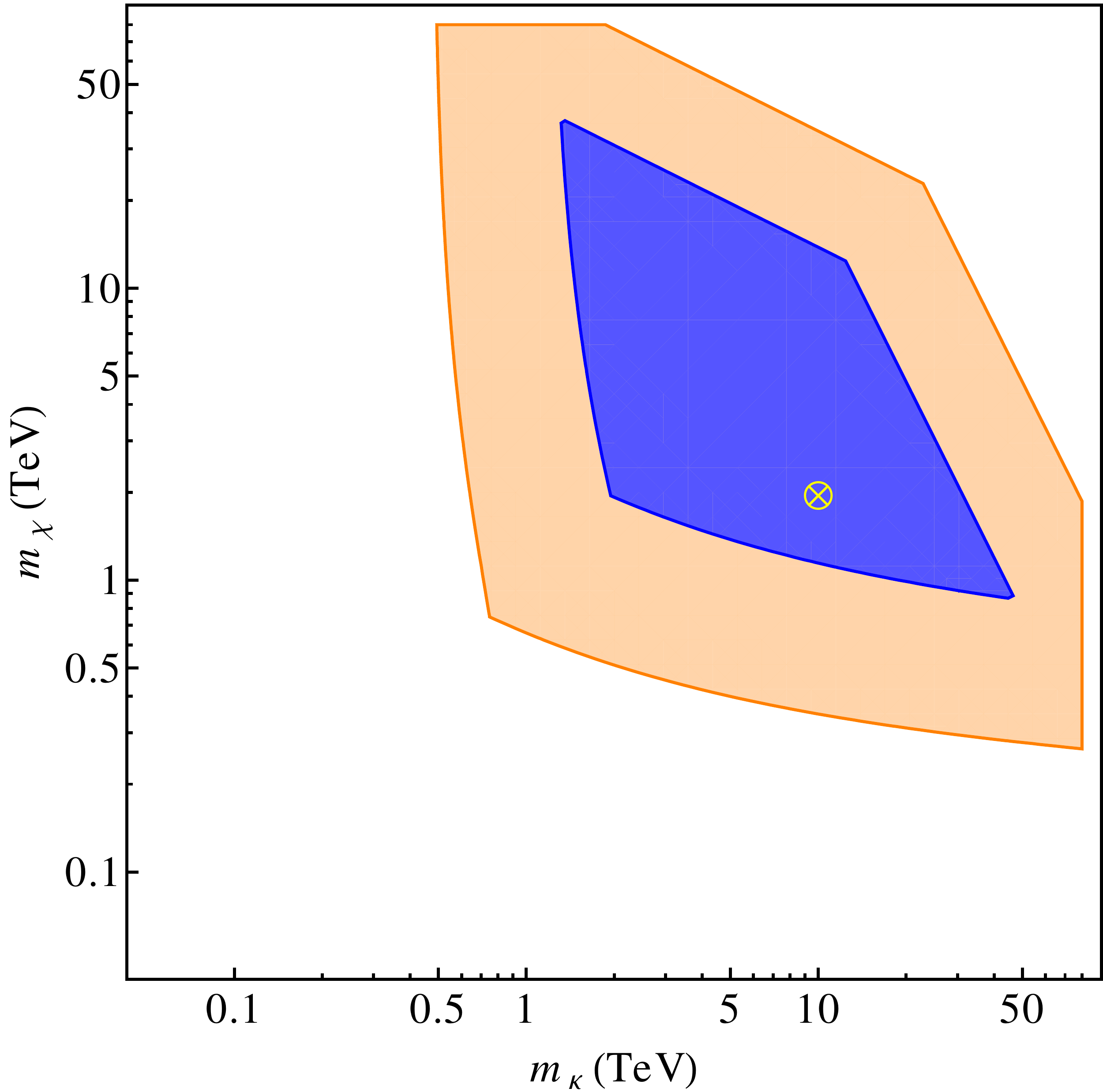}

	\raisebox{6cm}{\emph{b)}}
	\includegraphics[width=0.5\textwidth]{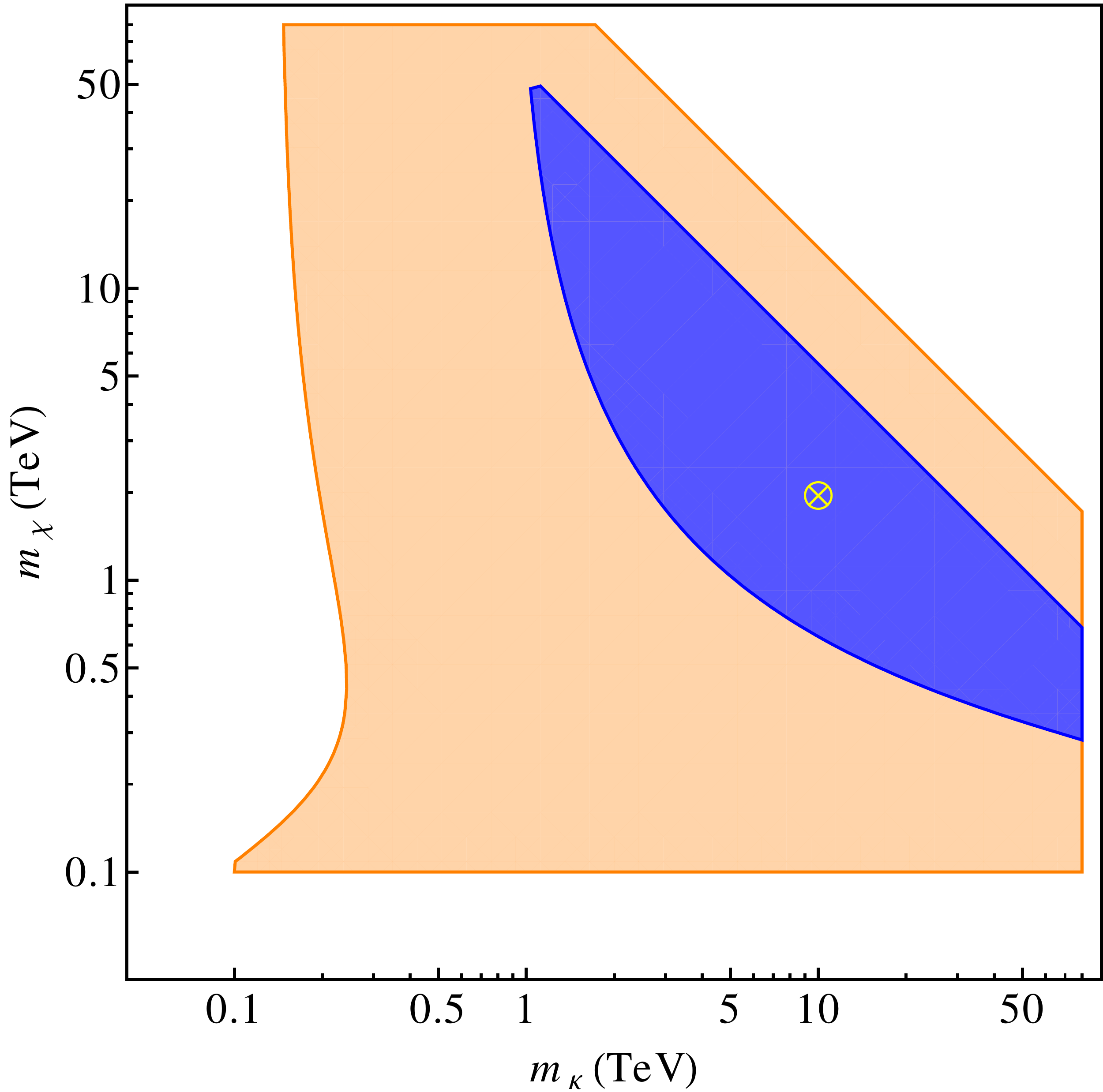}
	\caption{Two plots for the allowed region in the $m_\kappa - m_\chi$ plane.
			The regions are displayed after applying all the constraints. As
			the doubly-charged mixing angle is mainly small, we have implemented
			the approximation $m_{\kappa_1} \simeq m_\kappa, m_{\kappa_2} \simeq
			m_\chi$. The blue areas represent the allowed region if the triplet VEV
			is forced to be $v_\chi < 2 \; \mathrm{GeV}$; the orange areas
			display the allowed region if the bound is relaxed to $v_\chi <
			5 \; \mathrm{GeV}$.
			The plot labeled as \emph{a)} implements the naturality
			limit for $\mu_\kappa$, equation \eqref{WWee-model-muk-bound-naturality},
			whereas \emph{b)} implements just an upper bound on the LN-breaking
			scale, $\mu_\kappa < 20 \; \mathrm{TeV}$. The reference point 
			described in equation \eqref{WWee-model-reference-point}
			is presented in both plots as a yellow crossed circle.
			} \label{fig:WWee-model-mk-mchi-plots}
\end{figure}

Let us express the coalescence of the constraints in terms of the scalar masses
$m_\kappa$ and $m_\chi$, which are relevant for the appearance of 
the new particles in collider experiments. Throughout
this section we will be assuming that the doubly-charged mixing is small
and thus $m_{\kappa_1} \simeq m_\kappa$ and $m_{\kappa_2} \simeq m_\chi$. 
Equation \eqref{WWee-model-0nu2beta-bounds} can be our starting point, as
it imposes both an upper and a lower bound. Note first that as $v_\chi$ is small, 
$m_\kappa$ and $m_\chi$ are rather large, and $g_{e e}$ can be made small
without disturbing neutrino masses, the upper bound is easy to satisfy.
The lower bound, however, poses more problems: the product $\mu_\kappa 
v_\chi^2 |g_{e e}|$ is upper-bounded by unitarity, naturality and the $\rho$
parameter as seen in equations \eqref{WWee-model-g-bound-unitarity},
\eqref{WWee-model-muk-bound-naturality} and \eqref{WWee-model-vchi-bound};
this presses the scalar masses to light values, potentially leaving the
new scalars at the reach of the LHC. In figure 
\ref{fig:WWee-model-mk-mchi-plots}\emph{a)}
we depict the allowed region in the $m_\kappa - m_\chi$ plane with all the
constraints considered. The blue area represents the allowed region if the
triplet VEV is forced to be $v_\chi < 2 \; \mathrm{GeV}$, and the orange
zone corresponds to the more conservative bound $v_\chi < 5 \; \mathrm{GeV}$.

As we see, the permitted region is rather narrow, and the prospects are not
utterly optimistic about a potential LHC discovery. The issue in the 
`light' region is an undesirable interaction between neutrino masses,
naturality and LFV constraints which comes about as follows: as we see for
instance in equation \eqref{WWee-model-tautoe2mu-bound}, if the scalar
masses are light LFV forces the $g$ Yukawas to be smaller. Then we turn 
to neutrino masses; see the mass matrix \eqref{WWee-model-mass-matrix-example}:
if the $m_{e e}$ and $m_{e \mu}$ elements are to be small, some
other elements must carry the weight of the `heavy' eigenvalue: 
$m_{\mu \mu}$, for instance, has to be $\mathcal{O} (10^{-2} \; \mathrm{eV})$.
But look at equation \eqref{WWee-model-generated-numass}: if the $g$'s
are small and the $m$'s must be large, there is only one way to move: make
$\mu_\kappa$ larger. But this cannot happen, for $\mu_\kappa$ is forced by
\eqref{WWee-model-muk-bound-naturality} not to be much larger than $m_\kappa$ and
$m_\chi$, which are light. Conclusion: the whole `light' region for 
$m_\kappa$ and $m_\chi$ is forbidden. Let us work out this situation numerically:
from \eqref{WWee-model-generated-numass} we obtain that
\begin{displaymath}
	| g_{e \tau} g_{\mu \mu} | = \frac{\left( 2 (2\pi)^4 \, v_\phi^4 \right)^2}
			{m_e m_\tau m_\mu^2} \, \frac{|m_{e \tau}| \, |m_{\mu \mu}|}
			{\mu_\kappa^2 \, v_\chi^4 \, I_\nu^2} 
			> 
			\frac{0.065}{I_\nu^2} \left(\frac{\mathrm{TeV}}
			{\min (m_{\kappa_1}, m_{\kappa_2})} \right)^2 \, ,
\end{displaymath}
where in the last step we used the upper limits on $\mu_\kappa$ and $v_\chi$
(2 GeV in this case) and we made $| m_{e \tau} | \sim 10^{-3} \; \mathrm{eV}$ 
and $| m_{\mu \mu} | \sim 10^{-2} \; \mathrm{eV}$, according to equation
\eqref{WWee-model-mass-matrix-example}. Then, combining this result with the
bound for $\tau^- \rightarrow e^+ \mu^- \mu^-$, equation 
\eqref{WWee-model-tautoe2mu-bound}, we obtain $m_\kappa > 1.2 \; \mathrm{TeV}$,
limit that can be also observed in figure 
\ref{fig:WWee-model-mk-mchi-plots}\emph{a)}.

Of the three constraints that close the light scalar region, two of \mbox{them 
--neutrino} masses and \mbox{LFV bounds-- are not} negotiable; the naturality
bound on $\mu_\kappa$, however, is an estimate based on what is considered 
to be `natural' for a theory. We can relax this latter constraint and see
if the light scalar region remains closed; what we will do is to fix a maximum
scale for lepton number violation, that is to say, an upper bound on $\mu_\kappa$
irrespective of the masses of $\kappa_1$ and $\kappa_2$. In figure 
\ref{fig:WWee-model-mk-mchi-plots}\emph{b)} we depict the allowed region
substituting the naturality bound for $\mu_\kappa < 20 \; \mathrm{TeV}$,
and indeed we observe that most of the light scalar region is now open. As
a conclusion we might say that a detection of the new scalars at the LHC 
is not definitely ruled out, but it appears that the phenomenological 
constraints rather favour a scenario with the scalars at the order of several
TeV.

Something to note about the graphs in figure \ref{fig:WWee-model-mk-mchi-plots}
is that all observables violating LN are proportional to $\mu_\kappa v_\chi^2$; 
hence, an increase in $v_\chi$ can be traded by the corresponding increase 
in $\mu_\kappa$, and vice versa. So, the orange areas in the plots can be also 
interpreted as the allowed regions for $v_\chi = 2 \; \mathrm{GeV}$ and 
\mbox{$\mu_\kappa < 25 \pi \, \min (m_{\kappa_1}, m_{\kappa_2})$ --in \emph{a)}--
or} \mbox{$\mu_\kappa < 125 \; \mathrm{TeV}$ --in \emph{b)}--.
At the} same time, one may wonder why we chose 20 TeV for the bound on
$\mu_\kappa$ in \emph{b)} or, equivalently, what is the effect of varying
such a value. The answer is simple: the blue, $v_\chi < 2 \; \mathrm{GeV}$
region in figure \ref{fig:WWee-model-mk-mchi-plots}\emph{b)} 
disappears for $\mu_\kappa \lesssim 8 \; \mathrm{TeV}$, which is a reflection of 
the narrowness of the range allowed by equation \eqref{WWee-model-0nu2beta-bounds}, 
as required by our main working assumption that $0\nu\beta\beta$
should be observed in the next round of experiments.
The allowed regions in figure \ref{fig:WWee-model-mk-mchi-plots} are 
appreciably enlarged by reducing the lower limit in this equation.
This can also be achieved by dropping the issues with naturality and
further increasing $\mu_\kappa$.

Finally, we provide in figure \ref{fig:WWee-model-mk-mchi-plots} a 
`benchmark point', denoted by a cross in the figures, where all constraints are 
satisfied, including the observation of $0 \nu \beta \beta$ and a VEV triplet 
under 2 GeV. The coordinates of this point in the parameter space of the model are
\begin{align} \label{WWee-model-reference-point}
	m_\kappa &= m_{\kappa_1} = 10 \; \mathrm{TeV}
		& m_\chi &= m_{\kappa_2} = 2 \; \mathrm{TeV}
	\nonumber \\
	v_\chi &= 2 \; \mathrm{GeV} 
		& \mu_\kappa &= 15 \; \mathrm{TeV}
	\\
	g_{e e} &= 1 
	& g_{e \mu} &= 0.001 \, .
	\nonumber
\end{align}
The point is used in the plots throughout this chapter to show that, although
under tight pressure, the model can accommodate all the phenomenological and
theoretical considerations that we have exposed. It also helps to follow the
track of the predictions of the model along their various relevant phenomenological
aspects.

\section{The model at colliders} \label{sec:WWee-model-colliders}

Direct evidence for this type of models would be the 
discovery of the new scalars at a large collider together with a demonstration
of the presence of LNV-ing interactions. This latter goal seems difficult;
as we have seen throughout the discussion, violation of LN in this model involves 
several \mbox{couplings --typically} expected to be \mbox{small--, which yields}
small LNV-ing production and decay rates. In general, the dominant production 
mechanisms are standard and LN-conserving, or otherwise they are suppressed.
As for decays, not all the decay channels of the new particles produce 
LNV-ing signals, 
though in some cases they might be dominant. Generally, the idea is 
to look first for the new resonances in the most sensitive channels and 
only afterwards to address the observation of LNV-ing events. 
Doubly-charged scalars, for example, have fixed couplings to photons and 
are produced at colliders with known cross sections. 
In addition, their decay into leptons offers a very clean signal, 
which is particularly important at hadronic machines. 
Therefore, doubly-charged scalars seem a priori the best signal to look
for, if they are light enough to be produced.

Studies to search for doubly-charged scalars at colliders 
have been performed in the past, in many cases motivated by mass-generating
mechanisms like seesaw type II, which are different from the one we 
discussed in this chapter but can have a similar particle content
\cite{Gunion:1989in,Huitu:1996su,Gunion:1996pq,Akeroyd:2005gt,Azuelos:2005uc,delAguila:2008cj,Akeroyd:2010ip};
model-independent studies have  also been carried out in 
the literature, like for example \cite{Dion:1998pw,Cuypers:1996ia}. 
The general conclusion is that the LHC discovery limit reaches masses
over $600 \; \mathrm{GeV}$ for a center of mass energy of $14 \; \mathrm{TeV}$
and an integrated luminosity of $30 \; \mathrm{fb}^{-1}$ 
\cite{delAguila:2008cj,delAguila:2009bb};
however, the actual limits achieved may be way better given 
the outstanding LHC performance, which almost matches the 
most favourable expectations for a center-of-mass energy of $7 \; \mathrm{TeV}$ 
\cite{delAguila:2010uw}. See, for instance, for a review \cite{Nath:2010zj}.

Recently, the first results from CMS have been presented at a CM energy of 
$7 \; \mathrm{TeV}$ with an integrated luminosity of $0.89 \; \mathrm{fb}^{-1}$
\cite{cms-pas-hig-11-007:2011}.
The analysis  assumed a scalar triplet coupled
to leptons, with $100\%$ branching ratio to each leptonic channel. 
No excess over the background was observed, leading to a lower bound on 
the mass of the doubly-charged scalar  
of about $250 \; \mathrm{GeV}$ if the main decay
involves $\tau$ leptons, of about $300 \; \mathrm{GeV}$ 
if the main decay product is electrons and muons, and extending up to 
$375 \; \mathrm{GeV}$ if only muons are produced.
Weaker limits were obtained previously by LEP and the Tevatron;
the absence at LEP of a pair-production signal of the type
$e^+ e^- \rightarrow \gamma^\ast, Z^\ast \rightarrow \kappa^{+ +} \kappa^{- -}$
yields the constraint $m_\kappa > 100 \; \mathrm{GeV}$ 
\cite{Abdallah:2002qj,Abbiendi:2001cr,Achard:2003mv}.
Single production via $e^+ e^- \rightarrow \kappa ee$, as well as the
$u$-channel contribution of $\kappa$ to Bhabha scattering have also been studied 
at LEP \cite{Achard:2003mv,Abbiendi:2003pr}, but
the corresponding bounds depend on the unknown values of the
Yukawa couplings and are not much informative. Limits on this
sort of scalars have been also derived using Tevatron data
\cite{Abazov:2004au,Acosta:2004uj,Acosta:2005np}, leading to a
limit $m_\kappa > 100-150 \; \mathrm{GeV}$, depending on the details of the model.

In our model the triplet does not directly couple to fermions, while the 
doubly-charged singlet does not couple to $W$ pairs; however, triplet and 
singlet mix. Of the resulting mass eigenstates, $\kappa_1$ is mainly a singlet 
and decays dominantly to lepton pairs, while the other, $\kappa_{2}$, is mainly a 
triplet and will decay to gauge bosons if kinematically allowed.
Both of them can be produced at LHC via the Drell-Yan mechanism,
$q \bar{q} \rightarrow \gamma^*, Z^* \rightarrow \kappa^{+ +} \kappa^{- -}$,
with full strength. Since this is the main production process considered
by CMS, the former limits apply directly  to $\kappa_1$ if the mixing can be
safely neglected: $m_{\kappa_1} > 300 \; \mathrm {GeV}$. Limits on $m_{\kappa_2}$
will be more difficult to obtain because the process  
$q \bar{q} \rightarrow \gamma^*, Z^* \rightarrow \kappa_2^{+ +} \kappa_2^{- -}
\rightarrow W^+ W^+ W^- W^-$ is much more complicated to deal with, due to its 
large backgrounds and the generally difficult reconstruction of 
several leptonic $W$ decays. 

Notice that there are other production processes that are more specific
of our model; in particular, the process that yields $0 \nu \beta \beta$, 
shown in figure \ref{fig:WWee-model-0nu2beta}, can be reverted and act 
as a $WW$-fusion production channel for $\kappa_1$ and $\kappa_2$.
$\kappa_2$ will be the main product of such processes, as it is
mostly a triplet, and it will decay again to $W$'s which are difficult
to spot. However, a number of $\kappa_1$'s could be produced, depending
on the value of the doubly-charged mixing angle; the production amplitude
will in general be suppressed by the triplet VEV, $v_\chi$, and by
$\sin \theta_{+ +}$. Nonetheless, this could prove to be the
dominant production channel at LHC if $v_\chi > 1 \; \mathrm{GeV}$ and 
$m_{\kappa_{1,2}} > 500\; \mathrm{GeV}$ \cite{Huitu:1996su}. 
This is especially relevant for our model since the various constraints,
as discussed in section \ref{sec:WWee-model-constraints-parameters},
favour a relatively large $v_\chi$ as well as large scalar masses,
unlike other triplet models with tree-level neutrino masses, like
type II seesaw. A thorough study of the various possibilities is somewhat 
involved \cite{Melfo:2011nx}, but might deserve future investigation.

\section{Domain walls and alternatives} \label{sec:WWee-model-drawbacks}

The model that we have discussed in this chapter is an example of how to 
obtain $0 \nu \beta \beta$ and neutrino masses through the $WWee$ mechanism.
It is, however, not unique: it belongs to a class of models that share 
many features, and in particular the suppression of neutrino masses
with respect to $0 \nu \beta \beta$. It is also not speckless: containing as it does
a scalar that breaks spontaneously a discrete symmetry, it may trigger 
the creation
of domain walls in the early universe, which is forbidden by cosmological 
observations \cite{Zeldovich:1974uw,Kibble:1976sj,Vilenkin:1984ib}. 
Essentially, the problem is that domain walls have their own
energy density and should exhibit gravitational effects which have not been
detected; if the appropriate circumstances meet they could also yield
other effects that have likewise not been observed. 
The model, therefore, can be regarded as it stands as problematic; we
have preferred however to discuss in detail this particular version 
because we find
that the particle physics motivation is neater here, but in this section
we proceed to describe two possible variations that would tackle the
domain wall problem.

A first way to avoid the creation of domain walls is to promote the $\sigma$
field to a complex field, as already commented in section 
\ref{sec:WWee-model-presentation}. In this version of the model lepton number
is not explicitly broken, as it can be transmitted from $\kappa$ to $\chi$ 
through the $\mu_\kappa$ interaction and then again to $\sigma$ in the
$\lambda_6$ vertex. Specifically, we would have 
\begin{align*}
	L (\kappa) &= +2 & L (\chi) &= +1 & L (\sigma) &= -1 \, .
\end{align*}
Within this extension no discrete symmetry is needed: the enforcement of lepton
number suffices to forbid the terms that generate neutrino masses at tree level;
in fact, one can view the $Z_2$ symmetry in the model with $\sigma$ real
as a remnant of LN from a extended, high-energy theory which broke spontaneously
somehow. The complex-$\sigma$ extension, however, is not free of complications
either: lepton number is now a good global symmetry of the theory, and
as $\sigma$ acquires \mbox{a VEV --and it must} happen in order to have 
tree-\mbox{level $0 \nu \beta \beta$-- a Majoron} is created which requires
extra phenomenological considerations. Fortunately, the Majoron will be mainly 
a singlet and the constraints won't be very severe: its couplings to ordinary
matter will be small and its coupling to the Higgs field is essentially free
\cite{Chikashige:1980ui}. Even in these favorable circumstances, however, 
some constraints would need to be considered: the Majorons can be produced
inside a supernova and then escape, yielding a new source of supernova cooling
that can be bounded \cite{Choi:1989hi}; they can provide a mechanism for neutrino
decay or annihilation that would affect the neutrino relic abundance 
\cite{Chikashige:1980qk,Choi:1991aa}. One can choose to compute all these 
restrictions and then consider the region of parameter space where the model
is still viable. Another possibility is to gauge the lepton number 
sym\-\mbox{metry -- or rather} baryon minus lepton number, $B-L$, as $B+L$ is 
broken in the SM by nonperturbative effects \cite{Rubakov:1996vz}. This would
be an interesting possibility, and with renewed experimental interest 
(see, for instance, \cite{Basso:2011hn} for a review), but it requires the
addition of right-handed neutrinos, which provide new sources of neutrino
masses and somewhat complicate the scenario. Thus, we consider that it's not 
so appealing for the pure investigation of the $WWee$ mechanism.

A second option to evade the domain wall problem would be to keep the real
$\sigma$ field but engineer the model so that $v_\sigma$ is very heavy. The 
domain walls are created during a phase transition in the early universe 
that occurs roughly at a temperature $T_c \simeq v_\sigma$; if we manage to 
place this temperature before the inflationary epoch it is reasonable to 
think that the domain walls exist but are beyond our observable universe.
This requires, of course, \emph{huge} values for $v_\sigma$, but the phenomenology
of the model can be kept more or less the same just by choosing very small
values \mbox{for $\lambda_6$ -- so small,} actually, that some might consider that
a hierarchy issue is in order. We do not pursue this discussion because within 
this class of models there is an even simpler one, with the same neutrino physics 
at low energy and none of these potential drawbacks:
it is the same model presented throughout this chapter but without the $\sigma$
field; instead, we replace it by its vacuum expectation value, $\sigma \to
v_\sigma$. Up to coupling constant redefinition this yields the same potential 
as equation \eqref{WWee-model-scalar-potential} but without the terms containing 
$\sigma$, \emph{except} for the $\lambda_6$ term, which becomes 
\begin{equation} \label{WWee-model-muchi}
	\mu_\chi \, \phi^\dagger \chi \tilde{\phi} \, , \; \mathrm{with} \quad
		\mu_\chi = \lambda_6 \, v_\sigma \, . 
\end{equation}
This term is nothing but a explicit breaking of the $Z_2$ symmetry, and could
be regarded as a remnant of a high-energy model with a very heavy $\sigma$.
The $\mu_\chi$ coupling can be taken as $\mathcal{O} (\mathrm{TeV})$, and then
the resulting renormalizable Lagrangian has the same quantum behaviour than ours; 
in particular, neutrino masses are finite and generated at two loops,
and can be obtained from our results by eliminating $\lambda_6$
using \eqref{WWee-model-muchi}. Note that within this variation lepton number
is still explicitly broken, by the $\mu_\kappa$ and $\mu_\chi$ interactions, 
but none of them allows to assign $L(\chi) = 2$. The vertex
$\overline{\tilde{\ell}_{\mathrm{L}}} \chi \ell_{\mathrm{L}}$
is no longer forbidden, because $Z_2$ is broken, but it can be absent from
the \mbox{Lagrangian --if the} model is so \mbox{devised--, and it}
will be finitely generated by \mbox{loops -- by the}
analogous of the diagrams in figure 
\ref{fig:WWee-model-numass-WudkaGauge}, and we could check that the resulting
neutrino masses will be proportional to $v_\chi^2$, thus ensuring that the
$WWee$ mechanism is in action, and not a type-II-seesaw-like one.

\chapter{Conclusions}

In this doctoral dissertation we have presented our research on nonstandard 
properties of neutrinos through the interplay between effective operators
and models. The two approaches offer a different set of advantages and 
drawbacks: effective theories are very general and allow the study of
wide classes of models, but have a limited predictive ability; for some time
the community stressed this feature and disregarded effective theories as
not very powerful tools, but the attitude has changed in the past two decades
and now effective operators are actively utilised in theoretical particle
physics, including in the exploration of the boundaries of known physics.
Models, on the other hand, are complete frameworks which aim to full
predictivity; if renormalisable they can yield results up to arbitrary
energies and they include all the new particles, whose properties can
be investigated with no theoretical impediment. However, precisely because
of this, models need to include all the details that make them phenomenologically
viable, and these details can be complicated, even to the point of obscuring
the relevant mechanism and yielding too specific predictions. Our philosophy
is that it is in the combination of these two approaches that we can
benefit from the advantages of both while buffering their drawbacks.
In our investigation we have applied this principle to the study of new
properties of the neutrinos.

This work can be naturally divided into two distinct parts. The first covers 
chapters \ref{chap:nuR-magmo-eff} and \ref{chap:nuR-magmo-model}, and it is
centered on effective operators involving right-handed neutrinos, and especially
right-handed neutrino magnetic moments. It is rather unusual that the
$\nu_{\mathrm{R}}$'s are considered among the low-energy fields when constructing 
an effective theory; this is motivated by the fact that their role in neutrino mass
generation is most diaphanously understood when they are heavy and implement
a seesaw mechanism to yield small masses for the $\nu_{\mathrm{L}}$'s. However,
this is not the only possibility: for instance, neutrinos could be Dirac particles;
or, being Majorana, some of the $\nu_{\mathrm{R}}$ degrees of freedom might 
be lighter than the electroweak scale. Including right-handed neutrinos in an
effective theory as low-energy fields allows to explore their relation to 
high-energy particles, to use them as a window to new physics. 
With this aim in mind, we examined in chapter \ref{chap:nuR-magmo-eff} the 
lowest-order effective operators that involved Standard Model fields and 
right-handed neutrinos, and we found, as soon as at dimension five, two 
operators that have not been previously discussed in the literature. 

Both of them were lepton-number-violating operators. The first,
\begin{equation} \label{conclusions-nuR-nuR-Higgs}
	\overline{\nu_{\mathrm{R}}^\mathrm{c}} \nu_{\mathrm{R}} \, 
		\phi^\dagger \phi \, ,
\end{equation}
connects a pair of 
$\nu_{\mathrm{R}}$'s to a pair of Higgs fields, and it provides an additional
contribution to the Majorana mass of right-handed neutrinos once electroweak
symmetry breaks spontaneously. It also yields a Yukawa-like interaction
$\overline{\nu_{\mathrm{R}}^\mathrm{c}} \nu_{\mathrm{R}} \, H$ which triggers
decays of the Higgs boson to two right-handed neutrinos; in section 
\ref{sec:nuR-eff-Higgs-to-NN}
we discuss the issues related to these decays, which may include invisible
Higgs boson decays in certain zones of the parameter space.

The second operator is a magnetic-moment-like interaction between a pair 
of right-handed neutrinos and the hypercharge gauge boson,
\begin{equation} \label{conclusions-electroweak-moments}
	\overline{\nu_{\mathrm{R}}^\mathrm{c}} \sigma^{\mu \nu} \nu_{\mathrm{R}} \,
			B_{\mu \nu} \, ,
\end{equation}
which in turn yields proper magnetic moments for the $\nu_{\mathrm{R}}$'s plus
similar interactions with the $Z$ boson. The consequences of this set of interactions
would be remarkable: right-handed neutrinos, often regarded as `sterile', could
be produced anywhere a photon \mbox{is present -- that is,} everywhere. As we
have not \mbox{gathered --yet-- evidence} of the existence of $\nu_{\mathrm{R}}$'s
we can only conclude that either they are rather heavy or the effective interactions
are very \mbox{suppressed -- of course,} given the optimistic assumption that they 
both exist. Throughout chapter \ref{chap:nuR-magmo-eff} we discuss the phenomenology
of these electroweak-moment interactions; in section \ref{sec:nuR-eff-conclusions}
we summarise the constraints and prospects. Some basic conclusions include that if
the $\nu_{\mathrm{R}}$'s are very light astrophysical bounds push the new physics
responsible for the electroweak moments to very high energies; if the right-handed
neutrinos are heavier than $100 \; \mathrm{MeV}$ the best constraint is provided
by LEP and the new particles cannot be lighter than some TeV's,
but there's still room to be explored by the LHC.

If the new physics responsible for the right-handed neutrino effective operators
is weakly coupled it is pretty likely that the new particles are easier to
detect than the effective interactions. This is an additional reason to 
explore models that realise the operators \eqref{conclusions-nuR-nuR-Higgs} and
\eqref{conclusions-electroweak-moments}. In chapter \ref{chap:nuR-magmo-model} 
we provide such a model. It is devised as a minimal extension
and contains two new $SU(2)$ singlets: a scalar, $\omega$, and a vector-like 
fermion, $E$, both of which need to be charged so that the $B$
field couples to them while generating the electroweak moments.
The new particles must be endowed at least with two trilinear couplings,
$\overline{\nu_{\mathrm{R}}} \, E \,\omega^\dagger$ and
$\overline{\nu_{\mathrm{R}}^\mathrm{c}} \:  E \, \omega^\dagger$,
which combined break lepton number and thus allow to generate the 
operators we are interested in. $\omega$ and $E$ can also have a variety of
couplings to the Standard Model fields, but in order to simplify the 
model we only allow a mixing term between $E$ and the light right-handed 
charged leptons. 
This mixing induces the decay of the heavy charged
particles and avoids the problem of $E$ and $\omega$ accumulating as
charged dark matter, which occurs if we only allow for the trilinear terms.
We then discuss the phenomenology of the production and decay of the new
particles in collider experiments, and the constraints that can be set on 
the mixing of the heavy fermions. The model is
altogether phenomenologically viable, but from a wider view it may appear as an
\emph{ad hoc} mechanism. It would be interesting to consider in future works
whether this class of additions, and their LNV-ing trilinear couplings,
can be embedded in a more complex framework where their features are
more deeply understood.

The second part of the thesis comprises chapters \ref{chap:0nu2beta-eff},
\ref{chap:Wnue-model} and \ref{chap:WWee-model}, and in them we discuss
the relation
between neutrinoless double beta decay and neutrino masses for a certain class
of effective operators. The idea that even effective pictures of $0 \nu \beta \beta$
offer information about the generation of neutrino masses is far from new;
the same Schechter-Valle \mbox{theorem --that we} depict in \mbox{figure 
\ref{fig:intro-0nu2beta-Schechter-Valle}-- is an ex}\-ample of this sort of argument.
In chapter \ref{chap:0nu2beta-eff} we consider a family of effective interactions
that yield $0 \nu \beta \beta$: those that provide two charged leptons with 
violation of lepton number and with no quarks. We realised that the effective
lepton-number-violating interactions involving $W$'s and leptons had not received 
much attention in the literature; 
two such interactions yield $0 \nu \beta \beta$ at tree
level in the effective theory, as we see in figure \ref{fig:0nu2beta-eff-7-and-9}: 
the first with one $W$, one electron and one neutrino ($W \nu e$ vertex) and the 
second with two $W$'s and two electrons ($WWee$ vertex). The effective theory
also suggests that the $W \nu e$ interaction generates neutrino masses suppressed
by one loop with respect to $0 \nu \beta \beta$ and the $WWee$ vertex yields
neutrino masses suppressed by two loops (figure 
\ref{fig:0nu2beta-eff-numass-7-and-9}); however, this hierarchy will only be realised
if neutrino masses are not independently generated at tree \mbox{level -- that is,}
if the Weinberg operator is suppressed with respect to the $W \nu e$ and $WWee$
interactions. In order to isolate the various mass-generating operators we notice
that the chirality of the leptons involved in the effective interaction allows
some sort of selection on the dominant mechanism: indeed, the lowest-order operator
in this class with one left-handed and one right-handed lepton is the
dimension-seven
\begin{equation} \label{conclusions-0nu2beta-7}
	\overline{e_{\mathrm{R}}} \, \gamma^\mu \left( \phi^\dagger
			\tilde \ell_{\mathrm{L}} \right) \left( \phi^\dagger
			D_\mu \tilde \phi \right) \, ,
\end{equation}
which upon spontaneous symmetry breaking realises the $W \nu e$ interaction.
Similarly, the lowest-order operator with two right-handed leptons is the
dimension-nine
\begin{equation} \label{conclusions-0nu2beta-9}
	\overline{e_{\mathrm{R}}} \, e_{\mathrm{R}}^\mathrm{c} \left( \phi^\dagger
			D^\mu \tilde \phi \right) \left( \phi^\dagger D_\mu \tilde \phi 
			\right) \, ,
\end{equation}
which yields a $WWee$-like interaction. Chirality, thus, may be used in
model-building for favouring the desired mechanism.

Still in chapter \ref{chap:0nu2beta-eff}, we take the operators 
\eqref{conclusions-0nu2beta-7} and \eqref{conclusions-0nu2beta-9} as paradigmatic
of the $W \nu e$ and $WWee$ mechanisms and discuss the implications of the
current $0 \nu \beta \beta$ bounds on the mass of the required new particles. 
We obtain for the operator \eqref{conclusions-0nu2beta-7} that the new physics
scale should lie roughly above $100 \; \mathrm{TeV}$, with some dependence on
the effective \mbox{coefficient -- see} equation \eqref{bound-0nu2beta-O7}; 
essentially, if the coefficient is small the scale may be lowered to energies 
at the reach of present or planned experiments. For the operator 
\eqref{conclusions-0nu2beta-9} we obtain a lower bound on the new physics scale
of just $3 \; \mathrm{TeV}$, which can be further lowered if the effective 
coefficient \mbox{is small -- see} equation \eqref{bound-0nu2beta-O9}. In section 
\ref{sec:0nu2beta-eff-preeminence} we also address the question of the 
dominant contribution to neutrinoless double beta decay: the operators 
\eqref{conclusions-0nu2beta-7} and \eqref{conclusions-0nu2beta-9} induce
$0 \nu \beta \beta$, but so do the neutrino masses that they generate
through loops. Which one and under which circumstances will dominate?
The effective theory allows us to estimate that if the new particles are
relatively \mbox{light --in particular,} if the new physics scale is below 
\mbox{$30 \; \mathrm{TeV}$-- the effective} operators will dominate; however,
if the new physics is heavy $0 \nu \beta \beta$ will be dominated by neutrino
masses.

The possibility of inducing a hierarchy between $0 \nu \beta \beta$ and
neutrino masses that enhances the former and suppresses the latter is our 
main concern when considering the $W \nu e$ and $WWee$ interactions.
However, the mass-generating loop diagrams in figure 
\ref{fig:0nu2beta-eff-numass-7-and-9}, which are the source of our argument,
are actually not calculable in the effective theory, as it is not
renormalisable. In order to see if the argument can be realised we must turn 
to concrete models, and that is what we do in chapters \ref{chap:Wnue-model}
and \ref{chap:WWee-model}. Chapter \ref{chap:Wnue-model} presents a rapid
examination of a model that realises the $W \nu e$ mechanism and chapter
\ref{chap:WWee-model} reports more extensively on a model that yields
the $WWee$ mechanism and has prospects of observability in several near
future experiments. In both cases a main concern is to suppress any alternative
mass-generating mechanism so that neutrino masses are indeed induced by the
$W \nu e$ and $WWee$ interactions; this is achieved by invoking discrete
symmetries that forbid the undesired terms in the Lagrangian. In both cases
we check that $0 \nu \beta \beta$ can be induced at tree level and suppressed
neutrino masses are generated, at one loop for the $W \nu e$ mechanism and
at two loops for the $WWee$ mechanism.

Chapter \ref{chap:Wnue-model} proposes a realisation of the $W \nu e$ mechanism.
For that aim, we extend the Standard Model with a heavy scalar triplet,
a heavy vector-like fermion doublet and a light scalar doublet; therefore,
the model resembles an extended two-Higgs-doublet model. The operator
\eqref{conclusions-0nu2beta-7} can be generated either at tree level,
by spontaneous breaking of lepton number, or at one loop, by explicit breaking.
Neutrino masses, as we expect from the $W \nu e$ mechanism, can be generated
either at one loop by spontaneous breaking or at two loops by explicit breaking.
The preeminence of spontaneous or explicit breaking depends on the balance
of several parameters, but especially the magnitude of a quartic scalar 
\mbox{coupling --the one} that breaks explicitly lepton \mbox{number-- and the} 
VEV of the second Higgs doublet, which should be of the order of $v_\phi$ to ensure
the dominance of spontaneous breaking. Our phenomenological analysis yields
that explicit breaking should dominate if the masses of the heavy particles
lie roughly above $100 \; \mathrm{TeV}$, whereas spontaneous breaking is 
the leading mechanism for light masses at the reach of present or near future 
experiments. All in all, we have verified that the $W \nu e$ mechanism can be 
realised,
and $0 \nu \beta \beta$ dominated by the operator \eqref{conclusions-0nu2beta-7}
with one-loop suppressed neutrino masses is feasible, especially if the masses
of the heavy particles are below $35 \; \mathrm{TeV}$. However, our analysis
has not been exhaustive and many issues remain to be addressed; to cite just a
few, the violation of lepton family numbers should be examined in detail, and
variants of the model should be considered in order to determine if the
second Higgs doublet is a necessary ingredient of this class of models.

In chapter \ref{chap:WWee-model} we discuss a model that realises the $WWee$
mechanism and has the ability to produce signals in several ongoing or near-future
experiments. The model requires the addition of a heavy scalar triplet and a
heavy doubly-charged singlet; several variants of this setup are possible, and
we choose also to add a heavy real scalar singlet. The model yields $0 \nu \beta
\beta$ at tree level and neutrino masses at two loops. It also induces 
lepton-flavour-violating processes such as $\mu \rightarrow 3 e$ or
$\tau \rightarrow e + 2 \mu$. The generated neutrino masses are not only 
suppressed by loop factors, but also by factors of the masses of the charged
leptons, which induce a very definite texture on the mass matrix; as a consequence,
the model predicts a nonzero $\theta_{13}$ compatible with the recent measurement
of the Daya Bay experiment. The rate
of $0 \nu \beta \beta$, the rate of LFV-ing processes and the elements of the
mass matrix are all controlled by the same set of couplings, which induces
tensions among the various phenomenological and theoretical constraints;
in particular, if one demands that the model provides a positive signal in 
the present or near-future $0 \nu \beta \beta$ experiments, the allowed
range for the parameters gets drastically narrowed, though it remains open.
The constraints are discussed in detail in section 
\ref{sec:WWee-model-constraints-parameters}; we may conclude from them
that the model is definitely viable from the phenomenological point of view.
It can comply with the experimental limits, it can provide signals in $0 \nu \beta
\beta$ searches and lepton flavour violation experiments, and the heavy particles
can have masses around the TeV or even \mbox{lighter -- but if} we demand 
all these features at the same time the situation gets tight. Relaxing the
requirement of an imminent $0 \nu \beta \beta$ signal helps to get more
space in the heavy-mass range, and relaxing the theoretical constraint of
naturality yields more room for light new particles. All in all, this model
and its cognates form a family with a rich phenomenology and realise satisfactorily
the $WWee$ mechanism, providing rather suppressed neutrino masses while 
keeping the $0 \nu \beta \beta$ rate near observable levels. These models will be
tested in the next years by the forthcoming round of searches for lepton number
and lepton flavour violation.

To end up, along this work we have explored several ways in which effective
field theories may yield insight about neutrino properties and provide fruitful
ideas for model building. We hope that the next years provide us with new 
experimental data to tune up our aims, separate the promising ideas from those
that were just fun to explore and inspire new directions of research. It's been
a long way, but the road goes ever on.

\chapter{En la lengua de Cervantes}

El trabajo doctoral que acabamos de exponer presenta \mbox{nuestras 
--modes}\-\mbox{tas-- contribuciones} a la física de los neutrinos. 
En este capítulo ofrecemos un resumen en castellano, como la legislación 
requiere; aprovecharemos este obligado \emph{impasse} para enmarcar nuestros
trabajos con un lenguaje no técnico, pensando en el público no experto.
En el trayecto cometeré, no lo dudo, alguna imprecisión; espero que esos
defectos, imposibles de subsanar en un texto breve como éste, se vean compensados
por algún otro acierto, y que al final el lector termine con la impresión de
que sabe algo más que al principio.

\section{El título}

Parece lógico empezar por el principio, o sea, por explicar el título de este
trabajo: \emph{Exotic properties of neutrinos using effective Lagrangians
and specific models}, o lo que es lo mismo: ``Propiedades exóticas de los
neutrinos a través de lagrangianos efectivos y modelos concretos''. Esta
tesis tiene un protagonista absoluto: el neutrino; de él daré alguna cuenta
en la sección \ref{sec:cervantes-neutrinos} de este mismo capítulo. Nuestro
trabajo ha consistido en estudiar propiedades \emph{exóticas} de esta partícula,
es decir, propiedades que uno \emph{no} esperaría encontrar en los neutrinos;
punto número uno, pues: nuestra investigación se centra en propiedades hipotéticas, 
que nunca han sido observadas. Si el lector se está preguntando ``¿y cómo se estudia
algo que nadie ha visto?'', la respuesta es simple: con lápiz, papel y ordenadores.
Aún nos quedan muchas cosas que aprender sobre los neutrinos, pero una que hemos
aprendido es que su comportamiento viene excelentemente descrito por la 
teoría cuántica de campos, un fabuloso conjunto de herramientas matemáticas que
nos permiten predecir cómo se comportan las partículas.
Los neutrinos, junto con el resto de partículas
conocidas, se encuadran en una gran teoría llamada Modelo Estándar, que
básicamente plantea las reglas del \mbox{juego --las masas,} las cargas, 
las relaciones entre las diferentes partí\-\mbox{culas-- que luego} vamos a usar 
cuando pongamos en marcha la máquina de la teoría cuántica de campos. Pero el 
Modelo Estándar no son las únicas reglas posibles; son unas que funcionan muy
bien, pero que probablemente necesiten alguna corrección o algún añadido.

En ese punto entramos nosotros, y lo que hacemos es proponer \emph{extensiones}
del Modelo Estándar, añadidos que ayuden a entender las cosas que el Modelo Estándar
``puro'' no deja del todo \mbox{claras -- alguno} de estos ``defectos'' lo comentamos
en la sección \ref{sec:cervantes-neutrinos} de este mismo capítulo. Estas extensiones,
sin embargo, no salen gratis: aparte de resolver tus problemas pueden crear
algunos nuevos (!). Nuestro trabajo es estudiar esas propuestas, comprobar cuáles
son sus consecuencias y asegurarnos de que todas están bajo control; por ejemplo,
asegurarnos de que cierta extensión que ayuda a entender la masa de los neutrinos
no destruye todos los átomos del universo. Por el camino, estas extensiones dan
lugar a fenómenos exóticos: cosas que no esperaríamos encontrar ahí y que, si
llegásemos a observar, supondrían un punto a favor de esa extensión en concreto,
y quizá el primer paso hacia el descubrimiento de algo nuevo. Esos fenómenos exóticos
pueden ser muy variados: desde lo más escandaloso, como que se observe una
partícula desconocida en un acelerador, hasta lo más sutil, como que la
probabilidad de cierta desintegración nuclear sea un poco mayor o un poco menor
de lo esperado.

Así pues, ya tenemos las \emph{Exotic properties of neutrinos}; el resto
del título sólo explica el \emph{cómo}: en ocasiones hemos usado lagrangianos
efectivos y en ocasiones hemos usado modelos. ¿Cuál es la diferencia? Pues hay
una diferencia de profundidad y de generalidad. Cuando
uno dice que usa una teoría \emph{efectiva} lo que está haciendo es admitir
que sólo aspira a describir el nuevo fenómeno, pero no a explicarlo. Explicar
un fenómeno implica decir qué nuevas partículas hacen falta, cuáles son exactamente
sus propiedades, cómo se relacionan con las partículas que ya conocemos\dots 
$\;$especificar todos estos detalles es construir un \emph{modelo}. Por ejemplo,
digamos que el fenómeno que queremos estudiar es que los neutrinos forman cuerpos
sólidos del tamaño de una persona%
\footnote{Esto es completamente ciencia ficción. Prueba de ello es que
Stanis{\l}aw Lem, uno de los maestros del género, se plantea este fenómeno en su 
novela \emph{Solaris} \cite{lem:solaris}.
};
estudiarlo de manera efectiva consistiría en estudiar estos cuerpos sólidos,
su dureza, su estabilidad, cómo interaccionan con la luz, pero nunca plantearse
qué es lo que los mantiene unidos. Estudiar este fenómeno mediante un modelo 
sería describir por qué los neutrinos se unen, cómo sucede esto exactamente,
qué nuevas partículas harían falta para ello\dots

Los lagrangianos efectivos son una estupenda manera de centrarse en el fenómeno
relevante; cualquier conclusión que uno extrae
de un análisis efectivo debe ser válida para todos los modelos que aspiren a
explicar el fenómeno, así que el enfoque efectivo es también muy general.
Por contra, es incompleto; tal vez una versión realista de tu fenómeno requiere
condiciones que tu teoría efectiva no llega a ver. Por eso es útil construir
un modelo: para ver en qué medida te encuentras con dificultades en el momento 
en que quieres hacer algo realista y consistente. Ambos enfoques se complementan,
y ésa, la de buscar esa complementariedad, es la filosofía que hemos seguido
en este trabajo.

\section{Neutrinos} \label{sec:cervantes-neutrinos}

En esta sección queremos explicar qué es exactamente un neutrino y por qué nos
interesa lo suficiente como para que cada año centenares de personas en todo
el mundo escriban sus tesis sobre ellos. Para decirlo en una sola frase, 
los neutrinos son un tipo de partícula elemental que se caracteriza muy 
especialmente por sus débiles interacciones. Aunque se conocen decenas de 
partículas diferentes%
\footnote{El lector interesado puede consultar las listas que aparecen en 
Wikpedia \cite{wiki:list-particles}, o mejor, las exhaustivas tablas del 
Particle Data Group \cite{Beringer:1900zz,pdg:tables}.
},
sólo unas pocas son \emph{elementales}, esto es, 
indivisibles; la física moderna lista por centenares las partículas 
\emph{compuestas}, formadas por asociación de otras partículas, como pequeños 
enjambres o sistemas planetarios, pero apenas un \mbox{puñado --una} 
\mbox{veintena-- parecen} estar formadas sólo por ellas mismas: son elementales. 
Los neutrinos son
tres de los miembros de ese selecto club, caracterizados por interaccionar
de manera muy limitada con el resto de las especies; esta propiedad las hace
únicas entre las partículas conocidas: sean elementales o compuestas, todas
las demás partículas interaccionan más que los protagonistas de este trabajo.

¿Pero qué significa exactamente ``interaccionar débilmente''? En pocas palabras,
que los neutrinos son capaces de atravesar ingentes cantidades de materia
sin darse cuenta de que está ahí. Un cálculo sencillo nos arroja rápidamente
que para tener buenas chances de frenar un neutrino de baja energía es necesario 
interponerle \emph{un año-luz} de plomo. Lo repetiré: si a un neutrino le pones
delante ``sólo'' medio año-luz de plomo es bien probable que lo atraviese 
sin inmutarse. Este dato es sólo un dato, una anécdota, pero extendámoslo al 
quehacer científico:
imaginemos lo difícil que es estudiar un objeto, el que sea, cuando para estudiarlo
hace falta que deje alguna información en nuestras máquinas\dots $\;$y para el
99,999\dots \% de los neutrinos nuestras máquinas son tan transparentes como 
el aire.

Las razones para este comportamiento están codificadas en las reglas del juego
que nos da el Modelo Estándar: los neutrinos no tienen carga eléctrica, así que
no interaccionan con los fotones, las partículas que forman la luz. Sí pueden
interaccionar con los electrones, protones y neutrones que hay en los átomos, pero
siempre con la ayuda de terceras partículas mucho más pesadas, como los bosones
$W$ y $Z$; la mediación necesaria de estas partículas pesadas hace, por las
reglas de la teoría cuántica de campos, que la interacción con electrones,
protones y neutrones sea muy improbable, y de ahí el carácter ``fantasmal'' de los
neutrinos. Sin embargo, todo esto es conocimiento asentado: lo entendemos y
lo tenemos controlado desde hace varias décadas. ¿Qué es, pues, lo que hace
que los neutrinos sigan siendo interesantes para la comunidad científica \emph{hoy}?

Hay varias respuestas a esta pregunta: la más sencilla quizá sea que debido a 
su carácter huidizo aún son relativamente desconocidos; otra posible respuesta
es que este carácter permite que los neutrinos tengan comportamientos curiosos,
como la oscilación, que hace los tres tipos de neutrino cambien de
una variedad a otra sólo por el hecho de moverse. Pero la que nos interesa
aquí es una tercera razón: su \emph{masa}. Desde que, allá por el año 2000,
quedó claro que los neutrinos que se producen en el Sol no desaparecen a medio
camino, sino que se transforman en otro tipo de \mbox{neutrino --\emph{oscilan}} 
entre las tres \mbox{variedades--, las} ecuaciones nos exigían admitir que 
los neutrinos habían de tener masa. Hasta entonces no habíamos tenido una sola
prueba fehaciente de que no fueran partículas sin masa, como los fotones: se movían 
a altas velocidades, muy cercanas a la de la luz, y cuando intervenían en reacciones
nucleares no dejaban constancia de su carácter masivo. Todas las evidencias previas
apuntaban a que, si tenían masa realmente, había de ser muy pequeña, tan pequeña
que quedaba eclipsada por su energía cinética, la energía de su movimiento%
\footnote{Es un hecho poco conocido que éste es un principio general: la masa
es una forma de energía y el movimiento, otra. En el mundo en que vivimos los
objetos tienen mucha más energía acumulada en su masa que en su movimiento, y
sería un poco tonto decir que tenemos dudas de si un sofá tiene masa o no: es
evidente que la tiene. En el mundo de las partículas, sin embargo, a veces sucede 
lo contrario: el movimiento aporta mucha más energía que su masa. Entonces se da
el caso contrario: es evidente que está en movimiento, pero ¿tiene masa? Las
partículas con mucha más energía cinética que masa son prácticamente indistinguibles
de una partícula sin masa, y sólo cuando las frenamos empezamos a ``darnos cuenta''
de que son masivas. Los neutrinos, sin embargo, con su dificultad para interaccionar
con la materia también son difíciles de frenar, así que la faena está servida:
¿cómo medimos sus masas?
}.
A día de hoy seguimos intentando observar neutrinos más y más lentos, con la
esperanza de distinguir en su comportamiento alguna pista sobre su pequeñísima masa, 
pero hasta ahora seguimos a ciegas.

Así pues, aquí está el quid de la cuestión: sabemos que los neutrinos tienen masa,
y sabemos que es muy muy \mbox{pequeña -- es,} al menos, un millón de veces menor que
la masa del electrón, que ya de por sí tiene una masa pequeña entre todas las
demás partículas. Es más, los neutrinos están solos como partículas con masas
ridículamente pequeñas: las demás se mueven entre los 500 keV del electrón y los 175
GeV del quark top, una horquilla de un factor un millón; por debajo del electrón
hay otra horquilla de un factor un millón aparentemente \emph{vacía}, sin ninguna
partícula conocida; y debajo de ésta, en el sótano de las masas, los tres neutrinos,
haciendo gala de su singularidad. Pues bien, esa singularidad es la que los hace
interesantes: sabemos que las masas del resto de partículas se pueden explicar
gracias al mecanismo \mbox{de Higgs --ahora ya} podemos decirlo, \emph{¡al fin!}, 
después de que en 2012 descubriéramos al deseado \mbox{bosón--, pero} a nadie se 
le escapa que tres individuos un factor un millón por debajo resultan sospechosos:
toda la comunidad piensa que lo lógico es que sus masas vengan de otro sitio,
de otro mecanismo.

¿Cuál es ese mecanismo? Bueno, ésa es la cosa; hay decenas de propuestas, decenas
de candidaturas para explicar que los neutrinos funcionan de otra manera y por qué
su manera de funcionar les da masas tan \mbox{pequeñas -- podría} ser que los neutrinos
se relacionen con una nueva clase de partícula, hasta ahora desconocida, que 
de alguna manera les ``obligue'' a tener masas pequeñas; podría ser que alguna
nueva ley de la física exija que sus masas, y no otras, tienen que ser pequeñas.
Pero más allá de esa lista de
candidaturas no tenemos nada: no hemos podido medir aún el valor de las masas,
y los experimentos en los que esperábamos ver alguna pista sobre el origen de la
masa de los neutrinos han dado resultados negativos. En este contexto se inscribe
mi tesis; hay toda una comunidad de físicos teóricos pensando en maneras para que
los neutrinos tengan masas pequeñas y, sobre todo, tratando de relacionar esos 
nuevos mecanismos con experimentos actuales, para que podamos ponerlos a prueba.
En esa dirección transitan los trabajos que aquí hemos presentado, y que ahora
describiremos brevemente.

\section{Momentos magnéticos de neutrinos dextrógiros} \label{sec:cervantes-magmo}

Todos los mecanismos que buscan explicar la pequeñez de la masa de los neutrinos
pasan por añadir partículas nuevas al elenco que conocemos actualmente. Una de las
incorporaciones más populares es añadir, por cada uno de los neutrinos que tenemos%
\footnote{O sea, en principio tres, pero luego se pueden hacer variaciones: añadir
más, añadir menos, hacer que las neuvas partículas se fusionen con los neutrinos
convencionales\dots $\;$toda esa zoología modelística requeriría mucho más espacio
del que disponemos hoy aquí.
},
una nueva partícula, similar a éstos pero con una interacción todavía más débil
con el resto de la fauna subatómica. Esas nuevas partículas reciben el nombre de
\emph{neutrinos dextró}\-\mbox{\emph{giros} --aunque} a muchos les sonará más la 
terminología inglesa: right-handed \mbox{neutrinos--, o,} a veces, neutrinos 
\emph{estériles}, donde esta desagradable palabra hace referencia precisamente
a que nuestros nuevos inquilinos parecen sentir cierta aversión a relacionarse con 
el resto de partículas. El carácter estéril de los neutrinos dextrógiros, sin
embargo, no contribuye particularmente a explicar las masas de los neutrinos 
\mbox{``normales'' -- la} verdadera utilidad de los recién llegados reside en otras
de sus propiedades, que no corresponde comentar aquí.

Así que fue hasta cierto punto una sorpresa descubrir, mientras estudiábamos algo
ligeramente diferente, que nadie había considerado que los neutrinos dextrógiros
pudieran tener un momento magnético. El momento magnético es una interacción entre
una pareja de partículas y un fotón pero, a diferencia del acoplamiento ``habitual''
de los \mbox{fotones --vectorial} y \mbox{gauge--, el} momento magnético involucra 
necesariamente a terceras partículas. Con el acoplamiento vectorial gauge, las 
partículas que interaccionan con el fotón han de tener necesariamente carga eléctrica;
con el momento magnético basta con que las terceras partículas implicadas la tengan%
\footnote{El ejemplo paradigmático es un átomo. Los átomos son neutros porque tienen
el mismo número de protones y de electrones. Como tales objetos neutros no deberían
interaccionar con fotones, pero como en su interior hay electrones y protones que sí
tienen carga, los átomos como conjunto pueden interaccionar con los fotones a 
través de un momento \mbox{magnético -- que,} en última instancia, involucra a los 
electrones y protones de su interior.
}.
Conclusión: los neutrinos dextrógiros, paradigmáticamente estériles y sin carga
eléctrica, pueden interaccionar con los fotones siempre que, además, interaccionen
con otras partículas que sí tengan carga.

Como digo, esto es conocimiento estándar, nada nuevo hasta aquí. Lo que fue 
sorprendente fue darnos cuenta, mientras estudiábamos posibles interacciones
exóticas de los neutrinos dextrógiros, que nadie había estudiado una relativamente
poco exótica: los momentos magnéticos. Esto es lo que nos proponemos en las 
referencias \cite{Aparici:2009fh} y \cite{Aparici:2009mj}. Hay que decir por 
adelantado que a día de hoy no tenemos una sola evidencia de la existencia
de neutrinos dextrógiros, y por ende tampoco de sus momentos magnéticos; es más,
si los fotones andaran por ahí produciendo neutrinos deberíamos habernos dado cuenta
ya. El hecho de que no lo hayamos observado señala, genéricamente hablando, que
o bien la interacción no existe, o bien es muy débil, o bien los neutrinos
dextrógiros tienen masas muy grandes y requieren de fenómenos muy energéticos
para manifestarse.

En la referencia \cite{Aparici:2009fh} ponemos números a toda esta casuística.
En ella estudiamos los momentos magnéticos de 
neutrinos dextrógiros desde un punto de vista efectivo, es decir, olvidándonos
de cómo se origina la \mbox{interacción --quiénes} son las partículas cargadas 
con las que el neutrino se \mbox{relaciona-- y}
concentrándonos simplemente en cuáles serían las consecuencias de que los fotones
se relacionaran con neutrinos dextrógiros a través de un momento magnético.
Y de estas consecuencias hay un buen puñado: el universo está 
lleno de fotones, y nuestros experimentos están llenos de fotones; las estrellas
producen ingentes cantidades de fotones, los aceleradores de partículas también;
si este momento magnético fuese un hecho, los neutrinos dextrógiros deberían 
haber dejado su huella en todos estos sitios. Como no hemos encontrado esa 
huella, nuestro corpus de observaciones debería decirnos cómo \emph{no es}
el momento magnético: si es muy débil, cuán debil ha de ser para que no lo hayamos
visto; si los neutrinos han de ser pesados, cuán pesados. En la sección 
\ref{sec:nuR-eff-colliders} analizamos las consecuencias que los momentos magnéticos
tendrían en aceleradores de partículas; en la sección \ref{sec:nuR-eff-astro-cosmo}
analizamos cómo los neutrinos dextrógiros afectarían a diversos escenarios
astrofísicos y cosmológicos. La figura \ref{fig:nuR-eff-summary-bounds} resume
las conclusiones del estudio: en el eje de abscisas representamos la masa de los
neutrinos dextrógiros y en el de ordenadas la ``escala de nueva física'', que 
básicamente debería corresponderse con la masa de las nuevas partículas que
generan los momentos magnéticos. Las regiones sombreadas marcadas como ``$N$ magnetic
moment'', ``$N - \nu$ transition'' y ``LEP'' representan regiones excluidas:
valores de las masas que entrarían en contradicción con lo que hemos observado.

Por ejemplo, para masas pequeñas, por debajo de $10 \; \mathrm{keV}$, la región
sombreada azul sitúa las nuevas partículas por encima de $10^9 \; 
\mathrm{GeV}$, lo que implica un momento magnético pequeñísimo, prácticamente
inobservable. Esta cota proviene de la física de estrellas gigantes rojas, cuya 
evolución y propiedades en cada etapa de su existencia depende, entre otras cosas,
de cuán rápido se enfrían; si los fotones que hay dentro de la estrella
pudieran transformarse en neutrinos dextrógiros, éstos escaparían de la estrella
llevándose parte de su energía, y por tanto enfriándola. Nuestras observaciones
de este tipo de estrella nos dan información sobre cuán rápido se enfrían, 
y nos permiten deducir que la práctica totalidad de su enfriamiento procede por
los cauces \mbox{usuales -- fotones} y neutrinos convencionales; por tanto, se
deduce que los momentos magnéticos, de existir, han de ser muy pequeños para
que el número de neutrinos dextrógiros que se produzca sea muy limitado.

En la referencia \cite{Aparici:2009mj} analizamos un modelo muy simple que 
proporciona momentos magnéticos a los neutrinos dextrógiros. El modelo requiere
dos partículas nuevas: un fermión pesado, que sería similar a un electrón pero
con una masa mucho mayor, y un escalar \mbox{cargado -- digamos,} una suerte de bosón
de Higgs, pero a diferencia de éste, con carga eléctrica. El modelo es viable
fenomenológicamente, y arrojaría señales en experimentos tales como el LHC de 
Ginebra, aunque en el tiempo que lleva funcionando no ha encontrado rastro
de partículas exóticas como éstas. Por lo demás, el modelo demuestra que es 
relativamente sencillo generar los momentos magnéticos de neutrinos dextrógiros
sin necesidad de extensiones complicadas del Modelo Estándar, y las predicciones
del modelo coinciden esencialmente con las que se obtienen de la teoría efectiva
en la referencia \cite{Aparici:2009fh}.

\section{Desintegración doble beta sin neutrinos} \label{sec:cervantes-0nu2beta}

La desintegración doble beta es un tipo muy raro de desintegración nuclear, en la
que dos neutrones del núcleo atómico se transforman en dos protones y emiten dos
electrones. Este proceso ha sido observado en una decena de núcleos, y en todos
los casos se ha producido con la emisión de dos antineutrinos; por ejemplo, 
el núcleo de $\tensor[^{136}]{\mathrm{Xe}}{}$ se desintegra a 
$\tensor[^{136}]{\mathrm{Ba}}{}$,
\begin{equation} \label{cervantes-2nu2beta-Xe}
	\tensor[^{136}]{\mathrm{Xe}}{} \longrightarrow 
			\tensor[^{136}]{\mathrm{Ba}}{} + 2 e^- + 2 \bar \nu_e \, ,
\end{equation}
con una vida media de $2,4 \times 10^{21}$ años%
\footnote{Hemos dicho que estos procesos son en extremo raros: efectivamente, la
vida media del $\tensor[^{136}]{\mathrm{Xe}}{}$ es billones de veces más larga
que la vida del universo. Para poder observar esta desintegración en un experimento
hay que reunir grandes cantidades de xenón radiactivo, y aun entonces apenas
se observa la desintegración de un puñado de núcleos.
}.
La emisión de los dos antineutrinos no es baladí: las ecuaciones del
Modelo Estándar nos dan unas cuantas normas sobre cómo puede o no puede producirse
una partícula. Los electrones pertenecen a una familia llamada \emph{leptones},
y una de las normas del Modelo Estándar es que el número total de leptones
se mantiene constante en todos los procesos físicos usuales (hay unos pocos 
en los que no, pero sólo se dan a altísimas temperaturas); esta constancia del número 
de leptones se conoce como \emph{conservación del número leptónico}. Aplicado
a la desintegración doble beta, esto quiere decir que si se han producido dos 
electrones (leptones) hay que producir dos \emph{anti}leptones para compensar; 
esos dos antileptones son los dos antineutrinos. La desintegración del 
$\tensor[^{136}]{\mathrm{Xe}}{}$ que hemos escrito en \eqref{cervantes-2nu2beta-Xe} 
respeta la conservación del número leptónico, como es de esperar.

Ahora bien, volviendo por un instante a nuestra cruzada por explicar la pequeñez 
de las masas de los neutrinos, resulta que la gran mayoría de las propuestas
exige que el número leptónico no se conserve. O, dicho con un poco más de precisión,
resulta que hay muchas maneras de justificar que la masa del neutrino sea pequeña
si éste es su propia antipartícula. Si esto confunde al lector, que piense lo 
siguiente: con un electrón es fácil distinguir entre partícula y antipartícula;
el electrón tiene carga $-1$ y el positrón, su antipartícula, la contraria, $+1$.
Pero con el neutrino, que no tiene carga eléctrica, ¿cómo aplicamos ese
criterio? Hay que diseñar pruebas más sutiles, que pasan por separar el 
supuesto neutrino del supuesto antineutrino y ver si se comportan de manera
diferente; pero dado que los neutrinos se resisten tan tozudamente a ser detectados
y manipulados esas pruebas no son factibles. El hecho, a día de hoy, es que no 
sabemos si cuando hablamos de ``neutrinos'' y ``antineutrinos'' estamos hablando
de dos partículas diferentes, ambas fantasmales, ambas sin carga eléctrica, o de la 
misma, que pasa por delante de nuestros ojos travestida de una guisa o de otra
sin que nosotros nos demos cuenta.

Hay, sin embargo, una manera de poner a prueba este ``carácter'' del neutrino:
si el neutrino es su propia antipartícula querría decir que es leptón y antileptón
al mismo tiempo. Tal cosa no es posible, es un sinsentido; lo es hasta el punto
de que hace que el número leptónico deje también de tener sentido: ¿cómo vamos
a contar el número de leptones que hay en una habitación si hay unas cuantas
partículas que son, a la vez, un $+1$ y un $-1$? Si el neutrino es su propia 
antipartícula inmediatamente el número leptónico deja de ser una vara de medir
válida; inmediatamente deja de ser constante; e inmediatamente procesos como
\eqref{cervantes-2nu2beta-Xe} pueden ocurrir de otras maneras. Los detalles en
cuanto a
de qué maneras se pueden producir y cuán probable es cada camino pueden \mbox{llenar
--y llenan-- mono}\-grafías enteras y hoy no los comentaremos, pero sí vamos a
señalar una alternativa: la desintegración del $\tensor[^{136}]{\mathrm{Xe}}{}$
podría también producirse de la siguiente manera:
\begin{equation} \label{cervantes-0nu2beta-Xe}
	\tensor[^{136}]{\mathrm{Xe}}{} \longrightarrow 
			\tensor[^{136}]{\mathrm{Ba}}{} + 2 e^- \, ,
\end{equation}
es decir, con producción de dos electrones sin ningún antileptón para compensar,
en franca violación de la conservación de número leptónico. Si los neutrinos
fuesen su propia antipartícula, procesos como \eqref{cervantes-0nu2beta-Xe}
deberían ocurrir; pero es más: un célebre artículo del año 1982 
\cite{Schechter:1981bd} demostró que también lo contrario es verdad: si se observa
un proceso como \eqref{cervantes-0nu2beta-Xe}, entonces el neutrino es su 
propia antipartícula.

Procesos como \eqref{cervantes-0nu2beta-Xe} reciben el nombre de \emph{desintegración
doble beta sin neutrinos} (abreviado, $0 \nu \beta \beta$) y, como confío que 
el lector comprenda ahora, son de la máxima relevancia para todos los que 
investigamos las propiedades de estas partículas%
\footnote{Resulta irónico: para aprender sobre las propiedades del neutrino
debemos diseñar complejos experimentos buscando una desintegración en la que
el neutrino no aparece.
}.
Nuestra contribución a este campo ha sido analizar una familia de interacciones
que violan la conservación del número leptónico, centrándonos en cómo dan lugar
a $0 \nu \beta \beta$ y cómo nos pueden ayudar a entender las masas de los 
neutrinos. 

Las interacciones que hemos estudiado se dividen en dos clases: las que involucran
a un electrón, un neutrino y un bosón $W$, a las que llamamos ``tipo $W \nu e$'', 
y las que involucran a dos electrones y dos bosones $W$, que llamaremos ``de tipo
$WW ee$''. En la referencia \cite{delAguila:2012nu} y el capítulo 
\ref{chap:0nu2beta-eff} las analizamos desde un punto de vista estrictamente
efectivo, es decir, olvidándonos de qué las origina y pensando únicamente
en sus consecuencias para $0 \nu \beta \beta$ y masas de neutrinos. La primera
cuestión que nos ocupa es cómo evaluar si nuestras interacciones constituyen
la principal contribución a la desintegración doble beta; habitualmente, incluso 
en modelos sencillos, uno va a tener más de un proceso físico que dé lugar a
$0 \nu \beta \beta$, pero nosotros querremos construir modelos en los que 
la contribución más importante venga de las interacciones $W \nu e$ o $WWee$.
A lo largo de nuestro análisis desarrollamos un criterio para separar las
interacciones que nos interesan de otras que a menudo las acompañan y que también
producen $0 \nu \beta \beta$; de esta manera el físico que está construyendo
un modelo sabe dónde tiene que mirar y en qué puntos ha de ser cuidadoso si quiere
que las interacciones $W \nu e$ y $WWee$ sean las más relevantes. Este análisis 
culmina con una estimación de que las partículas que generan las nuevas
interacciones deberían tener masas por debajo de $35 \; \mathrm{TeV}$ para que
las interacciones $W \nu e$ y $WW ee$ sean las dominantes en $0 \nu \beta \beta$.

Seguidamente analizamos, siempre desde el punto de vista efectivo, qué información
nos da acerca de las nuevas interacciones lo que
sabemos sobre desintegración doble beta y masas de neutrinos. $0 \nu \beta \beta$
nos da cotas bastante definidas sobre las masas de las partículas que generan
las nuevas interacciones: las responsables de la interacción $W \nu e$ deberían
tener masas por encima de $100 \; \mathrm{TeV}$, y las asociadas con la interacción
$WWee$ deberían estar por encima de los $2 \; \mathrm{TeV}$, aunque en ambos
casos las masas podrían ser algo más ligeras dependiendo de los detalles del
modelo subyacente. El caso de la interacción $WWee$, pues, nos ofrece una
perspectiva bastante optimista, pues partículas de $2 \; \mathrm{TeV}$ o algo
más ligeras podrían ser vistas en el LHC. Las masas de neutrinos nos ofrecen
una información en principio menos fiable: no es posible, desde la teoría
efectiva, calcular las masas de los neutrinos sin comprometerse con un modelo 
\mbox{concreto -- cosa} que no queremos hacer, porque parte del poder de la teoría
efectiva es estar por encima de los modelos. Además, nuestro conocimiento de las
masas de los neutrinos es muy incompleto, y es difícil usarlas para obtener
información sobre las nuevas interacciones. En cualquier caso, sí es posible hacer
estimaciones, aunque éstas están necesariamente sometidas a mucha incertidumbre;
tales estimaciones arrojan que las masas de las nuevas partículas deberían estar,
para la interacción $W \nu e$, alrededor de 40 millones de TeV, una cifra 
estratosférica que las haría imposible de descubrir en cualquier experimento 
presente o del futuro cercano. La misma estimación ligada a masas de neutrinos
sitúa las partículas asociadas con la interacción $WW ee$ en masas de $1000 \;
\mathrm{TeV}$, más razonables pero aún demasiado altas para cualquier experimento
que podamos imaginar. Estas estimaciones, pues, enfrían un poco el optimismo de
las cifras anteriores, aunque sabemos que sólo son estimaciones. Para dilucidar
la cuestión lo mejor es usar los criterios que antes hemos descrito para construir
modelos concretos de las interacciones $W \nu e$ y $WW ee$; puesto que los modelos
contienen toda la información, éstos sí nos darán información fiable sobre
las masas de los neutrinos y sobre cuán ligeras pueden ser las nuevas partículas.

En el capítulo \ref{chap:Wnue-model} describimos brevemente un modelo que
da lugar a la interacción $W \nu e$. Este modelo requiere numerosas nuevas partículas,
incluyendo nuevos fermiones pesados, que se comportarían de manera similar a
electrones y neutrinos pero con masas mucho mayores, y nuevos escalares pesados,
en particular uno con carga $+2$ que podría ser fácil de detectar en un
acelerador. El modelo es complicado y ofrece varias maneras de generar la 
interacción $W \nu e$ pero, a cambio, es más optimista de lo que cabría esperar
del análisis efectivo: las nuevas partículas podrían ser relativamente ligeras
y producir alguna señal en el LHC o en otros experimentos que exploran nueva
física en las cercanías del TeV. Sin embargo, nuestro análisis de este modelo
ha tenido que ser somero por cuestiones de tiempo, y sería deseable volver sobre
él para estudiar otros aspectos de su fenomenología, como sus contribuciones a
procesos con violación de sabor leptónico.

En el capítulo \ref{chap:WWee-model} y la referencia \cite{delAguila:2011gr} 
analizamos con detalle un modelo que da lugar a la interacción $WW ee$. El modelo
requiere una cantidad de nuevos \mbox{escalares -- partículas} similares al bosón
de Higgs, en este caso con cargas eléctricas 0, 1 y 2. Se trata de un modelo 
bastante prometedor, porque permite generar un espectro de masas de neutrinos
compatible con lo observado y, a la vez, podría dar señales en una variedad
de experimentos, desde el LHC a experimentos de $0 \nu \beta \beta$. Nuestra
discusión se centra en si es posible obtener el pleno total: proporcionar masa
a los neutrinos, ser compatible con las observaciones que tenemos a día de hoy
y arrojar también una señal en la nueva generación de experimentos de desintegración
doble beta que está empezando a tomar datos actualmente. La respuesta es sí, 
aunque al exigir que el modelo cumpla en tantos frentes los valores posibles
para las masas de las nuevas partículas quedan bastante restringidos: si
queremos observar una señal en los experimentos de $0 \nu \beta \beta$ 
querremos que las masas de las nuevas partículas no sean muy pesadas, lo cual es
bueno, porque las pone al alcance del LHC; sin embargo, masas \emph{demasiado}
ligeras para las nuevas partículas hacen más difícil ajustar las masas de los
neutrinos, además de entrar en tensión con otras cotas experimentales. Teniéndolo
todo en cuenta, las nuevas partículas habrían de tener masas entre unos pocos TeV
y algunas decenas de TeV, lo cual las situaría un poco por encima de lo que el
LHC va a poder observar. Si renunciamos a alguna de nuestras exigencias, como
por ejemplo que se observe una señal de $0 \nu \beta \beta$ inminentemente,
la situación se vuelve más laxa: las nuevas partículas pueden ser más pesadas,
quedando igualmente lejos del LHC pero dando más margen en la generación de masas
de neutrinos.

En el capítulo \ref{chap:WWee-model}, pues, vemos un buen ejemplo de lo que 
requiere el análisis de modelos: considerar todas las consecuencias observables
de las nuevas partículas que estás analizando y ver hasta qué punto pueden
cumplir todas tus expectativas. Habitualmente hay que renunciar a alguna de esas
expectativas para que el modelo resulte viable; en este caso ha habido suerte
y hemos dado con un modelo versátil que puede acomodar, aunque con estrecheces,
un escenario bastante optimista. Ahora, para que el optimismo sea completo,
sólo faltaría que el modelo fuera cierto.

\chapter{Agradecimientos}

Esto no es un adiós. Pero sería deshonesto pretender que no se le parece, aunque
sólo sea un poco. No recuerdo cuándo supe que la ciencia iba a ser mi vida.
Recuerdo el día que en el colegio nos enseñaron la tabla \mbox{del dos -- 
``multi}\-plicar'', me dije, ``guau, esto son matemáticas de verdad''; es
uno de mis recuerdos emocionales más antiguos. Recuerdo también haber estado
rodeado siempre de libros de zoología: interminables catálogos de insectos,
aves rapaces, anfibios, \mbox{reptiles -- y tortugas;} muchas tortugas, claro.
No descarto que mi afición por lo enciclopédico derive de haber devorado con
afán insensato listas y listas de bichos diversos, descripciones de sus libreas,
especificaciones sobre sus hábitos, consejos para combatir sus enfermedades.
Finalmente, recuerdo el día en que en la
biblioteca descubrí, más bien por azar, \emph{El electrón es zurdo} de Isaac Asimov; 
era un volumen vetusto, de desvencijada tapa dura, con el olor a rancio y el color
moreno que sólo los buenos jamones conocen. Si usted no lo ha leído, entonces yo
poco le puedo decir; sólo que un descubrimiento le está esperando, y que no 
se arrepentirá si da el paso. Después de \emph{El electrón...} vino
el inimitable \emph{El Universo}, y luego un irracionalmente exhaustivo 
volumen que repasaba la ciencia de los últimos 400 años y cuyo título no consigo 
recordar. Al maestro Asimov le debo sin duda el estar escribiendo estas
líneas. Pero no sólo a él; también se lo debo a otros maestros, gente que
por aquel entonces acertó a enseñarme que esto de la ciencia tenía su gracia.
Recuerdo con especial cariño el estilo impulsivo de Asun, que me
enseñó mi primer teorema; los chistes de Elvira, la polvorienta geología de Tere.
Y la física con María Dolores. Lo hacía bien. Para ser química. También por esa
época, recuerdo, el padre Javier me descubrió a un señor llamado Albert Camus, 
que daría para otra historia, y conocí \emph{Cosmos}, de Carl Sagan.
Recuerdo que no le presté mucha \mbox{atención entonces.} Cosas de niños, supongo.

Luego vino la carrera. Es sorprendente que aquello saliera tan bien; elegí Física
casi a ciegas, sin saber dónde me \mbox{metía --tan lejana} es la enseñanza de 
instituto de la \mbox{física real--, y no} estaba en absoluto seguro de que ésta 
fuese la carrera para mí. Unos meses después, sin embargo, no podía imaginarme
en otro sitio ni rodeado de otra gente. En esta facultad me han
enseñado a pensar, a ser crítico y humilde, a buscar en los libros las respuestas 
que las
personas no tienen, y \mbox{sobre todo --porque} éste es el trabajo de \mbox{un 
físico-- que} cuando los problemas son grandes y complejos
podemos aspirar a atacarlos si los sabemos transformar en otras cosas, más
pequeñas, que sí conocemos.
También he jugado mucho al mus y he aprendido que puede haber mentes muy brillantes
en personas muy oscuras. Destacar a toda la gente que ha sido importante para mí en
esta facultad sería imposible, son demasiados años. Pero aun así me enfangaré
y citaré a una selección; después de tantas páginas de tesis las cosas no pueden 
ir a peor\dots

Entre el profesorado no olvidaré a Pepe Bernabéu, que daba clase con pasión y nos 
enseñó aquello de ``nunca hay que empezar un cálculo cuyo resultado no conoces 
de antemano''; a Pepe Navarro, cuyas lecciones fueron para mí el comienzo de las 
altas matemáticas; a Adolfo de Azcárraga, que enseñaba con un detallismo casi 
erudito;
a Toni Pich, que jugaba en sus clases con lo que sabíamos y lo que queríamos saber,
como los buenos \emph{thrillers} hacen. Y en fin, sólo me faltan
Eugenio y Germán, con los que di mis primeros bocados a la investigación
y aprendí cosas que aún uso, tanto sobre la ciencia como sobre las personas.

And speaking about science and people, there's a bunch of scientists who are also 
a little bit 
responsible for this story: my collaborators, Kyungwook, Subha, Paco, Nuria and 
Juan; Sacha, Vincenzo and Michael, who so kindly hosted me at Lyon and Los Alamos 
and tried to extract some science out of me. And maybe especially José Wudka,
whose ideas and momentum were fundamental for the birth of these works.
To all of them I feel indebted, and this work is also somewhat theirs.

En el mismo orden de cosas no puedo dejar de mencionar a las personas que han
evaluado esta tesis, colegas y amigos que han dedicado una parte de su apretada
agenda a leer este texto que ha quedado tan largo y a aportarle lo mejor que han
podido. Muchas gracias por vuestro esfuerzo y generosidad.

Ésta ha sido una etapa de maduración científica, pero \mbox{también --me}
atrevería a decir que \mbox{sobre todo-- humana.} Los testigos más cualificados
para dar fe de esto son mi familia y amigos. A mis padres y mis hermanas les
debo que se hayan mantenido al pie del cañón a pesar de mi humor insoportable
y mi tendencia a preferir la soledad de mi habitación al salón de casa.
Tengo la sensación de que todo esto puede haber pasado por delante de vuestros
ojos sin que hayáis entendido mucho qué estaba sucediendo, ni por qué ni cómo;
me tranquiliza un poco pensar que en realidad tampoco necesitabais entenderlo:
lo que queríais era que yo fuera feliz y que pudiera compartir esa alegría con
vosotros. Temo en este punto no haber estado siempre a la altura. Sólo puedo daros
las gracias por vuestra paciencia y el tozudo cariño que siempre me dais.
No desfallezcáis, ya veis que hasta yo puedo hacerme mayor.

Hay amigos que son casi como si fueran familia; gente que si durante una invasión
zombi te dijeran ``tírate por esa ventana'' no tendrías otra que hacerles caso.
Gente que, a menudo, te conoce mejor que tú mismo. Pienso en 
Carl, Apa, David, Sergio, Marta, Juan, Joaquín, Antonio. Para vosotros tengo
lo único que vale la pena dar: todo mi cariño.
Mirando atrás creo que sois lo que ha hecho 
que esta andadura valga la pena; con vosotros he aprendido a vivir y por vosotros 
me he convertido en alguien mejor, y ahora me miro al espejo y me alegro del paso 
de los años. No voy a repasar nuestros mejores momentos, no voy a elevar grandes
deseos de que esto dure para siempre; pienso que lo nuestro está por encima
de esas cosas. Os diré simplemente que gracias. Que me mantenéis unido a este
mundo. Que mañana nos veremos, y que probablemente será un día mejor que hoy.

Junto a ellos está la legión; gente y más gente que han sido, fueron, son o serán
importantes. Sus méritos han sido reír, montar telescopios, comprar botellón,
ver películas, hablar de política y de física, comer mucho y bien, copiarse, 
dejar copiar, escuchar siempre, ser sinceros junto a una lumbre o una cerveza.
Hoy me siento especialmente incompetente intentando haceros justicia. 
A mis amigos de Castellón les diré que han sido los cimientos de todo; os
recordaré las tardes hablando de agujeros negros en algún parque, lo extraño
que me sentía entrando a una discoteca de estrangis, lo a gusto que me siento
ahora. 
A Carlos, Paula, Luis y Juan les diré que les quiero; que es por vosotros
que soy un poco de letras; que hemos pasado mucho frío en ese puto portal, 
leches; y que sólo consigo acordarme de buenos momentos. 
A Bernat le agradeceré ser mi familia durante seis años, las tertulias por
la noche, los partidos de Champions, aguantar mis rarezas y ser práctico
con los problemas; para Pepe y para Berta irá una parte alícuota de estas
mismas cosas.
A los Caballeros de la 
Orden les diré que me hicisteis sentir en casa por primera vez, que será
difícil encontrar otra gente con la que sea tan sencillo ser yo mismo;
también que creo que nunca he cumplido tantos sueños de la infancia como
la noche del Hyundai. 
A Pepón le diré que no conozco a nadie que vea las matemáticas como tú lo haces,
y que si la ciencia va a echar de menos tus teoremas me alegro de que al menos
pueda disfrutar tus poemas; al fin y al cabo, como todo el mundo sabe, 
la poesía y las matemáticas son la misma cosa.
A la gente de la Asociación de Astronomía les agradeceré la paciencia, la
generosidad que hace fácil lo difícil, el entusiasmo por enseñar; y les diré
que este punto azul pálido es demasiado pequeño para que no estemos, de una
manera o de otra, cerca.
A Avelino, Rika, los Javis, las Susanas, Ferrándiz, Víctor y todos los demás les 
confesaré que si no os hubiera conocido no habría logrado recuperarle el pulso 
a la carrera, que la habría terminado de cualquier manera; sé que pensaréis que 
exagero, pero eso es porque siempre fuisteis demasiado generosos. 
Puedo decir sin inmutarme que sois el mejor grupo humano que he conocido, y que 
formar parte de él ha \mbox{sido --sigue} \mbox{siendo-- una} verdadera alegría.
A toda la gente de iGEM Valencia 2006 les contaré lo que ya saben: que ese verano 
fue mágico, que hicimos muchas más cosas que ciencia, que lo que construimos
ha sido duradero; y añadiré que si hoy hubiese de decidir cuándo
comenzó el cambio creo que lo situaría en aquel verano extraño en el que aprendí
que existe un mundo más allá de la física. 
A Francesc, Manu, Paco, Vicent, 
María y el resto les diré que siempre sonrío con vosotros; que nos hemos juntado
poco y que somos un grupo heterogéneo, pero que todas nuestras reuniones 
son un buen momento. 
A toda la gente del Departamento, desde a mis compañeros de despacho a los
profesores pasando por la gente de secretaría les agradeceré que me hayan hecho
las cosas tan fáciles; que pudiendo ser mis compañeros de trabajo se hayan
convertido en mis amigos; y les diré que si he tenido que huir no ha sido 
porque estuviera a disgusto, sino porque es \emph{imposible} escribir una
tesis estando tan a gusto. 
A Dani, Gustavo, Paco, Álex y Clarilla\dots $\;$no os puedo decir nada, porque
no me habéis visto el pelo! Hablando en serio, ya lo sabéis: mi vida es un 
completo desorden, y quizá por eso pongo tanto esfuerzo en que al menos la
casa esté ordenada; haciendo gala de estas costumbres que conocéis tan bien,
estas líneas las estoy escribiendo a las ocho y pico de la noche en un edificio
semidesierto, y mañana es fiesta. Realmente quizá haya cosas que nunca cambian.
A los últimos en llegar: David, Clari, Oliver, Javi, Beni, Diana, os
diré que habéis conocido una versión depauperada de mí mismo y que, pese a ello,
habéis hecho lo posible por sacarme a flote; os diré también que ha funcionado,
está funcionando. Por ello, pues, gracias; nada me hará más feliz que el día en
que \emph{no} tenga que devolveros el favor porque todo os vaya bien.

Todas las cosas se acaban, incluso las malas. Ésta, sin embargo, no puede
terminar hasta que haya hablado de Arcadi. Escoger un director de tesis no 
es necesariamente un asunto fácil, y menos cuando hay mucha oferta: todos
los proyectos, en el fondo, suenan bien, porque en realidad no tienes ni idea
de cómo evaluarlos, y no puedes comprender cuáles son los aspectos relevantes hasta
que estés dentro. Cuando escogí a Arcadi lo hice porque me gustaba la física que
me ofrecía,
pero también porque parecía un hombre de fiar, una buena persona. En lo que
respecta a la física, Arcadi, tú la ves con mucha mayor claridad de lo que
yo nunca la veré. Has intentado enseñarme, señalarme cuáles eran las estrategias
más inteligentes, y no ha sido un esfuerzo infructuoso porque creo que algo
he aprendido; sin embargo, no puedo dejar de pensar que te he tenido 
desaprovechado, porque en serio, creo que eres uno de los físicos más
brillantes que he conocido. Ha sido un placer descubrir cosas a tu lado.
En lo humano\dots $\;$bueno, sé que va a sonar a coba, pero el nivel ha sido
igual o mejor. Eres mi director de tesis, sí, pero sobre todo eres mi amigo,
y te he contado cosas que creo que la mayor parte de gente no le contaría a 
su jefe. Eso es relevante; eso tiene peso. Eso, aunque me suspendan el día
de la defensa, seguirá \mbox{existiendo -- bueno vale,} entenderé que te enfades 
si me suspenden, tampoco vamos a exagerar. En fin, que muchas gracias por este
viaje; espero que, como yo, pienses que el viaje no ha sido en vano.

Es hora de echar el cierre. Antes he dicho que los últimos tiempos han visto 
una versión depauperada de mí mismo. Lo sigo pensando. La redacción de este
texto que aquí termina y otras tantas complicaciones que ahora no vienen al
caso han hecho de estos dos últimos años un periodo difícil. La suerte es que
junto con esas dificultades han venido otras cosas que no dejan de empujar
para seguir adelante; quizá la más sorprendente y la más importante sea
la gente de La Brújula, a la que mando un saludo desde aquí.
He dicho antes también que esto no era un adiós, y es cierto.
Tiendo a pensar en la vida como en una partida de cartas: jugamos bien o jugamos
mal, pero al final siempre acabamos con unas cuantas cartas en la mano.
¿Cómo vas a decir adiós si al final sigues jugando, como al principio?
Yo no siempre he jugado bien; no me hacen feliz los errores, pero sí atesoro
unos cuantos aciertos. Si estás leyendo esto es porque eres uno de esos aciertos.
Hasta siempre.


\cleardoublepage
\phantomsection
\addcontentsline{toc}{chapter}{Bibliography}

\bibliographystyle{apsrev4.1-mod-long}
\bibliography{aparici-phd-dissertation}

\begin{thebibliography}{100}%
\makeatletter
\providecommand \@ifxundefined [1]{%
 \ifx #1\undefined \expandafter \@firstoftwo
 \else \expandafter \@secondoftwo
\fi
}%
\providecommand \@ifnum [1]{%
 \ifnum #1\expandafter \@firstoftwo
 \else \expandafter \@secondoftwo
\fi
}%
\providecommand \enquote [1]{``#1''}%
\providecommand \bibnamefont  [1]{#1}%
\providecommand \bibfnamefont [1]{#1}%
\providecommand \citenamefont [1]{#1}%
\providecommand\href[0]{\@sanitize\@href}%
\providecommand\@href[1]{\endgroup\@@startlink{#1}\endgroup\@@href}%
\providecommand\@@href[1]{#1\@@endlink}%
\providecommand \@sanitize [0]{\begingroup\catcode`\&12\catcode`\#12\relax}%
\@ifxundefined \pdfoutput {\@firstoftwo}{%
 \@ifnum{\z@=\pdfoutput}{\@firstoftwo}{\@secondoftwo}%
}{%
 \providecommand\@@startlink[1]{\leavevmode\special{html:<a href="#1">}}%
 \providecommand\@@endlink[0]{\special{html:</a>}}%
}{%
 \providecommand\@@startlink[1]{%
  \leavevmode
  \pdfstartlink
   attr{/Border[0 0 1 ]/H/I/C[0 1 1]}%
   user{/Subtype/Link/A<</Type/Action/S/URI/URI(#1)>>}%
  \relax
 }%
 \providecommand\@@endlink[0]{\pdfendlink}%
}%
\providecommand \url  [0]{\begingroup\@sanitize \@url }%
\providecommand \@url [1]{\endgroup\@href {#1}{\urlprefix}}%
\providecommand \urlprefix [0]{URL }%
\providecommand \Eprint[0]{\href }%
\@ifxundefined \urlstyle {%
  \providecommand \doi [1]{doi:\discretionary{}{}{}#1}%
}{%
  \providecommand \doi [0]{doi:\discretionary{}{}{}\begingroup
  \urlstyle{rm}\Url }%
}%
\providecommand \doibase [0]{http://dx.doi.org/}%
\providecommand \Doi[1]{\href{\doibase#1}}%
\providecommand \bibAnnote [3]{%
  \BibitemShut{#1}%
  \begin{quotation}\noindent
    \textsc{Key:}\ #2\\\textsc{Annotation:}\ #3%
  \end{quotation}%
}%
\providecommand \bibAnnoteFile [2]{%
  \IfFileExists{#2}{\bibAnnote {#1} {#2} {\input{#2}}}{}%
}%
\providecommand \typeout [0]{\immediate \write \m@ne }%
\providecommand \selectlanguage [0]{\@gobble}%
\providecommand \bibinfo [0]{\@secondoftwo}%
\providecommand \bibfield [0]{\@secondoftwo}%
\providecommand \translation [1]{[#1]}%
\providecommand \BibitemOpen[0]{}%
\providecommand \bibitemStop [0]{}%
\providecommand \bibitemNoStop [0]{.\EOS\space}%
\providecommand \EOS [0]{\spacefactor3000\relax}%
\providecommand \BibitemShut [1]{\csname bibitem#1\endcsname}%
\bibitem{Aparici:2009fh}%
  \BibitemOpen
  \bibfield{author}{%
  \bibinfo {author} {\bibfnamefont{Alberto}\ \bibnamefont{Aparici}}, \bibinfo
  {author} {\bibfnamefont{Kyungwook}\ \bibnamefont{Kim}}, \bibinfo {author}
  {\bibfnamefont{Arcadi}\ \bibnamefont{Santamaria}},\ and\ \bibinfo {author}
  {\bibfnamefont{Jose}\ \bibnamefont{Wudka}},\ }%
  \bibfield{title}{%
  \enquote{\bibinfo {title} {{Right-handed neutrino magnetic moments}},}\ }%
  \bibfield{journal}{%
  \Doi{10.1103/PhysRevD.80.013010}{\bibinfo {journal} {Phys.Rev.}}\ }%
  \textbf{\bibinfo {volume} {D80}},\ \bibinfo {pages} {013010} (\bibinfo {year}
  {2009}),\ \Eprint{http://arxiv.org/abs/0904.3244}{arXiv:0904.3244 [hep-ph]}%
  \bibAnnoteFile{NoStop}{Aparici:2009fh}%

\bibitem{Aparici:2009mj}%
  \BibitemOpen
  \bibfield{author}{%
  \bibinfo {author} {\bibfnamefont{Alberto}\ \bibnamefont{Aparici}}, \bibinfo
  {author} {\bibfnamefont{Arcadi}\ \bibnamefont{Santamaria}},\ and\ \bibinfo
  {author} {\bibfnamefont{Jose}\ \bibnamefont{Wudka}},\ }%
  \bibfield{title}{%
  \enquote{\bibinfo {title} {{A model for right-handed neutrino magnetic
  moments}},}\ }%
  \bibfield{journal}{%
  \Doi{10.1088/0954-3899/37/7/075012}{\bibinfo {journal} {J.Phys.}}\ }%
  \textbf{\bibinfo {volume} {G37}},\ \bibinfo {pages} {075012} (\bibinfo {year}
  {2010}),\ \Eprint{http://arxiv.org/abs/0911.4103}{arXiv:0911.4103 [hep-ph]}%
  \bibAnnoteFile{NoStop}{Aparici:2009mj}%

\bibitem{Aparici:2010zz}%
  \BibitemOpen
  \bibfield{author}{%
  \bibinfo {author} {\bibfnamefont{Alberto}\ \bibnamefont{Aparici}}, \bibinfo
  {author} {\bibfnamefont{Kyung-Wook}\ \bibnamefont{Kim}}, \bibinfo {author}
  {\bibfnamefont{Arcadi}\ \bibnamefont{Santamaria}},\ and\ \bibinfo {author}
  {\bibfnamefont{Jose}\ \bibnamefont{Wudka}},\ }%
  \bibfield{title}{%
  \enquote{\bibinfo {title} {{Right-handed neutrino magnetic moments}},}\ }%
  \bibfield{journal}{%
  \Doi{10.1088/1742-6596/259/1/012089}{\bibinfo {journal} {J.Phys.Conf.Ser.}}\
  }%
  \textbf{\bibinfo {volume} {259}},\ \bibinfo {pages} {012089} (\bibinfo {year}
  {2010})%
  \bibAnnoteFile{NoStop}{Aparici:2010zz}%

\bibitem{delAguila:2011gr}%
  \BibitemOpen
  \bibfield{author}{%
  \bibinfo {author} {\bibfnamefont{Francisco}\ \bibnamefont{del Aguila}},
  \bibinfo {author} {\bibfnamefont{Alberto}\ \bibnamefont{Aparici}}, \bibinfo
  {author} {\bibfnamefont{Subhaditya}\ \bibnamefont{Bhattacharya}}, \bibinfo
  {author} {\bibfnamefont{Arcadi}\ \bibnamefont{Santamaria}},\ and\ \bibinfo
  {author} {\bibfnamefont{Jose}\ \bibnamefont{Wudka}},\ }%
  \bibfield{title}{%
  \enquote{\bibinfo {title} {{A realistic model of neutrino masses with a large
  neutrinoless double beta decay rate}},}\ }%
  \bibfield{journal}{%
  \Doi{10.1007/JHEP05(2012)133}{\bibinfo {journal} {JHEP}}\ }%
  \textbf{\bibinfo {volume} {1205}},\ \bibinfo {pages} {133} (\bibinfo {year}
  {2012}),\ \Eprint{http://arxiv.org/abs/1111.6960}{arXiv:1111.6960 [hep-ph]}%
  \bibAnnoteFile{NoStop}{delAguila:2011gr}%

\bibitem{delAguila:2012nu}%
  \BibitemOpen
  \bibfield{author}{%
  \bibinfo {author} {\bibfnamefont{Francisco}\ \bibnamefont{del Aguila}},
  \bibinfo {author} {\bibfnamefont{Alberto}\ \bibnamefont{Aparici}}, \bibinfo
  {author} {\bibfnamefont{Subhaditya}\ \bibnamefont{Bhattacharya}}, \bibinfo
  {author} {\bibfnamefont{Arcadi}\ \bibnamefont{Santamaria}},\ and\ \bibinfo
  {author} {\bibfnamefont{Jose}\ \bibnamefont{Wudka}},\ }%
  \bibfield{title}{%
  \enquote{\bibinfo {title} {{Effective Lagrangian approach to neutrinoless
  double beta decay and neutrino masses}},}\ }%
  \bibfield{journal}{%
  \Doi{10.1007/JHEP06(2012)146}{\bibinfo {journal} {JHEP}}\ }%
  \textbf{\bibinfo {volume} {1206}},\ \bibinfo {pages} {146} (\bibinfo {year}
  {2012}),\ \Eprint{http://arxiv.org/abs/1204.5986}{arXiv:1204.5986 [hep-ph]}%
  \bibAnnoteFile{NoStop}{delAguila:2012nu}%

\bibitem{delAguila:2013zba}%
  \BibitemOpen
  \bibfield{author}{%
  \bibinfo {author} {\bibfnamefont{F.}~\bibnamefont{del Águila}}, \bibinfo
  {author} {\bibfnamefont{A.}~\bibnamefont{Aparici}}, \bibinfo {author}
  {\bibfnamefont{S.}~\bibnamefont{Bhattacharya}}, \bibinfo {author}
  {\bibfnamefont{A.}~\bibnamefont{Santamaria}},\ and\ \bibinfo {author}
  {\bibfnamefont{J.}~\bibnamefont{Wudka}},\ }%
  \bibfield{title}{%
  \enquote{\bibinfo {title} {{Neutrinoless double $\beta$ decay with small
  neutrino masses}},}\ }%
  \bibfield{journal}{%
  \bibinfo {journal} {PoS}\ }%
  \textbf{\bibinfo {volume} {Corfu2012}},\ \bibinfo {pages} {028} (\bibinfo
  {year} {2013}),\ \Eprint{http://arxiv.org/abs/1305.4900}{arXiv:1305.4900
  [hep-ph]}%
  \bibAnnoteFile{NoStop}{delAguila:2013zba}%

\bibitem{Glashow:1961tr}%
  \BibitemOpen
  \bibfield{author}{%
  \bibinfo {author} {\bibfnamefont{S.L.}\ \bibnamefont{Glashow}},\ }%
  \bibfield{title}{%
  \enquote{\bibinfo {title} {Partial symmetries of weak interactions},}\ }%
  \bibfield{journal}{%
  \Doi{10.1016/0029-5582(61)90469-2}{\bibinfo {journal} {Nucl.Phys.}}\ }%
  \textbf{\bibinfo {volume} {22}},\ \bibinfo {pages} {579--588} (\bibinfo
  {year} {1961})%
  \bibAnnoteFile{NoStop}{Glashow:1961tr}%

\bibitem{Higgs:1964pj}%
  \BibitemOpen
  \bibfield{author}{%
  \bibinfo {author} {\bibfnamefont{Peter~W.}\ \bibnamefont{Higgs}},\ }%
  \bibfield{title}{%
  \enquote{\bibinfo {title} {Broken symmetries and the masses of gauge
  bosons},}\ }%
  \bibfield{journal}{%
  \Doi{10.1103/PhysRevLett.13.508}{\bibinfo {journal} {Phys.Rev.Lett.}}\ }%
  \textbf{\bibinfo {volume} {13}},\ \bibinfo {pages} {508--509} (\bibinfo
  {year} {1964})%
  \bibAnnoteFile{NoStop}{Higgs:1964pj}%

\bibitem{Englert:1964et}%
  \BibitemOpen
  \bibfield{author}{%
  \bibinfo {author} {\bibfnamefont{F.}~\bibnamefont{Englert}}\ and\ \bibinfo
  {author} {\bibfnamefont{R.}~\bibnamefont{Brout}},\ }%
  \bibfield{title}{%
  \enquote{\bibinfo {title} {{Broken symmetry and the mass of gauge vector
  mesons}},}\ }%
  \bibfield{journal}{%
  \Doi{10.1103/PhysRevLett.13.321}{\bibinfo {journal} {Phys.Rev.Lett.}}\ }%
  \textbf{\bibinfo {volume} {13}},\ \bibinfo {pages} {321--323} (\bibinfo
  {year} {1964})%
  \bibAnnoteFile{NoStop}{Englert:1964et}%

\bibitem{Guralnik:1964eu}%
  \BibitemOpen
  \bibfield{author}{%
  \bibinfo {author} {\bibfnamefont{G.S.}\ \bibnamefont{Guralnik}}, \bibinfo
  {author} {\bibfnamefont{C.R.}\ \bibnamefont{Hagen}},\ and\ \bibinfo {author}
  {\bibfnamefont{T.W.B.}\ \bibnamefont{Kibble}},\ }%
  \bibfield{title}{%
  \enquote{\bibinfo {title} {Global conservation laws and massless
  particles},}\ }%
  \bibfield{journal}{%
  \Doi{10.1103/PhysRevLett.13.585}{\bibinfo {journal} {Phys.Rev.Lett.}}\ }%
  \textbf{\bibinfo {volume} {13}},\ \bibinfo {pages} {585--587} (\bibinfo
  {year} {1964})%
  \bibAnnoteFile{NoStop}{Guralnik:1964eu}%

\bibitem{Higgs:1966ev}%
  \BibitemOpen
  \bibfield{author}{%
  \bibinfo {author} {\bibfnamefont{Peter~W.}\ \bibnamefont{Higgs}},\ }%
  \bibfield{title}{%
  \enquote{\bibinfo {title} {Spontaneous symmetry breakdown without massless
  bosons},}\ }%
  \bibfield{journal}{%
  \Doi{10.1103/PhysRev.145.1156}{\bibinfo {journal} {Phys.Rev.}}\ }%
  \textbf{\bibinfo {volume} {145}},\ \bibinfo {pages} {1156--1163} (\bibinfo
  {year} {1966})%
  \bibAnnoteFile{NoStop}{Higgs:1966ev}%

\bibitem{Kibble:1967sv}%
  \BibitemOpen
  \bibfield{author}{%
  \bibinfo {author} {\bibfnamefont{T.W.B.}\ \bibnamefont{Kibble}},\ }%
  \bibfield{title}{%
  \enquote{\bibinfo {title} {{Symmetry breaking in non-Abelian gauge
  theories}},}\ }%
  \bibfield{journal}{%
  \Doi{10.1103/PhysRev.155.1554}{\bibinfo {journal} {Phys.Rev.}}\ }%
  \textbf{\bibinfo {volume} {155}},\ \bibinfo {pages} {1554--1561} (\bibinfo
  {year} {1967})%
  \bibAnnoteFile{NoStop}{Kibble:1967sv}%

\bibitem{Weinberg:1967tq}%
  \BibitemOpen
  \bibfield{author}{%
  \bibinfo {author} {\bibfnamefont{Steven}\ \bibnamefont{Weinberg}},\ }%
  \bibfield{title}{%
  \enquote{\bibinfo {title} {A model of leptons},}\ }%
  \bibfield{journal}{%
  \Doi{10.1103/PhysRevLett.19.1264}{\bibinfo {journal} {Phys.Rev.Lett.}}\ }%
  \textbf{\bibinfo {volume} {19}},\ \bibinfo {pages} {1264--1266} (\bibinfo
  {year} {1967})%
  \bibAnnoteFile{NoStop}{Weinberg:1967tq}%

\bibitem{Salam:1968rm}%
  \BibitemOpen
  \bibfield{author}{%
  \bibinfo {author} {\bibfnamefont{Abdus}\ \bibnamefont{Salam}},\ }%
  \bibfield{title}{%
  \enquote{\bibinfo {title} {Weak and electromagnetic interactions},}\ }%
  \bibfield{journal}{%
  \bibinfo {journal} {Conf.Proc.}\ }%
  \textbf{\bibinfo {volume} {C680519}},\ \bibinfo {pages} {367--377} (\bibinfo
  {year} {1968})%
  \bibAnnoteFile{NoStop}{Salam:1968rm}%

\bibitem{Han:1965pf}%
  \BibitemOpen
  \bibfield{author}{%
  \bibinfo {author} {\bibfnamefont{M.Y.}\ \bibnamefont{Han}}\ and\ \bibinfo
  {author} {\bibfnamefont{Yoichiro}\ \bibnamefont{Nambu}},\ }%
  \bibfield{title}{%
  \enquote{\bibinfo {title} {Three-triplet model with double $su(3)$
  symmetry},}\ }%
  \bibfield{journal}{%
  \Doi{10.1103/PhysRev.139.B1006}{\bibinfo {journal} {Phys.Rev.}}\ }%
  \textbf{\bibinfo {volume} {139}},\ \bibinfo {pages} {B1006--B1010} (\bibinfo
  {year} {1965})%
  \bibAnnoteFile{NoStop}{Han:1965pf}%

\bibitem{Fritzsch:1973pi}%
  \BibitemOpen
  \bibfield{author}{%
  \bibinfo {author} {\bibfnamefont{H.}~\bibnamefont{Fritzsch}}, \bibinfo
  {author} {\bibfnamefont{Murray}\ \bibnamefont{Gell-Mann}},\ and\ \bibinfo
  {author} {\bibfnamefont{H.}~\bibnamefont{Leutwyler}},\ }%
  \bibfield{title}{%
  \enquote{\bibinfo {title} {Advantages of the color octet gluon picture},}\ }%
  \bibfield{journal}{%
  \Doi{10.1016/0370-2693(73)90625-4}{\bibinfo {journal} {Phys.Lett.}}\ }%
  \textbf{\bibinfo {volume} {B47}},\ \bibinfo {pages} {365--368} (\bibinfo
  {year} {1973})%
  \bibAnnoteFile{NoStop}{Fritzsch:1973pi}%

\bibitem{GellMann:1981ph}%
  \BibitemOpen
  \bibfield{author}{%
  \bibinfo {author} {\bibfnamefont{Murray}\ \bibnamefont{Gell-Mann}},\ }%
  \bibfield{title}{%
  \enquote{\bibinfo {title} {{Quarks}},}\ }%
  \bibfield{journal}{%
  \bibinfo {journal} {Acta Phys.Austriaca Suppl.}\ }%
  \textbf{\bibinfo {volume} {9}},\ \bibinfo {pages} {733--761} (\bibinfo {year}
  {1972})%
  \bibAnnoteFile{NoStop}{GellMann:1981ph}%

\bibitem{Fritzsch:1972jv}%
  \BibitemOpen
  \bibfield{author}{%
  \bibinfo {author} {\bibfnamefont{Harald}\ \bibnamefont{Fritzsch}}\ and\
  \bibinfo {author} {\bibfnamefont{Murray}\ \bibnamefont{Gell-Mann}},\ }%
  \bibfield{title}{%
  \enquote{\bibinfo {title} {{Current algebra: Quarks and what else?}}.}\ }%
  \bibfield{journal}{%
  \bibinfo {journal} {eConf}\ }%
  \textbf{\bibinfo {volume} {C720906V2}},\ \bibinfo {pages} {135--165}
  (\bibinfo {year} {1972}),\
  \Eprint{http://arxiv.org/abs/hep-ph/0208010}{arXiv:hep-ph/0208010 [hep-ph]}%
  \bibAnnoteFile{NoStop}{Fritzsch:1972jv}%

\bibitem{Pich:2012sx}%
  \BibitemOpen
  \bibfield{author}{%
  \bibinfo {author} {\bibfnamefont{Antonio}\ \bibnamefont{Pich}},\ }%
  \bibfield{title}{%
  \enquote{\bibinfo {title} {The standard model of electroweak interactions},}\
  }%
   (\bibinfo {year} {2012}),\
  \Eprint{http://arxiv.org/abs/1201.0537}{arXiv:1201.0537 [hep-ph]}%
  \bibAnnoteFile{NoStop}{Pich:2012sx}%

\bibitem{Pich:1999yz}%
  \BibitemOpen
  \bibfield{author}{%
  \bibinfo {author} {\bibfnamefont{Antonio}\ \bibnamefont{Pich}},\ }%
  \bibfield{title}{%
  \enquote{\bibinfo {title} {{Aspects of quantum chromodynamics}},}\ }%
   (\bibinfo {year} {1999}),\
  \Eprint{http://arxiv.org/abs/hep-ph/0001118}{arXiv:hep-ph/0001118 [hep-ph]}%
  \bibAnnoteFile{NoStop}{Pich:1999yz}%

\bibitem{Fermi:1934sk}%
  \BibitemOpen
  \bibfield{author}{%
  \bibinfo {author} {\bibfnamefont{E.}~\bibnamefont{Fermi}},\ }%
  \bibfield{title}{%
  \enquote{\bibinfo {title} {Trends to a theory of beta radiation. (in
  {Italian})},}\ }%
  \bibfield{journal}{%
  \Doi{10.1007/BF02959820}{\bibinfo {journal} {Nuovo Cim.}}\ }%
  \textbf{\bibinfo {volume} {11}},\ \bibinfo {pages} {1--19} (\bibinfo {year}
  {1934})%
  \bibAnnoteFile{NoStop}{Fermi:1934sk}%

\bibitem{Fermi:1934hr}%
  \BibitemOpen
  \bibfield{author}{%
  \bibinfo {author} {\bibfnamefont{E.}~\bibnamefont{Fermi}},\ }%
  \bibfield{title}{%
  \enquote{\bibinfo {title} {{An attempt of a theory of beta radiation. 1.}}.}\
  }%
  \bibfield{journal}{%
  \Doi{10.1007/BF01351864}{\bibinfo {journal} {Z.Phys.}}\ }%
  \textbf{\bibinfo {volume} {88}},\ \bibinfo {pages} {161--177} (\bibinfo
  {year} {1934})%
  \bibAnnoteFile{NoStop}{Fermi:1934hr}%

\bibitem{wilson:1150}%
  \BibitemOpen
  \bibfield{author}{%
  \bibinfo {author} {\bibfnamefont{Fred~L.}\ \bibnamefont{Wilson}},\ }%
  \bibfield{title}{%
  \enquote{\bibinfo {title} {Fermi's theory of beta decay},}\ }%
  \bibfield{journal}{%
  \Doi{10.1119/1.1974382}{\bibinfo {journal} {American Journal of Physics}}\ }%
  \textbf{\bibinfo {volume} {36}},\ \bibinfo {pages} {1150--1160} (\bibinfo
  {year} {1968})%
  \bibAnnoteFile{NoStop}{wilson:1150}%

\bibitem{Wu:1957my}%
  \BibitemOpen
  \bibfield{author}{%
  \bibinfo {author} {\bibfnamefont{C.S.}\ \bibnamefont{Wu}}, \bibinfo {author}
  {\bibfnamefont{E.}~\bibnamefont{Ambler}}, \bibinfo {author}
  {\bibfnamefont{R.W.}\ \bibnamefont{Hayward}}, \bibinfo {author}
  {\bibfnamefont{D.D.}\ \bibnamefont{Hoppes}},\ and\ \bibinfo {author}
  {\bibfnamefont{R.P.}\ \bibnamefont{Hudson}},\ }%
  \bibfield{title}{%
  \enquote{\bibinfo {title} {Experimental test of parity conservation in beta
  decay},}\ }%
  \bibfield{journal}{%
  \Doi{10.1103/PhysRev.105.1413}{\bibinfo {journal} {Phys.Rev.}}\ }%
  \textbf{\bibinfo {volume} {105}},\ \bibinfo {pages} {1413--1414} (\bibinfo
  {year} {1957})%
  \bibAnnoteFile{NoStop}{Wu:1957my}%

\bibitem{Lee:1956qn}%
  \BibitemOpen
  \bibfield{author}{%
  \bibinfo {author} {\bibfnamefont{T.D.}\ \bibnamefont{Lee}}\ and\ \bibinfo
  {author} {\bibfnamefont{Chen-Ning}\ \bibnamefont{Yang}},\ }%
  \bibfield{title}{%
  \enquote{\bibinfo {title} {Question of parity conservation in weak
  interactions},}\ }%
  \bibfield{journal}{%
  \Doi{10.1103/PhysRev.104.254}{\bibinfo {journal} {Phys.Rev.}}\ }%
  \textbf{\bibinfo {volume} {104}},\ \bibinfo {pages} {254--258} (\bibinfo
  {year} {1956})%
  \bibAnnoteFile{NoStop}{Lee:1956qn}%

\bibitem{Cheng:1985bj}%
  \BibitemOpen
  \bibfield{author}{%
  \bibinfo {author} {\bibfnamefont{T.P.}\ \bibnamefont{Cheng}}\ and\ \bibinfo
  {author} {\bibfnamefont{L.F.}\ \bibnamefont{Li}},\ }%
  \emph{\bibinfo {title} {Gauge theory of elementary particle physics}}\
  (\bibinfo {publisher} {{Oxford University Press}},\ \bibinfo {year} {1984})\
  ISBN \bibinfo {isbn} {978-0198519614}%
  \bibAnnoteFile{NoStop}{Cheng:1985bj}%

\bibitem{Jackiw:1979wf}%
  \BibitemOpen
  \bibfield{author}{%
  \bibinfo {author} {\bibfnamefont{R.}~\bibnamefont{Jackiw}}\ and\ \bibinfo
  {author} {\bibfnamefont{A.}~\bibnamefont{Kerman}},\ }%
  \bibfield{title}{%
  \enquote{\bibinfo {title} {Time-dependent variational principle and the
  effective action},}\ }%
  \bibfield{journal}{%
  \Doi{10.1016/0375-9601(79)90151-8}{\bibinfo {journal} {Phys.Lett.}}\ }%
  \textbf{\bibinfo {volume} {A71}},\ \bibinfo {pages} {158--162} (\bibinfo
  {year} {1979})%
  \bibAnnoteFile{NoStop}{Jackiw:1979wf}%

\bibitem{Jackiw:1974cv}%
  \BibitemOpen
  \bibfield{author}{%
  \bibinfo {author} {\bibfnamefont{R.}~\bibnamefont{Jackiw}},\ }%
  \bibfield{title}{%
  \enquote{\bibinfo {title} {{Functional evaluation of the effective
  potential}},}\ }%
  \bibfield{journal}{%
  \Doi{10.1103/PhysRevD.9.1686}{\bibinfo {journal} {Phys.Rev.}}\ }%
  \textbf{\bibinfo {volume} {D9}},\ \bibinfo {pages} {1686} (\bibinfo {year}
  {1974})%
  \bibAnnoteFile{NoStop}{Jackiw:1974cv}%

\bibitem{Cornwall:1974vz}%
  \BibitemOpen
  \bibfield{author}{%
  \bibinfo {author} {\bibfnamefont{John~M.}\ \bibnamefont{Cornwall}}, \bibinfo
  {author} {\bibfnamefont{R.}~\bibnamefont{Jackiw}},\ and\ \bibinfo {author}
  {\bibfnamefont{E.}~\bibnamefont{Tomboulis}},\ }%
  \bibfield{title}{%
  \enquote{\bibinfo {title} {Effective action for composite operators},}\ }%
  \bibfield{journal}{%
  \Doi{10.1103/PhysRevD.10.2428}{\bibinfo {journal} {Phys.Rev.}}\ }%
  \textbf{\bibinfo {volume} {D10}},\ \bibinfo {pages} {2428--2445} (\bibinfo
  {year} {1974})%
  \bibAnnoteFile{NoStop}{Cornwall:1974vz}%

\bibitem{Georgi:weak}%
  \BibitemOpen
  \bibfield{author}{%
  \bibinfo {author} {\bibfnamefont{Howard}\ \bibnamefont{Georgi}},\ }%
  \emph{\bibinfo {title} {Lecture notes on weak interactions}},\
  \url{http://www.people.fas.harvard.edu/~hgeorgi/weak.pdf}%
  \bibAnnoteFile{NoStop}{Georgi:weak}%

\bibitem{Coleman:aspects}%
  \BibitemOpen
  \bibfield{author}{%
  \bibinfo {author} {\bibfnamefont{Sydney}\ \bibnamefont{Coleman}},\ }%
  \emph{\bibinfo {title} {Aspects of symmetry}}\ (\bibinfo {publisher}
  {Cambridge University Press},\ \bibinfo {year} {1985})%
  \bibAnnoteFile{NoStop}{Coleman:aspects}%

\bibitem{Beringer:1900zz}%
  \BibitemOpen
  \bibfield{author}{%
  \bibinfo {author} {\bibfnamefont{J.}~\bibnamefont{Beringer}} \emph{et~al.}
  (\bibinfo {collaboration} {Particle Data Group}),\ }%
  \bibfield{title}{%
  \enquote{\bibinfo {title} {{Review of particle physics (RPP)}},}\ }%
  \bibfield{journal}{%
  \Doi{10.1103/PhysRevD.86.010001}{\bibinfo {journal} {Phys.Rev.}}\ }%
  \textbf{\bibinfo {volume} {D86}},\ \bibinfo {pages} {010001} (\bibinfo {year}
  {2012})%
  \bibAnnoteFile{NoStop}{Beringer:1900zz}%

\bibitem{Aad:2012tfa}%
  \BibitemOpen
  \bibfield{author}{%
  \bibinfo {author} {\bibfnamefont{Georges}\ \bibnamefont{Aad}} \emph{et~al.}
  (\bibinfo {collaboration} {ATLAS Collaboration}),\ }%
  \bibfield{title}{%
  \enquote{\bibinfo {title} {{Observation of a new particle in the search for
  the Standard Model Higgs boson with the ATLAS detector at the LHC}},}\ }%
  \bibfield{journal}{%
  \Doi{10.1016/j.physletb.2012.08.020}{\bibinfo {journal} {Phys.Lett.}}\ }%
  \textbf{\bibinfo {volume} {B716}},\ \bibinfo {pages} {1--29} (\bibinfo {year}
  {2012}),\ \Eprint{http://arxiv.org/abs/1207.7214}{arXiv:1207.7214 [hep-ex]}%
  \bibAnnoteFile{NoStop}{Aad:2012tfa}%

\bibitem{Chatrchyan:2012ufa}%
  \BibitemOpen
  \bibfield{author}{%
  \bibinfo {author} {\bibfnamefont{Serguei}\ \bibnamefont{Chatrchyan}}
  \emph{et~al.} (\bibinfo {collaboration} {CMS Collaboration}),\ }%
  \bibfield{title}{%
  \enquote{\bibinfo {title} {{Observation of a new boson at a mass of 125 GeV
  with the CMS experiment at the LHC}},}\ }%
  \bibfield{journal}{%
  \Doi{10.1016/j.physletb.2012.08.021}{\bibinfo {journal} {Phys.Lett.}}\ }%
  \textbf{\bibinfo {volume} {B716}},\ \bibinfo {pages} {30--61} (\bibinfo
  {year} {2012}),\ \Eprint{http://arxiv.org/abs/1207.7235}{arXiv:1207.7235
  [hep-ex]}%
  \bibAnnoteFile{NoStop}{Chatrchyan:2012ufa}%

\bibitem{Abers:1973qs}%
  \BibitemOpen
  \bibfield{author}{%
  \bibinfo {author} {\bibfnamefont{E.S.}\ \bibnamefont{Abers}}\ and\ \bibinfo
  {author} {\bibfnamefont{B.W.}\ \bibnamefont{Lee}},\ }%
  \bibfield{title}{%
  \enquote{\bibinfo {title} {Gauge theories},}\ }%
  \bibfield{journal}{%
  \Doi{10.1016/0370-1573(73)90027-6}{\bibinfo {journal} {Phys.Rept.}}\ }%
  \textbf{\bibinfo {volume} {9}},\ \bibinfo {pages} {1--141} (\bibinfo {year}
  {1973})%
  \bibAnnoteFile{NoStop}{Abers:1973qs}%

\bibitem{Santamaria:1993ah}%
  \BibitemOpen
  \bibfield{author}{%
  \bibinfo {author} {\bibfnamefont{Arcadi}\ \bibnamefont{Santamaria}},\ }%
  \bibfield{title}{%
  \enquote{\bibinfo {title} {{Masses, mixings, Yukawa couplings and their
  symmetries}},}\ }%
  \bibfield{journal}{%
  \Doi{10.1016/0370-2693(93)91110-9}{\bibinfo {journal} {Phys.Lett.}}\ }%
  \textbf{\bibinfo {volume} {B305}},\ \bibinfo {pages} {90--97} (\bibinfo
  {year} {1993}),\
  \Eprint{http://arxiv.org/abs/hep-ph/9302301}{arXiv:hep-ph/9302301 [hep-ph]}%
  \bibAnnoteFile{NoStop}{Santamaria:1993ah}%

\bibitem{Weinberg:1978kz}%
  \BibitemOpen
  \bibfield{author}{%
  \bibinfo {author} {\bibfnamefont{Steven}\ \bibnamefont{Weinberg}},\ }%
  \bibfield{title}{%
  \enquote{\bibinfo {title} {{Phenomenological Lagrangians}},}\ }%
  \bibfield{journal}{%
  \bibinfo {journal} {Physica}\ }%
  \textbf{\bibinfo {volume} {A96}},\ \bibinfo {pages} {327} (\bibinfo {year}
  {1979})%
  \bibAnnoteFile{NoStop}{Weinberg:1978kz}%

\bibitem{Georgi:1994qn}%
  \BibitemOpen
  \bibfield{author}{%
  \bibinfo {author} {\bibfnamefont{H.}~\bibnamefont{Georgi}},\ }%
  \bibfield{title}{%
  \enquote{\bibinfo {title} {{Effective field theory}},}\ }%
  \bibfield{journal}{%
  \bibinfo {journal} {Ann.Rev.Nucl.Part.Sci.}\ }%
  \textbf{\bibinfo {volume} {43}},\ \bibinfo {pages} {209--252} (\bibinfo
  {year} {1993})%
  \bibAnnoteFile{NoStop}{Georgi:1994qn}%

\bibitem{Kaplan:1995uv}%
  \BibitemOpen
  \bibfield{author}{%
  \bibinfo {author} {\bibfnamefont{David~B.}\ \bibnamefont{Kaplan}},\ }%
  \bibfield{title}{%
  \enquote{\bibinfo {title} {{Effective field theories}},}\ }%
   (\bibinfo {year} {1995}),\
  \Eprint{http://arxiv.org/abs/nucl-th/9506035}{arXiv:nucl-th/9506035
  [nucl-th]}%
  \bibAnnoteFile{NoStop}{Kaplan:1995uv}%

\bibitem{Burgess:1998nm}%
  \BibitemOpen
  \bibfield{author}{%
  \bibinfo {author} {\bibfnamefont{C.P.}\ \bibnamefont{Burgess}},\ }%
  \bibfield{title}{%
  \enquote{\bibinfo {title} {An ode to effective {Lagrangians}},}\ }%
  \bibfield{journal}{%
  \bibinfo {journal} {Proceedings of RADCOR 98}}%
   (\bibinfo {year} {1998}),\
  \Eprint{http://arxiv.org/abs/hep-ph/9812470}{arXiv:hep-ph/9812470 [hep-ph]}%
  \bibAnnoteFile{NoStop}{Burgess:1998nm}%

\bibitem{Manohar:1996cq}%
  \BibitemOpen
  \bibfield{author}{%
  \bibinfo {author} {\bibfnamefont{Aneesh~V.}\ \bibnamefont{Manohar}},\ }%
  \enquote{\bibinfo {title} {{Effective field theories}},}\ in\
  \Doi{10.1007/BFb0104294}{\emph{\bibinfo {booktitle} {Perturbative and
  Nonperturbative Aspects of Quantum Field Theory}}},\ \bibinfo {series}
  {Lecture Notes in Physics}, Vol.\ \bibinfo {volume} {479},\ \bibinfo {editor}
  {edited by\ \bibinfo {editor} {\bibfnamefont{H.}~\bibnamefont{Latal}}\ and\
  \bibinfo {editor} {\bibfnamefont{W.}~\bibnamefont{Schweiger}}}\ (\bibinfo
  {publisher} {Springer Berlin Heidelberg},\ \bibinfo {year} {1997})\ pp.\
  \bibinfo {pages} {311--362},\ ISBN \bibinfo {isbn} {978-3-540-62478-3},\
  \Eprint{http://arxiv.org/abs/hep-ph/9606222}{arXiv:hep-ph/9606222 [hep-ph]}%
  \bibAnnoteFile{NoStop}{Manohar:1996cq}%

\bibitem{Pich:1998xt}%
  \BibitemOpen
  \bibfield{author}{%
  \bibinfo {author} {\bibfnamefont{Antonio}\ \bibnamefont{Pich}},\ }%
  \bibfield{title}{%
  \enquote{\bibinfo {title} {{Effective field theory}},}\ ,\ \bibinfo {pages}
  {949--1049}}%
   (\bibinfo {year} {1998}),\
  \Eprint{http://arxiv.org/abs/hep-ph/9806303}{arXiv:hep-ph/9806303 [hep-ph]}%
  \bibAnnoteFile{NoStop}{Pich:1998xt}%

\bibitem{Wudka:1994ny}%
  \BibitemOpen
  \bibfield{author}{%
  \bibinfo {author} {\bibfnamefont{Jose}\ \bibnamefont{Wudka}},\ }%
  \bibfield{title}{%
  \enquote{\bibinfo {title} {{Electroweak effective Lagrangians}},}\ }%
  \bibfield{journal}{%
  \Doi{10.1142/S0217751X94000959}{\bibinfo {journal} {Int.J.Mod.Phys.}}\ }%
  \textbf{\bibinfo {volume} {A9}},\ \bibinfo {pages} {2301--2362} (\bibinfo
  {year} {1994}),\
  \Eprint{http://arxiv.org/abs/hep-ph/9406205}{arXiv:hep-ph/9406205 [hep-ph]}%
  \bibAnnoteFile{NoStop}{Wudka:1994ny}%

\bibitem{Mohapatra:2005wg}%
  \BibitemOpen
  \bibfield{author}{%
  \bibinfo {author} {\bibfnamefont{R.N.}\ \bibnamefont{Mohapatra}}, \bibinfo
  {author} {\bibfnamefont{S.}~\bibnamefont{Antusch}}, \bibinfo {author}
  {\bibfnamefont{K.S.}\ \bibnamefont{Babu}}, \bibinfo {author}
  {\bibfnamefont{G.}~\bibnamefont{Barenboim}}, \bibinfo {author}
  {\bibfnamefont{Mu-Chun}\ \bibnamefont{Chen}}, \emph{et~al.},\ }%
  \bibfield{title}{%
  \enquote{\bibinfo {title} {Theory of neutrinos: A white paper},}\ }%
  \bibfield{journal}{%
  \Doi{10.1088/0034-4885/70/11/R02}{\bibinfo {journal} {Rept.Prog.Phys.}}\ }%
  \textbf{\bibinfo {volume} {70}},\ \bibinfo {pages} {1757--1867} (\bibinfo
  {year} {2007}),\
  \Eprint{http://arxiv.org/abs/hep-ph/0510213}{arXiv:hep-ph/0510213 [hep-ph]}%
  \bibAnnoteFile{NoStop}{Mohapatra:2005wg}%

\bibitem{Mohapatra:1998rq}%
  \BibitemOpen
  \bibfield{author}{%
  \bibinfo {author} {\bibfnamefont{R.N.}\ \bibnamefont{Mohapatra}}\ and\
  \bibinfo {author} {\bibfnamefont{P.B.}\ \bibnamefont{Pal}},\ }%
  \emph{\bibinfo {title} {Massive neutrinos in physics and astrophysics}},\
  \bibinfo {edition} {3rd}\ ed.,\ \bibinfo {series} {Lecture Notes in Physics},
  Vol.~\bibinfo {volume} {72}\ (\bibinfo {publisher} {World Scientific},\
  \bibinfo {year} {2004})\ ISBN \bibinfo {isbn} {978-9812380708}%
  \bibAnnoteFile{NoStop}{Mohapatra:1998rq}%

\bibitem{Giunti:2007ry}%
  \BibitemOpen
  \bibfield{author}{%
  \bibinfo {author} {\bibfnamefont{Carlo}\ \bibnamefont{Giunti}}\ and\ \bibinfo
  {author} {\bibfnamefont{Chung~W.}\ \bibnamefont{Kim}},\ }%
  \emph{\bibinfo {title} {Fundamentals of Neutrino Physics and Astrophysics}}\
  (\bibinfo {publisher} {Oxford University Press},\ \bibinfo {year} {2007})\
  ISBN \bibinfo {isbn} {978-0198508717}%
  \bibAnnoteFile{NoStop}{Giunti:2007ry}%

\bibitem{Xing:2011zza}%
  \BibitemOpen
  \bibfield{author}{%
  \bibinfo {author} {\bibfnamefont{Zhi-zhong}\ \bibnamefont{Xing}}\ and\
  \bibinfo {author} {\bibfnamefont{Shun}\ \bibnamefont{Zhou}},\ }%
  \emph{\bibinfo {title} {{Neutrinos in particle physics, astronomy and
  cosmology}}}\ (\bibinfo {publisher} {Springer},\ \bibinfo {year} {2011})\
  ISBN \bibinfo {isbn} {978-3642175596}%
  \bibAnnoteFile{NoStop}{Xing:2011zza}%

\bibitem{Raffelt:1996wa}%
  \BibitemOpen
  \bibfield{author}{%
  \bibinfo {author} {\bibfnamefont{G.G.}\ \bibnamefont{Raffelt}},\ }%
  \emph{\bibinfo {title} {{Stars as laboratories for fundamental physics: the
  astrophysics of neutrinos, axions, and other weakly interacting particles}}}\
  (\bibinfo {publisher} {University of Chicago Press},\ \bibinfo {year}
  {1996})\ ISBN \bibinfo {isbn} {978-0226702728}%
  \bibAnnoteFile{NoStop}{Raffelt:1996wa}%

\bibitem{Strumia:2006db}%
  \BibitemOpen
  \bibfield{author}{%
  \bibinfo {author} {\bibfnamefont{Alessandro}\ \bibnamefont{Strumia}}\ and\
  \bibinfo {author} {\bibfnamefont{Francesco}\ \bibnamefont{Vissani}},\ }%
  \bibfield{title}{%
  \enquote{\bibinfo {title} {{Neutrino masses and mixings and...}}.}\ }%
   (\bibinfo {year} {2006}),\
  \Eprint{http://arxiv.org/abs/hep-ph/0606054}{arXiv:hep-ph/0606054 [hep-ph]}%
  \bibAnnoteFile{NoStop}{Strumia:2006db}%

\bibitem{GonzalezGarcia:2007ib}%
  \BibitemOpen
  \bibfield{author}{%
  \bibinfo {author} {\bibfnamefont{M.C.}\ \bibnamefont{Gonzalez-Garcia}}\ and\
  \bibinfo {author} {\bibfnamefont{Michele}\ \bibnamefont{Maltoni}},\ }%
  \bibfield{title}{%
  \enquote{\bibinfo {title} {Phenomenology with massive neutrinos},}\ }%
  \bibfield{journal}{%
  \Doi{10.1016/j.physrep.2007.12.004}{\bibinfo {journal} {Phys.Rept.}}\ }%
  \textbf{\bibinfo {volume} {460}},\ \bibinfo {pages} {1--129} (\bibinfo {year}
  {2008}),\ \Eprint{http://arxiv.org/abs/0704.1800}{arXiv:0704.1800 [hep-ph]}%
  \bibAnnoteFile{NoStop}{GonzalezGarcia:2007ib}%

\bibitem{Dirac:1928hu}%
  \BibitemOpen
  \bibfield{author}{%
  \bibinfo {author} {\bibfnamefont{Paul}\ \bibnamefont{Dirac}},\ }%
  \bibfield{title}{%
  \enquote{\bibinfo {title} {{The quantum theory of electron}},}\ }%
  \bibfield{journal}{%
  \Doi{10.1098/rspa.1928.0023}{\bibinfo {journal} {Proc. Roy. Soc. Lond.}}\ }%
  \textbf{\bibinfo {volume} {A117}},\ \bibinfo {pages} {610--624} (\bibinfo
  {year} {1928})%
  \bibAnnoteFile{NoStop}{Dirac:1928hu}%

\bibitem{Dirac:1928ej}%
  \BibitemOpen
  \bibfield{author}{%
  \bibinfo {author} {\bibfnamefont{Paul}\ \bibnamefont{Dirac}},\ }%
  \bibfield{title}{%
  \enquote{\bibinfo {title} {{The quantum theory of electron. 2.}}.}\ }%
  \bibfield{journal}{%
  \Doi{10.1098/rspa.1928.0056}{\bibinfo {journal} {Proc. Roy. Soc. Lond.}}\ }%
  \textbf{\bibinfo {volume} {A118}},\ \bibinfo {pages} {351} (\bibinfo {year}
  {1928})%
  \bibAnnoteFile{NoStop}{Dirac:1928ej}%

\bibitem{Majorana:1937vz}%
  \BibitemOpen
  \bibfield{author}{%
  \bibinfo {author} {\bibfnamefont{Ettore}\ \bibnamefont{Majorana}},\ }%
  \bibfield{title}{%
  \enquote{\bibinfo {title} {Theory of the symmetry of electrons and
  positrons},}\ }%
  \bibfield{journal}{%
  \Doi{10.1007/BF02961314}{\bibinfo {journal} {Nuovo Cim.}}\ }%
  \textbf{\bibinfo {volume} {14}},\ \bibinfo {pages} {171--184} (\bibinfo
  {year} {1937})%
  \bibAnnoteFile{NoStop}{Majorana:1937vz}%

\bibitem{Weinberg:1979sa}%
  \BibitemOpen
  \bibfield{author}{%
  \bibinfo {author} {\bibfnamefont{Steven}\ \bibnamefont{Weinberg}},\ }%
  \bibfield{title}{%
  \enquote{\bibinfo {title} {Baryon and lepton nonconserving processes},}\ }%
  \bibfield{journal}{%
  \Doi{10.1103/PhysRevLett.43.1566}{\bibinfo {journal} {Phys.Rev.Lett.}}\ }%
  \textbf{\bibinfo {volume} {43}},\ \bibinfo {pages} {1566--1570} (\bibinfo
  {year} {1979})%
  \bibAnnoteFile{NoStop}{Weinberg:1979sa}%

\bibitem{Weldon:1980gi}%
  \BibitemOpen
  \bibfield{author}{%
  \bibinfo {author} {\bibfnamefont{H.A.}\ \bibnamefont{Weldon}}\ and\ \bibinfo
  {author} {\bibfnamefont{A.}~\bibnamefont{Zee}},\ }%
  \bibfield{title}{%
  \enquote{\bibinfo {title} {Operator analysis of new physics},}\ }%
  \bibfield{journal}{%
  \Doi{10.1016/0550-3213(80)90218-7}{\bibinfo {journal} {Nucl.Phys.}}\ }%
  \textbf{\bibinfo {volume} {B173}},\ \bibinfo {pages} {269} (\bibinfo {year}
  {1980})%
  \bibAnnoteFile{NoStop}{Weldon:1980gi}%

\bibitem{Mohapatra:1980yp}%
  \BibitemOpen
  \bibfield{author}{%
  \bibinfo {author} {\bibfnamefont{Rabindra~N.}\ \bibnamefont{Mohapatra}}\ and\
  \bibinfo {author} {\bibfnamefont{Goran}\ \bibnamefont{Senjanovic}},\ }%
  \bibfield{title}{%
  \enquote{\bibinfo {title} {Neutrino masses and mixings in gauge models with
  spontaneous parity violation},}\ }%
  \bibfield{journal}{%
  \Doi{10.1103/PhysRevD.23.165}{\bibinfo {journal} {Phys.Rev.}}\ }%
  \textbf{\bibinfo {volume} {D23}},\ \bibinfo {pages} {165} (\bibinfo {year}
  {1981})%
  \bibAnnoteFile{NoStop}{Mohapatra:1980yp}%

\bibitem{Cheng:1980qt}%
  \BibitemOpen
  \bibfield{author}{%
  \bibinfo {author} {\bibfnamefont{T.P.}\ \bibnamefont{Cheng}}\ and\ \bibinfo
  {author} {\bibfnamefont{Ling-Fong}\ \bibnamefont{Li}},\ }%
  \bibfield{title}{%
  \enquote{\bibinfo {title} {Neutrino masses, mixings and oscillations in
  {$SU(2) \otimes U(1)$} models of electroweak interactions},}\ }%
  \bibfield{journal}{%
  \Doi{10.1103/PhysRevD.22.2860}{\bibinfo {journal} {Phys.Rev.}}\ }%
  \textbf{\bibinfo {volume} {D22}},\ \bibinfo {pages} {2860} (\bibinfo {year}
  {1980})%
  \bibAnnoteFile{NoStop}{Cheng:1980qt}%

\bibitem{Zee:1985id}%
  \BibitemOpen
  \bibfield{author}{%
  \bibinfo {author} {\bibfnamefont{A.}~\bibnamefont{Zee}},\ }%
  \bibfield{title}{%
  \enquote{\bibinfo {title} {Quantum numbers of {Majorana} neutrino masses},}\
  }%
  \bibfield{journal}{%
  \Doi{10.1016/0550-3213(86)90475-X}{\bibinfo {journal} {Nucl.Phys.}}\ }%
  \textbf{\bibinfo {volume} {B264}},\ \bibinfo {pages} {99} (\bibinfo {year}
  {1986})%
  \bibAnnoteFile{NoStop}{Zee:1985id}%

\bibitem{SilvaMarcos:1999wy}%
  \BibitemOpen
  \bibfield{author}{%
  \bibinfo {author} {\bibfnamefont{J.I.}\ \bibnamefont{Silva-Marcos}},\ }%
  \bibfield{title}{%
  \enquote{\bibinfo {title} {{Alternative to the seesaw mechanism}},}\ }%
  \bibfield{journal}{%
  \Doi{10.1103/PhysRevD.59.091301}{\bibinfo {journal} {Phys.Rev.}}\ }%
  \textbf{\bibinfo {volume} {D59}},\ \bibinfo {pages} {091301} (\bibinfo {year}
  {1999})%
  \bibAnnoteFile{NoStop}{SilvaMarcos:1999wy}%

\bibitem{Ma:2006km}%
  \BibitemOpen
  \bibfield{author}{%
  \bibinfo {author} {\bibfnamefont{Ernest}\ \bibnamefont{Ma}},\ }%
  \bibfield{title}{%
  \enquote{\bibinfo {title} {{Verifiable radiative seesaw mechanism of neutrino
  mass and dark matter}},}\ }%
  \bibfield{journal}{%
  \Doi{10.1103/PhysRevD.73.077301}{\bibinfo {journal} {Phys.Rev.}}\ }%
  \textbf{\bibinfo {volume} {D73}},\ \bibinfo {pages} {077301} (\bibinfo {year}
  {2006}),\ \Eprint{http://arxiv.org/abs/hep-ph/0601225}{arXiv:hep-ph/0601225
  [hep-ph]}%
  \bibAnnoteFile{NoStop}{Ma:2006km}%

\bibitem{Ibanez:2009du}%
  \BibitemOpen
  \bibfield{author}{%
  \bibinfo {author} {\bibfnamefont{D.}~\bibnamefont{Ibanez}}, \bibinfo {author}
  {\bibfnamefont{S.}~\bibnamefont{Morisi}},\ and\ \bibinfo {author}
  {\bibfnamefont{J.W.F.}\ \bibnamefont{Valle}},\ }%
  \bibfield{title}{%
  \enquote{\bibinfo {title} {{Inverse tri-bimaximal type III seesaw and lepton
  flavor violation}},}\ }%
  \bibfield{journal}{%
  \Doi{10.1103/PhysRevD.80.053015}{\bibinfo {journal} {Phys.Rev.}}\ }%
  \textbf{\bibinfo {volume} {D80}},\ \bibinfo {pages} {053015} (\bibinfo {year}
  {2009}),\ \Eprint{http://arxiv.org/abs/0907.3109}{arXiv:0907.3109 [hep-ph]}%
  \bibAnnoteFile{NoStop}{Ibanez:2009du}%

\bibitem{Aparici:2011nu}%
  \BibitemOpen
  \bibfield{author}{%
  \bibinfo {author} {\bibfnamefont{Alberto}\ \bibnamefont{Aparici}}, \bibinfo
  {author} {\bibfnamefont{Juan}\ \bibnamefont{Herrero-Garcia}}, \bibinfo
  {author} {\bibfnamefont{Nuria}\ \bibnamefont{Rius}},\ and\ \bibinfo {author}
  {\bibfnamefont{Arcadi}\ \bibnamefont{Santamaria}},\ }%
  \bibfield{title}{%
  \enquote{\bibinfo {title} {{Neutrino masses from new generations}},}\ }%
  \bibfield{journal}{%
  \Doi{10.1007/JHEP07(2011)122}{\bibinfo {journal} {JHEP}}\ }%
  \textbf{\bibinfo {volume} {1107}},\ \bibinfo {pages} {122} (\bibinfo {year}
  {2011}),\ \Eprint{http://arxiv.org/abs/1104.4068}{arXiv:1104.4068 [hep-ph]}%
  \bibAnnoteFile{NoStop}{Aparici:2011nu}%

\bibitem{Peinado:2012tp}%
  \BibitemOpen
  \bibfield{author}{%
  \bibinfo {author} {\bibfnamefont{E.}~\bibnamefont{Peinado}}\ and\ \bibinfo
  {author} {\bibfnamefont{A.}~\bibnamefont{Vicente}},\ }%
  \bibfield{title}{%
  \enquote{\bibinfo {title} {Neutrino masses from {R}-parity violation with a
  {$Z_3$} symmetry},}\ }%
  \bibfield{journal}{%
  \Doi{10.1103/PhysRevD.86.093024}{\bibinfo {journal} {Phys.Rev.}}\ }%
  \textbf{\bibinfo {volume} {D86}},\ \bibinfo {pages} {093024} (\bibinfo {year}
  {2012}),\ \Eprint{http://arxiv.org/abs/1207.6641}{arXiv:1207.6641 [hep-ph]}%
  \bibAnnoteFile{NoStop}{Peinado:2012tp}%

\bibitem{Davis:1964hf}%
  \BibitemOpen
  \bibfield{author}{%
  \bibinfo {author} {\bibfnamefont{R.}~\bibnamefont{Davis}},\ }%
  \bibfield{title}{%
  \enquote{\bibinfo {title} {{Solar neutrinos. II: Experimental}},}\ }%
  \bibfield{journal}{%
  \Doi{10.1103/PhysRevLett.12.303}{\bibinfo {journal} {Phys.Rev.Lett.}}\ }%
  \textbf{\bibinfo {volume} {12}},\ \bibinfo {pages} {303--305} (\bibinfo
  {year} {1964})%
  \bibAnnoteFile{NoStop}{Davis:1964hf}%

\bibitem{Cleveland:1998nv}%
  \BibitemOpen
  \bibfield{author}{%
  \bibinfo {author} {\bibfnamefont{B.T.}\ \bibnamefont{Cleveland}}, \bibinfo
  {author} {\bibfnamefont{Timothy}\ \bibnamefont{Daily}}, \bibinfo {author}
  {\bibfnamefont{Jr.}\ \bibnamefont{Davis}, \bibfnamefont{Raymond}}, \bibinfo
  {author} {\bibfnamefont{James~R.}\ \bibnamefont{Distel}}, \bibinfo {author}
  {\bibfnamefont{Kenneth}\ \bibnamefont{Lande}}, \emph{et~al.},\ }%
  \bibfield{title}{%
  \enquote{\bibinfo {title} {{Measurement of the solar electron neutrino flux
  with the Homestake chlorine detector}},}\ }%
  \bibfield{journal}{%
  \Doi{10.1086/305343}{\bibinfo {journal} {Astrophys.J.}}\ }%
  \textbf{\bibinfo {volume} {496}},\ \bibinfo {pages} {505--526} (\bibinfo
  {year} {1998})%
  \bibAnnoteFile{NoStop}{Cleveland:1998nv}%

\bibitem{Fukuda:1996sz}%
  \BibitemOpen
  \bibfield{author}{%
  \bibinfo {author} {\bibfnamefont{Y.}~\bibnamefont{Fukuda}} \emph{et~al.}
  (\bibinfo {collaboration} {Kamiokande Collaboration}),\ }%
  \bibfield{title}{%
  \enquote{\bibinfo {title} {{Solar neutrino data covering solar cycle 22}},}\
  }%
  \bibfield{journal}{%
  \Doi{10.1103/PhysRevLett.77.1683}{\bibinfo {journal} {Phys.Rev.Lett.}}\ }%
  \textbf{\bibinfo {volume} {77}},\ \bibinfo {pages} {1683--1686} (\bibinfo
  {year} {1996})%
  \bibAnnoteFile{NoStop}{Fukuda:1996sz}%

\bibitem{Hampel:1998xg}%
  \BibitemOpen
  \bibfield{author}{%
  \bibinfo {author} {\bibfnamefont{W.}~\bibnamefont{Hampel}} \emph{et~al.}
  (\bibinfo {collaboration} {GALLEX Collaboration}),\ }%
  \bibfield{title}{%
  \enquote{\bibinfo {title} {{GALLEX solar neutrino observations: Results for
  GALLEX IV}},}\ }%
  \bibfield{journal}{%
  \Doi{10.1016/S0370-2693(98)01579-2}{\bibinfo {journal} {Phys.Lett.}}\ }%
  \textbf{\bibinfo {volume} {B447}},\ \bibinfo {pages} {127--133} (\bibinfo
  {year} {1999})%
  \bibAnnoteFile{NoStop}{Hampel:1998xg}%

\bibitem{Fukuda:1998fd}%
  \BibitemOpen
  \bibfield{author}{%
  \bibinfo {author} {\bibfnamefont{Y.}~\bibnamefont{Fukuda}} \emph{et~al.}
  (\bibinfo {collaboration} {Super-Kamiokande Collaboration}),\ }%
  \bibfield{title}{%
  \enquote{\bibinfo {title} {{Measurements of the solar neutrino flux from
  Super-Kamiokande's first 300 days}},}\ }%
  \bibfield{journal}{%
  \Doi{10.1103/PhysRevLett.81.1158}{\bibinfo {journal} {Phys.Rev.Lett.}}\ }%
  \textbf{\bibinfo {volume} {81}},\ \bibinfo {pages} {1158--1162} (\bibinfo
  {year} {1998}),\
  \Eprint{http://arxiv.org/abs/hep-ex/9805021}{arXiv:hep-ex/9805021 [hep-ex]}%
  \bibAnnoteFile{NoStop}{Fukuda:1998fd}%

\bibitem{Ahmad:2001an}%
  \BibitemOpen
  \bibfield{author}{%
  \bibinfo {author} {\bibfnamefont{Q.R.}\ \bibnamefont{Ahmad}} \emph{et~al.}
  (\bibinfo {collaboration} {SNO Collaboration}),\ }%
  \bibfield{title}{%
  \enquote{\bibinfo {title} {Measurement of the rate of $\nu_e + d \rightarrow
  p + p + e^-$ interactions produced by {$\tensor[^8]{\mathrm{B}}{}$ solar
  neutrinos at the Sudbury Neutrino Observatory}},}\ }%
  \bibfield{journal}{%
  \Doi{10.1103/PhysRevLett.87.071301}{\bibinfo {journal} {Phys.Rev.Lett.}}\ }%
  \textbf{\bibinfo {volume} {87}},\ \bibinfo {pages} {071301} (\bibinfo {year}
  {2001}),\ \Eprint{http://arxiv.org/abs/nucl-ex/0106015}{arXiv:nucl-ex/0106015
  [nucl-ex]}%
  \bibAnnoteFile{NoStop}{Ahmad:2001an}%

\bibitem{Abdurashitov:2002nt}%
  \BibitemOpen
  \bibfield{author}{%
  \bibinfo {author} {\bibfnamefont{J.N.}\ \bibnamefont{Abdurashitov}}
  \emph{et~al.} (\bibinfo {collaboration} {SAGE Collaboration}),\ }%
  \bibfield{title}{%
  \enquote{\bibinfo {title} {{Solar neutrino flux measurements by the
  Soviet-American Gallium Experiment (SAGE) for half the 22 year solar
  cycle}},}\ }%
  \bibfield{journal}{%
  \Doi{10.1134/1.1506424}{\bibinfo {journal} {J.Exp.Theor.Phys.}}\ }%
  \textbf{\bibinfo {volume} {95}},\ \bibinfo {pages} {181--193} (\bibinfo
  {year} {2002}),\
  \Eprint{http://arxiv.org/abs/astro-ph/0204245}{arXiv:astro-ph/0204245
  [astro-ph]}%
  \bibAnnoteFile{NoStop}{Abdurashitov:2002nt}%

\bibitem{Altmann:2005ix}%
  \BibitemOpen
  \bibfield{author}{%
  \bibinfo {author} {\bibfnamefont{M.}~\bibnamefont{Altmann}} \emph{et~al.}
  (\bibinfo {collaboration} {GNO COLLABORATION}),\ }%
  \bibfield{title}{%
  \enquote{\bibinfo {title} {{Complete results for five years of GNO solar
  neutrino observations}},}\ }%
  \bibfield{journal}{%
  \Doi{10.1016/j.physletb.2005.04.068}{\bibinfo {journal} {Phys.Lett.}}\ }%
  \textbf{\bibinfo {volume} {B616}},\ \bibinfo {pages} {174--190} (\bibinfo
  {year} {2005}),\
  \Eprint{http://arxiv.org/abs/hep-ex/0504037}{arXiv:hep-ex/0504037 [hep-ex]}%
  \bibAnnoteFile{NoStop}{Altmann:2005ix}%

\bibitem{Reines:1965qk}%
  \BibitemOpen
  \bibfield{author}{%
  \bibinfo {author} {\bibfnamefont{F.}~\bibnamefont{Reines}}, \bibinfo {author}
  {\bibfnamefont{M.F.}\ \bibnamefont{Crouch}}, \bibinfo {author}
  {\bibfnamefont{T.L.}\ \bibnamefont{Jenkins}}, \bibinfo {author}
  {\bibfnamefont{W.R.}\ \bibnamefont{Kropp}}, \bibinfo {author}
  {\bibfnamefont{H.S.}\ \bibnamefont{Gurr}}, \emph{et~al.},\ }%
  \bibfield{title}{%
  \enquote{\bibinfo {title} {{Evidence for high-energy cosmic ray neutrino
  interactions}},}\ }%
  \bibfield{journal}{%
  \Doi{10.1103/PhysRevLett.15.429}{\bibinfo {journal} {Phys.Rev.Lett.}}\ }%
  \textbf{\bibinfo {volume} {15}},\ \bibinfo {pages} {429--433} (\bibinfo
  {year} {1965})%
  \bibAnnoteFile{NoStop}{Reines:1965qk}%

\bibitem{Achar1965196}%
  \BibitemOpen
  \bibfield{author}{%
  \bibinfo {author} {\bibfnamefont{C.V.}\ \bibnamefont{Achar}} \emph{et~al.},\
  }%
  \bibfield{title}{%
  \enquote{\bibinfo {title} {Detection of muons produced by cosmic ray
  neutrinos deep underground},}\ }%
  \bibfield{journal}{%
  \Doi{10.1016/0031-9163(65)90712-2}{\bibinfo {journal} {Physics Letters}}\ }%
  \textbf{\bibinfo {volume} {18}},\ \bibinfo {pages} {196 -- 199} (\bibinfo
  {year} {1965}),\ ISSN \bibinfo {issn} {0031-9163}%
  \bibAnnoteFile{NoStop}{Achar1965196}%

\bibitem{Aglietta:1988be}%
  \BibitemOpen
  \bibfield{author}{%
  \bibinfo {author} {\bibfnamefont{M.}~\bibnamefont{Aglietta}} \emph{et~al.}
  (\bibinfo {collaboration} {NUSEX Collaboration}),\ }%
  \bibfield{title}{%
  \enquote{\bibinfo {title} {{Experimental study of atmospheric neutrino flux
  in the NUSEX experiment}},}\ }%
  \bibfield{journal}{%
  \Doi{10.1209/0295-5075/8/7/005}{\bibinfo {journal} {Europhys.Lett.}}\ }%
  \textbf{\bibinfo {volume} {8}},\ \bibinfo {pages} {611--614} (\bibinfo {year}
  {1989})%
  \bibAnnoteFile{NoStop}{Aglietta:1988be}%

\bibitem{Daum:1994bf}%
  \BibitemOpen
  \bibfield{author}{%
  \bibinfo {author} {\bibfnamefont{K.}~\bibnamefont{Daum}} \emph{et~al.}
  (\bibinfo {collaboration} {Frejus Collaboration.}),\ }%
  \bibfield{title}{%
  \enquote{\bibinfo {title} {{Determination of the atmospheric neutrino spectra
  with the Fréjus detector}},}\ }%
  \bibfield{journal}{%
  \Doi{10.1007/BF01556368}{\bibinfo {journal} {Z.Phys.}}\ }%
  \textbf{\bibinfo {volume} {C66}},\ \bibinfo {pages} {417--428} (\bibinfo
  {year} {1995})%
  \bibAnnoteFile{NoStop}{Daum:1994bf}%

\bibitem{BeckerSzendy:1992hq}%
  \BibitemOpen
  \bibfield{author}{%
  \bibinfo {author} {\bibfnamefont{R.}~\bibnamefont{Becker-Szendy}}, \bibinfo
  {author} {\bibfnamefont{C.B.}\ \bibnamefont{Bratton}}, \bibinfo {author}
  {\bibfnamefont{D.}~\bibnamefont{Casper}}, \bibinfo {author}
  {\bibfnamefont{S.T.}\ \bibnamefont{Dye}}, \bibinfo {author}
  {\bibfnamefont{W.}~\bibnamefont{Gajewski}}, \emph{et~al.},\ }%
  \bibfield{title}{%
  \enquote{\bibinfo {title} {The electron-neutrino and muon-neutrino content of
  the atmospheric flux},}\ }%
  \bibfield{journal}{%
  \Doi{10.1103/PhysRevD.46.3720}{\bibinfo {journal} {Phys.Rev.}}\ }%
  \textbf{\bibinfo {volume} {D46}},\ \bibinfo {pages} {3720--3724} (\bibinfo
  {year} {1992})%
  \bibAnnoteFile{NoStop}{BeckerSzendy:1992hq}%

\bibitem{Fukuda:1994mc}%
  \BibitemOpen
  \bibfield{author}{%
  \bibinfo {author} {\bibfnamefont{Y.}~\bibnamefont{Fukuda}} \emph{et~al.}
  (\bibinfo {collaboration} {Kamiokande Collaboration}),\ }%
  \bibfield{title}{%
  \enquote{\bibinfo {title} {{Atmospheric muon-neutrino / electron-neutrino
  ratio in the multiGeV energy range}},}\ }%
  \bibfield{journal}{%
  \Doi{10.1016/0370-2693(94)91420-6}{\bibinfo {journal} {Phys.Lett.}}\ }%
  \textbf{\bibinfo {volume} {B335}},\ \bibinfo {pages} {237--245} (\bibinfo
  {year} {1994})%
  \bibAnnoteFile{NoStop}{Fukuda:1994mc}%

\bibitem{Ahmad:2002jz}%
  \BibitemOpen
  \bibfield{author}{%
  \bibinfo {author} {\bibfnamefont{Q.R.}\ \bibnamefont{Ahmad}} \emph{et~al.}
  (\bibinfo {collaboration} {SNO Collaboration}),\ }%
  \bibfield{title}{%
  \enquote{\bibinfo {title} {{Direct evidence for neutrino flavor
  transformation from neutral current interactions in the Sudbury Neutrino
  Observatory}},}\ }%
  \bibfield{journal}{%
  \Doi{10.1103/PhysRevLett.89.011301}{\bibinfo {journal} {Phys.Rev.Lett.}}\ }%
  \textbf{\bibinfo {volume} {89}},\ \bibinfo {pages} {011301} (\bibinfo {year}
  {2002}),\ \Eprint{http://arxiv.org/abs/nucl-ex/0204008}{arXiv:nucl-ex/0204008
  [nucl-ex]}%
  \bibAnnoteFile{NoStop}{Ahmad:2002jz}%

\bibitem{Giunti:2000kw}%
  \BibitemOpen
  \bibfield{author}{%
  \bibinfo {author} {\bibfnamefont{Carlo}\ \bibnamefont{Giunti}}\ and\ \bibinfo
  {author} {\bibfnamefont{Chung~W.}\ \bibnamefont{Kim}},\ }%
  \bibfield{title}{%
  \enquote{\bibinfo {title} {{Quantum mechanics of neutrino oscillations}},}\
  }%
  \bibfield{journal}{%
  \Doi{10.1023/A:1012230026160}{\bibinfo {journal} {Found.Phys.Lett.}}\ }%
  \textbf{\bibinfo {volume} {14}},\ \bibinfo {pages} {213--229} (\bibinfo
  {year} {2001}),\
  \Eprint{http://arxiv.org/abs/hep-ph/0011074}{arXiv:hep-ph/0011074 [hep-ph]}%
  \bibAnnoteFile{NoStop}{Giunti:2000kw}%

\bibitem{Giunti:2008cf}%
  \BibitemOpen
  \bibfield{author}{%
  \bibinfo {author} {\bibfnamefont{Carlo}\ \bibnamefont{Giunti}},\ }%
  \bibfield{title}{%
  \enquote{\bibinfo {title} {Neutrino flavor states and the quantum theory of
  neutrino oscillations},}\ }%
  \bibfield{journal}{%
  \Doi{10.1063/1.2965075}{\bibinfo {journal} {AIP Conf.Proc.}}\ }%
  \textbf{\bibinfo {volume} {1026}},\ \bibinfo {pages} {3--19} (\bibinfo {year}
  {2008}),\ \Eprint{http://arxiv.org/abs/0801.0653}{arXiv:0801.0653 [hep-ph]}%
  \bibAnnoteFile{NoStop}{Giunti:2008cf}%

\bibitem{Waltham:2003tf}%
  \BibitemOpen
  \bibfield{author}{%
  \bibinfo {author} {\bibfnamefont{Chris}\ \bibnamefont{Waltham}},\ }%
  \bibfield{title}{%
  \enquote{\bibinfo {title} {{Teaching neutrino oscillations}},}\ }%
  \bibfield{journal}{%
  \Doi{10.1119/1.1646132}{\bibinfo {journal} {Am.J.Phys.}}\ }%
  \textbf{\bibinfo {volume} {72}},\ \bibinfo {pages} {742--752} (\bibinfo
  {year} {2004}),\
  \Eprint{http://arxiv.org/abs/physics/0303116}{arXiv:physics/0303116
  [physics.ed-ph]}%
  \bibAnnoteFile{NoStop}{Waltham:2003tf}%

\bibitem{Araki:2004mb}%
  \BibitemOpen
  \bibfield{author}{%
  \bibinfo {author} {\bibfnamefont{T.}~\bibnamefont{Araki}} \emph{et~al.}
  (\bibinfo {collaboration} {KamLAND Collaboration}),\ }%
  \bibfield{title}{%
  \enquote{\bibinfo {title} {{Measurement of neutrino oscillation with KamLAND:
  Evidence of spectral distortion}},}\ }%
  \bibfield{journal}{%
  \Doi{10.1103/PhysRevLett.94.081801}{\bibinfo {journal} {Phys.Rev.Lett.}}\ }%
  \textbf{\bibinfo {volume} {94}},\ \bibinfo {pages} {081801} (\bibinfo {year}
  {2005}),\ \Eprint{http://arxiv.org/abs/hep-ex/0406035}{arXiv:hep-ex/0406035
  [hep-ex]}%
  \bibAnnoteFile{NoStop}{Araki:2004mb}%

\bibitem{Pontecorvo:1957cp}%
  \BibitemOpen
  \bibfield{author}{%
  \bibinfo {author} {\bibfnamefont{B.}~\bibnamefont{Pontecorvo}},\ }%
  \bibfield{title}{%
  \enquote{\bibinfo {title} {{Mesonium and anti-mesonium}},}\ }%
  \bibfield{journal}{%
  \bibinfo {journal} {Sov.Phys.JETP}\ }%
  \textbf{\bibinfo {volume} {6}},\ \bibinfo {pages} {429} (\bibinfo {year}
  {1957}),\
  \url{http://puhep1.princeton.edu/~mcdonald/examples/EP/pontecorvo_spjetp_6_4%
29_57.pdf}%
  \bibAnnoteFile{NoStop}{Pontecorvo:1957cp}%

\bibitem{Pontecorvo:1957qd}%
  \BibitemOpen
  \bibfield{author}{%
  \bibinfo {author} {\bibfnamefont{B.}~\bibnamefont{Pontecorvo}},\ }%
  \bibfield{title}{%
  \enquote{\bibinfo {title} {{Inverse beta processes and nonconservation of
  lepton charge}},}\ }%
  \bibfield{journal}{%
  \bibinfo {journal} {Sov.Phys.JETP}\ }%
  \textbf{\bibinfo {volume} {7}},\ \bibinfo {pages} {172--173} (\bibinfo {year}
  {1958})%
  \bibAnnoteFile{NoStop}{Pontecorvo:1957qd}%

\bibitem{Maki:1962mu}%
  \BibitemOpen
  \bibfield{author}{%
  \bibinfo {author} {\bibfnamefont{Ziro}\ \bibnamefont{Maki}}, \bibinfo
  {author} {\bibfnamefont{Masami}\ \bibnamefont{Nakagawa}},\ and\ \bibinfo
  {author} {\bibfnamefont{Shoichi}\ \bibnamefont{Sakata}},\ }%
  \bibfield{title}{%
  \enquote{\bibinfo {title} {{Remarks on the unified model of elementary
  particles}},}\ }%
  \bibfield{journal}{%
  \Doi{10.1143/PTP.28.870}{\bibinfo {journal} {Prog.Theor.Phys.}}\ }%
  \textbf{\bibinfo {volume} {28}},\ \bibinfo {pages} {870--880} (\bibinfo
  {year} {1962})%
  \bibAnnoteFile{NoStop}{Maki:1962mu}%

\bibitem{Nakagawa:1963uw}%
  \BibitemOpen
  \bibfield{author}{%
  \bibinfo {author} {\bibfnamefont{M.}~\bibnamefont{Nakagawa}}, \bibinfo
  {author} {\bibfnamefont{H.}~\bibnamefont{Okonogi}}, \bibinfo {author}
  {\bibfnamefont{S.}~\bibnamefont{Sakata}},\ and\ \bibinfo {author}
  {\bibfnamefont{A.}~\bibnamefont{Toyoda}},\ }%
  \bibfield{title}{%
  \enquote{\bibinfo {title} {{Possible existence of a neutrino with mass and
  partial conservation of muon charge}},}\ }%
  \bibfield{journal}{%
  \Doi{10.1143/PTP.30.727}{\bibinfo {journal} {Prog.Theor.Phys.}}\ }%
  \textbf{\bibinfo {volume} {30}},\ \bibinfo {pages} {727--729} (\bibinfo
  {year} {1963})%
  \bibAnnoteFile{NoStop}{Nakagawa:1963uw}%

\bibitem{Schechter:1980gr}%
  \BibitemOpen
  \bibfield{author}{%
  \bibinfo {author} {\bibfnamefont{J.}~\bibnamefont{Schechter}}\ and\ \bibinfo
  {author} {\bibfnamefont{J.W.F.}\ \bibnamefont{Valle}},\ }%
  \bibfield{title}{%
  \enquote{\bibinfo {title} {Neutrino masses in {$SU(2) \otimes U(1)$}
  theories},}\ }%
  \bibfield{journal}{%
  \Doi{10.1103/PhysRevD.22.2227}{\bibinfo {journal} {Phys.Rev.}}\ }%
  \textbf{\bibinfo {volume} {D22}},\ \bibinfo {pages} {2227} (\bibinfo {year}
  {1980})%
  \bibAnnoteFile{NoStop}{Schechter:1980gr}%

\bibitem{Harrison:2002er}%
  \BibitemOpen
  \bibfield{author}{%
  \bibinfo {author} {\bibfnamefont{P.F.}\ \bibnamefont{Harrison}}, \bibinfo
  {author} {\bibfnamefont{D.H.}\ \bibnamefont{Perkins}},\ and\ \bibinfo
  {author} {\bibfnamefont{W.G.}\ \bibnamefont{Scott}},\ }%
  \bibfield{title}{%
  \enquote{\bibinfo {title} {{Tri-bimaximal mixing and the neutrino oscillation
  data}},}\ }%
  \bibfield{journal}{%
  \Doi{10.1016/S0370-2693(02)01336-9}{\bibinfo {journal} {Phys.Lett.}}\ }%
  \textbf{\bibinfo {volume} {B530}},\ \bibinfo {pages} {167} (\bibinfo {year}
  {2002}),\ \Eprint{http://arxiv.org/abs/hep-ph/0202074}{arXiv:hep-ph/0202074
  [hep-ph]}%
  \bibAnnoteFile{NoStop}{Harrison:2002er}%

\bibitem{Antonelli:2012qu}%
  \BibitemOpen
  \bibfield{author}{%
  \bibinfo {author} {\bibfnamefont{V.}~\bibnamefont{Antonelli}}, \bibinfo
  {author} {\bibfnamefont{L.}~\bibnamefont{Miramonti}}, \bibinfo {author}
  {\bibfnamefont{C.}~\bibnamefont{Pena~Garay}},\ and\ \bibinfo {author}
  {\bibfnamefont{A.}~\bibnamefont{Serenelli}},\ }%
  \bibfield{title}{%
  \enquote{\bibinfo {title} {Solar neutrinos},}\ }%
  \bibfield{journal}{%
  \Doi{10.1155/2013/351926}{\bibinfo {journal} {Adv.High Energy Phys.}}\ }%
  \textbf{\bibinfo {volume} {2013}},\ \bibinfo {pages} {351926} (\bibinfo
  {year} {2013}),\ \Eprint{http://arxiv.org/abs/1208.1356}{arXiv:1208.1356
  [hep-ex]}%
  \bibAnnoteFile{NoStop}{Antonelli:2012qu}%

\bibitem{Wolfenstein:1977ue}%
  \BibitemOpen
  \bibfield{author}{%
  \bibinfo {author} {\bibfnamefont{L.}~\bibnamefont{Wolfenstein}},\ }%
  \bibfield{title}{%
  \enquote{\bibinfo {title} {Neutrino oscillations in matter},}\ }%
  \bibfield{journal}{%
  \Doi{10.1103/PhysRevD.17.2369}{\bibinfo {journal} {Phys.Rev.}}\ }%
  \textbf{\bibinfo {volume} {D17}},\ \bibinfo {pages} {2369--2374} (\bibinfo
  {year} {1978})%
  \bibAnnoteFile{NoStop}{Wolfenstein:1977ue}%

\bibitem{Wolfenstein:1979ni}%
  \BibitemOpen
  \bibfield{author}{%
  \bibinfo {author} {\bibfnamefont{L.}~\bibnamefont{Wolfenstein}},\ }%
  \bibfield{title}{%
  \enquote{\bibinfo {title} {Neutrino oscillations and stellar collapse},}\ }%
  \bibfield{journal}{%
  \Doi{10.1103/PhysRevD.20.2634}{\bibinfo {journal} {Phys.Rev.}}\ }%
  \textbf{\bibinfo {volume} {D20}},\ \bibinfo {pages} {2634--2635} (\bibinfo
  {year} {1979})%
  \bibAnnoteFile{NoStop}{Wolfenstein:1979ni}%

\bibitem{Mikheev:1986wj}%
  \BibitemOpen
  \bibfield{author}{%
  \bibinfo {author} {\bibfnamefont{S.P.}\ \bibnamefont{Mikheev}}\ and\ \bibinfo
  {author} {\bibfnamefont{A.~Yu.}\ \bibnamefont{Smirnov}},\ }%
  \bibfield{title}{%
  \enquote{\bibinfo {title} {{Resonant amplification of neutrino oscillations
  in matter and solar neutrino spectroscopy}},}\ }%
  \bibfield{journal}{%
  \Doi{10.1007/BF02508049}{\bibinfo {journal} {Nuovo Cim.}}\ }%
  \textbf{\bibinfo {volume} {C9}},\ \bibinfo {pages} {17--26} (\bibinfo {year}
  {1986})%
  \bibAnnoteFile{NoStop}{Mikheev:1986wj}%

\bibitem{Mikheev:1986gs}%
  \BibitemOpen
  \bibfield{author}{%
  \bibinfo {author} {\bibfnamefont{S.P.}\ \bibnamefont{Mikheev}}\ and\ \bibinfo
  {author} {\bibfnamefont{A.~Yu.}\ \bibnamefont{Smirnov}},\ }%
  \bibfield{title}{%
  \enquote{\bibinfo {title} {Resonance amplification of oscillations in matter
  and spectroscopy of solar neutrinos},}\ }%
  \bibfield{journal}{%
  \bibinfo {journal} {Sov.J.Nucl.Phys.}\ }%
  \textbf{\bibinfo {volume} {42}},\ \bibinfo {pages} {913--917} (\bibinfo
  {year} {1985})%
  \bibAnnoteFile{NoStop}{Mikheev:1986gs}%

\bibitem{Smirnov:2004zv}%
  \BibitemOpen
  \bibfield{author}{%
  \bibinfo {author} {\bibfnamefont{A.~Yu.}\ \bibnamefont{Smirnov}},\ }%
  \bibfield{title}{%
  \enquote{\bibinfo {title} {{The MSW effect and matter effects in neutrino
  oscillations}},}\ }%
  \bibfield{journal}{%
  \Doi{10.1088/0031-8949/2005/T121/008}{\bibinfo {journal} {Phys.Scripta}}\ }%
  \textbf{\bibinfo {volume} {T121}},\ \bibinfo {pages} {57--64} (\bibinfo
  {year} {2005}),\
  \Eprint{http://arxiv.org/abs/hep-ph/0412391}{arXiv:hep-ph/0412391 [hep-ph]}%
  \bibAnnoteFile{NoStop}{Smirnov:2004zv}%

\bibitem{Kajita:2012vc}%
  \BibitemOpen
  \bibfield{author}{%
  \bibinfo {author} {\bibfnamefont{Takaaki}\ \bibnamefont{Kajita}},\ }%
  \bibfield{title}{%
  \enquote{\bibinfo {title} {{Atmospheric neutrinos}},}\ }%
  \bibfield{journal}{%
  \Doi{10.1155/2012/504715}{\bibinfo {journal} {Adv.High Energy Phys.}}\ }%
  \textbf{\bibinfo {volume} {2012}},\ \bibinfo {pages} {504715} (\bibinfo
  {year} {2012})%
  \bibAnnoteFile{NoStop}{Kajita:2012vc}%

\bibitem{Fukuda:1998mi}%
  \BibitemOpen
  \bibfield{author}{%
  \bibinfo {author} {\bibfnamefont{Y.}~\bibnamefont{Fukuda}} \emph{et~al.}
  (\bibinfo {collaboration} {Super-Kamiokande Collaboration}),\ }%
  \bibfield{title}{%
  \enquote{\bibinfo {title} {{Evidence for oscillation of atmospheric
  neutrinos}},}\ }%
  \bibfield{journal}{%
  \Doi{10.1103/PhysRevLett.81.1562}{\bibinfo {journal} {Phys.Rev.Lett.}}\ }%
  \textbf{\bibinfo {volume} {81}},\ \bibinfo {pages} {1562--1567} (\bibinfo
  {year} {1998}),\
  \Eprint{http://arxiv.org/abs/hep-ex/9807003}{arXiv:hep-ex/9807003 [hep-ex]}%
  \bibAnnoteFile{NoStop}{Fukuda:1998mi}%

\bibitem{Ambrosio:2001je}%
  \BibitemOpen
  \bibfield{author}{%
  \bibinfo {author} {\bibfnamefont{M.}~\bibnamefont{Ambrosio}} \emph{et~al.}
  (\bibinfo {collaboration} {MACRO Collaboration}),\ }%
  \bibfield{title}{%
  \enquote{\bibinfo {title} {{Matter effects in upward-going muons and sterile
  neutrino oscillations}},}\ }%
  \bibfield{journal}{%
  \Doi{10.1016/S0370-2693(01)00992-3}{\bibinfo {journal} {Phys.Lett.}}\ }%
  \textbf{\bibinfo {volume} {B517}},\ \bibinfo {pages} {59--66} (\bibinfo
  {year} {2001}),\
  \Eprint{http://arxiv.org/abs/hep-ex/0106049}{arXiv:hep-ex/0106049 [hep-ex]}%
  \bibAnnoteFile{NoStop}{Ambrosio:2001je}%

\bibitem{Sanchez:2003rb}%
  \BibitemOpen
  \bibfield{author}{%
  \bibinfo {author} {\bibfnamefont{Mayly~C.}\ \bibnamefont{Sanchez}}
  \emph{et~al.} (\bibinfo {collaboration} {Soudan 2 Collaboration}),\ }%
  \bibfield{title}{%
  \enquote{\bibinfo {title} {{Measurement of the L/E distributions of
  atmospheric neutrinos in Soudan 2 and their interpretation as neutrino
  oscillations}},}\ }%
  \bibfield{journal}{%
  \Doi{10.1103/PhysRevD.68.113004}{\bibinfo {journal} {Phys.Rev.}}\ }%
  \textbf{\bibinfo {volume} {D68}},\ \bibinfo {pages} {113004} (\bibinfo {year}
  {2003}),\ \Eprint{http://arxiv.org/abs/hep-ex/0307069}{arXiv:hep-ex/0307069
  [hep-ex]}%
  \bibAnnoteFile{NoStop}{Sanchez:2003rb}%

\bibitem{AdrianMartinez:2012ph}%
  \BibitemOpen
  \bibfield{author}{%
  \bibinfo {author} {\bibfnamefont{S.}~\bibnamefont{Adrian-Martinez}}
  \emph{et~al.} (\bibinfo {collaboration} {ANTARES collaboration}),\ }%
  \bibfield{title}{%
  \enquote{\bibinfo {title} {{Measurement of atmospheric neutrino oscillations
  with the ANTARES neutrino telescope}},}\ }%
  \bibfield{journal}{%
  \Doi{10.1016/j.physletb.2012.07.002}{\bibinfo {journal} {Phys.Lett.}}\ }%
  \textbf{\bibinfo {volume} {B714}},\ \bibinfo {pages} {224--230} (\bibinfo
  {year} {2012}),\ \Eprint{http://arxiv.org/abs/1206.0645}{arXiv:1206.0645
  [hep-ex]}%
  \bibAnnoteFile{NoStop}{AdrianMartinez:2012ph}%

\bibitem{Aartsen:2013jza}%
  \BibitemOpen
  \bibfield{author}{%
  \bibinfo {author} {\bibfnamefont{M.G.}\ \bibnamefont{Aartsen}} \emph{et~al.}
  (\bibinfo {collaboration} {IceCube Collaboration}),\ }%
  \bibfield{title}{%
  \enquote{\bibinfo {title} {{Measurement of atmospheric neutrino oscillations
  with IceCube}},}\ }%
  \bibfield{journal}{%
  \Doi{10.1103/PhysRevLett.111.081801}{\bibinfo {journal} {Phys.Rev.Lett.}}\ }%
  \textbf{\bibinfo {volume} {111}},\ \bibinfo {pages} {081801} (\bibinfo {year}
  {2013}),\ \Eprint{http://arxiv.org/abs/1305.3909}{arXiv:1305.3909 [hep-ex]}%
  \bibAnnoteFile{NoStop}{Aartsen:2013jza}%

\bibitem{Petcov:2005rv}%
  \BibitemOpen
  \bibfield{author}{%
  \bibinfo {author} {\bibfnamefont{S.T.}\ \bibnamefont{Petcov}}\ and\ \bibinfo
  {author} {\bibfnamefont{T.}~\bibnamefont{Schwetz}},\ }%
  \bibfield{title}{%
  \enquote{\bibinfo {title} {{Determining the neutrino mass hierarchy with
  atmospheric neutrinos}},}\ }%
  \bibfield{journal}{%
  \Doi{10.1016/j.nuclphysb.2006.01.020}{\bibinfo {journal} {Nucl.Phys.}}\ }%
  \textbf{\bibinfo {volume} {B740}},\ \bibinfo {pages} {1--22} (\bibinfo {year}
  {2006}),\ \Eprint{http://arxiv.org/abs/hep-ph/0511277}{arXiv:hep-ph/0511277
  [hep-ph]}%
  \bibAnnoteFile{NoStop}{Petcov:2005rv}%

\bibitem{Mena:2008rh}%
  \BibitemOpen
  \bibfield{author}{%
  \bibinfo {author} {\bibfnamefont{Olga}\ \bibnamefont{Mena}}, \bibinfo
  {author} {\bibfnamefont{Irina}\ \bibnamefont{Mocioiu}},\ and\ \bibinfo
  {author} {\bibfnamefont{Soebur}\ \bibnamefont{Razzaque}},\ }%
  \bibfield{title}{%
  \enquote{\bibinfo {title} {{Neutrino mass hierarchy extraction using
  atmospheric neutrinos in ice}},}\ }%
  \bibfield{journal}{%
  \Doi{10.1103/PhysRevD.78.093003}{\bibinfo {journal} {Phys.Rev.}}\ }%
  \textbf{\bibinfo {volume} {D78}},\ \bibinfo {pages} {093003} (\bibinfo {year}
  {2008}),\ \Eprint{http://arxiv.org/abs/0803.3044}{arXiv:0803.3044 [hep-ph]}%
  \bibAnnoteFile{NoStop}{Mena:2008rh}%

\bibitem{Lasserre:2005qw}%
  \BibitemOpen
  \bibfield{author}{%
  \bibinfo {author} {\bibfnamefont{T.}~\bibnamefont{Lasserre}}\ and\ \bibinfo
  {author} {\bibfnamefont{H.W.}\ \bibnamefont{Sobel}},\ }%
  \bibfield{title}{%
  \enquote{\bibinfo {title} {{Reactor neutrinos}},}\ }%
  \bibfield{journal}{%
  \Doi{10.1016/J.CRHY.2005.08.002}{\bibinfo {journal} {Comptes Rendus
  Physique}}\ }%
  \textbf{\bibinfo {volume} {6}},\ \bibinfo {pages} {749--757} (\bibinfo {year}
  {2005}),\ \Eprint{http://arxiv.org/abs/nucl-ex/0601013}{arXiv:nucl-ex/0601013
  [nucl-ex]}%
  \bibAnnoteFile{NoStop}{Lasserre:2005qw}%

\bibitem{Zacek:1986cu}%
  \BibitemOpen
  \bibfield{author}{%
  \bibinfo {author} {\bibfnamefont{G.}~\bibnamefont{Zacek}} \emph{et~al.}
  (\bibinfo {collaboration} {CALTECH-SIN-TUM COLLABORATION}),\ }%
  \bibfield{title}{%
  \enquote{\bibinfo {title} {Neutrino oscillation experiments at the {Gosgen}
  nuclear power reactor},}\ }%
  \bibfield{journal}{%
  \Doi{10.1103/PhysRevD.34.2621}{\bibinfo {journal} {Phys.Rev.}}\ }%
  \textbf{\bibinfo {volume} {D34}},\ \bibinfo {pages} {2621--2636} (\bibinfo
  {year} {1986})%
  \bibAnnoteFile{NoStop}{Zacek:1986cu}%

\bibitem{Vidyakin:1994ut}%
  \BibitemOpen
  \bibfield{author}{%
  \bibinfo {author} {\bibfnamefont{G.S.}\ \bibnamefont{Vidyakin}}, \bibinfo
  {author} {\bibfnamefont{V.N.}\ \bibnamefont{Vyrodov}}, \bibinfo {author}
  {\bibfnamefont{Yu.V.}\ \bibnamefont{Kozlov}}, \bibinfo {author}
  {\bibfnamefont{A.V.}\ \bibnamefont{Martemyanov}}, \bibinfo {author}
  {\bibfnamefont{V.P.}\ \bibnamefont{Martemyanov}}, \emph{et~al.},\ }%
  \bibfield{title}{%
  \enquote{\bibinfo {title} {{Limitations on the characteristics of neutrino
  oscillations}},}\ }%
  \bibfield{journal}{%
  \bibinfo {journal} {JETP Lett.}\ }%
  \textbf{\bibinfo {volume} {59}},\ \bibinfo {pages} {390--393} (\bibinfo
  {year} {1994})%
  \bibAnnoteFile{NoStop}{Vidyakin:1994ut}%

\bibitem{Declais:1994su}%
  \BibitemOpen
  \bibfield{author}{%
  \bibinfo {author} {\bibfnamefont{Y.}~\bibnamefont{Declais}}, \bibinfo
  {author} {\bibfnamefont{J.}~\bibnamefont{Favier}}, \bibinfo {author}
  {\bibfnamefont{A.}~\bibnamefont{Metref}}, \bibinfo {author}
  {\bibfnamefont{H.}~\bibnamefont{Pessard}}, \bibinfo {author}
  {\bibfnamefont{B.}~\bibnamefont{Achkar}}, \emph{et~al.},\ }%
  \bibfield{title}{%
  \enquote{\bibinfo {title} {{Search for neutrino oscillations at 15-meters,
  40-meters, and 95-meters from a nuclear power reactor at Bugey}},}\ }%
  \bibfield{journal}{%
  \Doi{10.1016/0550-3213(94)00513-E}{\bibinfo {journal} {Nucl.Phys.}}\ }%
  \textbf{\bibinfo {volume} {B434}},\ \bibinfo {pages} {503--534} (\bibinfo
  {year} {1995})%
  \bibAnnoteFile{NoStop}{Declais:1994su}%

\bibitem{Apollonio:1999ae}%
  \BibitemOpen
  \bibfield{author}{%
  \bibinfo {author} {\bibfnamefont{M.}~\bibnamefont{Apollonio}} \emph{et~al.}
  (\bibinfo {collaboration} {CHOOZ Collaboration}),\ }%
  \bibfield{title}{%
  \enquote{\bibinfo {title} {{Limits on neutrino oscillations from the CHOOZ
  experiment}},}\ }%
  \bibfield{journal}{%
  \Doi{10.1016/S0370-2693(99)01072-2}{\bibinfo {journal} {Phys.Lett.}}\ }%
  \textbf{\bibinfo {volume} {B466}},\ \bibinfo {pages} {415--430} (\bibinfo
  {year} {1999}),\
  \Eprint{http://arxiv.org/abs/hep-ex/9907037}{arXiv:hep-ex/9907037 [hep-ex]}%
  \bibAnnoteFile{NoStop}{Apollonio:1999ae}%

\bibitem{Piepke:2002ju}%
  \BibitemOpen
  \bibfield{author}{%
  \bibinfo {author} {\bibfnamefont{A.}~\bibnamefont{Piepke}} (\bibinfo
  {collaboration} {Palo Verde Collaboration}),\ }%
  \bibfield{title}{%
  \enquote{\bibinfo {title} {{Final results from the Palo Verde neutrino
  oscillation experiment}},}\ }%
  \bibfield{journal}{%
  \Doi{10.1016/S0146-6410(02)00117-5}{\bibinfo {journal}
  {Prog.Part.Nucl.Phys.}}\ }%
  \textbf{\bibinfo {volume} {48}},\ \bibinfo {pages} {113--121} (\bibinfo
  {year} {2002})%
  \bibAnnoteFile{NoStop}{Piepke:2002ju}%

\bibitem{Eguchi:2002dm}%
  \BibitemOpen
  \bibfield{author}{%
  \bibinfo {author} {\bibfnamefont{K.}~\bibnamefont{Eguchi}} \emph{et~al.}
  (\bibinfo {collaboration} {KamLAND Collaboration}),\ }%
  \bibfield{title}{%
  \enquote{\bibinfo {title} {{First results from KamLAND: Evidence for reactor
  anti-neutrino disappearance}},}\ }%
  \bibfield{journal}{%
  \Doi{10.1103/PhysRevLett.90.021802}{\bibinfo {journal} {Phys.Rev.Lett.}}\ }%
  \textbf{\bibinfo {volume} {90}},\ \bibinfo {pages} {021802} (\bibinfo {year}
  {2003}),\ \Eprint{http://arxiv.org/abs/hep-ex/0212021}{arXiv:hep-ex/0212021
  [hep-ex]}%
  \bibAnnoteFile{NoStop}{Eguchi:2002dm}%

\bibitem{Abe:2011fz}%
  \BibitemOpen
  \bibfield{author}{%
  \bibinfo {author} {\bibfnamefont{Y.}~\bibnamefont{Abe}} \emph{et~al.}
  (\bibinfo {collaboration} {DOUBLE-CHOOZ Collaboration}),\ }%
  \bibfield{title}{%
  \enquote{\bibinfo {title} {{Indication for the disappearance of reactor
  electron antineutrinos in the Double Chooz experiment}},}\ }%
  \bibfield{journal}{%
  \Doi{10.1103/PhysRevLett.108.131801}{\bibinfo {journal} {Phys.Rev.Lett.}}\ }%
  \textbf{\bibinfo {volume} {108}},\ \bibinfo {pages} {131801} (\bibinfo {year}
  {2012}),\ \Eprint{http://arxiv.org/abs/1112.6353}{arXiv:1112.6353 [hep-ex]}%
  \bibAnnoteFile{NoStop}{Abe:2011fz}%

\bibitem{An:2012eh}%
  \BibitemOpen
  \bibfield{author}{%
  \bibinfo {author} {\bibfnamefont{F.P.}\ \bibnamefont{An}} \emph{et~al.}
  (\bibinfo {collaboration} {DAYA-BAY Collaboration}),\ }%
  \bibfield{title}{%
  \enquote{\bibinfo {title} {{Observation of electron-antineutrino
  disappearance at Daya Bay}},}\ }%
  \bibfield{journal}{%
  \Doi{10.1103/PhysRevLett.108.171803}{\bibinfo {journal} {Phys.Rev.Lett.}}\ }%
  \textbf{\bibinfo {volume} {108}},\ \bibinfo {pages} {171803} (\bibinfo {year}
  {2012}),\ \Eprint{http://arxiv.org/abs/1203.1669}{arXiv:1203.1669 [hep-ex]}%
  \bibAnnoteFile{NoStop}{An:2012eh}%

\bibitem{Ahn:2012nd}%
  \BibitemOpen
  \bibfield{author}{%
  \bibinfo {author} {\bibfnamefont{J.K.}\ \bibnamefont{Ahn}} \emph{et~al.}
  (\bibinfo {collaboration} {RENO collaboration}),\ }%
  \bibfield{title}{%
  \enquote{\bibinfo {title} {Observation of reactor electron antineutrino
  disappearance in the {RENO} experiment},}\ }%
  \bibfield{journal}{%
  \Doi{10.1103/PhysRevLett.108.191802}{\bibinfo {journal} {Phys.Rev.Lett.}}\ }%
  \textbf{\bibinfo {volume} {108}},\ \bibinfo {pages} {191802} (\bibinfo {year}
  {2012}),\ \Eprint{http://arxiv.org/abs/1204.0626}{arXiv:1204.0626 [hep-ex]}%
  \bibAnnoteFile{NoStop}{Ahn:2012nd}%

\bibitem{Feldman:2012qt}%
  \BibitemOpen
  \bibfield{author}{%
  \bibinfo {author} {\bibfnamefont{G.J.}\ \bibnamefont{Feldman}}, \bibinfo
  {author} {\bibfnamefont{J.}~\bibnamefont{Hartnell}},\ and\ \bibinfo {author}
  {\bibfnamefont{T.}~\bibnamefont{Kobayashi}},\ }%
  \bibfield{title}{%
  \enquote{\bibinfo {title} {{Long-baseline neutrino oscillation
  experiments}},}\ }%
  \bibfield{journal}{%
  \Doi{10.1155/2013/475749}{\bibinfo {journal} {Adv.High Energy Phys.}}\ }%
  \textbf{\bibinfo {volume} {2013}},\ \bibinfo {pages} {475749} (\bibinfo
  {year} {2013}),\ \Eprint{http://arxiv.org/abs/1210.1778}{arXiv:1210.1778
  [hep-ex]}%
  \bibAnnoteFile{NoStop}{Feldman:2012qt}%

\bibitem{Agafonova:2012zz}%
  \BibitemOpen
  \bibfield{author}{%
  \bibinfo {author} {\bibfnamefont{N.}~\bibnamefont{Agafonova}} \emph{et~al.}
  (\bibinfo {collaboration} {OPERA Collaboration}),\ }%
  \bibfield{title}{%
  \enquote{\bibinfo {title} {{Search for $\nu_{\mu} \to \nu_{\tau}$ oscillation
  with the OPERA experiment in the CNGS beam}},}\ }%
  \bibfield{journal}{%
  \Doi{10.1088/1367-2630/14/3/033017}{\bibinfo {journal} {New J.Phys.}}\ }%
  \textbf{\bibinfo {volume} {14}},\ \bibinfo {pages} {033017} (\bibinfo {year}
  {2012})%
  \bibAnnoteFile{NoStop}{Agafonova:2012zz}%

\bibitem{DiMarco:2013qxa}%
  \BibitemOpen
  \bibfield{author}{%
  \bibinfo {author} {\bibfnamefont{N.}~\bibnamefont{Di~Marco}} (\bibinfo
  {collaboration} {OPERA Collaboration}),\ }%
  \bibfield{title}{%
  \enquote{\bibinfo {title} {{Recent results of the OPERA experiment}},}\ }%
  \bibfield{journal}{%
  \Doi{10.1016/j.nuclphysbps.2013.04.086}{\bibinfo {journal}
  {Nucl.Phys.Proc.Suppl.}}\ }%
  \textbf{\bibinfo {volume} {237-238}},\ \bibinfo {pages} {187--189} (\bibinfo
  {year} {2013})%
  \bibAnnoteFile{NoStop}{DiMarco:2013qxa}%

\bibitem{Ahn:2006zza}%
  \BibitemOpen
  \bibfield{author}{%
  \bibinfo {author} {\bibfnamefont{M.H.}\ \bibnamefont{Ahn}} \emph{et~al.}
  (\bibinfo {collaboration} {K2K Collaboration}),\ }%
  \bibfield{title}{%
  \enquote{\bibinfo {title} {Measurement of neutrino oscillation by the {K2K}
  experiment},}\ }%
  \bibfield{journal}{%
  \Doi{10.1103/PhysRevD.74.072003}{\bibinfo {journal} {Phys.Rev.}}\ }%
  \textbf{\bibinfo {volume} {D74}},\ \bibinfo {pages} {072003} (\bibinfo {year}
  {2006}),\ \Eprint{http://arxiv.org/abs/hep-ex/0606032}{arXiv:hep-ex/0606032
  [hep-ex]}%
  \bibAnnoteFile{NoStop}{Ahn:2006zza}%

\bibitem{deJong:2013/04/19pya}%
  \BibitemOpen
  \bibfield{author}{%
  \bibinfo {author} {\bibfnamefont{J.K.}\ \bibnamefont{de~Jong}} (\bibinfo
  {collaboration} {MINOS Collaboration}),\ }%
  \bibfield{title}{%
  \enquote{\bibinfo {title} {{Near-to-final MINOS oscillation results}},}\ }%
  \bibfield{journal}{%
  \Doi{10.1016/j.nuclphysbps.2013.04.080}{\bibinfo {journal}
  {Nucl.Phys.Proc.Suppl.}}\ }%
  \textbf{\bibinfo {volume} {237-238}},\ \bibinfo {pages} {166--169} (\bibinfo
  {year} {2013})%
  \bibAnnoteFile{NoStop}{deJong:2013/04/19pya}%

\bibitem{Yamamoto:2006ty}%
  \BibitemOpen
  \bibfield{author}{%
  \bibinfo {author} {\bibfnamefont{S.}~\bibnamefont{Yamamoto}} \emph{et~al.}
  (\bibinfo {collaboration} {K2K Collaboration}),\ }%
  \bibfield{title}{%
  \enquote{\bibinfo {title} {{An improved search for $\nu_\mu \rightarrow
  \nu_e$ oscillation in a long-baseline accelerator experiment}},}\ }%
  \bibfield{journal}{%
  \Doi{10.1103/PhysRevLett.96.181801}{\bibinfo {journal} {Phys.Rev.Lett.}}\ }%
  \textbf{\bibinfo {volume} {96}},\ \bibinfo {pages} {181801} (\bibinfo {year}
  {2006}),\ \Eprint{http://arxiv.org/abs/hep-ex/0603004}{arXiv:hep-ex/0603004
  [hep-ex]}%
  \bibAnnoteFile{NoStop}{Yamamoto:2006ty}%

\bibitem{Adamson:2011qu}%
  \BibitemOpen
  \bibfield{author}{%
  \bibinfo {author} {\bibfnamefont{P.}~\bibnamefont{Adamson}} \emph{et~al.}
  (\bibinfo {collaboration} {MINOS Collaboration}),\ }%
  \bibfield{title}{%
  \enquote{\bibinfo {title} {{Improved search for muon-neutrino to
  electron-neutrino oscillations in MINOS}},}\ }%
  \bibfield{journal}{%
  \Doi{10.1103/PhysRevLett.107.181802}{\bibinfo {journal} {Phys.Rev.Lett.}}\ }%
  \textbf{\bibinfo {volume} {107}},\ \bibinfo {pages} {181802} (\bibinfo {year}
  {2011}),\ \Eprint{http://arxiv.org/abs/1108.0015}{arXiv:1108.0015 [hep-ex]}%
  \bibAnnoteFile{NoStop}{Adamson:2011qu}%

\bibitem{Catanesi:2013fxa}%
  \BibitemOpen
  \bibfield{author}{%
  \bibinfo {author} {\bibfnamefont{M.G.}\ \bibnamefont{Catanesi}} (\bibinfo
  {collaboration} {T2K Collaboration}),\ }%
  \bibfield{title}{%
  \enquote{\bibinfo {title} {{T2K results and perspectives}},}\ }%
  \bibfield{journal}{%
  \Doi{10.1016/j.nuclphysbps.2013.04.074}{\bibinfo {journal}
  {Nucl.Phys.Proc.Suppl.}}\ }%
  \textbf{\bibinfo {volume} {237-238}},\ \bibinfo {pages} {129--134} (\bibinfo
  {year} {2013})%
  \bibAnnoteFile{NoStop}{Catanesi:2013fxa}%

\bibitem{Agafonova:2013xsk}%
  \BibitemOpen
  \bibfield{author}{%
  \bibinfo {author} {\bibfnamefont{N.}~\bibnamefont{Agafonova}} \emph{et~al.}
  (\bibinfo {collaboration} {OPERA collaboration}),\ }%
  \bibfield{title}{%
  \enquote{\bibinfo {title} {{Search for $\nu_\mu \rightarrow \nu_e$
  oscillations with the OPERA experiment in the CNGS beam}},}\ }%
  \bibfield{journal}{%
  \Doi{10.1007/JHEP07(2013)004, 10.1007/JHEP07(2013)085}{\bibinfo {journal}
  {JHEP}}\ }%
  \textbf{\bibinfo {volume} {1307}},\ \bibinfo {pages} {004} (\bibinfo {year}
  {2013}),\ \Eprint{http://arxiv.org/abs/1303.3953}{arXiv:1303.3953 [hep-ex]}%
  \bibAnnoteFile{NoStop}{Agafonova:2013xsk}%

\bibitem{Shanahan:2011zz}%
  \BibitemOpen
  \bibfield{author}{%
  \bibinfo {author} {\bibfnamefont{P.}~\bibnamefont{Shanahan}} (\bibinfo
  {collaboration} {NOVA Collaboration}),\ }%
  \bibfield{title}{%
  \enquote{\bibinfo {title} {{Status and prospects of the NOVA experiment}},}\
  }%
  \bibfield{journal}{%
  \Doi{10.1016/j.nuclphysbps.2011.09.013}{\bibinfo {journal}
  {Nucl.Phys.Proc.Suppl.}}\ }%
  \textbf{\bibinfo {volume} {221}},\ \bibinfo {pages} {254--259} (\bibinfo
  {year} {2011})%
  \bibAnnoteFile{NoStop}{Shanahan:2011zz}%

\bibitem{Aguilar:2001ty}%
  \BibitemOpen
  \bibfield{author}{%
  \bibinfo {author} {\bibfnamefont{A.}~\bibnamefont{Aguilar-Arevalo}}
  \emph{et~al.} (\bibinfo {collaboration} {LSND Collaboration}),\ }%
  \bibfield{title}{%
  \enquote{\bibinfo {title} {{Evidence for neutrino oscillations from the
  observation of $\bar \nu_e$ appearance in a $\bar \nu_\mu$ beam}},}\ }%
  \bibfield{journal}{%
  \Doi{10.1103/PhysRevD.64.112007}{\bibinfo {journal} {Phys.Rev.}}\ }%
  \textbf{\bibinfo {volume} {D64}},\ \bibinfo {pages} {112007} (\bibinfo {year}
  {2001}),\ \Eprint{http://arxiv.org/abs/hep-ex/0104049}{arXiv:hep-ex/0104049
  [hep-ex]}%
  \bibAnnoteFile{NoStop}{Aguilar:2001ty}%

\bibitem{Armbruster:2002mp}%
  \BibitemOpen
  \bibfield{author}{%
  \bibinfo {author} {\bibfnamefont{B.}~\bibnamefont{Armbruster}} \emph{et~al.}
  (\bibinfo {collaboration} {KARMEN Collaboration}),\ }%
  \bibfield{title}{%
  \enquote{\bibinfo {title} {Upper limits for neutrino oscillations $\bar
  \nu_\mu \rightarrow \bar \nu_e$ from muon decay at rest},}\ }%
  \bibfield{journal}{%
  \Doi{10.1103/PhysRevD.65.112001}{\bibinfo {journal} {Phys.Rev.}}\ }%
  \textbf{\bibinfo {volume} {D65}},\ \bibinfo {pages} {112001} (\bibinfo {year}
  {2002}),\ \Eprint{http://arxiv.org/abs/hep-ex/0203021}{arXiv:hep-ex/0203021
  [hep-ex]}%
  \bibAnnoteFile{NoStop}{Armbruster:2002mp}%

\bibitem{AguilarArevalo:2007it}%
  \BibitemOpen
  \bibfield{author}{%
  \bibinfo {author} {\bibfnamefont{A.A.}\ \bibnamefont{Aguilar-Arevalo}}
  \emph{et~al.} (\bibinfo {collaboration} {MiniBooNE Collaboration}),\ }%
  \bibfield{title}{%
  \enquote{\bibinfo {title} {{A search for electron neutrino appearance at the
  $\Delta m^{2} \sim 1 \; \mathrm{eV}^2$ scale}},}\ }%
  \bibfield{journal}{%
  \Doi{10.1103/PhysRevLett.98.231801}{\bibinfo {journal} {Phys.Rev.Lett.}}\ }%
  \textbf{\bibinfo {volume} {98}},\ \bibinfo {pages} {231801} (\bibinfo {year}
  {2007}),\ \Eprint{http://arxiv.org/abs/0704.1500}{arXiv:0704.1500 [hep-ex]}%
  \bibAnnoteFile{NoStop}{AguilarArevalo:2007it}%

\bibitem{VandeWater:2012tsa}%
  \BibitemOpen
  \bibfield{author}{%
  \bibinfo {author} {\bibfnamefont{R.G.}\ \bibnamefont{Van~de Water}} (\bibinfo
  {collaboration} {MiniBooNE Collaboration}),\ }%
  \bibfield{title}{%
  \enquote{\bibinfo {title} {{MiniBooNE search for $\bar{\nu}_{\mu} \to
  \bar{\nu}_{e}$ oscillations}},}\ }%
  \bibfield{journal}{%
  \Doi{10.1016/j.nuclphysbps.2012.09.007}{\bibinfo {journal}
  {Nucl.Phys.Proc.Suppl.}}\ }%
  \textbf{\bibinfo {volume} {229-232}},\ \bibinfo {pages} {45--49} (\bibinfo
  {year} {2012})%
  \bibAnnoteFile{NoStop}{VandeWater:2012tsa}%

\bibitem{King:2013eh}%
  \BibitemOpen
  \bibfield{author}{%
  \bibinfo {author} {\bibfnamefont{Stephen~F.}\ \bibnamefont{King}}\ and\
  \bibinfo {author} {\bibfnamefont{Christoph}\ \bibnamefont{Luhn}},\ }%
  \bibfield{title}{%
  \enquote{\bibinfo {title} {Neutrino mass and mixing with discrete
  symmetry},}\ }%
  \bibfield{journal}{%
  \Doi{10.1088/0034-4885/76/5/056201}{\bibinfo {journal} {Rept.Prog.Phys.}}\ }%
  \textbf{\bibinfo {volume} {76}},\ \bibinfo {pages} {056201} (\bibinfo {year}
  {2013}),\ \Eprint{http://arxiv.org/abs/1301.1340}{arXiv:1301.1340 [hep-ph]}%
  \bibAnnoteFile{NoStop}{King:2013eh}%

\bibitem{Tortola:2012te}%
  \BibitemOpen
  \bibfield{author}{%
  \bibinfo {author} {\bibfnamefont{D.V.}\ \bibnamefont{Forero}}, \bibinfo
  {author} {\bibfnamefont{M.}~\bibnamefont{Tortola}},\ and\ \bibinfo {author}
  {\bibfnamefont{J.W.F.}\ \bibnamefont{Valle}},\ }%
  \bibfield{title}{%
  \enquote{\bibinfo {title} {{Global status of neutrino oscillation parameters
  after Neutrino-2012}},}\ }%
  \bibfield{journal}{%
  \Doi{10.1103/PhysRevD.86.073012}{\bibinfo {journal} {Phys.Rev.}}\ }%
  \textbf{\bibinfo {volume} {D86}},\ \bibinfo {pages} {073012} (\bibinfo {year}
  {2012}),\ \Eprint{http://arxiv.org/abs/1205.4018}{arXiv:1205.4018 [hep-ph]}%
  \bibAnnoteFile{NoStop}{Tortola:2012te}%

\bibitem{Aseev:2011dq}%
  \BibitemOpen
  \bibfield{author}{%
  \bibinfo {author} {\bibfnamefont{V.N.}\ \bibnamefont{Aseev}} \emph{et~al.}
  (\bibinfo {collaboration} {Troitsk Collaboration}),\ }%
  \bibfield{title}{%
  \enquote{\bibinfo {title} {{An upper limit on electron antineutrino mass from
  Troitsk experiment}},}\ }%
  \bibfield{journal}{%
  \Doi{10.1103/PhysRevD.84.112003}{\bibinfo {journal} {Phys.Rev.}}\ }%
  \textbf{\bibinfo {volume} {D84}},\ \bibinfo {pages} {112003} (\bibinfo {year}
  {2011}),\ \Eprint{http://arxiv.org/abs/1108.5034}{arXiv:1108.5034 [hep-ex]}%
  \bibAnnoteFile{NoStop}{Aseev:2011dq}%

\bibitem{Ade:2013zuv}%
  \BibitemOpen
  \bibfield{author}{%
  \bibinfo {author} {\bibfnamefont{P.A.R.}\ \bibnamefont{Ade}} \emph{et~al.}
  (\bibinfo {collaboration} {Planck Collaboration}),\ }%
  \bibfield{title}{%
  \enquote{\bibinfo {title} {{Planck 2013 results. XVI. Cosmological
  parameters}},}\ }%
   (\bibinfo {year} {2013}),\
  \Eprint{http://arxiv.org/abs/1303.5076}{arXiv:1303.5076 [astro-ph.CO]}%
  \bibAnnoteFile{NoStop}{Ade:2013zuv}%

\bibitem{Auger:2012ar}%
  \BibitemOpen
  \bibfield{author}{%
  \bibinfo {author} {\bibfnamefont{M.}~\bibnamefont{Auger}} \emph{et~al.}
  (\bibinfo {collaboration} {EXO Collaboration}),\ }%
  \bibfield{title}{%
  \enquote{\bibinfo {title} {Search for neutrinoless double-beta decay in
  {$\tensor[^{136}]{\mathrm{Xe}}{}$ with EXO-200}},}\ }%
  \bibfield{journal}{%
  \Doi{10.1103/PhysRevLett.109.032505}{\bibinfo {journal} {Phys.Rev.Lett.}}\ }%
  \textbf{\bibinfo {volume} {109}},\ \bibinfo {pages} {032505} (\bibinfo {year}
  {2012}),\ \Eprint{http://arxiv.org/abs/1205.5608}{arXiv:1205.5608 [hep-ex]}%
  \bibAnnoteFile{NoStop}{Auger:2012ar}%

\bibitem{Agostini:2013mzu}%
  \BibitemOpen
  \bibfield{author}{%
  \bibinfo {author} {\bibfnamefont{M.}~\bibnamefont{Agostini}} \emph{et~al.}
  (\bibinfo {collaboration} {GERDA Collaboration}),\ }%
  \bibfield{title}{%
  \enquote{\bibinfo {title} {{Results on neutrinoless double beta decay of
  $\tensor[^{76}]{\mathrm{Ge}}{}$ from GERDA Phase I}},}\ }%
  \bibfield{journal}{%
  \Doi{10.1103/PhysRevLett.111.122503}{\bibinfo {journal} {Phys.Rev.Lett.}}\ }%
  \textbf{\bibinfo {volume} {111}},\ \bibinfo {pages} {122503} (\bibinfo {year}
  {2013}),\ \Eprint{http://arxiv.org/abs/1307.4720}{arXiv:1307.4720 [nucl-ex]}%
  \bibAnnoteFile{NoStop}{Agostini:2013mzu}%

\bibitem{Drexlin:2013lha}%
  \BibitemOpen
  \bibfield{author}{%
  \bibinfo {author} {\bibfnamefont{G.}~\bibnamefont{Drexlin}}, \bibinfo
  {author} {\bibfnamefont{V.}~\bibnamefont{Hannen}}, \bibinfo {author}
  {\bibfnamefont{S.}~\bibnamefont{Mertens}},\ and\ \bibinfo {author}
  {\bibfnamefont{C.}~\bibnamefont{Weinheimer}},\ }%
  \bibfield{title}{%
  \enquote{\bibinfo {title} {{Current direct neutrino mass experiments}},}\ }%
  \bibfield{journal}{%
  \Doi{10.1155/2013/293986}{\bibinfo {journal} {Adv.High Energy Phys.}}\ }%
  \textbf{\bibinfo {volume} {2013}},\ \bibinfo {pages} {293986} (\bibinfo
  {year} {2013}),\ \Eprint{http://arxiv.org/abs/1307.0101}{arXiv:1307.0101
  [physics.ins-det]}%
  \bibAnnoteFile{NoStop}{Drexlin:2013lha}%

\bibitem{Lesgourgues:2012uu}%
  \BibitemOpen
  \bibfield{author}{%
  \bibinfo {author} {\bibfnamefont{Julien}\ \bibnamefont{Lesgourgues}}\ and\
  \bibinfo {author} {\bibfnamefont{Sergio}\ \bibnamefont{Pastor}},\ }%
  \bibfield{title}{%
  \enquote{\bibinfo {title} {Neutrino mass from cosmology},}\ }%
  \bibfield{journal}{%
  \Doi{10.1155/2012/608515}{\bibinfo {journal} {Adv.High Energy Phys.}}\ }%
  \textbf{\bibinfo {volume} {2012}},\ \bibinfo {pages} {608515} (\bibinfo
  {year} {2012}),\ \Eprint{http://arxiv.org/abs/1212.6154}{arXiv:1212.6154
  [hep-ph]}%
  \bibAnnoteFile{NoStop}{Lesgourgues:2012uu}%

\bibitem{Pastor:2013book}%
  \BibitemOpen
  \bibfield{author}{%
  \bibinfo {author} {\bibfnamefont{J.}~\bibnamefont{Lesgourgues}}, \bibinfo
  {author} {\bibfnamefont{G.}~\bibnamefont{Mangano}}, \bibinfo {author}
  {\bibfnamefont{G.}~\bibnamefont{Miele}},\ and\ \bibinfo {author}
  {\bibfnamefont{S.}~\bibnamefont{Pastor}},\ }%
  \Doi{10.1017/CBO9781139012874}{\emph{\bibinfo {title} {Neutrino Cosmology}}}\
  (\bibinfo {publisher} {Cambridge University Press},\ \bibinfo {year} {2013})\
  ISBN \bibinfo {isbn} {9781107013957}%
  \bibAnnoteFile{NoStop}{Pastor:2013book}%

\bibitem{Barabash:2005sh}%
  \BibitemOpen
  \bibfield{author}{%
  \bibinfo {author} {\bibfnamefont{A.S.}\ \bibnamefont{Barabash}},\ }%
  \bibfield{title}{%
  \enquote{\bibinfo {title} {{Average and recommended half-life values for
  two-neutrino double beta decay: Upgrade'05}},}\ }%
  \bibfield{journal}{%
  \Doi{10.1007/s10582-006-0106-6}{\bibinfo {journal} {Czech.J.Phys.}}\ }%
  \textbf{\bibinfo {volume} {56}},\ \bibinfo {pages} {437--445} (\bibinfo
  {year} {2006}),\
  \Eprint{http://arxiv.org/abs/nucl-ex/0602009}{arXiv:nucl-ex/0602009
  [nucl-ex]}%
  \bibAnnoteFile{NoStop}{Barabash:2005sh}%

\bibitem{Ackerman:2011gz}%
  \BibitemOpen
  \bibfield{author}{%
  \bibinfo {author} {\bibfnamefont{N.}~\bibnamefont{Ackerman}} \emph{et~al.}
  (\bibinfo {collaboration} {EXO-200 Collaboration}),\ }%
  \bibfield{title}{%
  \enquote{\bibinfo {title} {Observation of two-neutrino double-beta decay in
  {$\tensor[^{136}]{\mathrm{Xe}}{}$ with EXO-200}},}\ }%
  \bibfield{journal}{%
  \Doi{10.1103/PhysRevLett.107.212501}{\bibinfo {journal} {Phys.Rev.Lett.}}\ }%
  \textbf{\bibinfo {volume} {107}},\ \bibinfo {pages} {212501} (\bibinfo {year}
  {2011}),\ \Eprint{http://arxiv.org/abs/1108.4193}{arXiv:1108.4193 [nucl-ex]}%
  \bibAnnoteFile{NoStop}{Ackerman:2011gz}%

\bibitem{Furry:1939qr}%
  \BibitemOpen
  \bibfield{author}{%
  \bibinfo {author} {\bibfnamefont{W.H.}\ \bibnamefont{Furry}},\ }%
  \bibfield{title}{%
  \enquote{\bibinfo {title} {{On transition probabilities in double beta
  disintegration}},}\ }%
  \bibfield{journal}{%
  \Doi{10.1103/PhysRev.56.1184}{\bibinfo {journal} {Phys.Rev.}}\ }%
  \textbf{\bibinfo {volume} {56}},\ \bibinfo {pages} {1184--1193} (\bibinfo
  {year} {1939})%
  \bibAnnoteFile{NoStop}{Furry:1939qr}%

\bibitem{Avignone:2007fu}%
  \BibitemOpen
  \bibfield{author}{%
  \bibinfo {author} {\bibfnamefont{Frank~T.}\ \bibnamefont{Avignone}}, \bibinfo
  {author} {\bibfnamefont{Steven~R.}\ \bibnamefont{Elliott}},\ and\ \bibinfo
  {author} {\bibfnamefont{Jonathan}\ \bibnamefont{Engel}},\ }%
  \bibfield{title}{%
  \enquote{\bibinfo {title} {Double beta decay, {Majorana} neutrinos, and
  neutrino mass},}\ }%
  \bibfield{journal}{%
  \Doi{10.1103/RevModPhys.80.481}{\bibinfo {journal} {Rev.Mod.Phys.}}\ }%
  \textbf{\bibinfo {volume} {80}},\ \bibinfo {pages} {481--516} (\bibinfo
  {year} {2008}),\ \Eprint{http://arxiv.org/abs/0708.1033}{arXiv:0708.1033
  [nucl-ex]}%
  \bibAnnoteFile{NoStop}{Avignone:2007fu}%

\bibitem{GomezCadenas:2011it}%
  \BibitemOpen
  \bibfield{author}{%
  \bibinfo {author} {\bibfnamefont{J.J.}\ \bibnamefont{Gomez-Cadenas}},
  \bibinfo {author} {\bibfnamefont{J.}~\bibnamefont{Martin-Albo}}, \bibinfo
  {author} {\bibfnamefont{M.}~\bibnamefont{Mezzetto}}, \bibinfo {author}
  {\bibfnamefont{F.}~\bibnamefont{Monrabal}},\ and\ \bibinfo {author}
  {\bibfnamefont{M.}~\bibnamefont{Sorel}},\ }%
  \bibfield{title}{%
  \enquote{\bibinfo {title} {{The Search for neutrinoless double beta
  decay}},}\ }%
  \bibfield{journal}{%
  \Doi{10.1393/ncr/i2012-10074-9}{\bibinfo {journal} {Riv.Nuovo Cim.}}\ }%
  \textbf{\bibinfo {volume} {35}},\ \bibinfo {pages} {29--98} (\bibinfo {year}
  {2012}),\ \Eprint{http://arxiv.org/abs/1109.5515}{arXiv:1109.5515 [hep-ex]}%
  \bibAnnoteFile{NoStop}{GomezCadenas:2011it}%

\bibitem{Bilenky:2012qi}%
  \BibitemOpen
  \bibfield{author}{%
  \bibinfo {author} {\bibfnamefont{S.M.}\ \bibnamefont{Bilenky}}\ and\ \bibinfo
  {author} {\bibfnamefont{Carlo}\ \bibnamefont{Giunti}},\ }%
  \bibfield{title}{%
  \enquote{\bibinfo {title} {Neutrinoless double beta decay: A brief review},}\
  }%
  \bibfield{journal}{%
  \Doi{10.1142/S0217732312300157}{\bibinfo {journal} {Mod.Phys.Lett.}}\ }%
  \textbf{\bibinfo {volume} {A27}},\ \bibinfo {pages} {1230015} (\bibinfo
  {year} {2012}),\ \Eprint{http://arxiv.org/abs/1203.5250}{arXiv:1203.5250
  [hep-ph]}%
  \bibAnnoteFile{NoStop}{Bilenky:2012qi}%

\bibitem{Deppisch:2012nb}%
  \BibitemOpen
  \bibfield{author}{%
  \bibinfo {author} {\bibfnamefont{Frank~F.}\ \bibnamefont{Deppisch}}, \bibinfo
  {author} {\bibfnamefont{Martin}\ \bibnamefont{Hirsch}},\ and\ \bibinfo
  {author} {\bibfnamefont{Heinrich}\ \bibnamefont{Päs}},\ }%
  \bibfield{title}{%
  \enquote{\bibinfo {title} {{Neutrinoless Double Beta Decay and Physics Beyond
  the Standard Model}},}\ }%
  \bibfield{journal}{%
  \Doi{10.1088/0954-3899/39/12/124007}{\bibinfo {journal} {J.Phys.}}\ }%
  \textbf{\bibinfo {volume} {G39}},\ \bibinfo {pages} {124007} (\bibinfo {year}
  {2012}),\ \Eprint{http://arxiv.org/abs/1208.0727}{arXiv:1208.0727 [hep-ph]}%
  \bibAnnoteFile{NoStop}{Deppisch:2012nb}%

\bibitem{Schechter:1981bd}%
  \BibitemOpen
  \bibfield{author}{%
  \bibinfo {author} {\bibfnamefont{J.}~\bibnamefont{Schechter}}\ and\ \bibinfo
  {author} {\bibfnamefont{J.W.F.}\ \bibnamefont{Valle}},\ }%
  \bibfield{title}{%
  \enquote{\bibinfo {title} {Neutrinoless double beta decay in {$SU(2) \otimes
  U(1)$} theories},}\ }%
  \bibfield{journal}{%
  \Doi{10.1103/PhysRevD.25.2951}{\bibinfo {journal} {Phys.Rev.}}\ }%
  \textbf{\bibinfo {volume} {D25}},\ \bibinfo {pages} {2951} (\bibinfo {year}
  {1982})%
  \bibAnnoteFile{NoStop}{Schechter:1981bd}%

\bibitem{Pas:1999fc}%
  \BibitemOpen
  \bibfield{author}{%
  \bibinfo {author} {\bibfnamefont{H.}~\bibnamefont{Päs}}, \bibinfo {author}
  {\bibfnamefont{M.}~\bibnamefont{Hirsch}}, \bibinfo {author}
  {\bibfnamefont{H.V.}\ \bibnamefont{Klapdor-Kleingrothaus}},\ and\ \bibinfo
  {author} {\bibfnamefont{S.G.}\ \bibnamefont{Kovalenko}},\ }%
  \bibfield{title}{%
  \enquote{\bibinfo {title} {{Towards a superformula for neutrinoless double
  beta decay}},}\ }%
  \bibfield{journal}{%
  \Doi{10.1016/S0370-2693(99)00330-5}{\bibinfo {journal} {Phys.Lett.}}\ }%
  \textbf{\bibinfo {volume} {B453}},\ \bibinfo {pages} {194--198} (\bibinfo
  {year} {1999})%
  \bibAnnoteFile{NoStop}{Pas:1999fc}%

\bibitem{Pas:2000vn}%
  \BibitemOpen
  \bibfield{author}{%
  \bibinfo {author} {\bibfnamefont{H.}~\bibnamefont{Päs}}, \bibinfo {author}
  {\bibfnamefont{M.}~\bibnamefont{Hirsch}}, \bibinfo {author}
  {\bibfnamefont{H.V.}\ \bibnamefont{Klapdor-Kleingrothaus}},\ and\ \bibinfo
  {author} {\bibfnamefont{S.G.}\ \bibnamefont{Kovalenko}},\ }%
  \bibfield{title}{%
  \enquote{\bibinfo {title} {{A superformula for neutrinoless double beta
  decay. 2. The short-range part}},}\ }%
  \bibfield{journal}{%
  \Doi{10.1016/S0370-2693(00)01359-9}{\bibinfo {journal} {Phys.Lett.}}\ }%
  \textbf{\bibinfo {volume} {B498}},\ \bibinfo {pages} {35--39} (\bibinfo
  {year} {2001}),\
  \Eprint{http://arxiv.org/abs/hep-ph/0008182}{arXiv:hep-ph/0008182 [hep-ph]}%
  \bibAnnoteFile{NoStop}{Pas:2000vn}%

\bibitem{KlapdorKleingrothaus:2000sn}%
  \BibitemOpen
  \bibfield{author}{%
  \bibinfo {author} {\bibfnamefont{H.V.}\ \bibnamefont{Klapdor-Kleingrothaus}},
  \bibinfo {author} {\bibfnamefont{A.}~\bibnamefont{Dietz}}, \bibinfo {author}
  {\bibfnamefont{L.}~\bibnamefont{Baudis}}, \bibinfo {author}
  {\bibfnamefont{G.}~\bibnamefont{Heusser}}, \bibinfo {author}
  {\bibfnamefont{I.V.}\ \bibnamefont{Krivosheina}}, \emph{et~al.},\ }%
  \bibfield{title}{%
  \enquote{\bibinfo {title} {{Latest results from the Heidelberg-Moscow double
  beta decay experiment}},}\ }%
  \bibfield{journal}{%
  \Doi{10.1007/s100500170022}{\bibinfo {journal} {Eur.Phys.J.}}\ }%
  \textbf{\bibinfo {volume} {A12}},\ \bibinfo {pages} {147--154} (\bibinfo
  {year} {2001}),\
  \Eprint{http://arxiv.org/abs/hep-ph/0103062}{arXiv:hep-ph/0103062 [hep-ph]}%
  \bibAnnoteFile{NoStop}{KlapdorKleingrothaus:2000sn}%

\bibitem{KlapdorKleingrothaus:2001ke}%
  \BibitemOpen
  \bibfield{author}{%
  \bibinfo {author} {\bibfnamefont{H.V.}\ \bibnamefont{Klapdor-Kleingrothaus}},
  \bibinfo {author} {\bibfnamefont{A.}~\bibnamefont{Dietz}}, \bibinfo {author}
  {\bibfnamefont{H.L.}\ \bibnamefont{Harney}},\ and\ \bibinfo {author}
  {\bibfnamefont{I.V.}\ \bibnamefont{Krivosheina}},\ }%
  \bibfield{title}{%
  \enquote{\bibinfo {title} {{Evidence for neutrinoless double beta decay}},}\
  }%
  \bibfield{journal}{%
  \Doi{10.1142/S0217732301005825}{\bibinfo {journal} {Mod.Phys.Lett.}}\ }%
  \textbf{\bibinfo {volume} {A16}},\ \bibinfo {pages} {2409--2420} (\bibinfo
  {year} {2001}),\
  \Eprint{http://arxiv.org/abs/hep-ph/0201231}{arXiv:hep-ph/0201231 [hep-ph]}%
  \bibAnnoteFile{NoStop}{KlapdorKleingrothaus:2001ke}%

\bibitem{KlapdorKleingrothaus:2006ff}%
  \BibitemOpen
  \bibfield{author}{%
  \bibinfo {author} {\bibfnamefont{H.V.}\ \bibnamefont{Klapdor-Kleingrothaus}}\
  and\ \bibinfo {author} {\bibfnamefont{I.V.}\ \bibnamefont{Krivosheina}},\ }%
  \bibfield{title}{%
  \enquote{\bibinfo {title} {{The evidence for the observation of $0 \nu \beta
  \beta$ decay: the identification of $0 \nu \beta \beta$ events from the full
  spectra}},}\ }%
  \bibfield{journal}{%
  \Doi{10.1142/S0217732306020937}{\bibinfo {journal} {Mod.Phys.Lett.}}\ }%
  \textbf{\bibinfo {volume} {A21}},\ \bibinfo {pages} {1547--1566} (\bibinfo
  {year} {2006})%
  \bibAnnoteFile{NoStop}{KlapdorKleingrothaus:2006ff}%

\bibitem{GellMann:1980vs}%
  \BibitemOpen
  \bibfield{author}{%
  \bibinfo {author} {\bibfnamefont{Murray}\ \bibnamefont{Gell-Mann}}, \bibinfo
  {author} {\bibfnamefont{Pierre}\ \bibnamefont{Ramond}},\ and\ \bibinfo
  {author} {\bibfnamefont{Richard}\ \bibnamefont{Slansky}},\ }%
  \bibfield{title}{%
  \enquote{\bibinfo {title} {Complex spinors and unified theories},}\ }%
  \bibfield{journal}{%
  \bibinfo {journal} {Conf.Proc.}\ }%
  \textbf{\bibinfo {volume} {C790927}},\ \bibinfo {pages} {315--321} (\bibinfo
  {year} {1979}),\ \bibinfo {note} {published in Supergravity, P. van
  Nieuwenhuizen and D.Z. Freedman (eds.), North Holland Publ. Co., 1979},\
  \Eprint{http://arxiv.org/abs/1306.4669}{arXiv:1306.4669 [hep-th]},\
  \url{http://tuvalu.santafe.edu/~mgm/Site/Publications_files/MGM\%2086.pdf}%
  \bibAnnoteFile{NoStop}{GellMann:1980vs}%

\bibitem{Ramond:1979py}%
  \BibitemOpen
  \bibfield{author}{%
  \bibinfo {author} {\bibfnamefont{Pierre}\ \bibnamefont{Ramond}},\ }%
  \bibfield{title}{%
  \enquote{\bibinfo {title} {The family group in grand unified theories},}\ ,\
  \bibinfo {pages} {265--280}}%
   (\bibinfo {year} {1979}),\ \bibinfo {note} {retro-Preprint (1979 unpublished
  Caltech preprint no: CALT-68-709). Invited talk at the Sanibel Symposia, Feb
  1979},\ \Eprint{http://arxiv.org/abs/hep-ph/9809459}{arXiv:hep-ph/9809459
  [hep-ph]}%
  \bibAnnoteFile{NoStop}{Ramond:1979py}%

\bibitem{Minkowski:1977sc}%
  \BibitemOpen
  \bibfield{author}{%
  \bibinfo {author} {\bibfnamefont{Peter}\ \bibnamefont{Minkowski}},\ }%
  \bibfield{title}{%
  \enquote{\bibinfo {title} {$\mu \rightarrow e \gamma$ at a rate of one out of
  1-billion muon decays?}.}\ }%
  \bibfield{journal}{%
  \Doi{10.1016/0370-2693(77)90435-X}{\bibinfo {journal} {Phys.Lett.}}\ }%
  \textbf{\bibinfo {volume} {B67}},\ \bibinfo {pages} {421} (\bibinfo {year}
  {1977})%
  \bibAnnoteFile{NoStop}{Minkowski:1977sc}%

\bibitem{Yanagida:1979as}%
  \BibitemOpen
  \bibfield{author}{%
  \bibinfo {author} {\bibfnamefont{Tsutomu}\ \bibnamefont{Yanagida}},\ }%
  \bibfield{title}{%
  \enquote{\bibinfo {title} {Horizontal symmetry and masses of neutrinos},}\ }%
  \bibfield{journal}{%
  \Doi{10.1143/PTP.64.1103}{\bibinfo {journal} {Prog. Theor. Phys.}}\ }%
  \textbf{\bibinfo {volume} {64}},\ \bibinfo {pages} {1103--1105} (\bibinfo
  {year} {1980})%
  \bibAnnoteFile{NoStop}{Yanagida:1979as}%

\bibitem{Mohapatra:1979ia}%
  \BibitemOpen
  \bibfield{author}{%
  \bibinfo {author} {\bibfnamefont{Rabindra~N.}\ \bibnamefont{Mohapatra}}\ and\
  \bibinfo {author} {\bibfnamefont{Goran}\ \bibnamefont{Senjanovic}},\ }%
  \bibfield{title}{%
  \enquote{\bibinfo {title} {Neutrino mass and spontaneous parity violation},}\
  }%
  \bibfield{journal}{%
  \Doi{10.1103/PhysRevLett.44.912}{\bibinfo {journal} {Phys.Rev.Lett.}}\ }%
  \textbf{\bibinfo {volume} {44}},\ \bibinfo {pages} {912} (\bibinfo {year}
  {1980})%
  \bibAnnoteFile{NoStop}{Mohapatra:1979ia}%

\bibitem{Wyler:1982dd}%
  \BibitemOpen
  \bibfield{author}{%
  \bibinfo {author} {\bibfnamefont{D.}~\bibnamefont{Wyler}}\ and\ \bibinfo
  {author} {\bibfnamefont{L.}~\bibnamefont{Wolfenstein}},\ }%
  \bibfield{title}{%
  \enquote{\bibinfo {title} {Massless neutrinos in left-right-symmetric
  models},}\ }%
  \bibfield{journal}{%
  \Doi{10.1016/0550-3213(83)90482-0}{\bibinfo {journal} {Nucl.Phys.}}\ }%
  \textbf{\bibinfo {volume} {B218}},\ \bibinfo {pages} {205} (\bibinfo {year}
  {1983})%
  \bibAnnoteFile{NoStop}{Wyler:1982dd}%

\bibitem{Mohapatra:1986aw}%
  \BibitemOpen
  \bibfield{author}{%
  \bibinfo {author} {\bibfnamefont{R.N.}\ \bibnamefont{Mohapatra}},\ }%
  \bibfield{title}{%
  \enquote{\bibinfo {title} {Mechanism for understanding small neutrino mass in
  superstring theories},}\ }%
  \bibfield{journal}{%
  \Doi{10.1103/PhysRevLett.56.561}{\bibinfo {journal} {Phys.Rev.Lett.}}\ }%
  \textbf{\bibinfo {volume} {56}},\ \bibinfo {pages} {561--563} (\bibinfo
  {year} {1986})%
  \bibAnnoteFile{NoStop}{Mohapatra:1986aw}%

\bibitem{Mohapatra:1986bd}%
  \BibitemOpen
  \bibfield{author}{%
  \bibinfo {author} {\bibfnamefont{R.N.}\ \bibnamefont{Mohapatra}}\ and\
  \bibinfo {author} {\bibfnamefont{J.W.F.}\ \bibnamefont{Valle}},\ }%
  \bibfield{title}{%
  \enquote{\bibinfo {title} {Neutrino mass and baryon number nonconservation in
  superstring models},}\ }%
  \bibfield{journal}{%
  \Doi{10.1103/PhysRevD.34.1642}{\bibinfo {journal} {Phys.Rev.}}\ }%
  \textbf{\bibinfo {volume} {D34}},\ \bibinfo {pages} {1642} (\bibinfo {year}
  {1986})%
  \bibAnnoteFile{NoStop}{Mohapatra:1986bd}%

\bibitem{Konetschny:1977bn}%
  \BibitemOpen
  \bibfield{author}{%
  \bibinfo {author} {\bibfnamefont{W.}~\bibnamefont{Konetschny}}\ and\ \bibinfo
  {author} {\bibfnamefont{W.}~\bibnamefont{Kummer}},\ }%
  \bibfield{title}{%
  \enquote{\bibinfo {title} {Nonconservation of total lepton number with scalar
  bosons},}\ }%
  \bibfield{journal}{%
  \Doi{10.1016/0370-2693(77)90407-5}{\bibinfo {journal} {Phys.Lett.}}\ }%
  \textbf{\bibinfo {volume} {B70}},\ \bibinfo {pages} {433} (\bibinfo {year}
  {1977})%
  \bibAnnoteFile{NoStop}{Konetschny:1977bn}%

\bibitem{Lazarides:1980nt}%
  \BibitemOpen
  \bibfield{author}{%
  \bibinfo {author} {\bibfnamefont{George}\ \bibnamefont{Lazarides}}, \bibinfo
  {author} {\bibfnamefont{Q.}~\bibnamefont{Shafi}},\ and\ \bibinfo {author}
  {\bibfnamefont{C.}~\bibnamefont{Wetterich}},\ }%
  \bibfield{title}{%
  \enquote{\bibinfo {title} {Proton lifetime and fermion masses in an $so(10)$
  model},}\ }%
  \bibfield{journal}{%
  \Doi{10.1016/0550-3213(81)90354-0}{\bibinfo {journal} {Nucl.Phys.}}\ }%
  \textbf{\bibinfo {volume} {B181}},\ \bibinfo {pages} {287} (\bibinfo {year}
  {1981})%
  \bibAnnoteFile{NoStop}{Lazarides:1980nt}%

\bibitem{Magg:1980ut}%
  \BibitemOpen
  \bibfield{author}{%
  \bibinfo {author} {\bibfnamefont{M.}~\bibnamefont{Magg}}\ and\ \bibinfo
  {author} {\bibfnamefont{C.}~\bibnamefont{Wetterich}},\ }%
  \bibfield{title}{%
  \enquote{\bibinfo {title} {Neutrino mass problem and gauge hierarchy},}\ }%
  \bibfield{journal}{%
  \Doi{10.1016/0370-2693(80)90825-4}{\bibinfo {journal} {Phys.Lett.}}\ }%
  \textbf{\bibinfo {volume} {B94}},\ \bibinfo {pages} {61} (\bibinfo {year}
  {1980})%
  \bibAnnoteFile{NoStop}{Magg:1980ut}%

\bibitem{Foot:1988aq}%
  \BibitemOpen
  \bibfield{author}{%
  \bibinfo {author} {\bibfnamefont{R.}~\bibnamefont{Foot}}, \bibinfo {author}
  {\bibfnamefont{H.}~\bibnamefont{Lew}}, \bibinfo {author}
  {\bibfnamefont{X.G.}\ \bibnamefont{He}},\ and\ \bibinfo {author}
  {\bibfnamefont{G.~C.}\ \bibnamefont{Joshi}},\ }%
  \bibfield{title}{%
  \enquote{\bibinfo {title} {{Seesaw neutrino masses induced by a triplet of
  leptons}},}\ }%
  \bibfield{journal}{%
  \Doi{10.1007/BF01415558}{\bibinfo {journal} {Z.Phys.}}\ }%
  \textbf{\bibinfo {volume} {C44}},\ \bibinfo {pages} {441} (\bibinfo {year}
  {1989})%
  \bibAnnoteFile{NoStop}{Foot:1988aq}%

\bibitem{Ma:2002pf}%
  \BibitemOpen
  \bibfield{author}{%
  \bibinfo {author} {\bibfnamefont{Ernest}\ \bibnamefont{Ma}}\ and\ \bibinfo
  {author} {\bibfnamefont{D.P.}\ \bibnamefont{Roy}},\ }%
  \bibfield{title}{%
  \enquote{\bibinfo {title} {{Heavy triplet leptons and new gauge boson}},}\ }%
  \bibfield{journal}{%
  \Doi{10.1016/S0550-3213(02)00815-5}{\bibinfo {journal} {Nucl.Phys.}}\ }%
  \textbf{\bibinfo {volume} {B644}},\ \bibinfo {pages} {290--302} (\bibinfo
  {year} {2002}),\
  \Eprint{http://arxiv.org/abs/hep-ph/0206150}{arXiv:hep-ph/0206150 [hep-ph]}%
  \bibAnnoteFile{NoStop}{Ma:2002pf}%

\bibitem{Ross:1985ai}%
  \BibitemOpen
  \bibfield{author}{%
  \bibinfo {author} {\bibfnamefont{Graham~G.}\ \bibnamefont{Ross}},\ }%
  \emph{\bibinfo {title} {Grand unified theories}},\ Frontiers in Physics\
  (\bibinfo {publisher} {Westview Press},\ \bibinfo {year} {2003})\ ISBN
  \bibinfo {isbn} {978-0805369687}%
  \bibAnnoteFile{NoStop}{Ross:1985ai}%

\bibitem{Fukugita:2003en}%
  \BibitemOpen
  \bibfield{author}{%
  \bibinfo {author} {\bibfnamefont{M.}~\bibnamefont{Fukugita}}\ and\ \bibinfo
  {author} {\bibfnamefont{T.}~\bibnamefont{Yanagida}},\ }%
  \emph{\bibinfo {title} {{Physics of neutrinos and applications to
  astrophysics}}}\ (\bibinfo {publisher} {Springer},\ \bibinfo {year} {2003})\
  ISBN \bibinfo {isbn} {978-3540438007}%
  \bibAnnoteFile{NoStop}{Fukugita:2003en}%

\bibitem{Kayser:1989iu}%
  \BibitemOpen
  \bibfield{author}{%
  \bibinfo {author} {\bibfnamefont{Boris}\ \bibnamefont{Kayser}}, \bibinfo
  {author} {\bibfnamefont{F.}~\bibnamefont{Gibrat-Debu}},\ and\ \bibinfo
  {author} {\bibfnamefont{F.}~\bibnamefont{Perrier}},\ }%
  \emph{\bibinfo {title} {The Physics of massive neutrinos}},\ \bibinfo
  {series} {Lecture Notes in Physics}, Vol.~\bibinfo {volume} {25}\ (\bibinfo
  {publisher} {World Scientific},\ \bibinfo {year} {1989})\ ISBN \bibinfo
  {isbn} {978-9971506612}%
  \bibAnnoteFile{NoStop}{Kayser:1989iu}%

\bibitem{Davidson:2005cs}%
  \BibitemOpen
  \bibfield{author}{%
  \bibinfo {author} {\bibfnamefont{Sacha}\ \bibnamefont{Davidson}}, \bibinfo
  {author} {\bibfnamefont{Martin}\ \bibnamefont{Gorbahn}},\ and\ \bibinfo
  {author} {\bibfnamefont{Arcadi}\ \bibnamefont{Santamaria}},\ }%
  \bibfield{title}{%
  \enquote{\bibinfo {title} {From transition magnetic moments to {Majorana}
  neutrino masses},}\ }%
  \bibfield{journal}{%
  \Doi{10.1016/j.physletb.2005.08.086}{\bibinfo {journal} {Phys.Lett.}}\ }%
  \textbf{\bibinfo {volume} {B626}},\ \bibinfo {pages} {151--160} (\bibinfo
  {year} {2005}),\
  \Eprint{http://arxiv.org/abs/hep-ph/0506085}{arXiv:hep-ph/0506085 [hep-ph]}%
  \bibAnnoteFile{NoStop}{Davidson:2005cs}%

\bibitem{Bell:2005kz}%
  \BibitemOpen
  \bibfield{author}{%
  \bibinfo {author} {\bibfnamefont{Nicole~F.}\ \bibnamefont{Bell}}, \bibinfo
  {author} {\bibfnamefont{Vincenzo}\ \bibnamefont{Cirigliano}}, \bibinfo
  {author} {\bibfnamefont{Michael~J.}\ \bibnamefont{Ramsey-Musolf}}, \bibinfo
  {author} {\bibfnamefont{Petr}\ \bibnamefont{Vogel}},\ and\ \bibinfo {author}
  {\bibfnamefont{Mark~B.}\ \bibnamefont{Wise}},\ }%
  \bibfield{title}{%
  \enquote{\bibinfo {title} {{How magnetic is the Dirac neutrino?}}.}\ }%
  \bibfield{journal}{%
  \Doi{10.1103/PhysRevLett.95.151802}{\bibinfo {journal} {Phys.Rev.Lett.}}\ }%
  \textbf{\bibinfo {volume} {95}},\ \bibinfo {pages} {151802} (\bibinfo {year}
  {2005}),\ \Eprint{http://arxiv.org/abs/hep-ph/0504134}{arXiv:hep-ph/0504134
  [hep-ph]}%
  \bibAnnoteFile{NoStop}{Bell:2005kz}%

\bibitem{Raffelt:1999tx}%
  \BibitemOpen
  \bibfield{author}{%
  \bibinfo {author} {\bibfnamefont{Georg~G.}\ \bibnamefont{Raffelt}},\ }%
  \bibfield{title}{%
  \enquote{\bibinfo {title} {{Particle physics from stars}},}\ }%
  \bibfield{journal}{%
  \Doi{10.1146/annurev.nucl.49.1.163}{\bibinfo {journal}
  {Ann.Rev.Nucl.Part.Sci.}}\ }%
  \textbf{\bibinfo {volume} {49}},\ \bibinfo {pages} {163--216} (\bibinfo
  {year} {1999}),\
  \Eprint{http://arxiv.org/abs/hep-ph/9903472}{arXiv:hep-ph/9903472 [hep-ph]}%
  \bibAnnoteFile{NoStop}{Raffelt:1999tx}%

\bibitem{Raffelt:1990pj}%
  \BibitemOpen
  \bibfield{author}{%
  \bibinfo {author} {\bibfnamefont{G.G.}\ \bibnamefont{Raffelt}},\ }%
  \bibfield{title}{%
  \enquote{\bibinfo {title} {{New bound on neutrino dipole moments from
  globular cluster stars}},}\ }%
  \bibfield{journal}{%
  \Doi{10.1103/PhysRevLett.64.2856}{\bibinfo {journal} {Phys.Rev.Lett.}}\ }%
  \textbf{\bibinfo {volume} {64}},\ \bibinfo {pages} {2856--2858} (\bibinfo
  {year} {1990})%
  \bibAnnoteFile{NoStop}{Raffelt:1990pj}%

\bibitem{Weinberg:1980wa}%
  \BibitemOpen
  \bibfield{author}{%
  \bibinfo {author} {\bibfnamefont{Steven}\ \bibnamefont{Weinberg}},\ }%
  \bibfield{title}{%
  \enquote{\bibinfo {title} {Effective gauge theories},}\ }%
  \bibfield{journal}{%
  \Doi{10.1016/0370-2693(80)90660-7}{\bibinfo {journal} {Phys.Lett.}}\ }%
  \textbf{\bibinfo {volume} {B91}},\ \bibinfo {pages} {51} (\bibinfo {year}
  {1980})%
  \bibAnnoteFile{NoStop}{Weinberg:1980wa}%

\bibitem{Veltman:1980mj}%
  \BibitemOpen
  \bibfield{author}{%
  \bibinfo {author} {\bibfnamefont{M.J.G.}\ \bibnamefont{Veltman}},\ }%
  \bibfield{title}{%
  \enquote{\bibinfo {title} {The infrared-ultraviolet connection},}\ }%
  \bibfield{journal}{%
  \bibinfo {journal} {Acta Phys.Polon.}\ }%
  \textbf{\bibinfo {volume} {B12}},\ \bibinfo {pages} {437} (\bibinfo {year}
  {1981})%
  \bibAnnoteFile{NoStop}{Veltman:1980mj}%

\bibitem{deGouvea:2006gz}%
  \BibitemOpen
  \bibfield{author}{%
  \bibinfo {author} {\bibfnamefont{Andre}\ \bibnamefont{de~Gouvea}}, \bibinfo
  {author} {\bibfnamefont{James}\ \bibnamefont{Jenkins}},\ and\ \bibinfo
  {author} {\bibfnamefont{Nirmala}\ \bibnamefont{Vasudevan}},\ }%
  \bibfield{title}{%
  \enquote{\bibinfo {title} {Neutrino phenomenology of very-low-energy
  seesaws},}\ }%
  \bibfield{journal}{%
  \Doi{10.1103/PhysRevD.75.013003}{\bibinfo {journal} {Phys.Rev.}}\ }%
  \textbf{\bibinfo {volume} {D75}},\ \bibinfo {pages} {013003} (\bibinfo {year}
  {2007}),\ \Eprint{http://arxiv.org/abs/hep-ph/0608147}{arXiv:hep-ph/0608147
  [hep-ph]}%
  \bibAnnoteFile{NoStop}{deGouvea:2006gz}%

\bibitem{Ibarra:2010xw}%
  \BibitemOpen
  \bibfield{author}{%
  \bibinfo {author} {\bibfnamefont{A.}~\bibnamefont{Ibarra}}, \bibinfo {author}
  {\bibfnamefont{E.}~\bibnamefont{Molinaro}},\ and\ \bibinfo {author}
  {\bibfnamefont{S.T.}\ \bibnamefont{Petcov}},\ }%
  \bibfield{title}{%
  \enquote{\bibinfo {title} {{TeV}-scale seesaw mechanisms of neutrino mass
  generation, the {Majorana} nature of the heavy singlet neutrinos and
  $(\beta\beta)_{0\nu}$ decay},}\ }%
  \bibfield{journal}{%
  \Doi{10.1007/JHEP09(2010)108}{\bibinfo {journal} {JHEP}}\ }%
  \textbf{\bibinfo {volume} {1009}},\ \bibinfo {pages} {108} (\bibinfo {year}
  {2010}),\ \Eprint{http://arxiv.org/abs/1007.2378}{arXiv:1007.2378 [hep-ph]}%
  \bibAnnoteFile{NoStop}{Ibarra:2010xw}%

\bibitem{Manohar:1983md}%
  \BibitemOpen
  \bibfield{author}{%
  \bibinfo {author} {\bibfnamefont{Aneesh}\ \bibnamefont{Manohar}}\ and\
  \bibinfo {author} {\bibfnamefont{Howard}\ \bibnamefont{Georgi}},\ }%
  \bibfield{title}{%
  \enquote{\bibinfo {title} {Chiral quarks and the nonrelativistic quark
  model},}\ }%
  \bibfield{journal}{%
  \Doi{10.1016/0550-3213(84)90231-1}{\bibinfo {journal} {Nucl.Phys.}}\ }%
  \textbf{\bibinfo {volume} {B234}},\ \bibinfo {pages} {189} (\bibinfo {year}
  {1984})%
  \bibAnnoteFile{NoStop}{Manohar:1983md}%

\bibitem{Georgi:1992dw}%
  \BibitemOpen
  \bibfield{author}{%
  \bibinfo {author} {\bibfnamefont{Howard}\ \bibnamefont{Georgi}},\ }%
  \bibfield{title}{%
  \enquote{\bibinfo {title} {{Generalized dimensional analysis}},}\ }%
  \bibfield{journal}{%
  \Doi{10.1016/0370-2693(93)91728-6}{\bibinfo {journal} {Phys.Lett.}}\ }%
  \textbf{\bibinfo {volume} {B298}},\ \bibinfo {pages} {187--189} (\bibinfo
  {year} {1993}),\
  \Eprint{http://arxiv.org/abs/hep-ph/9207278}{arXiv:hep-ph/9207278 [hep-ph]}%
  \bibAnnoteFile{NoStop}{Georgi:1992dw}%

\bibitem{Lee:1977eg}%
  \BibitemOpen
  \bibfield{author}{%
  \bibinfo {author} {\bibfnamefont{Benjamin~W.}\ \bibnamefont{Lee}}, \bibinfo
  {author} {\bibfnamefont{C.}~\bibnamefont{Quigg}},\ and\ \bibinfo {author}
  {\bibfnamefont{H.B.}\ \bibnamefont{Thacker}},\ }%
  \bibfield{title}{%
  \enquote{\bibinfo {title} {Weak interactions at very high-energies: The role
  of the {Higgs} boson mass},}\ }%
  \bibfield{journal}{%
  \Doi{10.1103/PhysRevD.16.1519}{\bibinfo {journal} {Phys.Rev.}}\ }%
  \textbf{\bibinfo {volume} {D16}},\ \bibinfo {pages} {1519} (\bibinfo {year}
  {1977})%
  \bibAnnoteFile{NoStop}{Lee:1977eg}%

\bibitem{Cornwall:1974km}%
  \BibitemOpen
  \bibfield{author}{%
  \bibinfo {author} {\bibfnamefont{John~M.}\ \bibnamefont{Cornwall}}, \bibinfo
  {author} {\bibfnamefont{David~N.}\ \bibnamefont{Levin}},\ and\ \bibinfo
  {author} {\bibfnamefont{George}\ \bibnamefont{Tiktopoulos}},\ }%
  \bibfield{title}{%
  \enquote{\bibinfo {title} {Derivation of gauge invariance from high-energy
  unitarity bounds on the {S} matrix},}\ }%
  \bibfield{journal}{%
  \Doi{10.1103/PhysRevD.10.1145, 10.1103/PhysRevD.11.972}{\bibinfo {journal}
  {Phys.Rev.}}\ }%
  \textbf{\bibinfo {volume} {D10}},\ \bibinfo {pages} {1145} (\bibinfo {year}
  {1974})%
  \bibAnnoteFile{NoStop}{Cornwall:1974km}%

\bibitem{Terwort:2008ii}%
  \BibitemOpen
  \bibfield{author}{%
  \bibinfo {author} {\bibfnamefont{Mark}\ \bibnamefont{Terwort}} (\bibinfo
  {collaboration} {ATLAS Collaboration, CMS Collaboration}),\ }%
  \bibfield{title}{%
  \enquote{\bibinfo {title} {{Searches for GMSB at the LHC}},}\ }%
  \bibfield{journal}{%
  \Doi{10.3360/dis.2008.110}{\bibinfo {journal} {Proceedings of DIS 2008}},\
  \bibinfo {pages} {110--113}}%
   (\bibinfo {year} {2008}),\
  \Eprint{http://arxiv.org/abs/0805.2524}{arXiv:0805.2524 [hep-ex]}%
  \bibAnnoteFile{NoStop}{Terwort:2008ii}%

\bibitem{Zalewski:2007up}%
  \BibitemOpen
  \bibfield{author}{%
  \bibinfo {author} {\bibfnamefont{Piotr}\ \bibnamefont{Zalewski}},\ }%
  \bibfield{title}{%
  \enquote{\bibinfo {title} {{Search for GMSB NLSPs at LHC}},}\ }%
  \bibfield{journal}{%
  \bibinfo {journal} {Proceedings of SUSY 2007},\ \bibinfo {pages} {314--317}}%
   (\bibinfo {year} {2007}),\
  \Eprint{http://arxiv.org/abs/0710.2647}{arXiv:0710.2647 [hep-ph]},\
  \url{http://www-ekp.physik.uni-karlsruhe.de/~susy07/Proceedings/proceedings/%
susy07.pdf}%
  \bibAnnoteFile{NoStop}{Zalewski:2007up}%

\bibitem{Prieur:2005xv}%
  \BibitemOpen
  \bibfield{author}{%
  \bibinfo {author} {\bibfnamefont{Damien}\ \bibnamefont{Prieur}},\ }%
  \bibfield{title}{%
  \enquote{\bibinfo {title} {{GMSB SUSY models with non pointing photons
  signatures in ATLAS at the LHC}},}\ }%
   (\bibinfo {year} {2005}),\
  \Eprint{http://arxiv.org/abs/hep-ph/0507083}{arXiv:hep-ph/0507083 [hep-ph]}%
  \bibAnnoteFile{NoStop}{Prieur:2005xv}%

\bibitem{Abreu:1996pa}%
  \BibitemOpen
  \bibfield{author}{%
  \bibinfo {author} {\bibfnamefont{P.}~\bibnamefont{Abreu}} \emph{et~al.}
  (\bibinfo {collaboration} {DELPHI Collaboration}),\ }%
  \bibfield{title}{%
  \enquote{\bibinfo {title} {{Search for neutral heavy leptons produced in $Z$
  decays}},}\ }%
  \bibfield{journal}{%
  \Doi{10.1007/s002880050370}{\bibinfo {journal} {Z.Phys.}}\ }%
  \textbf{\bibinfo {volume} {C74}},\ \bibinfo {pages} {57--71} (\bibinfo {year}
  {1997})%
  \bibAnnoteFile{NoStop}{Abreu:1996pa}%

\bibitem{Adriani:1992pq}%
  \BibitemOpen
  \bibfield{author}{%
  \bibinfo {author} {\bibfnamefont{O.}~\bibnamefont{Adriani}} \emph{et~al.}
  (\bibinfo {collaboration} {L3 Collaboration}),\ }%
  \bibfield{title}{%
  \enquote{\bibinfo {title} {{Search for isosinglet neutral heavy leptons in
  $Z^0$ decays}},}\ }%
  \bibfield{journal}{%
  \Doi{10.1016/0370-2693(92)91579-X}{\bibinfo {journal} {Phys.Lett.}}\ }%
  \textbf{\bibinfo {volume} {B295}},\ \bibinfo {pages} {371--382} (\bibinfo
  {year} {1992})%
  \bibAnnoteFile{NoStop}{Adriani:1992pq}%

\bibitem{Akrawy:1990zq}%
  \BibitemOpen
  \bibfield{author}{%
  \bibinfo {author} {\bibfnamefont{M.Z.}\ \bibnamefont{Akrawy}} \emph{et~al.}
  (\bibinfo {collaboration} {OPAL Collaboration}),\ }%
  \bibfield{title}{%
  \enquote{\bibinfo {title} {{Limits on neutral heavy lepton production from
  $Z^0$ decay}},}\ }%
  \bibfield{journal}{%
  \Doi{10.1016/0370-2693(90)90924-U}{\bibinfo {journal} {Phys.Lett.}}\ }%
  \textbf{\bibinfo {volume} {B247}},\ \bibinfo {pages} {448--457} (\bibinfo
  {year} {1990})%
  \bibAnnoteFile{NoStop}{Akrawy:1990zq}%

\bibitem{Dittmar:1989yg}%
  \BibitemOpen
  \bibfield{author}{%
  \bibinfo {author} {\bibfnamefont{M.}~\bibnamefont{Dittmar}}, \bibinfo
  {author} {\bibfnamefont{A.}~\bibnamefont{Santamaria}}, \bibinfo {author}
  {\bibfnamefont{M.C.}\ \bibnamefont{Gonzalez-Garcia}},\ and\ \bibinfo {author}
  {\bibfnamefont{J.W.F.}\ \bibnamefont{Valle}},\ }%
  \bibfield{title}{%
  \enquote{\bibinfo {title} {Production mechanisms and signatures of isosinglet
  neutral heavy leptons in {$Z^0$} decays},}\ }%
  \bibfield{journal}{%
  \Doi{10.1016/0550-3213(90)90028-C}{\bibinfo {journal} {Nucl.Phys.}}\ }%
  \textbf{\bibinfo {volume} {B332}},\ \bibinfo {pages} {1} (\bibinfo {year}
  {1990})%
  \bibAnnoteFile{NoStop}{Dittmar:1989yg}%

\bibitem{Abdallah:2003np}%
  \BibitemOpen
  \bibfield{author}{%
  \bibinfo {author} {\bibfnamefont{J.}~\bibnamefont{Abdallah}} \emph{et~al.}
  (\bibinfo {collaboration} {DELPHI Collaboration}),\ }%
  \bibfield{title}{%
  \enquote{\bibinfo {title} {{Photon events with missing energy in $e^+ e^-$
  collisions at $\sqrt{s} = 130$ to 209 GeV}},}\ }%
  \bibfield{journal}{%
  \Doi{10.1140/epjc/s2004-02051-8}{\bibinfo {journal} {Eur.Phys.J.}}\ }%
  \textbf{\bibinfo {volume} {C38}},\ \bibinfo {pages} {395--411} (\bibinfo
  {year} {2005}),\
  \Eprint{http://arxiv.org/abs/hep-ex/0406019}{arXiv:hep-ex/0406019 [hep-ex]}%
  \bibAnnoteFile{NoStop}{Abdallah:2003np}%

\bibitem{Achard:2003tx}%
  \BibitemOpen
  \bibfield{author}{%
  \bibinfo {author} {\bibfnamefont{P.}~\bibnamefont{Achard}} \emph{et~al.}
  (\bibinfo {collaboration} {L3 Collaboration}),\ }%
  \bibfield{title}{%
  \enquote{\bibinfo {title} {{Single photon and multiphoton events with missing
  energy in $e^{+} e^{-}$ collisions at LEP}},}\ }%
  \bibfield{journal}{%
  \Doi{10.1016/j.physletb.2004.01.010}{\bibinfo {journal} {Phys.Lett.}}\ }%
  \textbf{\bibinfo {volume} {B587}},\ \bibinfo {pages} {16--32} (\bibinfo
  {year} {2004}),\
  \Eprint{http://arxiv.org/abs/hep-ex/0402002}{arXiv:hep-ex/0402002 [hep-ex]}%
  \bibAnnoteFile{NoStop}{Achard:2003tx}%

\bibitem{Abbiendi:1998yu}%
  \BibitemOpen
  \bibfield{author}{%
  \bibinfo {author} {\bibfnamefont{G.}~\bibnamefont{Abbiendi}} \emph{et~al.}
  (\bibinfo {collaboration} {OPAL Collaboration}),\ }%
  \bibfield{title}{%
  \enquote{\bibinfo {title} {{Search for anomalous photonic events with missing
  energy in $e^+ e^-$ collisions at $\sqrt{s} = 130$ GeV, 136 GeV and 183
  GeV}},}\ }%
  \bibfield{journal}{%
  \Doi{10.1007/s100520050442}{\bibinfo {journal} {Eur.Phys.J.}}\ }%
  \textbf{\bibinfo {volume} {C8}},\ \bibinfo {pages} {23--40} (\bibinfo {year}
  {1999}),\ \Eprint{http://arxiv.org/abs/hep-ex/9810021}{arXiv:hep-ex/9810021
  [hep-ex]}%
  \bibAnnoteFile{NoStop}{Abbiendi:1998yu}%

\bibitem{Feldman:1997qc}%
  \BibitemOpen
  \bibfield{author}{%
  \bibinfo {author} {\bibfnamefont{Gary~J.}\ \bibnamefont{Feldman}}\ and\
  \bibinfo {author} {\bibfnamefont{Robert~D.}\ \bibnamefont{Cousins}},\ }%
  \bibfield{title}{%
  \enquote{\bibinfo {title} {{A unified approach to the classical statistical
  analysis of small signals}},}\ }%
  \bibfield{journal}{%
  \Doi{10.1103/PhysRevD.57.3873}{\bibinfo {journal} {Phys.Rev.}}\ }%
  \textbf{\bibinfo {volume} {D57}},\ \bibinfo {pages} {3873--3889} (\bibinfo
  {year} {1998}),\
  \Eprint{http://arxiv.org/abs/physics/9711021}{arXiv:physics/9711021
  [physics.data-an]}%
  \bibAnnoteFile{NoStop}{Feldman:1997qc}%

\bibitem{Abreu:1996vd}%
  \BibitemOpen
  \bibfield{author}{%
  \bibinfo {author} {\bibfnamefont{P.}~\bibnamefont{Abreu}} \emph{et~al.}
  (\bibinfo {collaboration} {DELPHI Collaboration}),\ }%
  \bibfield{title}{%
  \enquote{\bibinfo {title} {{Search for new phenomena using single photon
  events in the DELPHI detector at LEP}},}\ }%
  \bibfield{journal}{%
  \Doi{10.1007/s002880050421}{\bibinfo {journal} {Z.Phys.}}\ }%
  \textbf{\bibinfo {volume} {C74}},\ \bibinfo {pages} {577--586} (\bibinfo
  {year} {1997})%
  \bibAnnoteFile{NoStop}{Abreu:1996vd}%

\bibitem{Campbell:2006wx}%
  \BibitemOpen
  \bibfield{author}{%
  \bibinfo {author} {\bibfnamefont{John~M.}\ \bibnamefont{Campbell}}, \bibinfo
  {author} {\bibfnamefont{J.W.}\ \bibnamefont{Huston}},\ and\ \bibinfo {author}
  {\bibfnamefont{W.J.}\ \bibnamefont{Stirling}},\ }%
  \bibfield{title}{%
  \enquote{\bibinfo {title} {Hard interactions of quarks and gluons: A primer
  for {LHC} physics},}\ }%
  \bibfield{journal}{%
  \Doi{10.1088/0034-4885/70/1/R02}{\bibinfo {journal} {Rept.Prog.Phys.}}\ }%
  \textbf{\bibinfo {volume} {70}},\ \bibinfo {pages} {89} (\bibinfo {year}
  {2007}),\ \Eprint{http://arxiv.org/abs/hep-ph/0611148}{arXiv:hep-ph/0611148
  [hep-ph]}%
  \bibAnnoteFile{NoStop}{Campbell:2006wx}%

\bibitem{Pumplin:2002vw}%
  \BibitemOpen
  \bibfield{author}{%
  \bibinfo {author} {\bibfnamefont{J.}~\bibnamefont{Pumplin}}, \bibinfo
  {author} {\bibfnamefont{D.R.}\ \bibnamefont{Stump}}, \bibinfo {author}
  {\bibfnamefont{J.}~\bibnamefont{Huston}}, \bibinfo {author}
  {\bibfnamefont{H.L.}\ \bibnamefont{Lai}}, \bibinfo {author}
  {\bibfnamefont{Pavel~M.}\ \bibnamefont{Nadolsky}}, \emph{et~al.},\ }%
  \bibfield{title}{%
  \enquote{\bibinfo {title} {{New generation of parton distributions with
  uncertainties from global QCD analysis}},}\ }%
  \bibfield{journal}{%
  \bibinfo {journal} {JHEP}\ }%
  \textbf{\bibinfo {volume} {0207}},\ \bibinfo {pages} {012} (\bibinfo {year}
  {2002}),\ \Eprint{http://arxiv.org/abs/hep-ph/0201195}{arXiv:hep-ph/0201195
  [hep-ph]}%
  \bibAnnoteFile{NoStop}{Pumplin:2002vw}%

\bibitem{Pukhov:1999gg}%
  \BibitemOpen
  \bibfield{author}{%
  \bibinfo {author} {\bibfnamefont{A.}~\bibnamefont{Pukhov}}, \bibinfo {author}
  {\bibfnamefont{E.}~\bibnamefont{Boos}}, \bibinfo {author}
  {\bibfnamefont{M.}~\bibnamefont{Dubinin}}, \bibinfo {author}
  {\bibfnamefont{V.}~\bibnamefont{Edneral}}, \bibinfo {author}
  {\bibfnamefont{V.}~\bibnamefont{Ilyin}}, \emph{et~al.},\ }%
  \bibfield{title}{%
  \enquote{\bibinfo {title} {{CompHEP: a package for evaluation of Feynman
  diagrams and integration over multiparticle phase space}},}\ }%
   (\bibinfo {year} {1999}),\
  \Eprint{http://arxiv.org/abs/hep-ph/9908288}{arXiv:hep-ph/9908288 [hep-ph]}%
  \bibAnnoteFile{NoStop}{Pukhov:1999gg}%

\bibitem{Boos:2004kh}%
  \BibitemOpen
  \bibfield{author}{%
  \bibinfo {author} {\bibfnamefont{E.}~\bibnamefont{Boos}} \emph{et~al.}
  (\bibinfo {collaboration} {CompHEP Collaboration}),\ }%
  \bibfield{title}{%
  \enquote{\bibinfo {title} {{CompHEP 4.4: Automatic computations from
  Lagrangians to events}},}\ }%
  \bibfield{journal}{%
  \Doi{10.1016/j.nima.2004.07.096}{\bibinfo {journal} {Nucl.Instrum.Meth.}}\ }%
  \textbf{\bibinfo {volume} {A534}},\ \bibinfo {pages} {250--259} (\bibinfo
  {year} {2004}),\
  \Eprint{http://arxiv.org/abs/hep-ph/0403113}{arXiv:hep-ph/0403113 [hep-ph]}%
  \bibAnnoteFile{NoStop}{Boos:2004kh}%

\bibitem{Ghosh:2012ep}%
  \BibitemOpen
  \bibfield{author}{%
  \bibinfo {author} {\bibfnamefont{Diptimoy}\ \bibnamefont{Ghosh}}, \bibinfo
  {author} {\bibfnamefont{Rohini}\ \bibnamefont{Godbole}}, \bibinfo {author}
  {\bibfnamefont{Monoranjan}\ \bibnamefont{Guchait}}, \bibinfo {author}
  {\bibfnamefont{Kirtimaan}\ \bibnamefont{Mohan}},\ and\ \bibinfo {author}
  {\bibfnamefont{Dipan}\ \bibnamefont{Sengupta}},\ }%
  \bibfield{title}{%
  \enquote{\bibinfo {title} {{Looking for an invisible Higgs signal at the
  LHC}},}\ }%
  \bibfield{journal}{%
  \Doi{10.1016/j.physletb.2013.07.042}{\bibinfo {journal} {Phys. Lett.}}\ }%
  \textbf{\bibinfo {volume} {B 725}},\ \bibinfo {pages} {344} (\bibinfo {year}
  {2013}),\ \Eprint{http://arxiv.org/abs/1211.7015}{arXiv:1211.7015 [hep-ph]}%
  \bibAnnoteFile{NoStop}{Ghosh:2012ep}%

\bibitem{Belanger:2013kya}%
  \BibitemOpen
  \bibfield{author}{%
  \bibinfo {author} {\bibfnamefont{G.}~\bibnamefont{Belanger}}, \bibinfo
  {author} {\bibfnamefont{B.}~\bibnamefont{Dumont}}, \bibinfo {author}
  {\bibfnamefont{U.}~\bibnamefont{Ellwanger}}, \bibinfo {author}
  {\bibfnamefont{J.F.}\ \bibnamefont{Gunion}},\ and\ \bibinfo {author}
  {\bibfnamefont{S.}~\bibnamefont{Kraml}},\ }%
  \bibfield{title}{%
  \enquote{\bibinfo {title} {{Status of invisible Higgs decays}},}\ }%
  \bibfield{journal}{%
  \Doi{10.1016/j.physletb.2013.05.024}{\bibinfo {journal} {Phys.Lett.}}\ }%
  \textbf{\bibinfo {volume} {B723}},\ \bibinfo {pages} {340--347} (\bibinfo
  {year} {2013}),\ \Eprint{http://arxiv.org/abs/1302.5694}{arXiv:1302.5694
  [hep-ph]}%
  \bibAnnoteFile{NoStop}{Belanger:2013kya}%

\bibitem{Chatrchyan:2013zna}%
  \BibitemOpen
  \bibfield{author}{%
  \bibinfo {author} {\bibfnamefont{Serguei}\ \bibnamefont{Chatrchyan}}
  \emph{et~al.} (\bibinfo {collaboration} {CMS Collaboration}),\ }%
  \bibfield{title}{%
  \enquote{\bibinfo {title} {{Search for the standard model Higgs boson
  produced in association with a W or a Z boson and decaying to bottom
  quarks}},}\ }%
   (\bibinfo {year} {2013}),\
  \Eprint{http://arxiv.org/abs/1310.3687}{arXiv:1310.3687 [hep-ex]}%
  \bibAnnoteFile{NoStop}{Chatrchyan:2013zna}%

\bibitem{Aad:2013wqa}%
  \BibitemOpen
  \bibfield{author}{%
  \bibinfo {author} {\bibfnamefont{Georges}\ \bibnamefont{Aad}} \emph{et~al.}
  (\bibinfo {collaboration} {ATLAS Collaboration}),\ }%
  \bibfield{title}{%
  \enquote{\bibinfo {title} {{Measurements of Higgs boson production and
  couplings in diboson final states with the ATLAS detector at the LHC}},}\ }%
  \bibfield{journal}{%
  \Doi{10.1016/j.physletb.2013.08.010}{\bibinfo {journal} {Phys.Lett.}}\ }%
  \textbf{\bibinfo {volume} {B726}},\ \bibinfo {pages} {88--119} (\bibinfo
  {year} {2013}),\ \Eprint{http://arxiv.org/abs/1307.1427}{arXiv:1307.1427
  [hep-ex]}%
  \bibAnnoteFile{NoStop}{Aad:2013wqa}%

\bibitem{Castellani:1993hs}%
  \BibitemOpen
  \bibfield{author}{%
  \bibinfo {author} {\bibfnamefont{V.}~\bibnamefont{Castellani}}\ and\ \bibinfo
  {author} {\bibfnamefont{S.}~\bibnamefont{Degl'Innocenti}},\ }%
  \bibfield{title}{%
  \enquote{\bibinfo {title} {{Stellar evolution as a probe of neutrino
  properties}},}\ }%
  \bibfield{journal}{%
  \Doi{10.1086/172159}{\bibinfo {journal} {Astrophys.J.}}\ }%
  \textbf{\bibinfo {volume} {402}},\ \bibinfo {pages} {574--578} (\bibinfo
  {year} {1993})%
  \bibAnnoteFile{NoStop}{Castellani:1993hs}%

\bibitem{Catelan:1996-461}%
  \BibitemOpen
  \bibfield{author}{%
  \bibinfo {author} {\bibfnamefont{M.}~\bibnamefont{Catelan}}, \bibinfo
  {author} {\bibfnamefont{J.~A.}\ \bibnamefont{{de Freitas Pacheco}}},\ and\
  \bibinfo {author} {\bibfnamefont{J.~E.}\ \bibnamefont{{Horvath}}},\ }%
  \bibfield{title}{%
  \enquote{\bibinfo {title} {The helium-core mass at the helium flash in
  low-mass red giant stars: Observations and theory},}\ }%
  \bibfield{journal}{%
  \Doi{10.1086/177051}{\bibinfo {journal} {Astrophys. J.}}\ }%
  \textbf{\bibinfo {volume} {461}},\ \bibinfo {pages} {231} (\bibinfo {year}
  {1996}),\
  \Eprint{http://arxiv.org/abs/arXiv:astro-ph/9509062}{arXiv:astro-ph/9509062}%
  \bibAnnoteFile{NoStop}{Catelan:1996-461}%

\bibitem{Haft:1993jt}%
  \BibitemOpen
  \bibfield{author}{%
  \bibinfo {author} {\bibfnamefont{Martin}\ \bibnamefont{Haft}}, \bibinfo
  {author} {\bibfnamefont{Georg}\ \bibnamefont{Raffelt}},\ and\ \bibinfo
  {author} {\bibfnamefont{Achim}\ \bibnamefont{Weiss}},\ }%
  \bibfield{title}{%
  \enquote{\bibinfo {title} {{Standard and nonstandard plasma neutrino emission
  revisited}},}\ }%
  \bibfield{journal}{%
  \Doi{10.1086/173978}{\bibinfo {journal} {Astrophys.J.}}\ }%
  \textbf{\bibinfo {volume} {425}},\ \bibinfo {pages} {222--230} (\bibinfo
  {year} {1994}),\
  \Eprint{http://arxiv.org/abs/astro-ph/9309014}{arXiv:astro-ph/9309014
  [astro-ph]}%
  \bibAnnoteFile{NoStop}{Haft:1993jt}%

\bibitem{Raffelt:1989xu}%
  \BibitemOpen
  \bibfield{author}{%
  \bibinfo {author} {\bibfnamefont{Georg~G.}\ \bibnamefont{Raffelt}},\ }%
  \bibfield{title}{%
  \enquote{\bibinfo {title} {Core mass at the helium flash from observations
  and a new bound on neutrino electromagnetic properties},}\ }%
  \bibfield{journal}{%
  \Doi{10.1086/169510}{\bibinfo {journal} {Astrophys.J.}}\ }%
  \textbf{\bibinfo {volume} {365}},\ \bibinfo {pages} {559} (\bibinfo {year}
  {1990})%
  \bibAnnoteFile{NoStop}{Raffelt:1989xu}%

\bibitem{Raffelt:1992pi}%
  \BibitemOpen
  \bibfield{author}{%
  \bibinfo {author} {\bibfnamefont{Georg}\ \bibnamefont{Raffelt}}\ and\
  \bibinfo {author} {\bibfnamefont{Achim}\ \bibnamefont{Weiss}},\ }%
  \bibfield{title}{%
  \enquote{\bibinfo {title} {{Nonstandard neutrino interactions and the
  evolution of red giants}},}\ }%
  \bibfield{journal}{%
  \bibinfo {journal} {Astron.Astrophys.}\ }%
  \textbf{\bibinfo {volume} {264}},\ \bibinfo {pages} {536--546} (\bibinfo
  {year} {1992}),\ \url{http://adsabs.harvard.edu/abs/1992A&A...264..536R}%
  \bibAnnoteFile{NoStop}{Raffelt:1992pi}%

\bibitem{Heger:2008er}%
  \BibitemOpen
  \bibfield{author}{%
  \bibinfo {author} {\bibfnamefont{Alexander}\ \bibnamefont{Heger}}, \bibinfo
  {author} {\bibfnamefont{Alexander}\ \bibnamefont{Friedland}}, \bibinfo
  {author} {\bibfnamefont{Maurizio}\ \bibnamefont{Giannotti}},\ and\ \bibinfo
  {author} {\bibfnamefont{Vincenzo}\ \bibnamefont{Cirigliano}},\ }%
  \bibfield{title}{%
  \enquote{\bibinfo {title} {The impact of neutrino magnetic moments on the
  evolution of massive stars},}\ }%
  \bibfield{journal}{%
  \Doi{10.1088/0004-637X/696/1/608}{\bibinfo {journal} {Astrophys.J.}}\ }%
  \textbf{\bibinfo {volume} {696}},\ \bibinfo {pages} {608--619} (\bibinfo
  {year} {2009}),\ \Eprint{http://arxiv.org/abs/0809.4703}{arXiv:0809.4703
  [astro-ph]}%
  \bibAnnoteFile{NoStop}{Heger:2008er}%

\bibitem{Iwamoto:1994zd}%
  \BibitemOpen
  \bibfield{author}{%
  \bibinfo {author} {\bibfnamefont{Naoki}\ \bibnamefont{Iwamoto}}, \bibinfo
  {author} {\bibfnamefont{Letao}\ \bibnamefont{Qin}}, \bibinfo {author}
  {\bibfnamefont{Masataka}\ \bibnamefont{Fukugita}},\ and\ \bibinfo {author}
  {\bibfnamefont{Sachiko}\ \bibnamefont{Tsuruta}},\ }%
  \bibfield{title}{%
  \enquote{\bibinfo {title} {{Neutrino magnetic moment and neutron star
  cooling}},}\ }%
  \bibfield{journal}{%
  \Doi{10.1103/PhysRevD.51.348}{\bibinfo {journal} {Phys.Rev.}}\ }%
  \textbf{\bibinfo {volume} {D51}},\ \bibinfo {pages} {348--352} (\bibinfo
  {year} {1995})%
  \bibAnnoteFile{NoStop}{Iwamoto:1994zd}%

\bibitem{Raffelt:1999gv}%
  \BibitemOpen
  \bibfield{author}{%
  \bibinfo {author} {\bibfnamefont{G.G.}\ \bibnamefont{Raffelt}},\ }%
  \bibfield{title}{%
  \enquote{\bibinfo {title} {{Limits on neutrino electromagnetic properties: an
  update}},}\ }%
  \bibfield{journal}{%
  \Doi{10.1016/S0370-1573(99)00074-5}{\bibinfo {journal} {Phys.Rept.}}\ }%
  \textbf{\bibinfo {volume} {320}},\ \bibinfo {pages} {319--327} (\bibinfo
  {year} {1999})%
  \bibAnnoteFile{NoStop}{Raffelt:1999gv}%

\bibitem{Steigman:1976ev}%
  \BibitemOpen
  \bibfield{author}{%
  \bibinfo {author} {\bibfnamefont{G.}~\bibnamefont{Steigman}},\ }%
  \bibfield{title}{%
  \enquote{\bibinfo {title} {{Observational tests of antimatter
  cosmologies}},}\ }%
  \bibfield{journal}{%
  \Doi{10.1146/annurev.aa.14.090176.002011}{\bibinfo {journal}
  {Ann.Rev.Astron.Astrophys.}}\ }%
  \textbf{\bibinfo {volume} {14}},\ \bibinfo {pages} {339--372} (\bibinfo
  {year} {1976})%
  \bibAnnoteFile{NoStop}{Steigman:1976ev}%

\bibitem{Cohen:1997ac}%
  \BibitemOpen
  \bibfield{author}{%
  \bibinfo {author} {\bibfnamefont{Andrew~G.}\ \bibnamefont{Cohen}}, \bibinfo
  {author} {\bibfnamefont{A.}~\bibnamefont{De~Rujula}},\ and\ \bibinfo {author}
  {\bibfnamefont{S.L.}\ \bibnamefont{Glashow}},\ }%
  \bibfield{title}{%
  \enquote{\bibinfo {title} {A matter-antimatter universe?}.}\ }%
  \bibfield{journal}{%
  \Doi{10.1086/305328}{\bibinfo {journal} {Astrophys.J.}}\ }%
  \textbf{\bibinfo {volume} {495}},\ \bibinfo {pages} {539--549} (\bibinfo
  {year} {1998}),\
  \Eprint{http://arxiv.org/abs/astro-ph/9707087}{arXiv:astro-ph/9707087
  [astro-ph]}%
  \bibAnnoteFile{NoStop}{Cohen:1997ac}%

\bibitem{Sakharov:1967dj}%
  \BibitemOpen
  \bibfield{author}{%
  \bibinfo {author} {\bibfnamefont{A.D.}\ \bibnamefont{Sakharov}},\ }%
  \bibfield{title}{%
  \enquote{\bibinfo {title} {Violation of {CP} invariance, {C} asymmetry and
  baryon asymmetry of the universe},}\ }%
  \bibfield{journal}{%
  \Doi{10.1070/PU1991v034n05ABEH002497}{\bibinfo {journal} {Pisma
  Zh.Eksp.Teor.Fiz.}}\ }%
  \textbf{\bibinfo {volume} {5}},\ \bibinfo {pages} {32--35} (\bibinfo {year}
  {1967})%
  \bibAnnoteFile{NoStop}{Sakharov:1967dj}%

\bibitem{Gavela:1993ts}%
  \BibitemOpen
  \bibfield{author}{%
  \bibinfo {author} {\bibfnamefont{M.B.}\ \bibnamefont{Gavela}}, \bibinfo
  {author} {\bibfnamefont{P.}~\bibnamefont{Hernandez}}, \bibinfo {author}
  {\bibfnamefont{J.}~\bibnamefont{Orloff}},\ and\ \bibinfo {author}
  {\bibfnamefont{O.}~\bibnamefont{Pene}},\ }%
  \bibfield{title}{%
  \enquote{\bibinfo {title} {{Standard model CP violation and baryon
  asymmetry}},}\ }%
  \bibfield{journal}{%
  \Doi{10.1142/S0217732394000629}{\bibinfo {journal} {Mod.Phys.Lett.}}\ }%
  \textbf{\bibinfo {volume} {A9}},\ \bibinfo {pages} {795--810} (\bibinfo
  {year} {1994}),\
  \Eprint{http://arxiv.org/abs/hep-ph/9312215}{arXiv:hep-ph/9312215 [hep-ph]}%
  \bibAnnoteFile{NoStop}{Gavela:1993ts}%

\bibitem{Gavela:1994dt}%
  \BibitemOpen
  \bibfield{author}{%
  \bibinfo {author} {\bibfnamefont{M.B.}\ \bibnamefont{Gavela}}, \bibinfo
  {author} {\bibfnamefont{P.}~\bibnamefont{Hernandez}}, \bibinfo {author}
  {\bibfnamefont{J.}~\bibnamefont{Orloff}}, \bibinfo {author}
  {\bibfnamefont{O.}~\bibnamefont{Pene}},\ and\ \bibinfo {author}
  {\bibfnamefont{C.}~\bibnamefont{Quimbay}},\ }%
  \bibfield{title}{%
  \enquote{\bibinfo {title} {{Standard model CP violation and baryon asymmetry.
  Part 2: Finite temperature}},}\ }%
  \bibfield{journal}{%
  \Doi{10.1016/0550-3213(94)00410-2}{\bibinfo {journal} {Nucl.Phys.}}\ }%
  \textbf{\bibinfo {volume} {B430}},\ \bibinfo {pages} {382--426} (\bibinfo
  {year} {1994}),\
  \Eprint{http://arxiv.org/abs/hep-ph/9406289}{arXiv:hep-ph/9406289 [hep-ph]}%
  \bibAnnoteFile{NoStop}{Gavela:1994dt}%

\bibitem{Huet:1994jb}%
  \BibitemOpen
  \bibfield{author}{%
  \bibinfo {author} {\bibfnamefont{Patrick}\ \bibnamefont{Huet}}\ and\ \bibinfo
  {author} {\bibfnamefont{Eric}\ \bibnamefont{Sather}},\ }%
  \bibfield{title}{%
  \enquote{\bibinfo {title} {{Electroweak baryogenesis and standard model CP
  violation}},}\ }%
  \bibfield{journal}{%
  \Doi{10.1103/PhysRevD.51.379}{\bibinfo {journal} {Phys.Rev.}}\ }%
  \textbf{\bibinfo {volume} {D51}},\ \bibinfo {pages} {379--394} (\bibinfo
  {year} {1995}),\
  \Eprint{http://arxiv.org/abs/hep-ph/9404302}{arXiv:hep-ph/9404302 [hep-ph]}%
  \bibAnnoteFile{NoStop}{Huet:1994jb}%

\bibitem{Fukugita:1986hr}%
  \BibitemOpen
  \bibfield{author}{%
  \bibinfo {author} {\bibfnamefont{M.}~\bibnamefont{Fukugita}}\ and\ \bibinfo
  {author} {\bibfnamefont{T.}~\bibnamefont{Yanagida}},\ }%
  \bibfield{title}{%
  \enquote{\bibinfo {title} {Baryogenesis without grand unification},}\ }%
  \bibfield{journal}{%
  \Doi{10.1016/0370-2693(86)91126-3}{\bibinfo {journal} {Phys.Lett.}}\ }%
  \textbf{\bibinfo {volume} {B174}},\ \bibinfo {pages} {45} (\bibinfo {year}
  {1986})%
  \bibAnnoteFile{NoStop}{Fukugita:1986hr}%

\bibitem{Flanz:1996fb}%
  \BibitemOpen
  \bibfield{author}{%
  \bibinfo {author} {\bibfnamefont{Marion}\ \bibnamefont{Flanz}}, \bibinfo
  {author} {\bibfnamefont{Emmanuel~A.}\ \bibnamefont{Paschos}}, \bibinfo
  {author} {\bibfnamefont{Utpal}\ \bibnamefont{Sarkar}},\ and\ \bibinfo
  {author} {\bibfnamefont{Jan}\ \bibnamefont{Weiss}},\ }%
  \bibfield{title}{%
  \enquote{\bibinfo {title} {{Baryogenesis through mixing of heavy Majorana
  neutrinos}},}\ }%
  \bibfield{journal}{%
  \Doi{10.1016/S0370-2693(96)01337-8}{\bibinfo {journal} {Phys.Lett.}}\ }%
  \textbf{\bibinfo {volume} {B389}},\ \bibinfo {pages} {693--699} (\bibinfo
  {year} {1996}),\
  \Eprint{http://arxiv.org/abs/hep-ph/9607310}{arXiv:hep-ph/9607310 [hep-ph]}%
  \bibAnnoteFile{NoStop}{Flanz:1996fb}%

\bibitem{Dolgov:1991fr}%
  \BibitemOpen
  \bibfield{author}{%
  \bibinfo {author} {\bibfnamefont{A.D.}\ \bibnamefont{Dolgov}},\ }%
  \bibfield{title}{%
  \enquote{\bibinfo {title} {{NonGUT baryogenesis}},}\ }%
  \bibfield{journal}{%
  \Doi{10.1016/0370-1573(92)90107-B}{\bibinfo {journal} {Phys.Rept.}}\ }%
  \textbf{\bibinfo {volume} {222}},\ \bibinfo {pages} {309--386} (\bibinfo
  {year} {1992})%
  \bibAnnoteFile{NoStop}{Dolgov:1991fr}%

\bibitem{Davidson:2008bu}%
  \BibitemOpen
  \bibfield{author}{%
  \bibinfo {author} {\bibfnamefont{Sacha}\ \bibnamefont{Davidson}}, \bibinfo
  {author} {\bibfnamefont{Enrico}\ \bibnamefont{Nardi}},\ and\ \bibinfo
  {author} {\bibfnamefont{Yosef}\ \bibnamefont{Nir}},\ }%
  \bibfield{title}{%
  \enquote{\bibinfo {title} {{Leptogenesis}},}\ }%
  \bibfield{journal}{%
  \Doi{10.1016/j.physrep.2008.06.002}{\bibinfo {journal} {Phys.Rept.}}\ }%
  \textbf{\bibinfo {volume} {466}},\ \bibinfo {pages} {105--177} (\bibinfo
  {year} {2008}),\ \Eprint{http://arxiv.org/abs/0802.2962}{arXiv:0802.2962
  [hep-ph]}%
  \bibAnnoteFile{NoStop}{Davidson:2008bu}%

\bibitem{Rubakov:1996vz}%
  \BibitemOpen
  \bibfield{author}{%
  \bibinfo {author} {\bibfnamefont{V.A.}\ \bibnamefont{Rubakov}}\ and\ \bibinfo
  {author} {\bibfnamefont{M.E.}\ \bibnamefont{Shaposhnikov}},\ }%
  \bibfield{title}{%
  \enquote{\bibinfo {title} {{Electroweak baryon number nonconservation in the
  early universe and in high-energy collisions}},}\ }%
  \bibfield{journal}{%
  \Doi{10.1070/PU1996v039n05ABEH000145}{\bibinfo {journal} {Usp.Fiz.Nauk}}\ }%
  \textbf{\bibinfo {volume} {166}},\ \bibinfo {pages} {493--537} (\bibinfo
  {year} {1996}),\
  \Eprint{http://arxiv.org/abs/hep-ph/9603208}{arXiv:hep-ph/9603208 [hep-ph]}%
  \bibAnnoteFile{NoStop}{Rubakov:1996vz}%

\bibitem{Pilaftsis:1997jf}%
  \BibitemOpen
  \bibfield{author}{%
  \bibinfo {author} {\bibfnamefont{Apostolos}\ \bibnamefont{Pilaftsis}},\ }%
  \bibfield{title}{%
  \enquote{\bibinfo {title} {{CP violation and baryogenesis due to heavy
  Majorana neutrinos}},}\ }%
  \bibfield{journal}{%
  \Doi{10.1103/PhysRevD.56.5431}{\bibinfo {journal} {Phys.Rev.}}\ }%
  \textbf{\bibinfo {volume} {D56}},\ \bibinfo {pages} {5431--5451} (\bibinfo
  {year} {1997}),\
  \Eprint{http://arxiv.org/abs/hep-ph/9707235}{arXiv:hep-ph/9707235 [hep-ph]}%
  \bibAnnoteFile{NoStop}{Pilaftsis:1997jf}%

\bibitem{Pilaftsis:2003gt}%
  \BibitemOpen
  \bibfield{author}{%
  \bibinfo {author} {\bibfnamefont{Apostolos}\ \bibnamefont{Pilaftsis}}\ and\
  \bibinfo {author} {\bibfnamefont{Thomas~E.J.}\ \bibnamefont{Underwood}},\ }%
  \bibfield{title}{%
  \enquote{\bibinfo {title} {{Resonant leptogenesis}},}\ }%
  \bibfield{journal}{%
  \Doi{10.1016/j.nuclphysb.2004.05.029}{\bibinfo {journal} {Nucl.Phys.}}\ }%
  \textbf{\bibinfo {volume} {B692}},\ \bibinfo {pages} {303--345} (\bibinfo
  {year} {2004}),\
  \Eprint{http://arxiv.org/abs/hep-ph/0309342}{arXiv:hep-ph/0309342 [hep-ph]}%
  \bibAnnoteFile{NoStop}{Pilaftsis:2003gt}%

\bibitem{DeRujula:1989fe}%
  \BibitemOpen
  \bibfield{author}{%
  \bibinfo {author} {\bibfnamefont{A.}~\bibnamefont{De~Rujula}}, \bibinfo
  {author} {\bibfnamefont{S.L.}\ \bibnamefont{Glashow}},\ and\ \bibinfo
  {author} {\bibfnamefont{U.}~\bibnamefont{Sarid}},\ }%
  \bibfield{title}{%
  \enquote{\bibinfo {title} {Charged dark matter},}\ }%
  \bibfield{journal}{%
  \Doi{10.1016/0550-3213(90)90227-5}{\bibinfo {journal} {Nucl.Phys.}}\ }%
  \textbf{\bibinfo {volume} {B333}},\ \bibinfo {pages} {173} (\bibinfo {year}
  {1990})%
  \bibAnnoteFile{NoStop}{DeRujula:1989fe}%

\bibitem{Dimopoulos:1989hk}%
  \BibitemOpen
  \bibfield{author}{%
  \bibinfo {author} {\bibfnamefont{Savas}\ \bibnamefont{Dimopoulos}}, \bibinfo
  {author} {\bibfnamefont{David}\ \bibnamefont{Eichler}}, \bibinfo {author}
  {\bibfnamefont{Rahim}\ \bibnamefont{Esmailzadeh}},\ and\ \bibinfo {author}
  {\bibfnamefont{Glenn~D.}\ \bibnamefont{Starkman}},\ }%
  \bibfield{title}{%
  \enquote{\bibinfo {title} {Getting a charge out of dark matter},}\ }%
  \bibfield{journal}{%
  \Doi{10.1103/PhysRevD.41.2388}{\bibinfo {journal} {Phys.Rev.}}\ }%
  \textbf{\bibinfo {volume} {D41}},\ \bibinfo {pages} {2388} (\bibinfo {year}
  {1990})%
  \bibAnnoteFile{NoStop}{Dimopoulos:1989hk}%

\bibitem{Basdevant:1989fh}%
  \BibitemOpen
  \bibfield{author}{%
  \bibinfo {author} {\bibfnamefont{J.L.}\ \bibnamefont{Basdevant}}, \bibinfo
  {author} {\bibfnamefont{R.}~\bibnamefont{Mochkovitch}}, \bibinfo {author}
  {\bibfnamefont{J.}~\bibnamefont{Rich}}, \bibinfo {author}
  {\bibfnamefont{M.}~\bibnamefont{Spiro}},\ and\ \bibinfo {author}
  {\bibfnamefont{A.}~\bibnamefont{Vidal-Madjar}},\ }%
  \bibfield{title}{%
  \enquote{\bibinfo {title} {Is there room for charged dark matter?}.}\ }%
  \bibfield{journal}{%
  \Doi{10.1016/0370-2693(90)91948-B}{\bibinfo {journal} {Phys.Lett.}}\ }%
  \textbf{\bibinfo {volume} {B234}},\ \bibinfo {pages} {395} (\bibinfo {year}
  {1990})%
  \bibAnnoteFile{NoStop}{Basdevant:1989fh}%

\bibitem{Hemmick:1989ns}%
  \BibitemOpen
  \bibfield{author}{%
  \bibinfo {author} {\bibfnamefont{T.K.}\ \bibnamefont{Hemmick}}, \bibinfo
  {author} {\bibfnamefont{D.}~\bibnamefont{Elmore}}, \bibinfo {author}
  {\bibfnamefont{T.}~\bibnamefont{Gentile}}, \bibinfo {author}
  {\bibfnamefont{P.W.}\ \bibnamefont{Kubik}}, \bibinfo {author}
  {\bibfnamefont{S.L.}\ \bibnamefont{Olsen}}, \emph{et~al.},\ }%
  \bibfield{title}{%
  \enquote{\bibinfo {title} {A search for anomalously heavy isotopes of low-{Z}
  nuclei},}\ }%
  \bibfield{journal}{%
  \Doi{10.1103/PhysRevD.41.2074}{\bibinfo {journal} {Phys.Rev.}}\ }%
  \textbf{\bibinfo {volume} {D41}},\ \bibinfo {pages} {2074--2080} (\bibinfo
  {year} {1990})%
  \bibAnnoteFile{NoStop}{Hemmick:1989ns}%

\bibitem{Yamagata:1993jq}%
  \BibitemOpen
  \bibfield{author}{%
  \bibinfo {author} {\bibfnamefont{T.}~\bibnamefont{Yamagata}}, \bibinfo
  {author} {\bibfnamefont{Y.}~\bibnamefont{Takamori}},\ and\ \bibinfo {author}
  {\bibfnamefont{H.}~\bibnamefont{Utsunomiya}},\ }%
  \bibfield{title}{%
  \enquote{\bibinfo {title} {{Search for anomalously heavy hydrogen in deep sea
  water at 4000 m}},}\ }%
  \bibfield{journal}{%
  \Doi{10.1103/PhysRevD.47.1231}{\bibinfo {journal} {Phys.Rev.}}\ }%
  \textbf{\bibinfo {volume} {D47}},\ \bibinfo {pages} {1231--1234} (\bibinfo
  {year} {1993})%
  \bibAnnoteFile{NoStop}{Yamagata:1993jq}%

\bibitem{Gould:1989gw}%
  \BibitemOpen
  \bibfield{author}{%
  \bibinfo {author} {\bibfnamefont{Andrew}\ \bibnamefont{Gould}}, \bibinfo
  {author} {\bibfnamefont{Bruce~T.}\ \bibnamefont{Draine}}, \bibinfo {author}
  {\bibfnamefont{Roger~W.}\ \bibnamefont{Romani}},\ and\ \bibinfo {author}
  {\bibfnamefont{Shmuel}\ \bibnamefont{Nussinov}},\ }%
  \bibfield{title}{%
  \enquote{\bibinfo {title} {Neutron stars: Graveyard of charged dark
  matter},}\ }%
  \bibfield{journal}{%
  \Doi{10.1016/0370-2693(90)91745-W}{\bibinfo {journal} {Phys.Lett.}}\ }%
  \textbf{\bibinfo {volume} {B238}},\ \bibinfo {pages} {337} (\bibinfo {year}
  {1990})%
  \bibAnnoteFile{NoStop}{Gould:1989gw}%

\bibitem{Chivukula:1989cc}%
  \BibitemOpen
  \bibfield{author}{%
  \bibinfo {author} {\bibfnamefont{R.~Sekhar}\ \bibnamefont{Chivukula}},
  \bibinfo {author} {\bibfnamefont{Andrew~G.}\ \bibnamefont{Cohen}}, \bibinfo
  {author} {\bibfnamefont{Savas}\ \bibnamefont{Dimopoulos}},\ and\ \bibinfo
  {author} {\bibfnamefont{Terry~P.}\ \bibnamefont{Walker}},\ }%
  \bibfield{title}{%
  \enquote{\bibinfo {title} {Bounds on halo particle interactions from
  interstellar calorimetry},}\ }%
  \bibfield{journal}{%
  \Doi{10.1103/PhysRevLett.65.957}{\bibinfo {journal} {Phys.Rev.Lett.}}\ }%
  \textbf{\bibinfo {volume} {65}},\ \bibinfo {pages} {957--959} (\bibinfo
  {year} {1990})%
  \bibAnnoteFile{NoStop}{Chivukula:1989cc}%

\bibitem{Wolfram:1978gp}%
  \BibitemOpen
  \bibfield{author}{%
  \bibinfo {author} {\bibfnamefont{Stephen}\ \bibnamefont{Wolfram}},\ }%
  \bibfield{title}{%
  \enquote{\bibinfo {title} {Abundances of stable particles produced in the
  early universe},}\ }%
  \bibfield{journal}{%
  \Doi{10.1016/0370-2693(79)90426-X}{\bibinfo {journal} {Phys.Lett.}}\ }%
  \textbf{\bibinfo {volume} {B82}},\ \bibinfo {pages} {65} (\bibinfo {year}
  {1979})%
  \bibAnnoteFile{NoStop}{Wolfram:1978gp}%

\bibitem{Chuzhoy:2008zy}%
  \BibitemOpen
  \bibfield{author}{%
  \bibinfo {author} {\bibfnamefont{Leonid}\ \bibnamefont{Chuzhoy}}\ and\
  \bibinfo {author} {\bibfnamefont{Edward~W.}\ \bibnamefont{Kolb}},\ }%
  \bibfield{title}{%
  \enquote{\bibinfo {title} {{Reopening the window on charged dark matter}},}\
  }%
  \bibfield{journal}{%
  \Doi{10.1088/1475-7516/2009/07/014}{\bibinfo {journal} {JCAP}}\ }%
  \textbf{\bibinfo {volume} {0907}},\ \bibinfo {pages} {014} (\bibinfo {year}
  {2009}),\ \Eprint{http://arxiv.org/abs/0809.0436}{arXiv:0809.0436
  [astro-ph]}%
  \bibAnnoteFile{NoStop}{Chuzhoy:2008zy}%

\bibitem{Jedamzik:2007qk}%
  \BibitemOpen
  \bibfield{author}{%
  \bibinfo {author} {\bibfnamefont{Karsten}\ \bibnamefont{Jedamzik}},\ }%
  \bibfield{title}{%
  \enquote{\bibinfo {title} {{Bounds on long-lived charged massive particles
  from Big Bang nucleosynthesis}},}\ }%
  \bibfield{journal}{%
  \Doi{10.1088/1475-7516/2008/03/008}{\bibinfo {journal} {JCAP}}\ }%
  \textbf{\bibinfo {volume} {0803}},\ \bibinfo {pages} {008} (\bibinfo {year}
  {2008}),\ \Eprint{http://arxiv.org/abs/0710.5153}{arXiv:0710.5153 [hep-ph]}%
  \bibAnnoteFile{NoStop}{Jedamzik:2007qk}%

\bibitem{Jedamzik:2007cp}%
  \BibitemOpen
  \bibfield{author}{%
  \bibinfo {author} {\bibfnamefont{Karsten}\ \bibnamefont{Jedamzik}},\ }%
  \bibfield{title}{%
  \enquote{\bibinfo {title} {{The cosmic $\tensor[^{6}]{\mathrm{Li}}{}$ and
  $\tensor[^7]{\mathrm{Li}}{}$ problems and BBN with long-lived charged massive
  particles}},}\ }%
  \bibfield{journal}{%
  \Doi{10.1103/PhysRevD.77.063524}{\bibinfo {journal} {Phys.Rev.}}\ }%
  \textbf{\bibinfo {volume} {D77}},\ \bibinfo {pages} {063524} (\bibinfo {year}
  {2008}),\ \Eprint{http://arxiv.org/abs/0707.2070}{arXiv:0707.2070
  [astro-ph]}%
  \bibAnnoteFile{NoStop}{Jedamzik:2007cp}%

\bibitem{Bilenky:1993bt}%
  \BibitemOpen
  \bibfield{author}{%
  \bibinfo {author} {\bibfnamefont{Mikhail~S.}\ \bibnamefont{Bilenky}}\ and\
  \bibinfo {author} {\bibfnamefont{Arcadi}\ \bibnamefont{Santamaria}},\ }%
  \bibfield{title}{%
  \enquote{\bibinfo {title} {{One loop effective Lagrangian for a Standard
  Model with a heavy charged scalar singlet}},}\ }%
  \bibfield{journal}{%
  \Doi{10.1016/0550-3213(94)90375-1}{\bibinfo {journal} {Nucl.Phys.}}\ }%
  \textbf{\bibinfo {volume} {B420}},\ \bibinfo {pages} {47--93} (\bibinfo
  {year} {1994}),\
  \Eprint{http://arxiv.org/abs/hep-ph/9310302}{arXiv:hep-ph/9310302 [hep-ph]}%
  \bibAnnoteFile{NoStop}{Bilenky:1993bt}%

\bibitem{Adam:2013mnn}%
  \BibitemOpen
  \bibfield{author}{%
  \bibinfo {author} {\bibfnamefont{J.}~\bibnamefont{Adam}} \emph{et~al.}
  (\bibinfo {collaboration} {MEG Collaboration}),\ }%
  \bibfield{title}{%
  \enquote{\bibinfo {title} {{New constraint on the existence of the $\mu
  \rightarrow e \gamma$ decay}},}\ }%
   (\bibinfo {year} {2013}),\
  \Eprint{http://arxiv.org/abs/1303.0754}{arXiv:1303.0754 [hep-ex]}%
  \bibAnnoteFile{NoStop}{Adam:2013mnn}%

\bibitem{Nishiguchi:2011zz}%
  \BibitemOpen
  \bibfield{author}{%
  \bibinfo {author} {\bibfnamefont{Hajime}\ \bibnamefont{Nishiguchi}} (\bibinfo
  {collaboration} {MEG Collaboration}),\ }%
  \bibfield{title}{%
  \enquote{\bibinfo {title} {{MEG experiment: new result and prospects}},}\ }%
  \bibfield{journal}{%
  \Doi{10.1063/1.3644322}{\bibinfo {journal} {AIP Conf.Proc.}}\ }%
  \textbf{\bibinfo {volume} {1382}},\ \bibinfo {pages} {239--241} (\bibinfo
  {year} {2011})%
  \bibAnnoteFile{NoStop}{Nishiguchi:2011zz}%

\bibitem{Dohmen:1993mp}%
  \BibitemOpen
  \bibfield{author}{%
  \bibinfo {author} {\bibfnamefont{C.}~\bibnamefont{Dohmen}} \emph{et~al.}
  (\bibinfo {collaboration} {SINDRUM II Collaboration.}),\ }%
  \bibfield{title}{%
  \enquote{\bibinfo {title} {{Test of lepton flavor conservation in $\mu - e$
  conversion on titanium}},}\ }%
  \bibfield{journal}{%
  \Doi{10.1016/0370-2693(93)91383-X}{\bibinfo {journal} {Phys.Lett.}}\ }%
  \textbf{\bibinfo {volume} {B317}},\ \bibinfo {pages} {631--636} (\bibinfo
  {year} {1993})%
  \bibAnnoteFile{NoStop}{Dohmen:1993mp}%

\bibitem{Kuno:2013mha}%
  \BibitemOpen
  \bibfield{author}{%
  \bibinfo {author} {\bibfnamefont{Yoshitaka}\ \bibnamefont{Kuno}} (\bibinfo
  {collaboration} {COMET Collaboration}),\ }%
  \bibfield{title}{%
  \enquote{\bibinfo {title} {{A search for muon-to-electron conversion at
  J-PARC: The COMET experiment}},}\ }%
  \bibfield{journal}{%
  \Doi{10.1093/ptep/pts089}{\bibinfo {journal} {PTEP}}\ }%
  \textbf{\bibinfo {volume} {2013}},\ \bibinfo {pages} {022C01} (\bibinfo
  {year} {2013})%
  \bibAnnoteFile{NoStop}{Kuno:2013mha}%

\bibitem{Pasternak:2013ksa}%
  \BibitemOpen
  \bibfield{author}{%
  \bibinfo {author} {\bibfnamefont{J.}~\bibnamefont{Pasternak}} (\bibinfo
  {collaboration} {PRISM Task Force}),\ }%
  \bibfield{title}{%
  \enquote{\bibinfo {title} {{Recent studies on the PRISM FFAG ring}},}\ }%
  \bibfield{journal}{%
  \Doi{10.1088/1742-6596/408/1/012080}{\bibinfo {journal} {J.Phys.Conf.Ser.}}\
  }%
  \textbf{\bibinfo {volume} {408}},\ \bibinfo {pages} {012080} (\bibinfo {year}
  {2013})%
  \bibAnnoteFile{NoStop}{Pasternak:2013ksa}%

\bibitem{Fairbairn:2006gg}%
  \BibitemOpen
  \bibfield{author}{%
  \bibinfo {author} {\bibfnamefont{M.}~\bibnamefont{Fairbairn}}, \bibinfo
  {author} {\bibfnamefont{A.C.}\ \bibnamefont{Kraan}}, \bibinfo {author}
  {\bibfnamefont{D.A.}\ \bibnamefont{Milstead}}, \bibinfo {author}
  {\bibfnamefont{T.}~\bibnamefont{Sjostrand}}, \bibinfo {author}
  {\bibfnamefont{Peter~Z.}\ \bibnamefont{Skands}}, \emph{et~al.},\ }%
  \bibfield{title}{%
  \enquote{\bibinfo {title} {{Stable massive particles at colliders}},}\ }%
  \bibfield{journal}{%
  \Doi{10.1016/j.physrep.2006.10.002}{\bibinfo {journal} {Phys.Rept.}}\ }%
  \textbf{\bibinfo {volume} {438}},\ \bibinfo {pages} {1--63} (\bibinfo {year}
  {2007}),\ \Eprint{http://arxiv.org/abs/hep-ph/0611040}{arXiv:hep-ph/0611040
  [hep-ph]}%
  \bibAnnoteFile{NoStop}{Fairbairn:2006gg}%

\bibitem{Chen:2009gu}%
  \BibitemOpen
  \bibfield{author}{%
  \bibinfo {author} {\bibfnamefont{Jie}\ \bibnamefont{Chen}}\ and\ \bibinfo
  {author} {\bibfnamefont{Todd}\ \bibnamefont{Adams}},\ }%
  \bibfield{title}{%
  \enquote{\bibinfo {title} {{Searching for high speed long-lived charged
  massive particles at the LHC}},}\ }%
  \bibfield{journal}{%
  \Doi{10.1140/epjc/s10052-010-1283-9}{\bibinfo {journal} {Eur.Phys.J.}}\ }%
  \textbf{\bibinfo {volume} {C67}},\ \bibinfo {pages} {335--342} (\bibinfo
  {year} {2010}),\ \Eprint{http://arxiv.org/abs/0909.3157}{arXiv:0909.3157
  [hep-ph]}%
  \bibAnnoteFile{NoStop}{Chen:2009gu}%

\bibitem{Franceschini:2008pz}%
  \BibitemOpen
  \bibfield{author}{%
  \bibinfo {author} {\bibfnamefont{Roberto}\ \bibnamefont{Franceschini}},
  \bibinfo {author} {\bibfnamefont{Thomas}\ \bibnamefont{Hambye}},\ and\
  \bibinfo {author} {\bibfnamefont{Alessandro}\ \bibnamefont{Strumia}},\ }%
  \bibfield{title}{%
  \enquote{\bibinfo {title} {{Type III seesaw at LHC}},}\ }%
  \bibfield{journal}{%
  \Doi{10.1103/PhysRevD.78.033002}{\bibinfo {journal} {Phys.Rev.}}\ }%
  \textbf{\bibinfo {volume} {D78}},\ \bibinfo {pages} {033002} (\bibinfo {year}
  {2008}),\ \Eprint{http://arxiv.org/abs/0805.1613}{arXiv:0805.1613 [hep-ph]}%
  \bibAnnoteFile{NoStop}{Franceschini:2008pz}%

\bibitem{deCampos:2008re}%
  \BibitemOpen
  \bibfield{author}{%
  \bibinfo {author} {\bibfnamefont{F.}~\bibnamefont{de~Campos}}, \bibinfo
  {author} {\bibfnamefont{O.J.P.}\ \bibnamefont{Eboli}}, \bibinfo {author}
  {\bibfnamefont{M.B.}\ \bibnamefont{Magro}},\ and\ \bibinfo {author}
  {\bibfnamefont{D.}~\bibnamefont{Restrepo}},\ }%
  \bibfield{title}{%
  \enquote{\bibinfo {title} {{Searching supersymmetry at the LHCb with
  displaced vertices}},}\ }%
  \bibfield{journal}{%
  \Doi{10.1103/PhysRevD.79.055008}{\bibinfo {journal} {Phys.Rev.}}\ }%
  \textbf{\bibinfo {volume} {D79}},\ \bibinfo {pages} {055008} (\bibinfo {year}
  {2009}),\ \Eprint{http://arxiv.org/abs/0809.0007}{arXiv:0809.0007 [hep-ph]}%
  \bibAnnoteFile{NoStop}{deCampos:2008re}%

\bibitem{Babu:2001ex}%
  \BibitemOpen
  \bibfield{author}{%
  \bibinfo {author} {\bibfnamefont{K.S.}\ \bibnamefont{Babu}}\ and\ \bibinfo
  {author} {\bibfnamefont{Chung~Ngoc}\ \bibnamefont{Leung}},\ }%
  \bibfield{title}{%
  \enquote{\bibinfo {title} {{Classification of effective neutrino mass
  operators}},}\ }%
  \bibfield{journal}{%
  \Doi{10.1016/S0550-3213(01)00504-1}{\bibinfo {journal} {Nucl.Phys.}}\ }%
  \textbf{\bibinfo {volume} {B619}},\ \bibinfo {pages} {667--689} (\bibinfo
  {year} {2001}),\
  \Eprint{http://arxiv.org/abs/hep-ph/0106054}{arXiv:hep-ph/0106054 [hep-ph]}%
  \bibAnnoteFile{NoStop}{Babu:2001ex}%

\bibitem{Choi:2002bb}%
  \BibitemOpen
  \bibfield{author}{%
  \bibinfo {author} {\bibfnamefont{Ki-woon}\ \bibnamefont{Choi}}, \bibinfo
  {author} {\bibfnamefont{Kwang~Sik}\ \bibnamefont{Jeong}},\ and\ \bibinfo
  {author} {\bibfnamefont{Wan~Young}\ \bibnamefont{Song}},\ }%
  \bibfield{title}{%
  \enquote{\bibinfo {title} {{Operator analysis of neutrinoless double beta
  decay}},}\ }%
  \bibfield{journal}{%
  \Doi{10.1103/PhysRevD.66.093007}{\bibinfo {journal} {Phys.Rev.}}\ }%
  \textbf{\bibinfo {volume} {D66}},\ \bibinfo {pages} {093007} (\bibinfo {year}
  {2002}),\ \Eprint{http://arxiv.org/abs/hep-ph/0207180}{arXiv:hep-ph/0207180
  [hep-ph]}%
  \bibAnnoteFile{NoStop}{Choi:2002bb}%

\bibitem{Engel:2003yr}%
  \BibitemOpen
  \bibfield{author}{%
  \bibinfo {author} {\bibfnamefont{J.}~\bibnamefont{Engel}}\ and\ \bibinfo
  {author} {\bibfnamefont{P.}~\bibnamefont{Vogel}},\ }%
  \bibfield{title}{%
  \enquote{\bibinfo {title} {{Effective operators for double beta decay}},}\ }%
  \bibfield{journal}{%
  \Doi{10.1103/PhysRevC.69.034304}{\bibinfo {journal} {Phys.Rev.}}\ }%
  \textbf{\bibinfo {volume} {C69}},\ \bibinfo {pages} {034304} (\bibinfo {year}
  {2004}),\ \Eprint{http://arxiv.org/abs/nucl-th/0311072}{arXiv:nucl-th/0311072
  [nucl-th]}%
  \bibAnnoteFile{NoStop}{Engel:2003yr}%

\bibitem{deGouvea:2007xp}%
  \BibitemOpen
  \bibfield{author}{%
  \bibinfo {author} {\bibfnamefont{Andre}\ \bibnamefont{de~Gouvea}}\ and\
  \bibinfo {author} {\bibfnamefont{James}\ \bibnamefont{Jenkins}},\ }%
  \bibfield{title}{%
  \enquote{\bibinfo {title} {A survey of lepton number violation via effective
  operators},}\ }%
  \bibfield{journal}{%
  \Doi{10.1103/PhysRevD.77.013008}{\bibinfo {journal} {Phys. Rev.}}\ }%
  \textbf{\bibinfo {volume} {D77}},\ \bibinfo {pages} {013008} (\bibinfo {year}
  {2008}),\ \Eprint{http://arxiv.org/abs/0708.1344}{arXiv:0708.1344 [hep-ph]}%
  \bibAnnoteFile{NoStop}{deGouvea:2007xp}%

\bibitem{Barabash:2011fg}%
  \BibitemOpen
  \bibfield{author}{%
  \bibinfo {author} {\bibfnamefont{A.S.}\ \bibnamefont{Barabash}},\ }%
  \bibfield{title}{%
  \enquote{\bibinfo {title} {{Double beta decay experiments}},}\ }%
  \bibfield{journal}{%
  \Doi{10.1134/S1063779611040022}{\bibinfo {journal} {Phys.Part.Nucl.}}\ }%
  \textbf{\bibinfo {volume} {42}},\ \bibinfo {pages} {613--627} (\bibinfo
  {year} {2011}),\ \Eprint{http://arxiv.org/abs/1107.5663}{arXiv:1107.5663
  [nucl-ex]}%
  \bibAnnoteFile{NoStop}{Barabash:2011fg}%

\bibitem{Muto:1989cd}%
  \BibitemOpen
  \bibfield{author}{%
  \bibinfo {author} {\bibfnamefont{K.}~\bibnamefont{Muto}}, \bibinfo {author}
  {\bibfnamefont{E.}~\bibnamefont{Bender}},\ and\ \bibinfo {author}
  {\bibfnamefont{H.V.}\ \bibnamefont{Klapdor}},\ }%
  \bibfield{title}{%
  \enquote{\bibinfo {title} {Nuclear structure effects on the neutrinoless
  double beta decay},}\ }%
  \bibfield{journal}{%
  \bibinfo {journal} {Z.Phys.}\ }%
  \textbf{\bibinfo {volume} {A334}},\ \bibinfo {pages} {187--194} (\bibinfo
  {year} {1989})%
  \bibAnnoteFile{NoStop}{Muto:1989cd}%

\bibitem{Lee:1973iz}%
  \BibitemOpen
  \bibfield{author}{%
  \bibinfo {author} {\bibfnamefont{T.D.}\ \bibnamefont{Lee}},\ }%
  \bibfield{title}{%
  \enquote{\bibinfo {title} {A theory of spontaneous {T} violation},}\ }%
  \bibfield{journal}{%
  \Doi{10.1103/PhysRevD.8.1226}{\bibinfo {journal} {Phys.Rev.}}\ }%
  \textbf{\bibinfo {volume} {D8}},\ \bibinfo {pages} {1226--1239} (\bibinfo
  {year} {1973})%
  \bibAnnoteFile{NoStop}{Lee:1973iz}%

\bibitem{Branco:2011iw}%
  \BibitemOpen
  \bibfield{author}{%
  \bibinfo {author} {\bibfnamefont{G.C.}\ \bibnamefont{Branco}}, \bibinfo
  {author} {\bibfnamefont{P.M.}\ \bibnamefont{Ferreira}}, \bibinfo {author}
  {\bibfnamefont{L.}~\bibnamefont{Lavoura}}, \bibinfo {author}
  {\bibfnamefont{M.N.}\ \bibnamefont{Rebelo}}, \bibinfo {author}
  {\bibfnamefont{Marc}\ \bibnamefont{Sher}}, \emph{et~al.},\ }%
  \bibfield{title}{%
  \enquote{\bibinfo {title} {{Theory and phenomenology of two-Higgs-doublet
  models}},}\ }%
  \bibfield{journal}{%
  \Doi{10.1016/j.physrep.2012.02.002}{\bibinfo {journal} {Phys.Rept.}}\ }%
  \textbf{\bibinfo {volume} {516}},\ \bibinfo {pages} {1--102} (\bibinfo {year}
  {2012}),\ \Eprint{http://arxiv.org/abs/1106.0034}{arXiv:1106.0034 [hep-ph]}%
  \bibAnnoteFile{NoStop}{Branco:2011iw}%

\bibitem{Blanke:2007db}%
  \BibitemOpen
  \bibfield{author}{%
  \bibinfo {author} {\bibfnamefont{Monika}\ \bibnamefont{Blanke}}, \bibinfo
  {author} {\bibfnamefont{Andrzej~J.}\ \bibnamefont{Buras}}, \bibinfo {author}
  {\bibfnamefont{Bjoern}\ \bibnamefont{Duling}}, \bibinfo {author}
  {\bibfnamefont{Anton}\ \bibnamefont{Poschenrieder}},\ and\ \bibinfo {author}
  {\bibfnamefont{Cecilia}\ \bibnamefont{Tarantino}},\ }%
  \bibfield{title}{%
  \enquote{\bibinfo {title} {Charged lepton flavour violation and $(g-2)_\mu$
  in the littlest {Higgs} model with {T}-parity: A clear distinction from
  supersymmetry},}\ }%
  \bibfield{journal}{%
  \Doi{10.1088/1126-6708/2007/05/013}{\bibinfo {journal} {JHEP}}\ }%
  \textbf{\bibinfo {volume} {0705}},\ \bibinfo {pages} {013} (\bibinfo {year}
  {2007}),\ \Eprint{http://arxiv.org/abs/hep-ph/0702136}{arXiv:hep-ph/0702136
  [hep-ph]}%
  \bibAnnoteFile{NoStop}{Blanke:2007db}%

\bibitem{delAguila:2008zu}%
  \BibitemOpen
  \bibfield{author}{%
  \bibinfo {author} {\bibfnamefont{F.}~\bibnamefont{del Aguila}}, \bibinfo
  {author} {\bibfnamefont{J.I.}\ \bibnamefont{Illana}},\ and\ \bibinfo {author}
  {\bibfnamefont{M.D.}\ \bibnamefont{Jenkins}},\ }%
  \bibfield{title}{%
  \enquote{\bibinfo {title} {{Precise limits from lepton-flavour-violating
  processes on the Littlest Higgs model with T-parity}},}\ }%
  \bibfield{journal}{%
  \Doi{10.1088/1126-6708/2009/01/080}{\bibinfo {journal} {JHEP}}\ }%
  \textbf{\bibinfo {volume} {0901}},\ \bibinfo {pages} {080} (\bibinfo {year}
  {2009}),\ \Eprint{http://arxiv.org/abs/0811.2891}{arXiv:0811.2891 [hep-ph]}%
  \bibAnnoteFile{NoStop}{delAguila:2008zu}%

\bibitem{Goto:2010sn}%
  \BibitemOpen
  \bibfield{author}{%
  \bibinfo {author} {\bibfnamefont{Toru}\ \bibnamefont{Goto}}, \bibinfo
  {author} {\bibfnamefont{Yasuhiro}\ \bibnamefont{Okada}},\ and\ \bibinfo
  {author} {\bibfnamefont{Yasuhiro}\ \bibnamefont{Yamamoto}},\ }%
  \bibfield{title}{%
  \enquote{\bibinfo {title} {{Tau and muon lepton flavor violations in the
  littlest Higgs model with T-parity}},}\ }%
  \bibfield{journal}{%
  \Doi{10.1103/PhysRevD.83.053011}{\bibinfo {journal} {Phys.Rev.}}\ }%
  \textbf{\bibinfo {volume} {D83}},\ \bibinfo {pages} {053011} (\bibinfo {year}
  {2011}),\ \Eprint{http://arxiv.org/abs/1012.4385}{arXiv:1012.4385 [hep-ph]}%
  \bibAnnoteFile{NoStop}{Goto:2010sn}%

\bibitem{delAguila:2010nv}%
  \BibitemOpen
  \bibfield{author}{%
  \bibinfo {author} {\bibfnamefont{F.}~\bibnamefont{del Aguila}}, \bibinfo
  {author} {\bibfnamefont{J.I.}\ \bibnamefont{Illana}},\ and\ \bibinfo {author}
  {\bibfnamefont{M.D.}\ \bibnamefont{Jenkins}},\ }%
  \bibfield{title}{%
  \enquote{\bibinfo {title} {{Muon to electron conversion in the Littlest Higgs
  model with T-parity}},}\ }%
  \bibfield{journal}{%
  \Doi{10.1007/JHEP09(2010)040}{\bibinfo {journal} {JHEP}}\ }%
  \textbf{\bibinfo {volume} {1009}},\ \bibinfo {pages} {040} (\bibinfo {year}
  {2010}),\ \Eprint{http://arxiv.org/abs/1006.5914}{arXiv:1006.5914 [hep-ph]}%
  \bibAnnoteFile{NoStop}{delAguila:2010nv}%

\bibitem{Glashow:1970gm}%
  \BibitemOpen
  \bibfield{author}{%
  \bibinfo {author} {\bibfnamefont{S.L.}\ \bibnamefont{Glashow}}, \bibinfo
  {author} {\bibfnamefont{J.}~\bibnamefont{Iliopoulos}},\ and\ \bibinfo
  {author} {\bibfnamefont{L.}~\bibnamefont{Maiani}},\ }%
  \bibfield{title}{%
  \enquote{\bibinfo {title} {Weak interactions with lepton-hadron symmetry},}\
  }%
  \bibfield{journal}{%
  \Doi{10.1103/PhysRevD.2.1285}{\bibinfo {journal} {Phys.Rev.}}\ }%
  \textbf{\bibinfo {volume} {D2}},\ \bibinfo {pages} {1285--1292} (\bibinfo
  {year} {1970})%
  \bibAnnoteFile{NoStop}{Glashow:1970gm}%

\bibitem{delAguila:1989rq}%
  \BibitemOpen
  \bibfield{author}{%
  \bibinfo {author} {\bibfnamefont{F.}~\bibnamefont{del Aguila}}, \bibinfo
  {author} {\bibfnamefont{L.}~\bibnamefont{Ametller}}, \bibinfo {author}
  {\bibfnamefont{Gordon~L.}\ \bibnamefont{Kane}},\ and\ \bibinfo {author}
  {\bibfnamefont{J.}~\bibnamefont{Vidal}},\ }%
  \bibfield{title}{%
  \enquote{\bibinfo {title} {Vector-like fermion and standard {Higgs}
  production at hadron colliders},}\ }%
  \bibfield{journal}{%
  \Doi{10.1016/0550-3213(90)90655-W}{\bibinfo {journal} {Nucl.Phys.}}\ }%
  \textbf{\bibinfo {volume} {B334}},\ \bibinfo {pages} {1} (\bibinfo {year}
  {1990})%
  \bibAnnoteFile{NoStop}{delAguila:1989rq}%

\bibitem{AguilarSaavedra:2009ik}%
  \BibitemOpen
  \bibfield{author}{%
  \bibinfo {author} {\bibfnamefont{J.A.}\ \bibnamefont{Aguilar-Saavedra}},\ }%
  \bibfield{title}{%
  \enquote{\bibinfo {title} {{Heavy lepton pair-production at LHC: model
  discrimination with multi-lepton signals}},}\ }%
  \bibfield{journal}{%
  \Doi{10.1016/j.nuclphysb.2009.11.021}{\bibinfo {journal} {Nucl.Phys.}}\ }%
  \textbf{\bibinfo {volume} {B828}},\ \bibinfo {pages} {289--316} (\bibinfo
  {year} {2010}),\ \Eprint{http://arxiv.org/abs/0905.2221}{arXiv:0905.2221
  [hep-ph]}%
  \bibAnnoteFile{NoStop}{AguilarSaavedra:2009ik}%

\bibitem{delAguila:2010es}%
  \BibitemOpen
  \bibfield{author}{%
  \bibinfo {author} {\bibfnamefont{Francisco}\ \bibnamefont{del Aguila}},
  \bibinfo {author} {\bibfnamefont{Adrian}\ \bibnamefont{Carmona}},\ and\
  \bibinfo {author} {\bibfnamefont{Jose}\ \bibnamefont{Santiago}},\ }%
  \bibfield{title}{%
  \enquote{\bibinfo {title} {{Tau custodian searches at the LHC}},}\ }%
  \bibfield{journal}{%
  \Doi{10.1016/j.physletb.2010.11.054}{\bibinfo {journal} {Phys.Lett.}}\ }%
  \textbf{\bibinfo {volume} {B695}},\ \bibinfo {pages} {449--453} (\bibinfo
  {year} {2011}),\ \Eprint{http://arxiv.org/abs/1007.4206}{arXiv:1007.4206
  [hep-ph]}%
  \bibAnnoteFile{NoStop}{delAguila:2010es}%

\bibitem{delAguila:2008ks}%
  \BibitemOpen
  \bibfield{author}{%
  \bibinfo {author} {\bibfnamefont{F.}~\bibnamefont{del Aguila}}, \bibinfo
  {author} {\bibfnamefont{J.A.}\ \bibnamefont{Aguilar-Saavedra}}, \bibinfo
  {author} {\bibfnamefont{J.}~\bibnamefont{de~Blas}},\ and\ \bibinfo {author}
  {\bibfnamefont{M.}~\bibnamefont{Perez-Victoria}},\ }%
  \bibfield{title}{%
  \enquote{\bibinfo {title} {{Electroweak constraints on seesaw messengers and
  their implications for LHC}},}\ }%
  \bibfield{journal}{%
  \bibinfo {journal} {Proceedings of the 43rd Rencontres de Moriond}\ }%
  \textbf{\bibinfo {volume} {I - Standard Model}},\ \bibinfo {pages} {45--52}
  (\bibinfo {year} {2008}),\
  \Eprint{http://arxiv.org/abs/0806.1023}{arXiv:0806.1023 [hep-ph]},\
  \url{https://indico.in2p3.fr/internalPage.py?pageId=10&confId=420}%
  \bibAnnoteFile{NoStop}{delAguila:2008ks}%

\bibitem{Einhorn:1981cy}%
  \BibitemOpen
  \bibfield{author}{%
  \bibinfo {author} {\bibfnamefont{M.B.}\ \bibnamefont{Einhorn}}, \bibinfo
  {author} {\bibfnamefont{D.R.T.}\ \bibnamefont{Jones}},\ and\ \bibinfo
  {author} {\bibfnamefont{M.J.G.}\ \bibnamefont{Veltman}},\ }%
  \bibfield{title}{%
  \enquote{\bibinfo {title} {{Heavy Particles and the rho Parameter in the
  Standard Model}},}\ }%
  \bibfield{journal}{%
  \Doi{10.1016/0550-3213(81)90292-3}{\bibinfo {journal} {Nucl.Phys.}}\ }%
  \textbf{\bibinfo {volume} {B191}},\ \bibinfo {pages} {146} (\bibinfo {year}
  {1981})%
  \bibAnnoteFile{NoStop}{Einhorn:1981cy}%

\bibitem{Golden:1986jn}%
  \BibitemOpen
  \bibfield{author}{%
  \bibinfo {author} {\bibfnamefont{Mitchell}\ \bibnamefont{Golden}},\ }%
  \bibfield{title}{%
  \enquote{\bibinfo {title} {An upper limit on the masses of the charged
  {Higgs} bosons in the {Gelmini-Roncadelli} model},}\ }%
  \bibfield{journal}{%
  \Doi{10.1016/0370-2693(86)90660-X}{\bibinfo {journal} {Phys.Lett.}}\ }%
  \textbf{\bibinfo {volume} {B169}},\ \bibinfo {pages} {248} (\bibinfo {year}
  {1986})%
  \bibAnnoteFile{NoStop}{Golden:1986jn}%

\bibitem{Kanemura:2012rs}%
  \BibitemOpen
  \bibfield{author}{%
  \bibinfo {author} {\bibfnamefont{Shinya}\ \bibnamefont{Kanemura}}\ and\
  \bibinfo {author} {\bibfnamefont{Kei}\ \bibnamefont{Yagyu}},\ }%
  \bibfield{title}{%
  \enquote{\bibinfo {title} {{Radiative corrections to electroweak parameters
  in the Higgs triplet model and implication with the recent Higgs boson
  searches}},}\ }%
  \bibfield{journal}{%
  \Doi{10.1103/PhysRevD.85.115009}{\bibinfo {journal} {Phys.Rev.}}\ }%
  \textbf{\bibinfo {volume} {D85}},\ \bibinfo {pages} {115009} (\bibinfo {year}
  {2012}),\ \Eprint{http://arxiv.org/abs/1201.6287}{arXiv:1201.6287 [hep-ph]}%
  \bibAnnoteFile{NoStop}{Kanemura:2012rs}%

\bibitem{Nebot:2007bc}%
  \BibitemOpen
  \bibfield{author}{%
  \bibinfo {author} {\bibfnamefont{Miguel}\ \bibnamefont{Nebot}}, \bibinfo
  {author} {\bibfnamefont{Josep~F.}\ \bibnamefont{Oliver}}, \bibinfo {author}
  {\bibfnamefont{David}\ \bibnamefont{Palao}},\ and\ \bibinfo {author}
  {\bibfnamefont{Arcadi}\ \bibnamefont{Santamaria}},\ }%
  \bibfield{title}{%
  \enquote{\bibinfo {title} {{Prospects for the Zee-Babu model at the CERN LHC
  and low energy experiments}},}\ }%
  \bibfield{journal}{%
  \Doi{10.1103/PhysRevD.77.093013}{\bibinfo {journal} {Phys.Rev.}}\ }%
  \textbf{\bibinfo {volume} {D77}},\ \bibinfo {pages} {093013} (\bibinfo {year}
  {2008}),\ \Eprint{http://arxiv.org/abs/0711.0483}{arXiv:0711.0483 [hep-ph]}%
  \bibAnnoteFile{NoStop}{Nebot:2007bc}%

\bibitem{Raidal:1997hq}%
  \BibitemOpen
  \bibfield{author}{%
  \bibinfo {author} {\bibfnamefont{Martti}\ \bibnamefont{Raidal}}\ and\
  \bibinfo {author} {\bibfnamefont{Arcadi}\ \bibnamefont{Santamaria}},\ }%
  \bibfield{title}{%
  \enquote{\bibinfo {title} {Muon-electron conversion in nuclei versus $\mu
  \rightarrow e \gamma$: An effective field theory point of view},}\ }%
  \bibfield{journal}{%
  \Doi{10.1016/S0370-2693(98)00020-3}{\bibinfo {journal} {Phys.Lett.}}\ }%
  \textbf{\bibinfo {volume} {B421}},\ \bibinfo {pages} {250--258} (\bibinfo
  {year} {1998}),\
  \Eprint{http://arxiv.org/abs/hep-ph/9710389}{arXiv:hep-ph/9710389 [hep-ph]}%
  \bibAnnoteFile{NoStop}{Raidal:1997hq}%

\bibitem{Weinberg:1995mt}%
  \BibitemOpen
  \bibfield{author}{%
  \bibinfo {author} {\bibfnamefont{Steven}\ \bibnamefont{Weinberg}},\ }%
  \emph{\bibinfo {title} {{The quantum theory of fields, Vol. 1:
  Foundations}}}\ (\bibinfo {publisher} {Cambridge University Press},\ \bibinfo
  {year} {2005})\ ISBN \bibinfo {isbn} {978-0521670531}%
  \bibAnnoteFile{NoStop}{Weinberg:1995mt}%

\bibitem{Lee:1977yc}%
  \BibitemOpen
  \bibfield{author}{%
  \bibinfo {author} {\bibfnamefont{Benjamin~W.}\ \bibnamefont{Lee}}, \bibinfo
  {author} {\bibfnamefont{C.}~\bibnamefont{Quigg}},\ and\ \bibinfo {author}
  {\bibfnamefont{H.B.}\ \bibnamefont{Thacker}},\ }%
  \bibfield{title}{%
  \enquote{\bibinfo {title} {The strength of weak interactions at very
  high-energies and the {Higgs} boson mass},}\ }%
  \bibfield{journal}{%
  \Doi{10.1103/PhysRevLett.38.883}{\bibinfo {journal} {Phys.Rev.Lett.}}\ }%
  \textbf{\bibinfo {volume} {38}},\ \bibinfo {pages} {883--885} (\bibinfo
  {year} {1977})%
  \bibAnnoteFile{NoStop}{Lee:1977yc}%

\bibitem{Chanowitz:1978mv}%
  \BibitemOpen
  \bibfield{author}{%
  \bibinfo {author} {\bibfnamefont{Michael~S.}\ \bibnamefont{Chanowitz}},
  \bibinfo {author} {\bibfnamefont{M.A.}\ \bibnamefont{Furman}},\ and\ \bibinfo
  {author} {\bibfnamefont{I.}~\bibnamefont{Hinchliffe}},\ }%
  \bibfield{title}{%
  \enquote{\bibinfo {title} {Weak interactions of ultraheavy fermions. 2.}.}\
  }%
  \bibfield{journal}{%
  \Doi{10.1016/0550-3213(79)90606-0}{\bibinfo {journal} {Nucl.Phys.}}\ }%
  \textbf{\bibinfo {volume} {B153}},\ \bibinfo {pages} {402} (\bibinfo {year}
  {1979})%
  \bibAnnoteFile{NoStop}{Chanowitz:1978mv}%

\bibitem{Schwetz:2011zk}%
  \BibitemOpen
  \bibfield{author}{%
  \bibinfo {author} {\bibfnamefont{Thomas}\ \bibnamefont{Schwetz}}, \bibinfo
  {author} {\bibfnamefont{Mariam}\ \bibnamefont{Tortola}},\ and\ \bibinfo
  {author} {\bibfnamefont{J.W.F.}\ \bibnamefont{Valle}},\ }%
  \bibfield{title}{%
  \enquote{\bibinfo {title} {{Where we are on $\theta_{13}$: addendum to
  `Global neutrino data and recent reactor fluxes: status of three-flavour
  oscillation parameters'}},}\ }%
  \bibfield{journal}{%
  \Doi{10.1088/1367-2630/13/10/109401}{\bibinfo {journal} {New J.Phys.}}\ }%
  \textbf{\bibinfo {volume} {13}},\ \bibinfo {pages} {109401} (\bibinfo {year}
  {2011}),\ \Eprint{http://arxiv.org/abs/1108.1376}{arXiv:1108.1376 [hep-ph]}%
  \bibAnnoteFile{NoStop}{Schwetz:2011zk}%

\bibitem{Gunion:1989in}%
  \BibitemOpen
  \bibfield{author}{%
  \bibinfo {author} {\bibfnamefont{J.F.}\ \bibnamefont{Gunion}}, \bibinfo
  {author} {\bibfnamefont{J.}~\bibnamefont{Grifols}}, \bibinfo {author}
  {\bibfnamefont{A.}~\bibnamefont{Mendez}}, \bibinfo {author}
  {\bibfnamefont{Boris}\ \bibnamefont{Kayser}},\ and\ \bibinfo {author}
  {\bibfnamefont{Fredrick~I.}\ \bibnamefont{Olness}},\ }%
  \bibfield{title}{%
  \enquote{\bibinfo {title} {Higgs bosons in left-right symmetric models},}\ }%
  \bibfield{journal}{%
  \Doi{10.1103/PhysRevD.40.1546}{\bibinfo {journal} {Phys.Rev.}}\ }%
  \textbf{\bibinfo {volume} {D40}},\ \bibinfo {pages} {1546} (\bibinfo {year}
  {1989})%
  \bibAnnoteFile{NoStop}{Gunion:1989in}%

\bibitem{Huitu:1996su}%
  \BibitemOpen
  \bibfield{author}{%
  \bibinfo {author} {\bibfnamefont{K.}~\bibnamefont{Huitu}}, \bibinfo {author}
  {\bibfnamefont{J.}~\bibnamefont{Maalampi}}, \bibinfo {author}
  {\bibfnamefont{A.}~\bibnamefont{Pietila}},\ and\ \bibinfo {author}
  {\bibfnamefont{M.}~\bibnamefont{Raidal}},\ }%
  \bibfield{title}{%
  \enquote{\bibinfo {title} {{Doubly charged Higgs at LHC}},}\ }%
  \bibfield{journal}{%
  \Doi{10.1016/S0550-3213(97)87466-4}{\bibinfo {journal} {Nucl.Phys.}}\ }%
  \textbf{\bibinfo {volume} {B487}},\ \bibinfo {pages} {27--42} (\bibinfo
  {year} {1997}),\
  \Eprint{http://arxiv.org/abs/hep-ph/9606311}{arXiv:hep-ph/9606311 [hep-ph]}%
  \bibAnnoteFile{NoStop}{Huitu:1996su}%

\bibitem{Gunion:1996pq}%
  \BibitemOpen
  \bibfield{author}{%
  \bibinfo {author} {\bibfnamefont{J.F.}\ \bibnamefont{Gunion}}, \bibinfo
  {author} {\bibfnamefont{C.}~\bibnamefont{Loomis}},\ and\ \bibinfo {author}
  {\bibfnamefont{K.T.}\ \bibnamefont{Pitts}},\ }%
  \bibfield{title}{%
  \enquote{\bibinfo {title} {{Searching for doubly charged Higgs bosons at
  future colliders}},}\ }%
  \bibfield{journal}{%
  \bibinfo {journal} {eConf}\ }%
  \textbf{\bibinfo {volume} {C960625}},\ \bibinfo {pages} {LTH096} (\bibinfo
  {year} {1996}),\
  \Eprint{http://arxiv.org/abs/hep-ph/9610237}{arXiv:hep-ph/9610237 [hep-ph]}%
  \bibAnnoteFile{NoStop}{Gunion:1996pq}%

\bibitem{Akeroyd:2005gt}%
  \BibitemOpen
  \bibfield{author}{%
  \bibinfo {author} {\bibfnamefont{A.G.}\ \bibnamefont{Akeroyd}}\ and\ \bibinfo
  {author} {\bibfnamefont{Mayumi}\ \bibnamefont{Aoki}},\ }%
  \bibfield{title}{%
  \enquote{\bibinfo {title} {{Single- and pair-production of doubly-charged
  Higgs bosons at hadron colliders}},}\ }%
  \bibfield{journal}{%
  \Doi{10.1103/PhysRevD.72.035011}{\bibinfo {journal} {Phys.Rev.}}\ }%
  \textbf{\bibinfo {volume} {D72}},\ \bibinfo {pages} {035011} (\bibinfo {year}
  {2005}),\ \Eprint{http://arxiv.org/abs/hep-ph/0506176}{arXiv:hep-ph/0506176
  [hep-ph]}%
  \bibAnnoteFile{NoStop}{Akeroyd:2005gt}%

\bibitem{Azuelos:2005uc}%
  \BibitemOpen
  \bibfield{author}{%
  \bibinfo {author} {\bibfnamefont{G.}~\bibnamefont{Azuelos}}, \bibinfo
  {author} {\bibfnamefont{K.}~\bibnamefont{Benslama}},\ and\ \bibinfo {author}
  {\bibfnamefont{J.}~\bibnamefont{Ferland}},\ }%
  \bibfield{title}{%
  \enquote{\bibinfo {title} {{Prospects for the search for a doubly-charged
  Higgs in the left-right symmetric model with ATLAS}},}\ }%
  \bibfield{journal}{%
  \Doi{10.1088/0954-3899/32/2/002}{\bibinfo {journal} {J.Phys.}}\ }%
  \textbf{\bibinfo {volume} {G32}},\ \bibinfo {pages} {73--92} (\bibinfo {year}
  {2006}),\ \Eprint{http://arxiv.org/abs/hep-ph/0503096}{arXiv:hep-ph/0503096
  [hep-ph]}%
  \bibAnnoteFile{NoStop}{Azuelos:2005uc}%

\bibitem{delAguila:2008cj}%
  \BibitemOpen
  \bibfield{author}{%
  \bibinfo {author} {\bibfnamefont{F.}~\bibnamefont{del Aguila}}\ and\ \bibinfo
  {author} {\bibfnamefont{J.A.}\ \bibnamefont{Aguilar-Saavedra}},\ }%
  \bibfield{title}{%
  \enquote{\bibinfo {title} {{Distinguishing seesaw models at LHC with
  multi-lepton signals}},}\ }%
  \bibfield{journal}{%
  \Doi{10.1016/j.nuclphysb.2008.12.029}{\bibinfo {journal} {Nucl.Phys.}}\ }%
  \textbf{\bibinfo {volume} {B813}},\ \bibinfo {pages} {22--90} (\bibinfo
  {year} {2009}),\ \Eprint{http://arxiv.org/abs/0808.2468}{arXiv:0808.2468
  [hep-ph]}%
  \bibAnnoteFile{NoStop}{delAguila:2008cj}%

\bibitem{Akeroyd:2010ip}%
  \BibitemOpen
  \bibfield{author}{%
  \bibinfo {author} {\bibfnamefont{A.G.}\ \bibnamefont{Akeroyd}}, \bibinfo
  {author} {\bibfnamefont{Cheng-Wei}\ \bibnamefont{Chiang}},\ and\ \bibinfo
  {author} {\bibfnamefont{Naveen}\ \bibnamefont{Gaur}},\ }%
  \bibfield{title}{%
  \enquote{\bibinfo {title} {{Leptonic signatures of doubly charged Higgs boson
  production at the LHC}},}\ }%
  \bibfield{journal}{%
  \Doi{10.1007/JHEP11(2010)005}{\bibinfo {journal} {JHEP}}\ }%
  \textbf{\bibinfo {volume} {1011}},\ \bibinfo {pages} {005} (\bibinfo {year}
  {2010}),\ \Eprint{http://arxiv.org/abs/1009.2780}{arXiv:1009.2780 [hep-ph]}%
  \bibAnnoteFile{NoStop}{Akeroyd:2010ip}%

\bibitem{Dion:1998pw}%
  \BibitemOpen
  \bibfield{author}{%
  \bibinfo {author} {\bibfnamefont{B.}~\bibnamefont{Dion}}, \bibinfo {author}
  {\bibfnamefont{T.}~\bibnamefont{Gregoire}}, \bibinfo {author}
  {\bibfnamefont{David}\ \bibnamefont{London}}, \bibinfo {author}
  {\bibfnamefont{L.}~\bibnamefont{Marleau}},\ and\ \bibinfo {author}
  {\bibfnamefont{H.}~\bibnamefont{Nadeau}},\ }%
  \bibfield{title}{%
  \enquote{\bibinfo {title} {{Bilepton production at hadron colliders}},}\ }%
  \bibfield{journal}{%
  \Doi{10.1103/PhysRevD.59.075006}{\bibinfo {journal} {Phys.Rev.}}\ }%
  \textbf{\bibinfo {volume} {D59}},\ \bibinfo {pages} {075006} (\bibinfo {year}
  {1999}),\ \Eprint{http://arxiv.org/abs/hep-ph/9810534}{arXiv:hep-ph/9810534
  [hep-ph]}%
  \bibAnnoteFile{NoStop}{Dion:1998pw}%

\bibitem{Cuypers:1996ia}%
  \BibitemOpen
  \bibfield{author}{%
  \bibinfo {author} {\bibfnamefont{Frank}\ \bibnamefont{Cuypers}}\ and\
  \bibinfo {author} {\bibfnamefont{Sacha}\ \bibnamefont{Davidson}},\ }%
  \bibfield{title}{%
  \enquote{\bibinfo {title} {{Bileptons: Present limits and future
  prospects}},}\ }%
  \bibfield{journal}{%
  \Doi{10.1007/s100520050157}{\bibinfo {journal} {Eur.Phys.J.}}\ }%
  \textbf{\bibinfo {volume} {C2}},\ \bibinfo {pages} {503--528} (\bibinfo
  {year} {1998}),\
  \Eprint{http://arxiv.org/abs/hep-ph/9609487}{arXiv:hep-ph/9609487 [hep-ph]}%
  \bibAnnoteFile{NoStop}{Cuypers:1996ia}%

\bibitem{delAguila:2009bb}%
  \BibitemOpen
  \bibfield{author}{%
  \bibinfo {author} {\bibfnamefont{F.}~\bibnamefont{del Aguila}}, \bibinfo
  {author} {\bibfnamefont{J.A.}\ \bibnamefont{Aguilar-Saavedra}},\ and\
  \bibinfo {author} {\bibfnamefont{J.}~\bibnamefont{de~Blas}},\ }%
  \bibfield{title}{%
  \enquote{\bibinfo {title} {{Trilepton signals: the golden channel for seesaw
  searches at LHC}},}\ }%
  \bibfield{journal}{%
  \bibinfo {journal} {Acta Phys.Polon.}\ }%
  \textbf{\bibinfo {volume} {B40}},\ \bibinfo {pages} {2901--2911} (\bibinfo
  {year} {2009}),\ \Eprint{http://arxiv.org/abs/0910.2720}{arXiv:0910.2720
  [hep-ph]}%
  \bibAnnoteFile{NoStop}{delAguila:2009bb}%

\bibitem{delAguila:2010uw}%
  \BibitemOpen
  \bibfield{author}{%
  \bibinfo {author} {\bibfnamefont{Francisco}\ \bibnamefont{del Aguila}},
  \bibinfo {author} {\bibfnamefont{Juan~Antonio}\
  \bibnamefont{Aguilar-Saavedra}},\ and\ \bibinfo {author}
  {\bibfnamefont{Jorge}\ \bibnamefont{de~Blas}},\ }%
  \bibfield{title}{%
  \enquote{\bibinfo {title} {{New neutrino interactions at large colliders}},}\
  }%
  \bibfield{journal}{%
  \bibinfo {journal} {PoS}\ }%
  \textbf{\bibinfo {volume} {ICHEP2010}},\ \bibinfo {pages} {296} (\bibinfo
  {year} {2010}),\ \Eprint{http://arxiv.org/abs/1012.1327}{arXiv:1012.1327
  [hep-ph]}%
  \bibAnnoteFile{NoStop}{delAguila:2010uw}%

\bibitem{Nath:2010zj}%
  \BibitemOpen
  \bibfield{author}{%
  \bibinfo {author} {\bibfnamefont{Pran}\ \bibnamefont{Nath}}, \bibinfo
  {author} {\bibfnamefont{Brent~D.}\ \bibnamefont{Nelson}}, \bibinfo {author}
  {\bibfnamefont{Hooman}\ \bibnamefont{Davoudiasl}}, \bibinfo {author}
  {\bibfnamefont{Bhaskar}\ \bibnamefont{Dutta}}, \bibinfo {author}
  {\bibfnamefont{Daniel}\ \bibnamefont{Feldman}}, \emph{et~al.},\ }%
  \bibfield{title}{%
  \enquote{\bibinfo {title} {The hunt for new physics at the large hadron
  collider},}\ }%
  \bibfield{journal}{%
  \Doi{10.1016/j.nuclphysbps.2010.03.001}{\bibinfo {journal}
  {Nucl.Phys.Proc.Suppl.}}\ }%
  \textbf{\bibinfo {volume} {200-202}},\ \bibinfo {pages} {185--417} (\bibinfo
  {year} {2010}),\ \Eprint{http://arxiv.org/abs/1001.2693}{arXiv:1001.2693
  [hep-ph]}%
  \bibAnnoteFile{NoStop}{Nath:2010zj}%

\bibitem{cms-pas-hig-11-007:2011}%
  \BibitemOpen
  \bibfield{author}{%
  \bibinfo {author} {\bibfnamefont{{CMS}}\ \bibnamefont{collaboration}},\ }%
  \enquote{\bibinfo {title} {{Inclusive search for doubly charged higgs in
  leptonic final states at $\sqrt{s}=7$ TeV}},}\ \bibinfo {howpublished}
  {{CMS-PAS-HIG-11-007}} (\bibinfo {year} {2011}),\
  \url{http://cdsweb.cern.ch/record/1369542}%
  \bibAnnoteFile{NoStop}{cms-pas-hig-11-007:2011}%

\bibitem{Abdallah:2002qj}%
  \BibitemOpen
  \bibfield{author}{%
  \bibinfo {author} {\bibfnamefont{J.}~\bibnamefont{Abdallah}} \emph{et~al.}
  (\bibinfo {collaboration} {DELPHI Collaboration}),\ }%
  \bibfield{title}{%
  \enquote{\bibinfo {title} {{Search for doubly charged Higgs bosons at
  LEP-2}},}\ }%
  \bibfield{journal}{%
  \Doi{10.1016/S0370-2693(02)03125-8}{\bibinfo {journal} {Phys.Lett.}}\ }%
  \textbf{\bibinfo {volume} {B552}},\ \bibinfo {pages} {127--137} (\bibinfo
  {year} {2003}),\
  \Eprint{http://arxiv.org/abs/hep-ex/0303026}{arXiv:hep-ex/0303026 [hep-ex]}%
  \bibAnnoteFile{NoStop}{Abdallah:2002qj}%

\bibitem{Abbiendi:2001cr}%
  \BibitemOpen
  \bibfield{author}{%
  \bibinfo {author} {\bibfnamefont{G.}~\bibnamefont{Abbiendi}} \emph{et~al.}
  (\bibinfo {collaboration} {OPAL Collaboration}),\ }%
  \bibfield{title}{%
  \enquote{\bibinfo {title} {{Search for doubly-charged Higgs bosons with the
  OPAL detector at LEP}},}\ }%
  \bibfield{journal}{%
  \Doi{10.1016/S0370-2693(01)01474-5}{\bibinfo {journal} {Phys.Lett.}}\ }%
  \textbf{\bibinfo {volume} {B526}},\ \bibinfo {pages} {221--232} (\bibinfo
  {year} {2002}),\
  \Eprint{http://arxiv.org/abs/hep-ex/0111059}{arXiv:hep-ex/0111059 [hep-ex]}%
  \bibAnnoteFile{NoStop}{Abbiendi:2001cr}%

\bibitem{Achard:2003mv}%
  \BibitemOpen
  \bibfield{author}{%
  \bibinfo {author} {\bibfnamefont{P.}~\bibnamefont{Achard}} \emph{et~al.}
  (\bibinfo {collaboration} {L3 Collaboration}),\ }%
  \bibfield{title}{%
  \enquote{\bibinfo {title} {{Search for doubly-charged Higgs bosons at
  LEP}},}\ }%
  \bibfield{journal}{%
  \Doi{10.1016/j.physletb.2003.09.082}{\bibinfo {journal} {Phys.Lett.}}\ }%
  \textbf{\bibinfo {volume} {B576}},\ \bibinfo {pages} {18--28} (\bibinfo
  {year} {2003}),\
  \Eprint{http://arxiv.org/abs/hep-ex/0309076}{arXiv:hep-ex/0309076 [hep-ex]}%
  \bibAnnoteFile{NoStop}{Achard:2003mv}%

\bibitem{Abbiendi:2003pr}%
  \BibitemOpen
  \bibfield{author}{%
  \bibinfo {author} {\bibfnamefont{G.}~\bibnamefont{Abbiendi}} \emph{et~al.}
  (\bibinfo {collaboration} {OPAL Collaboration}),\ }%
  \bibfield{title}{%
  \enquote{\bibinfo {title} {{Search for the single production of
  doubly-charged Higgs bosons and constraints on their couplings from Bhabha
  scattering}},}\ }%
  \bibfield{journal}{%
  \Doi{10.1016/j.physletb.2003.10.034}{\bibinfo {journal} {Phys.Lett.}}\ }%
  \textbf{\bibinfo {volume} {B577}},\ \bibinfo {pages} {93--108} (\bibinfo
  {year} {2003}),\
  \Eprint{http://arxiv.org/abs/hep-ex/0308052}{arXiv:hep-ex/0308052 [hep-ex]}%
  \bibAnnoteFile{NoStop}{Abbiendi:2003pr}%

\bibitem{Abazov:2004au}%
  \BibitemOpen
  \bibfield{author}{%
  \bibinfo {author} {\bibfnamefont{V.M.}\ \bibnamefont{Abazov}} \emph{et~al.}
  (\bibinfo {collaboration} {D0 Collaboration}),\ }%
  \bibfield{title}{%
  \enquote{\bibinfo {title} {{Search for doubly-charged Higgs boson
  pair-production in the decay to $\mu^+ \mu^+ \mu^- \mu^-$ in $p\bar{p}$
  collisions at $\sqrt{s} = 1.96$ TeV}},}\ }%
  \bibfield{journal}{%
  \Doi{10.1103/PhysRevLett.93.141801}{\bibinfo {journal} {Phys.Rev.Lett.}}\ }%
  \textbf{\bibinfo {volume} {93}},\ \bibinfo {pages} {141801} (\bibinfo {year}
  {2004}),\ \Eprint{http://arxiv.org/abs/hep-ex/0404015}{arXiv:hep-ex/0404015
  [hep-ex]}%
  \bibAnnoteFile{NoStop}{Abazov:2004au}%

\bibitem{Acosta:2004uj}%
  \BibitemOpen
  \bibfield{author}{%
  \bibinfo {author} {\bibfnamefont{D.}~\bibnamefont{Acosta}} \emph{et~al.}
  (\bibinfo {collaboration} {CDF Collaboration}),\ }%
  \bibfield{title}{%
  \enquote{\bibinfo {title} {{Search for doubly-charged Higgs bosons decaying
  to dileptons in $p\bar{p}$ collisions at $\sqrt{s} = 1.96$ TeV}},}\ }%
  \bibfield{journal}{%
  \Doi{10.1103/PhysRevLett.93.221802}{\bibinfo {journal} {Phys.Rev.Lett.}}\ }%
  \textbf{\bibinfo {volume} {93}},\ \bibinfo {pages} {221802} (\bibinfo {year}
  {2004}),\ \Eprint{http://arxiv.org/abs/hep-ex/0406073}{arXiv:hep-ex/0406073
  [hep-ex]}%
  \bibAnnoteFile{NoStop}{Acosta:2004uj}%

\bibitem{Acosta:2005np}%
  \BibitemOpen
  \bibfield{author}{%
  \bibinfo {author} {\bibfnamefont{D.}~\bibnamefont{Acosta}} \emph{et~al.}
  (\bibinfo {collaboration} {CDF Collaboration}),\ }%
  \bibfield{title}{%
  \enquote{\bibinfo {title} {{Search for long-lived doubly-charged Higgs bosons
  in $p\bar{p}$ collisions at $\sqrt{s} = 1.96$ TeV}},}\ }%
  \bibfield{journal}{%
  \Doi{10.1103/PhysRevLett.95.071801}{\bibinfo {journal} {Phys.Rev.Lett.}}\ }%
  \textbf{\bibinfo {volume} {95}},\ \bibinfo {pages} {071801} (\bibinfo {year}
  {2005}),\ \Eprint{http://arxiv.org/abs/hep-ex/0503004}{arXiv:hep-ex/0503004
  [hep-ex]}%
  \bibAnnoteFile{NoStop}{Acosta:2005np}%

\bibitem{Melfo:2011nx}%
  \BibitemOpen
  \bibfield{author}{%
  \bibinfo {author} {\bibfnamefont{Alejandra}\ \bibnamefont{Melfo}}, \bibinfo
  {author} {\bibfnamefont{Miha}\ \bibnamefont{Nemevsek}}, \bibinfo {author}
  {\bibfnamefont{Fabrizio}\ \bibnamefont{Nesti}}, \bibinfo {author}
  {\bibfnamefont{Goran}\ \bibnamefont{Senjanovic}},\ and\ \bibinfo {author}
  {\bibfnamefont{Yue}\ \bibnamefont{Zhang}},\ }%
  \bibfield{title}{%
  \enquote{\bibinfo {title} {{Type II seesaw at LHC: the roadmap}},}\ }%
  \bibfield{journal}{%
  \Doi{10.1103/PhysRevD.85.055018}{\bibinfo {journal} {Phys.Rev.}}\ }%
  \textbf{\bibinfo {volume} {D85}},\ \bibinfo {pages} {055018} (\bibinfo {year}
  {2012}),\ \Eprint{http://arxiv.org/abs/1108.4416}{arXiv:1108.4416 [hep-ph]}%
  \bibAnnoteFile{NoStop}{Melfo:2011nx}%

\bibitem{Zeldovich:1974uw}%
  \BibitemOpen
  \bibfield{author}{%
  \bibinfo {author} {\bibfnamefont{Ya.B.}\ \bibnamefont{Zeldovich}}, \bibinfo
  {author} {\bibfnamefont{I.~Yu.}\ \bibnamefont{Kobzarev}},\ and\ \bibinfo
  {author} {\bibfnamefont{L.B.}\ \bibnamefont{Okun}},\ }%
  \bibfield{title}{%
  \enquote{\bibinfo {title} {Cosmological consequences of the spontaneous
  breakdown of discrete symmetry},}\ }%
  \bibfield{journal}{%
  \bibinfo {journal} {Zh.Eksp.Teor.Fiz.}\ }%
  \textbf{\bibinfo {volume} {67}},\ \bibinfo {pages} {3--11} (\bibinfo {year}
  {1974})%
  \bibAnnoteFile{NoStop}{Zeldovich:1974uw}%

\bibitem{Kibble:1976sj}%
  \BibitemOpen
  \bibfield{author}{%
  \bibinfo {author} {\bibfnamefont{T.W.B.}\ \bibnamefont{Kibble}},\ }%
  \bibfield{title}{%
  \enquote{\bibinfo {title} {Topology of cosmic domains and strings},}\ }%
  \bibfield{journal}{%
  \Doi{10.1088/0305-4470/9/8/029}{\bibinfo {journal} {J.Phys.}}\ }%
  \textbf{\bibinfo {volume} {A9}},\ \bibinfo {pages} {1387--1398} (\bibinfo
  {year} {1976})%
  \bibAnnoteFile{NoStop}{Kibble:1976sj}%

\bibitem{Vilenkin:1984ib}%
  \BibitemOpen
  \bibfield{author}{%
  \bibinfo {author} {\bibfnamefont{Alexander}\ \bibnamefont{Vilenkin}},\ }%
  \bibfield{title}{%
  \enquote{\bibinfo {title} {Cosmic strings and domain walls},}\ }%
  \bibfield{journal}{%
  \Doi{10.1016/0370-1573(85)90033-X}{\bibinfo {journal} {Phys.Rept.}}\ }%
  \textbf{\bibinfo {volume} {121}},\ \bibinfo {pages} {263} (\bibinfo {year}
  {1985})%
  \bibAnnoteFile{NoStop}{Vilenkin:1984ib}%

\bibitem{Chikashige:1980ui}%
  \BibitemOpen
  \bibfield{author}{%
  \bibinfo {author} {\bibfnamefont{Y.}~\bibnamefont{Chikashige}}, \bibinfo
  {author} {\bibfnamefont{Rabindra~N.}\ \bibnamefont{Mohapatra}},\ and\
  \bibinfo {author} {\bibfnamefont{R.D.}\ \bibnamefont{Peccei}},\ }%
  \bibfield{title}{%
  \enquote{\bibinfo {title} {Are there real {Goldstone} bosons associated with
  broken lepton number?}.}\ }%
  \bibfield{journal}{%
  \Doi{10.1016/0370-2693(81)90011-3}{\bibinfo {journal} {Phys.Lett.}}\ }%
  \textbf{\bibinfo {volume} {B98}},\ \bibinfo {pages} {265} (\bibinfo {year}
  {1981})%
  \bibAnnoteFile{NoStop}{Chikashige:1980ui}%

\bibitem{Choi:1989hi}%
  \BibitemOpen
  \bibfield{author}{%
  \bibinfo {author} {\bibfnamefont{Kiwoon}\ \bibnamefont{Choi}}\ and\ \bibinfo
  {author} {\bibfnamefont{A.}~\bibnamefont{Santamaria}},\ }%
  \bibfield{title}{%
  \enquote{\bibinfo {title} {Majorons and supernova cooling},}\ }%
  \bibfield{journal}{%
  \Doi{10.1103/PhysRevD.42.293}{\bibinfo {journal} {Phys.Rev.}}\ }%
  \textbf{\bibinfo {volume} {D42}},\ \bibinfo {pages} {293--306} (\bibinfo
  {year} {1990})%
  \bibAnnoteFile{NoStop}{Choi:1989hi}%

\bibitem{Chikashige:1980qk}%
  \BibitemOpen
  \bibfield{author}{%
  \bibinfo {author} {\bibfnamefont{Y.}~\bibnamefont{Chikashige}}, \bibinfo
  {author} {\bibfnamefont{Rabindra~N.}\ \bibnamefont{Mohapatra}},\ and\
  \bibinfo {author} {\bibfnamefont{R.D.}\ \bibnamefont{Peccei}},\ }%
  \bibfield{title}{%
  \enquote{\bibinfo {title} {Spontaneously broken lepton number and
  cosmological constraints on the neutrino mass spectrum},}\ }%
  \bibfield{journal}{%
  \Doi{10.1103/PhysRevLett.45.1926}{\bibinfo {journal} {Phys.Rev.Lett.}}\ }%
  \textbf{\bibinfo {volume} {45}},\ \bibinfo {pages} {1926} (\bibinfo {year}
  {1980})%
  \bibAnnoteFile{NoStop}{Chikashige:1980qk}%

\bibitem{Choi:1991aa}%
  \BibitemOpen
  \bibfield{author}{%
  \bibinfo {author} {\bibfnamefont{Kiwoon}\ \bibnamefont{Choi}}\ and\ \bibinfo
  {author} {\bibfnamefont{A.}~\bibnamefont{Santamaria}},\ }%
  \bibfield{title}{%
  \enquote{\bibinfo {title} {{17-KeV neutrino in a singlet - triplet majoron
  model}},}\ }%
  \bibfield{journal}{%
  \Doi{10.1016/0370-2693(91)90900-B}{\bibinfo {journal} {Phys.Lett.}}\ }%
  \textbf{\bibinfo {volume} {B267}},\ \bibinfo {pages} {504--508} (\bibinfo
  {year} {1991})%
  \bibAnnoteFile{NoStop}{Choi:1991aa}%

\bibitem{Basso:2011hn}%
  \BibitemOpen
  \bibfield{author}{%
  \bibinfo {author} {\bibfnamefont{Lorenzo}\ \bibnamefont{Basso}},\ }%
  \emph{\bibinfo {title} {{Phenomenology of the minimal B-L extension of the
  Standard Model at the LHC}}},\ Ph.D. thesis,\ \bibinfo {school} {University
  of Southampton} (\bibinfo {year} {2011}),\
  \Eprint{http://arxiv.org/abs/1106.4462}{arXiv:1106.4462 [hep-ph]}%
  \bibAnnoteFile{NoStop}{Basso:2011hn}%

\bibitem{lem:solaris}%
  \BibitemOpen
  \bibfield{author}{%
  \bibinfo {author} {\bibfnamefont{Stanis{\l}aw}\ \bibnamefont{Lem}},\ }%
  \emph{\bibinfo {title} {Solaris}}\ (\bibinfo {publisher} {Impedimenta},\
  \bibinfo {year} {2011})\ ISBN \bibinfo {isbn} {9788415130093},\ \bibinfo
  {note} {edición en español; traducción del polaco por Joanna Orzechowska}%
  \bibAnnoteFile{NoStop}{lem:solaris}%

\bibitem{wiki:list-particles}%
  \BibitemOpen
  \bibfield{author}{%
  \bibinfo {author} {\bibnamefont{{W}ikipedia{,} The Free~Encyclopedia}},\ }%
  \enquote{\bibinfo {title} {List of particles},}\ \bibinfo {note} {[Online;
  accessed 30-September-2013]},\
  \url{http://en.wikipedia.org/wiki/List_of_subatomic_particles}%
  \bibAnnoteFile{NoStop}{wiki:list-particles}%

\bibitem{pdg:tables}%
  \BibitemOpen
  \bibfield{author}{%
  \bibinfo {author} {\bibfnamefont{Particle~Data}\ \bibnamefont{Group}},\ }%
  \enquote{\bibinfo {title} {Summary tables of the {PDG}},}\ \bibinfo {note}
  {[Online; accessed 30-September-2013]},\
  \url{http://pdg.lbl.gov/2013/tables/contents_tables.html}%
  \bibAnnoteFile{NoStop}{pdg:tables}%

\end{thebibliography}%

\newpage
\thispagestyle{empty}
\vspace*{15cm}

\centering
\includegraphics[width=0.1\textwidth]{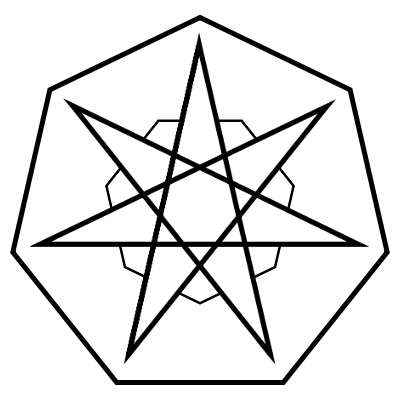}

\end{document}